\documentclass[12pt,twoside]{mitthesis}
\usepackage{lgrind}
\usepackage{cmap}
\usepackage[T1]{fontenc}
\usepackage{fontawesome}
\usepackage{CJKutf8}
\usepackage{mathptmx}
\usepackage{titlesec}
\usepackage{xspace}
\usepackage{expl3}

\newcommand{\chapternumberbox}[1]{%
	\makebox[0pt][l]{
		\fontsize{60}{70}\selectfont
		\color{lightgray}
		\hspace{6.1in}
		\raisebox{.2\height}[0pt][0pt]{#1}
	}%
}
\titleformat{\chapter}[display]
{\normalfont\LARGE\bfseries}
{\chapternumberbox{\thechapter}}
{0pt}
{\titlerule\vspace{1ex}\sc\color{Maroon}
}
[\vspace{.5ex}
\color{black}\titlerule]

\usepackage[numbers]{natbib}

\usepackage{amssymb}
\usepackage{amsmath}
\usepackage{mathtools}
\usepackage{braket}

\usepackage{tabularx}
\usepackage{multirow}
\usepackage{dcolumn}
\usepackage{makecell}
\usepackage{adjustbox}

\usepackage{float}
\usepackage{epsfig}
\usepackage{graphicx}
\usepackage{grffile}
\usepackage{subfigure}
\usepackage[dvipsnames]{xcolor}
\usepackage[colorlinks=true,
urlcolor=Maroon,
anchorcolor=WildStrawberry,
citecolor=Maroon,
filecolor=WildStrawberry,
linkcolor=Mahogany,
menucolor=WildStrawberry,
hyperfootnotes=false,
linktocpage=true,
pdfproducer=medialab,
pdfa=true
]{hyperref}

\usepackage{tikz}
\usetikzlibrary{shapes,arrows}
\usepackage[compat=1.1.0]{tikz-feynman}

\usepackage{siunitx}
\DeclareSIUnit \h {\ensuremath{\mathit{h}}}
\DeclareSIUnit\eV{e\kern-.05em V}
\DeclareSIUnit\electronvolt{e\kern-.05em V}
\DeclareSIUnit\parsec{pc}
\DeclareSIUnit\year{yr}
\DeclareSIUnit\erg{erg}
\DeclareSIUnit\sr{sr}

\titleformat{\paragraph}
{\normalfont\normalsize\bfseries}{\theparagraph}{1em}{}
\titlespacing*{\paragraph}
{0pt}{3.25ex plus 1ex minus .2ex}{1.5ex plus .2ex}

\newcommand\beq{\begin{equation}}
\newcommand\eeq{\end{equation}}
\def\be{\begin{equation}}
\def\ee{\end{equation}}
\def\figs/B{B}
\def\bea{\begin{eqnarray}}
\def\eea{\end{eqnarray}}
\def\bg{\begin{eqnarray}}
\def\nd{\end{eqnarray}}
\def\sin{{\rm sin}}
\def\cos{{\rm cos}}

\def\log{{\rm log}}
\def\ln{{\rm log}}

\newcommand{\dhis}{\texttt{DarkHistory}\xspace}
\newcommand{\n}{\nonumber \\}

\newcommand*\bbar[1]{%
	\vbox{%
		\hrule height 0.5pt
		\kern-0.4ex
		\hbox{%
			\kern-0.2em
			\ifmmode#1\else\ensuremath{#1}\fi
			\kern-0.1em
		}
	}
}
\newcommand{\githubmaster}{\href{https://github.com/hongwanliu/DarkHistory/tree/lowengelec_upgrade}{\faGithub}\xspace}
\newcommand{\githubZeus}{\href{https://github.com/JulianBMunoz/Zeus21}{\faGithub}} 
\newcolumntype{C}[1]{>{\centering\let\newline\\\arraybackslash\hspace{0pt}}m{#1}}
\defcitealias{paperI}{Paper~I}
\defcitealias{paperII}{Paper~II}

\newcommand\Msun{\text{M}_{\odot}}

\newcommand{\dbar}{d\hspace*{-0.08em}\bar{}\hspace*{0.1em}}
\newcommand{\thesan}{\textsc{thesan}\xspace}

\def\aj{\rm{AJ}}                   
\def\apj{\rm{ApJ}}                 
\def\apjl{\rm{ApJ}}                
\def\aap{\rm{A\&A}}                

\def\jcap{\rm{J. Cosmology Astropart. Phys.}}
\def\mnras{\rm{MNRAS}}             
\def\prd{\rm{Phys.~Rev.~D}}        
\def\prl{\rm{Phys.~Rev.~Lett.}}    
\def\pasa{\rm{PASA}}               
\def\pasp{\rm{PASP}}               
\def\nat{\rm{Nature}}              

\def\all{all}
\ifx\files\all  \else 

\begin{document}

\title{Illuminating the Cosmos:\\
	Dark matter, primordial black holes, and cosmic dawn}

\author{Wenzer Qin}
\department{Department of Physics}

 \degree{DOCTOR OF PHILOSOPHY IN PHYSICS}

\degreemonth{May}
\degreeyear{2024}
\thesisdate{April 30, 2024}


\supervisor{Tracy R. Slatyer}{Professor of Physics}

\chairman{Lindley Winslow}{Associate Department Head of Physics}

\maketitle



\cleardoublepage
\setcounter{savepage}{\thepage}
\begin{abstractpage}
%
%
%

The $\Lambda$-CDM model of cosmology has done much to clarify our picture of the early universe. 
However, there are still some questions that $\Lambda$-CDM does not necessarily answer; questions such as what is the fundamental nature of dark matter? 
What is its origin? 
And what causes the intriguing measurements that we are seeing from cosmic dawn? 
In this thesis, I will describe three directions in which I have pushed forward our understanding of how fundamental physics manifests in cosmology.
First, I have studied the signatures of exotic energy injection in various astrophysical and cosmological probes, including the Lyman-$\alpha$ forest, the blackbody spectrum of the cosmic microwave background, the power spectrum of the cosmic microwave background, and the formation of the earliest stars in our universe.
Second, I have investigated the formation of primordial black hole dark matter in a general model for inflation with multiple scalar fields.
I have identified the space of models that can generate primordial black holes while remaining in compliance with observational constraints using a Markov Chain Monte Carlo, and also showed that future gravitational wave observatories will be able to further constrain these models.
Finally, I have developed an analytic description of signals from 21\,cm cosmology using methods inspired by effective field theory.
This method includes realistic observational effects and has been validated against state-of-the-art radiation hydrodynamic simulations, including those with alternative dark matter scenarios.
With these recent efforts, we are advancing the frontiers of dark matter phenomenology and cosmology, thereby paving the way towards illuminating the remaining mysteries of our cosmos and drawing closer to a comprehensive understanding of the universe.
\end{abstractpage}


\cleardoublepage

\section*{Acknowledgments}

\setstretch{1}
The words I'm writing here are woefully inadequate to acknowledge the many people who have helped me get to this point, but hopefully this section can convey some of my gratitude.

To my advisor, Tracy Slatyer: if all physicists were half the scientist that you are, I think we would have found dark matter by now.
Five years ago, when I was going around graduate school open houses trying to figure out where I was going to do my PhD, I remember encountering something quite amazing---no matter where I was, if I mentioned that I was thinking about going to MIT to join your group, I would always hear people say,\,``Tracy is the best.''
It didn't matter if we were talking about research, advising, or collaborating, this was the response that I always got.
In the time that we've worked together since then, I have to say that I think it's true; I've learned so much from you and I am continually inspired by the thoughtfulness you put into the many many things that you do and how you seem to bring the best out of everyone around you.
I hope to emulate your qualities as a physicist as I navigate becoming an independent scientist.

I would not have made it to this point without an incredible network of mentors and collaborators: my academic big brothers, Greg Ridgway and Hongwan Liu; my first graduate school role model, Katelin Schutz; my fellow groupmates, Yitian Sun and Marianne Moore; my sage alter-advisors and thesis committee, Dave Kaiser, Kiyo Masui, and Jackie Hewitt; and everyone else that I have learned so much from, Sarah Geller, Evan McDonough, Shyam Balaji, Josh Foster, Julian Mu\~{n}oz, Adrian Liu, Kai-Feng Chen, and Ben Lehmann.
I also have to thank all of my undergraduate teachers and mentors who helped me start down this path: Charles Hussong, Petar Maksimovic, Candice You, Andrei Gritsan, David Nataf, Nadia Zakamska, Kim Boddy, and Marc Kamionkowski.

This Ph.D. thesis would not have been possible if not for some amazing administrators, without whom I would have ended up very lost in this program and very poor: Scott Morley, Charles Suggs, Sydney Miller, Cathy Modica, and Shannon Larkin.

Life would be insufferable without friends to suffer through it with, including my moral support crew, Patrick Oare, Artur Avkhadiev, and Sam Alipour-Fard, as well as Zhiquan Sun, Rahul Jayaraman; steadfast roommates who have been with me through my best and worst times, Ben Reichelt, Bhaamati Borkhetaria, Charlie Shvartsman, Kaliro\"{e} Pappas, Cammie Farrugio, Enid Cruz Col\'{o}n, Christine Cho, Abby Berk, and Miranda Grenville; my feeziks buddies, Kat Xiang, Robert Barr, Lalit Varada; people who've known me for way too long, Angela Lim, Natalie Wigger, Molly Vornholt, Amanda Sun, Sherry Xie, and Stephanie Zhang; and too many others to name here.

To Nick Kamp: graduate school is hard, but it would have been so much harder without you. 
Our time together has helped me grow so much as a person and work towards the best version of myself, and I still have so much to learn from your charm, generosity, and zest for life.
Here's to many more jam sessions and adventures together.

To Betelgeuse: Mrrrrowww \faPaw.

Finally, to the people who started it all: my family.
Boiar, you were my first role model and I'm trying really really hard but I'll never be as cool as you.
\begin{CJK}{UTF8}{gkai}爸爸妈妈\end{CJK}, you've always pushed me to do my best and I would not be the person I am today without you both.
Thank you for your support and for the freedom to choose my own way through life.


\pagestyle{plain}
\tableofcontents
\newpage
\listoffigures
\newpage
\listoftables


\chapter{Introduction}
\label{sec:thesis_intro}

In spite of the huge advances of the past few decades in physical cosmology and the development of the $\Lambda$-CDM model, there are still a number of open questions about the early universe.
For example, the nature of dark matter is still a mystery. 
Although there is a wealth of gravitational evidence pointing to its existence, we still do not have a microphysical description of what dark matter is. 
We do not know its mass, we do not know where it came from, we do not know if its non-gravitational interactions are nonexistent or just very feeble.
Although terrestrial experiments allow us to control the environments in which we search for dark matter, cosmology remains especially promising for investigating these questions since we can leverage both pristine initial conditions and effects that build up over the longest possible timescales. 

Following a summary of our modern understanding of cosmology and dark matter, as well as prerequisite knowledge for the remainder of this thesis, I will describe the ways in which I have worked to answer the above questions.
In Chapter~\ref{sec:DH}, I will discuss various constraints and predicted signals from exotic energy injection.
In Chapter~\ref{sec:PBHandMFI}, I will demonstrate that a range of multifield inflation models are capable of producing primordial black hole dark matter while remaining in compliance with observational constraints.
In Chapter~\ref{sec:EFTof21cm}, I will develop an analytic description for the 21\,cm cosmology signal based on methods from effective field theory.
I will then conclude in Chapter~\ref{sec:thesis-conclusion}.
Throughout this thesis, I will work in natural units, where $\hslash = c = k_B = 1$.

\section{A brief historical review}
\label{sec:history}

The necessity and impact of new work cannot be recognized without also knowing the foundations upon which our current understanding of cosmology is built.
Although I cannot hope to review the entire history of the field, I will attempt to give some context here.
For a more comprehensive summary of the history of dark matter and cosmology, see Refs.~\cite{Bertone:2016nfn,deSwart:2017heh,deSwart2022dark}.

\begin{figure}
	\centering
	\includegraphics[width=0.6\textwidth]{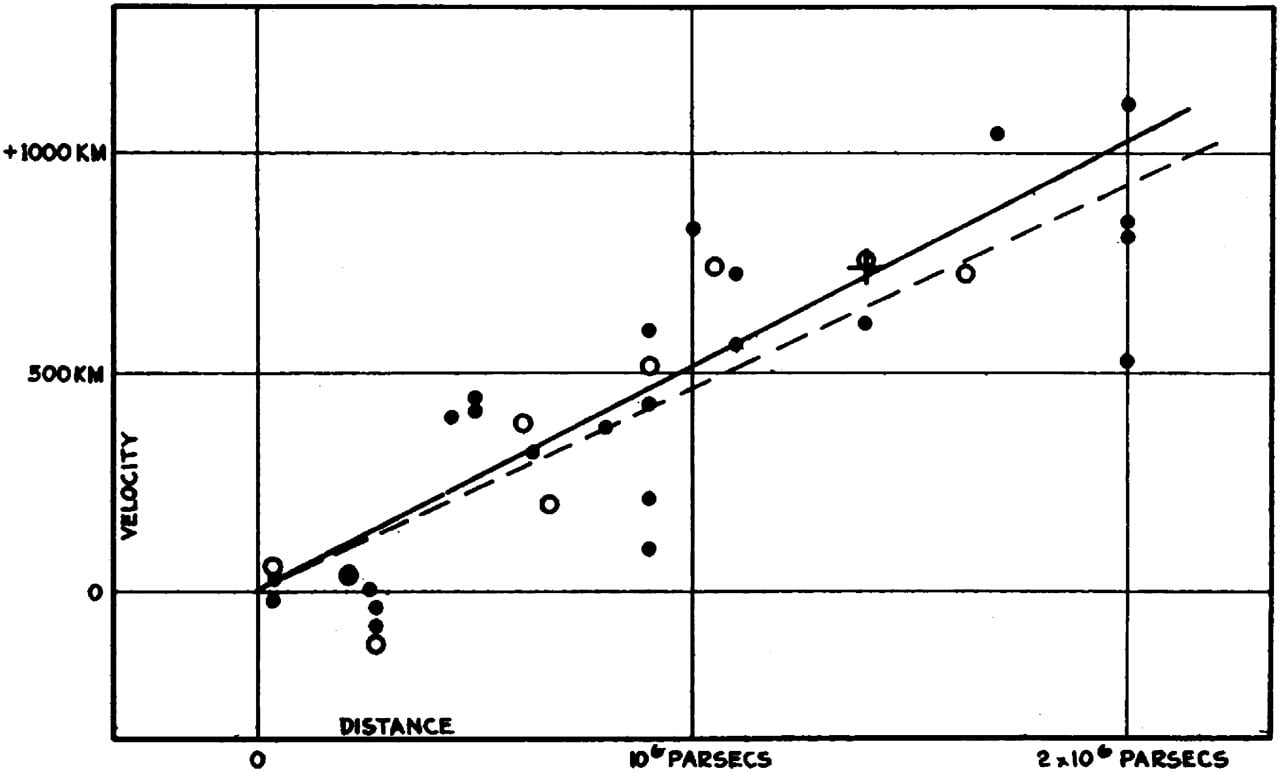}
	\caption{
		The radial velocities of galaxies as a function of their distance, and Edwin Hubble's best fit for a linear relationship between the two.
		Reproduced from Ref.~\cite{Hubble:1929ig}.
	}
	\label{fig:hubble-diagram}
\end{figure}
At the beginning of the 20th century, physicists were already toying with ideas that would form the basis of modern cosmology.
For example, in 1912, Vesto M. Slipher measured the Doppler shift of spiral galaxies and inferred that almost all of them appeared to be receding from us~\cite{1915PA.....23...21S}.
However, it was not until much later that the cosmological implications were understood.
George Lema\^{i}tre proposed in 1927 that the universe was expanding~\cite{Lemaitre:1927zz} and later that it may have began from an ``explosion"~\cite{Lemaitre:1931zzb}---this was the origin of the ``Big Bang theory".
Observational evidence for Lema\^{i}tre's theory was provided in 1929 by Edwin Hubble using Cepheid variable stars~\cite{Hubble:1929ig}; the famous Hubble diagram from this work is reproduced in Fig.~\ref{fig:hubble-diagram}.

In addition, there were discussions of unseen or ``meteoric" matter within the Milky Way.
It was generally accepted that there may exist matter out in space that could not be observed by telescopes; however, astronomers thought that such matter would likely take the form of stars that were too faint to be seen by existing means.
Drawing analogies between stars in the galaxy and gases of particles, scientists such as Lord Kelvin, Henri Poincar\'{e}, Ernst \"{O}pik, Jacobus Kapteyn, and Jan Oort realized that the velocities of stars could be used to infer the local density of matter, and they could therefore constrain the amount of invisible matter through its gravitational influence~\cite{Bertone:2016nfn}.
They concluded that the existence of this unseen matter was unlikely, or at least its abundance was less than that of visible stars.

\begin{figure}
	\centering
	\includegraphics[width=0.6\textwidth]{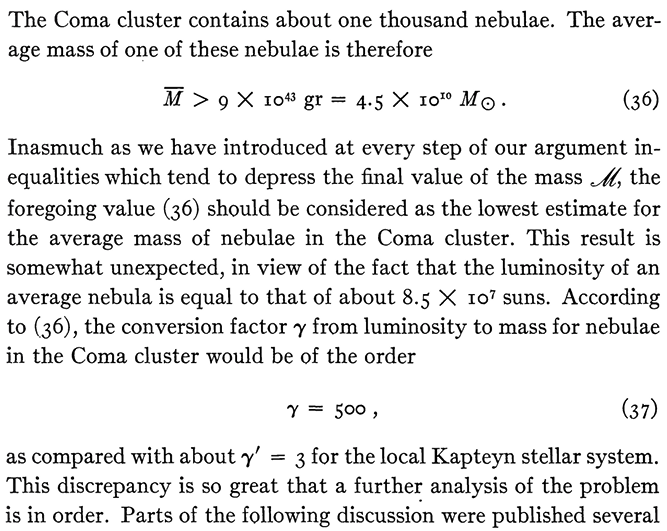}
	\caption{
		Excerpt from Ref.~\cite{Zwicky:1937zza} showing Zwicky's conclusion that the Coma Cluster has a remarkably large mass-to-light ratio.
	}
	\label{fig:Zwicky}
\end{figure}
These studies set the stage for Fritz Zwicky's study of the Coma Cluster in the 1933, which is one of the most well-known pieces of evidence for dark matter even today.
Zwicky applied the virial theorem to the observed galaxies within the cluster and inferred that the velocity dispersion of the galaxies should be 80 km/s~\cite{Zwicky:1933gu}.
In contrast, the observed velocity dispersion was closer to 1000 km/s.
In 1937, Zwicky refined this estimate and concluded that the mass-to-light ratio was nearly 500, and even accounting for the overly large value of the Hubble constant that he used at the time, his calculation still implied the existence of some non-luminous matter~\cite{Zwicky:1937zza}; I include an excerpt from this paper in Fig.~\ref{fig:Zwicky}.
Sinclair Smith performed a similar estimate on the Virgo Cluster and found an average mass per galaxy of about $10^{11} \,M_\odot$, which was about a hundred times larger than Hubble's estimate for the average galaxy mass~\cite{1936ApJ....83...23S}.
The astonishing mass estimates by Zwicky and Smith triggered a flurry of discussions and new experimental efforts, and would continue to puzzle scientists for the decades to come.

In the meantime, clues about the missing mass problem were also beginning to emerge in the rotational motion of galaxies.
Throughout the early 1900s, measurements of the rotation curves of galaxies were steadily improving, in part due to repurposing radio technology from World War II for astronomy and the first detection of the 21\,cm line in 1951, which serves as a precise reference for measuring the rotational velocities of galaxies and is still an important observable for modern cosmology~\cite{Bertone:2016nfn,deSwart:2017heh}.
The rotation curves measured in that era occasionally yielded unexpectedly large mass-to-light ratios.
However, at the time these anomalous measurements were not yet considered a crisis.

\begin{figure}
	\centering
	\includegraphics[width=0.8\textwidth]{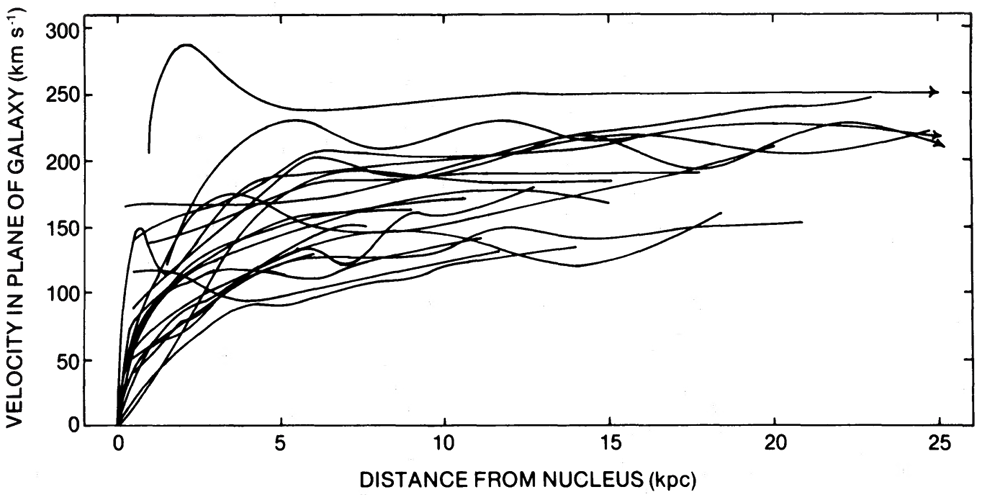}
	\caption{
		Rotation curves for twenty-one galaxies showing a gradual increase or nearly constant velocity out to large radii.
		Reproduced from Ref.~\cite{1980ApJ...238..471R}.
	}
	\label{fig:rotation_curves}
\end{figure}
Around the 1970s, further study of galactic rotation curves began to lead scientists to conclude that there was significant unseen mass in the outer regions of galaxies.
The unexpected flatness of rotation curves was seen in both optical data published by e.g. Vera Rubin and Kent Ford~\cite{1970ApJ...159..379R} and radio data, as obtained by Morton Roberts and Arnold Rots~\cite{1973A&A....26..483R}.
In 1974, two papers by Jaan Einasto, Ants Kaasik, and Enn Saar~\cite{1974Natur.250..309E}, as well as Jerry Ostriker, Jim Peebles, and Amos Yahil connected the missing mass problem in galaxy clusters and galactic rotation curves~\cite{1974ApJ...193L...1O}.
Additional publications in subsequent years firmly established the flatness of rotation curves, including the widely cited paper by Vera Rubin, Kent Ford, and Norbert Thonnard~\cite{1980ApJ...238..471R}.
With this, the scientific community began to take seriously the problem of missing matter.

During this time, the field of physical cosmology as we know it today also began to emerge~\cite{Kragh_cosmo_contro,Peebles:2022bya}.
From the 1940's to the 1960's, astronomers were divided between the Big Bang theory and ``steady-state" universe by Hermann Bondi, Thomas Gold, and Fred Hoyle in which the universe was expanding but matter was also continuously created such that the universe appeared the same at any point in time~\cite{Bondi:1948qk,Hoyle:1948zz}.
However, evidence in support of the Big Bang emerged with the first quasars, which were discovered at large distances in the 1950s and indicated that the universe must have looked very different in the past.
The key discovery came from the cosmic microwave background (CMB), which was predicted to be relic radiation from the Big Bang by Ralph Alpher and Robert Herman in 1948~\cite{Alpher:1948srz} and first observed in 1964 by Arno Penzias and Robert Wilson~\cite{Penzias:1965wn}.
The blackbody spectrum of the CMB is difficult to explain in the steady-state model, leading to the theory's decline.

Although the universe was known to be expanding, its geometry was still unclear.
From the Friedmann equations, the energy density of the universe determines its curvature and also its eventual fate.
At a critical value for the density, the universe is flat; for a density smaller than this, the universe is open and will continue to expand forever; larger, and the universe is closed and will eventually collapse.
An accounting of the luminous matter in galaxies yielded an average density which was smaller than the critical density, although it was within a couple orders of magnitude.
Hence, if the universe is not open, then this points to other components contributing to the universe's density.

One can also show from the Friedmann equations that departures from the critical density will rapidly increase with time; therefore, a density that is couple orders of magnitude from the critical value today implies that the universe was exceedingly close to flat in the distant past.
This presented a fine-tuning problem for cosmology, sometimes called the ``flatness" problem.
At the time there were also two other conundrums: the ``monopole" problem, i.e. why there are no magnetic monopoles in the observable universe, and the ``horizon problem", i.e. why our universe is so homogeneous when it appears to be made up of causally disconnected patches.
These problems were solved by the theory of inflation, which was first proposed by Alan Guth in 1981~\cite{Guth:1980zm}.
Slow-roll inflation was then developed by Andrei Linde~\cite{Linde:1981mu} and independently by Andreas Albrecht and Paul Steinhardt~\cite{Albrecht:1982wi}, and is still used to define standard inflationary dynamics today.

Another key event in the history of cosmology was the first observational evidence for dark energy in 1998. 
The teams led by Saul Perlmutter, Brian Schmidt, and Adam Riess used Type 1A supernovae to show that the universe was expanding at an accelerating rate, pointing to the existence of dark energy~\cite{SupernovaSearchTeam:1998fmf,SupernovaCosmologyProject:1998vns}.
This dark energy could take the form of a cosmological constant, which had been first proposed by Albert Einstein.

\begin{figure}
	\centering
	\includegraphics[width=0.8\textwidth]{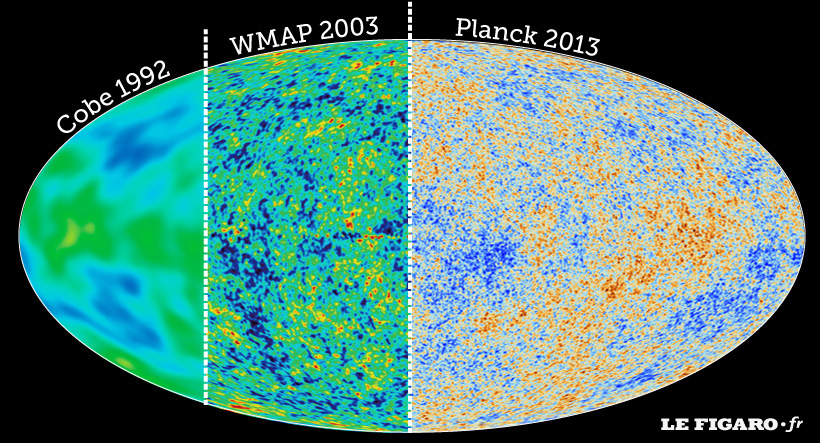}
	\caption{
		Maps of the CMB as measured by COBE, WMAP, and \textit{Planck}.
		Reproduced from Ref.~\cite{Vey_2013}.
	}
	\label{fig:cmb_maps}
\end{figure}
Measurements of the composition of the universe's energy budget began in the 1990s around the time of the Cosmic Background Explorer (COBE). 
In addition to confirming the near-perfect blackbody spectrum of the CMB, COBE detected faint anisotropies in the CMB~\cite{COBE:1992syq,Bennett:1996ce}.
Subsequent CMB experiments began to show strong evidence for a flat universe, particularly the Wilkinson Microwave Anisotropy Probe (WMAP) which constrained the energy content of the universe to be about 5\% ordinary ``baryonic" matter, 24\% cold dark matter, and 71\% dark energy~\cite{WMAP:2003elm}.
These results were in agreement with measurements of the matter density from galaxy surveys such as 2dFGRS and SDSS.
Today, some of the most precise measurements of cosmological parameters and the abundance of dark matter and dark energy come from the \textit{Planck} observatory~\cite{Planck:2018vyg}.
Fig.~\ref{fig:cmb_maps} shows a comparison of maps of the CMB as measured by these experiments.

Finally, an alternative to dark matter includes modified Newtonian dynamics (MOND), which describes a class of theories in which Newton's second law is modified to scale as $F = m a^2 / a_0$ in the limit of very small accelerations.
The strongest evidence against MOND was published in 2006 using observations of the Bullet Cluster~\cite{Clowe:2006eq}.
X-ray and lensing observations show that the plasma distribution is offset from the gravitational potential, which is difficult to explain in MOND but consistent with collisionless dark matter.

Altogether, the progress of the 1990s and early 2000s established the $\Lambda$-CDM paradigm of cosmology~\cite{Peebles:2022bya}.
This is the model widely accepted by physicists today, and much of the current research in cosmology is focused on refining observational evidence and our understanding of this model, including determining the microphysical details of dark matter, the mechanisms that generated its abundance, and confirming the predictions of $\Lambda$-CDM in yet unobserved redshifts and unexamined regimes.

\section{Status of dark matter searches: \textit{What are we?}}
\label{sec:DM_status}

Today, we know that dark matter needs to have a few key properties:
\begin{enumerate}
	\item Other than through gravity, its interactions with Standard Model particles is very weak,
	
	\item It is stable on cosmological timescales,
	
	\item Its mass must be at least large enough that its Compton wavelength fits within a galaxy~\footnote{There are other constraints on the mass range such as the Tremaine-Gunn bound or unitarity limits, but this is the most model-independent statement.},
	
	\item It is cold, i.e. moving non-relativistically, otherwise structure in our universe would be too washed out,
	
	\item It can be produced in nearly the same abundance as baryonic matter in the early universe.
\end{enumerate}
Specifying any other details about dark matter requires choosing a particular model.

This last property of dark matter gives the closest thing to a ``target" that we have in searches for dark matter.
The fact that the density in dark matter is only an $\mathcal{O} (1)$ factor away from the density of baryonic matter implies that the two sectors may have been interacting in the early universe; otherwise, their nearly equal abundances would be a startling coincidence.
Hence, dark matter and Standard Model particles may have interactions that were efficient in the far past and very feeble today.
Additionally, for a particular dark matter model and production mechanism, the parameter space that produces the correct abundance presents a theoretically-motivated target for experimental searches.
We will return to types of production mechanisms for dark matter in the next section.

Given this argument that dark matter likely has some interactions with Standard Model particles, we can now classify ways to search for such interactions.
\begin{itemize}
	\item Colliders: if dark matter is a new particle, then colliding Standard Model particles at high enough energies may result in the production of dark matter.
	This would manifest as missing energy and momentum in the resulting jets.
	
	\item Direct Detection: dark matter may be able to scatter off of ordinary matter and be detected through e.g. nuclear recoils.
	
	\item Indirect Detection: Standard Model particles may be produced from the dark sector, e.g. through decays or annihilations by dark matter particles.
	Depending on the mass of the dark matter, the resulting particles can be very energetic and leave distinct signatures in astrophysical searches.
\end{itemize}
These three types of searches can be summarized in the diagram below.
For the rest of this section, I will focus on indirect detection.
\begin{figure}[h]
	\centering
	\hspace{1in}
	\begin{tikzpicture}
		\begin{feynman}
		\vertex (a);
		\vertex [above right=of a] (f1) {\(\text{SM}\)};
		\vertex [below right=of a] (f2) {\(\text{SM}\)};
		\vertex [above left=of a] (i1) {\(\chi\)};
		\vertex [below left=of a] (i2) {\(\chi\)};
		\vertex[blob, fill=black!20, minimum size=30pt] at (a) {};
		
		\diagram* {
			(i1) -- [plain] (a) -- [plain] (f1),
			(i2) -- [plain] (a) -- [plain] (f2),
		};
		
		\draw[->,thick] ([shift={(0.5,0.5)}]i1.north west) -- node[above, midway] {Indirect} ([shift={(-0.5,0.5)}]f1.north east);
		\draw[->,thick] ([shift={(0.7,0.7)}]f1.south east) -- node[right, midway] {Direct} ([shift={(0.7,-0.7)}]f2.north east);
		\draw[->,thick] ([shift={(0.2,-0.3)}]f2.south west) -- node[below, midway] {Colliders} ([shift={(-0.2,-0.3)}]i2.south east);
		\end{feynman}
	\end{tikzpicture}
\end{figure}

With indirect detection, one can search for the injection of energy into electromagnetic observables by processes not readily explained by $\Lambda$-CDM or the Standard Model of particle physics, e.g. \textit{exotic energy injection}.
Decaying dark matter is one such example, and the amount of energy injected per unit volume per time goes as
\begin{equation}
	\frac{dE}{dV \, dt} = \frac{\rho_\chi (z)}{\tau},
\end{equation}
where $V$ is physical volume, $\rho_\chi$ is the dark matter energy density as a function of the redshift $z$, and $\tau$ is the decay lifetime.
Another example is annihilating dark matter, where the injection rate is 
\begin{equation}
	\frac{dE}{dV \, dt} = \frac{\rho_\chi^2 (z) \langle \sigma v \rangle}{m_\chi},
\end{equation}
where $\langle \sigma v \rangle$ is the thermally-averaged annihilation cross-section and $m_\chi$ is the mass of the dark matter particle.
Often times, the cross-section is expanded in terms of velocities.
\begin{equation}
	\langle \sigma v \rangle = \langle a + b v^2 + c v^4 + \cdots \rangle 
\end{equation}
When expanding the cross-section in terms of partial waves, the velocity-independent term on receives contributions from $s$-wave annihilation, whereas the next-to-leading order term receives contributions from both $s$ and $p$-wave annihilation.
Other examples of exotic energy injection include evaporation and accretion onto primordial black holes.

From examining the injection rates, we can already see that different types of surveys will be sensitive to different forms of energy injection.
For example, since the dark matter density scales with redshift as $\rho_\chi (z) \propto (1+z)^3$, then the energy injection rate for decays also goes as $(1+z)^3$.
For $s$-wave annihilation, the only redshift dependence of the injection rate again comes from the dark matter density, so the injection rate scales as $(1+z)^6$ until structure formation begins, at which point the annihilation is boosted due to the fact that $\langle \rho_\chi^2 \rangle > \langle \rho_\chi \rangle^2$.
At redshifts where the enhancement from structure formation is small, the contributions from $s$-wave annihilation are more heavily weighted towards earlier redshifts compared to decays, which means that later redshift probes such as 21\,cm or Lyman-$\alpha$ can be better for constraining decays compared to e.g. the CMB.
The relative advantage of different energy injection probes also depends, of course, on experimental precision as well as the types of products (photons, electrons, hadrons, etc.) resulting from the injection events.

\begin{figure}
	\centering
	\includegraphics[width=0.8\textwidth]{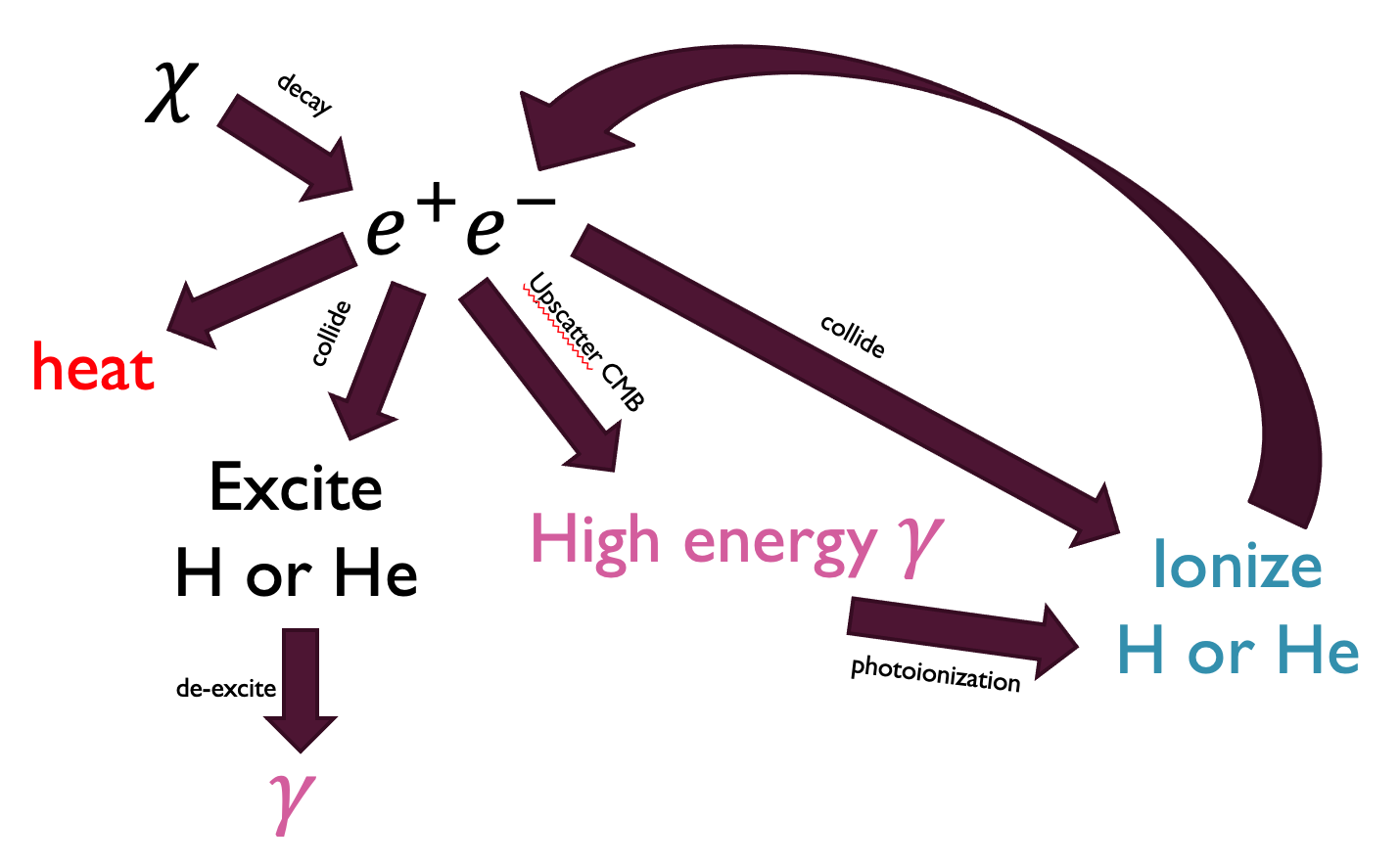}
	\caption{
		Diagram depicting the cooling cascade for dark matter decaying to $e^+ e^-$ pairs.
	}
	\label{fig:cascade}
\end{figure}
The injected Standard Model particles may interact with the thermal bath of the universe and deposit their energy in a number of different ways.
Fig.~\ref{fig:cascade} depicts a cooling cascade for dark matter decaying to pairs of electrons and positrons.
Depending on the dark matter mass, the leptons may carry significant kinetic energy and thermalize with existing charged particles in the universe, thus depositing their energy as heat.
Alternatively, the electrons and positrons can collide with atoms in our universe and excite them.
When these atoms de-excite, they will emit line photons that we can search for on top of background radiation.
The charged particles can also upscatter CMB very photons to higher energies, further distorting the CMB spectrum, and if the upscattered photons have a high enough energy, they can go on to photoionize atoms.
Ionizations can also be caused by collisions with the primary $e^+ e^-$, and the secondary electrons resulting from the ionizations can go through this cascade all over again.
Hence, \textit{exotic energy injection can manifest in observables as additional heat, ionization, or radiation.}

We can parametrize the amount of energy deposited into a channel $c$ such as heating, excitation, or ionization as
\begin{equation}
	\left( \frac{dE}{dV dt} \right)^\mathrm{dep}_c = f_c (z) \left( \frac{dE}{dV dt} \right)^\mathrm{inj} ,
\end{equation}
where these $f_c$'s are often estimated using the prescription of Ref.~\cite{Chen:2003gz},
\begin{equation}
	f_\mathrm{ion} \sim f_\mathrm{exc} \sim \frac{1 - x_e}{3}, \qquad f_\mathrm{heat} \sim \frac{1 + 2 x_e}{3} ,
\end{equation}
where $x_e = n_e / n_\mathrm{H}$ is the free electron fraction.
The justification for this prescription is that in a totally neutral medium, energy deposition is nearly equally divided between heating, ionization, and excitation, whereas in a fully ionized medium, there are no atoms to ionize or excite and any energy injected is deposited into heat.
The form of the $f_c$'s above smoothly interpolates between these limits.
However, the $f_c$'s are more accurately solved for using codes such as \texttt{DarkHistory}~\cite{DarkHistory}.

Once the $f_c$'s are known, they can be incorporated into evolution equations for the temperature and ionization fraction.
For example, to get to a heating rate from $\left( \frac{dE}{dV dt} \right)^\mathrm{dep}_\mathrm{heat}$, we have to divide by the number density to get the rate of change of energy per baryon and by the heat capacity to convert to a temperature change.
Then the evolution equation for the matter temperature is given by
\begin{equation}
	\dot{T}_m = - 2 H T_m + \Gamma_C (T_\mathrm{CMB} - T_m) + \dot{T}_m^\mathrm{re} + \frac{2 f_\mathrm{heat}}{3 (1 + x_e + \chi) n_\mathrm{H}} \left( \frac{dE}{dV dt} \right)^\mathrm{inj} ,
\end{equation}
where the first term corresponds to the adiabatic cooling term from the universe's expansion, the second term is the Compton cooling rate, where $\Gamma_C$ will be defined in App.~\ref{app:Lya_supp}, the third term represents heating from sources of reionization, and the last term is the exotic heating term, with $\chi =  n_\mathrm{He} / n_\mathrm{H}$ being the helium fraction.

Similarly, we can write the evolution equation for the fraction of ionized hydrogen.
Both exotic ionizations and exotic excitations will affect $x_\mathrm{HII}$, since excited hydrogen may be ionized if it does not decay back to the ground state first.
\footnote{
	We will use the standard notation in astrophysics for representing states of ionization, where an element symbol followed by `I' denotes a neutral atom, `II' a singly ionized atom, and so on and so forth.
}
Assuming that the CMB is a perfect blackbody and all hydrogen energy levels above the first excited state are in thermal equilibrium (i.e. so that hydrogen can be treated as a ``three-level atom"), then 
\begin{align}
	\dot{x}_\mathrm{HII} = &- \mathcal{C} \left[ n_\mathrm{H} x_e x_\mathrm{HII} \alpha_\mathrm{B} - 4 (1 - x_\mathrm{HII}) \beta_\mathrm{B} e^{-E_\alpha / T_\mathrm{CMB}} \right] + \dot{x}_\mathrm{HII}^\mathrm{re} \n
	&+ \left[ \frac{f_\mathrm{ion}}{\mathcal{R} n_\mathrm{H}} + \frac{(1-\mathcal{C}) f_\mathrm{exc}}{E_\alpha n_\mathrm{H}} \right] \left( \frac{dE}{dV dt} \right)^\mathrm{inj}
\end{align}
where $\mathcal{C}$ is the Peebles-$\mathcal{C}$ factor representing the probability of a hydrogen atom in the first excited state decaying to the ground state before being photoionized, $\alpha_\mathrm{B}$ and $\beta_\mathrm{B}$ are the case-B recombination and photoionization coefficients for hydrogen, $\mathcal{R} = 13.6$ eV is the ionization energy of hydrogen, e.g. a Rydberg, and $E_\alpha = 3 \mathcal{R} / 4$ is the energy corresponding to the Lyman-$\alpha$ transition~\cite{Peebles:1968ja,Seager:1999km}.
In this equation, the terms in the first line correspond to recombination, photoionization, and reionization, whereas the terms in the last line represent the exotic contributions.

With these ideas at hand, we can use any measurement of the global temperature or ionization fraction to constrain exotic energy injection, and thus search for signatures of dark matter in a relatively model-independent manner.
In Section~\ref{sec:Lya}, I will show how to use Lyman-$\alpha$ forest measurements of the gas temperature to constrain exotic energy injection.
Section~\ref{sec:DHv2_tech} and ~\ref{sec:DHv2_apps} will describe refinements to this calculation where we relax the assumptions of the three-level atom and improve the deposition calculation at low energies.
Finally, in Section~\ref{sec:first_stars}, I will show how exotic energy injection impacts the formation of the first stars in the universe.

\section{Production of dark matter: \textit{Where do we come from?}}
\label{sec:DM_production}

As mentioned previously, one of the few things that we can say with great certainty about dark matter is that it is about five times as abundant as ordinary baryonic matter.
Hence, for any proposed dark matter model, a key consideration is how well the model can produce the correct relic abundance and how constrained (or within reach of future experiments) is the corresponding parameter space.
Here, I will review a few popular dark matter candidates and their respective production mechanisms.

\subsection{Weakly Interacting Massive Particles}

Consider a dark matter particle $\chi$ that can annihilate into light Standard Model particles, and assume that these interactions were initially in thermal equilibrium.
If we recall that the abundance of a non-relativistic species with mass $m$ in thermal equilibrium at temperature $T$ is given by
\begin{equation}
	n_\mathrm{eq} \sim \left( m T \right)^{3/2} e^{-m /T}, 
	\label{eqn:therm_equib_n}
\end{equation}
then the Boltzmann equation determining the evolution of the dark matter abundance is
\begin{equation}
	\frac{d n_\chi}{dt} = - 3 H(z) n_\chi - \langle \sigma v \rangle [n_\chi^2 - n_{\chi, \mathrm{eq}}^2], 
\end{equation}
where $n_\chi$ is the number density of dark matter and $H(z)$ is the Hubble parameter.
In this equation, the first term represents dilution of the dark matter from the expansion of the universe, the second corresponds to depletion of dark matter from annihilation, and the last is reverse annihilations.

If the annihilation rate $\Gamma_\mathrm{ann} = n_\chi \langle \sigma v \rangle$ is much larger than $H(z)$, then we can neglect the expansion term and the dark matter abundance is driven towards $n_\chi \sim n_{\chi, \mathrm{eq}}$.
However, when $\Gamma_\mathrm{ann} < H(z)$, then dark matter decouples from Standard Model species and annihilations no longer affect the abundance of dark matter.
Another way to think about this is that the universe is expanding fast enough that particles cannot even find each other to annihilate.
We often call this process ``freeze-out" of dark matter.

From Eqn.~\ref{eqn:therm_equib_n}, the number abundance of dark matter drops exponentially once $T \ll m_\chi$, which will cause the annihilation rate to drop below the Hubble expansion rate, so we will take $T \sim m_\chi / x$ as the freeze-out temperature, where $x \sim 10$ for the sake of an estimate.
In addition, the Friedmann equations tell us that the Hubble parameter during radiation domination can be approximated by $H(z) \sim T^2 / M_\mathrm{Pl}$, where $M_\mathrm{Pl}$ is the Planck mass.
Hence the abundance of dark matter today is approximately given by
\begin{equation}
	\Gamma_\mathrm{ann} = H(z) \qquad \rightarrow \qquad n_{\chi, \mathrm{fr}} \langle \sigma v \rangle = \frac{m_\chi^2}{x^2 M_\mathrm{Pl}} .
	\label{eqn:freeze-out}
\end{equation}

While the number density of dark matter dilutes with the expansion of the universe, $n_\chi / n_\gamma$ is constant after dark matter freeze-out, since both scale as $(1+z)^3$.
Knowing that $n_\chi \sim T^3$, $\rho_\chi \approx 5 \rho_b$, and the baryon-to-photon ratio is $\eta \equiv n_b / n_\gamma \approx 5 \times 10^{-10}$, we can also write the freeze-out abundance of dark matter as 
\begin{equation}
	\frac{n_{\chi, \mathrm{fr}}}{n_{\gamma, \mathrm{fr}}} = \frac{n_{\chi, 0}}{n_{\gamma, 0}} = \frac{5 \eta m_\mathrm{H}}{m_\chi}
	\label{eqn:dm_abundance}
\end{equation}

Putting Eqns.~\ref{eqn:freeze-out} and \ref{eqn:dm_abundance} together, we find 
\begin{equation}
	\langle \sigma v \rangle = \frac{x}{5 \eta m_\mathrm{H} M_\mathrm{Pl}} \approx 10^{-26} \,\mathrm{cm}^3 / \mathrm{s} .
\end{equation}
Remarkably, the cross-section needed to get the correct relic abundance is independent of the dark matter mass!
Hence, this value is used as a benchmark for experimental searches, although it is already ruled out for several mass ranges~\cite{ParticleDataGroup:2022pth}.

We can take one more step to determine what a theoretically motivated mass range for this model might be.
From dimensional analysis, the annihilation cross-section should scale as $\langle \sigma v \rangle = g^2 / m_\chi^2$, where $g$ is the dimensionless coupling constant characterizing the interaction.
Substituting this into the previous expression, we find
\begin{equation}
	g \sim 0.01 \times \frac{m_\chi}{1000 \, \mathrm{GeV}} .
\end{equation}
Therefore, a new particle with a mass and coupling constant near the electroweak scale gives the correct relic abundance for dark matter.
Particles fitting this description are therefore called \textit{weakly interacting massive particle}, or WIMPs for short, and this coincidence is sometimes called the ``WIMP miracle".
Because WIMPs give rise so neatly to the right dark matter density and are also predicted to exist in well-motivated theories like supersymmetry, they were considered to be the canonical dark matter candidate for many years.

\subsection{The QCD axion}

The QCD axion is a hypothetical pseudoscalar particle motivated as a solution to the strong CP problem, where C refers to the symmetry of charge conjugation and P refers to parity.
The strong CP problem is a question of why the CP violating term of QCD,
\begin{equation}
	\mathcal{L} \supset \theta \frac{g_s^2}{32 \pi^2} G^a_{\mu \nu} \tilde{G}^{a\mu\nu},
\end{equation}
is so small when CP violation exists in other sectors of the Standard Model.
In this expression, $\theta$ is a dimensionless parameter characterizing the level of CP violation, $g_s$ is the strong coupling constant and $G^a$ is the gluon field strength tensor.
Measurements of the neutron electric dipole moment limit $\theta \lesssim 10^{-10}$~\cite{Abel:2020pzs}, which seems anomalously fine-tuned if $\theta$ can in principal take on any value between $-\pi$ and $\pi$.
The solution is to introduce a spontaneously broken $U(1)$ symmetry, called the Peccei-Quinn symmetry after the first physicists who postulated this idea~\cite{Peccei:1977hh}, and the resulting pseudo-Nambu-Goldstone boson is the axion field $a(x)$ with a coupling to gluons, 
\begin{equation}
	\mathcal{L} \supset \left(\theta + \frac{a(x)}{f_a} \right) \frac{g_s^2}{32 \pi^2} G^a_{\mu \nu} \tilde{G}^{a\mu\nu}.
\end{equation}
Here, $f_a$ is the axion decay constant.
The introduction of the axion field allows this term to dynamically relax to zero, thus solving the strong CP problem~\cite{Hook:2018dlk}.

In order for the axion to be a good dark matter candidate, it has to be cosmologically stable, cold, and have a production mechanism in the early universe.
For the QCD axion, one can show that $m_a f_a \approx (100\,\mathrm{MeV})^2$~\cite{ParticleDataGroup:2022pth}.
If we assume the axion can decay into two photons, then its lifetime is given by
\begin{equation}
	\tau = \frac{64 \pi}{g_{a\gamma\gamma}^2 m_a^3} = \frac{256 \pi^2}{\alpha^2 m_a^3} f_a^2,
\end{equation}
where $g_{a\gamma\gamma}$ is the axion-photon coupling and $\alpha$ is the fine structure constant, and requiring the lifetime to be longer than the age of the universe means that $m_a < 24$ eV.

Then there are two main ways by which axions can have a residual abundance today that is comparable to dark matter.
First, the axion could have been in thermal equilibrium with the rest of the universe
At these masses, if the axion was in thermal equilibrium with the rest of the universe, then it would contribute to the total dark matter abundance as a \textit{hot} dark matter component, which is strongly constrained by the CMB to be a very small fraction.
Since the relic energy density of axions scales linearly with the mass, then CMB limits require $m_a \lesssim 1$ eV~\cite{DiValentino:2015wba}.
In addition, past studies have shown that axions with masses below $\sim 0.1$ eV decouple before the QCD phase transition and therefore have their abundance significantly diluted by entropy production---hence, in order to get a significant cosmological thermal abundance, the axion must have a mass above this threshold~\cite{Archidiacono:2015mda}.

Second, if the axion was never in thermal contact with the Standard Model, but was initially `misaligned' with its potential minimum, then after it begins to roll into the minimum, the resulting oscillations may give rise to a significant axion abundance.
If the axion begins oscillating near the QCD scale, $T \sim \Lambda_\mathrm{QCD} \approx 200$ MeV, then its number density today is approximately
\begin{equation}
	n_a \approx \frac{1}{2} \theta_0^2 f_a^2 \frac{\Lambda_\mathrm{QCD}^2}{M_\mathrm{Pl}} \times \left( \frac{a_\mathrm{QCD}}{a_0} \right)^3 ,
\end{equation}
where $\theta_0$ is the initial misalignment angle and $a$ is the scale factor of the universe.
The ratio of the scale factors is equal to the temperature today over $\Lambda_\mathrm{QCD}$, and $n_\gamma \approx T^3$, so we can rewrite this as 
\begin{equation}
	n_a \approx \frac{1}{2} \theta_0^2 (100 \,\mathrm{MeV})^4 \frac{1}{m_a^2 \Lambda_\mathrm{QCD} M_\mathrm{Pl}} n_\gamma .
\end{equation}
Again, since we know the relative abundance of dark matter and baryons, as well as the baryon-to-photon ratio, then the energy density in axions relative to the measured energy density of dark matter is 
\begin{equation}
	\frac{\Omega_a}{\Omega_\chi} \approx \theta_0^2 \frac{10^{-5} \,\mathrm{eV}}{m_a}.
\end{equation}
Assuming $\theta_0$ is a $\mathcal{O} (1)$ number, then $m_a \lesssim 10^{-5}$ eV in order for axions to make up the entire dark matter abundance.

%
\vspace{3mm}
\begin{tikzpicture}[]

\newcount\nOne; \nOne=-7
\def\w{14}      
\def\n{9}      
\def\noffset{0} 
\def\nskip{2}   
\def\la{2.00}   
\def\lt{0.20}   
\def\ls{0.15}   

\def\myx(#1){{(#1-\nOne)*\w/\n}}
\def\arrowLabel(#1,#2,#3,#4){
	\def\xy{(#1-\nOne)*\w/\n}; \pgfmathparse{int(#2*100)};
	\ifnum \pgfmathresult<0
	\def\yyp{{(\lt*(-0.10+#2))}}; \def\yyw{{(\yyp-\la*\lt*#3)}}
	\draw[<-,thick,black!50!blue,align=center]
	(\myx(#1),\yyp) -- (\myx(#1),\yyw)
	node[below,black!80!blue] {#4}; 
	\else
	\def\yyp{{(\lt*(0.10+#2)}}; \def\yyw{{(\yyp+\la*\lt*#3)}}
	\draw[<-,thick,black!50!blue,align=center]
	(\myx(#1),\yyp) -- (\myx(#1),\yyw)
	node[above,black!80!blue] {#4};
	\fi}

\draw[->,thick] (-\w*0.03,0) -- (\w*1.06,0)
node[right=4pt,below=6pt] {$m_a$ [eV]};

\foreach \tick in {0,1,...,\n}{
	\def\x{{\tick*\w/\n}}
	\def\dec{\the\numexpr \nOne+\tick \relax}
	\pgfmathparse{Mod(\tick-\noffset,\nskip)==0?1:0}
	\ifnum\pgfmathresult>0
	\draw[thick] (\x,\lt) -- (\x,-\lt) 
	node[below] {$10^{\dec}$}; 
	\else
	\draw[thick] (\x,\ls) -- (\x,-\ls); 
	\fi
}

\draw[|->,thick,black!20!black]
({(1.4-\nOne)*\w/\n},0.5) -- ({(2.5-\nOne)*\w/\n},0.5)
node[midway,above=1pt] {unstable};

\draw[<->,thick,black!20!black]
({(-1-\nOne)*\w/\n},0.95) -- ({(1-\nOne)*\w/\n},0.95)
node[midway,above=1pt] {allowed thermal production};

\draw[<-|,thick,black!20!black]
({(-7.5-\nOne)*\w/\n},0.5) -- ({(-5-\nOne)*\w/\n},0.5)
node[midway,above=1pt] {misalignment};

\end{tikzpicture}
\begin{tikzpicture}[]

\newcount\nOne; \nOne=14
\def\w{14}      
\def\n{9}      
\def\noffset{0} 
\def\nskip{2}   
\def\la{2.00}   
\def\lt{0.20}   
\def\ls{0.15}   

\def\myx(#1){{(#1-\nOne)*\w/\n}}
\def\arrowLabel(#1,#2,#3,#4){
	\def\xy{(#1-\nOne)*\w/\n}; \pgfmathparse{int(#2*100)};
	\ifnum \pgfmathresult<0
	\def\yyp{{(\lt*(-0.10+#2))}}; \def\yyw{{(\yyp-\la*\lt*#3)}}
	\draw[<-,thick,black!50!blue,align=center]
	(\myx(#1),\yyp) -- (\myx(#1),\yyw)
	node[below,black!80!blue] {#4}; 
	\else
	\def\yyp{{(\lt*(0.10+#2)}}; \def\yyw{{(\yyp+\la*\lt*#3)}}
	\draw[<-,thick,black!50!blue,align=center]
	(\myx(#1),\yyp) -- (\myx(#1),\yyw)
	node[above,black!80!blue] {#4};
	\fi}

 axis
\draw[<-,thick] (-\w*0.1,0) -- (\w*1.06,0)
node[right=4pt,below=6pt] {$f_a$ [GeV]};

\foreach \tick in {0,1,...,\n}{
	\def\x{{\tick*\w/\n}}
	\def\dec{\the\numexpr \nOne-\tick \relax}
	\pgfmathparse{Mod(\tick-\noffset,\nskip)==0?1:0}
	\ifnum\pgfmathresult>0
	\draw[thick] (\x,\lt) -- (\x,-\lt) 
	node[below] {$10^{\dec}$}; 
	\else
	\draw[thick] (\x,\ls) -- (\x,-\ls); 
	\fi
}
\end{tikzpicture}

These production mechanisms are summarized in the diagram above.
Note that this is specifically for the QCD axion---axion-like particles such as those predicted from string theory are subject to different constraints.

The phenomenology of axions is also determined by whether or not the Peccei-Quinn symmetry is broken before or after the end of inflation.
When the symmetry is broken, the axion field will randomly settle into different potential minima, with causally disconnected patches of the universe taking on different values and sometimes forming topological defects.
If this occurs after inflation, then as the universe expands, modes will reenter the horizon and the universe will contain many patches with different initial values for the axion field, whereas if the breaking occurs before, then the patches will be inflated such that the axion field only takes on one value throughout the universe.

\subsection{Primordial black holes}

Primordial black holes are black holes that formed in the very early universe.
They are a particularly interesting candidate since new physics is not necessarily required to explain their existence but only their production, and we know that such new physics may already exist at inflationary scales.

\begin{figure}
	\centering
	\includegraphics[width=0.8\textwidth]{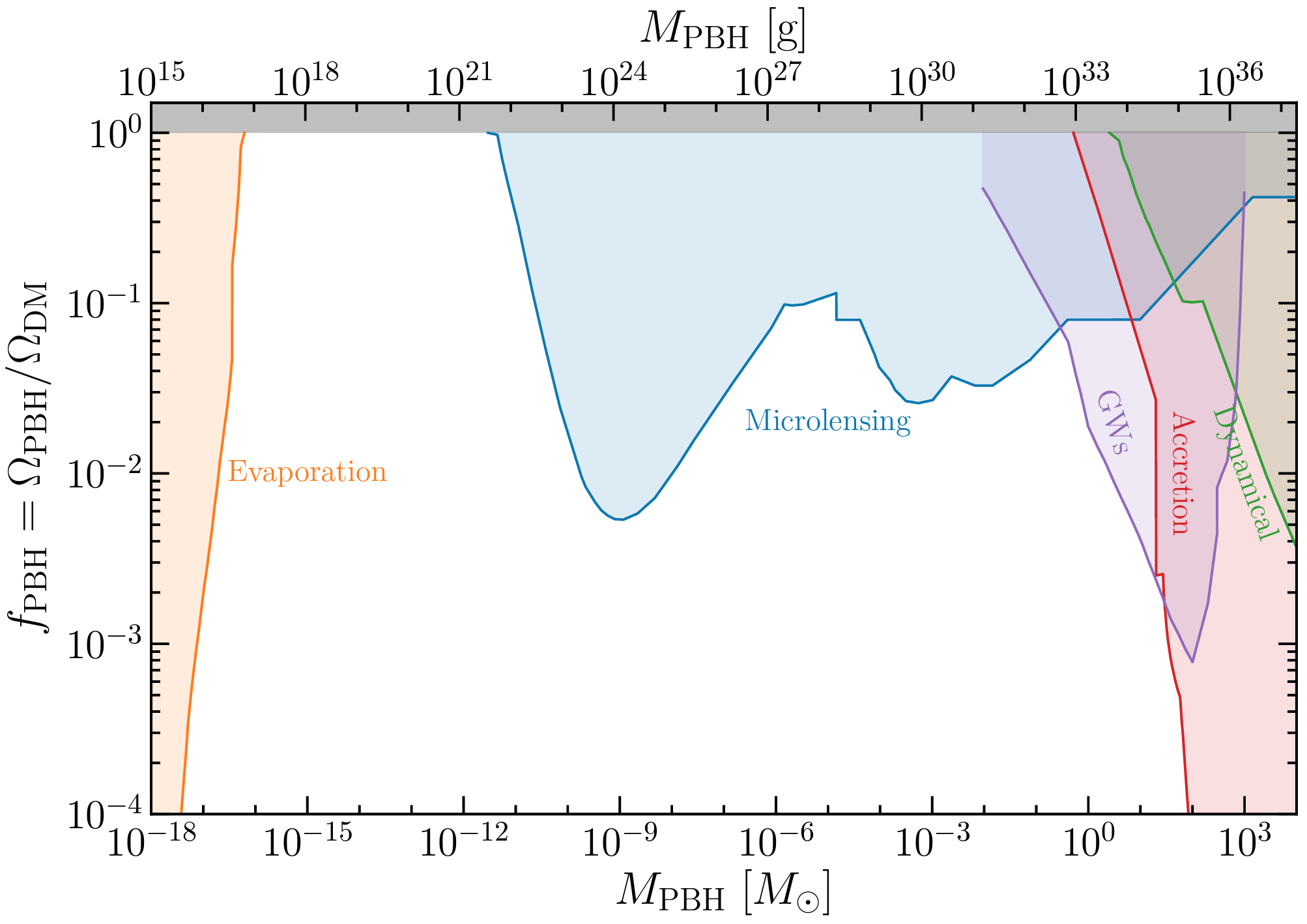}
	\caption{
		Constraints on the abundance of primordial black holes, reproduced from Ref.~\cite{Green:2020jor}.
	}
	\label{fig:PBH_constraints}
\end{figure}
At the moment, the only mass range in which primordial black holes remain a viable dark matter candidate is the asteroid mass range corresponding to about $10^{-17} M_\odot \leq M_\mathrm{PBH} \leq 10^{-12} M_\odot$ or $10^{17} \,\mathrm{g} \leq M_\mathrm{PBH} \leq 10^{22} \,\mathrm{g}$.
In other ranges, strong constraints can be placed on $f_\mathrm{PBH}$, the fraction of dark matter that could be made up of primordial black holes, and I list some existing limits below from the lightest mass PBHs to the heaviest:
\begin{itemize}
	\item \textit{Evaporation}: Black holes evaporate through the emission of Hawking radiation~\cite{Hawking:1975vcx}, with the lifetime of the black hole scaling as $\tau \sim M^3$.
	For primordial black holes, this implies that if their mass is less than $M \lesssim 5 \times 10^{14}$ g, then they would have evaporated away by the present day.
	At masses slightly larger than this, the abundance of black holes can be strongly constrained by e.g. their $\gamma$-ray emission~\cite{Carr:2009jm}, their contribution to the positron flux~\cite{Boudaud:2018hqb}, and the 511 keV line from positron annihilation~\cite{DeRocco:2019fjq,Laha:2019ssq}.
	\footnote{Recently, there has been new discussion about whether the standard Hawking emission formula can be trusted all the way to the total evaporation of a black hole, since the formula is a semi-classical result~\cite{Dvali:2020wft,Alexandre:2024nuo,Thoss:2024hsr}.}
	
	\item \textit{Lensing}: If the black holes are massive enough, they can significantly lens luminous sources. 
	For example, constraints on PBHs have been set by looking at microlensing of stars, quasars, typa 1A supernovae, and strong lensing of fast radio bursts~\cite{Green:2020jor}.
	
	Previously, it was proposed that lensing could cause detectable interference between the multiple lensed images of a $\gamma$-ray burst~\cite{1992ApJ...386L...5G}, and the non-detection of of this effect was used to set constraints on PBHs in the mass range of $5\times10^{17} - 10^{20}$ g~\cite{2012PhRvD..86d3001B}.
	However, it was later shown that $\gamma$-ray bursts cannot be modeled as point sources~\cite{Katz:2018zrn} and wave-optics effects also need to be taken into account~\cite{Ulmer:1994ij}.
	Hence, femtolensing cannot be used to set constraints and this ``asteroid-mass" window for PBHs remains open.
	
	\item \textit{Gravitational waves}: Binary mergers of primordial black holes may give rise to gravitational waves that are detectable in existing observatories.
	LIGO-Virgo has constrained $f_\mathrm{PBH} < \mathcal{O} (10^{-3})$ in the mass range $10 M_\odot \leq M_\mathrm{PBH} \leq 300 M_\odot$~\cite{Sasaki:2016jop,Ali-Haimoud:2017rtz,Kavanagh:2018ggo}, and $f_\mathrm{PBH} < \mathcal{O} (10^{-1})$ down to $M_\mathrm{PBH} \sim 0.2 M_\odot$~\cite{LIGOScientific:2019kan}, although there are uncertainties from the impact of clustering and binary survival.
	
	\item \textit{Dynamical}: There exist a variety of constraints from various dynamical effects of primordial black holes disrupting astronomical systems, including heating and expanding ultra-faint dwarf galaxies~\cite{Brandt:2016aco}, as well as destroying wide stellar binaries~\cite{2014ApJ...790..159M}. Collectively, these limits rule out PBHs as dark matter from $M_\mathrm{PBH} \sim 10 M_\odot$ up to about $M_\mathrm{PBH} \sim 10^4 M_\odot$.
	
	\item \textit{Accretion}: If large black holes exist in the Milky Way, the can accrete interstellar gas and shine brightly in X-ray and radio frequencies.
	This has been used to constrain $f_\mathrm{PBH}$ between black holes masses of $30 M_\odot$ up to $10^7 M_\odot$.
	
	\item \textit{Early universe constraints}: The large perturbations that give rise to PBHs can dissipate and give rise to CMB spectral distortions~\cite{Kohri:2014lza}, as well as modify the light element abundances from BBN~\cite{Inomata:2016uip}.
\end{itemize}
These limits are also compiled in Fig.~\ref{fig:PBH_constraints}, which is reproduced from Ref.~\cite{Green:2020jor}.
Additional constraints at even higher black holes masses are listed in Ref.~\cite{Carr:2020xqk}.

In the next two sections, I will review how to produce the correct abundance of PBHs, particularly from models of inflation.

\subsubsection{Mass and abundance}

Consider a mode of wavenumber $k$ entering the horizon at some time $t \approx 1/H$.
The local overdensity is given by $\delta (x) = \frac{\rho(x) - \overline{\rho}}{\overline{\rho}}$, and if this value exceeds some critical value $\delta_c$, then the density perturbation can collapse to form a black hole~\cite{1975ApJ...201....1C}.
The mass of the resulting black hole will be approximately given by the mass inside the horizon at the horizon crossing time,
\begin{equation}
	M_\mathrm{PBH} \sim M_H = \frac{4}{3} \pi R_H^3 \rho = \frac{4}{3} \pi t^3 \times \frac{3 H^2}{8 \pi G} (1+\delta) = \frac{t}{2 G} (1+\delta) .
\end{equation}
Dropping the $\mathcal{O} (1)$ factors, we find that
\begin{equation}
	M_\mathrm{PBH} \sim 10^{15} \,\mathrm{g} \frac{t}{10^{-23} \,\mathrm{s}} .
\end{equation}
Hence, a PBH which forms around the time of the QCD phase transition ($t \sim 10^{-5}$ s) will be close to a solar mass ($M_\odot \sim 2 \times 10^{30}$ kg).

More precisely, since gravitational collapse into a PBH is a critical phenomenon controlled by the overdensity, then observables obey the scaling relation $\kappa (\delta - \delta_c)^\gamma$, which for the black hole mass we can write as
\begin{equation}
	M_\mathrm{PBH} = \kappa M_H (\delta - \delta_c)^\gamma
\end{equation}
such that $\kappa$ and $\gamma$ are dimensionless constants that depend on the shape of the initial perturbation and background equation of state~\cite{Choptuik:1992jv,Niemeyer:1997mt,Niemeyer:1999ak}.
For example, for PBHs formed from wine-bottle-shaped perturbations during radiation domination, $\gamma = 0.357$ and $\kappa = 4.02$~\cite{Musco:2008hv}.
Calculations of the collapse threshold depends on the shape of the density perturbation but typically yield values around $\delta_c \sim 0.5$, and although the mapping between density and curvature perturbations $\mathcal{R}$ is non-linear and also depends on the shape of overdensities, this threshold roughly corresponds to $\Delta_\mathcal{R}^2 (k) \geq 10^{-3}$ to form PBHs at a particular scale~\cite{Young:2019yug}.

Assuming a monochromatic mass distribution where all PBHs form with the same mass, and the PBH mass remains constant (i.e. there is not significant evaporation or accretion), then the fraction of the universe's energy density in PBHs is given by 
\begin{equation}
	\Omega_\mathrm{PBH} = \frac{\rho_\mathrm{PBH}}{\rho_c} = \frac{3 H^2 M_\mathrm{PBH}}{8 \pi G} n_\mathrm{PBH, 0}
\end{equation}
where $\rho_c$ is the critical energy density and $n_\mathrm{PBH, 0}$ is the number density of black holes today.
Recall that the entropy density of the universe is given by~\cite{Rubakov:2017xzr}
\begin{equation}
	s (t) = \frac{2\pi}{45} g_{*S} (t) T^3 (t),
\end{equation}
where $g_{*S}$ is the effective number of entropy degrees of freedom.
Since both $n_\mathrm{PBH, 0}$ and the entropy density $s$ scale with the expansion of the universe as $a^{-3}$, then $n_\mathrm{PBH, 0}/s$ is constant and we can rewrite the abundance of PBHs as 
\begin{equation}
	\Omega_\mathrm{PBH} = \frac{3 H^2 M_\mathrm{PBH}}{8 \pi G} \times s(t_0) \frac{n_\mathrm{PBH} (t_\mathrm{PBH})}{s (t_\mathrm{PBH})} 
	= \frac{3 H^2 M_\mathrm{PBH}}{8 \pi G} \frac{3.38 T_\mathrm{CMB}^3}{g_{*S} (t_\mathrm{PBH}) T^3 (t_\mathrm{PBH})} n_\mathrm{PBH} (t_\mathrm{PBH}) ,
\end{equation}
where we have used $t_\mathrm{PBH}$ to denote the time of formation.

The only remaining unknown is the number density of black holes at the time of their formation.
This can be estimated using the Press-Schechter formalism as~\cite{Press:1973iz,1991ApJ...379..440B}
\begin{equation}
	\frac{M_\mathrm{PBH} n_\mathrm{PBH} (t_\mathrm{PBH})}{\rho_c (t_\mathrm{PBH})} = 2 \int_{\delta_c}^{\infty} \frac{M_\mathrm{PBH}}{M_H} P (\delta (R)) \,d \delta(R) ,
\end{equation}
where $\delta(R)$ is the overdensity field smoothed over a radius $R$ and $P(\delta)$ is the probability distribution of the fluctuations.
However, the results are strongly dependent on the choice of window function used to smooth the density field.

\subsubsection{Formation from inflation}

In the previous section, I stated that in order to collapse and form a black hole, there must be density perturbations large enough such that either $\delta > \delta_c \sim 0.5$ or $\Delta_\mathcal{R}^2 (k) \geq 10^{-3}$~\cite{Young:2019yug}.
Assuming that the initial curvature power spectrum is nearly scale-invariant and therefore of the form
\begin{equation}
	\Delta_\mathcal{R}^2 (k) = \frac{k^3}{2\pi^2} P_\mathcal{R} (k) = A_s \left( \frac{k}{k_*} \right)^{n_s - 1} ,
\end{equation}
then at the pivot scale $k_* = 0.05$ Mpc$^{-1}$, \textit{Planck} 2018 constrains the amplitude of the initial curvature power spectrum to be $A_s \approx 2 \times 10^{-9}$ with a spectral index of $n_s \approx 0.97$~\cite{Planck2018}.
Hence, the curvature spectrum must be enhanced by \textit{six orders of magnitude} from what we would expect in order to generate PBHs.

Given that this prediction of a nearly scale-invariant primordial spectrum comes from vanilla slow-roll inflation, many have looked into modifying inflationary dynamics to produce small-scale enhancements to the primordial spectrum.
Below, I will briefly review how this occurs in single-field inflation.

Assuming that the universe is dominated by a new scalar field, which we call the \textit{inflaton}, then its equation of motion is given by
\begin{equation}
	\ddot{\phi} + 3 H \dot{\phi} + V' (\phi) = 0
\end{equation}
and the Friedmann equation becomes
\begin{equation}
	H^2 = \frac{8\pi G}{3} \left[ \frac{1}{2} \dot{\phi}^2 + V(\phi) \right] .
\end{equation}
The slow-roll parameters are defined as
\begin{equation}
	\epsilon = - \frac{\dot{H}}{H^2}, \qquad \eta = \epsilon - \frac{\ddot{\phi}}{H \dot{\phi}} .
\end{equation}

When the potential energy of the inflation field dominates over the kinetic energy, $\frac12 \dot{\phi}^2 \ll V(\phi)$, and the field is also accelerating slowly, $\vert \ddot{\phi} \vert \ll H \vert \dot{\phi} \vert$, then we say the universe is undergoing \textit{slow-roll} inflation, and the equation of motion is given by
\begin{equation}
	3H \dot{\phi} \simeq - V(\phi) .
\end{equation}
Under the slow-roll conditions, the parameters can be rewritten as
\begin{equation}
	\epsilon = \frac{1}{16 \pi G} \left( \frac{V'}{V}  \right)^2, \qquad \eta = \frac{1}{8\pi G} \frac{V''}{V} .
\end{equation}
Hence, an equivalent way to meet the conditions for slow-roll inflation is for both slow-roll parameters to be less than one.

During slow-roll inflation, the curvature power spectrum can be written as
\begin{equation}
	\Delta_\mathcal{R}^2 (k) = \frac{H^2}{8 \pi^2 M_\mathrm{Pl}^2 \epsilon} .
\end{equation}
From this expression, we can see that the power spectrum may be enhanced by many orders of magnitude if there exists a mechanism that exponentially suppresses $\epsilon$.

This could be caused by a slightly different phase of inflation called \textit{ultra-slow-roll} (USR). 
USR can occur if the inflaton potential suddenly becomes nearly flat, $V'(\phi) \sim 0$, such that the equation of motion instead becomes
\begin{equation}
	\ddot{\phi} + 3H \dot{\phi} \simeq 0.
\end{equation}
Then $\epsilon$ becomes exponentially small due to its dependence on $V'(\phi)$ and the second slow-roll parameter is given by~\cite{Kinney:2005vj}
\begin{equation}
	\eta \sim 3 .
\end{equation}

\begin{figure}
	\centering
	\includegraphics[width=0.9\textwidth]{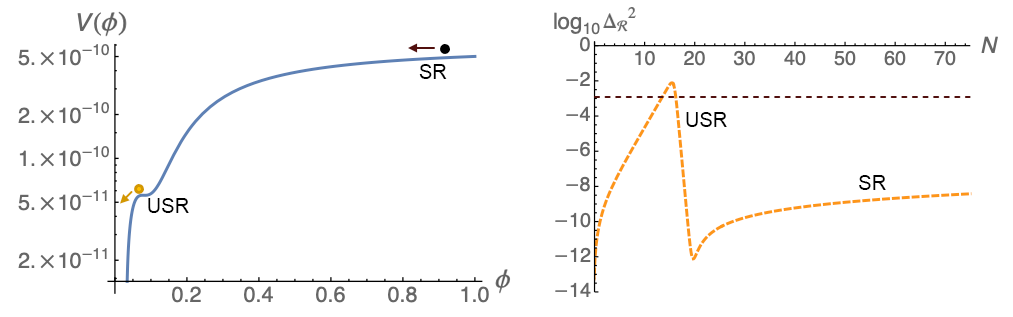}
	\caption{
		Example of a single-field inflation potential (\textit{left}) and the resulting power spectrum (\textit{right}) as a function of $e$-folds prior to the end of inflation.
	}
	\label{fig:SF_potential}
\end{figure}
Fig.~\ref{fig:SF_potential} shows an example of an inflaton potential that could give rise to a phase of USR inflation.
At large field values corresponding to large length scales that can be probed by CMB experiments (labeled ``SR" in Fig.~\ref{fig:SF_potential}), the inflaton undergoes the usual slow-roll inflation and generates a relatively scale-invariant primordial power spectrum.
As the inflaton rolls towards the global minimum, it falls into a shallow local minimum followed by a shallow local maximum (labeled ``USR" in Fig.~\ref{fig:SF_potential}).
The inflaton will undergo USR as it traverses this feature, thus generating a large enhancement in the power spectrum shown in the right panel, and crossing the $\Delta_\mathcal{R}^2 \geq 10^{-3}$ threshold needed to form PBHs.
Note that the presence of the USR feature slightly distorts the slow-roll plateau and in certain cases can pull $n_s (k_*)$ out of compliance with \textit{Planck} constraints.

In Section~\ref{sec:PBHs}, I will showcase one multifield inflation model with two scalar fields nonminimally coupled to gravity that can produce PBHs through a phase of ultra-slow-roll inflation while remaining in compliance with limits from \textit{Planck}.
In Section~\ref{sec:PBH_MCMC}, I will use Markov Chain Monte Carlo (MCMC) methods to show that this is possible with a range of different parameters in our model and also demonstrate that future gravitational wave observatories may be able to set new constraints on these models.

\section{Future probes and methods: \textit{Where are we going?}}
\label{sec:future_probes}

The next few decades will yield a wealth of new datasets, such as improved measurements of the CMB with the Simons Observatory~\cite{SimonsObservatory:2018koc} and CMB-S4~\cite{CMB-S4:2016ple,Abazajian:2019eic}, a new galaxy survey using the Vera C. Rubin Observatory~\cite{2019ApJ...873..111I}, observations of gamma-rays with unprecedented accuracy using the Cherenkov Telescope Array (CTA)~\cite{2019scta.book.....C}, multimessenger physics with IceCube Gen-2~\cite{IceCube-Gen2:2020qha}, and gravitational waves at new frequencies using instruments like the Laser Interferometer Space Antenna (LISA)~\cite{LISA:2017pwj}.
In particular, we may get our first look into the epoch of reionization and cosmic dawn from current 21\,cm radio interferometers such as the Hydrogen Epoch of Reionization Array (HERA)~\cite{DeBoer:2016tnn,HERA:2021bsv}, as well as planned telescopes like the Square Kilometer Array (SKA)~\cite{Weltman:2018zrl}.
21\,cm observations also hold great promise for cosmology because they could be used to map nearly the entire volume of our universe, although this is a more futuristic goal~\cite{2012RPPh...75h6901P}.
For the rest of this section, I will discuss predictions for the 21\,cm signal from the early universe and potential directions for extracting cosmology information from future measurements.

\subsubsection{Introduction to 21\,cm cosmology}

The 21\,cm line refers to the radiation corresponding to transitions between the two hyperfine levels of ground state neutral hydrogen.
The lifetime for hydrogen in the spin-1 state is about 10 million years, corresponding to a deexcitation rate of about $2.9 \times 10^{-15}$ s$^{-1}$. 
While the rate at which hydrogen decays from the excited triplet state to the singlet state may seem very low, the signal is still measurable simply due to the vast amount of hydrogen in the universe. 

The specific intensity $I_\nu = dE/dA d\Omega dt d\nu$ of radiation from a hydrogen cloud is often recast in terms of a ``brightness temperature" using the Rayleigh-Jeans law, $T_b = I_\nu / 2 \nu^2$. 
This facilitates comparison to other temperatures such as the CMB temperature $T_\mathrm{CMB}$, and a quantity known as the spin temperature, which is defined as
\begin{equation}
	\frac{n_1}{n_0} = 3 \exp \left( - \frac{T_*}{T_\mathrm{spin}} \right) ,
	\label{eqn:spintemp}
\end{equation}
where $n_1$ and $n_0$ are the number densities of the spin-1 triplet and spin-0 singlet states of ground state hydrogen and $T_* = 0.0681 \mathrm{K}$ is the temperature corresponding to the 21 cm wavelength. 
In other words, the spin temperature characterizes the relative occupation of the two energy levels.

The radiative transfer equation is given by
\begin{equation}
	\frac{d I_\nu}{d \tau_\nu} = - I_\nu + B_\nu (T_\mathrm{spin}),
\end{equation}
where $B_\nu$ is the specific intensity of background radiation and $\tau_\nu$ is the optical depth. 
This equation shows that as 21\,cm radiation passes through the cloud, some gets absorbed by the gas, but the cloud also emits radiation at the characteristic temperature $T_\mathrm{spin}$. 
Then, one can show that the brightness temperature is given by
\begin{equation}
	T_b (\nu) = T_{\mathrm{spin}} (1 - e^{-\tau_\nu}) + T_{CMB} (\nu) e^{-\tau_\nu}.
\end{equation}

In experiments, one can only detect the 21\,cm signal if there is a difference between the 21 cm brightness and the CMB. 
Thus, the quantity of interest is the \textit{differential} brightness temperature, which is given by
\begin{equation}
	\delta T_b = \frac{T_b(\nu) - T_\mathrm{CMB} (\nu)}{1+z} = \frac{(T_\mathrm{spin} - T_\mathrm{CMB} (\nu)) (1 - e^{-\tau_\nu})}{1+z}.
\end{equation}
If we solve for the optical depth (see ~\cite{Zaroubi_notes} for more detail), we find the full expression for the 21\,cm signal is given by
\begin{equation}
	\delta T_b = 28 (1+\delta) x_\mathrm{HI} \left[ 1 - \frac{T_\mathrm{CMB} (\nu)}{T_\mathrm{spin}} \right] \left(\frac{\Omega_b h^2}{0.0223}\right) \sqrt{\left(\frac{1+z}{10}\right) \left(\frac{0.24}{\Omega_m}\right)} \left(\frac{H(z)/(1+z)}{dv_\parallel / dr_\parallel}\right) \,\mathrm{mK},
	\label{eqn:signal}
\end{equation}
where $\Omega_b$ is the baryon density parameter, $h$ is the Hubble constant in units of $100 \mathrm{km} \mathrm{s}^{-1} \mathrm{Mpc}^{-1}$, $\Omega_m$ is the mass density parameter, and $dv_\parallel / dr_\parallel$ is the gradient of the proper velocity along the line of sight. 
From this expression, we can see the 21 cm signal depends on many cosmological parameters and is thus a complex, but very useful probe of the universe at high redshifts.

We can see that the sign of this expression is determined by the $ 1 - T_\mathrm{CMB} / T_\mathrm{spin}$ factor. When $T_\mathrm{spin} > T_\mathrm{CMB}$, this factor is positive, up to a maximum value of 1, and thus characterizes 21 cm emission. 
When $T_\mathrm{spin} < T_\mathrm{CMB}$, this factor has no lower bound, so the absorption signal can be very deep.

We know the temperature of the CMB very well, but what physics determines the value of $T_\mathrm{spin}$? Processes that contribute to the spontaneous emission of 21 cm radiation must be processes that excite hydrogen to the spin-1 triplet; these are mainly absorption of CMB photons, collisions with other hydrogen atoms, and excitation by Lyman-$\alpha$ and deexcitation (Wouthuysen-Field effect). 
Thus, the spin temperature is given by
\begin{equation}
T_\mathrm{spin} = \frac{T_{CMB} + y_\mathrm{kin} T_\mathrm{gas} + y_\alpha T_\mathrm{gas}}{1 + y_\mathrm{kin} + y_\alpha},
\end{equation}
where $T_\mathrm{gas}$ is the kinetic temperature of the hydrogen gas and the $y$'s characterize the strength of the different effects.

\begin{figure}[h]
	\centering
	\includegraphics[width=0.5\linewidth]{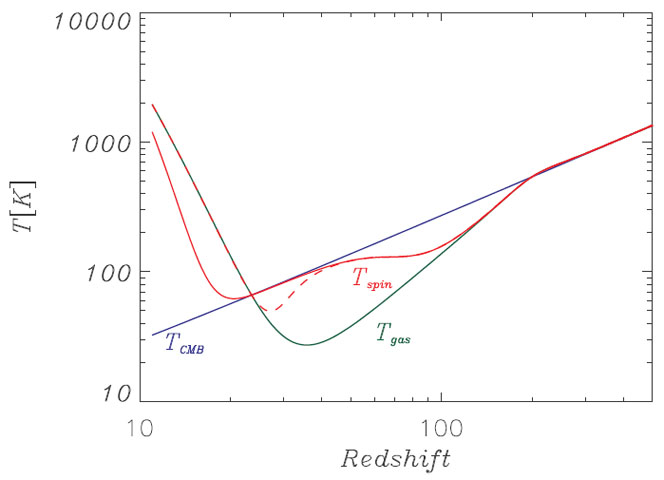}
	\caption{
		The CMB, gas, and spin temperatures as a function of redshift. 
		The red lines show two possible histories for the spin temperature: one where the spin temperature recouples to the gas temperature after it exceeds the CMB temperature (solid), and one where the temperatures recouple earlier and below the CMB temperature (dashed).
		Reproduced from Ref.~\cite{Zaroubi_notes}.
	}
	\label{fig:temperatures}
\end{figure}

To get a sense of what kind of signals we should expect in experiments, we can look at the CMB and gas temperatures over time (See Figure \ref{fig:temperatures}). 
$T_\mathrm{CMB}$ redshifts as $1+z$, and so decreases as time passes. 
The gas is initially coupled to the CMB by Compton scattering, but after decoupling begins to cool as $(1+z)^2$, which is faster than the CMB. 
Around $z \approx 30$, the first stars and black holes begin to form, thus reheating the gas until it is hotter than the CMB. 
The spin temperature is initially coupled to the gas, until the density of the gas decreases to the point where collisional excitations are no longer efficient, at which point the spin temperature returns to follow the CMB. 
As the gas begins to reheat and emit Lyman-$\alpha$ photons, the spin temperature once again recouples to the gas.

\begin{figure}[h]
	\centering
	\includegraphics[width=\linewidth]{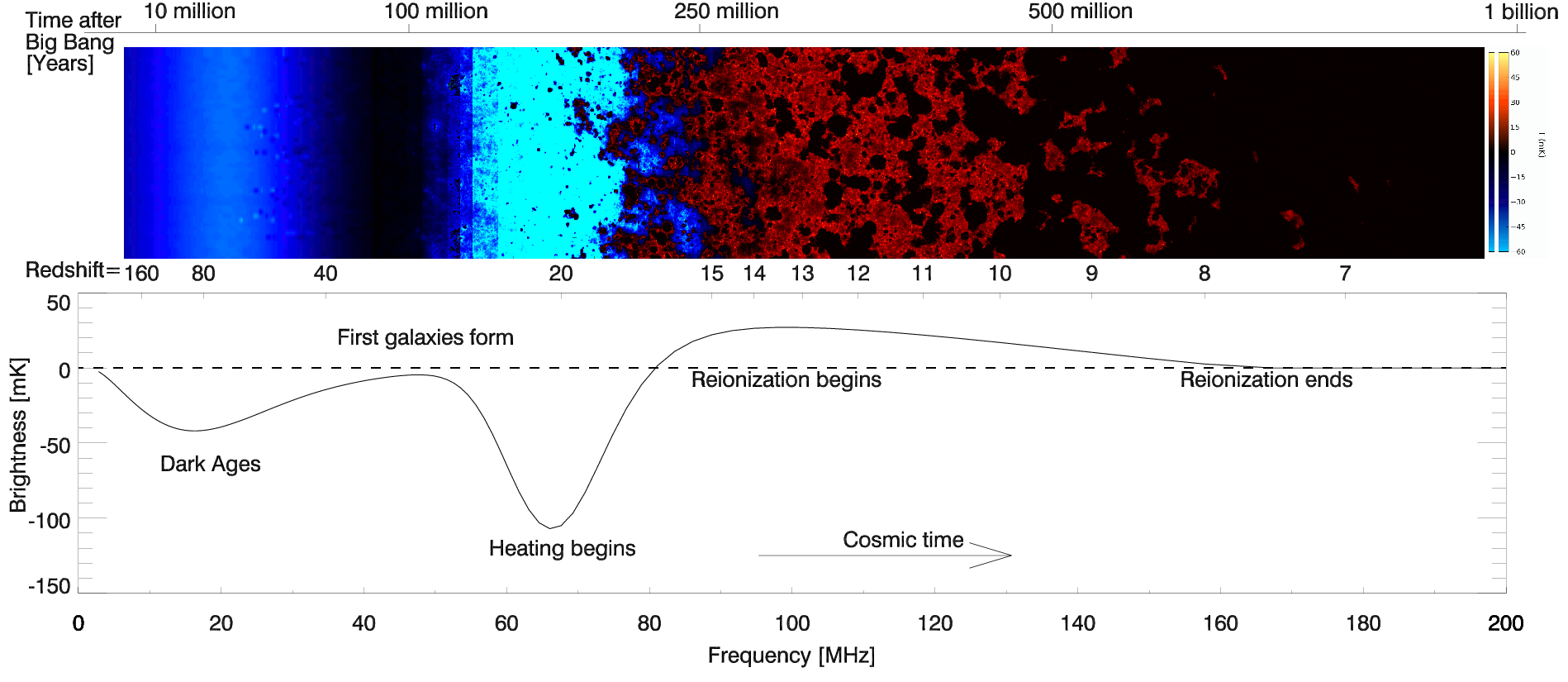}
	\caption{
		The expected brightness temperature global signal from a $\Lambda$-CDM universe. 
		Reproduced from Ref.~\cite{Pritchard:2011xb}.
	}
	\label{fig:brightness}
\end{figure}

Using the fact that $\delta T_b \propto T_\mathrm{spin} - T_\mathrm{CMB}$, we can predict what a global (sky-averaged signal) should look like. 
There will be two troughs: one from the dark ages and one from reheating. 
The brightness temperature may also be slightly positive during reionization, when the gas becomes hotter than the CMB. 
Figure \ref{fig:brightness} plots the expected signal from a standard cosmology as a function of redshift/time/frequency.

In Eqn.~\eqref{eqn:signal}, both the overdensity $\delta$ and neutral fraction $x_\mathrm{HI}$ vary throughout space; hence, the 21\,cm signal itself has spatial fluctuations.
Intensity mapping experiments such as HERA~\cite{DeBoer:2016tnn,HERA:2021bsv} and SKA~\cite{Weltman:2018zrl} aim to observe these fluctuations, which will not only sharpen our understanding of how the universe reionizes, but also give us an indirect measurement of the matter power spectrum at new wavenumbers.

\subsubsection{Cosmological perturbation theory}

Our theoretical understanding of the 21\,cm signal has traditionally been driven by simulations such as \texttt{THESAN}~\cite{thesan1} and semi-numerical methods such as \texttt{21cmFAST}~\cite{2020JOSS....5.2582M}.
The prevailing view has been that perturbative methods will not work for the epoch of reionization, given that the process is very patchy and nonlinear, hence, most methods for extracting cosmological information from 21\,cm measurements are computationally intensive.
However, recent works have shown that the effective field theory methods which have been employed with great success in galaxy surveys are applicable on redshifts and length scales that are within the sensitivity of experiments like HERA.
In this section, we will give a brief introduction to standard perturbation theory and effective field theory; some of the following material is drawn from Refs.~\cite{Bertolini:2016bmt,Ivanov:2022mrd}.
Here and throughout this dissertation, we will consider a spatially flat, isotropic, and homogeneous background spacetime.

Given the phase space distribution of an ensemble of particles, $f(t, \mathbf{x}, \mathbf{p})$, i.e. the probability distribution for a particle to be at comoving coordinates $\mathbf{x}$ with comoving momentum $\mathbf{p}$ at time $t$, we can simplify this information by taking moments of the distribution to obtain
\begin{itemize}
	\item the density, $\rho (t, \mathbf{x}) = m \int \dbar^3 \mathbf{p} \, f(t, \mathbf{x}, \mathbf{p})$,
	
	\item the momentum density, $\pi^i (t, \mathbf{x}) = \int \dbar^3 \mathbf{p} \, p^i f(t, \mathbf{x}, \mathbf{p})$,
	
	\item and the velocity dispersion tensor, $\sigma^{ij} (t, \mathbf{x}) = \frac{1}{m^2} \int \dbar^3 \mathbf{p} \, p^i p^j f(t, \mathbf{x}, \mathbf{p}) - \frac{\pi^i \pi^j}{m \rho}$,
\end{itemize}
where $m$ denotes the mass of the particles and $\dbar \mathbf{p} = d^3 \mathbf{p} / (2\pi)^3$.

One can show from Liouville's theorem that the equations of motion for these particles are given by the continuity equation,
\begin{equation}
	0 = \frac{d \rho}{d \tau} + \frac{1}{a} \partial_i \pi^i ,
\end{equation}
and Euler's equation,
\begin{equation}
	0 = \frac{d \pi_i}{d \tau} + \frac{1}{a} \partial^j \frac{\pi_i \pi_j}{\rho} + a \rho \partial_i \phi .
\end{equation}
In these expressions, we have switched to using the conformal time which is given by $dt = a d\tau$, $\partial_i$ denotes the partial derivative $\partial / \partial x^i$, and $\phi$ is the gravitational potential.

Additionally, it is more common to work in terms of the overdensity $\delta$ and velocity $v^i$ instead of the momentum density, since velocities are more observable, and the two quantities are related by $\pi^i = a \rho v^i$.
Then the eqns. of motion become
\begin{align}
	0 &= \partial_\tau \delta + \partial_i (v^i (1 + \delta)) , \\
	0 &= \partial_\tau v_i + \mathcal{H} v_i + v_j \partial^j v_i + \partial_i \phi ,
\end{align}
where $\mathcal{H} = \frac{1}{a} \frac{da}{d\tau}$ is the conformal Hubble parameter
The velocity can be further decomposed into its divergence, $\theta = \partial_i v^i$, and its curl or vorticity, $\omega^i = \epsilon^{ijk} \partial_j v_k$, with $\epsilon^{ijk}$ denoting the Levi-Civita tensor.
Then the continuity equation can be rewritten as
\begin{equation}
	\partial_\tau \delta + \theta = - \left( \delta \theta + \frac{\partial^i}{\partial^2} \theta \partial_i \delta \right) + \epsilon^{ijk} \frac{\partial^j}{\partial^2} \omega_k \partial_i \delta 
	\label{eqn:delta_ev}
\end{equation}
and the Euler equation can be decomposed into evolution equations for $\theta$ and $\omega$,
\begin{align}
	\partial_\tau \theta + \mathcal{H} \theta + \partial^2 \phi &=  -\left( \frac{\partial_i \partial_j}{\partial^2} \theta \frac{\partial^j \partial^i}{\partial^2} \theta + \frac{\partial_j}{\partial^2} \theta \partial^j \theta \right) \n
	&\qquad + \left( 2\frac{\partial_i \partial_j}{\partial^2} \theta \epsilon^{imn} \frac{\partial^j \partial_m}{\partial^2} \omega_n + {\epsilon_j}^{kl} \frac{\partial_k}{\partial^2} \omega_l \partial^j \theta \right) \n
	&\qquad - {\epsilon_j}^{kl} \frac{\partial_i \partial_k}{\partial^2} \omega_l \epsilon^{imn} \frac{\partial^j \partial_m}{\partial^2} \omega_n , 
	\label{eqn:theta_ev} \\
	\partial_\tau \omega^i + \mathcal{H} \omega^i &= - \epsilon^{ijk} \partial_j (v_l \partial^l v_k) .
	\label{eqn:omega_ev}
\end{align}

The linearized version of Eqn.~\eqref{eqn:omega_ev} is simply given by $0 = \partial_\tau \omega^i + \mathcal{H} \omega^i$ and has the solution
\begin{equation}
	\omega^i \propto \frac1a .
\end{equation}
Hence, the vorticity quickly decays away in the early universe and is often dropped from the equations of motion.

We can take combinations of Eqns.~\eqref{eqn:delta_ev} and \eqref{eqn:theta_ev} in Fourier space to diagonalize them in terms of $\delta$ and $\theta$, obtaining
\begin{align}
	\left( a^2 \partial_a^2 + \frac{3}{2} a \partial_a - \frac{3}{2} \right) \delta &= \frac{1 + a \partial_a}{\mathcal{H}} I_1 - \frac{1}{\mathcal{H}^2} I_2 , \\
	\left( a^2 \partial_a^2 + \frac{5}{2} a \partial_a -1 \right) \theta &= \frac{1 + a \partial_a}{\mathcal{H}} I_2 - \frac{3}{2} I_1 ,
\end{align}
where $I_1$ and $I_2$ stand for the integrals
\begin{align}
	I_1 &= - \int \dbar^3 \mathbf{q} \, \alpha(\mathbf{q}, \mathbf{k}-\mathbf{q}) \theta(\mathbf{q}) \delta(\mathbf{k}-\mathbf{q}) , \\
	I_2 &= - \int \dbar^3 \mathbf{q} \, \beta(\mathbf{q}, \mathbf{k}-\mathbf{q}) \theta(\mathbf{q}) \theta(\mathbf{k}-\mathbf{q}) ,
\end{align}
with the kernels given by
\begin{equation}
	\alpha(\mathbf{q}, \mathbf{p}) = 1 + \frac{\mathbf{q} \cdot \mathbf{p}}{q^2} , 
	\qquad \beta(\mathbf{q}, \mathbf{p}) = \frac{(\mathbf{q} \cdot \mathbf{p}) (\mathbf{q} + \mathbf{p})^2}{2 q^2 p^2} .
\end{equation}

We now impose the series ansatz
\begin{equation}
	\delta (\tau, \mathbf{k}) = \sum_n D^n (\tau) \delta_n (\mathbf{k}) ,
	\qquad \theta (\tau, \mathbf{k}) = \sum_n D^n (\tau) \theta_n (\mathbf{k}) ,
\end{equation}
where, $D(\tau)$ is the time-dependent growth factor that can be solved for in linear theory, and the $n$-th order perturbative solutions are given by
\begin{align}
	\delta_n (\mathbf{k}) &= \int \dbar^3 \mathbf{q}_1 \cdots \int \dbar^3 \mathbf{q}_n \, (2\pi)^3 \delta^D (\mathbf{k} - \sum_i \mathbf{q}_i) F_n (\mathbf{q}_1, ..., \mathbf{q}_n) \delta_1 (\mathbf{q}_1) \cdots \delta_1 (\mathbf{q}_n) , \\
	\theta_n (\mathbf{k}) &= \int \dbar^3 \mathbf{q}_1 \cdots \int \dbar^3 \mathbf{q}_n \, (2\pi)^3 \delta^D (\mathbf{k} - \sum_i \mathbf{q}_i) G_n (\mathbf{q}_1, ..., \mathbf{q}_n) \delta_1 (\mathbf{q}_1) \cdots \delta_1 (\mathbf{q}_n) .
\end{align}
In these equations, $\delta_1$ represents the initial density fluctuation.
Requiring that these ansatzes solve the equations of motion then yields a set of recursive relations for the $F_n$ and $G_n$ kernels.
\begin{align}
	F_n (\mathbf{q}_1,...,\mathbf{q}_n) = - \sum_m \frac{G_m(\mathbf{q}_1,...,\mathbf{q}_m)}{2n^2 + n - 3} &\left[ \left( 2n+ 1 \right) \alpha \left(\sum_{i=1}^{m} \mathbf{q}_i, \sum_{j=m+1}^{n} \mathbf{q}_j \right)  F_{n-m} (\mathbf{q}_{m+1},...,\mathbf{q}_n) \right. \n
	&\left.\qquad + 2 \beta \left(\sum_{i=1}^{m} \mathbf{q}_i , \sum_{j=m+1}^{n} \mathbf{q}_j \ \right) G_{n-m} (\mathbf{q}_{m+1},...,\mathbf{q}_n) \right] \\
	G_n (\mathbf{q}_1,...,\mathbf{q}_n) = - \sum_m \frac{G_m(\mathbf{q}_1,...,\mathbf{q}_m)}{2n^2 + n - 3} &\left[ 3\alpha \left(\sum_{i=1}^{m} \mathbf{q}_i, \sum_{j=m+1}^{n} \mathbf{q}_j \right) F_{n-m} (\mathbf{q}_{m+1},...,\mathbf{q}_n) \right. \n
	&\left.\qquad + 2n \beta \left(\sum_{i=1}^{m} \mathbf{q}_i , \sum_{j=m+1}^{n} \mathbf{q}_j \ \right) G_{n-m} (\mathbf{q}_{m+1},...,\mathbf{q}_n) \right]
\end{align}
The first and second order kernels are given by $F_1 = G_1 = 1$ and
\begin{align}
	F_2(\mathbf{q}_1, \mathbf{q}_2) &= \frac{5}{7} + \frac{2}{7} \frac{(\mathbf{q}_1 \cdot \mathbf{q}_2)^2}{q_1^2 q_2^2} + \frac{\mathbf{q}_1 \cdot \mathbf{q}_2}{2} \left( \frac{1}{q_1^2} + \frac{1}{q_2^2} \right), \\
	G_2(\mathbf{q}_1, \mathbf{q}_2) &= \frac{3}{7} + \frac{4}{7} \frac{(\mathbf{q}_1 \cdot \mathbf{q}_2)^2}{q_1^2 q_2^2} + \frac{\mathbf{q}_1 \cdot \mathbf{q}_2}{2} \left( \frac{1}{q_1^2} + \frac{1}{q_2^2} \right).
\end{align}

Just as in quantum field theory, we can calculate observables from $N$-point correlation functions of the fields and introduce a diagrammatic language to represent these calculations.
The rules are as follows:

\begin{enumerate}
	\item Every $\delta_n (\mathbf{k})$ and $\theta_n (\mathbf{k})$ is represented by a vertex with one external leg of wavenumber $\mathbf{k}$ that is coupled to $n$ internal legs representing the factors of the linear density $\delta_1 (\mathbf{q}_i)$ by either the $F_n (\mathbf{q}_1, \dots, \mathbf{q}_n)$ or $G_n (\mathbf{q}_1, \dots, \mathbf{q}_n)$ kernel, depending on which field is involved. 
	In order to conserve momentum, each vertex also carries a factor of $(2\pi)^3 \delta^D \left( \mathbf{k} - \sum_{i=1}^n \mathbf{q}_i \right)$.
	Filled dots represent the density field and open dots represent the velocity field.
	\begin{equation}
	\delta_n (\mathbf{k}) \quad \rightarrow \quad
	\begin{tikzpicture}[baseline=(current bounding box.center)]
	\begin{feynman}
	\vertex (i);
	\vertex [right=1.5cm of i, scale=1.5, dot] (j) {};
	
	\vertex [above right=2cm of j] (a) {};
	\vertex [below=0.5cm of a] (b) {};
	\vertex [below right=2cm of j] (c) {};
	
	\vertex [below=1cm of b, scale=0.3, dot] (l) {};
	\vertex [below=0.2cm of l, scale=0.3, dot] (m) {};
	\vertex [below=0.2cm of m, scale=0.3, dot] (n) {};
	
	\diagram*{
		(i) -- [edge label=\(\mathbf{k}\)] (j) ,
		(j) -- [scalar, edge label=\(\mathbf{q_1}\)] {(a)},
		(j) -- [scalar] {(b)},
		(j) -- [scalar, edge label'=\(\mathbf{q_n}\)] {(c)}
	};
	\end{feynman}
	\end{tikzpicture}
	\quad = \quad (2\pi)^3 \delta^D \left( \mathbf{k} - \sum_{i=1}^n \mathbf{q}_i \right) F_n (\mathbf{q}_1, \dots, \mathbf{q}_n)
	\end{equation}
	
	\begin{equation}
	\theta_n (\mathbf{k}) \quad \rightarrow \quad
	\begin{tikzpicture}[baseline=(current bounding box.center)]
	\begin{feynman}
	\vertex (i);
	\vertex [right=1.5cm of i, scale=1.5, empty dot] (j) {};
	
	\vertex [above right=2cm of j] (a) {};
	\vertex [below=0.5cm of a] (b) {};
	\vertex [below right=2cm of j] (c) {};
	
	\vertex [below=1cm of b, scale=0.3, dot] (l) {};
	\vertex [below=0.2cm of l, scale=0.3, dot] (m) {};
	\vertex [below=0.2cm of m, scale=0.3, dot] (n) {};
	
	\diagram*{
		(i) -- [edge label=\(\mathbf{k}\)] (j) ,
		(j) -- [scalar, edge label=\(\mathbf{q_1}\)] {(a)},
		(j) -- [scalar] {(b)},
		(j) -- [scalar, edge label'=\(\mathbf{q_n}\)] {(c)}
	};
	\end{feynman}
	\end{tikzpicture}
	\quad = \quad (2\pi)^3 \delta^D \left( \mathbf{k} - \sum_{i=1}^n \mathbf{q}_i \right) G_n (\mathbf{q}_1, \dots, \mathbf{q}_n)
	\end{equation}
	
	\item To compute a correlation function, draw all connected diagrams that can be made by contracting the internal $\delta_1$ legs. 
	When using the symmetrized $F_n$ and $G_n$ kernels, permuting the $\delta_1$ legs on each $\delta_n$ vertex will give rise to a symmetry factor of $n!$, and one must also track combinatoric factors from loops to prevent double-counting of diagrams.
	
	\item For each internal leg carrying wavenumber $\mathbf{p}$, write down a factor of $P_L (\mathbf{p})$, the linear matter power spectrum. This is analogous to the propagator of the linear, ``free'' density fields of the internal legs, since the power spectrum is related to the two-point correlation function.
	\begin{equation}
	\begin{tikzpicture}
	\begin{feynman}
	\vertex [dot, scale=1.5] (i) {};
	\vertex [right=2cm of i, empty dot, scale=1.5] (j) {};
	\diagram*{
		(i) --[scalar, momentum=\(\mathbf{p}\)] (j)
	};
	\end{feynman}
	\end{tikzpicture}
	\quad = \quad P_L (\mathbf{p})
	\end{equation}
	The vertices on the ends of the propagator can correspond to both $\delta$, both $\theta$, or one of each; the factor of $P_L (\mathbf{p})$ associated with the propagator is the same in any case, since $\delta^{(1)}$ and $\theta^{(1)}$ are spatially the same up to time-dependent factors.
	
	\item Integrate over internal wavenumbers $\mathbf{q}$ with $\int \dbar^3 q$.
	
	\item We assume momentum is conserved such that we can remove Dirac delta functions over external wavenumbers and their corresponding factors of $2\pi$, i.e. $$\langle \delta_1 (\mathbf{k})  \delta_1 (\mathbf{k'}) \rangle = (2\pi)^3 \delta(\mathbf{k} - \mathbf{k'}) P_L (\mathbf{k}) \quad \rightarrow \quad P_L (\mathbf{k}). $$
\end{enumerate}

For example, the matter power spectrum to lowest order in the perturbative series will be given by the ``tree-level" diagram,
\begin{equation}
	P_\delta^\mathrm{tree} (\mathbf{k}) \quad = \quad
	\begin{tikzpicture}
	\begin{feynman}
	\vertex (i);
	\vertex [right=1cm of i, dot, scale=1.5] (j) {};
	\vertex [right=1.2cm of j, dot, scale=1.5] (k) {};
	\vertex [right=1cm of k] (l) {};
	\diagram*{
		(i) -- (j) --[scalar, momentum=\(\mathbf{k}\)] (k) -- (l)
	};
	\end{feynman}
	\end{tikzpicture}
	\quad = \quad P_L (\mathbf{k}) ,
\end{equation}
and the next order corrections are the ``one-loop" diagrams,
\begin{align}
	P_\delta^\mathrm{1-loop} (\mathbf{k}) \quad &= \quad
	\begin{tikzpicture}[baseline=(i.base)]
	\begin{feynman}
	\vertex (i);
	\vertex [right=1cm of i, dot, scale=1.5] (j) {};
	\vertex [right=1.2cm of j, dot, scale=1.5] (k) {};
	\vertex [right=1cm of k] (l) {};
	\diagram*{
		(i) -- [momentum=\(\mathbf{k}\)] (j) --[scalar, half left, momentum=\(\mathbf{k} - \mathbf{q}\)] (k) --  [momentum=\(\mathbf{k}\)] (l),
		(k) --[scalar, half left, momentum=\(\mathbf{q}\)] (j)
	};
	\end{feynman}
	\end{tikzpicture} 
	+ 2 \times
	\begin{tikzpicture}[baseline=(i.base)]
	\begin{feynman}
	\vertex (i);
	\vertex [right=1cm of i, dot, scale=1.5] (j) {};
	\vertex [right=1.2cm of j, dot, scale=1.5] (k) {};
	\vertex [right=1cm of k] (l) {};
	\diagram*{
		(i) -- (j) --[scalar] (k) --[momentum=\(\mathbf{k}\)] (l);
		j --[scalar, loop, out=135, in=45, min distance=2cm, momentum=\(\mathbf{q}\)] j
	};
	\end{feynman}
	\end{tikzpicture}
	\n
	&= \quad 2 \int \dbar^3 q F_2^2 (\mathbf{q}, \mathbf{k}-\mathbf{q})  P_L (\mathbf{q}) P_L (\mathbf{k}-\mathbf{q}) + 6 P_L(\mathbf{k}) \int \dbar^3 \mathbf{q} \,F_3 (-\mathbf{k}, \mathbf{q}, - \mathbf{q}) P_L(\mathbf{q}) .
\end{align}
Similarly, using cycle notation to denote permutations of the external momenta, the three-point function or bispectrum is given at tree-level by
\footnote{To be explicit, the permutation denoted by $(1\,2\,3)$ means that in a particular expression, one should substitute $k_1$ with $k_2$, and $k_2$ should be substituted by $k_3$, and $k_3$ by $k_1$.}
\begin{align}
B^\mathrm{tree} (\mathbf{k}_1, \mathbf{k}_2, \mathbf{k}_3) \quad &= \quad
\begin{tikzpicture}[baseline=(i.base)]
\begin{feynman}
\vertex (i);
\vertex [right=1.5cm of i, dot, scale=1.5] (j) {};
\vertex [above right=1cm of j, dot, scale=1.5] (k) {};
\vertex [above right=1cm of k] (l) {};
\vertex [below right=1cm of j, dot, scale=1.5] (m) {};
\vertex [below right=1cm of m] (n) {};
\diagram*{
	(i) -- [momentum=\(\mathbf{k}_1\)] (j);
	(l) -- (k) --[scalar, momentum=\(\mathbf{k}_2\)] (j);
	(n) --(m) --[scalar, momentum=\(\mathbf{k}_3\)]  (j)
};
\end{feynman}
\end{tikzpicture}
+ (1\,2\,3) + (1\,3\,2) \n
&= 2 P_L (\mathbf{k}_2) P_L (\mathbf{k}_3) F_2 (\mathbf{k}_2, \mathbf{k}_3) + (1\,2\,3) + (1\,3\,2)
\end{align}
and is corrected at next order by diagrams of the form 
\begin{align}
\begin{tikzpicture}[baseline=(i.base)]
	\begin{feynman}
	\vertex (i);
	\vertex [right=2cm of i] (center);
	\pgfmathsetmacro\vert{2 * sin(2 * pi / 6 r)};
	\pgfmathsetmacro\horz{2 * cos(2 * pi / 6 r)}
	\vertex [above right=\vert cm and \horz cm of center] (o1);
	\vertex [below right=\vert cm and \horz cm of center] (o2);
	\vertex [dot, scale=1.5] (v1) at ($(center)!0.5!(i)$) {};
	\vertex [dot, scale=1.5] (v2) at ($(center)!0.5!(o1)$) {};
	\vertex [dot, scale=1.5] (v3) at ($(center)!0.5!(o2)$) {};
	\diagram*{
		(i) --[momentum=\(\mathbf{k}_1\)] (v1) --[scalar, momentum'=\(\mathbf{q} + \mathbf{k}_1\)] (v3) --[scalar, momentum'=\(\mathbf{q} - \mathbf{k}_3\)] (v2) --[scalar, momentum'=\(\mathbf{q}\)] (v1),
		(o1) --[momentum=\(\mathbf{k}_2\)] (v2),
		(o2) --[momentum=\(\mathbf{k}_3\)] (v3),
	};
	\end{feynman}
\end{tikzpicture}
&= (2!)^3 \int \dbar^3 \mathbf{q} F_2 (\mathbf{q}, -(\mathbf{q} + \mathbf{k}_1)) F_2 (\mathbf{q} + \mathbf{k}_1, - \mathbf{q} + \mathbf{k}_3) F_2 (\mathbf{q} - \mathbf{k}_3, -\mathbf{q}) \n
&\qquad \times P_L(\mathbf{q}) P_L (\mathbf{q} + \mathbf{k}_1) P_L (\mathbf{q} - \mathbf{k}_3) ,
\end{align}
\begin{align}
\begin{tikzpicture}[baseline=(i.base)]
	\begin{feynman}
	\vertex (i);
	\vertex [right=1cm of i, dot, scale=1.5] (ii) {};
	\vertex [right=1cm of ii, dot, scale=1.5] (j) {};
	\vertex [above right=1cm of j, dot, scale=1.5] (k) {};
	\vertex [above right=1cm of k] (l) {};
	\vertex [below right=2cm of j] (m) {};
	\diagram*{
		(i) -- [momentum=\(\mathbf{k}_1\)] (ii) --[scalar, half left, momentum=\(\mathbf{q} + \mathbf{k}_1\)] (j) --[scalar, half left, momentum=\(\mathbf{q}\)] (ii);
		(l) -- (k) --[scalar, momentum=\(\mathbf{k}_2\)] (j) -- (m)
	};
	\end{feynman}
\end{tikzpicture}
&= \frac{2! 3!}{2} P_L(\mathbf{k}_2) \int \dbar^3 \mathbf{q} F_2 (\mathbf{q}, -(\mathbf{q} + \mathbf{k}_1)) F_3 (-\mathbf{q}, \mathbf{q}+ \mathbf{k}_1, \mathbf{k}_2) P_L (\mathbf{q}) P_L (\mathbf{q} + \mathbf{k}_1) \n
&\qquad+ 5\, \mathrm{permutations} ,
\end{align}
\begin{align}
\begin{tikzpicture}[baseline=(i.base)]
\begin{feynman}
\vertex (i);
\vertex [right=1cm of i, dot, scale=1.5] (ii) {};
\vertex [right=1cm of ii, dot, scale=1.5] (j) {};
\vertex [above right=1cm of j, dot, scale=1.5] (k) {};
\vertex [above right=1cm of k] (l) {};
\vertex [below right=2cm of j] (m) {};
\diagram*{
	(i) -- (ii) --[scalar, momentum'=\(\mathbf{k}_1\)] (j);
	ii --[scalar, loop, out=135, in=45, min distance=2cm, momentum=\(\mathbf{q}\)] ii;
	(l) -- (k) --[scalar, momentum=\(\mathbf{k}_2\)] (j) -- (m)
};
\end{feynman}
\end{tikzpicture}
&= \frac{2! 3!}{2} F_2 (\mathbf{k}_1, \mathbf{k}_2) P_L (\mathbf{k}_1) P_L(\mathbf{k}_2) \int \dbar^3 \mathbf{q} F_3 (-\mathbf{k}_1, \mathbf{q}, -\mathbf{q}) P_L(\mathbf{q})  \n
&\qquad + 5\, \mathrm{permutations} ,
\end{align}
\begin{align}
\begin{tikzpicture}[baseline=(i.base)]
\begin{feynman}
\vertex (i);
\vertex [right=1cm of i, dot, scale=1.5] (ii) {};
\vertex [right=1cm of ii, dot, scale=1.5] (j) {};
\vertex [above right=1cm of j, dot, scale=1.5] (k) {};
\vertex [above right=1cm of k] (l) {};
\vertex [below right=2cm of j] (m) {};
\diagram*{
	(i) -- (ii) --[scalar, momentum'=\(\mathbf{k}_1\)] (j);
	j --[scalar, loop, out=80, in=160, min distance=2cm, momentum'=\(\mathbf{q}\)] j;
	(l) -- (k) --[scalar, momentum=\(\mathbf{k}_2\)] (j) -- (m)
};
\end{feynman}
\end{tikzpicture}
&= \frac{4!}{2} P_L(\mathbf{k}_1) P_L(\mathbf{k}_2) \int \dbar^3 \mathbf{q} F_4 (\mathbf{k}_1, \mathbf{k}_2, \mathbf{q}, -\mathbf{q}) P_L(\mathbf{q}) + 2\, \mathrm{permutations} .
\end{align}
This concludes the introduction to \textit{standard perturbation theory}.

At small length scales, or large wavenumbers, perturbations have had more time to evolve and are therefore less perturbative than the large scale modes.
Hence, below some wavenumber $k_\mathrm{NL}$, the density field can no longer be treated perturbatively.
Instead, we would like to work in terms of the smoothed field $\delta_\Lambda (\mathbf{k})$, where the subscript indicates that the field has been smoothed over some characteristic length scale $1/\Lambda$.

However, to make things worse, in the integrals for the non-perturbative contributions to the correlation functions, nonperturbative modes become coupled to modes that we could otherwise treat perturbatively.
Hence, rewriting the higher-point statistics in terms of the smoothed field requires introducing so-called counterterms in order to account for the fact that for two fields $A$ and $B$, $(AB)_\Lambda \neq A_\Lambda B_\Lambda$.
A similar problem arises when considering biased tracers of the density field, e.g. a field that can be written in configuration space as
\begin{equation}
	\delta_\mathrm{tr} = b_1 \delta + b_2 \delta^2 + b_3 \delta^3 + \cdots
\end{equation}
The quadratic bias term convolves perturbative and nonperturbative modes in Fourier space, and therefore also must be regularized and renormalized.
A description of a tracer field and its statistics that has been regularized and renormalized in this manner is referred to as the \textit{effective field theory} for that probe, and for galaxies this is often called the effective field theory of large scale structure, or EFTofLSS for short.

In Chapter~\ref{sec:EFTof21cm}, I develop a perturbative description of the 21\,cm signal that includes renormalized bias and redshift space distortions.
This work is a step towards bringing analytic methods like effective field theory into the field of 21\,cm cosmology.

%
%
%
%
%
%
%
%
%
%
%

\chapter{Probes of Exotic Energy Injection}
\label{sec:DH}

Models for new or exotic physics generically feature interactions with Standard Model particles that can inject energetic particles into our universe, which then cool and deposit energy into astrophysical and cosmological observables.
We refer to this phenomenon as \textit{exotic energy injection}.
For example, weakly interacting massive particles (WIMPs) can annihilate into Standard Model particles through the electroweak force.
UV models for the QCD axion predict that the axion can decay into two photons.
Primordial black holes can also have an impact on the early universe by injecting energy through evaporation or accretion.

Throughout this dissertation, I will focus on injections of $e^+ e^-$ pairs or photons. 
Injections of heavier final states deposit much of their energy into electron, positron, and photon secondaries as they cool, and hence constraints on these channels can be estimated by taking appropriate linear combinations of the $e^+ e^-$ and photon results.

In this chapter, I will describe my work on studying signatures of exotic energy injection as well as upgrading the tools we use to predict these signals.
In Section~\ref{sec:Lya}, I will discuss constraints on dark matter energy injection from Lyman-$\alpha$ forest observations.
This work is based on Ref.~\cite{Liu:2020wqz} and was done in collaboration with Hongwan Liu, Greg Ridgway, and Tracy Slatyer.
While this work was led by Greg Ridgway, I contributed in setting the constraints and checked the effects of including HeIII (Appendix~\ref{sec:HeIII}) and compared our results to optical depth constraints (Appendix~\ref{sec:opdepth}).

In Section~\ref{sec:DHv2_tech}, I will describe major upgrades to the \texttt{DarkHistory} code to improve the treatment of low-energy depositon and to track the evolution of the spectrum of background radiation.
In Section~\ref{sec:DHv2_apps}, I will show predictions for CMB spectral distortions from exotic energy injection, and extend the constraints on dark matter decays to photons to lower masses than before.
Section~\ref{sec:DHv2_tech} is based on Ref.~\cite{Liu:2023fgu} and Section~\ref{sec:DHv2_apps} is based on the companion paper, Ref.~\cite{Liu:2023nct}.
Both of these works were done in collaboration with Hongwan Liu, Greg Ridgway, and Tracy Slatyer.
While Greg Ridgway significantly contributed to the beginning stages of these works, Hongwan Liu and I led them to completion.
In particular, I developed much of the new machinery for calculating spectral distortions from exotic energy injection and generated the results shown in Ref.~\cite{Liu:2023nct}.

Finally, in Section~\ref{sec:H2}, I will show the potential impact that dark matter energy injection can have on the formation of the earliest stars in the universe.
This work is based on Ref.~\cite{Qin:2023kkk} and was done in collaboration with Hongwan Liu, Julian Mu\~{n}oz, and Tracy Slatyer.
I led this analysis and wrote the code used to evolve various quantities in a spherical top-hat halo in the presence of energy injection.
I also generated most of the results presented in Ref.~\cite{Qin:2023kkk}.

\section{Lyman-$\alpha$ Constraints on Cosmic Heating from Dark Matter Annihilation and Decay}
\label{sec:Lya}

Dark matter (DM) interactions such as annihilation or decay can inject a significant amount of energy into the early Universe, producing observable changes in both its ionization and temperature histories.
Changes in the free electron fraction, 
for example, can alter the cosmic microwave background (CMB) anisotropy power spectrum \cite{Adams:1998nr, Chen:2003gz, Padmanabhan:2005es}, allowing constraints on the annihilation cross section \cite{Galli:2009zc,Slatyer:2009yq,Kanzaki:2009hf,Hisano:2011dc,Hutsi:2011vx,Galli:2011rz,Finkbeiner:2011dx,Slatyer:2012yq,Galli:2013dna,Madhavacheril:2013cna,Slatyer:2015jla,Slatyer:2015kla}
and the decay lifetime of DM~\cite{Zhang:2007zzh,Slatyer:2016qyl,Poulin:2016anj, Acharya:2019uba} to be set using Planck data~\cite{Planck2018}. 
Constraints based on modifications to the temperature history focus on two redshift ranges where measurement data is or will potentially be available: \textit{(i)} before hydrogen reionization at $z \sim 20$, and \textit{(ii)} during the reionization epoch at $2 \lesssim z \lesssim 6$. In the former redshift range, the 21-cm global signal~\cite{Poulin:2016anj,DAmico:2018sxd,Liu:2018uzy,Cheung:2018vww,Mitridate:2018iag,Clark:2018ghm} and power spectrum~\cite{Evoli:2014pva,Lopez-Honorez:2016sur} have been shown to be powerful probes of DM energy injection, and have the potential to be the leading constraint on the decay lifetime of sub-GeV DM~\cite{Liu:2018uzy}. In the latter range, measurements of the intergalactic medium (IGM) temperature derived from Lyman-$\alpha$ flux power spectra~\cite{Schaye:1999vr, Becker:2010cu} and Lyman-$\alpha$ absorption features in quasar spectra~\cite{Bolton:2011ck, Bolton:2010gr} have been used to constrain the $s$-wave annihilation cross section~\cite{Cirelli:2009bb}, the $p$-wave annihilation cross section, and the decay lifetime of DM~\cite{Diamanti:2013bia,Liu:2016cnk, Poulin:2016anj}. The IGM temperature can also be used to set limits on the kinetic mixing parameter for ultralight dark photon DM~\cite{Witte:2020rvb, Caputo:2020bdy, Caputo:2020rnx}, the strength of DM-baryon interactions~\cite{Munoz:2017qpy}, and the mass of primordial black hole DM~\cite{PBH2020}. 

In this section, we revisit the constraints on $p$-wave annihilating and decaying dark matter from the IGM temperature measurements during reionization.
This work is timely for two reasons. 
First and foremost, the development of \texttt{DarkHistory}~\cite{DarkHistory} allows us to improve on the results of Refs.~\cite{Cirelli:2009bb,Diamanti:2013bia,Liu:2016cnk} considerably.
We can now self-consistently take into account the positive feedback that increased ionization levels have on the IGM heating efficiency of DM energy injection processes.
This effect can give rise to large corrections in the predicted IGM temperature~\cite{DarkHistory} during reionization. 
Furthermore, \texttt{DarkHistory} can solve for the temperature evolution of the IGM in the presence of both astrophysical reionization sources and dark matter energy injection; previous work only set constraints assuming no reionization~\cite{Cirelli:2009bb}
or a rudimentary treatment of reionization and the energy deposition efficiency~\cite{Diamanti:2013bia, Liu:2016cnk}.
Second, experimental results published since Refs.~\cite{Cirelli:2009bb,Diamanti:2013bia,Liu:2016cnk} have considerably improved our knowledge of the Universe during and after reionization. These include: 
\begin{enumerate}
    \item \textit{Planck constraints on reionization}. 
    The low multipole moments of the Planck power spectrum provide information on the process of reionization~\cite{Planck2018}. 
    In particular, Planck provides 68th and 95th percentiles for the ionization fraction in the range $6 \lesssim z \lesssim 30$ using three different models~\cite{Hu:2003gh,Millea:2018bko}, arriving at qualitatively similar results.
    \item \textit{New determinations of the IGM temperature}. By comparing mock Lyman-$\alpha$ power spectra produced by a large grid of hydrodynamical simulations to power spectra calculated~\cite{Walther:2017cir} based on quasar spectra measured by BOSS~\cite{Palanque-Delabrouille:2013gaa}, HIRES~\cite{OMeara:2015lwd,2017AJ....154..114O}, MIKE~\cite{Viel:2013apy}, and XQ-100~\cite{Irsic:2017sop}, Ref.~\cite{Walther:2018pnn} (hereafter Walther\texttt{+}) determined the IGM temperature at mean density in the range $1.8 < z < 5.4$, overcoming a degeneracy between gas density and deduced temperature that hampered previous analyses \cite{Becker:2010cu,Boera:2014sia}. More recently, Ref.~\cite{Gaikwad:2020art} (hereafter Gaikwad\texttt{+}) fit the observed width distribution of the Ly$\alpha$ transmission spikes to simulation results, enabling a determination of the IGM temperature at mean density in the $5.4 < z < 5.8$ redshift range, again with only a weak dependence on the temperature-density relation.
\end{enumerate}

These improvements to both the understanding of energy deposition and the ionization/temperature histories are combined in our analysis into robust constraints on DM $p$-wave annihilation rates and decay lifetimes.
These constraints are competitive in the light DM mass regime ($\lesssim \SI{10}{\giga\eV}$) with existing limits on DM decay from the CMB anisotropy power spectrum~\cite{Slatyer:2016qyl} and are complementary to indirect detection limits~\cite{Essig:2013goa, Cohen:2016uyg, Boudaud:2016mos, Boudaud:2018oya}, being less sensitive to systematics associated with the galactic halo profile and interstellar cosmic ray propagation.

In the rest of this section,
we introduce the IGM ionization and temperature evolution equations,
discuss the data and statistical tests used, and finally present our new constraints.  
We also include supplemental materials in Appendix~\ref{app:Lya_supp} that provide additional details to support our main text. 

\subsection{Ionization and temperature histories}

Here, we write down the equations governing the evolution of the IGM temperature, $T_\text{m}$, and the IGM hydrogen ionization level, $x_\text{HII} \equiv n_\text{HII}/n_\text{H}$, where $n_\text{H}$ is the number density of both neutral and ionized hydrogen. The ionization evolution equation is:
\begin{alignat}{1}
    \dot{x}_\text{HII} = \dot{x}_\text{HII}^\text{atom} + \dot{x}_\text{HII}^\text{DM} + \dot{x}_\text{HII}^\star \,.
    \label{eq:ionization_diff_eq}
\end{alignat}
Here, $\dot{x}_\text{HII}^\text{atom}$ corresponds to atomic processes, i.e.\ recombination~\cite{Seager:1999bc, Seager:1999km, Chluba:2010ca, AliHaimoud:2010dx} and collisional ionization, which depend in a straightforward way on the ionization and temperature of the IGM, while $\dot{x}_\text{HII}^\text{DM}$ is the contribution to ionization from DM energy injection. These terms are discussed in detail in Ref.~\cite{DarkHistory}, and are given in full in Appendix~\ref{app:Lya_supp}, as well as a completely analogous HeII evolution equation.
The remaining term, $\dot{x}_\text{HII}^\star$, corresponds to the contribution to photoionization from astrophysical sources of reionization. This term will inevitably source photoheating, which will be important for the IGM temperature evolution equation (discussed below).
$\dot{x}_\text{HII}^\star$ can in principle be determined given a model of astrophysical sources of reionization, but there are large uncertainties associated with these sources. 
For example, the fraction of ionizing photons that escape into the IGM from their galactic sites of production is highly uncertain, ranging from essentially 0 to 1 depending on the model~\cite{McQuinn:2015icp}.

Instead, we rely on the Planck constraints on the process of reionization to fix the form of $\dot{x}_\text{e}$,
allowing us to fix $\dot{x}_\text{HII}^\star$ while remaining agnostic about astrophysical sources of reionization. Specifically, we begin by choosing a late time ionization history, $x_\text{e}^\text{Pl}(z)$ for $z<30$, within the 95\% confidence region determined using either the ``Tanh'' or ``FlexKnot'' model adopted by Planck~\cite{Planck2018}. We then make the common assumption that during hydrogen reionization HI and HeI have identical ionization fractions due to their similar ionizing potentials, but that helium remains only singly ionized due to HeII's deeper ionization potential~\cite{Onorbe2017}.
These assumptions allow us to set $x_\text{HII}^\text{Pl} = x_\text{e}^\text{Pl} / (1+\chi)$, where $\chi \equiv n_\text{He}/n_\text{H}$ is the primordial ratio of helium atoms to hydrogen atoms. 
Given a choice of $x_\text{e}^\text{Pl}(z)$ we can then rearrange Eq.~\eqref{eq:ionization_diff_eq} to set
\begin{alignat}{1}
    \dot{x}_\text{HII}^\star = \left(\frac{\dot{x}_\text{e}^\text{Pl}}{1 + \chi} - \dot{x}_\text{HII}^\text{atom} - \dot{x}_\text{HII}^\text{DM}\right) \theta(z^\star-z) \,,
    \label{eq:photoionization_rate}
\end{alignat}
where $\theta$ is a step function that enforces $\dot{x}_\text{HII}^\star = 0$ at sufficiently early redshifts when astrophysical reionization sources do not exist yet.
To fix $z^\star$, notice that at early times when $\dot{x}_\text{HII}^\star$ is turned off, ionization due to DM energy injection produces $x_\text{e}(z) \geq x_\text{e}^\text{Pl}(z)$.  Since DM cannot significantly reionize the universe~\cite{Liu:2016cnk}, there will exist a redshift past which $x_\text{e}(z) < x_\text{e}^\text{Pl}(z)$ if we do not turn on $\dot{x}_\text{HII}^\star$.  We define $z^\star$ to be this cross-over redshift where $x_\text{e}(z^\star) = x_\text{e}^\text{Pl}(z^\star)$. 

Thus, for any given DM model and $x_\text{e}^\text{Pl}$ we can use Eq.~\eqref{eq:photoionization_rate} to construct ionization histories that self-consistently include the effects of DM energy injection and reionization simultaneously.
We do not require the astrophysics that produces $\dot{x}_\text{HII}^\star$ to obey any constraint other than $\dot{x}_\text{HII}^\star \geq 0$, which maximizes freedom in the reionization model and leads to more conservative DM constraints.

The IGM temperature history can similarly be described by a differential equation:
\begin{alignat}{1}
    \dot{T}_\text{m} = \dot{T}_\text{adia} + \dot{T}_\text{C} + \dot{T}_\text{DM} + \dot{T}_\text{atom} + \dot{T}^\star \,,
    \label{eq:temp_diff_eq}
\end{alignat}
where $\dot{T}_\text{adia}$ is the adiabatic cooling term, $\dot{T}_\text{C}$ is the heating/cooling term from Compton scattering with the CMB, $\dot{T}_\text{DM}$ is the heating contribution from DM energy injection, and $\dot{T}_\text{atom}$ comprises all relevant atomic cooling processes.
These terms are also fully described in Ref.~\cite{DarkHistory}, and included in Appendix~\ref{app:Lya_supp} for completeness.
We stress that $\dot{T}_\text{DM}$ is computed, using \texttt{DarkHistory}~\cite{DarkHistory}, as a function of both redshift and ionization fraction $x_\text{e}$, self-consistently taking into account the strong dependence of $\dot{T}_\text{DM}$ on $x_\text{e}$, and strengthening the constraints we derive.

The remaining term, $\dot{T}^\star$, accounts for photoheating that accompanies the process of photoionization, as described in Eq.~\eqref{eq:photoionization_rate}. We adopt two different prescriptions for treating the photoheating rate, which we name `conservative' and `photoheated'. In the `conservative' treatment, we simply set $\dot{T}^\star = 0$. This treatment produces highly robust constraints on DM energy injection 
since the uncertainties of the reionization source modeling do not appear in our calculation. Any non-trivial model would only serve to increase the temperature of the IGM, strengthening our constraints. 

\begin{figure*}[!t]
    \centering
    \includegraphics[width=1.0\textwidth]{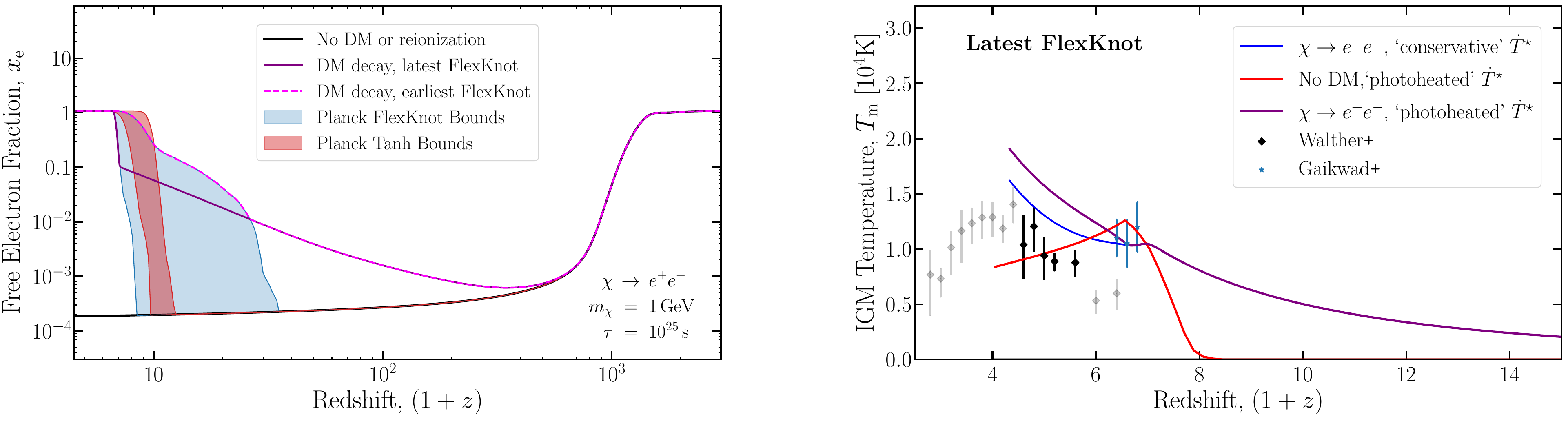}
    \caption{ The ionization history \textit{(Left)} and IGM temperature history \textit{(Right)} as functions of redshift.  The left plot shows the ionization history in the absence of DM energy injection and reionization sources (solid black), the 95\% confidence region for Planck's FlexKnot (shaded blue) and Tanh (shaded red) reionization histories, and the ionization history in the presence of both DM energy injection and reionization sources that produce Planck's latest (solid purple) and earliest (dashed magenta) FlexKnot histories at late times. The right plot shows the temperature history assuming 
\textit{(i)} DM decay and the `conservative' treatment of $\dot{T}^\star$ (blue), \textit{(ii)} the `photoheated' treatment and no DM energy injection (red), and \textit{(iii)} the `photoheated' treatment with DM decay (purple). 
\textit{(i)} and \textit{(iii)} assume a DM mass of $\SI{1}{\GeV}$ and decay to $e^+ e^-$ pairs with a lifetime of $\SI{e25}{\s}$ while 
\textit{(ii)} and \textit{(iii)} assume the latest FlexKnot reionization history and use 
parameter values $(\Delta T, \alpha_\text{bk}) = (\SI{24665}{\kelvin}, 0.57)$ and $(\SI{0}{\kelvin}, 1.5)$, respectively. Also included are the data from Ref.~\cite{Walther:2018pnn} (black diamonds) and Ref.~\cite{Gaikwad:2020art} (blue stars), where the solid data constitute our fiducial data set.
}
    \label{fig:example_history}
\end{figure*}

In the `photoheated' treatment, we implement a two-stage reionization model.  In the first stage --- prior to the completion of HI/HeI reionization --- we follow a simple 
parametrization adopted in e.g. Refs.~\cite{McQuinn:2015gda,Onorbe2017, Walther:2018pnn} and take $\dot{T}^\star = \dot{x}_\text{HII}^\star (1+\chi) \Delta T$ for some constant $\Delta T$.
This parameter is expected to be within the range \SIrange[range-phrase=--]{2e4}{3e4}{\kelvin} based on analytic arguments~\cite{Miralda_Escude_1994} and simulations~\cite{McQuinn:2012bq, Sanderbeck:2015bba}. 
We will either restrict $\Delta T \geq 0$ or impose a physical prior of $\Delta T\geq \SI{2e4}{K}$ in what we call our `photoheated-I' or `photoheated-II' constraints, respectively.

In the second stage --- after reionization is complete --- the IGM becomes optically thin. In this regime, reionization-only models find that the IGM is, to a good approximation, in photoionization equilibrium~\cite{Puchwein:2014zsa}. The photoheating rate in this limit is specified completely by the spectral index $\alpha_\text{bk}$ of the average specific intensity $J_\nu$ [with units \SI{}{\eV \per \second \per \hertz \per \steradian \per \centi\meter\squared}] of the ionizing background near the HI ionization threshold, i.e. $J_\nu \propto \nu^{-\alpha_\text{bk}}$~\cite{McQuinn:2015gda, Sanderbeck:2015bba}. By considering a range of reionization source models and using measurements of the column-density distribution of intergalactic hydrogen absorbers, the authors of Ref.~\cite{Sanderbeck:2015bba} bracketed the range of $\alpha_\text{bk}$ to be within $-0.5 < \alpha_\text{bk} < 1.5$, which we will use in our analysis.

In summary, the `photoheated' prescription is
\begin{alignat}{1}
    \dot{T}^\star = \begin{dcases}
        \dot{x}_\text{HII}^\star (1 + \chi) \Delta T \,, & x_\text{HII} < 0.99 \,, \\
        \sum_{i\in\{ \text{H}, \text{He} \}} \frac{E_{i\text{I}} x_i}{3(\gamma_{i\text{I}} - 1 + \alpha_\text{bk})} \alpha_{\text{A},i\text{I}} n_\text{H} \,, & x_\text{HII} \geq 0.99 \,,
    \end{dcases}
    \label{eq:modeled_term}
\end{alignat}
%
where $i$ runs over H and He (thus $x_i=$1, $n_\text{He}/n_\text{H}$), 
and for species $i$, $E_{i\text{I}}$ is the ionization potential, 
$\gamma_{i\text{I}}$ denotes the power-law index for the photoionization cross-section at threshold, 
and $\alpha_{A,i\text{I}}$ is the case-A recombination coefficient~\cite{Sanderbeck:2015bba}.
The `photoheated' model is therefore fully specified by two parameters, $\Delta T$ and $\alpha_\text{bk}$.  
Additionally, once HI/HeI reionization is complete, we set $1-x_\text{e} = \SI{4e-5}{}$, which is approximately its measured value~\cite{Bouwens:2015vha}.
This small fraction of neutral HI and HeI atoms 
dramatically decreases the photoionization rate relative to its pre-reionization value for photons of energy $\SI{13.6}{\eV} < E_\gamma < \SI{54.4}{\eV}$ injected by DM. 
Consequently, there is a non-negligible unabsorbed fraction of photons in each timestep, $\exp\left(- \sum_{i \in \{\text{HI}, \text{HeI}\}}n_i \sigma^\text{ion}_i(E_\gamma) \Delta t\right)$, where $\sigma_i^\text{ion}(E_\gamma)$ is the photoionization cross-section for species $i$ at photon energy $E_\gamma$. We modify \dhis to propagate these photons to the next timestep.

\begin{figure*}[t!]
\begin{tabular}{c}
\includegraphics[width=1.0\textwidth]{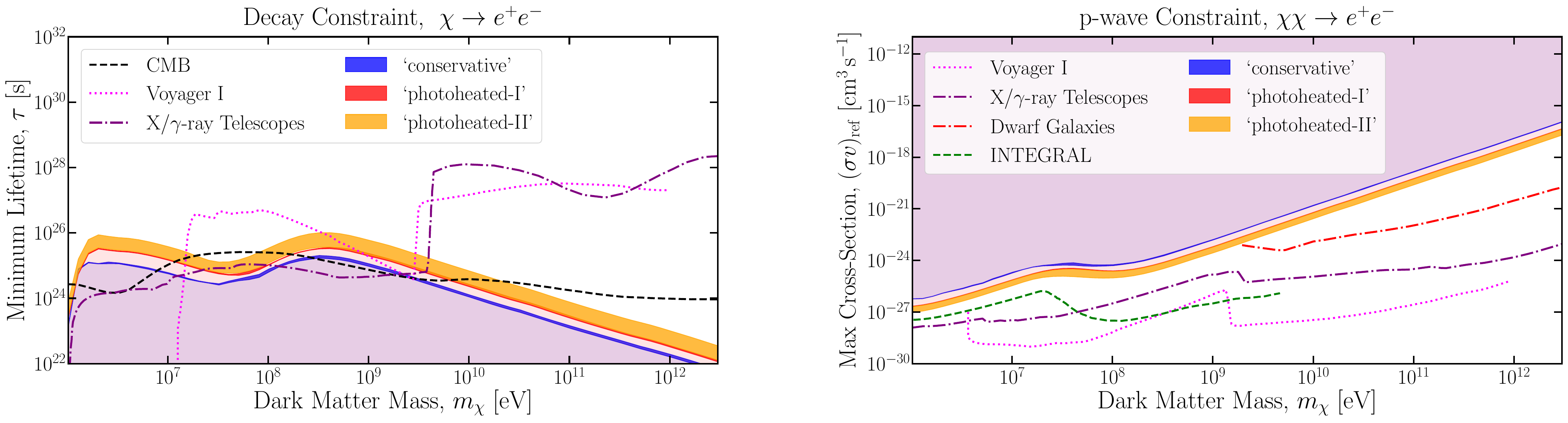}
\end{tabular}
\caption{Constraints for decay (left) or $p$-wave annihilation (right) to $e^+e^-$ pairs with $v_\text{ref} = \SI{100}{\km \, \s^{-1}}$.  We show our constraints using the `conservative' (blue band), `photoheated-I' (red band), or `photoheated-II' (orange band) treatment.
The darkly shaded bands show the variation of our constraints as we vary through the 95\% confidence regions of Planck's Tanh and FlexKnot reionization models.
We also include constraints from the CMB~\cite{Slatyer:2016qyl} (dashed-black), X/$\gamma$-ray telescopes~\cite{Essig:2013goa, Massari:2015xea, Cohen:2016uyg} (dot-dashed purple), INTEGRAL~\cite{Cirelli:2020bpc} where we have assumed $\left< v^2 \right> = \SI{220}{\km \, \s^{-1}}$ in the Milky Way, Voyager I~\cite{Boudaud:2016mos, Boudaud:2018oya} (dotted pink), and
gamma-ray observations of dwarf galaxies~\cite{Zhao:2016xie} (dot-dashed red).
}
\label{fig:constraints}
\end{figure*}

To demonstrate the effects of DM energy injection and our reionization modeling, we show in Fig.~\ref{fig:example_history} example histories obtained by integrating Eqs.~\eqref{eq:ionization_diff_eq} and~\eqref{eq:temp_diff_eq} for both the `conservative' and `photoheated' treatments, with and without DM decay.  
The left plot shows how our method can produce ionization histories that both take into account the extra ionization caused by DM energy injection and also vary over Planck's 95\% confidence region for the late-time ionization levels.
In the right panel, we assume the Planck FlexKnot curve with the latest reionization, and show in red our best fit temperature history assuming no DM energy injection, with the ‘photoheated’ treatment.
This history is a good fit to the fiducial data, with a total $\chi^2$ of about $5$. 
Additionally, once DM is added we show a model that is just consistent with our (95\% confidence) `conservative' constraints but ruled out by the `photoheated' constraints.

\subsection{Comparison with data} 
We compare our computed temperature histories with IGM temperature data obtained from Walther\texttt{+}~\cite{Walther:2018pnn} within the range $1.8 < z < 5.4$ and Gaikwad\texttt{+}~\cite{Gaikwad:2020art} within $5.4 < z < 5.8$.
To construct our fiducial IGM temperature dataset, we only consider data points with redshifts $z > 3.6$ (see Fig.~\ref{fig:example_history}, solid data points) since these redshifts are well separated from the redshift of full HeII reionization \cite{Becker:2010cu},
allowing us to safely use the transfer functions that \dhis currently uses, which assume $x_\text{HeIII}=0$. 
By neglecting HeII reionization and its significant heating of the IGM~\cite{McQuinn:2015icp} we derive more conservative constraints.
Additionally, the two Walther\texttt{+} data points above $z = 4.6$ are in tension with the Gaikwad\texttt{+} result; we discard them in favor of the higher $T_\text{m}$ values reported by Gaikwad\texttt{+}, since this results in less stringent limits. 

To assess the agreement between a computed temperature history and our fiducial temperature dataset using our `conservative' method, we perform a modified $\chi^2$ test. Specifically, our test statistic only penalizes DM models that overheat the IGM relative to the data, which accounts for the fact that any non-trivial photoheating model would only result in less agreement with the data, whereas DM models that underheat the IGM could be brought into agreement with the data given a specific photoheating model.
We define the following test statistic for the $i$th IGM temperature bin: 
\begin{alignat}{1}
    \text{TS}_i = \begin{dcases}
        0 \,, & T_{i,\text{pred}} < T_{i,\text{data}} \,, \\
        \left(\frac{T_{i,\text{pred}} - T_{i,\text{data}}}{\sigma_{i,\text{data}}}\right)^2 \,, & T_{i,\text{pred}} \geq T_{i,\text{data}} \,, 
    \end{dcases}
    \label{eq:one_sided_TS}
\end{alignat}
where $T_{i,\text{data}}$ is the fiducial IGM temperature measurement, $T_{i,\text{pred}}$ is the predicted IGM temperature given a DM model and photoheating prescription, and $\sigma_{i,\text{data}}$ is the $1\sigma$ upper error bar 
from the fiducial IGM temperature data. 
We then construct a global test statistic for all of the bins, simply given by $\text{TS} = \sum_i \text{TS}_i$. Assuming the data points $\{T_{i,\text{data}} \}$ are each independent, Gaussian random variables with standard deviation given by $\sigma_{i,\text{data}}$, the probability density function of $\text{TS}$ given some model $\{T_{i,\text{pred}} \}$ is given by
\begin{equation}
    f(\text{TS}|\{T_{i,\text{pred}}\}) = \frac{1}{2^N} \sum_{n=0}^N \frac{N!}{n! (N-n)!} f_{\chi^2}(\text{TS}; n) \,.
    \label{eq:TS_pdf}
\end{equation}
$N$ is the total number of temperature bins and $f_{\chi^2}(x;n)$ is the $\chi^2$-distribution with argument $x$ and number of degrees-of-freedom $n$, where the $n=0$ case is defined to be a Dirac delta function, $f_{\chi^2}(x;0) \equiv \delta(x)$. The hypothesis that the data $\{T_{i,\text{data}} \}$ is consistent with the $\{T_{i,\text{pred}} \}$ can then be accepted or rejected at the 95\% confidence level based on Eq.~\eqref{eq:TS_pdf}. 
See Appendix~\ref{app:Lya_supp} for more details.

For our `photoheated' constraints, we perform a standard $\chi^2$ goodness-of-fit test.  
For any given DM model we marginalize over the photoheating model parameters by finding the $\Delta T$ and $\alpha_\text{bk}$ values that minimize the total $\chi^2$ subject to the constraints $\Delta T \geq 0$ (`photoheated-I') or $\SI{2e4}{\K}$ (`photoheated-II') and $-0.5 < \alpha_\text{bk} < 1.5$.  We then accept or reject DM models at the 95\% confidence level using a $\chi^2$ test with 6 degrees of freedom (8 data points - 2 model parameters).

Fig.~\ref{fig:constraints} shows constraints for two classes of DM models: DM that decays or $p$-wave annihilates to $e^+e^-$. Our $p$-wave annihilation cross-section is defined by $\sigma v = (\sigma v)_\text{ref} \times (v/v_\text{ref})^2$ with $v_\text{ref} = \SI{100}{\kilo\meter\per\second}$. 
We also use the boost factor for Navarro-Frenk-White (NFW) dark matter density profiles for $p$-wave annihilation calculated in Ref.~\cite{Liu:2016cnk}.  
Although we only show constraints for $e^+e^-$ final states, our method applies to any other final state (see Appendix~\ref{app:Lya_supp}).
The blue, red, and orange regions are excluded by our `conservative,' `photoheated-I,' and `photoheated-II' constraints, respectively. 
The `photoheated' limits are generally a factor of $2-8$ times stronger than the `conservative' constraints. 

The thickness of the darkly shaded bands correspond to the variation in the constraints when we vary $x_\text{e}^\text{Pl}$ in Eq.~\eqref{eq:photoionization_rate} over the 95\% confidence region of Planck's FlexKnot and Tanh late-time ionization curves. 
The `conservative' and `photoheated-I' bands are narrow, demonstrating that the uncertainty in the late-time ionization curve is not an important uncertainty for these treatments. However, the `photoheated-II' treatment shows a larger spread, since the larger values of $\Delta T$ imposed by the prior significantly increase the rate of heating at $z \sim 6$, making the earliest temperature data points more constraining, and increasing the sensitivity to the ionization history at $z \simeq 6$. A better understanding of the process of reionization could therefore enhance our constraints significantly.

Our `conservative' constraints for decay to $e^+ e^-$ are the strongest constraints in the DM mass range $\sim\SI{1}{\MeV} - \SI{10}{\MeV}$ and competitive at around $\SI{1}{\GeV}$ while our $p$-wave constraints are competitive in the range $\sim\SI{1}{\MeV} - \SI{10}{\MeV}$.
For higher masses, constraints from Voyager I observations of interstellar cosmic rays are orders of magnitude stronger for both $p$-wave~\cite{Boudaud:2018oya} and decay~\cite{Boudaud:2016mos}. Constraints from X/$\gamma$-ray telescopes~\cite{Essig:2013goa, Massari:2015xea, Zhao:2016xie, Cohen:2016uyg} are stronger than ours for $m_\chi > \SI{1}{\GeV}$ and comparable for $m_\chi < \SI{1}{\GeV}$.

Importantly, all three types of constraints are affected by different systematics. The telescope constraints are affected by uncertainties in our galactic halo profile while Voyager's are affected by uncertainties in cosmic ray propagation.  
The $p$-wave boost factor is relatively insensitive to many details of structure formation, since it is dominated by the largest DM halos, which are well resolved in simulations (see Appendix~\ref{app:Lya_supp}). 
A more important systematic comes from our assumption of homogeneity. We assume that energy injected into the IGM spreads quickly and is deposited homogeneously, when in reality injected particles may be unable to efficiently escape their sites of production within halos~\cite{Schon:2014xoa, Schon:2017bvu}.
We leave a detailed exploration of these inhomogeneity effects for future work.

\subsection{Conclusion} 

We have described a method to self-consistently construct ionization and IGM temperature histories in the presence of reionization sources and DM energy injection by utilizing Planck's measurement of the late-time ionization level of the IGM.  We construct two types of constraints for models of DM decay and $p$-wave annihilation.  For the first `conservative' type of constraint, we assume that reionization sources can ionize the IGM but not heat it, resulting in constraints that are robust to the uncertainties of reionization. 
For the second `photoheated' type of constraint, we use a simple but well-motivated photoheating model that gives stronger limits than the `conservative' constraints by roughly a factor of $2-8$.  We expect that as the uncertainties on the IGM temperature measurements shrink, and as reionization and photoheating models become more constrained, these `photoheated' constraints will strengthen considerably. 
\section{
Exotic energy injection in the early universe I: a novel treatment for low-energy electrons and photons
}
\label{sec:DHv2_tech}

The early Universe is an excellent laboratory for new physics searches. It was relatively homogeneous, making it simple to treat; between $20 \lesssim z \lesssim 150$, the intergalactic medium (IGM) temperature in standard $\Lambda$CDM cosmology evolves only through adiabatic cooling, making it exceptionally sensitive to exotic sources of heating; finally, the effect of new physics processes on the Universe can accumulate over a timescale at least a billion times longer than any feasible terrestrial probe of new physics. As a result, energy injection from new physics that is otherwise undetectable terrestrially or in our local astrophysical neighborhood can both be accurately predicted and potentially observed with early-Universe probes.

High-energy Standard Model (SM) particles injected from new physics processes lose their energy as they interact with the IGM, ultimately depositing their energy into the following channels: 
\textit{1)} ionization, resulting in an increase in the free electron fraction $x_e \equiv n_e / n_\text{H}$, where $n_e$ and $n_\text{H}$ are the number densities of free electrons and hydrogen (both neutral and ionized) respectively; 
\textit{2)} heating of the IGM, resulting in an increase in the IGM temperature $T_m$; 
\textit{3)} the production of low-energy photons, a term we use to refer to photons below the ionization potential of hydrogen, $\mathcal{R} \equiv \SI{13.6}{\eV}$;
and potentially \textit{4)} the increased abundance of excited states of atoms.
Low-energy photons can show up as extragalactic background photons over a wide range of frequencies.\footnote{For generic unstable Standard Model particles, a significant portion of their energy can go into free-streaming neutrinos as well, but we will only focus on electromagnetic channels in this work.} 

Searches for new physics in all three of these channels have been studied extensively in the literature. The increase in ionization levels from dark matter (DM) annihilation and decay, and its impact on the cosmic microwave background (CMB) anisotropy power spectrum, provides some of the strongest limits on the rates of such processes for sub-GeV dark matter~\cite{Slatyer:2009yq, Kanzaki:2009hf, Slatyer:2015jla, Slatyer:2016qyl, Poulin:2016anj,Cang:2020exa}. 
Exotic heating of the IGM during the cosmic dark ages could modify the 21-cm brightness temperature significantly, so that future 21-cm observations could potentially lead to strong constraints on DM annihilation, decay, and primordial black holes (PBHs)~\cite{Evoli:2014pva, Lopez-Honorez:2016sur, DAmico:2018sxd, Liu:2018uzy, Cheung:2018vww, Mitridate:2018iag, Clark:2018ghm}. Heating of the IGM during reionization can also be probed by temperature measurements of the IGM using the Lyman-$\alpha$ forest~\cite{Hiss:2017qyw,Walther:2018pnn,Gaikwad:2020art,Gaikwad:2020eip}, setting constraints on DM velocity-dependent annihilation and decay~\cite{Cirelli:2009bb, Diamanti:2013bia, Liu:2016cnk,Liu:2020wqz}; dark photon dark matter can likewise convert into extremely low-frequency photons that heat the IGM~\cite{McDermott:2019lch,Caputo:2020bdy,Witte:2020rvb}, and may potentially provide sufficient heating to reconcile low- and high-redshift Lyman-$\alpha$ observations of the IGM~\cite{2206.13520}. 
Finally, the limits on CMB spectral distortions have been used to constrain dark matter scattering against Standard Model particles~\cite{Ali-Haimoud:2015pwa}.
In addition, there have been a number of studies that employ the method of Green's functions to study spectral distortions from heating~\cite{Chluba:2013vsa}, photon injection~\cite{Chluba:2015hma}, and electron injection~\cite{Acharya:2018iwh} prior to recombination.

The public \dhis code~\cite{DarkHistory} was designed to compute changes to the cosmic temperature and ionization history due to exotic energy injections, self-consistently taking into account energy injection from conventional sources (e.g. stars during the epoch of reionization).
In this section, we make significant improvements to the computation of the global free-electron fraction $x_e$, IGM temperature $T_m$, and the intensity of low-energy photons $I_\omega$ 
\footnote{Throughout this work, we will use angular frequency $\omega$ instead of frequency $\nu$, in accordance with our choice of natural units.}
in the presence of exotic energy injection, focusing on injection occurring during or after recombination, and provide publicly available tools for these computations.~\footnote{Our code is available at \githubmaster.}
The key advances are: 
\begin{enumerate}
    \item We improve our treatment of electron cooling by making significant upgrades to our treatment of low-energy ($< \SI{3}{\kilo\eV}$) electrons---especially in terms of collisional excitation into a range of hydrogen excited states---and inverse Compton scattering. This in turn allows us to track low-energy ($< \SI{10.2}{\eV}$) photons produced during electron cooling;
    \item We account for $y$-type spectral distortions produced by heating of baryons; 
    \item We track the population of a range of excited states in the hydrogen atom, as well as transitions between these states. We also allow low-energy photons produced from exotic energy injection to excite and de-excite hydrogen atoms, while also carefully tracking the low-energy photon spectrum produced by excitation and de-excitation. 
\end{enumerate}

Put together, these improvements result in two major achievements. 
First, they allow us to consistently calculate, for the first time, the frequency-dependent change in the photon background intensity $\Delta I_\omega$ over the CMB after recombination due to exotic sources of energy injection, enabling the use of spectral distortions to the CMB to look for such processes. 
Second, the occupation number of photons injected by dark matter with energies between the CMB temperature after recombination and the ionization potential of hydrogen significantly exceeds the occupation number of the CMB, and may lead to corrections to the ionization and thermal histories.
The ability to track $\Delta I_\omega$ enables us to quantify the influence of sub-\SI{10.2}{\eV} low-energy photons on the process of recombination, improving the reach of energy injection calculations to lower energies, as well as increasing the accuracy.

In a companion work, described in Section~\ref{sec:DHv2_apps}, we explore the applications of our improved calculation~\cite{paperII}. 
We provide the full space of predicted low-energy photon spectra---from radio to ultraviolet frequencies---for dark matter annihilation and decay into electron/positron pairs and photons, providing a new-physics benchmark for future CMB spectral distortion experiments. 
We show the corrections imparted by additional low-energy photons (injected by dark matter annihilation and decay) on the ionization and thermal histories. These corrections are at the level of a few percent; hence any results using the ionization histories calculated with \dhis \texttt{v1.0} are largely unchanged. 
We also show for the first time the ionization histories resulting from dark matter decaying to photons with masses less than \SI{10}{\kilo\eV}.
Finally, our improved treatment of low-energy electrons also allows us to extend existing CMB anisotropy power spectrum limits on the annihilation and decay of DM into photons to DM masses of less than \SI{10}{\kilo\eV}, giving rise to important limits on the coupling between axion-like particles (ALPs) and photons in this mass range that we present in Section~\ref{sec:DHv2_apps}. 

The complete contribution to the low-energy photon spectrum during and after recombination including exotic energy injection will be computed here for the first time.
In Section~\ref{sec:lowengelec}, we describe our improved treatment for low-energy electrons, including the ability to now track the spectrum of photons resulting from inverse Compton scattering (ICS) by these electrons. 
In Section~\ref{sec:y-type}, we describe the contributions of ICS and heating of the IGM to the low-energy photon spectrum.
Section~\ref{sec:evolution} outlines our method for evolving low-energy photons, by taking the contributions mentioned above and including their interactions with hydrogen atoms.
We conclude with Section~\ref{sec:tech_conclusion}.
In addition, we include a number of appendices which provide more detailed discussion to support the main text.

\subsection{Low-Energy Electrons}
\label{sec:lowengelec}

In this section we describe how an injected electron or positron loses its energy in the early universe as it cools. 
Since much of the following discussion builds directly on Ref.~\cite{DarkHistory}, 
we focus here on extensions and improvements we have made. 
In Ref.~\cite{DarkHistory}, electrons were artificially divided into high-energy ($> \SI{3}{\kilo\eV}$) and low-energy electrons ($\leq \SI{3}{\kilo\eV}$). 
For high-energy electrons, the particles undergo all possible cooling processes with some probability, and one can therefore solve a set of linear equations to determine the energy deposited by each process.
Previously in \dhis \texttt{v1.0}, when the electrons cool to kinetic energies below $\SI{3}{\keV}$, the interpolation tables provided by Ref.~\cite{MEDEAII} determined where the remaining energy was ultimately deposited. 
With this method, it is possible to track the total energy converted into photons with energy less than $E_\alpha =$ \SI{10.2}{\eV}, but not the \textit{spectrum} of these photons. 
In the subsequent section, we present the following changes to \dhis:
\begin{itemize}
	\item We update our collisional excitation cross-sections and extend our treatment of high-energy electrons to all electrons.
	\item In addition to tracking how low-energy electrons deposit their energy, we now simultaneously track the spectrum of photons they produce as they cool.
\end{itemize}

\subsubsection{Energy deposition from low-energy electrons}
\label{sec:eng_dep}

Electrons of all relevant energies injected into the Universe after recombination cool extremely efficiently, losing almost all of their energy by scattering off atoms, ions, electrons, and CMB photons over a timescale that is much shorter than a Hubble time~\cite{Slatyer:2009yq}. 
Consider an electron produced with energy $E'$. 
We would like to know how much energy this electron deposits, $R_c(E')$, into the IGM through a given channel $c$. 

As explained in Ref.~\cite{DarkHistory}, $R_c(E')$ can be calculated by considering the electron to cool over a timescale much shorter than the characteristic timescales of all cooling processes, and calculating the energy deposited promptly into all channels $c$, as well as the secondary electron spectrum $d N_e / dE$. Then $R_c(E')$ can be calculated recursively using the integral equation
\begin{alignat}{1}
    R_c(E') = \int dE \, R_c(E) \frac{d N_e}{dE} + P_c(E') \,,
    \label{eqn:elec_cooling_analytic}
\end{alignat}
where $P_c(E')$ is the energy promptly deposited into channel $c$. A similar equation can be written for the spectrum of photons produced through ICS. 
When discretized, this equation becomes a lower-triangular system of linear equations that we can solve numerically.

Previously in \dhis \texttt{v1.0}, this formula was only applied to ``high-energy electrons'' with $E'\geq\SI{3}{\keV}$, resolving them into (1) what was termed ``high-energy deposition'' with channels ionization, Lyman-$\alpha$ excitation, and heating, (2) a spectrum of ICS photons that was passed to the photon cooling part of the code, and (3) a low-energy $E'<\SI{3}{\keV}$ electron spectrum. 
This treatment was not extended to ``low-energy electrons'' due to approximations made in the cross-sections used in \dhis \texttt{v1.0}  that may not apply at lower energies. Instead, the energy deposited by these low-energy electrons was resolved by default using the results of the MEDEA code~\cite{MEDEAII}, which studied the cooling of low-energy electrons using Monte Carlo methods, ultimately resolving electrons into hydrogen ionization, helium ionization, Lyman-$\alpha$ excitation, heating, and continuum (sub-$E_\alpha$) photons for a range of different ionization levels, $x_e$. 
This division was also intended to make it easier for users to incorporate their own alternative models for the cooling of low-energy electrons (e.g.~accounting for updated cross-sections, the inclusion of additional states, or more sophisticated methodology) into \texttt{DarkHistory}, without needing to modify the rest of the code.

However, this division by energy is not a requirement;
 Eq.~\eqref{eqn:elec_cooling_analytic} applies to all electrons, regardless of their energy, and we can simply extend the previous high-energy electron treatment to cover all regimes so long as we have cross sections that apply across all relevant energies. 
 We note that the authors of Refs.~\cite{Acharya:2019uba} and~\cite{Jensen:2021mik} also treated electrons with no such division between high and low energies. 
 In our current work, we have updated the collisional ionization and excitation cross-sections of electrons with HI, HeI and HeII. For collisional ionization, we adopt the binary-encounter models described in Ref.~\cite{Kim_Rudd}, while for collisional excitation, we use results from Ref.~\cite{Stone_Kim_Desclaux} for all hydrogen $np$ states and HeI, and the CCC database~\cite{CCC} for all other hydrogen states. These results are extended to higher energies using the Bethe approximation~\cite{RevModPhys.43.297}. Further details on the cross sections can be found in Appendix~\ref{app:collisional_rates}. 

With these new cross sections, we now calculate the energy deposition of both low- and high-energy electrons using Eq.~\eqref{eqn:elec_cooling_analytic}, and extend the number of possible excitation transitions from the hydrogen and helium ground states to cover the following states: 
\begin{itemize}
	\item H excitation from $1s$ to any $nl$ state with $n\leq 4$, where the integers $n,l$ index the principal quantum number and angular momentum quantum number of the hydrogen atom,
	
	\item H excitation from $1s$ to any $np$ state with $n\leq 10$,
	
	\item HeII excitation to the first excited state and HeI excitation to the $n^{2S+1}L = 2^3s$ state. Here, $S$ and $L$ are the total spin and orbital angular momenta, respectively.
\end{itemize}
The default treatment in \dhis now includes all of these states, but users can change this as desired. 

At the end of our electron cooling calculation, the initial electron energy is subdivided between heating of the IGM, ionization of neutral hydrogen, ionization of neutral and singly ionized helium, excitation of neutral hydrogen, excitation of neutral helium, and a spectrum of photons. 
The spectrum of photons produced by electron cooling is then added to the propagating photon spectrum at the current timestep, and is resolved in the same manner as in \dhis \texttt{v1.0}. 
However, the excited states should also ultimately cascade down to the ground state, producing more photons that need to be treated as well. 
We have developed two methods to treat these deexcitation photons, under the simplifying assumption that deexcitation from helium can be neglected, given that the number density of helium nuclei $n_\text{He}$ is only 8\% of $n_\text{H}$. 

The first method involves tracking the population of a large number of excited states using the multilevel atom method described in Ref.~\cite{2010PhRvD..82f3521A}, and provides an accurate way of tracking photons that are emitted after deexcitation; we will describe this method in detail in Section~\ref{sec:evolution}. 
For calculations involving e.g. CMB spectral distortions, it is necessary to use this method.

The second method follows the prescription used in MEDEA~\cite{MEDEAII}; for any calculations that do not require accurately tracking the background spectrum of photons, this method is sufficient and faster than the first method.
Specifically, every excitation to an $nl$ state cascades down to either the $2s$ or the $2p$ state, emitting sub-$E_\alpha$ photons in the process. 
If the electron reaches the $2s$ state, it undergoes a two-photon forbidden decay process, further emitting sub-$E_\alpha$ photons; otherwise, the excitation is counted as an effective Lyman-$\alpha$ excitation. 
The probability for an $nl$ state cascading down to a $2s$ or $2p$ state is calculated in Ref.~\cite{Hirata:2005mz}. 
In this way, the initial energy of an electron is completely distributed among the following channels: \textit{1)} ionization of neutral hydrogen; \textit{2)} ionization of neutral and singly-ionized helium; \textit{3)} Lyman-$\alpha$ excitation; \textit{4)} heating of the IGM, and \textit{5)} sub-$E_\alpha$ continuum photons. 
We have also implemented this second method in \dhis and will use this method to compare our results with the existing literature in Section~\ref{sec:cross-checks}. 

\subsubsection{ICS of low-energy electrons}
\label{sec:ICS}

In addition, we obtain the resultant low-energy photon spectrum, $N^\text{ICS}_\omega$, produced during the cooling process through inverse Compton scattering of CMB photons. 
As explained in Ref.~\cite{DarkHistory}, we can use a similar strategy as in Section~\ref{sec:eng_dep} to determine the CMB distortion caused by ICS. 
The distortion is defined as the upscattered photon spectrum, subtracting off the original CMB photons that were scattered.
Previously, this method was applied to high-energy electrons, but ICS was neglected for electrons with energy less than 3 keV, since ICS is a subdominant process for these energies after recombination.
Now we extend the treatment to arbitrarily low energies. 

For low enough electron energies, the amount of energy gained by the upscattered CMB photons scales as $\beta^2$, where $\beta$ is the electron velocity; hence the upscattered photon spectrum becomes very similar to the original CMB spectrum as the electron's energy becomes small.
This leads to catastrophic cancellation between the two terms when determining the CMB distortion.
To solve this issue, we use an analytic formula for the difference of the two terms.

At low energies, the spectrum of photons produced per unit time due to ICS of low-energy electrons can be given by a Taylor expansion in $\beta$,
\begin{equation}
	\frac{dN^\text{ICS}_\omega}{dt} = n_\text{CMB}(\omega, T) \sigma_T + \frac{3 \sigma_T T^2}{32 \pi^2} \sum_{n=1}^\infty \sum_{j=1}^{2n}\frac{A_n \beta^{2n} x^3 P_{n,j}(x) e^{-jx}}{(1 - e^{-x})^{2n+1}} \, ,
    \label{eqn:thomson_scattered_phot_spec_expansion}
\end{equation}
where $N^\text{ICS}_\omega$ is the number of upscattered photons per energy $\omega$, $n_\text{CMB}(\omega, T)$ is the blackbody spectrum (number density of photons per unit energy), $\sigma_T$ is the Thomson cross-section, $A_n$ is a constant, $x \equiv \omega/T$, and $P_{n,j}(x)$ is a rational or polynomial function in $x$. 
Expressions for $A_n$ and $P_{n,j}(x)$ are given in the appendix of Ref.~\cite{DarkHistory}. 
The first term in the expansion is exactly the original spectrum of CMB photons that were upscattered. 
This is precisely the expected result as $\beta \to 0$, since we approach the Thomson scattering limit and photons are simply scattered elastically. 
Terms that are higher order in $\beta$ can therefore be regarded as distortions to the blackbody spectrum that occur due to scattering, and the energy loss can be directly computed from those terms alone. 
In fact, we can check that when integrated over $\omega$, 
\begin{equation}
  \int d\omega \sum_{n=1}^\infty \sum_{j=1}^{2n} \omega \frac{A_n \beta^{2n} x^3 P_{n,j}(x) e^{-jx}}{(1 - e^{-x})^{2n+1}} = \frac{4}{3} \sigma_T c \beta^2 (1 + \beta^2 + \beta^4 + \beta^6 + \cdots) u_\text{CMB}(T) \,,
\end{equation}
where $u_\text{CMB}(T)$ is the blackbody energy density with temperature $T$; this is simply the Taylor expansion of the energy loss rate of an electron scattering off a blackbody spectrum with temperature $T$ in the Thomson regime, $dE' / dt = (4/3) \sigma_T \beta^2 \gamma^2 u_\text{CMB}(T)$, where $\gamma \equiv (1 - \beta^2)^{-1/2}$. 
Therefore, we compute the distortion produced by low-energy electrons directly by computing only the $n \geq 1$ terms in Eq.~\eqref{eqn:thomson_scattered_phot_spec_expansion}; in practice, we include terms up to $n = 6$.

\subsubsection{Comparison to other calculations}
\label{sec:cross-checks}

%
\begin{figure*}
	\centering
	\includegraphics[scale=0.45]{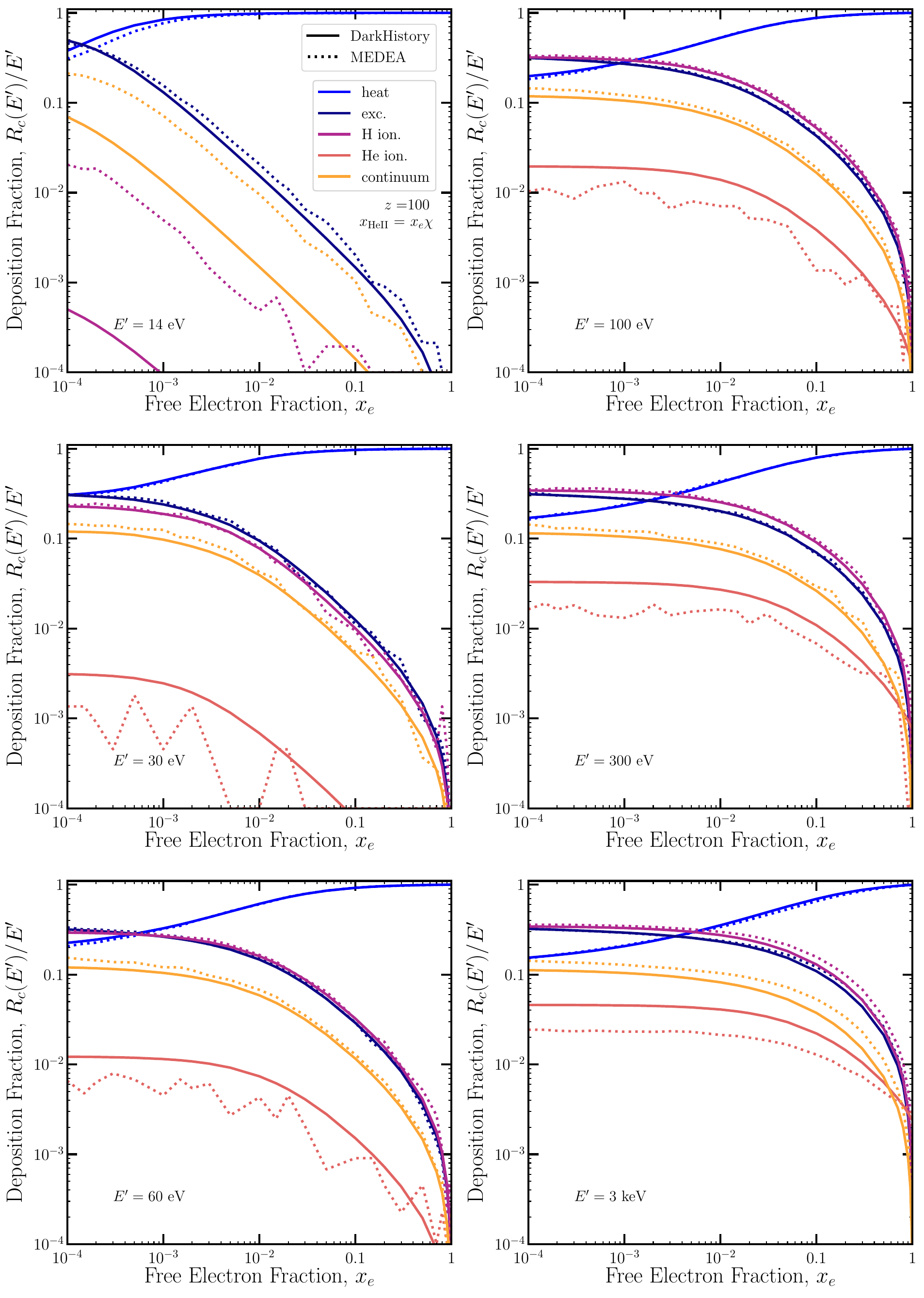}
	\caption{
	A comparison of the fraction of energy deposited into various channels by low-energy electrons, $R_c (E') / E'$, as calculated by \dhis (solid) and Ref.~\cite{MEDEAII} (dashed), where $E'$ is the energy of the electron. 
	The channels we include are heat (blue), Lyman-$\alpha$ excitation (indigo), hydrogen ionization (magenta), helium ionization (orange), and sub-$E_\alpha$ or `continuum' photons (yellow). 
	}
	\label{fig:MEDEA_xcheck}
\end{figure*}
\begin{figure*}
	\centering
    \includegraphics[scale=0.45]{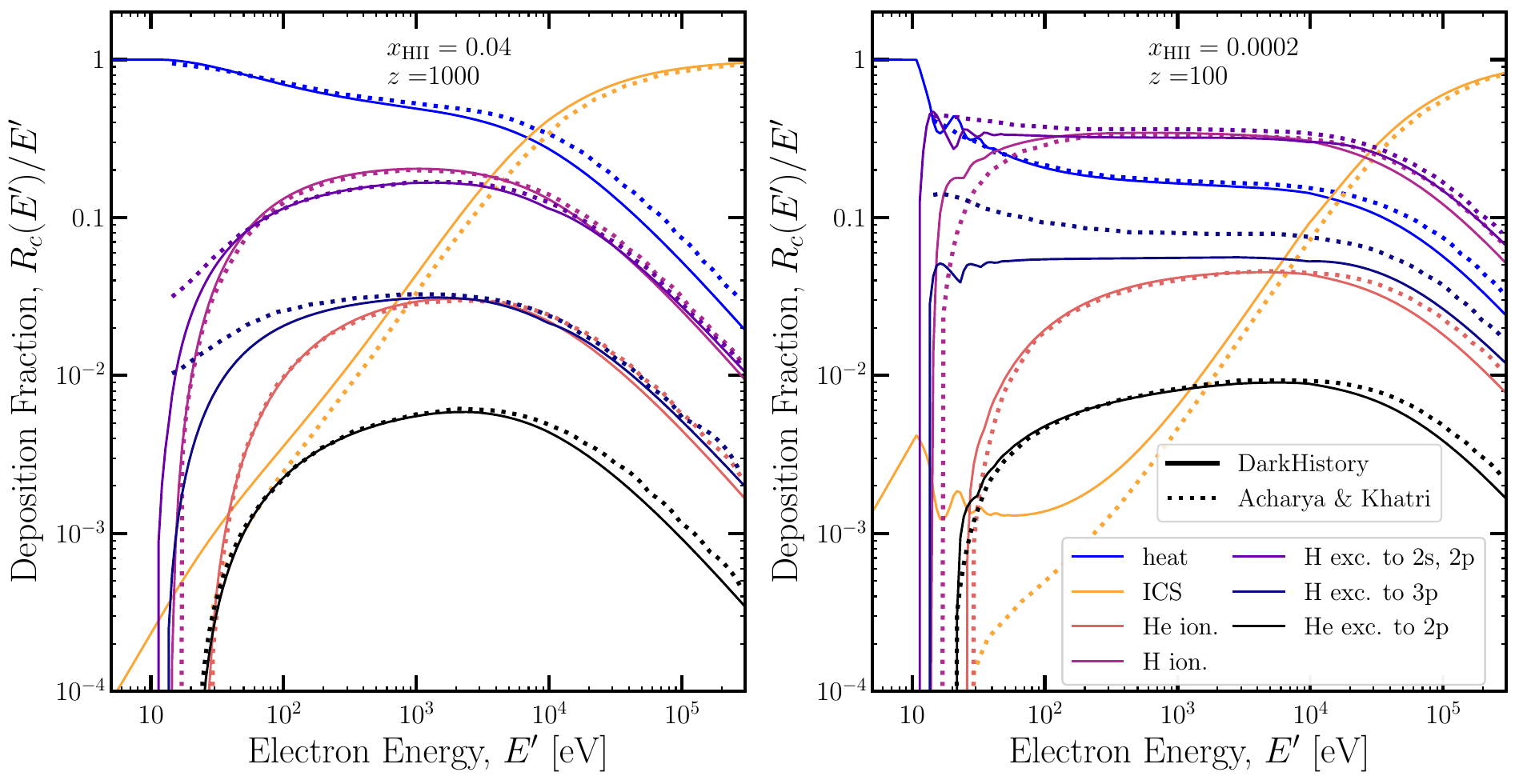}
	\caption{
	A comparison of the fraction of energy deposited into various channels by low-energy electrons, $R_c (E') / E'$, as calculated by \dhis (solid) and Ref.~\cite{Acharya:2019uba} (dashed). 
	The channels we include are heat (blue), ICS (yellow), helium ionization (orange), hydrogen ionization (magenta), 
	hydrogen excitation to the $n=2$ (purple) and $n=3$ state (indigo), and helium excitation to the $n^{2S+1}L = 2^1p$ state (black). 
	}
	\label{fig:AcharyaKhatri_xcheck}
\end{figure*}

In this section we compare our calculation of $R_c(E')$ to that of the MEDEA code~\cite{MEDEAII} and Ref.~\cite{Acharya:2019uba}. 
The former implements a Monte Carlo calculation of $R_c(E')$ while the latter solves a lower-triangular system of equations much like we do. 

To make the comparison with MEDEA, we use the second method described at the end of Section~\ref{sec:eng_dep}, restricting ourselves only to the $2s$ and $np$ states, in agreement with their method. 
In Fig.~\ref{fig:MEDEA_xcheck} we compare our results against the output of the MEDEA code~\cite{MEDEAII}, summarized in Ref.~\cite{Galli:2013dna}. 
Following Ref.~\cite{MEDEAII}, we plot $R_c(E') / E'$ against the free electron fraction $x_e$ for six different values of the input electron energy at $z=0$, and set the helium ionization level to linearly scale with the hydrogen ionization level, $x_\text{HeII} = x_\text{HII} \, \chi$, where $\chi \equiv n_\text{He}/n_\text{H}$ is the ratio of all helium to hydrogen nuclei by number. 

We find generally good agreement between the two methods.
However, compared to our method, relying on the MEDEA results has two disadvantages. First, Monte Carlo methods generally struggle to capture rare channels and continuous processes properly. 
The inability to capture rare channels accurately is evident in the noticeable differences between the helium ionization curves; our results are smoother across $x_e$ and tend to be slightly larger in amplitude.
Our \SI{14}{\eV} result also differs significantly from MEDEA's results, predicting a larger contribution to heating, and a suppression to energy deposited into continuum photons and hydrogen ionization (helium ionization is impossible for such low-energy electrons). This may be due to the fact that MEDEA models the heating loss process as discrete events, each leading to a loss of 5\% of the total kinetic energy of the electron. Since \SI{14}{\eV} electrons are close to the hydrogen ionization threshold, this discretization of a continuous process may be the cause of the discrepancy.  
The second disadvantage in relying on the MEDEA results is that we only have data for electrons with 14, 30, 60, 100, 300, and 3000 eV of kinetic energy; an interpolation must be performed over the kinetic energy for all other values. Our new method avoids this problem, and is able to compute $R_c(E')$ at any arbitrary energy. 

In Fig.~\ref{fig:AcharyaKhatri_xcheck} we compare our results for $R_c(E')$ against those of Ref.~\cite{Acharya:2019uba}. 
Again, the overall agreement is generally good, though there are a few qualitative differences. 
We find that the two results sometimes differ at excitation thresholds, with our results going smoothly to zero as energy decreases. This may be due to a difference in resolution for $E'$.

Additionally, some of our curves have oscillatory features while those of Ref.~\cite{Acharya:2019uba}  do not.
These oscillatory features are present in e.g. Refs.~\cite{1979ApJ...234..761S,2010MNRAS.404.1869F} and have a simple explanation. 
Consider the hydrogen excitation to the $n=2$ state in the right hand panel. 
Each minimum corresponds to a multiple of the Lyman-$\alpha$ energy $E_\alpha$. 
If the injected electron has energy just under $2 E_\alpha$, then after it excites an atom, it will have just under $E_\alpha$ energy left over, and therefore is unable to excite any more atoms.
As the injected electron energy increases above this threshold, then the secondary electrons will be energetic enough to also deposit energy into excitation.
The suppression of energy going into excitation means that energy must be deposited through other processes, hence we see maxima at kinetic energies of $2 E_\alpha$ in other channels.
The oscillations at the other multiples of $E_\alpha$ have a similar explanation. 

Finally, for completeness, we note that Ref.~\cite{Jensen:2021mik} proposed an analytic method to estimate the fraction of electron energy deposited directly into continuum photons; for sufficiently low-energy electrons, the remaining energy goes into heating, ionizations, and excitations. 
Within its regime of validity, this approach agrees very well with the results of \dhis \texttt{v1.0} presented in Ref.~\cite{DarkHistory}.

\subsection{Spectral distortions from Compton scattering}
\label{sec:y-type}

In addition to the ICS secondary photons described in Section~\ref{sec:ICS}, heating of the IGM from exotic sources also leads to a distortion of the CMB away from the blackbody spectrum, as photons Compton scatter off a thermal distribution of electrons at a different temperature. 
For redshifts $z \lesssim 5 \times 10^4$, the Compton scattering rate becomes smaller than Hubble expansion, and photon energies can no longer be efficiently redistributed to establish a Bose-Einstein distribution, leading to a $y$-distortion~\cite{Zeldovich:1969ff}. 
The shape of the $y$-type distortion is given by 
\begin{equation}
\Delta I_\omega = y \times \frac{\omega^3}{2 \pi^2} \frac{x e^x}{(e^x - 1)^2} \left[ x \coth \left( \frac{x}{2} \right) - 4 \right] .
\label{eqn:y-distortion}
\end{equation}
The amplitude of this distortion is controlled by the $y$-parameter, which is given by~\cite{Chluba:2018cww}
\begin{equation}
y = \int_0^t  dt \, \frac{T_m - T_\text{CMB}}{m_e} \sigma_T n_e \,,
\label{eqn:y-param}
\end{equation}
where $T_\text{CMB}$ is the CMB temperature and $m_e$ is the electron mass.
As defined, $y$ includes contributions from \textit{1)} the adiabatic cooling of baryons, which causes $T_m < T_\text{CMB}$ at $z \lesssim 155$; \textit{2)} photoheating during the process of reionization, and \textit{3)} exotic sources of energy injection. 
We define the contribution to $y$-distortions just from exotic energy injection as $y_\text{inj}$, which can be parametrized as
\begin{equation}
    y_\text{inj} = \int_0^t dt\, \frac{\Delta T}{m_e} \sigma_T n_e  \,, \qquad \Delta T = T_m - T_m^{(0)},
    \label{eqn:y_DM}
\end{equation}
where $T_m^{(0)}$ is the temperature history including reionization in the absence of exotic energy injection.
Throughout this section, we assume the default reionization model used in \dhis \texttt{v1.0}, which is based on Ref.~\cite{Puchwein:2018arm}; users may define their own reionization model in performing these calculations. 

At early enough times, Eq.~\eqref{eqn:y_DM} reduces to an integral over the heating rate, a form that is useful numerically when baryons are tightly coupled to the CMB with $T_m \approx T_\text{CMB}$, and is typically seen in the literature when discussing spectral distortions due to energy injection before recombination.
We give a novel derivation of this expression in Appendix~\ref{app:y_deriv}, following a similar calculation from Ref.~\cite{Hirata:2008ny}, and present the main results here.
At redshifts well before reionization, the evolution equation for the matter temperature is given by~\cite{DarkHistory}
\begin{equation}
	\dot{T}_m = -2 H T_m + \Gamma_C (T_\text{CMB} - T_m) + \dot{T}_m^\text{inj} ,
	\label{eq:general_Tm_evolution}
\end{equation}
where $H$ is the Hubble parameter, and $\Gamma_C$ is the Compton heating rate coefficient. The rate of change of the matter temperature due to exotic energy injection is defined by
\begin{alignat}{1}
    \dot{T}_m^\text{inj} \equiv \frac{2 \dot{Q}}{3 (1 + \chi + x_e) n_\text{H}} \,, \quad \dot{Q} \equiv f_\text{heat} \left(\frac{dE}{dV \, dt} \right)_\text{inj} \!\!\!.
\end{alignat}
$\dot{Q}$ is the usual exotic heating rate in terms of energy density per time and is parameterized in terms of the energy injection rate using the coefficient $f_\text{heat}$.
Define the quantity 
\begin{equation}
J \equiv \frac{8 \sigma_T u_\text{CMB} x_e}{3 (1 + \chi + x_e) m_e H} \,.
\end{equation}
The integrated $y$-parameter as defined in Eq.~\eqref{eqn:y-param}, including the $\Lambda$CDM contribution from $T_m^{(0)} \neq T_\text{CMB}$, reads in the limit where $J \gg 1$
\begin{alignat}{1}
    y \approx \int_0^t dt \, \left(\frac{\dot{T}_m^\text{inj}}{H J} - \frac{T_\text{CMB}}{J}\right) \frac{\sigma_T n_e}{m_e} \,,
\end{alignat}
an expression that we use for $J > 100$ to avoid numerical difficulties associated with $T_m$ being close to $T_\text{CMB}$. 
Here, $\Lambda$CDM refers to the standard cosmological model, within which dark matter neither decays nor annihilates.
We note that depending on the size of the energy injection, the $y$-distortion can take on negative values, as is expected in $\Lambda$CDM cosmology, since the matter temperature is always slightly colder than the CMB at early times, see e.g. Eq.~\eqref{eqn:analytic_Tm}. 
For $J \leq 100$ and after recombination, we use the more general Eq.~\eqref{eqn:y-distortion}, with $T_m$ solved using the usual machinery of \dhis to compute the $y$-distortion. 

\subsubsection{Validation}
\label{sec:y-validation}
\begin{figure}
	\centering
	\includegraphics[width=0.5\textwidth]{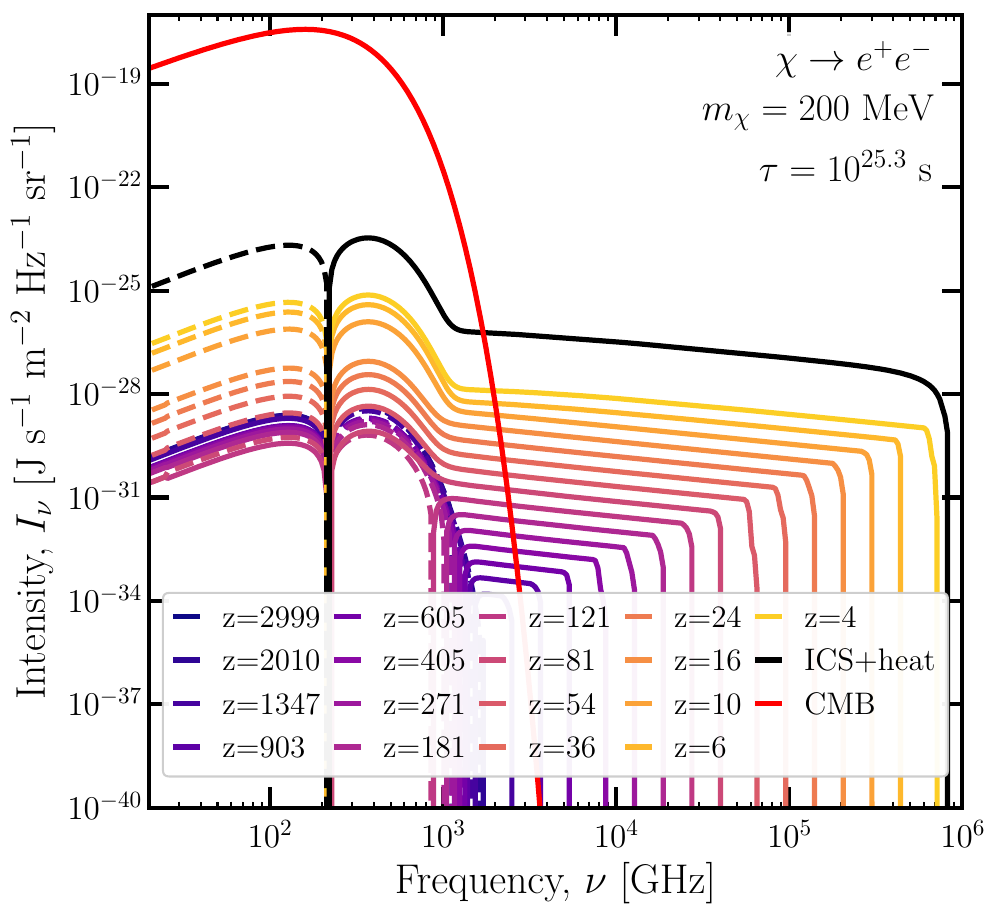}
	\caption{
		Contributions to the spectral distortion from ICS and heating at various redshifts as well as the total distortion.
		We also show the CMB spectrum in red for comparison.
		For visual clarity, we only show every 20th redshift step used to generate this spectral distortion and each contribution has been redshifted to $z=0$.
		The high energy cutoff of each contribution is due to photons with $\omega > \mathcal{R}$ getting absorbed through ionization.
	}
	\label{fig:redshift_dists}
\end{figure}

The spectral distortion from ICS and heating is calculated by first taking the low-energy photon spectra that are output at each redshift step, which we denote by, $J_\text{init} (\omega)$, and then adding in the component from heating using Eq.~\eqref{eqn:y-distortion}; we call their sum $J_\text{new} (\omega)$.
The total distortion from these cooling processes is then given by redshifting the contributions from each step to $z=0$ and summing up the resulting spectra.
Fig.~\ref{fig:redshift_dists} demonstrates this with the example of dark matter with a mass of 200 MeV and a lifetime of $10^{25.3}$ s decaying to $e^+ e^-$ pairs; the colorful lines ranging from blue to yellow show the contribution to the spectral distortion from each redshift step, with each spectrum redshifted to $z=0$.
The dominant component is the $y$-type distortion from the difference in the matter and radiation temperatures, which changes sign around $1+z=155$ when the matter and radiation temperatures decouple.
The black line then shows the sum of all these contributions.
The photons that are being upscattered by the high-energy electron decay products come from the CMB; hence, the troughs of the distortions are located near the same energy as the peak of the CMB blackbody spectrum.
In addition, a photon with initital energy $\omega_0$ will on average upscatter to an energy of $\langle \omega \rangle = \frac{4}{3} \gamma^2 \omega_0$, where $\gamma$ is the Lorentz boost factor for the relativistic electron; hence, the peak of the spectral distortion resulting from ICS is located at $\frac{4}{3} \gamma^2$ times the average energy of photons in the trough~\cite{Blumenthal:1970gc}.

We have also compared the spectral distortions calculated using our method to the Green's functions in Ref.~\cite{Acharya:2018iwh}.
A key difference between our work and Ref.~\cite{Acharya:2018iwh} is that they assume the ionization level is 100\%, so that there is no possibility for photoionization or excitation; at most of the early redshifts they are concerned with, this is a good assumption. 
However, at the relatively later redshifts that we are interested in, there is a small amount of neutral hydrogen prior to recombination, and secondary electrons resulting from the ionization of this hydrogen can deposit a non-negligible amount of energy into heating.
Hence, we find that our spectral distortions are generally larger in amplitude than the Green's functions from Ref.~\cite{Acharya:2018iwh}, see Appendix~\ref{app:cross_checks} for more details.

\subsection{An Improved Treatment of Low-Energy Photons}
\label{sec:evolution}

The final piece we must consider is that of emission and absorption of photons from atomic transitions.
Our goal in this section is to derive and numerically solve the full set of equations that govern the evolution of injected low-energy photons, and their interactions with atoms, as a function of time, and thereby determine their contribution to the CMB spectral distortion.
By ``injected" photons, we not only mean those directly injected through processes such as dark matter annihilating and decaying to photons, but also those produced as secondary particles from the cooling of high-energy particles, e.g. the spectra described in Sections~\ref{sec:ICS} and \ref{sec:y-type}.
These equations self-consistently solve for the evolution of the low-energy photons and atoms as they interact with one another.
\footnote{In principle, at redshifts low enough for molecules to form, interactions between photons and molecules could also become important; however, we defer this topic to future work.}

\subsubsection{Summary of previous work and new features}
\label{sec:past_work}

Our treatment here extends that of our previous work~\cite{DarkHistory}, which used many convenient approximations that are no longer sufficient when we seek to accurately predict the full spectrum of low-energy photons. Here, we discuss the previous approach and briefly highlight the differences, which we will discuss in more detail in the upcoming sections.

In both our previous work and our current treatment, it is true that at each redshift, exotic energy injection can give rise to a spectrum of high-energy photons, either through direct injection of photons or by the cooling of injected electrons.
We employ tables of precomputed transfer functions to determine how these high-energy photons deposit their energy through cooling processes or into low-energy electrons and photons, as well as how many photons propagate to the next redshift step without cooling.

In our previous work, the resulting low-energy photons are then deposited into different channels, depending on their energy.
We track the energy deposited by different processes using the functions $f_c$, defined by
\begin{equation}
	\left( \frac{dE}{dV dt} \right)_c \equiv f_c \left( \frac{dE}{dV dt} \right)_\text{inj},
\end{equation}
where $\left( \frac{dE}{dV dt} \right)_\text{inj}$ is the energy injected per volume per time and $\left( \frac{dE}{dV dt} \right)_c$ is the energy deposited into channel $c$ per volume per time.

Starting at the lowest energies, previously we assumed that non-CMB photons with energy below $E_\alpha$ redshift until today without further interactions, tracking the energy density of these photons with the function $f_\text{cont}$. 
In contrast, with the new treatment described in this work we can account for interactions these photons have with atoms and the resulting perturbations to their spectrum; 
for example, when a low-energy CMB photon of energy $E$ and an injected photon of energy $E_\alpha-E$ are both absorbed by an atom, exciting it to the $2s$ state.

At the next highest energy, we previously assumed that all photons with energy between $E_\alpha$ and $\mathcal{R}$ are instantly absorbed in exciting neutral hydrogen atoms; this absorbed energy was tracked in the function $f_\text{exc}$. 
These excited atoms can absorb a CMB photon and ionize, leading to an ionization rate of:
\begin{alignat}{1}
\frac1{n_\text{H}} \left( \frac{dE}{dVdt}\right)_\text{inj} (1-\mathcal{C}) \frac{f_\text{exc}}{E_\alpha},
\label{eq:exc_to_ion}
\end{alignat}
where $\mathcal{C}$ is the Peebles-$\mathcal{C}$ factor~\cite{Peebles:1968ja} and $\left( \frac{dE}{dVdt}\right)_\text{inj}$ is the exotic energy injection rate per unit volume and time. 
This treatment was also approximate, and did not allow us to keep track of the addition of photons through de-excitation of atoms, nor the subtraction of photons through photo-excitation. 
In the current work we expand the calculation to allow us to keep track of the detailed spectrum of photons absorbed and emitted through such excitations and de-excitations, and also refine the ionization rate of Eq.~\eqref{eq:exc_to_ion}.

Finally, low-energy photons with energies above $\mathcal{R}$ were assumed to all ionize hydrogen atoms (or helium atoms, depending on options set within \texttt{DarkHistory}) and their energy was tracked through the function $f_\text{H ion}$. 
This contributes to the ionization rate through the term:
\begin{alignat}{1}
	\frac{f_\text{H ion}}{\mathcal{R} n_\text{H}} \left( \frac{dE}{dVdt}\right)_\text{inj} .
	\label{eq:just_ion}
\end{alignat}

Our previous approach, based on the simple energy accounting described above, treated hydrogen atoms using the 
Three-Level Atom (TLA) model~\cite{Peebles:1968ja, Zeldovich:1969en}. 
This TLA-based treatment assumed that the background radiation bath was a blackbody, and that all excited states above $n=2$ were in perfect thermal equilibrium. 
In our current work, consistent with tracking the detailed interactions of low-energy photons with the gas, we account for the effect that a non-thermal background of photons and a non-equilibrium population of excited states have on the recombination and ionization rates. 

Specifically, to keep track of the higher-$n$ abundances, we drop the TLA approximation and generalize to the Multi-Level Atom (MLA) 
treatment~\cite{Seager:1999km, Chluba:2006bc,2010PhRvD..81h3005G,Chluba:2010fy}. 
We extend the MLA in two ways. 
First, we allow the phase space density of photons to be non-thermal, allowing changes in the number density of photons to alter the atomic transition rates.
Second, we add source terms due to excitations and ionizations caused by non-thermal injected particles.
In Appendix~\ref{app:cross_checks}, we show that at redshifts where the TLA assumptions hold, our extended MLA treatment and the earlier TLA method yield the same results.

\subsubsection{Notation and conventions}
\label{sec:dist_exc_intro}

Here, we define the CMB spectral distortion and describe our notation for atomic states and transition rates.
Since we use natural units, we will hereafter denote photon energy by $\omega$.
Denote the full photon phase space density as $f^\gamma(\omega, t)$, and assume that it is homogeneous and isotropic. 
We assume that inhomogeneities at early times are negligible, and ignore any inhomogeneities introduced at late times due to structure formation. 
It is convenient to separate out the dominant blackbody contribution from the rest of the phase space density, 
\begin{alignat}{1}
f^\gamma(\omega, t) = f^\text{CMB}(\omega,t) + \Delta f(\omega,t) \, ,
\label{eq:f_gamma} 
\end{alignat}
where $f^\text{CMB}(\omega, t) = \left[ \exp(\omega/T_\text{CMB}(t))-1 \right]^{-1}$ and $\Delta f(\omega,t)$ is the deviation from the blackbody distribution. 
From now on, we will refer to $\Delta f(\omega,t)$ as the CMB distortion (although note that it may have support at energies far from the peak of the CMB spectrum).
We also define the photon spectrum
\begin{alignat}{1}
N_\omega = \frac{g_\gamma \omega^2}{2 \pi^2 n_\text{B}} f^\gamma(\omega, t) \, ,
\label{eq:dNdE_to_f}
\end{alignat}
which is the number of photons per baryon per energy $\omega$, where $n_\text{B}$ is the baryon density and $g_\gamma=2$ is the number of degrees of freedom for the photon. 
One can check the above set of definitions by ensuring that the number density of photons is equal to
$g_\gamma \int \frac{d^3p}{(2\pi)^3} f^\gamma = n_\text{B} \int d\omega \, N_\omega$. 
The photon spectrum is also related to the spectral radiance by
\begin{equation}
N_\omega = \frac{4\pi}{n_b \omega} I_\omega .
\end{equation}

We will denote bound states of the hydrogen atom either by a subscript $i$ for generic states or subscript $nl$ when the quantum numbers are specified. 
We do not need to consider the magnetic quantum number $m$ in any of our calculations since any splitting effects would produce negligibly small effects at our working precision. 
The existence of $m$ substates is taken into account by multiplicity factors for each $nl$ state.

We denote the radiative transition rate from state $i$ to state $j$ as $R_{i \to j}$, the radiative recombination coefficient to state $i$ as $\alpha_i$, and the  photoionization rate from state $i$ as $\beta_i$; these quantities depend on the photon spectrum and $T_m$.
The energy difference corresponding to $R_{i \to j}$ is labeled as $\omega_{i \to j}$.
We have defined $R_{i \to j}$ to be the rate of transitions per second per hydrogen atom as in Ref.~\cite{2010PhRvD..82f3521A} without normalizing per volume as in Refs.~\cite{Wong:2005yr,Rubino-Martin:2006hng}, and we define $R_{i \to i} = 0$.
When calculating $R_{i \to j}$, we ignore numerous percent-level contributions that are included in precise recombination 
codes~\cite{Seager:1999bc, Seager:1999km, Chluba:2006bc,2009A&A...503..345C,2010MNRAS.402.1221C, AliHaimoud:2010dx}; the net effect of these is small, since we are able to reproduce the atomic lines from recombination at percent-level (see Section~\ref{sec:atom-cross-checks} for details).
For example, we ignore any collisional transitions from thermal electrons, focusing on only radiative transitions. 
While we include all dipole transitions and the single quadrupole transition $2s \to 1s$, we ignore all other quadrupole or higher-order transitions. 
We treat all resonances as perfect lines, ignoring any line broadening or other radiative transfer effects.
In Appendix~\ref{app:atomic_rates} we show how to calculate these rates given an arbitrary $f^\gamma(\omega)$ using the methods of Ref.~\cite{2010PhRvD..82f3521A}.

\subsubsection{General Evolution Equations}

The most general form of the equations that govern the population of each hydrogen atomic state $i = $ 1s, 2s, 2p, $\cdots$ up to states with principal quantum number $n = n_{\max}$ can be written as~\cite{AliHaimoud:2010dx}
\begin{align}
    \dot{x}_{1s} &= \sum_{k > 1s}^{n_{\max}} (x_k R_{k \to 1s} - x_{1s} R_{1s \to k}) + x_e^2 n_\text{H} \alpha_{1s} - x_{1s} \beta_{1s} - \dot{x}_\text{inj}^{\text{ion}} \,, \nonumber \\
    \dot{x}_{l > 1s} &= \sum_{j \neq l}^{n_{\max}} (x_j R_{j \to l} - x_l R_{l \to j}) + x_e^2 n_\text{H} \alpha_l - x_l \beta_l \,,
    \label{eq:general_xi_evolution}
\end{align}
where $x_i \equiv n_i / n_\text{H}$ is the ratio of the number density of hydrogen atoms in state $i$ to $n_\text{H}$; all summations over hydrogen states in this work are taken up to $n = n_{\max}$, and we will suppress indicating this maximum value in the summation for ease of notation. 
We will also reserve $i$, $j$ as indices to represent all states, while $k$ and $l$ represent only excited states. 
$\dot{x}_{\text{inj}}^{\text{ion}}$ is the net contribution to hydrogen ionization from exotic energy injection, including collisional ionization from secondary electrons, and photoionization from photons with $\omega \geq \mathcal{R}$; for $\omega < \mathcal{R}$, we include the effect of injected photons in $\alpha_i$, $\beta_i$ and $R_{i \to j}$, as we will detail below. Note that we assume that all exotic photoionization events occur with a hydrogen atom in the ground state.
In principle, we could attribute some of these ionizations to excited states, which would alter the amount of energy deposited into ionization; however, since the ground state is exponentially populated compared to the excited states throughout recombination, we choose to consider only ground state ionizations.
Differentiating the relation $x_e + \sum_i x_i = 1$ with respect to time, we can determine the evolution of the free electron fraction, 
\begin{alignat}{1}
    \dot{x}_e = - x_e^2 n_\text{H} \sum_i \alpha_i + \sum_i x_i \beta_i + \dot{x}_\text{inj}^\text{ion} \,.
    \label{eq:general_xe_evolution}
\end{alignat}
Since all of the atomic bound-bound rates and photoionization coefficients are functions of the photon spectrum, we also have to specify how the photon spectrum with $\omega \leq \mathcal{R}$ evolves. 
Throughout this section, we only track distortions to the CMB; any subsequent mention of a spectrum of photons should be understood as a distortion (both positive and negative) to the CMB. 
The evolution of the photon spectrum can be written as
\begin{alignat}{1}
    \dot{N}_\omega = - H \omega \frac{d N_\omega}{d \omega} + J(\omega) \,.
    \label{eq:general_photon_spec_evolution}
\end{alignat}
The first term on the right-hand side accounts for the redshifting of the spectrum, while $J(\omega)$ is the net number of photons injected per baryon, per energy, and per unit time. $J(\omega)$ includes the absorption or emission of photons from bound-bound transitions, recombination/photoionization, and exotic injection: 
\begin{alignat}{2}
    J(\omega) &=&& \frac{n_\text{H}}{n_\text{B}} \sum_i \sum_{j < i} (x_i R_{i \to j} - x_j R_{j \to i}) \delta_D(\omega - \omega_{i \to j}) \nonumber \\
    & &&+ \frac{n_\text{H}}{n_\text{B}}\sum_i \left[ x_e^2 n_\text{H} \gamma_i(\omega) - x_i \xi_i(\omega) \right] + J_\text{new}(\omega) \,. 
    \label{eqn:j_omega}
\end{alignat}
For bound-bound transitions in the first term, we take the spectrum of photons emitted/absorbed to be a line with $\delta_D$ being a Dirac-delta function, since the spectrum is much narrower than our binning. Next, every bound-free transition to/from state $i$ which produces or absorbs a photon of energy $\omega$ also produces or removes a free electron with kinetic energy defined as $\kappa^2 \mathcal{R}$, where $\kappa^2$ is a positive real number. The recombination spectrum is therefore proportional to $x_e^2 n_\text{H}$, with 
\begin{alignat}{1}
    \gamma_i(\omega) \equiv \int_0^\infty d \kappa^2 \, \frac{d \alpha_i}{d \kappa^2} \delta_D(\omega - (\kappa^2 + 1 / n_i^2) \mathcal{R}) \, ,
\end{alignat}
where $n_i$ is the principal quantum number of state $i$.
Similarly, the spectrum of photons absorbed in photoionization is proportional to $x_i$, with 
\begin{alignat}{1}
    \xi_i(\omega) \equiv \int_0^\infty d \kappa^2 \, \frac{d \beta_i}{d \kappa^2} \delta_D(\omega - (\kappa^2 + 1 / n_i^2) \mathcal{R}) \,.
\end{alignat}
$J_\text{new}(\omega)$ is the number of photons deposited per baryon, per energy, and per unit time; our upgraded version of \dhis computes precisely this low-energy photon spectrum given a source of injected energy. 
It receives contributions from ICS of injected high-energy electrons off the CMB (see Section~\ref{sec:ICS}) and from $y$-distortions due to heating of the IGM (see Section~\ref{sec:y-type}). 
We note that previous versions of \dhis only explicitly tracked the contribution from ICS, without including the backreaction of these distortions on bound-bound and bound-free transitions. 

To close these equations, we use Eq.~\eqref{eq:general_Tm_evolution} as the evolution equation for the IGM temperature.
Eqs.~\eqref{eq:general_xi_evolution},~\eqref{eq:general_xe_evolution},~\eqref{eq:general_photon_spec_evolution} and~\eqref{eq:general_Tm_evolution} are complete and can be solved to determine the joint evolution of hydrogen atoms and radiation in the presence of exotic energy injection. However, there are vast differences in timescales in the various terms that both make it difficult to solve this general set of equations, and also allow the use of well-known simplifications to reduce the equations into a more tractable form~\cite{Peebles:1968ja,2010PhRvD..82f3521A,AliHaimoud:2010dx}.

\subsubsection{Simplified Evolution Equations}
\label{sec:steady_state}

\textit{Rapid photoionization}
\vspace{2mm}

The first approximation arises from the fact that once an appreciable population of bound neutral hydrogen forms, which occurs at around $1+z \sim 1500$, the mean free time of a photoionizing photon is short compared to the Hubble timescale, leading to two effects, both outlined in Ref.~\cite{Peebles:1968ja}. 
First, the distribution of photons with $\omega > \mathcal{R}$ is driven strongly toward equilibrium with the IGM, which has a temperature $T_m \ll \mathcal{R}$, implying that all photoionizing photons produced by exotic injections are to a good approximation absorbed. This allows us to write
\begin{alignat}{1}
    \dot{x}_\text{inj}^\text{ion} = \frac{f_\text{H ion}(z, x_e)}{\mathcal{R} n_\text{H}} \left(\frac{dE}{dV \, dt}\right)_\text{inj} \,,
    \label{eqn:x_inj_term}
\end{alignat}
where the right-hand side gives the number of ionizations from exotic energy injections per H nucleus per unit time.
Second, for every recombination to the ground state, an ionizing photon is emitted that is immediately absorbed by another atom in the ground state, leading to no net recombination; this is known as case-B recombination. 
This allows us to drop $- x_e^2 n_\text{H} \alpha_{1s} + x_\text{1s} \beta_{1s}$ in Eq.~\eqref{eq:general_xi_evolution}, which cancel each other down to a residual much smaller than the Hubble rate~\cite{Peebles:1968ja}. 
This also simplifies the $x_e$ evolution equation to 
\begin{alignat}{1}
    \dot{x}_e &= - x_e^2 n_\text{H} \sum_{k > 1s} \alpha_k + \sum_{k > 1s} x_k \beta_k + \dot{x}_\text{inj}^\text{ion}, 
    \label{eq:simplified_intermediate_xe}
\end{alignat}
which depends on the properties of the IGM. 
Throughout, we assume that free electrons are distributed thermally with temperature $T_m$.

\vspace{3mm}
\noindent\textit{Rapid Lyman-series transitions}
\vspace{2mm}

Let us turn now to Lyman-series, i.e.\ $1s \to np$ photons. When an appreciable population of ground state hydrogen forms, the mean free path of Lyman series radiation is, as for photoionizing photons, much shorter than the Hubble length. As with photoionizing photons, Lyman-series photons are driven strongly toward equilibrium with the IGM, so that any exotic injections of Lyman-series photons are to a good approximation absorbed rapidly. Furthermore, any $np \to 1s$ transition produces a Lyman-series photon that is quickly reabsorbed, significantly suppressing the effectiveness of $np \to 1s$ processes in depleting the $np$ state. Unlike recombination photons, however, Lyman-series photons are emitted/absorbed in an extremely narrow frequency range, granting emitted photons the opportunity to redshift out of the narrow Lyman-series resonances via the first term in Eq.~\eqref{eq:general_photon_spec_evolution}. The net rate of transitions from $np$ to $1s$ can be properly accounted for by using the Sobolev approximation~\cite{Seager:1999km}, where the net rate is multiplied by the probability of a Lyman-series photon redshifting out of a line, $p_{np} \equiv [1 - \exp(-\tau_{np})] / \tau_{np}$, where $\tau_{np}$ is the Sobolev optical depth for a photon with energy corresponding to the $1s \to np$ transition. 

To take both effects into account in our calculation, we define
\begin{alignat}{1}
    \tilde{R}_{i \to j} \equiv \begin{cases}
        p_{np} R_{i \to j} \,, & i = 1s, j = np \text{ or } j=1s, i=np \,, \\
        R_{i \to j} \,, & \text{otherwise} \,,
    \end{cases}
\end{alignat}
and replace $R_{i \to j} \to \tilde{R}_{i \to j}$ in Eq.~\eqref{eq:general_xi_evolution}. We also assume that all Lyman-series photons $J_{np}$ from exotic injection are rapidly absorbed, leading directly to $1s \to np$ excitations. These photons are therefore not included in $J(\omega)$, but instead directly contribute to a term $\dot{x}_{\text{inj}, np}^\text{exc}$, where
\begin{alignat}{1}
    \dot{x}_{\text{inj},np}^\text{exc} \equiv \frac{n_\text{B}}{n_\text{H}} J_{np} + \dot{x}_{np}^{\text{coll. exc.}}\,. 
\end{alignat}
The term $\dot{x}_{np}^{\text{coll. exc.}}$ represents the contribution of collisional excitation from low-energy electrons.
Note that we do not include this correction for resonant photons that connect two excited states, since the abundance of excited states is extremely suppressed compared to the abundance of the ground state, and resonant lines between two excited states do not interact strongly enough to drive these photons to equilibrium with the IGM. Further details on this treatment can be found in Appendix~\ref{app:atomic_rates}. 

At this point, including both the approximations for fast photoionization and fast Lyman-series scattering, the evolution equations for $x_i$ read
\begin{alignat}{2}
    \dot{x}_{1s} &=&& \sum_{k > 1s} (x_k \tilde{R}_{k \to 1s} - x_{1s} \tilde{R}_{1s \to k} - \dot{x}_{\text{inj},k}^\text{exc}) - \dot{x}_\text{inj}^\text{ion} \,, \nonumber \\
    \dot{x}_{l>1s} &=&& \sum_{j \neq l} (x_j \tilde{R}_{j \to l} - x_l \tilde{R}_{l \to j}) + x_e^2 n_\text{H} \alpha_l - x_l \beta_l + \dot{x}_{\text{inj},l}^\text{exc}\,, 
    \label{eq:x_i_after_rapid_ion_exc_approx}
\end{alignat}
where we set $\dot{x}_{\text{inj},l \neq np}^\text{exc} \equiv 0$ for ease of notation. 

\vspace{3mm}
\noindent\textit{Steady state approximation}
\vspace{2mm}

The next significant approximation we make is the steady-state approximation for all excited states~\cite{2010PhRvD..81h3005G,2010PhRvD..82f3521A,Hirata:2008ny}. 
The total rate for transitioning out of an excited state $k$ is
\begin{alignat}{1}
    \tilde{R}_{k}^\text{out} \equiv \sum_j \tilde{R}_{k \to j} + \beta_k \,,
\end{alignat}
which is much faster than the Hubble rate at all redshifts~\cite{2010PhRvD..82f3521A,AliHaimoud:2010dx}.
The populations of the excited states in the hydrogen atom are therefore driven toward a fixed point for each $x_k$ set by $\dot{x}_k = 0$; any differences from these fixed points are rapidly erased on timescales much shorter than a Hubble time. This steady state approximation reduces the population of the atomic states to a set of algebraic equations of the form $x_k = M^{-1}_{kl} b_l$, with $M_{kl}$ and $b_l$ are objects indexed by excited states, evolving on Hubble timescales.  

Let us now explicitly write out the matrix $M$ and the inhomogeneous term $b$. Setting $\dot{x}_i = 0$ in Eq.~\eqref{eq:x_i_after_rapid_ion_exc_approx} and moving the negative terms to the right hand side, we obtain for the excited states
\begin{alignat}{1}
    x_k \tilde{R}_k^\text{out} = \sum_{l > 1s} x_l \tilde{R}_{l \to k} + x_{1s} \tilde{R}_{1s \to k} + x_e^2 n_\text{H} \alpha_k + \dot{x}_{\text{inj},k}^\text{exc} \,,
\end{alignat}
which simply equates the rate of leaving the state $i$ to the rate of all transitions into the state $i$. This is a linear system in excited states $x_k$, which we can now invert. To make the relation between this expression and previous results clear, however, we write the solution as
\begin{alignat}{1}
    x_k = \sum_{l > 1s} M^{-1}_{kl} (b_l^{1s} + b_l^\text{rec} + b_l^\text{inj}) \,,
    \label{eq:matrix_MLA}
\end{alignat}
where 
\begin{alignat}{1}
    M_{kl} &= \delta_{kl} \tilde{R}_k^\text{out} - \tilde{R}_{l \to k} \,, \nonumber \\
    b_l^{1s} = x_{1s} \tilde{R}_{1s \to l} \,, &\quad b_l^\text{rec} = x_e^2 n_\text{H} \alpha_l \,, \quad b_l^\text{inj} = \dot{x}_{\text{inj},l}^\text{exc} \,.
    \label{eq:M_and_b}
\end{alignat}
where $\delta_{kl}$ is the Kronecker delta function. The three source terms populate the excited state $l$ in three different ways: $b_l^{1s}$ via excitations from the ground state sourced by $f^\gamma (\omega,t)$, $b_l^\text{rec}$ via recombinations, and $b_l^\text{inj}$ via interactions with injected particles, including photoexcitations by Lyman-series photons and collisional excitations by injected electrons. 

With the aid of the identity
\begin{alignat}{1}
    \beta_k = \tilde{R}_k^\text{out} - \tilde{R}_{k \to 1s} - \sum_{l > 1s} \tilde{R}_{k \to l} = \sum_l M_{lk} - \tilde{R}_{k \to 1s} \,,
\end{alignat}
we can now also simplify Eq~\eqref{eq:simplified_intermediate_xe} to obtain 
\begin{alignat}{1}
    \dot{x}_e = - x_e^2 n_\text{H} \tilde{\alpha}_\text{B} + x_{1s} \tilde{\beta}_\text{B} + \dot{x}_\text{inj} \,,
    \label{eq:final_xe}
\end{alignat}
where
\begin{alignat}{1}
    \tilde{\alpha}_\text{B} &= \sum_{k > 1s}  Q_k \alpha_k \,, \nonumber \\
    \tilde{\beta}_\text{B} &= \sum_{k > 1s} (1 - Q_k) \tilde{R}_{1s \to k} \,, \nonumber \\
    \dot{x}_\text{inj} &= \sum_{k > 1s} (1 - Q_k) \dot{x}_{\text{inj}, k}^\text{exc} + \dot{x}_\text{inj}^\text{ion} \,, \nonumber \\
    Q_k &= \sum_{l} M_{lk}^{-1} \tilde{R}_{l \to 1s} \,. 
\end{alignat}
$Q_k$ is a weight that determines how effective recombination to the ground state is from state $k$; if $Q_k = 1$ for all $k$, then recombination to state $k$ ultimately always results in a ground state. 
$Q_k$ is related to the Peebles-$\mathcal{C}$ factor in the three-level atom model. 
Moreover, we use the suggestive notation $\tilde{\alpha}_\text{B}$ and $\tilde{\beta}_\text{B}$ because under the assumptions of the three-level atom, these quantities are closely related to the case-B recombination and photoionization rates.

Note that since $Q_k$ is very close to the identity, the quantity $1-Q_k$ can be very slow to compute.
In Appendix~\ref{app:matrix_method}, we outline an alternative procedure for calculating these quantities that is numerically faster and more in line with what is done in the code.
Further discussion of these equations and how they reduce to the three-level atom model can be found in Appendix~\ref{app:derivation_TLA}. 

To close these equations, we use the approximation $x_{1s} \approx 1 - x_e$, which is an excellent approximation, since we find that the excited states are typically populated at the level of $10^{-13}$ of the ground state shortly after recombination (See e.g. Fig.~\ref{fig:MLA_vs_TLA} in Appendix~\ref{app:cross_checks}). 
With this approximation, the only quantity we have to track explicitly is $x_e$, although given $x_e$ and the photon spectrum, the population of all hydrogen states can be computed. We stress that the effect of exotic particle injection on the ionization term is not solely contained in $\dot{x}_\text{inj}$, but also manifests itself in the fact that low-energy photons arising from the injection affect $\tilde{\alpha}_\text{B}$ and $\tilde{\beta}_\text{B}$, which both depend on the photon spectrum. Altogether, the equations we finally solve are
\begin{alignat}{1}
    \dot{x}_e &= - x_e^2 n_\text{H} \tilde{\alpha}_\text{B} + (1 - x_e) \tilde{\beta}_\text{B} + \dot{x}_\text{inj} + \dot{x}^\text{re} \,, \nonumber \\
    \dot{T}_m &= - 2 H T_m + \Gamma_C(T_\text{CMB} - T_m) + \dot{T}_m^\text{inj} + \dot{T}_m^\text{re} \,, \nonumber \\
    \dot{N}_\omega &= - H \omega \frac{d N_\omega}{d\omega} + J(\omega) \,.
    \label{eq:full_eqs}
\end{alignat}
We have now included $\dot{x}^\text{re}$ and $\dot{T}_m^\text{re}$ to account for ionization and heating from reionization sources at late times; the definition of these terms is unchanged relative to \dhis\texttt{v1.0}~\cite{DarkHistory}.
\footnote{In principle, one should also modify the Multi-Level Atom (MLA) treatment of the excited hydrogen states to include the radiation fields that cause reionization in the first place. Since we have not tested our numerical method described in Section~\ref{sec:numerical MLA} with these radiation fields, we leave including this source to future work.} 

Note that here we have omitted the helium ionization equations; they have also not changed from our previous treatment~\cite{DarkHistory}. While atomic transitions between states of helium could in principle also contribute to the spectral distortion, energy deposition into helium ionization and excitation is subdominant to the hydrogen contribution, so we expect the effect on the photon spectrum to be small, see e.g. Ref.~\cite{Rubino-Martin:2007tua}. To briefly summarize, the helium ionization equation is a sum of three contributions: 1) the expected contribution in the absence of energy injection, which is identical to the \texttt{Recfast} treatment~\cite{Seager:1999bc}, 2) a source term from processes that are active at reionization, and 3) a term from exotic energy injection which is analogous to the hydrogen ionization term given in Eq.\eqref{eqn:x_inj_term} but with $f_\text{H ion} \rightarrow f_\text{He ion}$, where $f_\text{He ion}$ is the fraction of injected energy that is deposited into helium ionization, and $\mathcal{R} \rightarrow \mathcal{R}_\text{He} = 24.6$ eV.
\dhis includes a few possible methods to treat the contributions of low-energy photons to $f_\text{He ion}$ that bracket the uncertainties on how helium ionization proceeds.
The default treatment is to ascribe the energy from all photoionizations to hydrogen, so the only helium ionizations occur through collisional ionizations by electrons.
While this may not be the most accurate treatment, it is the simplest and is valid well before reionization.

\subsubsection{Numerical Method}
\label{sec:numerical MLA}

%
\begin{figure*}
	\centering
	\includegraphics[width=\textwidth]{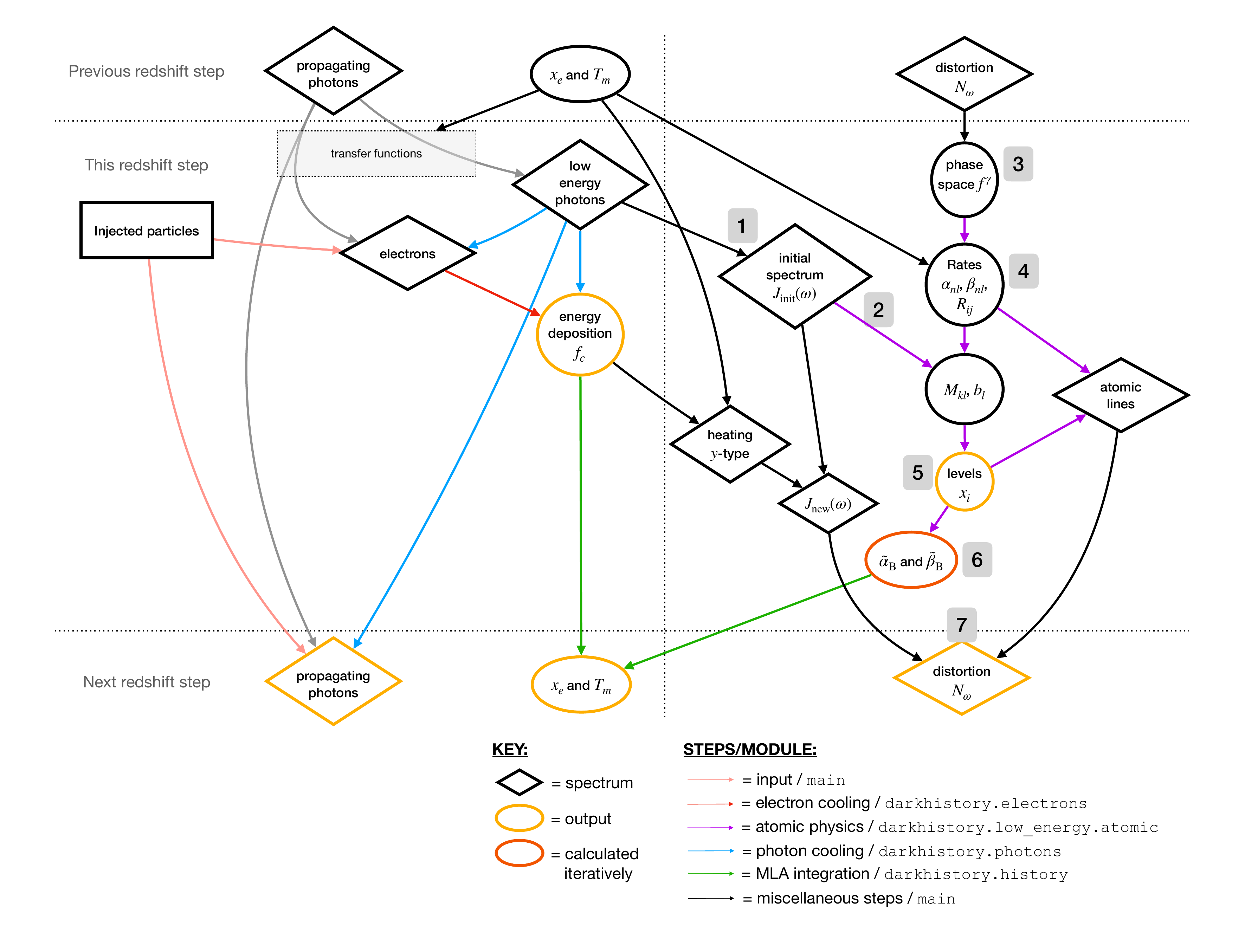}
	\caption{
		Flowchart summarizing our numerical procedure. 
		The outputs of \dhis are highlighted in the yellow shapes.
		We also highlight $\tilde{\alpha}_\text{B}$ and $\tilde{\beta}_\text{B}$ in orange-red to indicate that this quantity is calculated iteratively, as described at the end of Section~\ref{sec:numerical MLA}.
		The right side highlights the new quantities and atomic processes that we track in \texttt{DarkHistory}, as described in Section~\ref{sec:evolution}.
		The numbers label the steps outlined in Section~\ref{sec:numerical MLA}.
	}
	\label{fig:flowchart}
\end{figure*}

We now describe our numerical procedure for integrating the system of differential equations in Eq.~\eqref{eq:full_eqs}, and focus in particular on how we obtain the spectral distortion.
This procedure is implemented in \texttt{main.evolve()} provided the \texttt{distort} option is set to |True|.

Initially, the photon phase space density is set to its blackbody value, so $\Delta f(\omega,t) = 0$; the ionization level is set to its Saha equilibrium value; and the initial matter temperature is determined using the early time analytic formula, Eq.~\eqref{eqn:analytic_Tm}.
The initial redshift for the integration should be early enough to justify using the above initial conditions; by default, we use $1+z_\text{init} = 3000$.

To integrate from the initial redshift to today, we use a fixed step size in log redshift space, which by default is set to $\Delta \ln(1+z) = .001$. 
At each redshift step, we perform the following calculations.
\begin{enumerate}
	\item Calculate the initial spectrum of injected low-energy photons, $J_\text{init} (\omega)$.
	Note that this does not include directly injected photons, but rather photons that are deposited as result of the cooling of directly injected particles.
	\item Zero out the ionizing bins in the above spectrum with bin centers above $\mathcal{R}$ and the Lyman-series bins that contain energies $\mathcal{R} ( 1 - n^{-2})$, for $n \in \{1, 2,...,n_\text{max}\}$, accounting for the energy in these bins by modifying $f_\text{H ion}$ and the $b_l^\text{inj}$ term of Eq.~\eqref{eq:matrix_MLA}.
	\item Use the full spectrum of low-energy photons, $N_\omega$, to calculate the photon phase space density using Eq.~\eqref{eq:dNdE_to_f}.
	This spectrum includes the total spectral distortion accumulated up to this redshift.
	\item Compute the transition rates $\tilde{R}_{i \to j}$, $\alpha_i$, and $\beta_i$ using the equations within Appendix~\ref{app:atomic_rates} in the presence of a non-zero $\Delta f(\omega,t)$.
	\item Set $x_{1s} = 1-x_e$, solve for $M$ and $b$ using Eq.~\eqref{eq:M_and_b}, then determine the population levels $x_i(z)$ by inverting Eq.~\eqref{eq:matrix_MLA}.
	\item Use $M$, $\tilde{R}_{i \to j}$, $\alpha_i$, and $\beta_i$ to compute $\tilde{\alpha}_\text{B}$, $\tilde{\beta}_\text{B}$, and $\dot{x}_\text{inj}$,
	then solve for $x_e$ and $T_m$ at the end of the redshift step by integrating the first two equations of Eq.~\eqref{eq:full_eqs} through the redshift step,
	\item Using the occupation numbers $x_i(z)$ and rates already computed, calculate the photons produced from atomic processes using Eq.~\eqref{eqn:j_omega}.
	In addition, use $T_m$ to determine the amplitude of the $y$-type distortion contributed at this redshift step through Eq.~\eqref{eqn:y_DM}; this together with $J_\text{init} (\omega)$ constitutes $J_\text{new} (\omega)$.
	Add the atomic contribution and $J_\text{new} (\omega)$ to $N_\omega$.
	\item Redshift all spectra to the next redshift step.
\end{enumerate}
Looping through these steps evolves all quantities forward in time, gradually building up the spectral distortion $N_\omega$ until we reach the chosen ending redshift; in our analysis, we choose $1+z_\text{end}=4$ since our treatment of ionized helium breaks down past this point.
We then redshift the photon spectrum to today. 
The outputs of this version of the \dhis code are $T_m$ and $x_e$ as a function of time and the present day photon spectrum.
This procedure is summarized in Fig.~\ref{fig:flowchart}.

\begin{figure}
	\centering
	\includegraphics[width=0.5\textwidth]{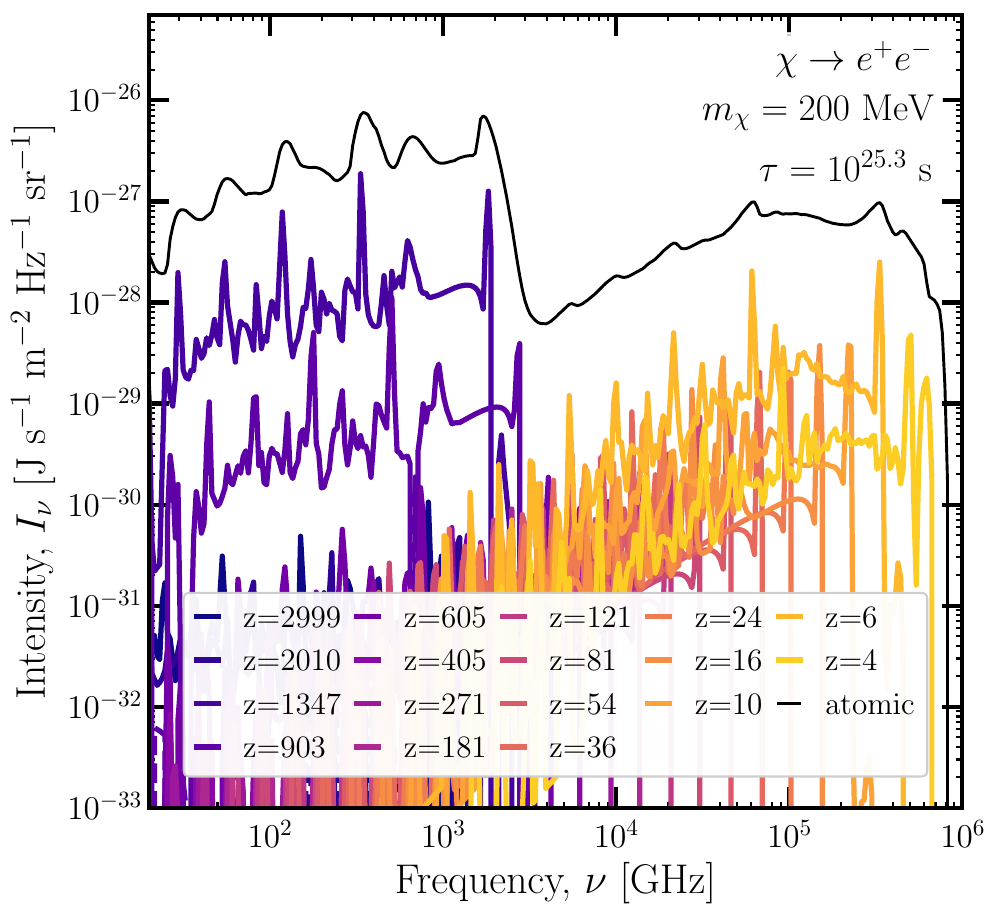}
	\caption{
		Changes to the spectral distortion from atomic transitions at each redshift step, as well as the total distortion.
		For visual clarity, we only show every 20th redshift step used to generate this spectral distortion and each contribution has been redshifted to $z=0$.
	}
	\label{fig:redshift_dists_atomic}
\end{figure}
\begin{figure}
	\centering
	\includegraphics[width=0.5\textwidth]{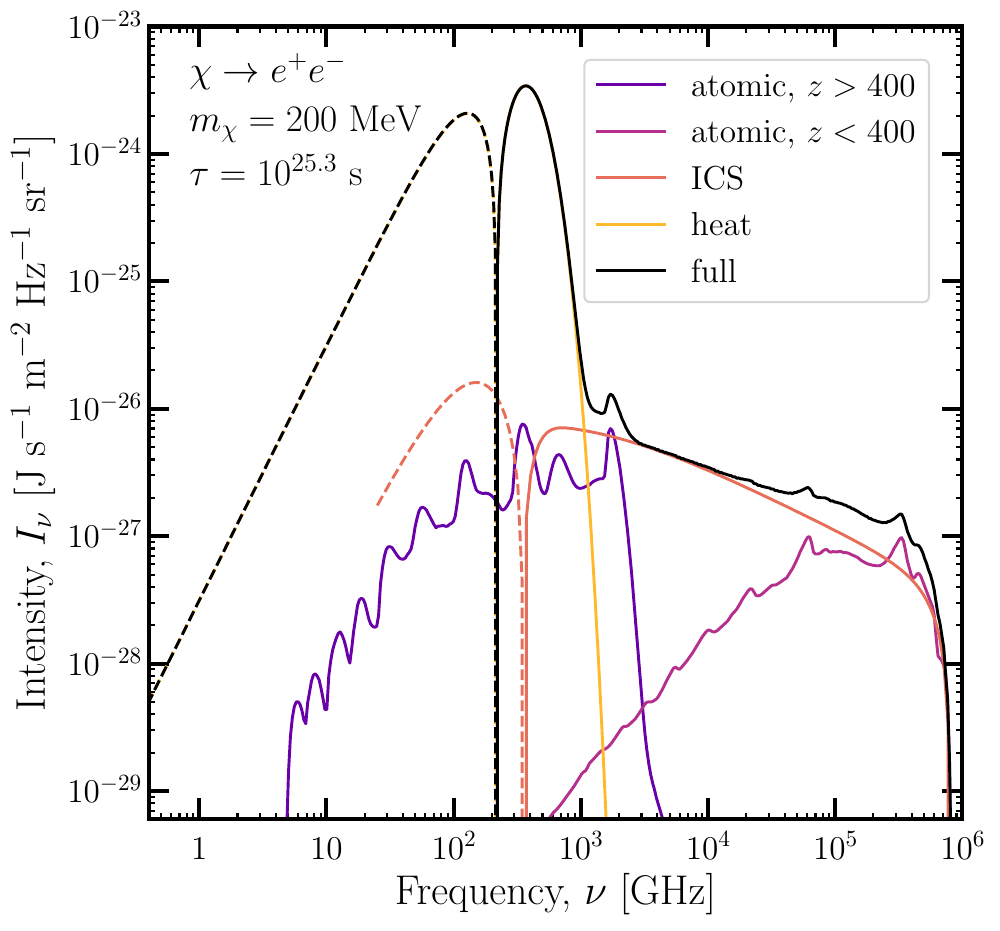}
	\caption{
		The different contributions to the spectral distortion; here we choose dark matter decaying to $e^+ e^-$ pairs, with a mass of $200$ MeV and a lifetime of $10^{25.3}$ s.
		The contributions we show are atomic lines from $z > 400$ which are dominated by the redshifts around recombination, the atomic lines from $z < 400$ which are dominated by the redshifts around reionization, the photons resulting from ICS, and the $y$-type distortion resulting from heating of the IGM.
		The sum of the ICS and heat contributions is the same as the black line in Fig.~\ref{fig:redshift_dists}.
	}
	\label{fig:components}
\end{figure}

In Fig.~\ref{fig:redshift_dists_atomic}, we show the change in the distortion resulting from atomic transitions at each redshift step, including the effect of the full distortion on the atomic states and their recombination/photoionization coefficients.
We also show the final distortion contributed by atomic lines in black.
All spectra shown are redshifted to $z=0$.
The dark matter model used is the same as that in Fig.~\ref{fig:redshift_dists}.

The total spectral distortion for the energy injection model used in Fig.~\ref{fig:redshift_dists_atomic} is shown in Fig.~\ref{fig:components} in black; we also break down the spectral distortion by the photon sources, including ICS, heating, atomic lines from before $z = 400$ (which are mostly from recombination), and atomic lines from after $z=400$ (mostly from reionization).
The component with the largest amplitude is generated by heating.
We note that many of these photons are generated by $\Lambda$CDM processes; in other words, some of this spectral distortion would still be present if we turned off exotic energy injections.

Let us pause to analyze how this procedure self-consistently captures the effect of the radiation on the evolution of the atoms, as well as the atoms' effect on the radiation.  
To capture the effect of radiation on the atoms, we add $\Delta f(\omega,t)$ to the black-body phase space density in our calculation of $\tilde{R}_{i \to j}$, altering the rates of de-excitation, excitation, recombination, and ionization. 
Since $\Delta f(\omega,t)$ can be negative or positive, these rates can be either diminished or enhanced.
The changes in these rates subsequently modifies $M_{kl}$ and the $b_l$ terms, and hence modifies the populations of atomic states and ionization level, $x_i$ and $x_e$.
Beyond the effects on the photon spectrum, we include exotic energy injections in the evolution equations via the term $\dot{x}_\text{inj}$, which captures ionization and excitation from the additional energy injection.

The atoms can in turn affect the radiation if the rates and populations are such that there is net absorption or emission between two atomic states.
This includes the absorption of photons from previous time steps which redshift into resonant lines.
The resulting photons or photon deficit are then added to $\Delta f(\omega,t)$ within the redshift step.

\begin{figure*}
	\centering
	\includegraphics[width=\textwidth]{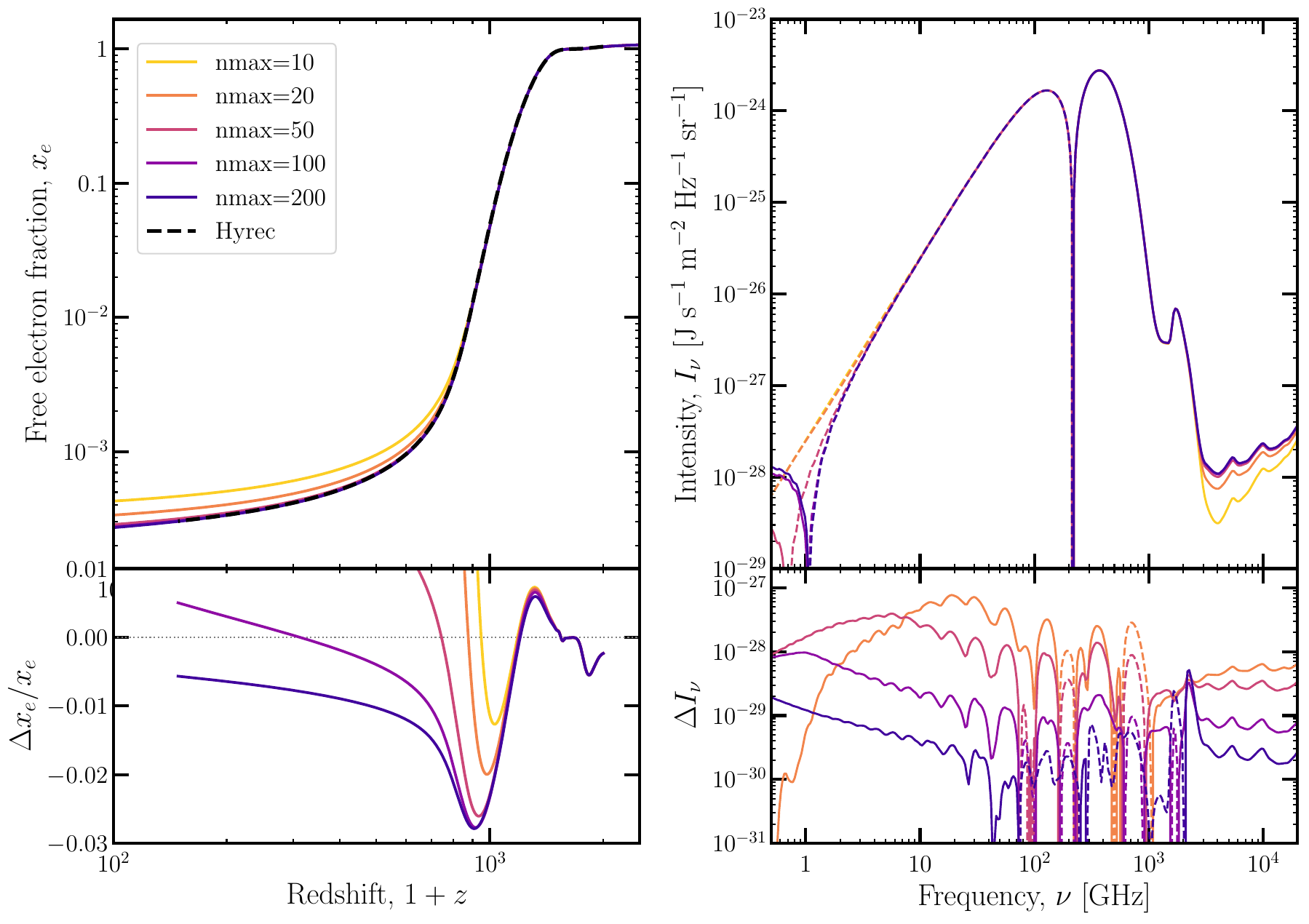}
	\caption{\textit{Left panel:} Convergence of the ionization history with $n_\text{max}$ calculated using our method, compared against the ionization calculated with \texttt{Hyrec}~\cite{AliHaimoud:2010dx} which is shown in the black dashed curve.
	The lower panel shows the percent difference between \dhis at the different values of $n_\text{max}$ and \texttt{Hyrec}. 
	The relative difference between the result with  $n_\text{max}=200$ and \texttt{Hyrec} is largest around recombination, but is still only at the level of a few percent.
	\textit{Right panel:} Convergence of the spectral distortion with $n_\text{max}$.
	The color coding is the same as in the left panel.
	The lower panel shows the change in the spectral distortion between using one value of $n_\text{max}$ and the next smallest value.
	Since this difference decreases as we increase $n_\text{max}$, the spectrum of low-energy photons is essentially converged by $n_\text{max}=100$.}
	\label{fig:nmax_convergence}
\end{figure*}
A number of additional practical points need to be addressed. 
One must choose a highest energy state at which to truncate the sum over excited states, $n_\text{max}$. 
We find that with $n_\text{max}=200$ the ionization level is essentially converged, as was also found in Ref.~\cite{Chluba:2010fy}, and with $n_\text{max}=100$ the photon spectrum $N_\omega$ is essentially converged.
In Fig.~\ref{fig:nmax_convergence}, we show the ionization histories at different values of $n_\text{max}$ on the left; the black dashed curve is reproduced from Fig. 3 of Ref.~\cite{AliHaimoud:2010dx}.
The lower panel shows the relative difference between the curves calculated with \dhis and the \texttt{Hyrec} result~\cite{AliHaimoud:2010dx}.
The right panel depicts the spectral distortion calculated at the same values for $n_\text{max}$; the lower panel shows the change in the spectral distortion as we increase $n_\text{max}$ from one value to the next largest value.
In both panels, we can see that the quantities are converged for $n_\text{max}$ greater than about 100.

Once $n_\text{max} \sim \mathcal{O}(10)$, the computation of $\tilde{R}_{i \to j}$, $\alpha_i$, $\beta_i$, and the matrix inversion in Eq.~\eqref{eq:matrix_MLA} become the most computationally expensive steps. 
To speed up our computations, we precompute any quantity that does not depend on redshift, like dipole operator matrix elements connecting bound-bound and bound-free states. 
In addition, $M_{kl}$ is a sparse matrix, meaning most of its elements are equal to zero.
There exist techniques and code packages for taking advantage of sparse structure to speed up linear algebra operations, so we employ \texttt{scipy.sparse} to more efficiently invert Eq.~\eqref{eq:matrix_MLA}.

Another issue is that the differential equations for $x_e$ and $T_m$ are stiff. 
At sufficiently early times the recombination, ionization, and Compton heating rates are so fast compared to a Hubble time that a numerical solution to the evolution equations 
yields $x_e = x_e^\text{Saha}$ and $T_m = T_\text{CMB}$ plus numerical noise. 
To combat this noise, for redshifts $z > 1555$\footnote{This value is close to the one used in Ref.~\cite{2010PhRvD..82f3521A}, but this is coincidental.} we simply set $x_e = x_e^\text{Saha}$ and $T_m$ to the analytic expression in Eq.~\eqref{eqn:analytic_Tm}.

\begin{figure}
	\centering
	\includegraphics[width=0.5\columnwidth]{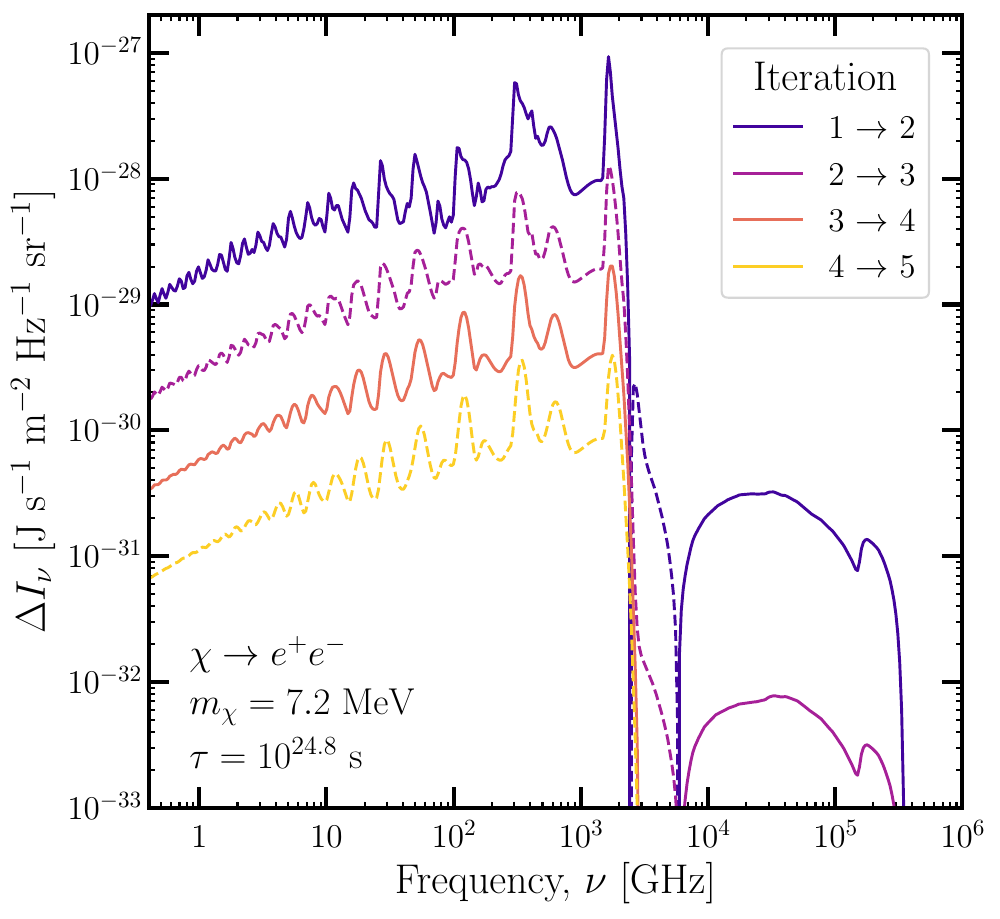}
	\caption{Change in the spectral distortion between one iteration of calculating $\tilde{\alpha}_\text{B}$ and $\tilde{\beta}_\text{B}$ over the full redshift range from $1+z_\text{init}$ to $1+z_\text{end}$ and the next iteration.
	With each iteration, the difference decreases by about an order of magnitude, indicating that the spectral distortion is rapidly converging.}
	\label{fig:iterations}
\end{figure}

Finally, we must be careful with the interpolation and extrapolation of the recombination and ionization rates used in the ionization evolution equation. 
Within each redshift step we calculate $\tilde{\alpha}_B$ and $\tilde{\beta}_\text{B}$. 
Other than Euler's method, which is insufficient for such stiff equations, any integration method requires multiple evaluations throughout the redshift step. 
Each of these evaluations would require recalculating $\tilde{\alpha}_B$ and $\tilde{\beta}_\text{B}$ at slightly different redshifts; however, since each evaluation of $\tilde{\alpha}_\text{B}$ and $\tilde{\beta}_\text{B}$ is so costly, we have devised an iterative method that avoids their computation during the integration loop.
For each step of the iterative method we integrate Eq.~\eqref{eq:full_eqs} over the full redshift range from $1+z_\text{init}$ to $1+z_\text{end}$ using the method described above, but using a precomputed formula functional form for $\tilde{\alpha}_\text{B}$ and $\tilde{\beta}_\text{B}$, instead of computing them during the loop. 
For the first iteration, we use the fitting functions used by \texttt{Recfast}~\cite{Seager:1999bc}, including the hydrogen fudge factor which we set to 1.125 and the double Gaussian function correction. 
At the end of this iteration, we have a new set of $\tilde{\alpha}_\text{B}$ and $\tilde{\beta}_\text{B}$ values associated with each redshift step that we computed. 
We linearly interpolate these new values with respect to redshift and proceed through the next iteration using these interpolation functions to determine $\tilde{\alpha}_\text{B}$ and $\tilde{\beta}_\text{B}$. 
We find that this process already converges after one iteration (see Fig.~\ref{fig:iterations}). 
%

\subsubsection{Comparison to Other Calculations}
\label{sec:atom-cross-checks}

In this section we perform the numerical procedure described in the previous section and compare our outputs to those of other codes. 
First we compare our calculation of the ionization history, $x_e(z)$, to the output of the recombination code, \texttt{Hyrec}. 
If we again look at Fig.~\ref{fig:nmax_convergence}, the dark purple line is the ionization history calculated using our code with $n_\text{max} = 200$; the black dashed line shows the same quantity calculated with \texttt{Hyrec}.
The second panel shows the relative difference between the two; the greatest deviations come from around the redshift of recombination, but are still only at the level of a few percent.
These differences may be due to a number of effects we neglected that are accounted for in \texttt{Hyrec}, including helium recombination, two-photon transitions from levels higher than 2s, and frequency diffusion in the Lyman-$\alpha$ line~\cite{AliHaimoud:2010dx}.
Hence, \texttt{Hyrec} is the more accurate recombination code, but few-percent-level accuracy is sufficient for our purposes.

\begin{figure}
	\centering
	\includegraphics[width=0.5\columnwidth]{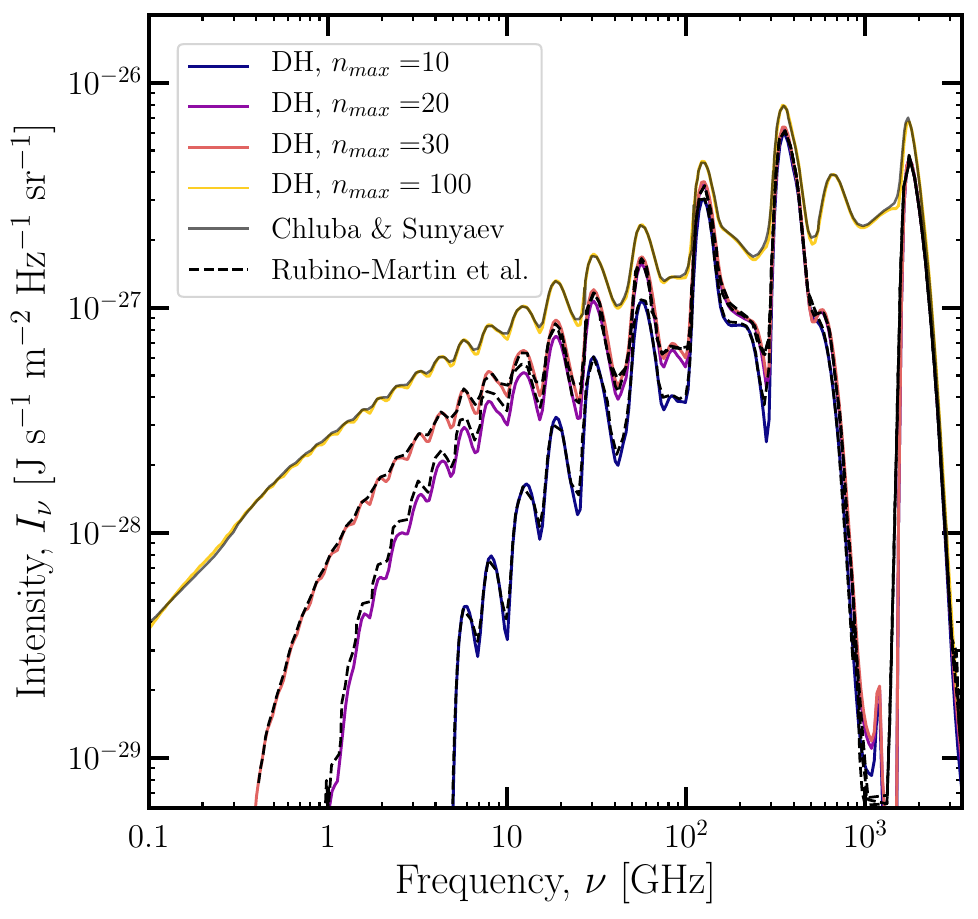}
	\caption{Spectral distortion due to atomic transitions at recombination, without exotic energy injection and tracking up to various $n_\text{max}$.
	For comparison, we also show the results of Ref.~\cite{Rubino-Martin:2006hng} in black dashed lines and Ref.~\cite{Chluba:2006xa} in the solid grey using the same values for $n_\text{max}$.}
	\label{fig:distortion_xcheck}
\end{figure}
Next, we compare our calculation of the CMB distortion due to atomic transitions to the outputs of Refs.~\cite{Rubino-Martin:2006hng, Chluba:2006xa}, calculated at different values for $n_\text{max}$.
In Fig.~\ref{fig:distortion_xcheck}, we show the hydrogen recombination spectrum calculated using our code for the same values of $n_\text{max}$ and no exotic energy injection.
For $n_\text{max} = 10$, 20, and 30, we are able to reproduce the recombination spectra from Ref.~\cite{Rubino-Martin:2006hng}, and for $n_\text{max} = 100$, we match the results from Ref.~\cite{Chluba:2006xa}.
In each case, we find only percent-level deviations compared to the results in the literature.

Lastly, the $R_{i \to j}$ transition rates are closely related to the probabilities for a hydrogen atom in the excited state to decay to the ground state with the emission of a Ly-$\alpha$ photon (e.g. by cascading through the $2p$ state, rather than the $2s$ state).
Table 1 of Ref.~\cite{Hirata:2005mz} lists these probabilities up to $n=30$, and we are able to reproduce the results to the last significant digit they report.

\subsection{Conclusion}
\label{sec:tech_conclusion}

In addition to modifying the global temperature and ionization history, exotic energy injections can change the spectrum of the universe's background photons, resulting in CMB spectral distortions apart from those expected in $\Lambda$CDM.
In this work, we have described major upgrades to the \dhis code which are necessary for self-consistently tracking these photons.
We summarize the main points below:
\begin{enumerate}
	\item We extend our treatment of high-energy electrons to lower energies, which allows us to track the spectrum of secondary photons from ICS of CMB photons off low-energy electrons and simultaneously provides a more precise treatment of heating and ionization from low-energy electrons; the latter is relevant to constraints on light DM.
	
	\item We calculate the $y$-type spectral distortion caused by gas heating.
	
	\item Instead of treating hydrogen as a TLA, we now track an arbitrary number of energy levels and the subsequent line emissions from transitions between these levels. 
	
	\item We can also account for the back-reaction of the altered background spectrum of photons on these transitions.
\end{enumerate}
At each step of the upgrade, we have cross-checked our results against various other codes.
This version of \dhis is publicly available on GitHub \githubmaster.

Throughout this work, we treat energy deposition as homogeneous.
This has also been a common assumption in previous works; however, structure formation in the late universe leads to inhomogeneous energy injection and deposition~\cite{Schon:2014xoa,Schon:2017bvu}.
We leave a detailed study of these inhomogeneities to future work.

Being able to track the late time spectrum of photons paves the way for future studies of observables from exotic energy injection.
In Section~\ref{sec:DHv2_apps}, we will demonstrate various directions that can be taken with this new technology, including examining the possibility of observing spectral distortions from yet unconstrained dark matter models using future CMB spectral distortion experiments, extending existing CMB anisotropy contraints, and setting limits on couplings between ALPs and photons.
Other possible follow-ups include studying various ways in which the modified photon background could affect signals from 21-cm cosmology and the extragalactic background light.
We leave this to future work.

\section{
Exotic energy injection in the early universe II: CMB spectral distortions and constraints on light dark matter
}
\label{sec:DHv2_apps}

Exotic sources of energy injection, such as dark matter (DM) annihilating or decaying into Standard Model (SM) particles, can inject a significant amount of energy into the universe.
This additional energy can manifest as modifications to the global ionization history $x_\text{HII} \equiv n_\text{HII} / n_\text{H}$, where $n_\text{HII}$ and $n_\text{H}$ are the number densities of ionized hydrogen and all hydrogen nuclei respectively, and the intergalactic medium (IGM) temperature history $T_m (z)$; thus, observations of these quantities can be used to constrain the properties of DM~\cite{Slatyer:2009yq, Cirelli:2009bb, Kanzaki:2009hf, Diamanti:2013bia, Evoli:2014pva, Slatyer:2015jla, Lopez-Honorez:2016sur,Liu:2016cnk, Slatyer:2016qyl, Poulin:2016anj,Hiss:2017qyw, DAmico:2018sxd, Liu:2018uzy, Cheung:2018vww, Mitridate:2018iag, Clark:2018ghm, Walther:2018pnn, McDermott:2019lch, Gaikwad:2020art, Caputo:2020bdy, Witte:2020rvb,Liu:2020wqz,Gaikwad:2020eip,2206.13520}.
Moreover, exotic energy injection can modify the background of low-energy photons.

Changes to the background spectrum of photons can have a number of interesting implications---for example, exotic energy injection could explain observed excesses, such as that of the cosmic optical background~\cite{Lauer:2022fgc, Bernal:2022wsu}.
In addition, excess Lyman-$\alpha$ radiation~\cite{Hirata:2005mz} can affect the global and inhomogeneous redshifted 21cm signal via the Wouthuysen-Field effect. 
Star formation is also affected by the background radiation; for example, the formation of molecular hydrogen, which is required to form the first stars, can be disrupted by photodetachment of intermediate states~\cite{Hirata:2006bt}, or photodissociation by photons in the Lyman-Werner band~\cite{1967ApJ...149L..29S,1992A&A...253..525A,1996ApJ...467..522H}.

The distortion of the Cosmic Microwave Background due to exotic energy injection is of particular interest.
COBE/FIRAS measured the CMB energy spectrum to be a perfect blackbody within about 
one part in $10^4$ over the frequency range of about 60 to 630 GHz~\cite{Mather:1993ij,Fixsen:1996nj}.
The balloon-borne ARCADE 2 experiment also measured the CMB spectrum between 3 to 90 GHz and reported a rise in the blackbody temperature at this low frequency tail of the spectrum~\cite{2011ApJ...734....5F,2011ApJ...734....6S}.
A proposed next-generation experiment that can improve upon these measurements is the Primordial Inflation Explorer (PIXIE)~\cite{2011JCAP...07..025K}; there are also a number of other proposed efforts which could improve sensitivity to the CMB energy spectrum by orders of magnitude~\cite{Chluba:2019nxa,Kogut:2019vqh,Maffei:2021xur,Chang:2022tzj}.

Distortions to the blackbody spectrum can arise when matter and radiation are driven out of thermal equilibrium after $z \lesssim 2 \times 10^6$, i.e. once thermalization is no longer efficient~\cite{Chluba:2019kpb}.
Spectral distortions from energy release in the early universe are often characterized by $\mu$ and $y$-type distortions. 
$\mu$-type distortions arise when Compton scattering is still fast enough to maintain the photons and electrons in kinetic equilibrium but photon number-changing processes are inefficient, causing the blackbody to develop a chemical potential~\cite{Illarionov_Sunyaev}.
$y$-type distortions form later, once Compton scattering is also inefficient and unable to hold the photon bath in full kinetic equilibrium~\cite{Zeldovich:1969ff}, and have magnitude controlled by the ``$y$-parameter'', which roughly describes the energy density exchanged by scattering off of electrons as a fraction of the total CMB energy density~\cite{Chluba:2018cww}.
The $y$-parameter can be non-zero even at very early redshifts, when the radiation and matter temperature are tightly coupled.
Even within the standard $\Lambda$CDM cosmological model, we expect such distortions to the CMB blackbody spectrum to arise due to reionization and structure formation; the overall size of the distortion from reionization could reach a $y$-parameter of up to $y \sim 10^{-6}$~\cite{Chluba:2016bvg, Chluba:2019kpb}. 

In addition to Compton scattering, distortions can be sourced by atomic transitions; in particular, such distortions arising from recombination have received much attention.
Earlier studies, such as Ref.~\cite{1993ASPC...51..548R} calculated these distortions using up to 10 excited states of hydrogen.
More recent studies, such as Refs.~\cite{Rubino-Martin:2006hng, Chluba:2006bc}, refined this calculation and included up to 100 energy levels of hydrogen.

Processes beyond $\Lambda$CDM such as DM annihilation or decay could lead to the injection of non-thermal, electromagnetically interacting particles into the early universe, potentially producing a large number of secondary photons as the injected particles cool and lose their energy. The effects of such energy injections, and DM interactions with the SM bath more broadly, on spectral distortions has been examined from a number of perspectives.
For example, limits on CMB distortions have been used to constrain DM scattering on SM particles~\cite{Ali-Haimoud:2015pwa}, energy injection from dark photons~\cite{McDermott:2019lch}, and axions and axion-like particles (ALPs)~\cite{Bolliet:2020ofj}.
Multiple studies have also used Green's functions to more generally study spectral distortions from heating~\cite{Chluba:2013vsa}, photon injection~\cite{Chluba:2015hma}, and electron injection~\cite{Acharya:2018iwh}. However, the focus of this latter set of studies has been the epoch prior to recombination, when the universe can be well-approximated as fully ionized, and photons scatter rapidly with electrons.

In addition to inducing spectral distortions, exotic energy injection can also affect anisotropies in the CMB by modifying the global ionization history~\cite{Adams:1998nr,Chen:2003gz,Padmanabhan:2005es}.
This has been studied in the context of both decaying~\cite{Zhang:2007zzh,Slatyer:2016qyl,Poulin:2016anj,Acharya:2019uba,Cang:2020exa} and annihilating DM~\cite{Galli:2009zc,Slatyer:2009yq,Kanzaki:2009hf,Hisano:2011dc,Hutsi:2011vx,Galli:2011rz,Finkbeiner:2011dx,Slatyer:2012yq,Galli:2013dna,Madhavacheril:2013cna,Slatyer:2015jla,Slatyer:2015kla}.
In many of these studies, the constraints were either applied only to a set of benchmark DM models; in studies where the constraints were calculated more generically, the limits only extended to DM masses corresponding to particle injection energies above a few keV. For example, in the work of Ref.~\cite{Slatyer:2015jla}, the limits were not extended to sub-keV DM on the basis of approximations made in the analysis that were likely to break down for sufficiently low DM masses (approaching the hydrogen and helium ionization thresholds).

In a companion paper, described in Section~\ref{sec:DHv2_tech}, we describe recent improvements made to the \dhis code package~\cite{DarkHistory} to better track the behavior of low-energy electrons and photons, either directly injected or arising as part of a secondary particle production cascade, in the universe at $z \lesssim 3000$. 
We treat the energy deposition as homogeneous, a common assumption also made in previous works~\cite{DarkHistory,Liu:2020wqz}.
In this work, we apply the tools that we developed in Section~\ref{sec:DHv2_tech} to obtain several immediate and significant results. 
In particular, we calculate the CMB spectral distortions arising from exotic energy injections during the post-recombination epoch for the first time, and extend CMB anisotropy constraints based on the ionization history to lower DM masses. 
We also demonstrate how additional low-energy photons produced by exotic energy injections back-react on the process of recombination itself, finding a small effect on the global ionization history.  

In Section~\ref{sec:distortions}, we show the spectral distortions resulting from energy injection between $3000 > 1+z > 4$ and compare their amplitude to the sensitivity of future experiments, as well as contributions to the distortion from other redshift ranges.
In Section~\ref{sec:ionization}, we show that although these spectral distortions could in principle modify the ionization history, changes to the ionization history compared to the \dhis \texttt{v1.0} calculation are not significant.
In Section~\ref{sec:anisotropy}, we extend the CMB anisotropy constraints on DM decaying to photons, and also translate this into a constraint on the ALP-photon coupling.
We conclude in Section~\ref{sec:apps_conclusion}. 

\begin{figure*}
	\centering
	\includegraphics[width=0.8\textwidth]{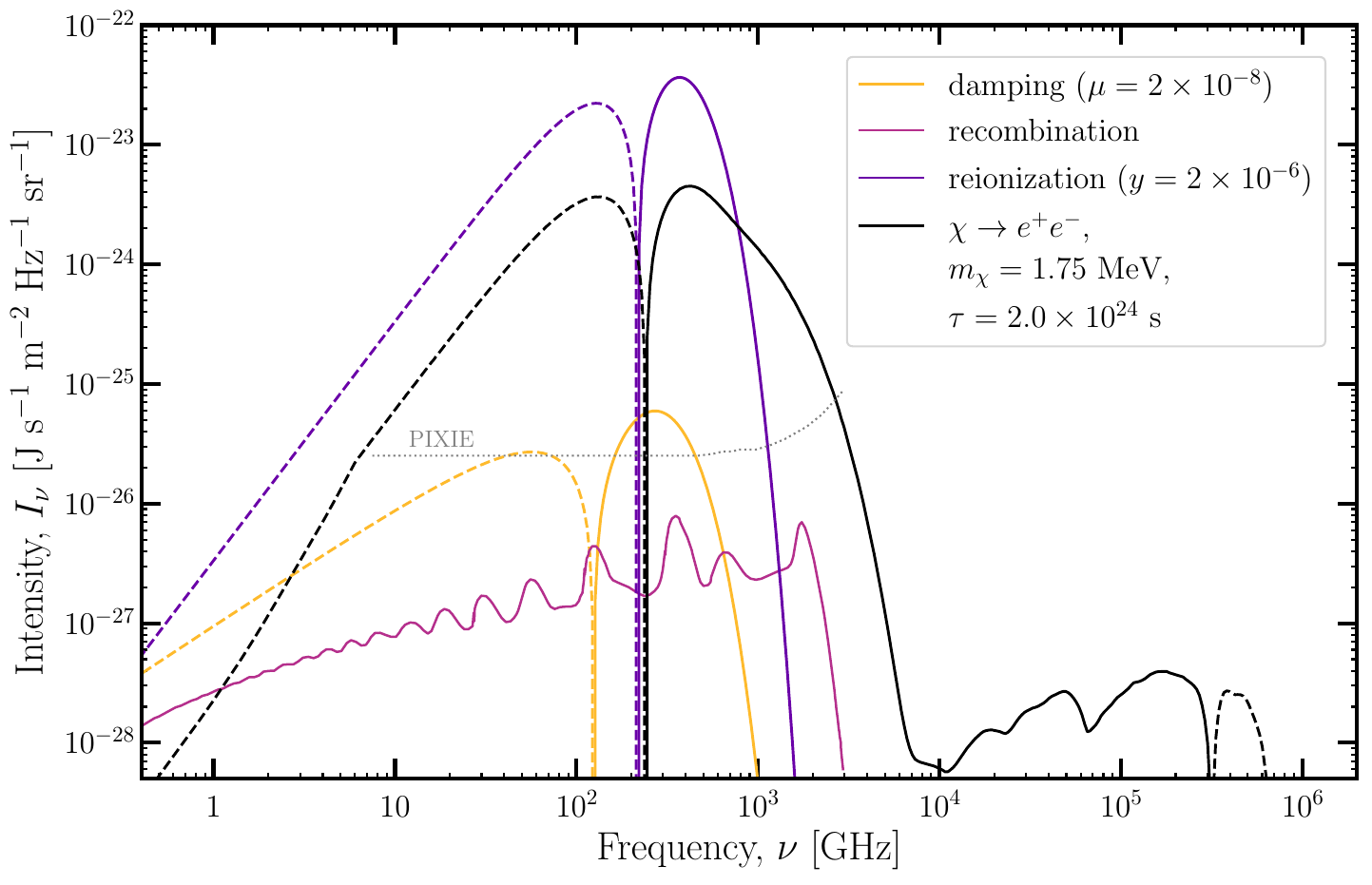}
	\caption{
		Predicted spectral distortions from $\Lambda$CDM, including the average $\mu$-distortion expected from Silk damping (yellow), the signal from recombination (magenta), and the $y$-distortion expected from reionization and structure formation (dark purple).
		Dashed lines indicate where the spectral distortions have negative values.
		The gray dotted line is the expected sensitivity of PIXIE~\cite{2011JCAP...07..025K, 2016SPIE.9904E..0WK}.
		Shown in black is the largest spectral distortion from DM energy injection in our study (the $\Lambda$CDM contributions, including contributions from a model of reionization, have been subtracted out from this line), restricting to scenarios that are not otherwise excluded.
	}
	\label{fig:LCDM}
\end{figure*}
%

\subsection{CMB Distortions}
\label{sec:distortions}

Hereafter, we define the CMB spectral distortion as the difference between the total photon intensity $I_\omega$ and the intensity of a blackbody at the CMB temperature $I_{\omega, \mathrm{CMB}}$:
\begin{equation}
	\Delta I_\omega = I_\omega - I_{\omega, \mathrm{CMB}} .
\end{equation}
Within $\Lambda$CDM cosmology, spectral distortions can be sourced by a number of processes~\cite{Chluba:2016bvg,Chluba:2019nxa}.
Prior to redshifts of $z \sim \text{few} \times 10^6$, any injected energy thermalizes very quickly and only changes the temperature of the CMB blackbody, leading to no discernible distortion.
Below this redshift, processes that do not conserve photon number become inefficient, while photons and electrons remain in thermal contact; energy injection therefore causes the photon bath to develop a chemical potential, $\mu$.
This leads to a $\mu$-type distortion, given by
\begin{equation}
	\Delta I_{\omega, \mu} = \frac{\mu \omega^3}{2\pi^2} \frac{e^{x}}{(e^{x}-1)^2} \left[ \frac{x}{\zeta} - 1 \right],
\end{equation}
where $x= \omega / T_\text{CMB}$ and $\zeta \simeq 2.1923$ is a constant~\cite{Chluba:2013vsa}.
As an example, dissipation of primordial density perturbations through Silk damping is expected to contribute a $\mu$-type distortion with parameter $\mu \simeq 2 \times 10^{-8}$~\cite{Chluba:2016bvg}, whereas data from COBE/FIRAS constrains $\mu$ to be less than several times $10^{-5}$~\cite{Fixsen:1996nj,Bolliet:2020ofj,Bianchini:2022dqh}.

Around $z \sim 5 \times 10^4$, Comptonization also becomes inefficient and energy injection typically results in $y$-type distortions, which are defined as
\begin{equation}
	\Delta I_{\omega, y} = \frac{y \omega^3}{2\pi^2} \frac{x e^{x}}{(e^{x}-1)^2} \left[ x \coth \left( \frac{x}{2} \right) - 4 \right],
\end{equation}
where the $y$ parameter characterizes the amplitude of the distortion.
For example, heating of the IGM during reionization and structure formation is expected to contribute a $y$-type distortion with amplitude of about $y \simeq 2 \times 10^{-6}$~\cite{Refregier:2000xz,Zhang:2004fh,Dolag:2015dta,Hill:2015tqa,DeZotti:2015awh,Chluba:2016bvg}.
COBE/FIRAS limits such distortions to $y \lesssim 10^{-5}$~\cite{Fixsen:1996nj,Bolliet:2020ofj}.

There can also exist distortions that are neither $\mu$-type nor $y$-type~\cite{Chluba:2011hw,Khatri:2012tw,Chluba:2013vsa,Acharya:2018iwh}.
For example, during recombination, as electrons are captured and cascade to the ground state, they emit out-of-equilibrium radiation that appear today as broadened and redshifted atomic lines (helium recombination also contributes a spectral distortion, but this is expected to be an order of magnitude smaller than the signal from hydrogen recombination, hence we only consider hydrogen~\cite{Wong:2005yr,Rubino-Martin:2007tua,Chluba:2019nxa}).
Fig.~\ref{fig:LCDM} shows these predicted signals, as well as the expected sensitivity of PIXIE, which can measure spectral distortions down to $\mu \sim 10^{-8}$ or $y \sim 2\times10^{-9}$~\cite{2011JCAP...07..025K, 2016SPIE.9904E..0WK}.

Exotic energy injections can similarly distort the CMB frequency spectrum.
For example, DM can decay or annihilate to photons, or to SM particles that promptly decay producing photons; these photons could contribute directly to the distortion, but they can also produce secondary electrons and photons through processes such as pair production, Compton scattering, photoionization, or photoexcitation, with subsequent de-excitation and recombination.
Electrons (either produced directly by the energy injection mechanism, or as secondaries) can produce more secondaries through collisional ionization and excitation, and inverse Compton scattering (ICS) off CMB photons.
Especially for higher-energy photons, the timescale for cooling processes can be comparable to or longer than a Hubble time, so it is important to take into account the expansion of the universe during the development of the secondary particle cascade.
Calculating these rich particle cascades with the necessary amount of detail to accurately track the resulting distortion is nontrivial; the numerical methods we have developed for this purpose are described in Ref.~\cite{DarkHistory} and Section~\ref{sec:DHv2_tech}.

For the rest of this section, we will focus on DM annihilation and decay as two main examples of exotic energy injection; we stress, however, that \texttt{DarkHistory} can handle general energy injection processes.
To include the effects of energy injection by DM in \texttt{DarkHistory}, one need only specify if DM decays or annihilates, the spectrum of decay/annihilation products, the DM mass $m_\chi$, and the lifetime $\tau$ or cross-section $\langle \sigma v \rangle$. 
When calculating the evolution of the temperature, ionization level, and spectral distortions, there are a number of options within \dhis that can be adjusted, depending on which effects one would like to include.
We describe a few of the most relevant options for this work below; see Ref.~\cite{DarkHistory} and Section~\ref{sec:DHv2_tech} for more details.

\begin{itemize}
	\item \texttt{backreaction}: If set to \texttt{False}, then the fraction of energy deposited into each energy deposition channel (e.g.\ heating, hydrogen ionization etc.) is computed assuming the baseline value of $x_e \equiv n_e / n_\text{H}$, the number density ratio of free electrons to all hydrogen nuclei, obtained without exotic energy injection. 
	Otherwise, if \texttt{True}, the effect on how energy is deposited due to modifications of the temperature and ionization histories away from the baseline solutions is taken into account. 
	This is crucial for an accurate determination of $T_m(z)$ for $z \lesssim 200$~\cite{DarkHistory}.
	
	\item \texttt{struct\_boost}: When structure formation begins, annihilation is enhanced due to the fact that the average of the squared density exceeds the square of the average density.
	With this option, the annihilation rate of DM is enhanced by a boost factor due to structure formation, based on the model first developed in Ref.~\cite{Liu:2016cnk}, and implemented in \dhis in Ref.~\cite{DarkHistory}. 
	
	\item \texttt{reion\_switch}: If \texttt{True}, then a model for reionization is included in the evolution.
	
	\begin{itemize}
		\item \texttt{reion\_method}: Specifies which reionization model to use.
		By default, we employ \texttt{`Puchwein'}, which is the model given in Ref.~\cite{Puchwein:2018arm}.
	\end{itemize}
	
	\item \texttt{distort}: If \texttt{True}, then \dhis will track the background spectrum of photons and its evolution.
	To accurately calculate the hydrogen atom transitions that contribute to spectral distortions, we treat hydrogen as a Multi-Level Atom (MLA).
	
	\begin{itemize}
		\item \texttt{nmax}: Sets the maximum hydrogen principal quantum number to track in the MLA.
		
		\item \texttt{iterations}: The MLA recombination and photoionization rates are the most computationally expensive quantities to compute; hence, we have developed an iterative method for their calculation, described in Section~\ref{sec:DHv2_tech}. 
		This option sets the number of iterations over which to improve the calculation.
		
		\item \texttt{reprocess\_distortion}: The atomic transition rates for the MLA depend on the state of the background radiation and affect subsequent emissions and absorptions by hydrogen atoms.
		If this option is set to \texttt{False}, then the rates are calculated assuming the CMB is a perfect blackbody; i.e. spectral distortions are not ``reprocessed'' through the MLA.
		If \texttt{True}, then we also account for the effect of spectral distortions on these rates.
		See Section~\ref{sec:DHv2_tech} for more details.
	\end{itemize}
\end{itemize}
For clarity, we will be explicit about how these flags are set in the results described below.

\label{sec:dist_results}

%
\begin{figure*}
	\centering
	\includegraphics[width=\textwidth]{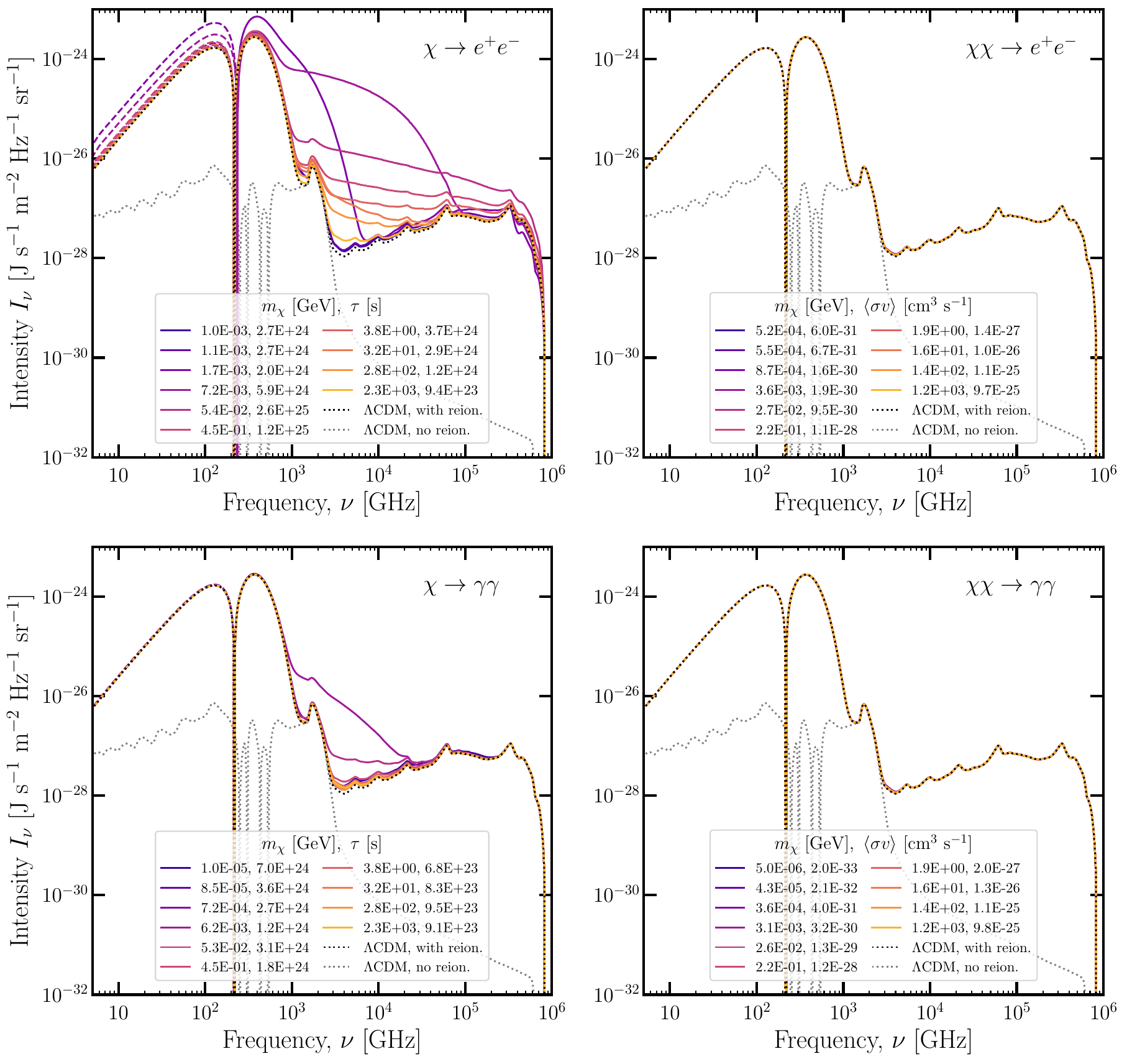}
	\caption{The distortion to the CMB spectrum including different types of DM energy injection; this distortion is relative to the CMB blackbody and hence includes components from $\Lambda$CDM processes.
		For each mass, the lifetime/cross-section is chosen to be at the edge of the CMB constraints given in Refs.~\cite{Slatyer:2016qyl,Slatyer:2015jla}.
		The left panels show the effect of decaying DM, while the right is from $s$-wave annihilation.
		In the top row, the decay/annihilation products are $e^+ e^-$ pairs, while the bottom row is for photon pairs.
		Negative values for the distortion are represented by colored dashed lines.
		The black dotted line in each panel shows the spectral distortion expected in the absence of exotic energy injection and including a reionization model; the grey dotted line shows the result when we also turn off reionization.}
	\label{fig:dist_grid}
\end{figure*}
\begin{figure*}
	\centering
	\includegraphics[width=\textwidth]{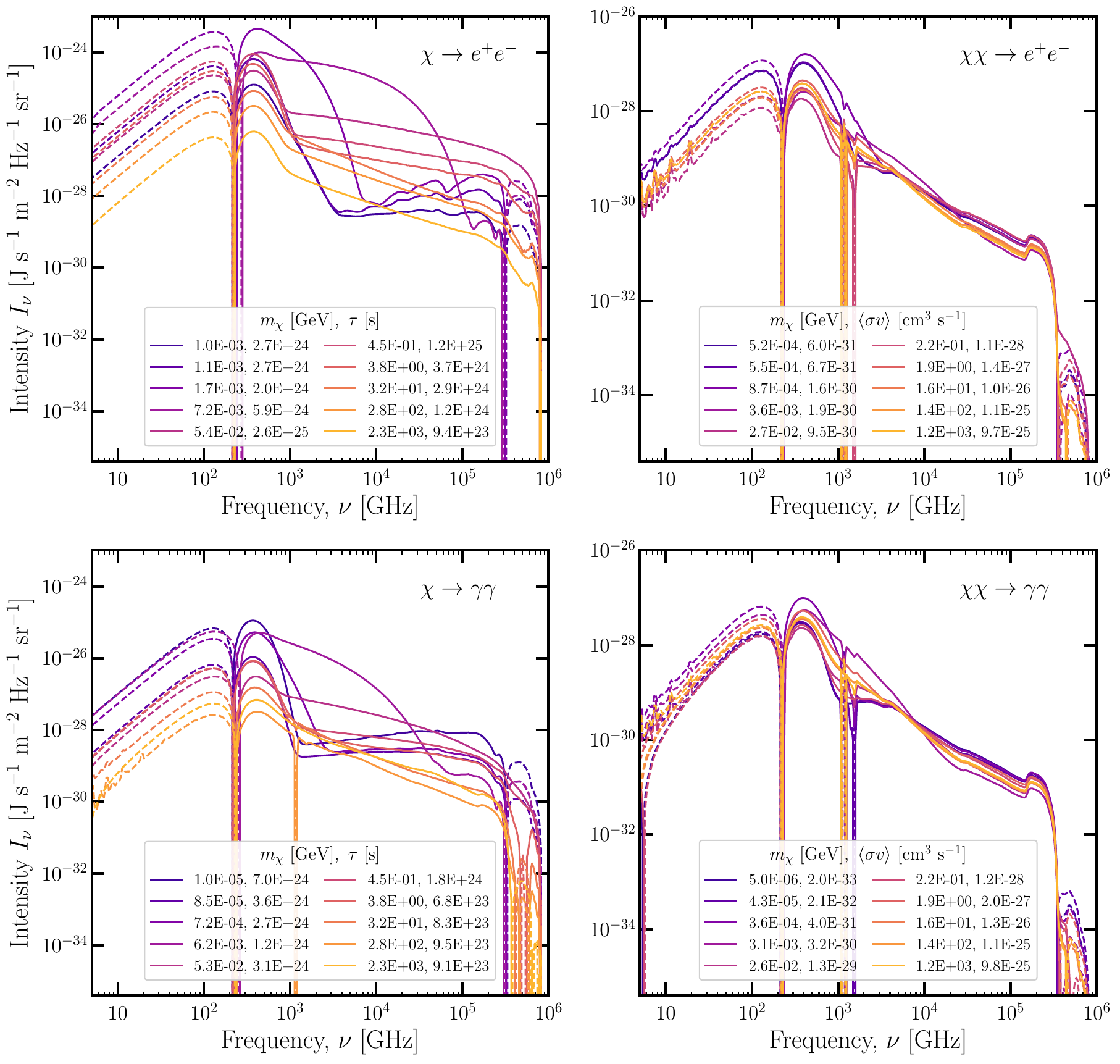}
	\caption{
		Same as Fig.~\ref{fig:dist_grid}, but subtracting the component of the distortion that would be present in the absence of exotic energy injection.
		In other words, here we show the spectral distortion relative to the expected $\Lambda$CDM contribution.
	}
	\label{fig:dist_grid_noLCDM}
\end{figure*}
Fig.~\ref{fig:dist_grid} shows the spectral distortion calculated by \dhis including the effects of DM energy injection for some sample DM models; we emphasize that the public code is capable of repeating these calculations for other final states, and more general exotic injection scenarios. 
We choose scenarios with annihilation or decay purely to photons or $e^+ e^-$ pairs because those are the simplest allowed channels for DM masses below the muon and pion thresholds, which is where these cosmological constraints are most competitive. 
Furthermore, even at higher masses, SM particles heavier than an electron promptly decay, producing electrons, positrons, and photons (as well as neutrinos, which are effectively free-streaming, and (anti)nuclei which typically have a significantly lower branching ratio) so that the results for a general scenario can be estimated by taking linear combinations of the results for photons and $e^+ e^-$.
\dhis includes a submodule than can calculate the electron and photon spectra from the injection of any arbitrary SM particle, based on the \textsc{pppc4dmid} results~\cite{Cirelli:2010xx}.

To be very explicit, the options we set for calculating these distortions are:
\begin{itemize}
	\item  \texttt{backreaction = True},
	
	\item  \texttt{struct\_boost = None},
	
	\item  \texttt{reion\_switch = True},
	
	\item  \texttt{reion\_method = `Puchwein'},
	
	\item  \texttt{distort = True},
	
	\item  \texttt{nmax = 200},
	
	\item  \texttt{iterations = 5},
	
	\item  \texttt{reprocess\_distortion = True}.
\end{itemize}
We conservatively choose not to include structure boost formation in our main results, and discuss the effect of structure formation later in this section.
Regarding \texttt{nmax} and \texttt{iterations}, we choose these values since the ionization level and spectral distortions are converged for \texttt{nmax = 200} and after only one iteration, see Section~\ref{sec:DHv2_tech}.

For decaying DM, we choose the lifetime at each mass point to be at the minimum lifetime allowed by the CMB anisotropy constraints from Ref.~\cite{Slatyer:2016qyl}; for $s$-wave annihilation, the CMB anisotropy constraints on the cross-section are taken from Ref.~\cite{Slatyer:2015jla}.
For comparison, we also plot on each panel a black dotted line, which is the spectral distortion calculated by \dhis in the absence of exotic energy injection but including the reionization model; the grey dotted line shows what is left when we also turn off reionization.
The largest distortions come from DM decaying to $e^+ e^-$ pairs at the masses that are least constrained, e.g. around \SIrange{1}{10}{\mega\eV}. 

For the annihilation models, exotic energy injection has a small effect on the CMB blackbody spectrum compared to the distortion that would be expected even in the absence of exotic injections; we call this latter component ``$\Lambda$CDM contributions'' since they are the spectral distortions that would still be present in a $\Lambda$CDM universe.
Hence, the spectral distortions from annihilations are indistinguishable by eye. This indicates that the CMB anisotropy constraints are strong enough to ensure that distortions from annihilation are small relative to the $\Lambda$CDM contribution.
Fig.~\ref{fig:dist_grid_noLCDM} shows the same spectral distortions as in Fig.~\ref{fig:dist_grid}, but with the $\Lambda$CDM contributions subtracted out.
At the lowest frequencies, this difference is very small and subject to numerical instabilities, hence we only show the distortion above a few GHz.
The largest of the distortions from Fig.~\ref{fig:dist_grid_noLCDM} is also plotted on Fig.~\ref{fig:LCDM} for comparison.

One can immediately distinguish contributions from various processes and redshifts to the distortions in Fig.~\ref{fig:dist_grid}.
All of the distortions show a trough-peak feature that resembles a $y$-type distortion; this is primarily due to Compton scattering of photons off of a hot IGM heated by DM and reionization.
There is also a narrow peak around 2000 GHz; this is the high energy end of the signal from atomic transitions at recombination, e.g. the magenta line in Fig.~\ref{fig:LCDM}.
Additionally, between about $3000$ and $10^6$ GHz, all the distortions show a spiky shoulder.
This feature contains a smooth component caused by ICS; the spikes are due to the enhancement of atomic line emission around reionization, as neutral hydrogen atoms are raised to excited and ionized states which can subsequently decay and emit line photons.
In Fig.~\ref{fig:components}, we show an example of these different contributions to the distortion resulting from DM decaying to $e^+ e^-$ pairs, with a mass of $200$ MeV and a lifetime of $10^{25.3}$ s.

By comparing the black and grey dotted lines in Fig.~\ref{fig:dist_grid}, we can clearly see that reionization has a large impact on the spectral distortion.
When we set \texttt{reion = True}, we include additional terms in the ionization and temperature evolution, which contribute in a number of different ways: the increase in temperature at late times leads to a large $y$-parameter, and the increasing ionization leads to additional emission of atomic lines.
In principle, our MLA treatment is also affected by the radiation fields that cause reionization in the first place, which could dominate over the late-time spectral distortion at certain frequencies.
One could add a model for these extra radiation fields as a new source of injected photons in \texttt{DarkHistory}; however, for this work we choose only to include the effect of reionization on the spectral distortions through the terms for the ionization and temperature equations (see e.g.\ Eq.~(36) in Section~\ref{sec:DHv2_tech}).

In addition, since the energy injection rate from $s$-wave annihilation goes as $\rho_\chi^2 \propto (1+z)^6$, where $\rho_\chi$ is the energy density in dark matter, and the same rate for decay goes as $\rho_\chi \propto (1+z)^3$, then decaying DM tends to modify the distortion more at late times compared to annihilation; hence, the $y$-distortion from late time heating is much larger for decay scenarios.
Currently unconstrained decaying DM models will generally give signals large enough to be detectable by future experiments such as PIXIE~\cite{2011JCAP...07..025K, 2016SPIE.9904E..0WK}. 

\begin{figure*}
	\centering
	\includegraphics[width=0.48\columnwidth]{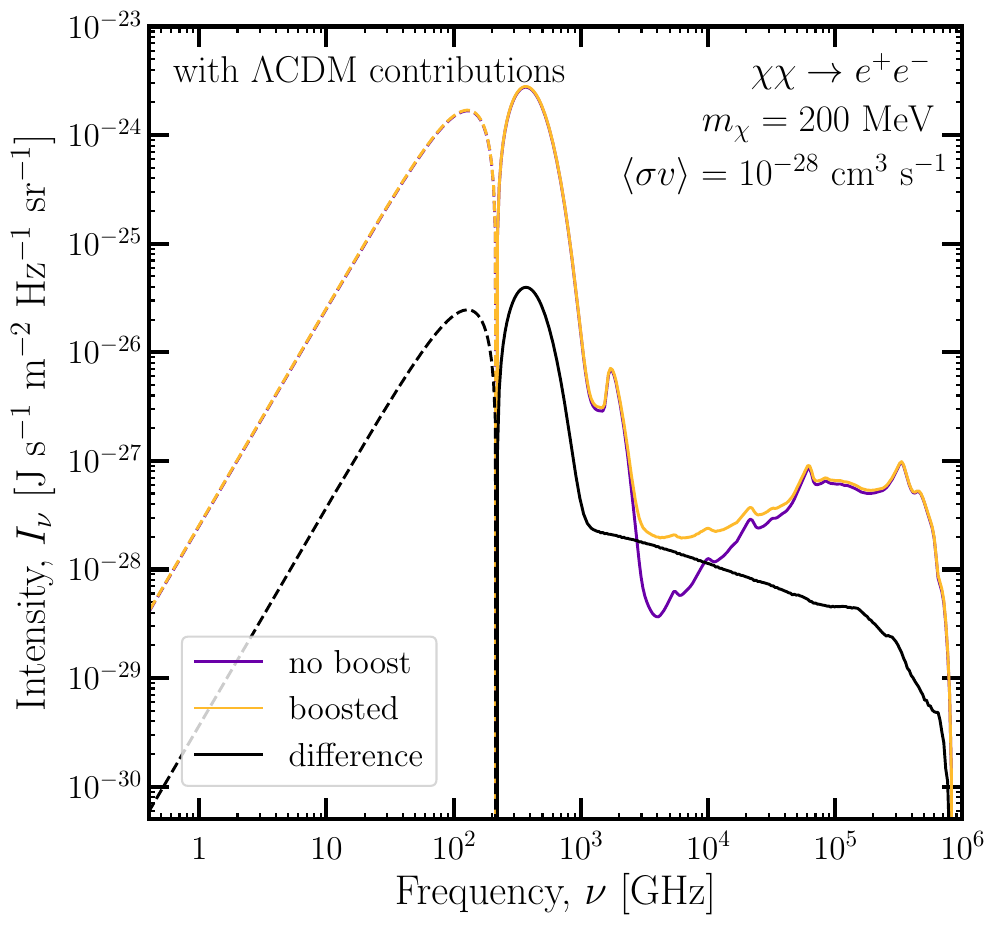}
	\includegraphics[width=0.48\columnwidth]{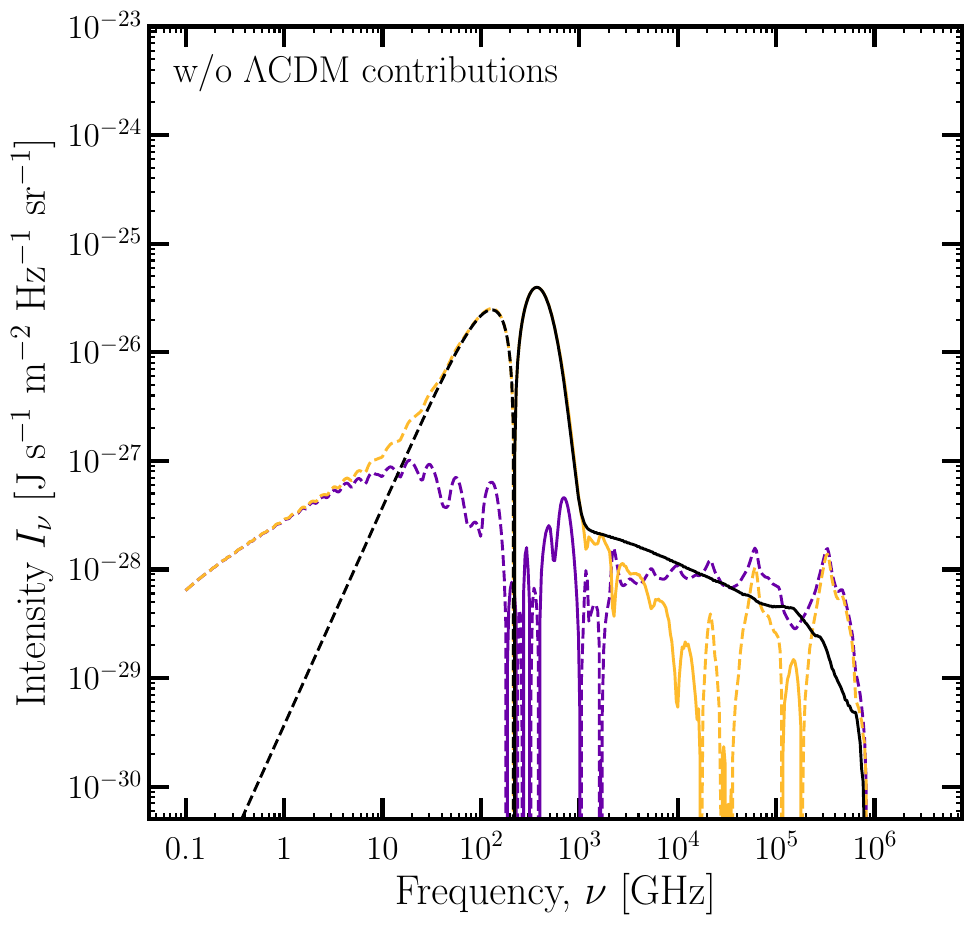}
	\caption{
		The effect of structure formation on the distortion.
		The left panel includes the $\Lambda$CDM and reionization contribution; the right panel has this contribution subtracted out.
		The purple line shows the distortion when we do not include the boost in power due to structure formation; yellow shows the distortion when we use the default model for structure formation in \dhis (see text for details).
		At late times, both the $y$-type distortion from heating and the ICS component of the distortion are enhanced.
		The difference between the two curves is shown by the black line.
	}
	\label{fig:struct_boost}
\end{figure*}
Small scale substructure can amplify spectral distortions from annihilating dark matter at late times; once dark matter halos begin to collapse, the annihilation rate can be significantly enhanced by the factor $\langle \rho_\chi^2 \rangle / \langle \rho_\chi \rangle^2$.
Fig.~\ref{fig:struct_boost} shows the effect of including structure formation; the boost factor is calculated assuming the halos have an Einasto profile and substructure, with properties as discussed in Ref.~\cite{Liu:2016cnk}.
The main difference is to enhance the heating $y$-type spectral distortion and the ICS contribution from later redshifts.
At frequencies less than a couple thousand GHz, the contribution of $\Lambda$CDM processes dominates over the spectral distortion solely from annihilations; hence, the only visible change to the spectral distortion in the left panel comes from the ICS component, which raises the high frequency shoulder around 4000 GHz.
For the models we considered, this effect is at most at the level of a few times $10^{-26}$ J s$^{-1}$ m$^{-2}$ Hz$^{-1}$ sr$^{-1}$, which may be just detectable by e.g. PIXIE~\cite{2011JCAP...07..025K, 2016SPIE.9904E..0WK}.
In the right panel, when we subtract out the $\Lambda$CDM contribution to the distortion, we can more clearly see the effect on the heating $y$-type component.

We reiterate that we only show contributions to the spectral distortions from between $1+z=3000$ and $1+z=4$.
We have estimated the contribution from other redshifts and find that it is subdominant to the contributions calculated here for the models we have studied; the degree to which the contribution from $3000 > 1+z > 4$ dominates depends on the channel for energy injection, see Appendix~\ref{app:other_rs} for details.
We also examine the rate of energy deposited into spectral distortions and find that it is consistent with the energy deposited into the continuum channel plus $y$-type distortion contributions, see Appendix~\ref{app:eng_rate} for details.

\subsection{Ionization Histories from Low Mass Dark Matter}
\label{sec:ionization}

Given the upgrades to the treatment of low-energy electrons described in Section~\ref{sec:DHv2_tech}, we can now accurately calculate the ionization histories with particles injected with energy less than a few keV. 
In this section, we present these results for the first time.
Moreover, since we have now implemented the capability for \dhis to track many excited states of hydrogen, transitions of these excited states can in principle modify the evolution of the free electron fraction, compared to the results derived by the previous version of \texttt{DarkHistory}.
The modified ionization rate and spectral distortion can affect each other: 
atomic transitions and recombination lead to the absorption or emission of low-energy photons,
while spectral distortions will affect the various atomic rates through the photon phase space density.
These modifications can in principle be large; however, given existing constraints on DM annihilation and decay, we will show that the effects on these processes are not significant enough to alter present constraints.

\subsubsection{Extending to lower energies}
\label{sec:ion_low_mass}

%
\begin{figure*}
	\includegraphics[width=\textwidth]{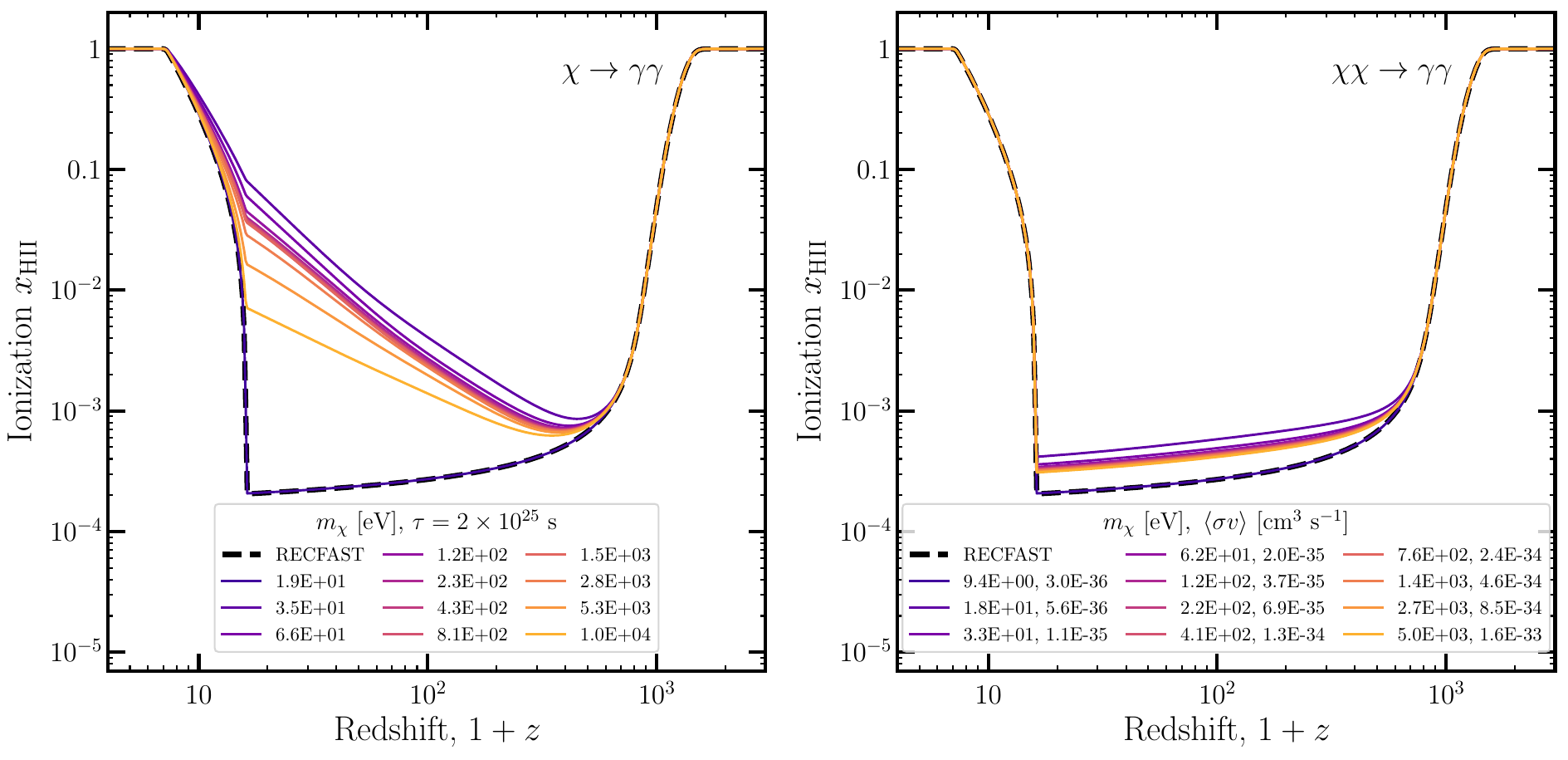}
	\caption{
		Ionization histories for dark matter decaying or annihilating to photons with masses $m_\chi <$ 10 keV. For decays, the lifetime is fixed to $\tau = 2 \times 10^{25}$ s; for annihilations, we choose the lifetime to be at the Planck 2018 constraint assuming 100\% efficiency of energy deposition, $\langle \sigma v \rangle / m_\chi = 3.2 \times 10^{-28}$ cm$^3$ s$^{-1}$ GeV$^{-1}$~\cite{Planck:2018vyg}.
		The black dashed line shows the ionization history calculated with \texttt{Recfast}~\cite{Seager:1999km,Seager:1999bc} in the absence of exotic energy injection.
	}
	\label{fig:xe_lowm}
\end{figure*}
A novel application of the method described in Section~\ref{sec:DHv2_tech} is the ability to accurately calculate the ionization histories resulting from dark matter with masses less than a few keV.
At energies lower than this, previous calculations relied on Monte Carlo results based on Ref.~\cite{MEDEAII}, while the applicability of the photoionization cross section used in earlier works as we approach the ionization potential of hydrogen and helium was unclear. 
With our improved low-energy treatment outlined in Section~\ref{sec:DHv2_tech}, we can now reliably calculate $x_\text{HII} (z)$ for arbitrarily low dark matter masses.

Fig.~\ref{fig:xe_lowm} shows the ionization histories from exotic energy injection.
The left panel shows dark matter decaying to photons, with masses between 19 eV and 10 keV and lifetime fixed at $\tau = 2 \times 10^{25}$ s.
At this lifetime, we see that the presence of exotic energy injection can significantly impact the ionization history prior to reionization; hence, these modifications can also be constrained by their impact on CMB anisotropies (see Section~\ref{sec:anisotropy}).
As $m_\chi$ drops below $2\mathcal{R}$, where $\mathcal{R} \equiv \SI{13.6}{\eV}$ denotes the hydrogen ionization potential, the resulting photons drop below the hydrogen ionization threshold; below this mass, exotic energy injection cannot directly ionize hydrogen and the curve calculated by \dhis almost exactly matches the standard \texttt{Recfast} ionization history; the difference arises from changes to recombination due to the presence of nonthermal photons below \SI{13.6}{\eV}, which we can calculate accurately, and will discuss further below.

The right panel of Fig.~\ref{fig:xe_lowm} is for dark matter annihilating to photons, with cross-section fixed to the Planck 2018 constraint assuming that all injected energy is transferred to the IGM, $\langle \sigma v \rangle / m_\chi = 3.2 \times 10^{-28}$ cm$^3$ s$^{-1}$ GeV$^{-1}$~\cite{Planck:2018vyg}.
We see that at this cross-section, the effect of exotic energy injection on ionization is at a level of less than $\Delta x_e < 10^{-3}$, which is smaller than the effect from decays but still potentially detectable by future experiments.
Again, we see that as the kinetic energy of the primary photons falls below $\mathcal{R}$, the effect on the ionization history is suppressed---although as we will discuss, it does not vanish entirely.

\subsubsection{Comparison to previous calculations}
\label{sec:old_ion}

%
\begin{figure*}
	\includegraphics[width=\textwidth]{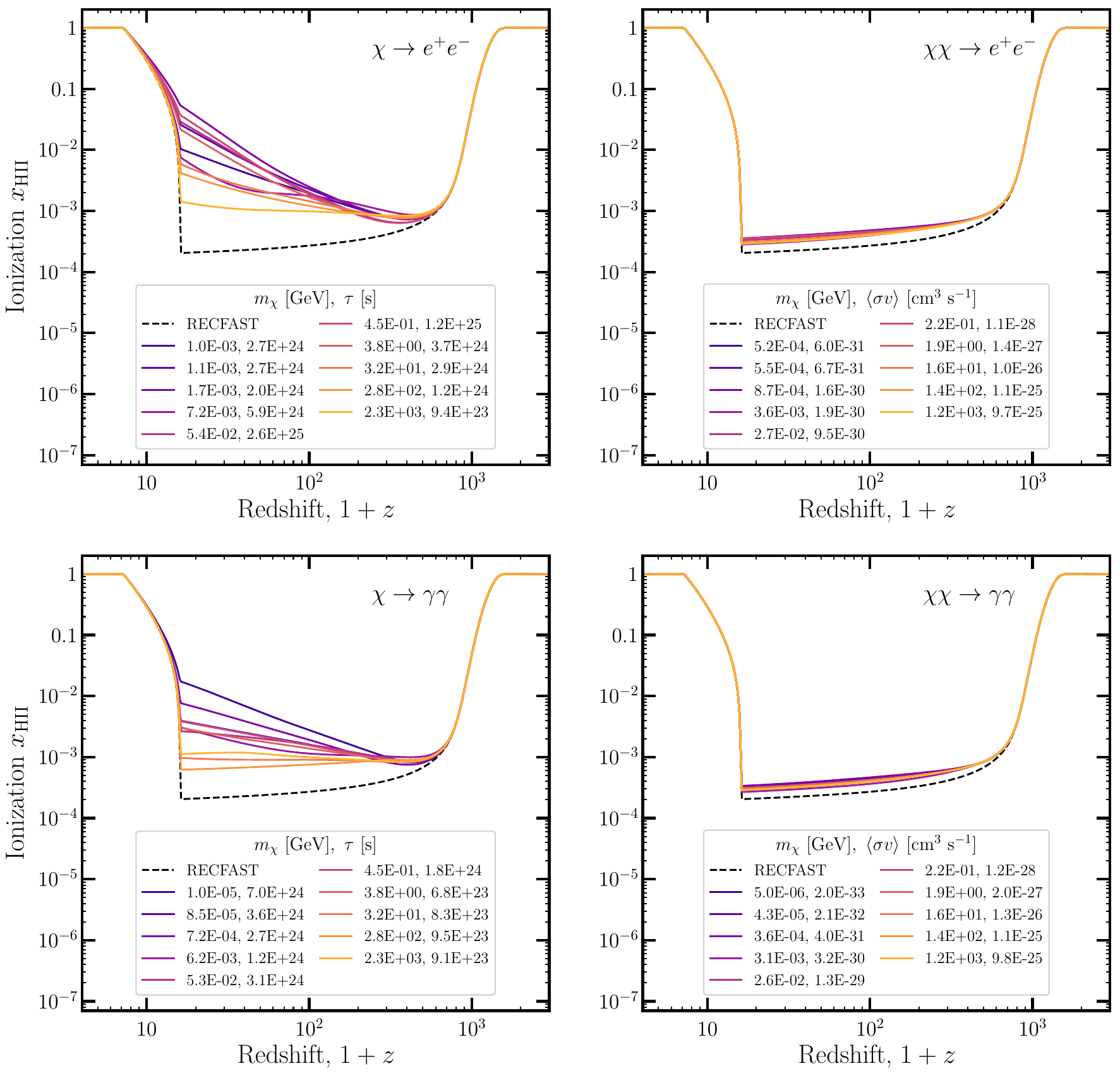}
	\caption{Ionization histories for the same DM models as in Fig.~\ref{fig:dist_grid}.
	The black dashed line shows the ionization history calculated with \texttt{Recfast}~\cite{Seager:1999km,Seager:1999bc} in the absence of exotic energy injection.
	While all models with exotic energy injection show an increase in the global ionization, the change in ionization is largest for the decaying DM models.}
	\label{fig:xe_grid}
\end{figure*}
\begin{figure*}
	\includegraphics[width=\textwidth]{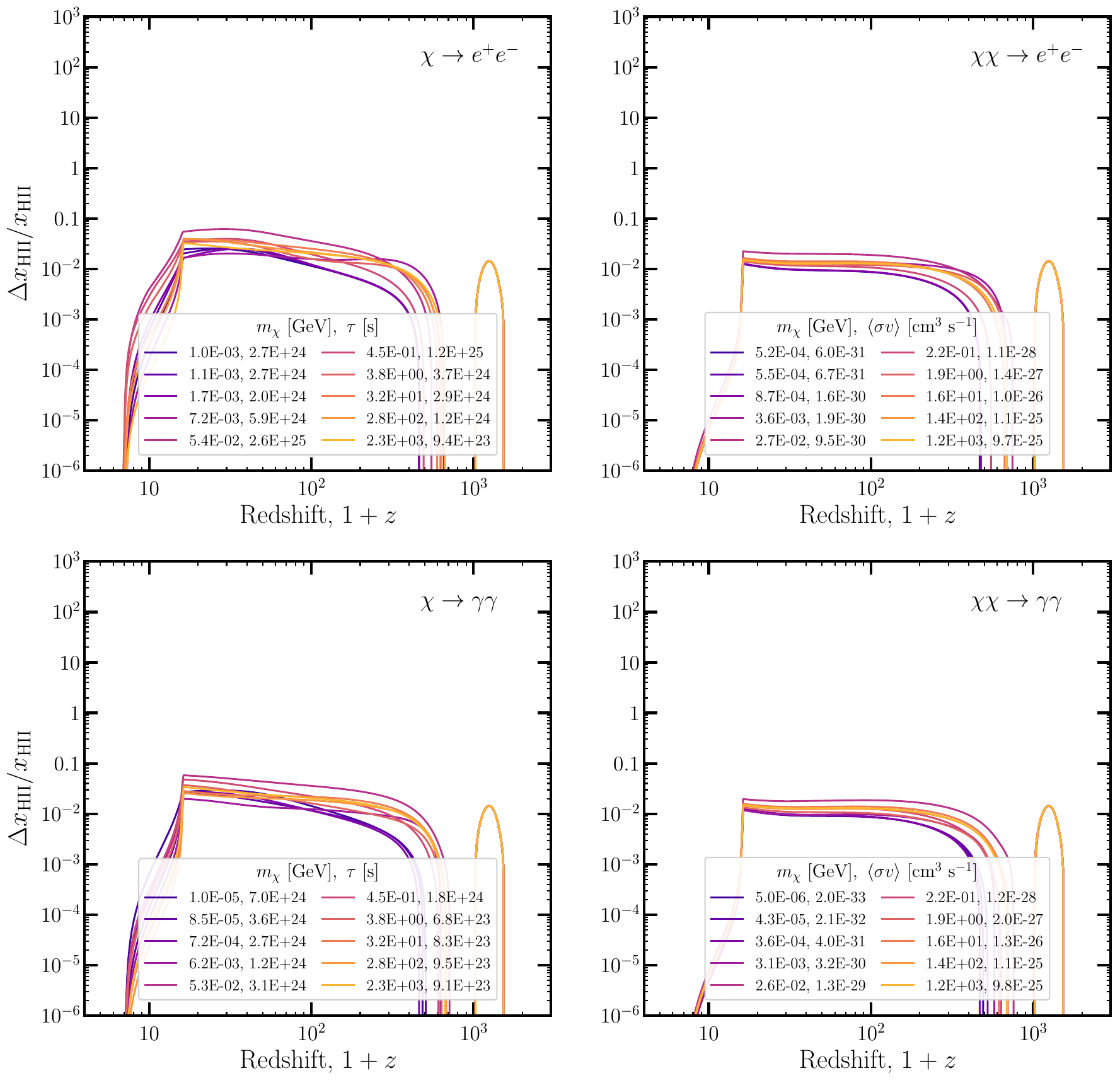}
	\caption{Difference in the ionization history relative to that calculated using \dhis \texttt{v1.0}.
	In other words, this is the difference in ionization between the two methods, divided by the history calculated with \dhis \texttt{v1.0}.
	The differences are at the level of less than 10 percent, which shows that tracking more excited states of hydrogen and the feedback between spectral distortions and ionization does not significantly modify the global ionization history.}
	\label{fig:delta_xe_grid_oldDH}
\end{figure*}
Due to the feedback between modifications to the global ionization history and spectral distortions, one might expect that accurately tracking the spectrum of background photons could change our calculation of the free electron fraction, $x_e$, relative to \dhis \texttt{v1.0} even from higher DM masses.
Here, we examine the size of this effect.

Fig.~\ref{fig:xe_grid} shows the ionization histories derived using the same DM models shown in Fig.~\ref{fig:dist_grid}.
Shown in the black dashed line is the ionization history calculated by \texttt{Recfast} in the absence of DM energy injection (including the hydrogen fudge factor, which we set to 1.125, and the double Gaussian function correction~\cite{Seager:1999km,Seager:1999bc}).
In all cases, as expected, including energy injections from DM annihilation/decay yields ionization histories that are larger than the \texttt{Recfast} history at all redshifts.
For DM annihilating to photons or $e^+ e^-$ pairs, we see that the spread in ionization histories is quite small.
For decaying DM, the spread in ionization histories is larger, and the histories can deviate more significantly from the \texttt{Recfast} calculation at late times (consistent with e.g. Ref.~\cite{Liu:2016cnk}).
We also show in Appendix~\ref{app:recfast_xcheck} the relative difference between the ionization history calculated with our new method versus that derived with \texttt{Recfast}. 

The reason for the difference in variation between annihilation and decay scenarios is due to the redshift dependence of the energy injection and the fact that we are using CMB-derived limits.
For decay, the constraint is mainly controlled by energy injection at redshift $1+z \sim 300$~\cite{Slatyer:2016qyl}, hence we would expect the ionization and difference relative to the \texttt{Recfast} curve at this redshift to be similar in the left panels of Figs.~\ref{fig:xe_grid}.
This can be clearly seen in the $\chi \rightarrow \gamma\gamma$ panels.
For annihilation constraints, the principal component comes from redshift $1+z \sim 600$~\cite{Slatyer:2015jla}, hence the ionization should be similar at this redshift in the right panels.
Since the power from annihilating DM decreases with redshift as $(1+z)^6$, then energy injection effectively shuts off after $1+z \sim 600$ and we see very little spread to the ionization histories, whereas the power from decaying DM only decreases as $(1+z)^3$ so there is much greater variation in the ionization histories at late times.

In order to determine how much the new machinery to track hydrogen levels and spectral distortions affects the ionization history, in Fig.~\ref{fig:delta_xe_grid_oldDH}, we show the relative difference between the ionization histories calculated using the new and original versions of \texttt{DarkHistory}; in Appendix~\ref{app:recfast_xcheck}, we also show the relative difference between the new \dhis and \texttt{Recfast}.
Again, we see that the spread in the histories is somewhat larger for decay scenarios than annihilation.
The largest differences are at less than the 10 percent level; hence, while feedback from the spectral distortions does modify the ionization history, this difference is not large for unconstrained values of $\tau$.
An important implication of this is that DM decay and annihilation results that depend on the ionization histories calculated by previous versions of this code remain largely unchanged.
This includes the constraints on decaying and annihilating DM derived using the effect of ionization on the CMB anisotropies~\cite{Slatyer:2015jla,Slatyer:2016qyl}.
We discuss the CMB anisotropy constraints in greater detail in the next section.

Finally, although the validity of the approximations made in \dhis \texttt{v1.0} was not clear for injections at dark matter masses less than 10 keV, we find that the differences between the ionization histories calculated with \dhis \texttt{v1.0} and with the improved treatment are also less than 10\% at all masses and redshifts.
The agreement between \dhis \texttt{v1.0} and the upgraded version shows that these approximations in fact yield accurate results for sub-keV dark matter, at least in the context of calculations of the modified ionization history and constraints that are determined by this modification.

\subsection{CMB anisotropy}
\label{sec:anisotropy}

Energy deposition into ionization from DM annihilation or decay modifies the process of recombination, which in turn modifies the CMB anisotropy power spectrum~\cite{Slatyer:2009yq, Kanzaki:2009hf}. 
Existing limits based on Planck data set strong limits on such processes, especially in the sub-GeV DM mass range~\cite{Slatyer:2015jla, Slatyer:2016qyl, Poulin:2016anj}. 
We can parametrize the power deposited into ionization relative to the power injected by DM as
\begin{equation}
	\left( \frac{dE}{dV dt} \right)_\text{H ion}= f_\text{H ion} \left( \frac{dE}{dV dt} \right)_\text{inj}.
\end{equation}
Properly computing the limits on DM interactions therefore relies on an accurate computation of $f_\text{H ion}(z)$ for annihilation or decay into $e^+e^-$ pairs and $\gamma \gamma$ pairs. 
Previous works on these limits only presented constraints down to a DM mass where the kinetic energy of each injected particle in these channel was $\SI{5}{\kilo\eV}$. 
As mentioned in Section~\ref{sec:ion_low_mass}, at lower energies, Monte Carlo results based on Ref.~\cite{MEDEAII} provided $f_\text{H ion}(z)$ for $e^+e^-$ injection, but required interpolation over a limited number of injection energies, and the applicability of cross-sections used in \dhis\texttt{v1.0} near the ionization potential of hydrogen and helium were unclear. 

With the method described in Section~\ref{sec:DHv2_tech}, we can now confidently extend the CMB power spectrum limits down to much lower kinetic energies of the injected particles. 
This is particularly important for DM particles such as ALPs decaying into photons, since these constraints are among the strongest available for $2 \mathcal{R} < m_\chi \lesssim \SI{10}{\kilo\eV}$, with CMB spectral distortions providing relevant limits for $m_\chi \geq 2 \mathcal{R}$~\cite{Bolliet:2020ofj}. 

We can obtain a good estimate for the constraint on DM decay by noting that the limit on the lifetime $\tau$ is approximately proportional to $f_\text{H ion}(z = 300)$~\cite{Slatyer:2016qyl}, since the impact of energy injection as a function of redshift on the CMB power spectrum is peaked at that redshift~\cite{Finkbeiner:2011dx}. 
We can therefore estimate limits for decay by simply computing $f_\text{H ion}(z = 300)$ as a function of $m_\chi$ for both channels, and rescaling that shape to match previous constraints on $\tau$. 
The options we use to calculate $f_\text{H ion}(z = 300)$ are:
\begin{itemize}
	\item  \texttt{backreaction = False},
	
	\item  \texttt{struct\_boost = None},
	
	\item  \texttt{reion\_switch = False},
	
	\item  \texttt{distort = True},
	
	\item  \texttt{nmax = 10},
	
	\item  \texttt{iterations = 1}
	
	\item  \texttt{reprocess\_distortion = False}.
\end{itemize}
Since we are only seeking an estimate of the CMB anisotropy constraint, we lower \texttt{nmax} to 10, \texttt{iterations} to 1, and set \texttt{reprocess\_distortion = False} to speed up the calculation.
This estimate is valid down to about $m_\chi = 2 \mathcal{R}$; for masses lower than this, the resulting photons are below the hydrogen ionization threshold; therefore, $f_\text{H ion}(z = 300)=0$.

\begin{figure}
	\centering
	\includegraphics[width=0.5\columnwidth]{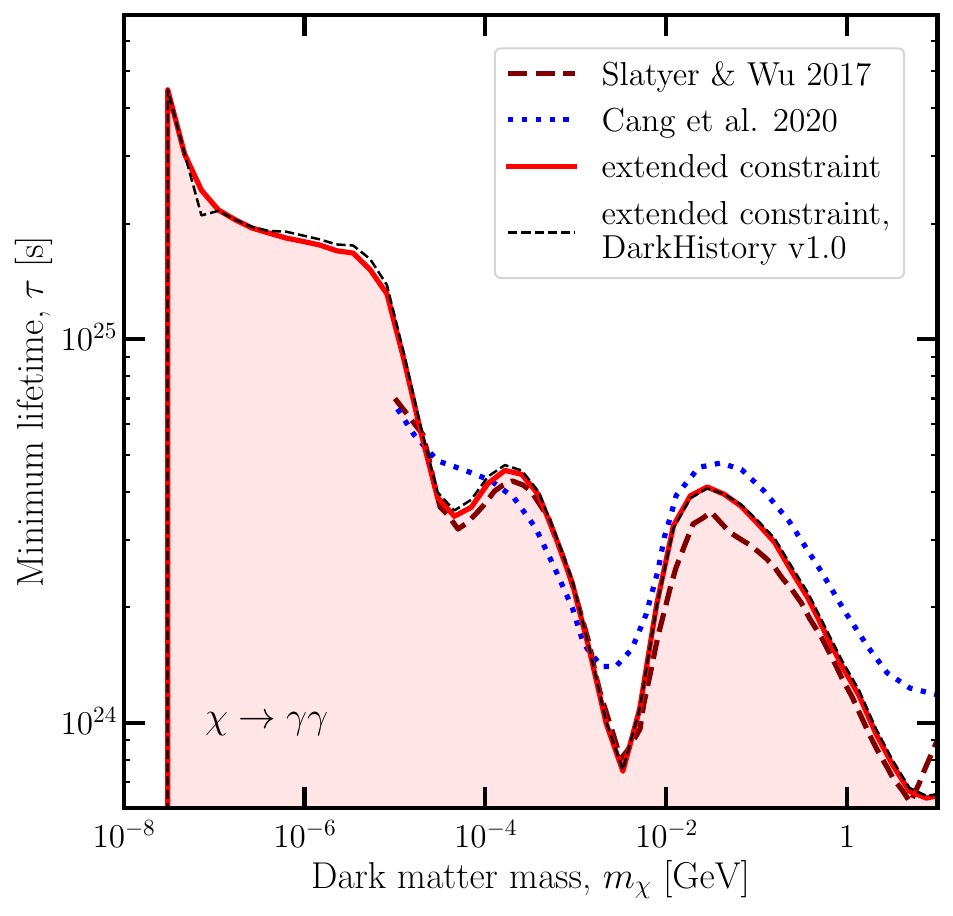}
	\caption{
		Estimated constraints on the lifetime for DM that decays to a pair photons, shown as the red contour. 
		These constraints were derived using the results of Ref.~\cite{Slatyer:2016qyl} which are shown as a dashed dark red line.
		We show the same estimate using \dhis \texttt{v1.0} in the thin dashed black line.
		Both of these results made use of Planck 2015 data.
		We also plot Ref.~\cite{Cang:2020exa}, which used Planck 2018 data, as a dotted blue line for comparison; we choose not to extend these constraints, since it is unclear if our estimation method applies to this data.
	}
	\label{fig:anisotropy_constraint}
\end{figure}
\begin{figure}
	\centering
	\includegraphics[width=0.5\columnwidth]{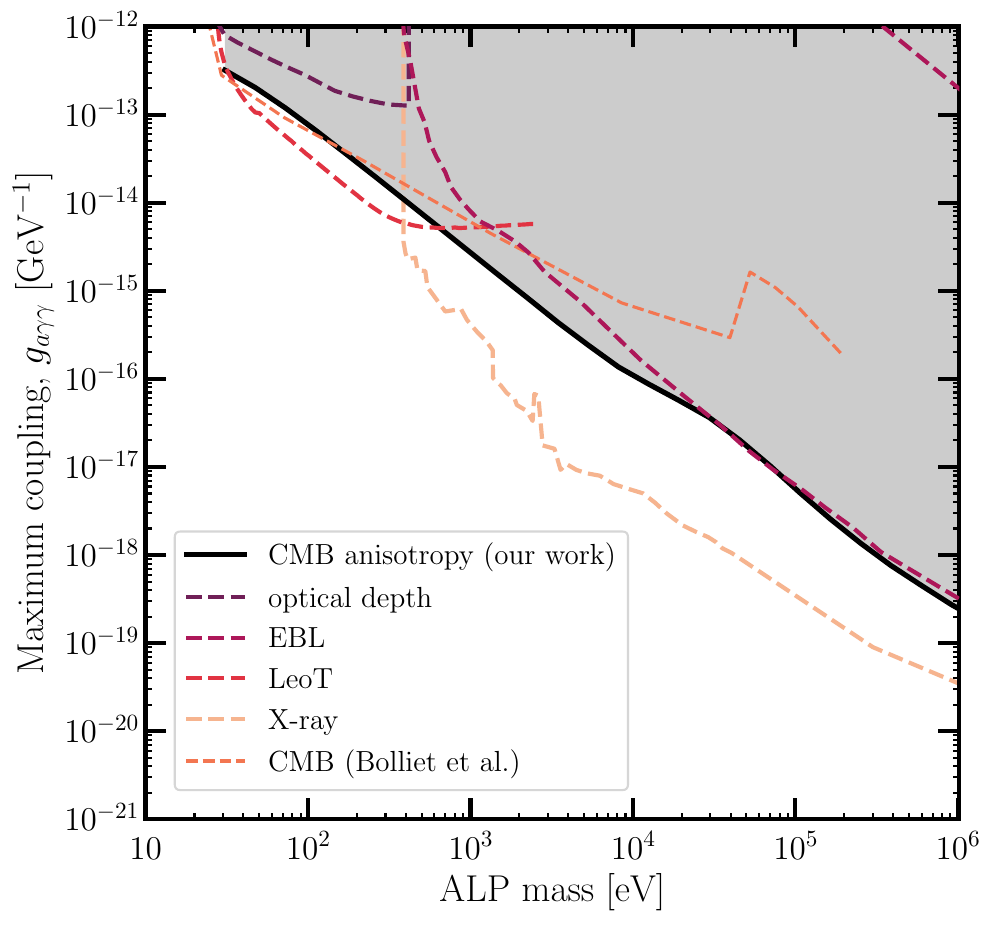}
	\caption{Constraints on the coupling of ALPs to photons.
	Our estimate of the constraint from modifications of CMB anisotropies is shown in the black contour.
	For comparison, we also show constraints from measurements of the ionization history/optical depth ($x_e$)~\cite{Cadamuro:2011fd}, extragalactic background light (EBL), heating in LeoT using the more conservative gas temperature of 7552 K~\cite{Wadekar:2021qae}, CMB spectral distortion and anisotropy constraints from Ref.~\cite{Bolliet:2020ofj}, and X-rays~\cite{Cadamuro:2011fd}.
	The constraints were plotted using the \texttt{AxionLimits} code~\cite{AxionLimits}.}
	\label{fig:ALP_constraint}
\end{figure}
\begin{figure}
	\centering
	\includegraphics[width=0.5\columnwidth]{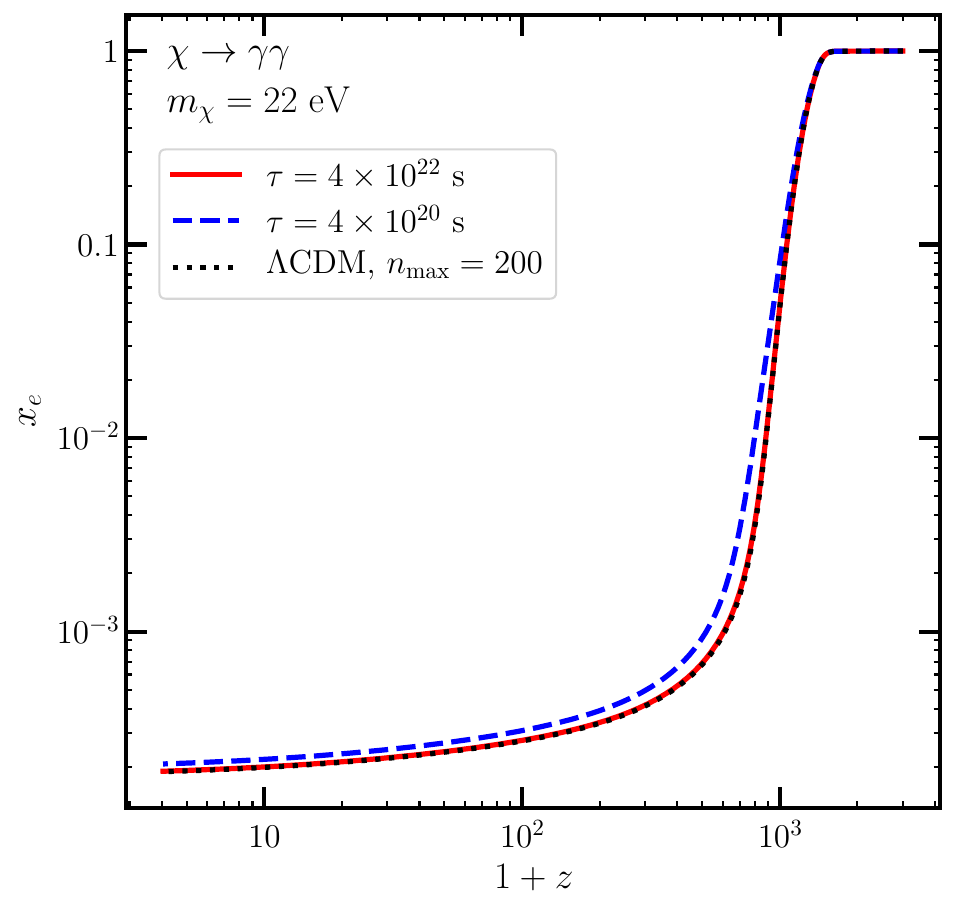}
	\caption{Ionization histories for DM decaying to photons at a mass of $m_\chi =$ 22 eV, i.e. for injection of photons with energies less than $\mathcal{R}$.
	At a DM lifetime of $\tau = 4 \times 10^{22}$ (red), which is taken from constraints based on measurements of the EBL~\cite{Nakayama:2022jza}, the ionization history is indistinguishable from the ionization history with no exotic energy injection (dotted black).
	In order to have an observable change to the ionization, we must decrease the lifetime quite a bit, e.g. to $\tau = 4 \times 10^{20}$ s (dashed blue).
	}
	\label{fig:subRydberg}
\end{figure}

Fig.~\ref{fig:anisotropy_constraint} shows our estimated lifetime constraints on DM decaying into a pair of photons from extending the bounds in Ref.~\cite{Slatyer:2016qyl}, which used the Planck 2015 dataset and likelihoods (as discussed in e.g.~\cite{Planck:2015bpv}). 
For $m_\chi > \SI{10}{\kilo\eV}$, we obtain excellent agreement between our current result and those of Ref.~\cite{Slatyer:2016qyl} as expected, but extend the anisotropy power spectrum limits down to $m_\chi = 2 \mathcal{R}$, where we find $\tau \gtrsim \SI{e25}{\second}$. 
We show the same estimate using \dhis \texttt{v1.0} in the thin dashed black line.
Per the discussion in Section\ref{sec:ionization}, we find that the approximations made in \dhis \texttt{v1.0} are in fact very good for this purpose, such that $f_\text{H ion}(z = 300)$ calculated using \dhis \texttt{v1.0} is remarkably close to the result using our improved low-energy treatment.
Hence, the constraints between the two versions of \dhis are nearly identical, although we are more confident in our control over theoretical uncertainties in the upgraded version. 
As we will discuss below, we can in principle also obtain constraints for $m_\chi < 2 \mathcal{R}$ coming from modifications to recombination due to the presence nonthermal low-energy photons, but these constraints appear to be weaker than existing limits, and not been computed carefully here.

We also show the constraints from Ref.~\cite{Cang:2020exa}, which used Planck 2018 + BAO data.
The shape of these constraints is smoother compared to Ref.~\cite{Slatyer:2016qyl}, since they used a different statistical approach (treating $m_\chi$ as a model parameter in their MCMC rather than performing a separate MCMC for each fixed $m_\chi$ value); hence it is not clear if the same principal component approach can be applied here.
As a result, we choose only to extend the estimated constraints using the results of Ref.~\cite{Slatyer:2016qyl} based on Planck 2015 data.
The most robust constraint would come from repeating the MCMC analysis for the Planck 2018 data using the latest version of \dhis presented here.

The limits on injected photons translate directly to a constraint on the ALP-photon coupling $g_{a \gamma \gamma}$, which we show in Fig.~\ref{fig:ALP_constraint}. 
Comparing to the limits derived using CMB anisotropy data in Ref.~\cite{Bolliet:2020ofj}, our constraints become stronger above $m_a \sim 100$ eV; this may at least in part be explained by simplifications in their treatment that are expected to break down at higher energies~\cite{Bolliet:private_comm}.
In addition, while weaker than limits based on heating of dwarf galaxies like LeoT~\cite{Wadekar:2021qae}, our constraints have completely complementary systematics.  
Note that this result is based on the full information provided by the Planck CMB power spectrum---across all available angular scales and using TT, TE and EE power spectra---instead of simply relying on the inferred optical depth to the surface of last scattering from WMAP7 to set a limit, as was done in Ref.~\cite{Cadamuro:2011fd}.
These results therefore supercede the estimate in Ref.~\cite{Cadamuro:2011fd} (labelled $x_e$ in their Fig. 11). 

For DM with $m_\chi < 2 \mathcal{R}$ decaying to photons, the primary photons cannot ionize anything, but the ionization history can still be affected: for example, injected photons can excite neutral hydrogen, and the resulting excited atoms have a higher probability of being ionized by a CMB photon.
Hence, although $f_\text{H ion} = 0$, there will still be some constraint on energy injection from the overall ionization history.
The current most competitive bound on DM decaying to photons at $m_\chi =$ 22 eV is given by measurements of the extragalactic background light (EBL) and corresponds to a lifetime of about $\tau \approx 4 \times 10^{22}$ s~\cite{Nakayama:2022jza}.
At this lifetime, we find the change to the ionization history is negligible.
In order to get a potentially observable change to the ionization, we would have to lower the lifetime by at least an order of magnitude or two, so any CMB anisotropy constraints for injection of sub-$\mathcal{R}$ photons are likely already ruled out by the EBL.
Fig.~\ref{fig:subRydberg} shows the ionization histories for $\Lambda$CDM and also including dark matter decaying to photons at $m_\chi =$ 22 eV and lifetimes of $\tau \approx 4 \times 10^{22}$ s and $4 \times 10^{20}$ s (to ensure this calculation is accurate, we again set \texttt{nmax = 200} and \texttt{iterations = 5} for this figure).
The curve with $\tau \approx 4 \times 10^{22}$ s is nearly identical to the $\Lambda$CDM curve; hence, it is unlikely that the resulting CMB anisotropy constraints would be competitive with current bounds in this low mass range.

Finally, Ref.~\cite{Langhoff:2022bij} pointed out that for sufficiently large couplings, a relic density of ALPs may freeze in from the thermal bath, forming an irreducible contribution to the DM density.
Although these ALPs may only comprise a small fraction of DM, DM searches can still be used to set very strong constraints on ALP parameter space using this relic abundance.
Ref.~\cite{Langhoff:2022bij} applied this argument to CMB constraints, using the anisotropy costraints in Ref.~\cite{Cang:2020exa} above masses of 10 keV and the spectral distortion information from Ref.~\cite{Bolliet:2020ofj} below 10 keV.
In principle, with the upgrades to \dhis presented here, we can now extend the CMB anisotropy constraints on such a population to arbitrarily low masses; we leave a detailed analysis to future work.


\subsection{Conclusion}
\label{sec:apps_conclusion}

In Section~\ref{sec:DHv2_tech}, we described a major upgrade to the capabilities of the \dhis package for modeling the effects of exotic energy injection in the early universe, including an improved treatment for energy deposition by low-energy electrons, tracking the full background spectrum of photons, and taking into account interactions between the photon bath and the hydrogen gas with an arbitrary number of excited states.
This treatment assumes that energy is deposited homogeneously, which may become inaccurate at late times once structure formation is well underway.
In this work, we describe several applications of this new machinery.

First, we demonstrated that we can now compute the CMB spectral distortions resulting from general exotic energy injections in the $z < 3000$ universe; specifically, we studied models of decaying and annihilating DM, with interaction rates at the exclusion limit from CMB anisotropy constraints, and predicted their signals in CMB spectral distortion.
The largest predicted signals come from light decaying DM, yielding spectral distortions up to about one part in $10^6$, and would be observable by future experiments such as PIXIE, which could detect spectral distortions down to about one part in $10^8$~\cite{2011JCAP...07..025K, 2016SPIE.9904E..0WK}. 
Other proposed experiments that may be able to detect this signal include PRISTINE~\cite{Chluba:2019nxa}, BISOU~\cite{Maffei:2021xur}, and FOSSIL~\cite{Chang:2022tzj}.
The signal from injection of high-energy particles can also be distinguished from other expected contributions by the shape of the high-frequency tail of the distortion, although we have not yet examined the impact of foregrounds on the detectability of these distortions.

Secondly, we extended the calculation of the modified ionization history from decaying DM to arbitrarily low masses.
We also showed that even though spectral distortions and modifications to the ionization history are coupled in the evolution equations, including spectral distortions (in models that are not currently excluded) does not significantly change the ionization history, with effects at the few-percent level and consistently below $10\%$.
This means that any constraints on DM annihilation and decay that were derived using the ionization history calculated with \dhis \texttt{v1.0} remain largely unchanged.

With these results in hand, we estimated the resulting CMB anisotropy constraints on low-mass scenarios. 
These results can also be translated into a limit on the coupling between ALPs and photons, which is competitive with other CMB-derived constraints and has complementary systematics to the leading constraints from e.g.~heating of dwarf galaxies.

There are numerous other possible applications of this improved technology, including studying the effects of the photons from exotic energy injection on the 21-cm signal or the EBL.
We leave these directions as topics for future study.

\section{Birth of the first stars amidst decaying and annihilating dark matter}
\label{sec:first_stars}

Exotic energy injection by e.g.\ decaying or annihilating dark matter (DM) is capable of heating and ionizing the baryonic gas in our universe, as well as altering the cosmic electromagnetic radiation background.
Such injection is a generic expectation of many proposed DM models, including weakly interacting massive particles (WIMPs).
Since the effects of such exotic energy injection can accumulate over cosmological timescales, interactions beyond the reach of terrestrial experiments can still lead to detectable changes in the global temperature and ionization level of the intergalactic medium (IGM), as well as in the cosmic electromagnetic background, detectable as e.g.\ distortions to the blackbody spectrum of the cosmic microwave background (CMB).

Many searches have been conducted to look for the effect of exotic energy injection on all three of these quantities.
For example, changes to the global ionization history can alter the cosmic microwave background anisotropies; measurements of the CMB power spectrum thus set strong limits on decaying or annihilating DM in the sub-GeV mass range~\cite{Adams:1998nr, Chen:2003gz, Padmanabhan:2005es, Slatyer:2009yq, Kanzaki:2009hf, Slatyer:2015jla, Slatyer:2016qyl, Poulin:2016anj, Cang:2020exa}.
Excess heating of the IGM can be constrained by measurements of the Lyman-$\alpha$ forest~\cite{Hiss:2017qyw, Walther:2018pnn, Gaikwad:2020art, Gaikwad:2020eip}, which has been used to set limits on DM velocity-dependent annihilation and decay~\cite{Cirelli:2009bb, Diamanti:2013bia, Liu:2016cnk, Liu:2020wqz}, as well as dark-photon DM~\cite{McDermott:2019lch, Caputo:2020bdy, Witte:2020rvb}; heating by dark photons has also been invoked to reconcile low- and high-redshift Lyman-$\alpha$ observations of the IGM~\cite{Bolton:2022hpt}. 
In addition, exotic heating has been shown to modify the 21\,cm brightness temperature, such that future observations could also set constraints on energy injection by decaying or annihilating DM, and primordial black holes~\cite{Evoli:2014pva, Lopez-Honorez:2016sur, DAmico:2018sxd, Liu:2018uzy, Cheung:2018vww, Mitridate:2018iag, Clark:2018ghm, 2022JCAP...03..030M}.
Lastly, new contributions to background radiation could be observed as distortions to the CMB blackbody spectrum~\cite{Liu:2023fgu,Liu:2023nct}, which has been used to set limits on a range of DM models before recombination~\cite{Chluba:2013vsa,Ali-Haimoud:2015pwa,Chluba:2015hma,Acharya:2018iwh}.

Current constraints still allow exotic energy injection at a level as large as \SI{1}{\eV} of energy per baryon (corresponding to a temperature of $\sim 10^4$ K) to the IGM at $2 \lesssim z \lesssim 5$~\cite{Liu:2020wqz}; however, such high gas temperatures may also affect gas collapse and subsequent star formation during the earlier epoch of cosmic dawn, when the gas temperature is expected to be much colder~\cite{Munoz:2018pzp}. 
The prospect of new observational probes of this epoch, e.g.\ 21\,cm telescopes, motivates understanding how the formation of the first stars will be impacted by exotic energy injection.

The first luminous objects are expected to have formed in pristine environments of primordial gas that have not yet been reprocessed by stars and galaxies.
For the small halos that collapse first, the only available coolants are atomic hydrogen, which is only effective at temperatures above $\sim\SI{e4}{\kelvin}$, and molecular hydrogen (H$_2$), which is thus thought to be crucial for early star formation.
The formation of H$_2$ depends on the temperature, ionization, and light irradiating the host halo; since exotic energy injection can affect all of these quantities, it may be possible to observe signatures of decaying or annihilating DM as we sharpen our understanding of the first stellar formation.
However, while we have made much progress in understanding a number of effects that are important for early star formation (such as the various channels by which H$_2$ forms, its dissociation by Lyman-Werner (LW) radiation, self-shielding of H$_2$, and baryon streaming velocities) there are still a number of astrophysical uncertainties regarding this process~\cite{2016MNRAS.462..601R,2018MNRAS.474..443G,Mebane:2020jwl,Munoz:2021psm,Hartwig:2022lon,Nebrin:2023yzm}.
Observations at high redshift by the James Webb Space Telescope (JWST)~\cite{Gardner:2006ky,2013MNRAS.429.3658R,2011ApJ...740...13Z,2022ApJ...937L...6R,2023AJ....165....2L}, Giant Magellan Telescope (GMT)~\cite{2012SPIE.8444E..1HJ}, or radio interferometers such as the Hydrogen Epoch of Reionization Array (HERA)~\cite{DeBoer:2016tnn,HERA:2021noe} and Square Kilometre Array (SKA)~\cite{Weltman:2018zrl,SKA:2018ckk} will improve our understanding of these effects and thus also help us constrain the impact of exotic energy injection. 

In this work, we seek a qualitative understanding of the effect of exotic energy injection on the first star-forming halos using a simple calculation for the molecular hydrogen formation rate and a toy model for collapsing halos.
A precise study will eventually require hydrodynamic simulations including all of the aforementioned effects (see e.g.\ Refs~\cite{Machacek:2000us,Yoshida:2003rw,Latif:2019zdi,Schauer:2018iig,2020MNRAS.492.4386S,Kulkarni:2020ovu,Park:2021sbs,Schauer:2020gvx,Schauer:2022cgf}). 
However, such simulations are prohibitively expensive to run for each DM scenario. 
Our work aims to pave the way for simulations by analytically finding regions of parameter space that will likely have the most interesting effects.

This section is structured as follows.
In Section~\ref{sec:H2}, we discuss the processes by which H$_2$ forms in the early universe.
Section~\ref{sec:injection} describes how we track the effects of exotic energy injection, and Section~\ref{sec:halo} outlines our toy halo model.
We then present our main results in Section~\ref{sec:results} before concluding with Section~\ref{sec:H2_conclusion}.
In addition, Appendix~\ref{app:LW} discusses the effect of self-shielding from LW radiation for certain energy injection channels, Appendix~\ref{app:IGM_vs_halo} estimates the relative contribution of decays and annihilations from within and beyond the halo, and Appendix~\ref{app:fs_halo} examines our assumptions on energy deposition within the halo.

\subsection{H$_2$ Formation}
\label{sec:H2}

The dominant pathway through which H$_2$ is formed in the first halos involves the formation of an intermediate H$^-$ ion. All other pathways are responsible for less than $\sim 2\%$ of the final abundance, and we neglect them in this analysis~\cite{Hirata:2006bt}. The H$^-$ mechanism for forming molecular hydrogen begins with radiative attachment of an electron to hydrogen:
\begin{equation}
    \mathrm{H} + e^- \rightarrow \mathrm{H}^- + \gamma \,.
    \label{eqn:H-_REA}
\end{equation}
H$_2$ is then formed through the detachment reaction
\begin{equation}
    \mathrm{H} + \mathrm{H}^- \rightarrow \mathrm{H}_2 + e^- \,.
    \label{eqn:H2_detach}
\end{equation}
The intermediate H$^-$ ion can be destroyed through mutual neutralization,
\begin{equation}
    \mathrm{H}^+ + \mathrm{H}^- \rightarrow 2\mathrm{H} \,,
\end{equation}
or through photodetachment, i.e.\ the reverse reaction of Eqn.~\eqref{eqn:H-_REA}.\footnote{We can neglect the reverse reaction of Eqn.~\eqref{eqn:H2_detach} since this is highly suppressed at temperatures below $10^4$ K~\cite{1998A&A...335..403G}.}
H$_2$ can also be destroyed by LW photons, which have energies between 11.2 and \SI{13.6}{\eV}~\cite{Haiman:1996rc,Tegmark:1996yt,Machacek:2000us,Yoshida:2003rw,Mesinger:2006pa,Wise:2007cf,OShea:2007ita,Schauer:2020gvx,Kulkarni:2020ovu}.

Using the same notation as in Ref.~\cite{Hirata:2006bt}, we denote the rates for each of these reactions by $k_1$, $k_2$, $k_3$, respectively; $k_{-1}$ for photodetachment; and $k_\mathrm{LW}$ for photodissociation by LW photons.
Then the abundances of H$^-$ and H$_2$ are given by 
\begin{alignat}{2}
    \frac{dx_{\mathrm{H}^-}}{dt} &=&& \, k_1 x_e n_\mathrm{HI} - k_{-1} x_{\mathrm{H}^-} - k_2 x_{\mathrm{H}^-} n_\mathrm{HI} - k_3 x_{\mathrm{H}^-} n_{\mathrm{HII}} \,, \label{eqn:dxHminusdt}\\
    \frac{dx_\mathrm{H_2}}{dt} &=&& \, k_2 x_\mathrm{H^-} n_\mathrm{HI} - k_\mathrm{LW} x_{\mathrm{H}_2} \,. \label{eqn:dxH2dt_general}
\end{alignat}
$x_i$ denotes the abundance of species $i$, given by its number density $n_i$ relative to the total number density of hydrogen nuclei (ionized, atomic, and in H$_2$). 
Note that we track the free-electron fraction $x_e$ and the ionized hydrogen fraction $x_\mathrm{HII}$ separately, since $x_e$ also receives contributions from ionized helium.
Ref.~\cite{1998ApJ...509....1S} provides fits for the rates $k_1$, $k_2$, and $k_3$ as a function of the gas temperature, $T$:
\begin{align}
    k_1 (T) &= \SI{3e-16}{\centi\meter\cubed\per\second} \left( \frac{T}{\SI{300}{\kelvin}} \right)^{0.95} \!\!\!\!\!\!\! \exp \left(- \frac{T}{\SI{9320}{\kelvin}} \right)  \,, \\
    k_2 (T) &= \SI{1.5e-9}{\centi\meter\cubed\per\second} \left( \frac{T}{\SI{300}{\kelvin}} \right)^{-0.1} \,, \\
    k_3 (T) &= \SI{4e-8}{\centi\meter\cubed\per\second} \left( \frac{T}{\SI{300}{\kelvin}} \right)^{-0.5} \,.
\end{align}

The photodetachment rate receives contributions from CMB photons, including a term from nonthermal radiation, i.e. radiation that is not part of the CMB blackbody~\cite{Hirata:2006bt,Coppola:2013qza},
\begin{equation}
    k_{-1} = 4 \left( \frac{m_e T_\mathrm{CMB}}{2\pi} \right)^{3/2} e^{-B(\mathrm{H}^-) / T_\mathrm{CMB}} k_1 (T_\mathrm{CMB}) + \int_{B(\mathrm{H}^-)}^{\infty} d\omega \frac{dn_\gamma}{d \omega} \sigma_{-1} (\omega) \,.
    \label{eqn:k-1}
\end{equation}
In the above equations, $m_e$ is the electron mass, $T_\mathrm{CMB}$ is the CMB temperature, $B(\mathrm{H}^-) = 0.754$ eV is the threshold energy for photodetachment, $dn_\gamma/d \omega$ is the number density of distortion photons per unit energy, and $\sigma_{-1}$ is the cross-section for photodetachment, which is well fit by the expression~\cite{Tegmark:1996yt}
\begin{equation}
    \sigma_{-1} (\omega) \approx 
    \SI{3.486e-16}{\centi\meter\squared}
    \times \frac{(\varpi-1)^{3/2}}{\varpi^{3.11}},
\end{equation}
where $\varpi = \omega / \SI{0.74}{\eV}$.

The LW photodissociation rate $k_{\rm LW}$ is highly dependent on self-shielding, i.e. whether or not the outermost shell of H$_2$ in a halo shields the H$_2$ at the center of the halo from the LW flux.
Given the current uncertainties on this process, we will bracket this effect by showing results for either complete self-shielding or no self-shielding.
However, recent hydrodynamical simulations find evidence for strong self-shielding~\cite{Schauer:2020gvx,Kulkarni:2020ovu,2020MNRAS.492.4386S}, so throughout the text we will focus on that limit by setting $k_\mathrm{LW} = 0$.
In the opposite limit of no self-shielding the photodissociation rate can be approximated by~\cite{Wolcott-Green:2016grm,Friedlander:2022ovf}
\begin{equation}
    k_\mathrm{LW} \approx\, \SI{1.39e-12}{\per\second} \times \left( \frac{J_\mathrm{LW}}{\SI{e-21}{\erg\per\second\per\hertz\per\centi\meter\squared\per\sr}} \right),
    \label{eqn:kLW}
\end{equation}
where $J_\mathrm{LW}$ is the average intensity in the LW band.

Finally, since the H$^-$ destruction rate is much faster than the Hubble rate, we can treat $x_{\mathrm{H}^-}$ assuming a steady-state approximation, i.e.\ $dx_{\mathrm{H}^-} / dt = 0$.
Setting the right-hand side of Eqn.~\eqref{eqn:dxHminusdt} to zero and substituting the resulting expression for $x_{\mathrm{H}^-}$ into Eqn.~\eqref{eqn:dxH2dt_general} gives~\cite{Hirata:2006bt} 
\begin{equation}
    \frac{d x_{\mathrm{H}_2}}{dt} = \frac{k_1 k_2 x_e n_\mathrm{HI}^2}{k_2 n_\mathrm{HI} + k_{-1} + k_3 n_\mathrm{HII}} - k_\mathrm{LW} x_{\mathrm{H}_2} \,.
    \label{eqn:dxh2dt}
\end{equation}
From the equations listed above, we see that the formation rate for H$_2$ depends on the matter temperature $T$, the spectrum of nonthermal radiation $dn_\gamma / d\omega$, the ionized hydrogen fraction $x_\mathrm{HII}$, and the free-electron fraction $x_e$.
The values for all of these quantities in the homogeneous IGM can be calculated in the presence of exotic energy injection using the \texttt{DarkHistory} code~\cite{DarkHistory,Liu:2023fgu,Liu:2023nct}, which we will describe in the next section.

There are a number of additional processes that affect H$_2$ formation~\cite{Nebrin:2023yzm}, such as the other pathways mentioned above, H$_2$ formation on dust grains~\cite{2020ApJ...905..151N}, and baryon streaming velocities~\cite{Tseliakhovich:2010bj,Stacy:2010gg,Greif:2011iv,Fialkov:2011iw,Naoz:2012fr,Fialkov:2014rba,Hirano:2017wbu,Schauer:2022cgf,Schauer:2020gvx,Schauer:2018iig,Kulkarni:2020ovu}.
These effects should be included in careful simulation studies; however, our focus here is the effect of exotic energy injection.
Moreover, given that the transfer functions used in \texttt{DarkHistory} rely on approximations that affect the changes to the temperature and ionization histories from energy injection at the level of about 10\%~\cite{Galli:2013dna,DarkHistory}, ignoring these effects is sufficient for an initial study of the effect of exotic energy injection on early star formation.

\subsection{Exotic energy injection in the IGM}
\label{sec:injection}

We now review our treatment of exotic energy injection, and the impact of these processes on the formation of H$_2$ in the IGM, which we treat as homogeneous and at mean cosmological energy density.
This will serve as a warmup to the calculation inside the first galaxies, which will follow.
Exotic energy injection, such as annihilating or decaying DM, can heat and ionize the universe more than would be expected in a universe without such injections~\cite{Adams:1998nr, Chen:2003gz, Padmanabhan:2005es, Zhang:2007zzh, Slatyer:2009yq, Cirelli:2009bb, Kanzaki:2009hf, Galli:2009zc, Hisano:2011dc, Hutsi:2011vx, Galli:2011rz, Finkbeiner:2011dx, Slatyer:2012yq, Galli:2013dna, Diamanti:2013bia, Madhavacheril:2013cna, Evoli:2014pva, Slatyer:2015jla, Slatyer:2015kla, Lopez-Honorez:2016sur, Liu:2016cnk, Slatyer:2016qyl, Poulin:2016anj, DAmico:2018sxd, Liu:2018uzy, Cheung:2018vww, Mitridate:2018iag, Clark:2018ghm, McDermott:2019lch, Caputo:2020bdy, Cang:2020exa, Witte:2020rvb, Liu:2020wqz, Bolton:2022hpt, 2022JCAP...03..030M}, and also modify the background radiation spectrum~\cite{Ali-Haimoud:2015pwa,Acharya:2018iwh,McDermott:2019lch,Bolliet:2020ofj,Bernal:2022wsu}.
\texttt{DarkHistory}\, \githubmaster\,~\cite{DarkHistory,Liu:2023fgu,Liu:2023nct} calculates the global temperature and ionization, and evolves background radiation while self-consistently including the effect of homogeneous energy injection; hence, we use \texttt{DarkHistory} to evolve the properties of the IGM and background radiation in which the first halos and stars will form.

\texttt{DarkHistory} tracks how injected particles cool and deposit their energy; the rate of energy deposited into a channel $c$ can be parametrized relative to the rate of energy injected as 
\begin{equation}
    \left( \frac{dE}{dV dt} \right)^\mathrm{dep}_c = f_c (z) \left( \frac{dE}{dV dt} \right)^\mathrm{inj} .
\end{equation}
These $f_c$'s can then be used in the equations determining the IGM temperature, ionization, and radiation spectrum to include the effects from injected particles. 
Since the $f_c$'s are calculated at each redshift step after having evolved the background equations, the $f_c$'s take backreaction into account; in other words, as exotic injections change the background ionization, and radiation spectrum, these changes modify $f_c$'s at later redshifts and subsequent energy deposition.

The evolution of the IGM gas temperature, $T_\mathrm{IGM}$, is given by
\begin{equation}
    \dot{T}_\mathrm{IGM} = 
    \dot{T}_\mathrm{IGM}^\mathrm{adia} + \dot{T}_\mathrm{IGM}^\mathrm{comp} + \dot{T}_\mathrm{IGM}^\mathrm{inj},
    \label{eqn:evol_Tm}
\end{equation}
where the adiabatic cooling term is given by
\begin{equation}
    \dot{T}^\mathrm{adia} = \frac{2}{3} \frac{\dot{n}_\mathrm{H}}{n_\mathrm{H}} T,
    \label{eqn:T_adia}
\end{equation}
and $n_\mathrm{H}$ is the hydrogen number density.\footnote{In principle, $n_\mathrm{H}$ should be replaced with the total number of particles in the system, but since $x_e$ remains small and we assume that helium evolves the same way as hydrogen, we treat $n_\mathrm{H}$ as a proxy for the total particle number evolution.}
Here, we drop the subscript on temperature since the expression is general and also applies for halos.
In the IGM, this expression simplifies to $\dot{T}_\mathrm{IGM}^\mathrm{adia} = - 2 H T_\mathrm{IGM}$, where $H$ is the Hubble parameter, since the background density evolves as $n_\mathrm{H} \propto (1+z)^3$.
The Compton scattering term is
\begin{equation}
    \dot{T}^\mathrm{comp} = \Gamma_C(T_\mathrm{CMB} - T) .
    \label{eqn:T_comp}
\end{equation}
In the above equation, $\Gamma_C$ is the Compton scattering rate (see e.g.\ Ref.~\cite{DarkHistory} for its definition).
$\dot{T}^\mathrm{inj}$ represents sources of exotic heating, and its homogeneous contribution can be written in terms of $f_\mathrm{heat}$ as
\begin{equation}
    \dot{T}_\mathrm{IGM}^\mathrm{inj} = \frac{2 f_\mathrm{heat}(z)}{3 \bar{n}_\mathrm{H} (1 + x_e + \mathcal{F}_\mathrm{He})} \left( \frac{dE}{dV dt} \right)^\mathrm{inj},
    \label{eqn:T_inj}
\end{equation}
where $\bar{n}_\mathrm{H}$ is the total number density of hydrogen nuclei(ionized, neutral, and molecular) in the IGM and $\mathcal{F}_\mathrm{He} = \overline{n}_\mathrm{He}/\bar{n}_\mathrm{H}$ is the relative abundance of helium nuclei by number.

To obtain the history of the global free-electron fraction $x_e$ and the spectrum of background radiation, we treat hydrogen as a modified Multi-Level Atom (MLA)~\cite{Seager:1999km,Chluba:2006bc,2010PhRvD..81h3005G,Rubino-Martin:2006hng} and track the energy levels of hydrogen up to the principal quantum number $n=200$.
The photon spectrum then takes into account the absorption and emission of photons from bound-bound transitions, recombination, photoionization, and exotic injection.
Details of how the MLA is implemented can be found in Ref.~\cite{Liu:2023fgu}.
Since we are concerned with the formation of the very first stars, we neglect astrophysical sources of radiation when calculating the spectrum of photons, e.g.\ for the LW background, we only consider the contribution from exotic energy injection.

\begin{figure}
    \includegraphics[scale=0.5]{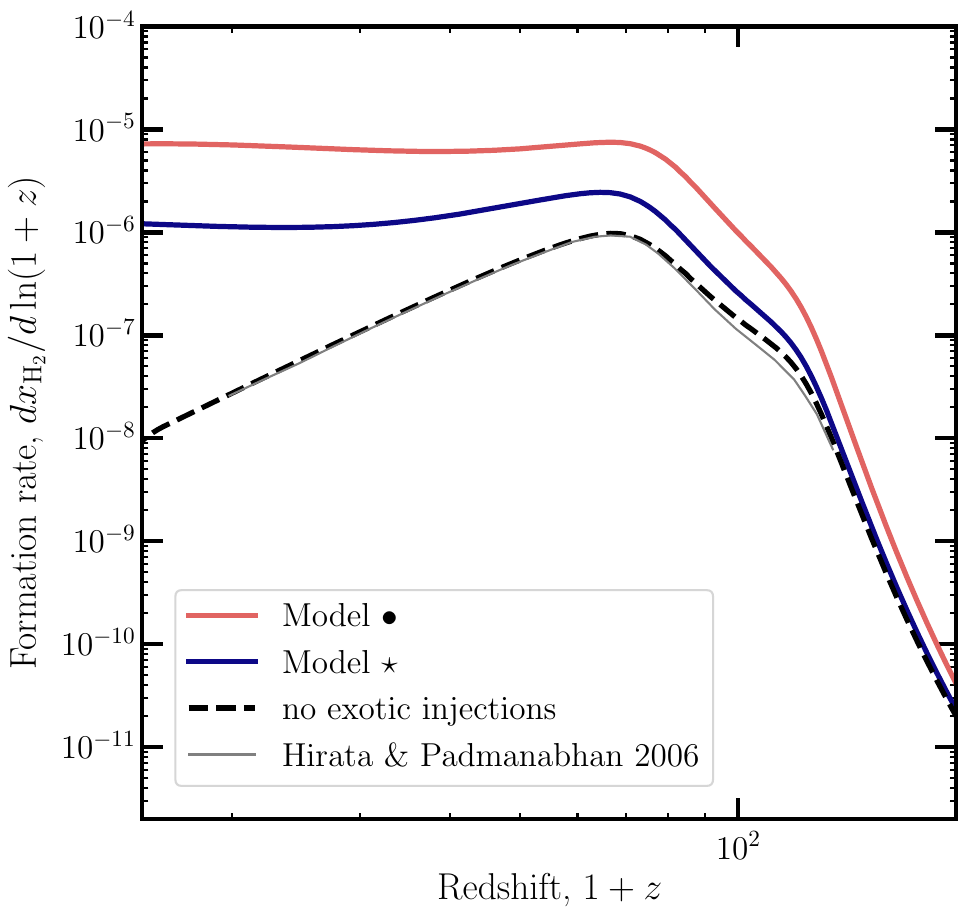}
    \hspace{5mm}
    \includegraphics[scale=0.5]{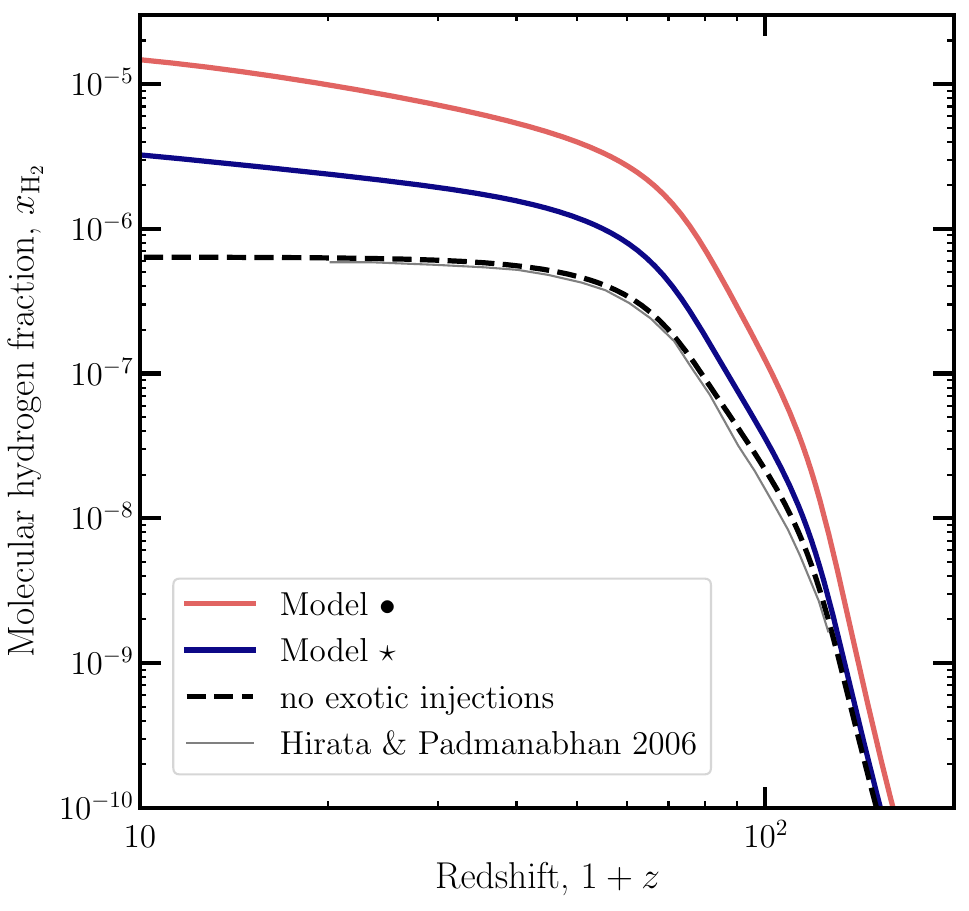}
    \caption{
    H$_2$ formation rate (left) and fraction (right) in the average IGM as a function of redshift.
    The black dashed curves show the results assuming standard temperature and ionization histories, including spectral distortions from standard cosmology and no exotic energy injection.
    For comparison, we show the results of Ref.~\cite{Hirata:2006bt} in the thin gray line.
    The blue and pink curves show the results if we include decays by DM to $e^+ e^-$ pairs; both models have $m_\chi = 185$ MeV, but different lifetimes.
    In both cases, the dominant effect is to \textit{enhance} the formation of H$_2$ due to the higher free-electron fraction caused by exotic ionizations.
    }
    \label{fig:global_H2}
\end{figure}

Together, Eqn.~\eqref{eqn:evol_Tm} and the MLA form a closed system of equations that \texttt{DarkHistory} solves to self-consistently determine the evolution of these quantities in the presence of sources of exotic energy injection, in particular decaying or annihilating DM.
The outputs of \texttt{DarkHistory} can then be used to calculate the abundance of H$_2$ in light of these effects.

\begin{table}[t!]
  \renewcommand{\arraystretch}{1.5}
  \centering
  \begin{tabular}{|c|c|c|c|}
    \hline
    & Channel & Mass [MeV] & $\log_{10} (\tau / [\mathrm{s}])$ \\ 
    \hline
    Model $\bullet$ & $\chi \rightarrow e^+ e^-$ & 185 & 25.6 \\
    Model $\star$ & $\chi \rightarrow e^+ e^-$ & 185 & 26.4 \\
    \hline
  \end{tabular} 
  \caption{
  Parameters for two fiducial DM models. 
  }
  \label{tab:fiducial}
\end{table}

As an initial validation of our molecular hydrogen treatment, Fig.~\ref{fig:global_H2} shows the formation rate and total abundance of H$_2$ in the IGM in terms of the molecular hydrogen fraction, $x_{\mathrm{H}_2}$; ultimately, for star formation, we will replace $T_\mathrm{IGM}$, $x_e$, and $x_{\mathrm{H}_2}$ with the analogous quantities in a halo in Section~\ref{sec:halo}, but for comparison to other works we only consider the IGM contribution here.
We also set $k_\mathrm{LW}$ to zero here for the purpose of comparison with results from the literature.
The black dashed curved shows our calculation in the absence of exotic energy injection; i.e. using the global temperature history, ionization history, and spectral distortions from standard cosmology.
We compare our results against Ref.~\cite{Hirata:2006bt}, finding excellent agreement.

We also show the results including two fiducial models of exotic energy injection, whose parameters are listed in Table~\ref{tab:fiducial}.
In both models, DM has a mass $m_\chi = 185$ MeV and decays to $e^+ e^-$ pairs (throughout this text, $\chi$ denotes the hypothetical DM particle).
We focus on sub-GeV decaying DM in this work, since this mass range is less constrained by existing probes, and the impact of the halo on energy deposition within the halo is expected to be more straightforward.
Two different lifetimes are chosen to reflect two distinct, opposite effects that energy injection can have on star formation.
Further discussion on our choice of models can be found in Sec.~\ref{sec:collapse}.
Model $\bullet$\, assumes a lifetime of $\log_{10} (\tau / [\mathrm{s}]) = 25.6$, which is excluded by Voyager constraints~\cite{Boudaud:2016mos, Boudaud:2018oya} but unconstrained by CMB anisotropies~\cite{Slatyer:2016qyl};\footnote{While Model $\bullet$\, is nominally ruled out, we include it for illustrative purposes because we show in Section~\ref{sec:halo} that the effect on star formation is in the opposite direction of Model $\star$.}
Model $\star$\, assumes an even longer lifetime $\log_{10} (\tau / [\mathrm{s}]) = 26.4$ is beyond current limits from Voyager and the CMB.
For the IGM, in both models, the H$_2$ abundance and formation rate is larger than for a standard cosmology; the effect is greater for Model $\bullet$\, because the lifetime is shorter.
This effect is due to the enhanced free-electron fraction from exotic ionization; the additional electrons serve to catalyze the formation of molecular hydrogen through the reaction in Eqn.~\eqref{eqn:H-_REA}.

\subsection{Halo evolution}
\label{sec:halo}

To study the effect of exotic energy injection on the first star-forming halos, we modify the simple procedure in Ref.~\cite{Tegmark:1996yt} to evolve the free-electron fraction $x_e$, the H$_2$ fraction $x_{\mathrm{H}_2}$, and the gas temperature $T_\mathrm{halo}$ inside a halo in the presence of energy injection from DM.
To summarize the method, we assume the halo has a mass $M_\mathrm{halo}$, and can be approximated as a spherical ``top-hat'' as it collapses, i.e. the halo has a uniform density $\rho$ within a spherical volume.
When the halo reaches either the virial density $\rho_\mathrm{vir}$ or virial temperature $T_\mathrm{vir}$, then the halo is considered virialized; we label the redshift at which this occurs as $z_\mathrm{vir}$.
Upon virialization, we assume the halo becomes an isothermal sphere; then the virial temperature is approximately related to the halo mass by~\cite{Barkana:2000fd}
\begin{equation}
    T_\mathrm{vir} \approx 224 \,\mathrm{K} \, \left( \frac{\Omega_m h^2}{0.14} \right)^{1/3}  \left( \frac{M_\mathrm{halo}}{10^4 M_\odot} \right)^{2/3} \left( \frac{1+z_\mathrm{vir}}{100} \right) .
    \label{eqn:Tvir}
\end{equation}
In the above equation, $h$ is defined by $H_0 = \SI[parse-numbers=false]{\num[parse-numbers=true]{100}h}{\kilo\meter\per\second\per\mega\parsec}$, where $H_0 = \SI{67.36}{\kilo\meter\per\second\per\mega\parsec}$ is the Hubble parameter today, $\Omega_m$ is the matter density parameter, and $M_\odot$ is a solar mass.
We take $\rho_\mathrm{vir} = 18 \pi^2 \rho_0 (1+z)^3$, where $\rho_0$ is the mean matter energy density today.
After the halo virializes, we hold its density fixed and continue to evolve the other quantities.
If the temperature of the gas falls quickly enough, then we infer that it is capable of collapsing to form stars; the precise conditions we employ are described in Sec.~\ref{sec:collapse}.
In the following sections, we describe the ingredients for this calculation in more detail.

In principle, decays and annihilations from within the star-forming halos themselves should be considered separately from the IGM contribution.
Moreover, one should account for the fact that energy deposition can be different in the overdense halo (even if illuminated with the background emitted from the average IGM).
Exotic energy injection from halos has been studied using Monte Carlo methods~\cite{Schon:2014xoa,Schon:2017bvu}, but is beyond the scope of this work.
Instead, we expect the energy deposited per particle inside halos to be comparable to that in the IGM for DM decay; we justify this assumption in two ways.

First, in Appendix~\ref{app:IGM_vs_halo}, we estimate the geometrical factors for decay and annihilation from early halos vs. from the IGM and find that the contributions from the IGM for decays dominate over the halo contribution for free-streaming decay/annihilation products.
Second, in Appendix~\ref{app:fs_halo}, we also follow simplified particle cascades to show that for most injected particles, the presence of the halo has a small impact on energy injection and deposition.
Intuitively, this happens because the path lengths of many of the particles produced in the cascade from the initial DM process tend to be much longer than the halo itself, leading to the intensity being dominated by the IGM contribution. 
If these long-path-length particles deposit their energy through interactions with the target gas particles, equal intensity of particles received in the halo and in the IGM translates into equal energy deposited \emph{per gas particle} in the halo, which is the relevant parameter for the effect of exotic energy injection on the free-electron fraction and temperature.
In the opposite limit, where products of decay/annihilation lose their energy promptly, for decays the enhanced injection in the halo is canceled by the higher density of targets, when computing the power deposited per target.
A more in-depth exploration of this question, while unnecessary for this study, would be an interesting subject for future work.

\subsubsection{Density}
\label{sec:halo_density}

Following Ref.~\cite{Tegmark:1996yt}, we take a simple approximation for the temporal evolution of the density of a collapsing spherical top-hat:\footnote{This expression corrects a sign error in Ref.~\cite{Tegmark:1996yt}, providing a good fit to the spherical collapse model.}
\begin{gather}
    \rho(z) \approx \rho_0 (1 + z)^3 \exp \left( \frac{1.9 A}{1 - 0.75A^2} \right), \quad A = \frac{1 + z_\mathrm{vir}}{1+z} \,,
    \label{eqn:rho_TH}
\end{gather}
At early times, the density of the halo is diluted by the expansion of the universe and closely follows the evolution of the IGM mean density; however, as $z$ approaches $z_\mathrm{vir}$, the halo begins to collapse and the density increases.
Once the halo passes the condition for virialization (see Sec.~\ref{sec:collapse}), the density is held constant at $\rho (z_\mathrm{vir})$.

The number density of hydrogen within the halo is similarly assumed to be given by
\begin{equation}
    n_\mathrm{H} (z) \approx \bar{n}_{\mathrm{H}, 0} (1 + z)^3 \exp \left( \frac{1.9 A}{1 - 0.75A^2} \right) ,
    \label{eqn:n_TH}
\end{equation}
where $\bar{n}_{\mathrm{H}, 0}$ is the mean number density of all hydrogen nuclei today.

\subsubsection{Temperature}
\label{sec:halo_temp}

The halo temperature is affected by adiabatic cooling/heating as the volume of the halo changes, as well as Compton scattering off the CMB, atomic hydrogen line cooling, molecular hydrogen cooling, and exotic energy injection.
Hence, we can write the temperature evolution of gas in a halo as
\begin{equation}
    \dot{T}_\mathrm{halo} = \dot{T}_\mathrm{halo}^\mathrm{adia} + \dot{T}_\mathrm{halo}^\mathrm{comp} + \dot{T}_\mathrm{halo}^\mathrm{line} + \dot{T}_\mathrm{halo}^{\mathrm{H}_2} + \dot{T}_\mathrm{halo}^\mathrm{inj},
    \label{eqn:dTmdt}
\end{equation}
which has two new terms compared to its IGM counterpart.
The adiabatic and Compton-scattering contributions are still given by Eqn.~\eqref{eqn:T_adia} and \eqref{eqn:T_comp}, respectively, where the number density is given in Eqn.~\eqref{eqn:n_TH}, thus accounting for the collapse of the halo.
All instances of the background density, temperature, or ionization fraction should be replaced with the corresponding quantities in the halo.
The terms for atomic-line cooling and molecular cooling are respectively given by
\begin{equation}
    \dot{T}_\mathrm{halo}^\mathrm{line} = -\SI{7.5e-19}{\erg\centi\meter\cubed\per\second} \times \frac{2 n_\mathrm{HI} n_e}{3 n_\mathrm{H} (1 + x_e + \mathcal{F}_\mathrm{He})} \exp \left(- \frac{\SI{118348}{\kelvin}}{T_\mathrm{halo}} \right) \,,
\end{equation}
where $n_\mathrm{H}$ denotes the total number density of hydrogen in the halo in this case, and
\begin{equation}
    \dot{T}_\mathrm{halo}^{\mathrm{H}_2} = - \Lambda_\mathrm{H_2} n_\mathrm{H_2},
\end{equation}
where the cooling rate $\Lambda_\mathrm{H_2}$ is given by Eqn.~(A.5) of Ref.~\cite{1998A&A...335..403G}. 

The exotic injection term $\dot{T}_\mathrm{halo}^\mathrm{inj}$ in Eqn.~\eqref{eqn:dTmdt} is taken to be identical to the expression in Eqn.~\eqref{eqn:T_inj}, with $\bar{n}_\mathrm{H}$ and the injection rate taking their homogeneous value, and $f_\mathrm{heat}(z)$ taken from \texttt{DarkHistory}, using the IGM temperature and ionization level. 
We justify this assumption in Appendix~\ref{app:fs_halo}, where we show that under reasonable simplifying assumptions, the intensity of particles received from exotic energy injection at the center of halos is roughly equal to the intensity in the homogeneous IGM for our fiducial models, and across much of the relevant parameter space for decaying DM.

\subsubsection{Ionization fraction}
\label{sec:halo_ion}

For the ionization fraction inside the halo, we use the following evolution equation. 
\begin{equation}
    \dot{x}_e = - \mathcal{C} \left[ x_e n_e \alpha_\mathrm{B} + 4 (1 - x_e) \beta_\mathrm{B} e^{-E_{21} / T_\mathrm{CMB}} \right] + \dot{x}_e^\mathrm{inj},
    \label{eqn:dxedt}
\end{equation}
In this equation, $\mathcal{C}$ is the Peebles-$\mathcal{C}$ factor, $\alpha_\mathrm{B}$ and $\beta_\mathrm{B}$ are the case-B recombination and photoionization coefficients, which are calculated using $T_\mathrm{halo}$, and $E_{21} = \SI{10.2}{\eV}$ is the energy of the Lyman-$\alpha$ transition\footnote{The case-B coefficients are appropriate to use both in the homogeneous medium~\cite{Peebles:1968ja,Zeldovich:1969en} and the denser halos~\cite{Nebrin:2023yzm}, since the gas is optically thick to Lyman limit photons}.
This is the equation for a Three-Level Atom (TLA) modified to include exotic energy injection; it is appropriate to use the TLA here instead of the MLA since we are not tracking the contribution of the halo to spectral distortions (see the discussion in Ref.~\cite{Liu:2023fgu}), and the MLA takes much longer to solve than the TLA.

Under the assumptions of the TLA, the term from exotic energy injection term, $\dot{x}_e^\mathrm{inj}$, is written as
\begin{equation}
    \dot{x}_e^\mathrm{inj} = \left[ \frac{f_\mathrm{H ion}}{\mathcal{R} n_\mathrm{H}} + \frac{(1-\mathcal{C}) f_\mathrm{exc}}{E_{21} n_\mathrm{H}} \right] \left( \frac{dE}{dV dt} \right)^\mathrm{inj},
    \label{eqn:x_inj}
\end{equation}
where $\mathcal{R} = \SI{13.6}{\eV}$ is the energy required to ionize hydrogen.
The first term represents photoionization and collisional ionization caused by exotic energy injection.
The second term represents excitations followed by ionization; the $(1 - \mathcal{C})$ factor ensures that this term does not include excitations that are immediately followed by recombination to the ground state.
Once again, we use the IGM result for $f_c / \bar{n}_\mathrm{H}$, since the energy deposited per baryon in the halo is expected to be similar to the same quantity in the homogeneous IGM for our fiducial models and for most of the parameter space for DM decay (as discussed in Appendix~\ref{app:fs_halo}).

\subsubsection{Full evolution and collapse}
\label{sec:collapse}

We simultaneously solve Eqns.~\eqref{eqn:dxh2dt},~\eqref{eqn:dTmdt}, and~\eqref{eqn:dxedt} to obtain the H$_2$ abundance, temperature, and ionization fraction inside the halo, while using Eqn.~\eqref{eqn:rho_TH} for the halo density.
There are two free parameters that parametrize the halo properties: \emph{1)} the virialization redshift $z_\mathrm{vir}$, and \emph{2)} the halo mass $M_\mathrm{halo}$, or equivalently the virial temperature $T_\mathrm{vir}$, which is related to $M_\mathrm{halo}$ via Eqn.~\eqref{eqn:Tvir}.
We also require the $f_c$'s and spectral distortions output by \texttt{DarkHistory} for a given DM model, which are calculated using the temperature and ionization from the homogeneous IGM.

For the initial conditions, we set $x_\mathrm{H_2} = 0$ and the halo temperature and ionization fraction to the IGM values at $1+z = 3000$.
Prior to $1+z = 900$, the system of differential equations is very stiff; since the halo density and temperature are very similar to the mean IGM density and temperature at these times, we fix $x_\mathrm{HII}$ and $T_\mathrm{halo}$ to the IGM values for $1+z > 900$, solving only the evolution equation for $x_\mathrm{H_2}$, and use the value of $x_\mathrm{H_2}$ at $1+z = 900$ as an initial condition for solving the full set of equations thereafter.

As we evolve the halo forward in time, the halo begins to collapse and increase in density. 
Eventually, at some redshift $z_\mathrm{vir}$, the halo is considered to be virialized when one of the following conditions is met:
\begin{itemize}
    \item the density reaches the value expected of a virialized halo, $\rho_\mathrm{vir} = 18\pi^2 \rho_0 (1+z)^3$, or

    \item $T_\mathrm{halo} = T_\mathrm{vir}$, and the halo is prevented from collapsing further by gas pressure~\cite{Tegmark:1996yt}.
\end{itemize}
Once one of these two conditions is met, we hold the gas density fixed to its collapsed value and raise the temperature to $T_\mathrm{vir}$ if it is below this value.
We continue to evolve the halo past $z_\mathrm{vir}$ down to $1+z=4$ in order to evaluate if it is cooling sufficiently quickly to collapse; however, we only use the results for studying halo cooling shortly after $z_\mathrm{vir}$, since our assumption that the halo density remains constant after virialization breaks down if the halo can cool and collapse further. 
During this collapsing phase, we only want to evaluate the ability of molecular and line cooling to lead to halo collapse, since these are crucial processes for star formation; hence, we neglect the Compton cooling term $\dot{T}_\mathrm{halo}^\mathrm{comp}$ after virialization.
This also ensures that the threshold we find is the minimum mass above which we expect \textit{all} halos to undergo runaway collapse.

We adopt the criterion in Ref.~\cite{Tegmark:1996yt}, and consider a halo to be cooling sufficiently quickly to collapse and form stars if
\begin{equation}
    T_\mathrm{halo} (\eta z_\mathrm{vir}) \leq \eta T_\mathrm{halo} (z_\mathrm{vir}),
    \label{eqn:criteria}
\end{equation}
where we choose $\eta = 0.75$.
As explained in Ref.~\cite{Tegmark:1996yt}, this condition comes from requiring the halo to cool by a significant amount within a Hubble time; since the scale factor is proportional to $t^{2/3}$ during matter domination, this roughly corresponds to the redshift dropping by $2^{2/3} \approx 0.63$.
We have checked that this condition is not particularly sensitive to the choice of $\eta$ by varying it between $0.6$ to $0.9$.

\subsection{Results}
\label{sec:results}

%
\begin{figure*}
	\centering
    \includegraphics[scale=0.5]{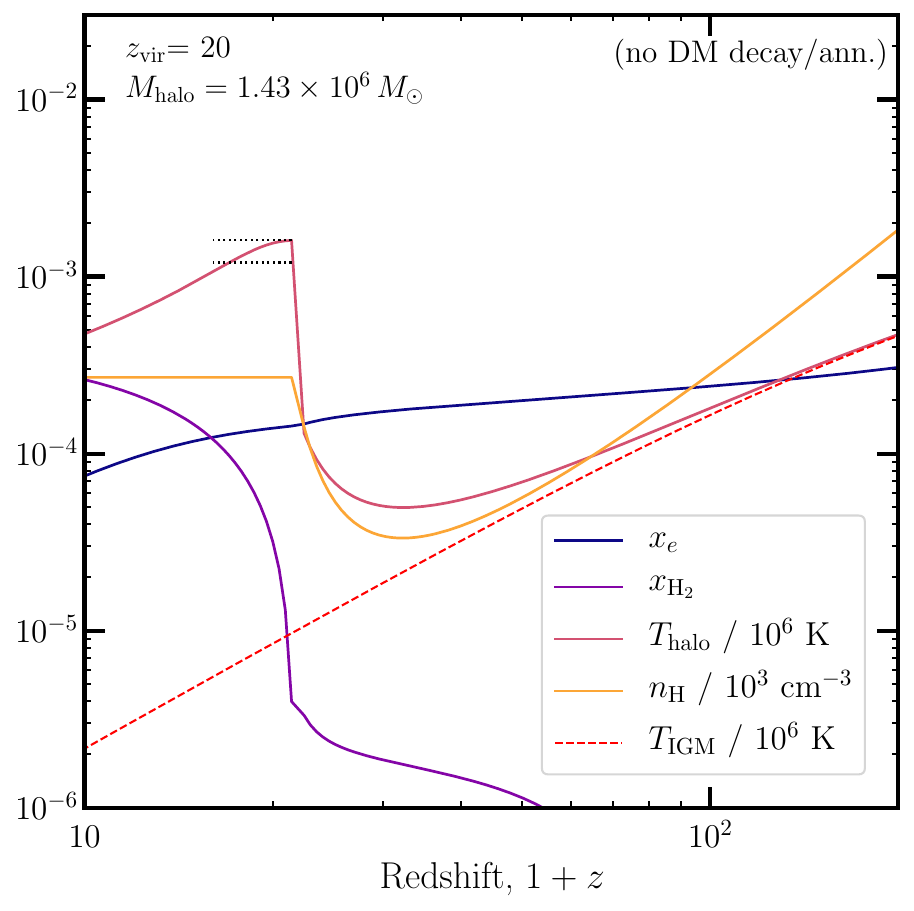}
    \\
    \includegraphics[scale=0.5]{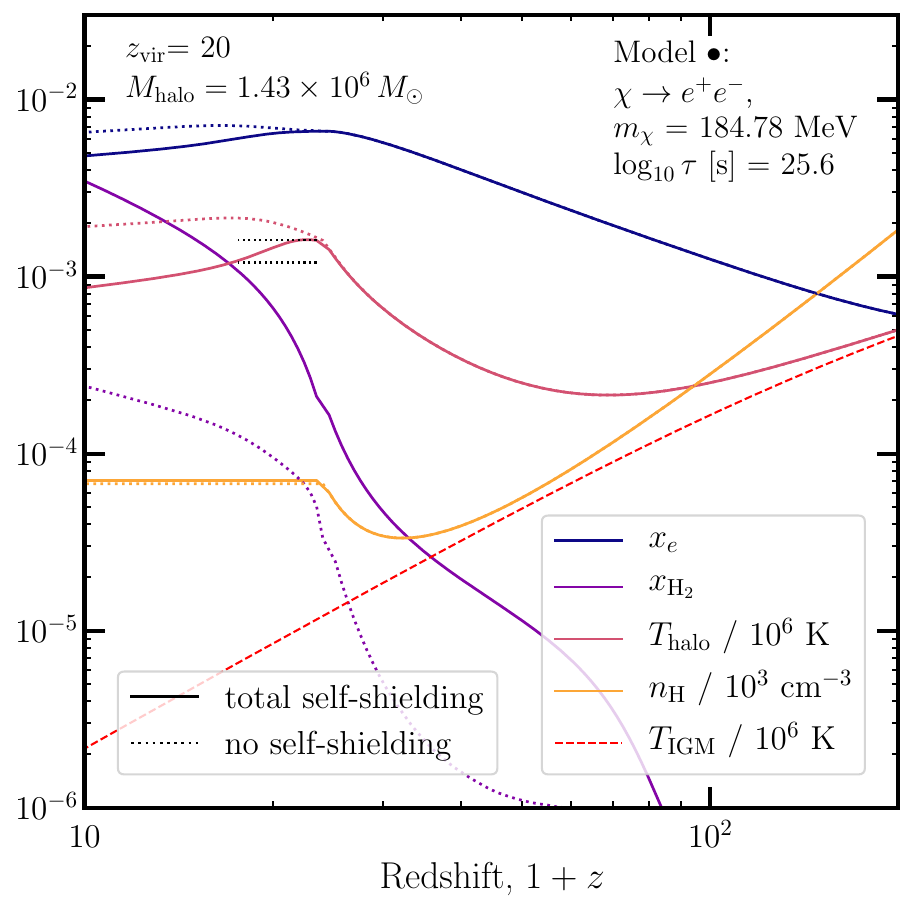}
    \hspace{5mm}
    \includegraphics[scale=0.5]{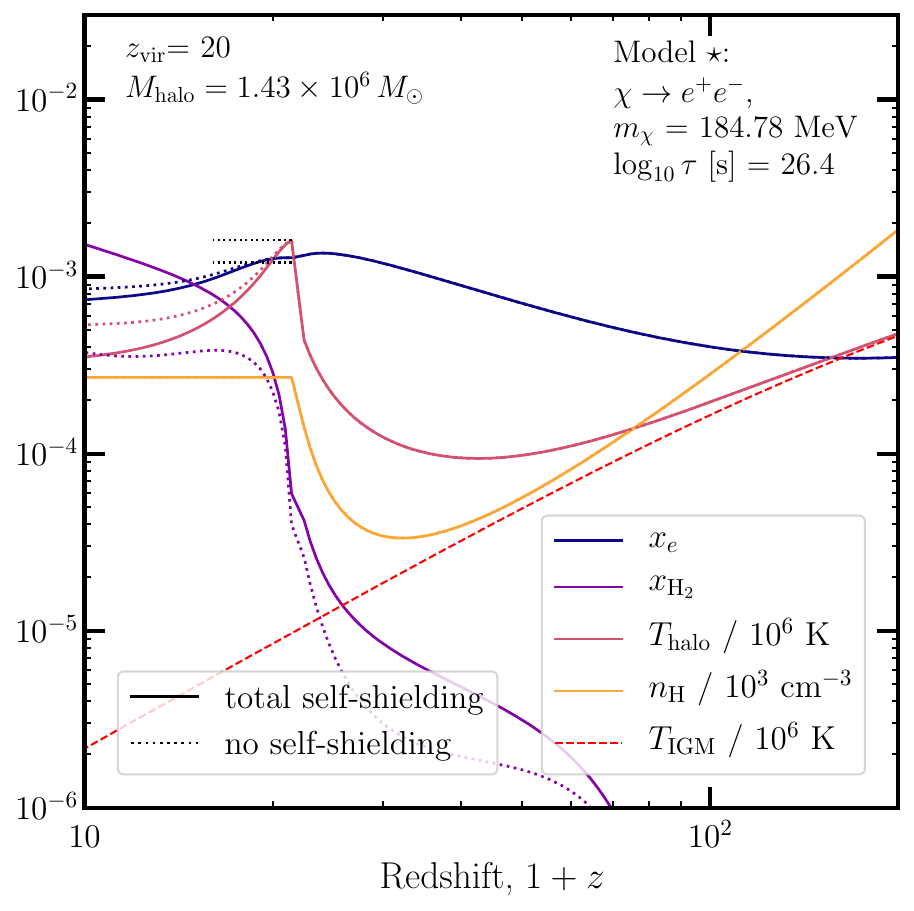}
    \caption{
    Three examples of halo evolution.
    In the top panel there is no DM decay or annihilation, and the halo virializes at $z_\mathrm{vir} = 20$ with a mass of $M_\mathrm{halo} = 1.43 \times 10^6 M_\odot$, corresponding to a virial temperature of $T_\mathrm{vir} = \SI{1600}{\kelvin}$.
    The bottom two panels include the effects of exotic energy injection, with Model $\bullet$\, on the left and Model $\star$\, on the right.
    In all plots, we show the free-electron fraction (blue), the molecular hydrogen fraction (purple), the halo temperature (magenta), and the number density of hydrogen nuclei (neutral or ionized, orange).
    Dotted lines include photodissociation by LW photons.
    The red dashed line indicates the average IGM temperature when we do not include exotic energy injection.
    The horizontal line segments on each panel indicate $T_\mathrm{vir}$ and $0.75 T_\mathrm{vir}$, and span the redshift range $(z_\mathrm{vir}, 0.75 z_\mathrm{vir})$; if the temperature curve crosses the lower line segment after virialization, the halo passes the criterion for collapsing and forming stars.
    In the top panel, the halo succeeds in collapsing.
    Model $\star$\, gives rise to even faster cooling, whereas Model $\bullet$\, cools more slowly, failing to meet our criterion for star formation.
    }
    \label{fig:halo_examples}
\end{figure*}

With the prescription described in Section~\ref{sec:collapse}, given $z_\mathrm{vir}$ and an exotic energy-injection model, we can determine the minimum $M_\mathrm{halo}$ or $T_\mathrm{vir}$ above which halos will collapse.
Fig.~\ref{fig:halo_examples} shows a few examples of evolving halos.
In each panel, we show $x_e$, $x_{\mathrm{H}_2}$, $T_\mathrm{halo}$, and $n_\mathrm{H}$ as a function of redshift, for both very efficient H$_2$ self-shielding and no self-shielding.
We also show two horizontal dotted line segments at $T_\mathrm{vir}$ and $0.75 T_\mathrm{vir}$.
The lines span the redshift range $z_\mathrm{vir}$ to $0.75 z_\mathrm{vir}$; hence, models where the temperature history crosses the lower line segment after virialization succeed in forming stars.
The top panel shows a halo that virializes at $z_\mathrm{vir} = 20$ with a mass of $M_\mathrm{halo} = 1.39 \times 10^6 M_\odot$, corresponding to a virial temperature of $T_\mathrm{vir} = 1600$ K.
The bottom two panels show halos with the same $z_\mathrm{vir}$ and $T_\mathrm{vir}$, but including the same energy injection models shown in Fig.~\ref{fig:global_H2}; Model $\bullet$\, is on the left and Model $\star$\, is shown on the right.

In the top panel at early times, the halo density decreases and closely follows the IGM density; at later times, as we approach $z_\mathrm{vir}$, the halo begins to collapse and the density increases until it reaches $\rho_\mathrm{vir}$.
At this point, the halo has virialized and the density is held fixed.
Similarly, at early times, the halo temperature follows the standard IGM temperature evolution, but starts to deviate around $z \sim 100$ as the overdensity begins to collapse.
Close to $z_\mathrm{vir}$, the temperature begins to increase and is raised to $T_\mathrm{vir}$ once the halo meets the conditions for virialization.
After this, the temperature decreases due to molecular cooling.
The curve crosses back through the lower line segment; hence, by the criteria given in Eqn.~\eqref{eqn:criteria}, this halo has succeeded in forming stars.

The bottom panels of Fig.~\ref{fig:halo_examples} demonstrate that exotic energy injection has multiple competing effects on halo collapse. 
On the one hand, the exotic heating impedes cooling, preventing halos from collapsing. 
If self-shielding of the halos is inefficient, then the effect of LW photons can further suppress cooling by reducing the H$_2$ abundance.
On the other hand, the increase in ionization levels catalyzes the formation of H$_2$, enhancing the molecular cooling rate. 
The effect of heating dominates for Model $\bullet$ \ in the lower-left plot, causing the halo temperature to reach $T_\mathrm{vir}$ before the density reaches $\rho_\mathrm{vir}$; after $z_\mathrm{vir}$, the exotic heating rate is also large enough that the halo cannot cool efficiently. 
The effect of ionization dominates for Model $\star$, leading to enhanced cooling relative to the standard cosmological expectation shown in the top plot.
This halo forms stars successfully by the criterion in Eqn.~\eqref{eqn:criteria}.
In both cases, inefficient self-shielding delays the cooling of the halo; the effect is larger for Model $\bullet$, where DM decays more quickly and therefore contributes a larger LW background.

\begin{figure*}
    \includegraphics[scale=0.5]{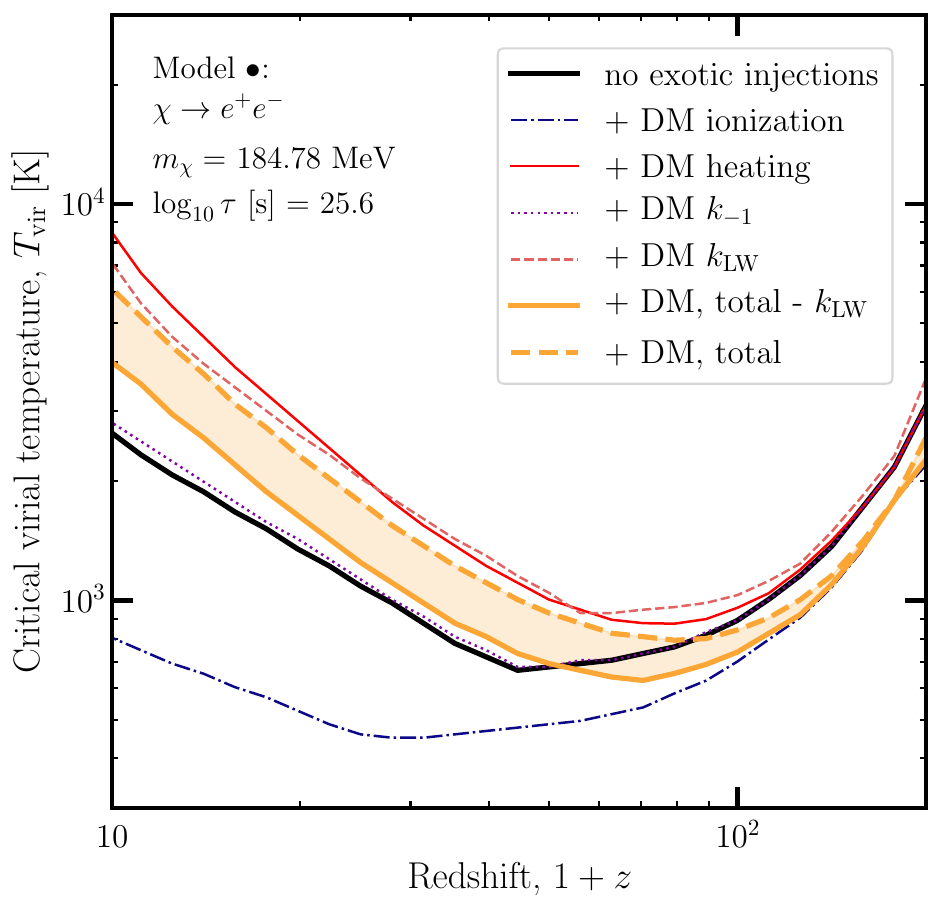}
    \hspace{5mm}
    \includegraphics[scale=0.5]{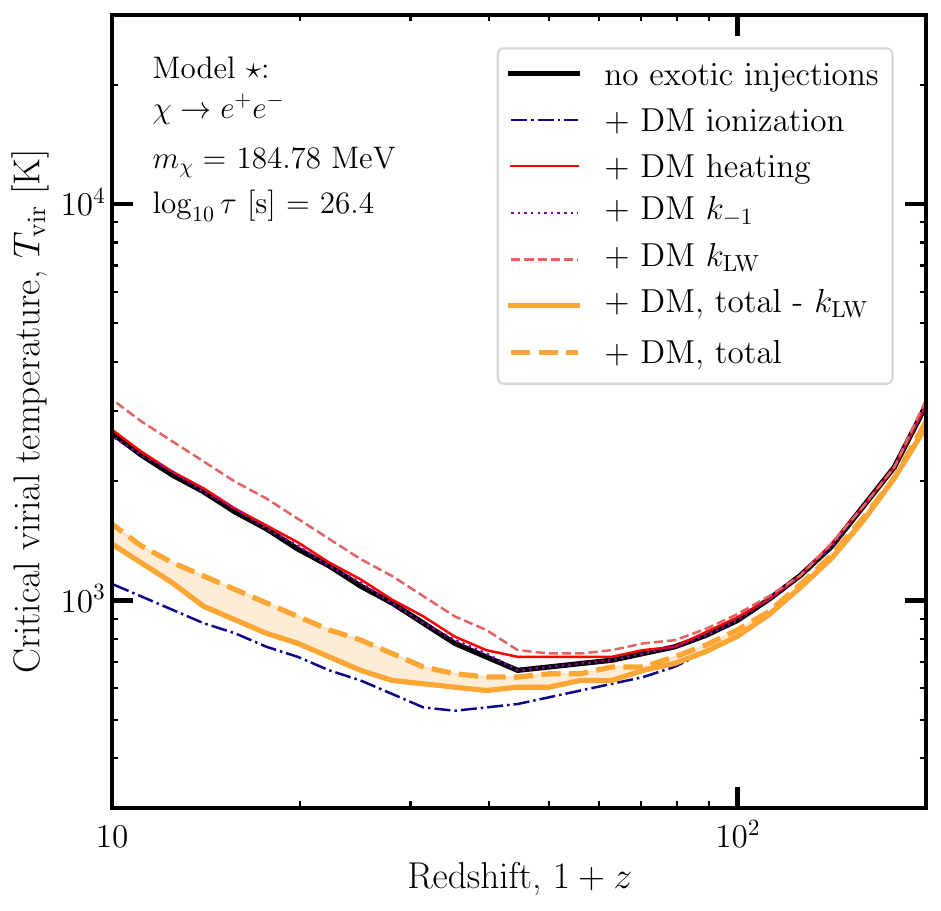}
    \caption{
    Critical virial temperature for collapse as a function of redshift, both in a standard cosmology without exotic injections (black) and including the fiducial DM energy injection models (gold); the contours bracket the effect of H$_2$ self-shielding.
    The left panel shows Model $\bullet$\, and the right shows Model $\star$.
    We also show the results from considering one effect at a time, including exotic ionization (blue dot dashed), exotic heating (red solid), H$^-$ photodetachment (purple dotted), and H$_2$ photodissociation (pink dashed).
    }
    \label{fig:critical_collapse_Tvir}
\end{figure*}
\begin{figure*}
    \includegraphics[scale=0.5]{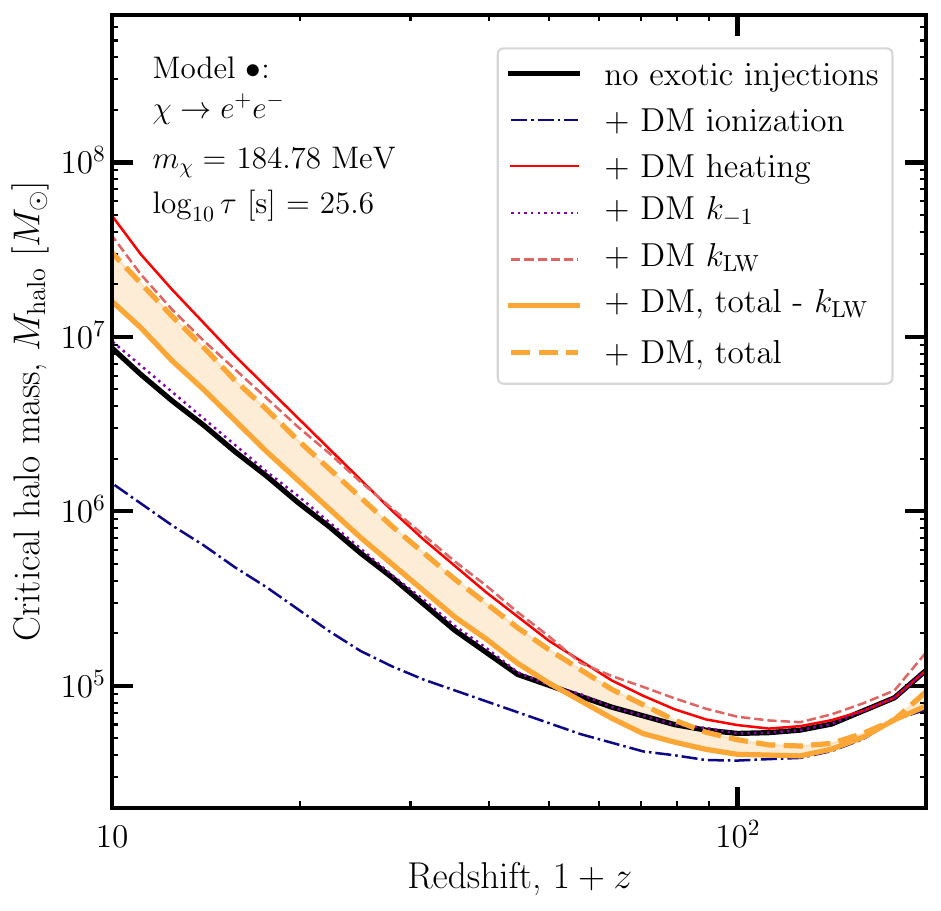}
    \hspace{5mm}
    \includegraphics[scale=0.5]{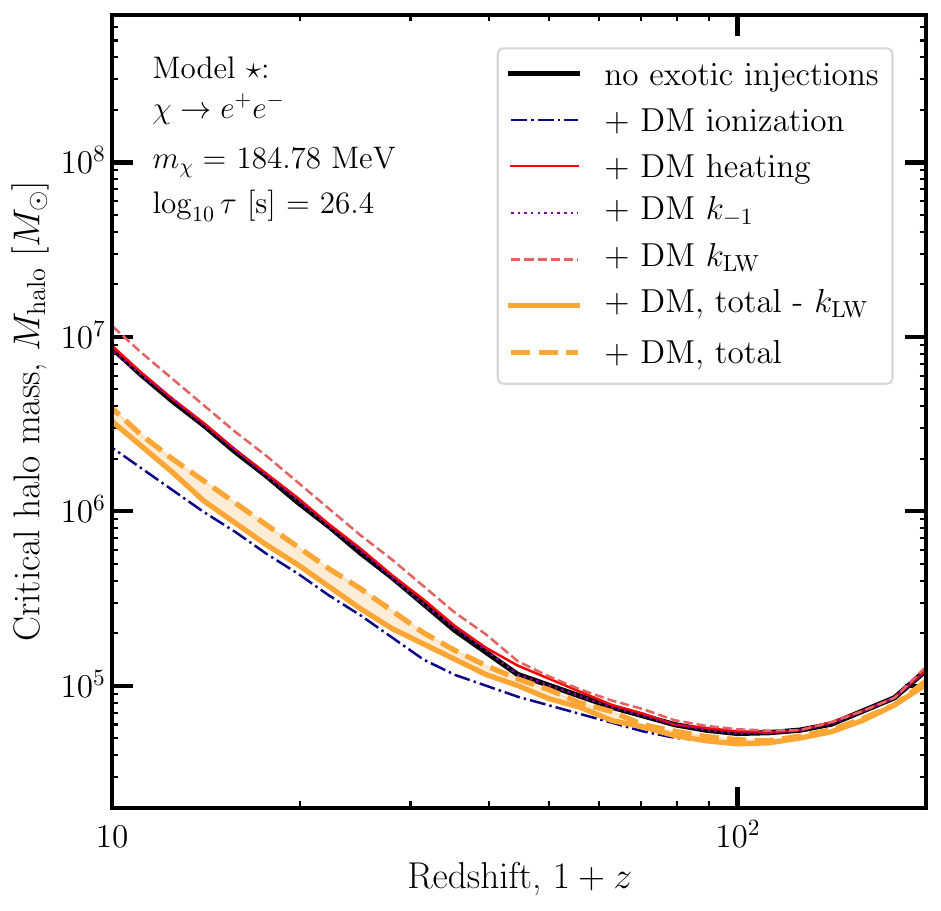}
    \caption{
    Same as Fig.~\ref{fig:critical_collapse_Tvir}, but showing results for the critical halo mass.
    }
    \label{fig:critical_collapse_Mhalo}
\end{figure*}

The threshold for halos to collapse is a redshift-dependent condition.
For example, Fig.~\ref{fig:critical_collapse_Tvir} shows the critical virial temperature as a function of the virialization redshift; Fig.~\ref{fig:critical_collapse_Mhalo} shows the same results, but in terms of the halo mass.
We show the critical values if we do not include exotic injections, and how the curves change if we include Model $\bullet$\, or Model $\star$, bracketing the effect of H$_2$ self-shielding.
We also show the resulting curves if in the evolution equations we only include one of the terms for exotic ionization (Eqn.~\eqref{eqn:x_inj}), exotic heating (Eqn.~\eqref{eqn:T_inj}), H$^-$ photodetachment (Eqn.~\eqref{eqn:k-1}), or H$_2$ photodissociation (Eqn.~\eqref{eqn:kLW}) from exotic injections.

From Figs.~\ref{fig:critical_collapse_Tvir} and~\ref{fig:critical_collapse_Mhalo}, we see that additional heating and LW photons generally raise the critical value of $T_\mathrm{vir}$ or equivalently $M_\mathrm{halo}$, making it more difficult for halos to collapse, whereas ionization caused by exotic energy injection lowers the critical value due to the enhanced production of H$_2$, allowing more halos to collapse.
Because of this interplay of effects, it is nontrivial to determine the direction of the effect on the critical virial temperature/halo mass for a given redshift and energy injection model.
For Model $\bullet$\, (left panels), it is easier for halos to collapse prior to $z \sim 50$ and harder after this, whereas for Model $\star$\, (right), the overall effect is to lower the critical masses or virial temperatures needed for collapse at all redshifts.
This complicated interplay of heating and ionization on collapsing halos has been noted in previous work studying energy injection by primordial black holes on the formation of supermassive black holes~\cite{Pandey:2018jun}; however, to our knowledge, our results are the first to show that the direction of the effect can change as one increases the energy injection rate. 

We can make a rough comparison to the results of studies with hydrodynamical simulations.
In Ref.~\cite{Schauer:2020gvx}, for the case of no LW background and no relative difference between baryons and DM, they find that the minimum halo mass of star-forming halos rises slightly from $3 \times 10^5 M_\odot$ at $1+z=23$ to $5 \times 10^5 M_\odot$ at $1+z=15$. 
For comparison, Ref.~\cite{Kulkarni:2020ovu} finds this threshold to rise from about $10^5 M_\odot$ at $1+z=28$ to $3 \times 10^5 M_\odot$ at $1+z=16$.
In both cases, this threshold is smaller than what we find at comparable redshifts, and the growth of this threshold as a function of redshift is much steeper in our study; however, we emphasize that our simple top-hat halo treatment agrees with the overall trends in Refs.~\cite{Schauer:2020gvx} and ~\cite{Kulkarni:2020ovu}.

\subsubsection{Scanning over DM parameter space}
\label{sec:scans}

Based on the results in the previous section, it is clear that different regions of DM parameter space can have opposite effects on halo collapse and star formation; moreover, the direction of the effect can also change with redshift.
We would like to now scan over the parameter space of various energy injection channels to understand where we get the strongest and most interesting signals.

In describing our methods up to this point, we have focused on the example of decaying DM; these largely apply to annihilations as well.
For $p$-wave annihilation, since the cross-section is velocity dependent, we define the cross-section using 
\begin{equation}
    \sigma v = (\sigma v)_\mathrm{ref} \left( \frac{v}{v_\mathrm{ref}} \right)^2
\end{equation}
and choose a reference velocity of $v_\mathrm{ref} = 100$ km/s, which is on the order of the dispersion velocity of DM in a Milky Way-like halo today.
We again assume that the $f_c$'s parametrizing energy deposition are the same in the halos as in the IGM.
This assumption is less likely to be valid for annihilations since the annihilation rate depends on density squared, and in particular may break down for $p$-wave annihilations since there is an additional velocity dependence.
Since the rate of energy injection from annihilations depends on density squared, this rate is boosted once structure formation begins; the average of the density squared exceeds the square of the average density.
We use the boost factor prescription included in \texttt{DarkHistory}~\cite{DarkHistory} under the assumption that halos have an NFW profile and no substructure; see Ref.~\cite{Liu:2016cnk} for more details about how this is calculated.

For both decays and annihilations, we will study photon and $e^+ e^-$ final states.
In principle, we could study decays or annihilations into other Standard Model particles; however, even when we inject other types of particles, the prompt decays of these particles primarily result in showers of secondary electrons, positrons, and photons.
Hence, one can usually understand the results for other final states by taking linear combinations of the results for injections of photons and $e^+ e^-$ pairs.

\subsubsection{Comparison to existing constraints}
\label{sec:constraints}

\begin{figure*}
	\centering
    \includegraphics[scale=0.39]{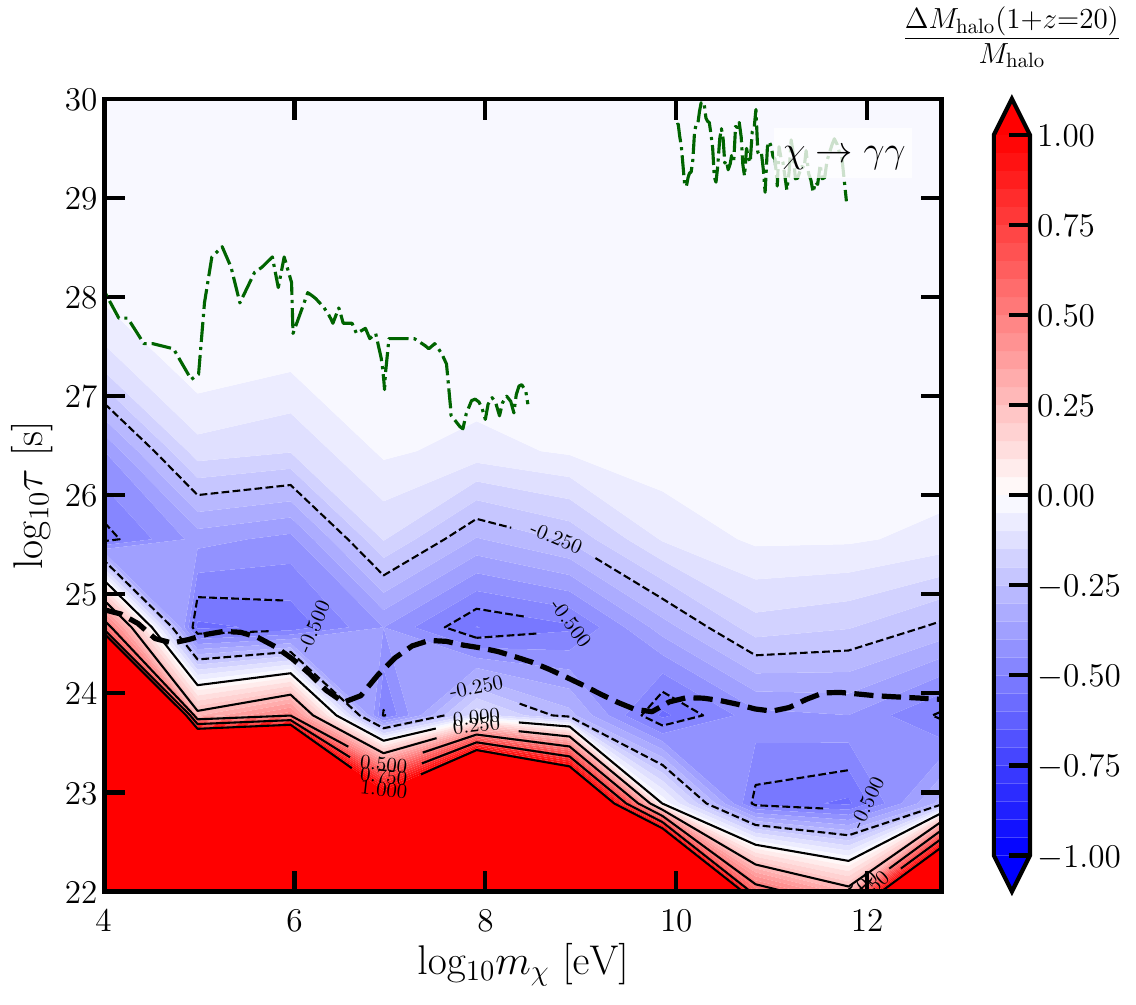}
    \includegraphics[scale=0.39]{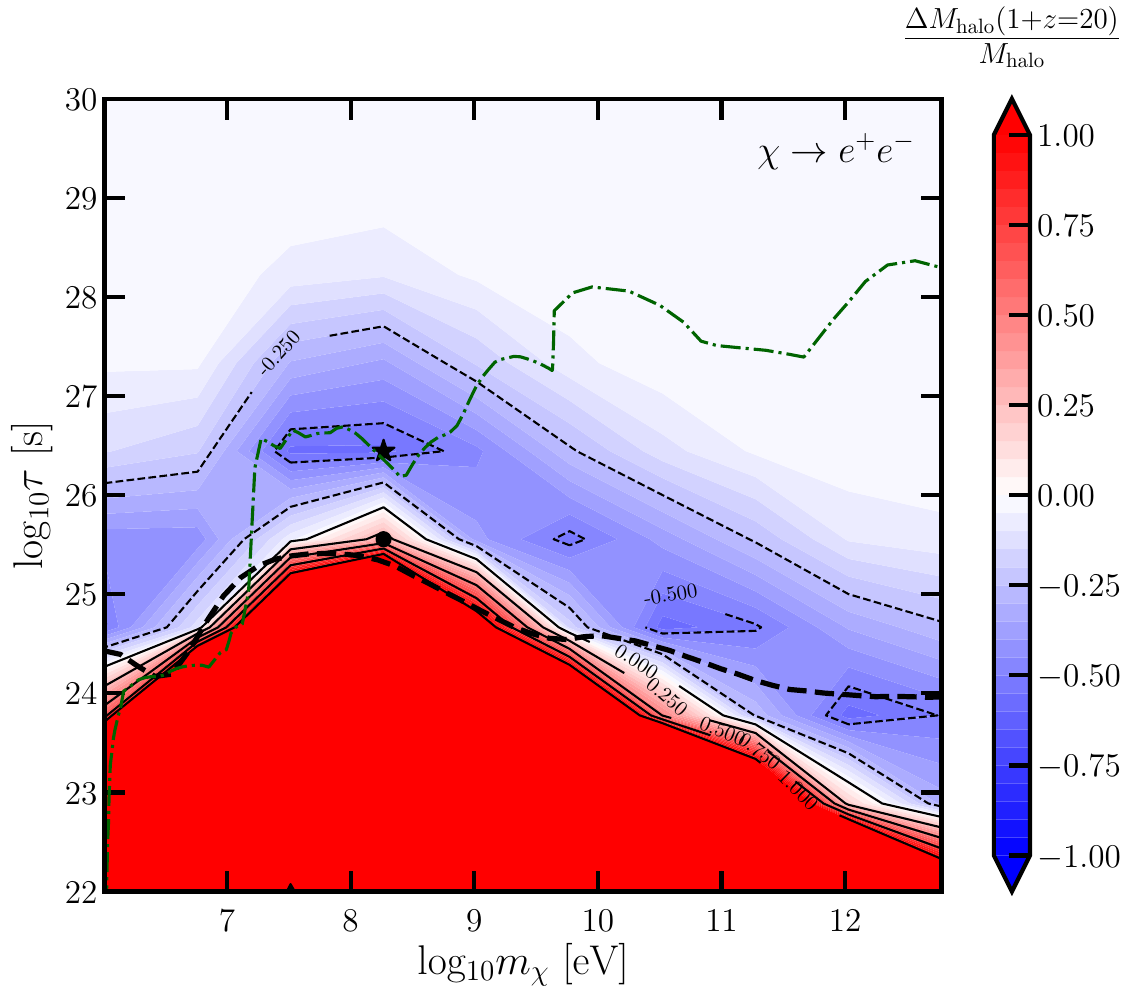} \\
    \includegraphics[scale=0.39]{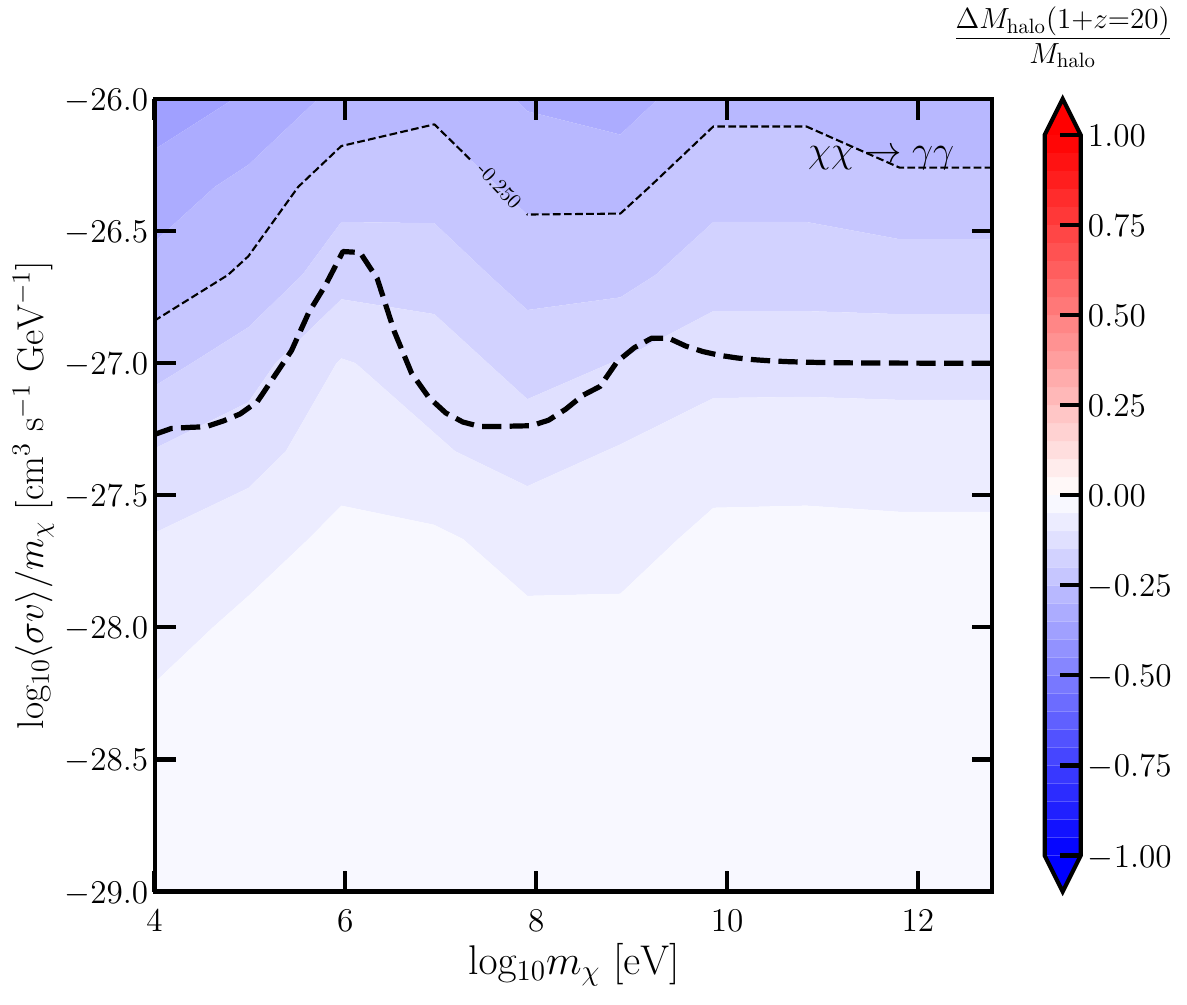}
    \includegraphics[scale=0.39]{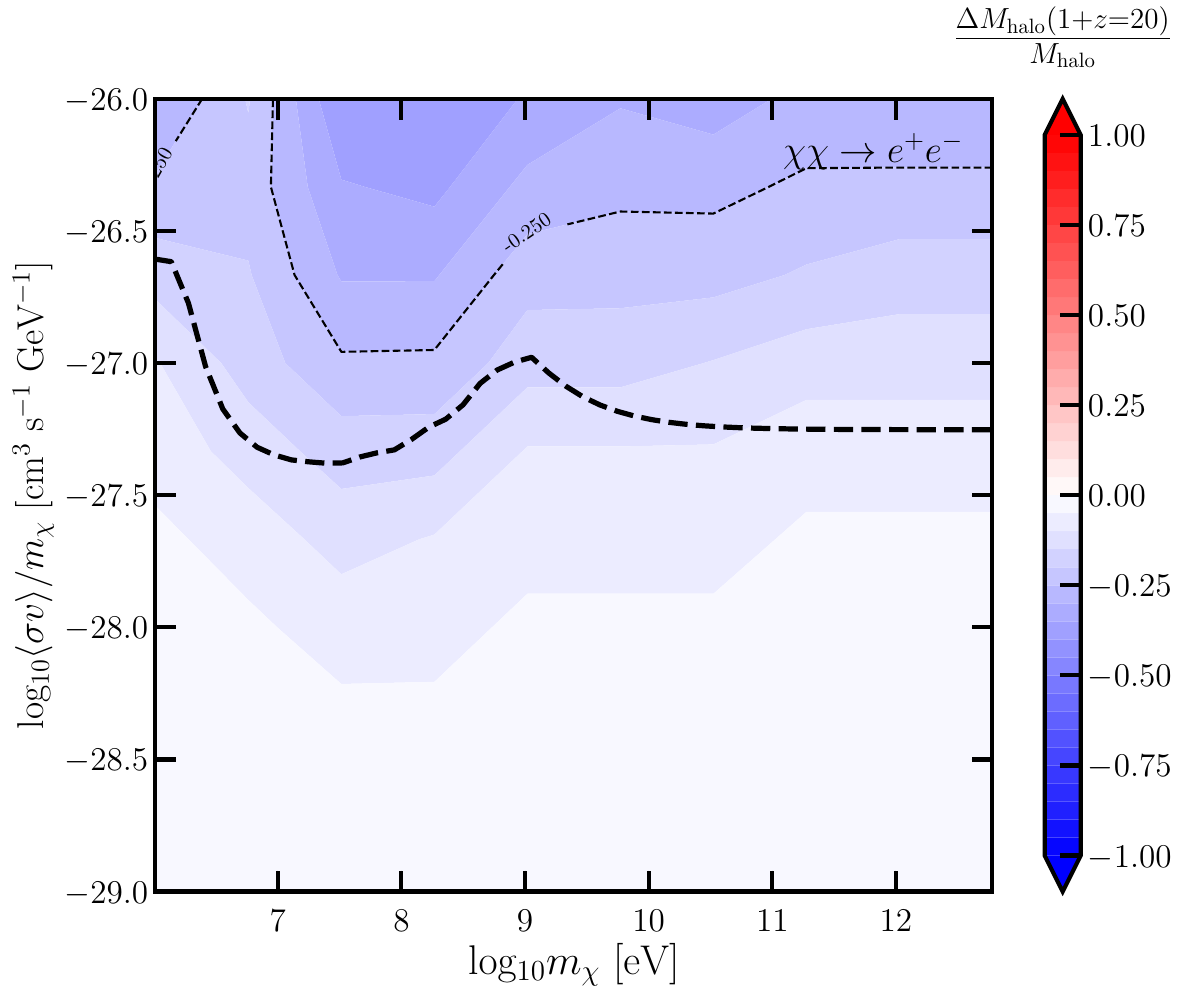} \\
    \includegraphics[scale=0.39]{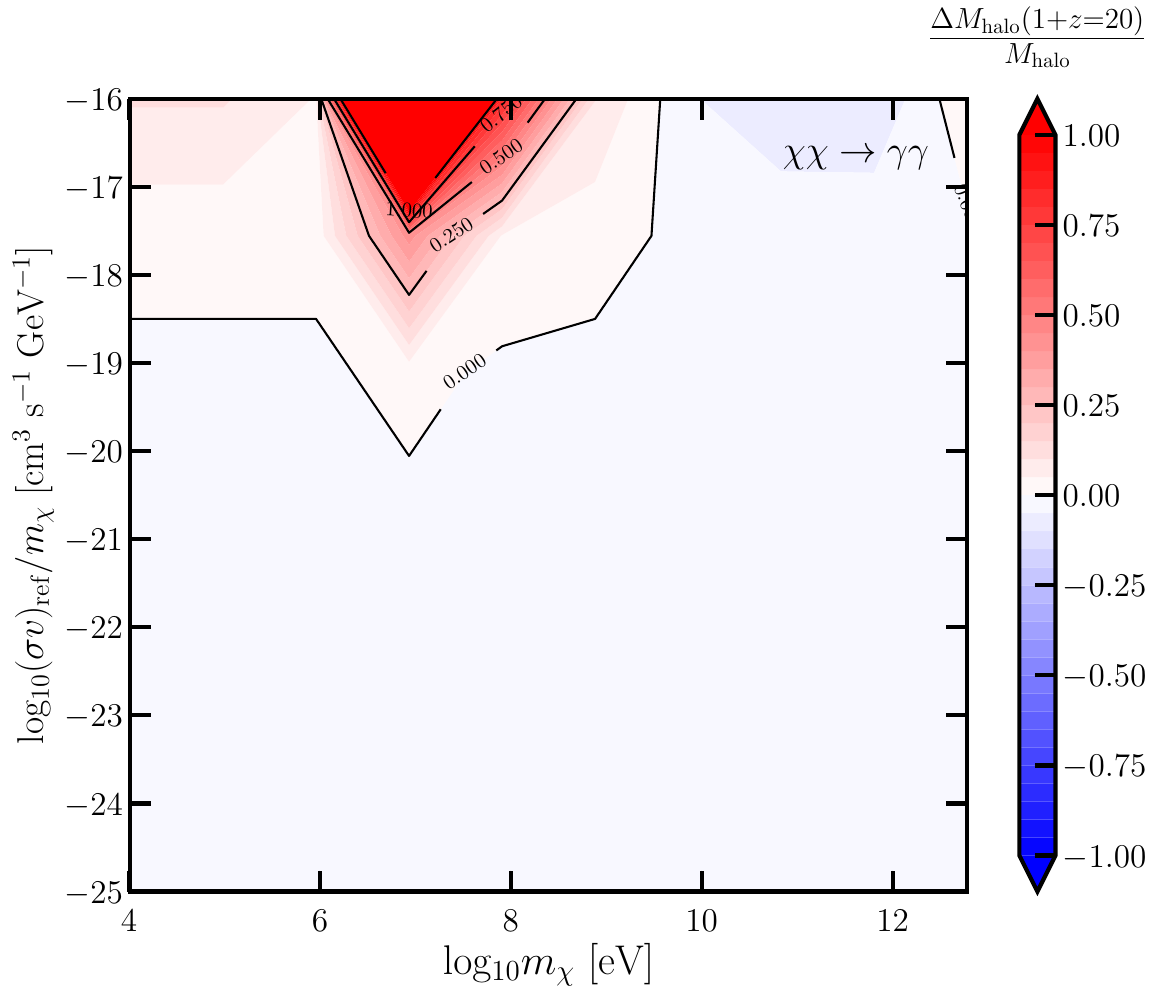}
    \includegraphics[scale=0.39]{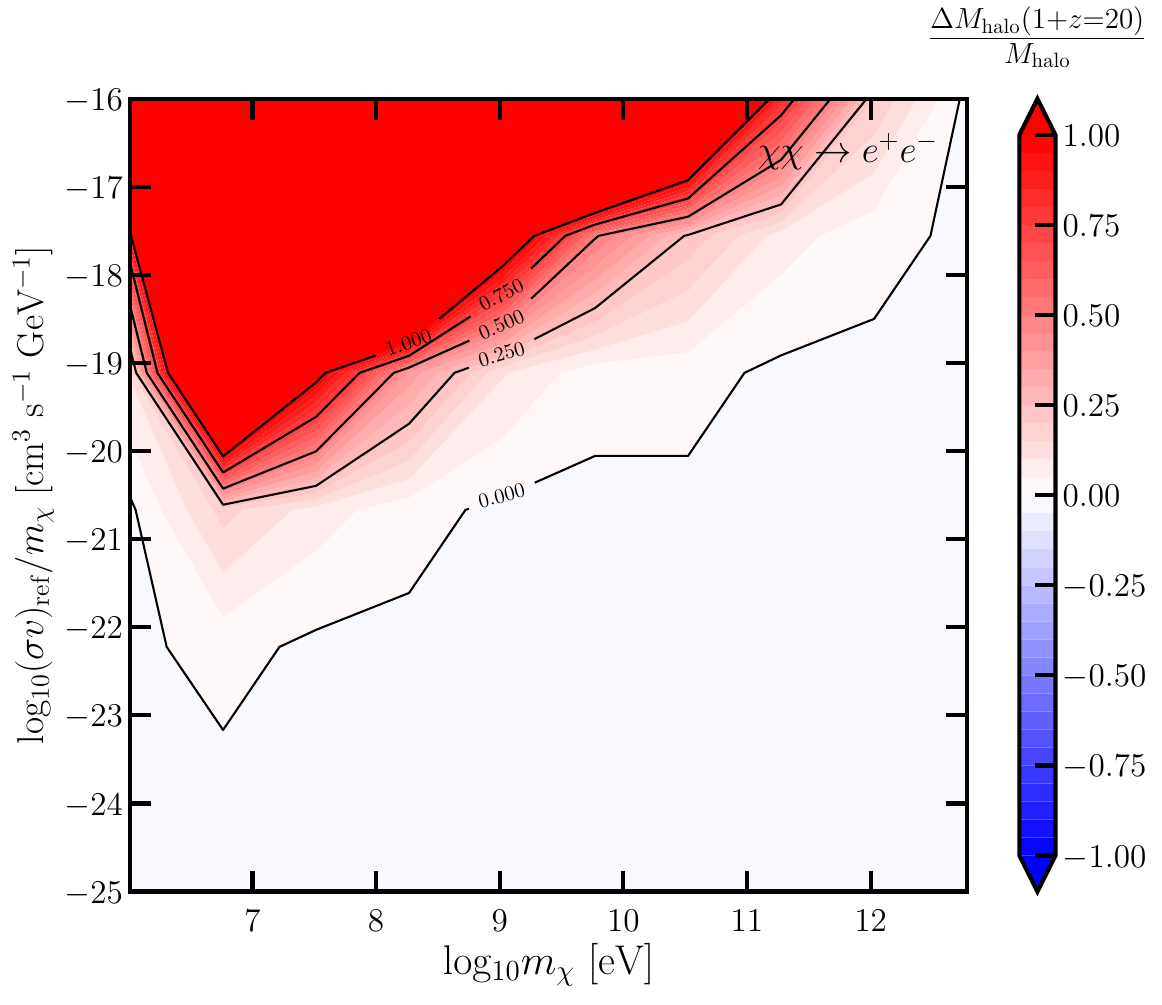}
    \caption{
    \footnotesize
    Change to the minimum halo mass $M_\mathrm{halo}$ necessary for star formation at $1+z=20$.
    From top row to bottom, the channels are decay, $s$-wave annihilation, and $p$-wave annihilation.
    For $p$-wave annihilation, the cross-section is defined using a reference velocity of 100 km/s as in Ref.~\cite{Liu:2016cnk}.
    In the left panels, the final state particles are photons; in the right, the final state particles are $e^+ e^-$ pairs.
    The thick black dashed lines show existing constraints from CMB data~\cite{Slatyer:2016qyl,Planck:2018vyg}.
    We also include constraints from X/$\gamma$-ray telescopes~\cite{Cadamuro:2011fd,Boddy:2015efa,Fermi-LAT:2013thd,Essig:2013goa, Massari:2015xea, Cohen:2016uyg,Laha:2020ivk,Cirelli:2020bpc,Cirelli:2023tnx}, where we have assumed $v = 220$ km s$^{-1}$ in the Milky Way, and Voyager I~\cite{Boudaud:2016mos, Boudaud:2018oya} (green dot-dashed). 
    For $p$-wave results, constraints lie below the bottom of the plot.
    }
    \label{fig:scans_z20}
\end{figure*}
Fig.~\ref{fig:scans_z20} shows the change in the minimum halo mass necessary for star formation at $1+z=20$ relative to the standard cosmology value over the parameter space for the energy injection channels described in Sec.~\ref{sec:scans}; here we assume self-shielding is very efficient and hence the effect of LW photons is suppressed, since this is closer to recent results from hydrodynamical simulations~\cite{Schauer:2020gvx,Kulkarni:2020ovu,2020MNRAS.492.4386S}.
We overlay existing constraints from CMB anisotropies~\cite{Slatyer:2016qyl,Planck:2018vyg} for decay and $s$-wave annihilation, as well as constraints from X/$\gamma$-ray telescopes~\cite{Cadamuro:2011fd,Boddy:2015efa,Fermi-LAT:2013thd,Essig:2013goa, Massari:2015xea, Cohen:2016uyg,Laha:2020ivk,Cirelli:2020bpc,Cirelli:2023tnx} and Voyager I~\cite{Boudaud:2016mos,Boudaud:2018oya} for decay, and mark Models $\bullet$\, and $\star$. 

For decays to photons, we show for illustration a selection of some of the strongest existing limits on photon lines from indirect detection~\cite{Cadamuro:2011fd,Boddy:2015efa,Fermi-LAT:2013thd}; we observe that these limits are generally markedly stronger than the CMB constraints. 
However, note that these bounds can only be applied directly to decays to $\gamma \gamma$ exclusively; the indirect constraints on photon-rich final states with continuum spectra are often considerably weaker. 
In contrast, we expect our parameter space to be sensitive primarily to the total injected energy (similar to the CMB limits), rather than the details of the photon spectrum; thus we expect the effects on star formation to be similar for the simple decay to $\gamma \gamma$ that we show and injection of continuum photons with similar total energy.

For $p$-wave annihilation, constraints lie below the bottom of the plot, and the $y$-axis on the bottom panels shows the value of $(\sigma v)_\mathrm{ref}$.
This velocity dependence means that the $p$-wave results are dominated by late-time structure formation.
Hence, these results should be used cautiously, since our assumption that the energy deposition fractions are equal between the IGM and the halo may be less reliable for annihilation, particularly $p$-wave annihilation.

Starting with the decay channels (top panels of Fig.~\ref{fig:scans_z20}), we see that for both photon and $e^+ e^-$ final states, most of the regions where the net effect is to raise the threshold for a halo to collapse are ruled out by current constraints; for decay to $e^+ e^-$ pairs, there still exist small regions just above the minimum allowed lifetime where the critical $M_\mathrm{halo}$ could be slightly raised; one is located around $m_\chi = 4$ MeV and the other just above $m_\chi = 1$ GeV.
There is also much unconstrained parameter space where the net effect is to \textit{lower} the threshold for collapse.
That is, DM can produce both positive and negative feedback in the first star formation, through its effect on the H$_2$ abundance of early galaxies.
The fiducial models studied earlier in this work were chosen such that one came from each of these regions.
For $s$-wave annihilation (middle panels), these models can only lower the threshold for collapse, but the regions with the largest effect are ruled out by CMB constraints.
For $p$-wave annihilation (bottom panels), these models only raise the threshold for collapse, but the regions where there is any significant effect are strongly ruled out by existing limits.

\begin{figure*}
	\centering
    \includegraphics[scale=0.4]{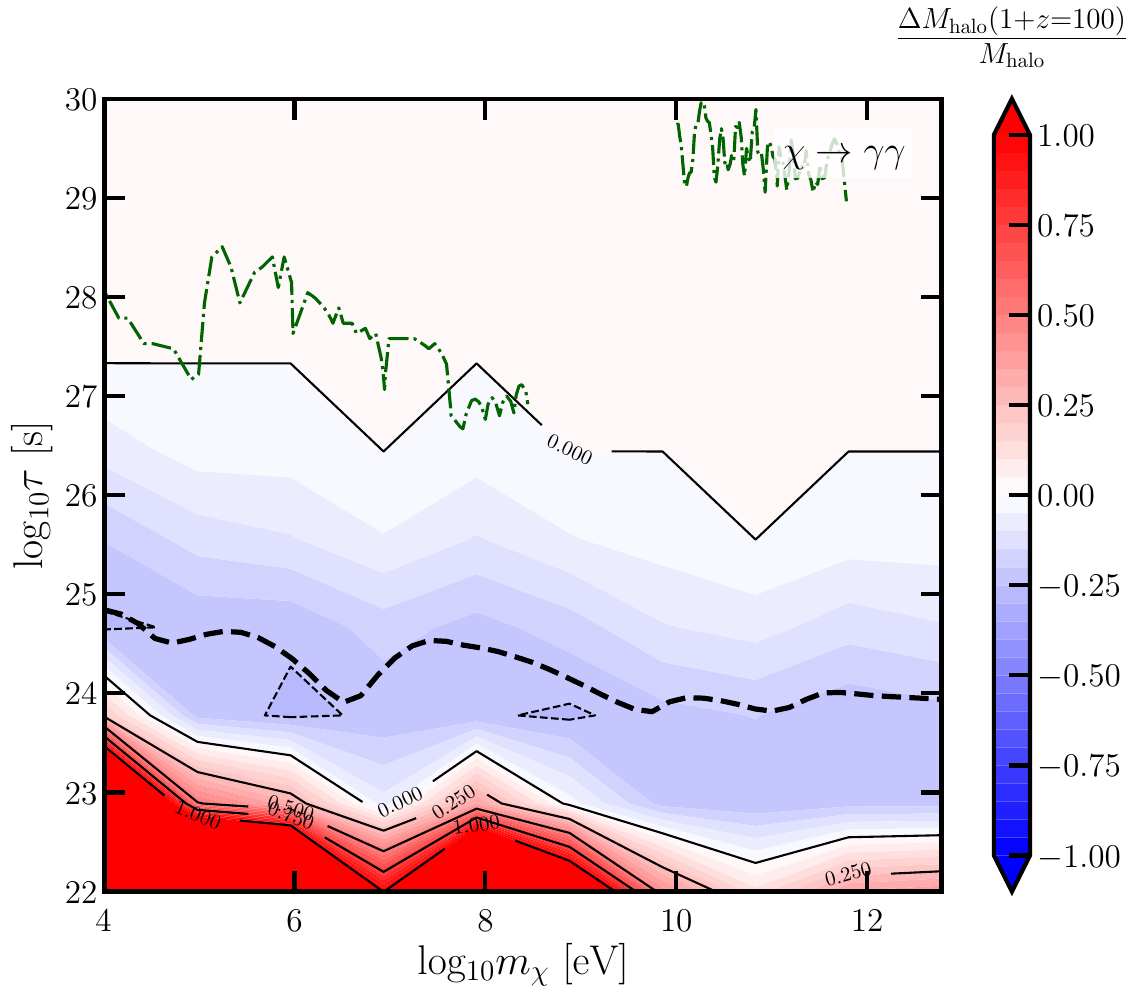}
    \includegraphics[scale=0.4]{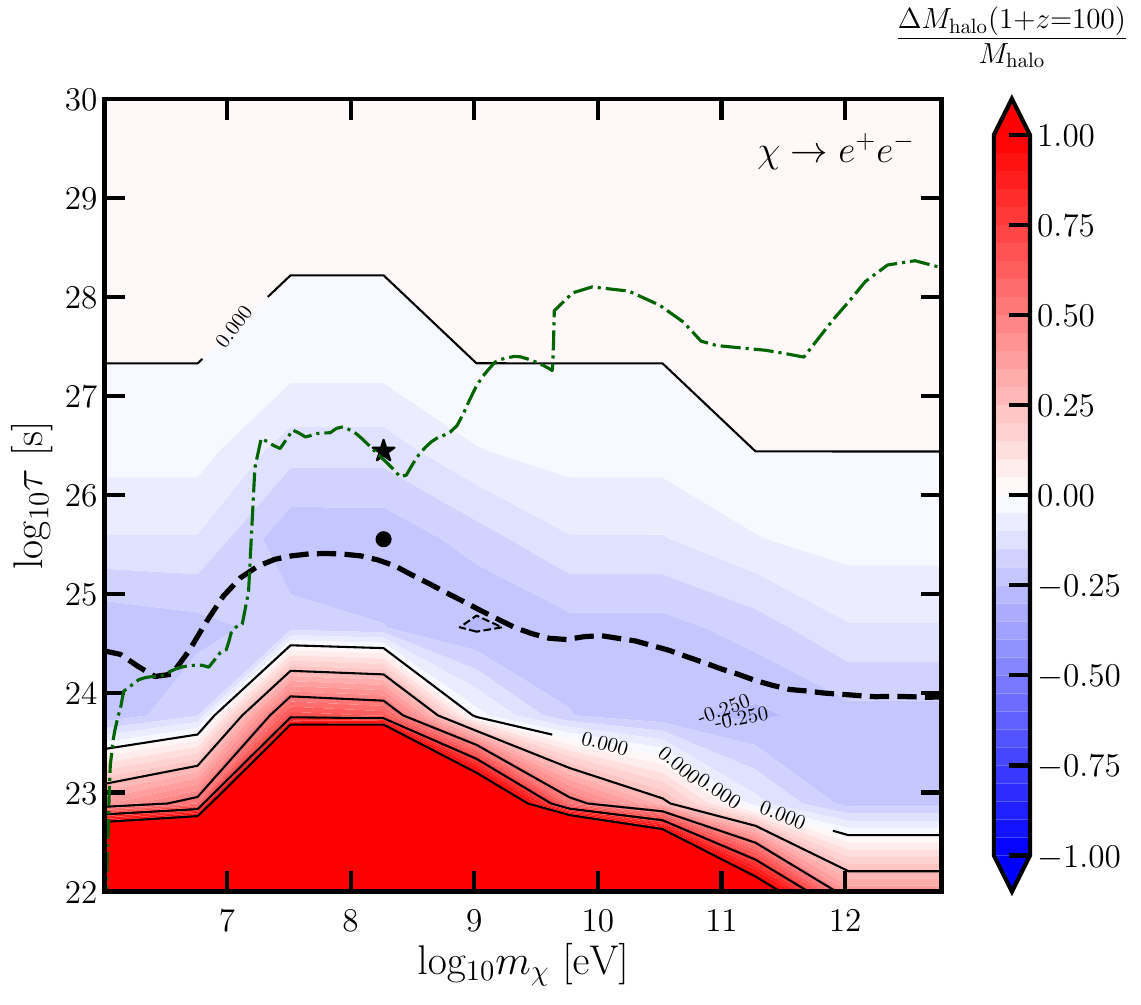} \\
    \includegraphics[scale=0.4]{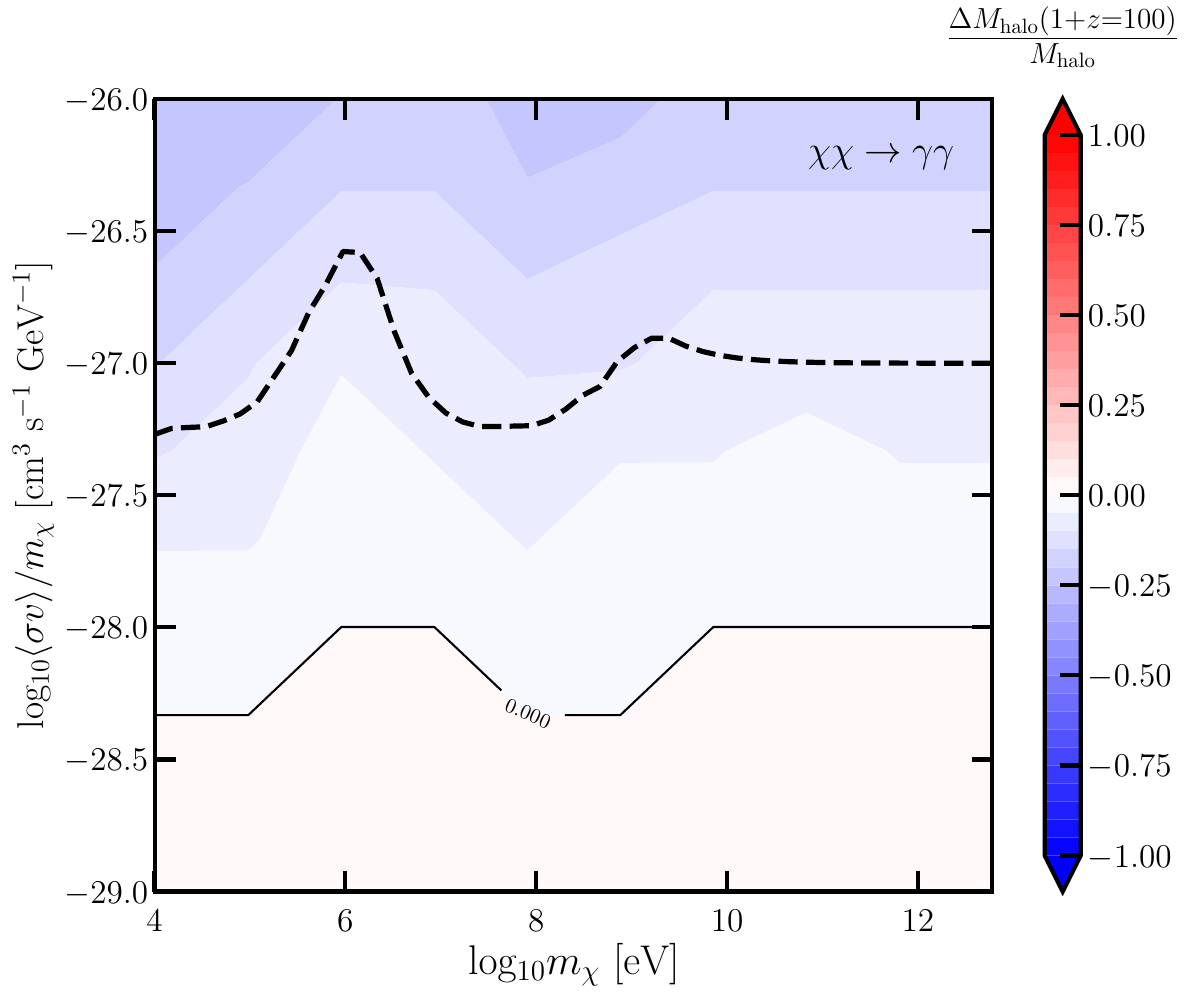}
    \includegraphics[scale=0.4]{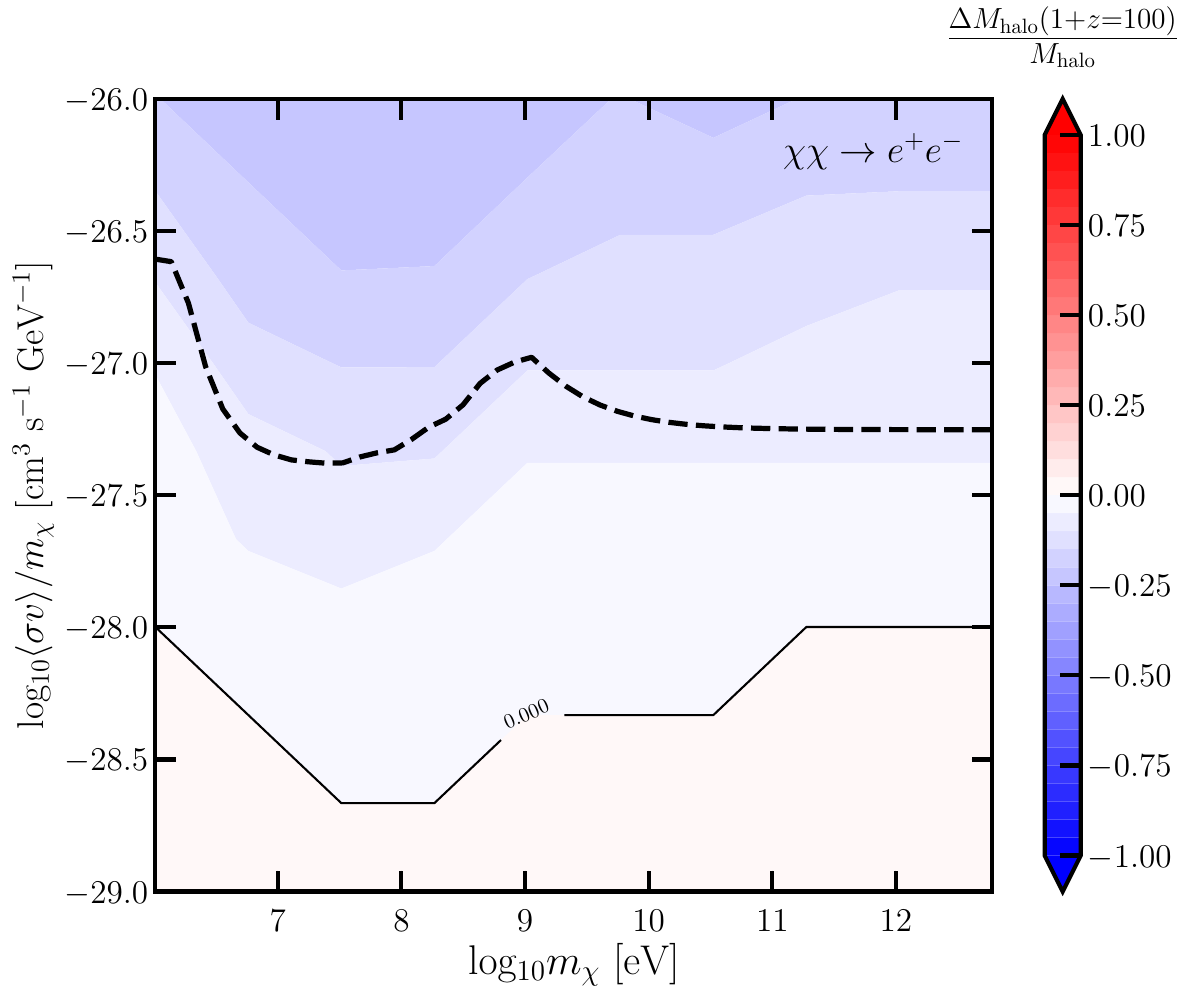} \\
    \includegraphics[scale=0.4]{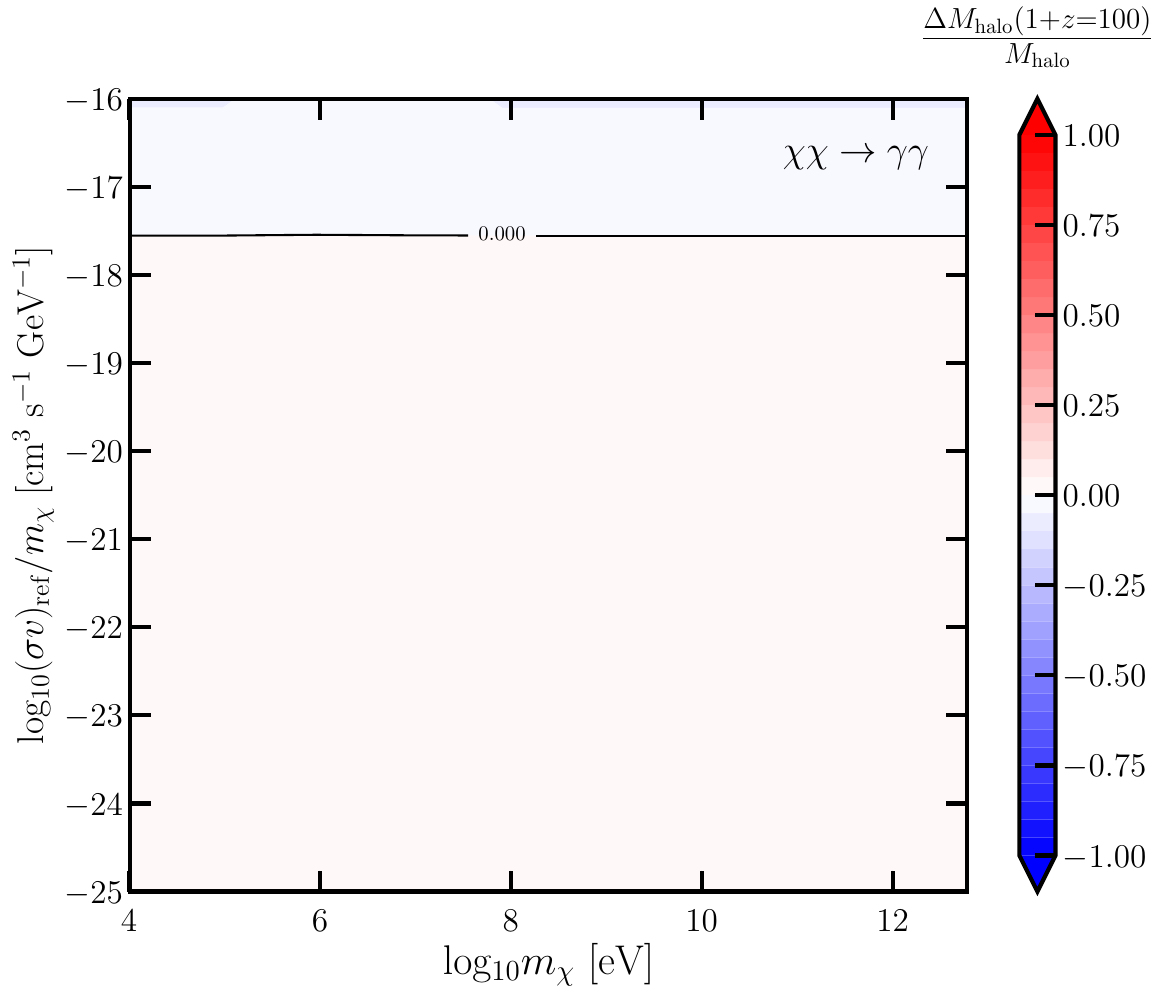}
    \includegraphics[scale=0.4]{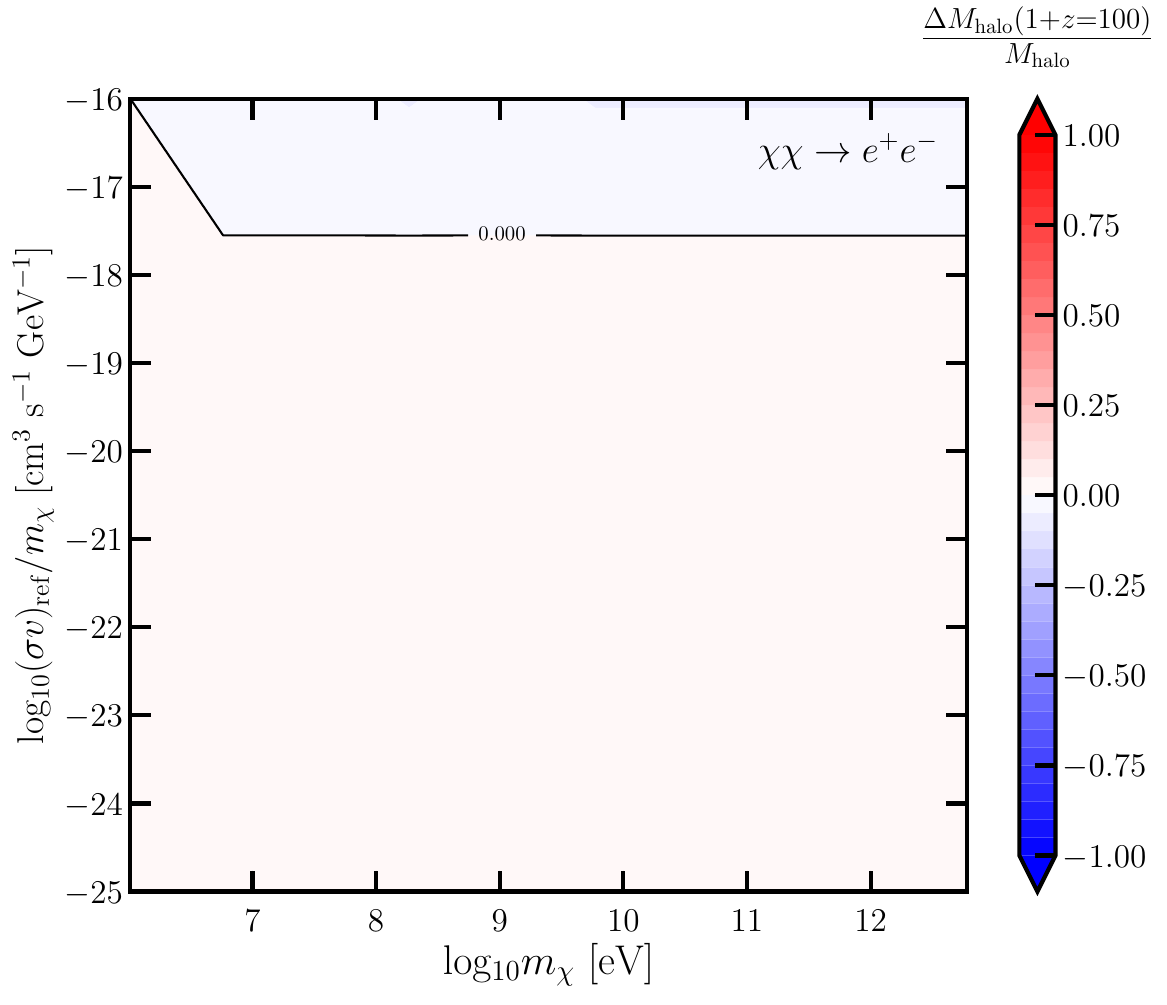}
    \caption{
    Same as Fig.~\ref{fig:scans_z20}, but showing the results at $1+z=100$.
    While the shape of the contours is broadly similar to Fig.~\ref{fig:scans_z20}, the contours are shifted to lower lifetimes/higher cross-sections, and the depth of the contours is reduced.
    }
    \label{fig:scans_z100}
\end{figure*}

The net effect of a particular energy injection model is redshift dependent.
Fig.~\ref{fig:scans_z100} shows an analogous plot to Fig.~\ref{fig:scans_z20}, but at a much earlier redshift of $1+z=100$; within the standard $\Lambda$ Cold Dark Matter ($\Lambda$CDM) cosmological model, we do not expect sufficient halos to form at this redshift, but we show these results for illustrative purposes.
The contours of the relative change to the critical $M_\mathrm{halo}$ change dramatically at this redshift.
For decaying DM, all regions that raise the critical $M_\mathrm{halo}$ at this redshift are ruled out by CMB constraints, and the region where $M_\mathrm{halo}$ can be lowered moves to smaller lifetimes.
Moreover, the shape of the contours at higher energies better matches the shape of the CMB constraints compared to the case at $1+z=20$.
This is because over time, the universe becomes increasingly transparent to photons with energy between 10 keV and 1 TeV~\cite{Chen:2003gz,Slatyer:2009yq}.
Prior to $1+z \sim 100$, most photons in this energy range will scatter and deposit their energy, so energy deposition results for these redshifts are relatively flat across the high mass range for decaying DM.
At lower redshifts, much of the particle cascade from high mass DM ends up in the transparency window and energy deposition becomes less efficient, hence the strength of the effects on star formation become weaker with increasing mass.

The panels showing the effect of $s$-wave annihilation at this redshift are similar to the case at $1+z=20$.
However, for $p$-wave annihilation, there is nearly no effect at all on the critical value for halo collapse; this is because the energy injection from $p$-wave annihilation scales even more steeply with velocity than for $s$-wave annihilation, hence the effects of $p$-wave annihilation are strongly suppressed for the redshifts just before structure formation.

\subsubsection{Bracketing Lyman-Werner effects}
\label{sec:shielding}

\begin{figure*}
	\centering
    \includegraphics[scale=0.4]{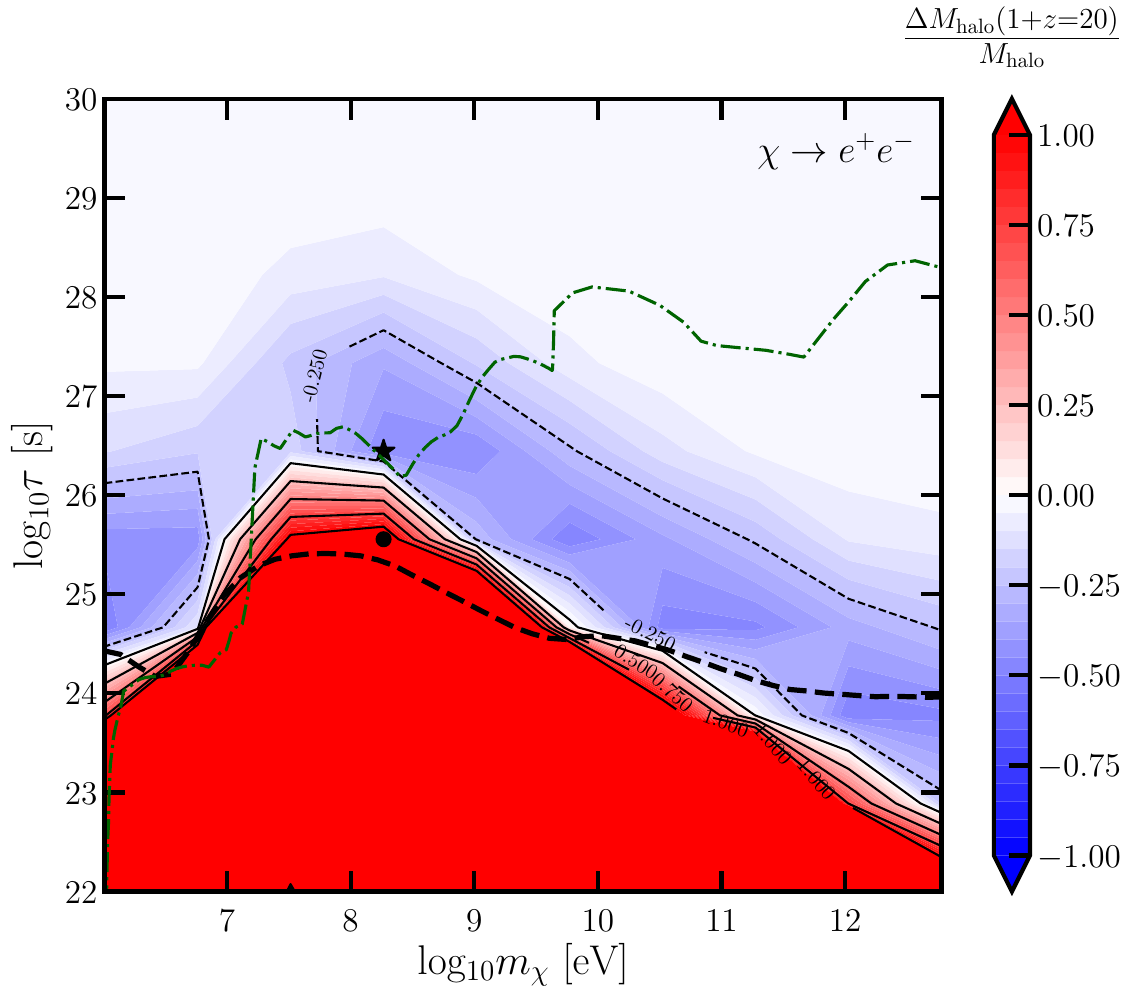}
    \includegraphics[scale=0.4]{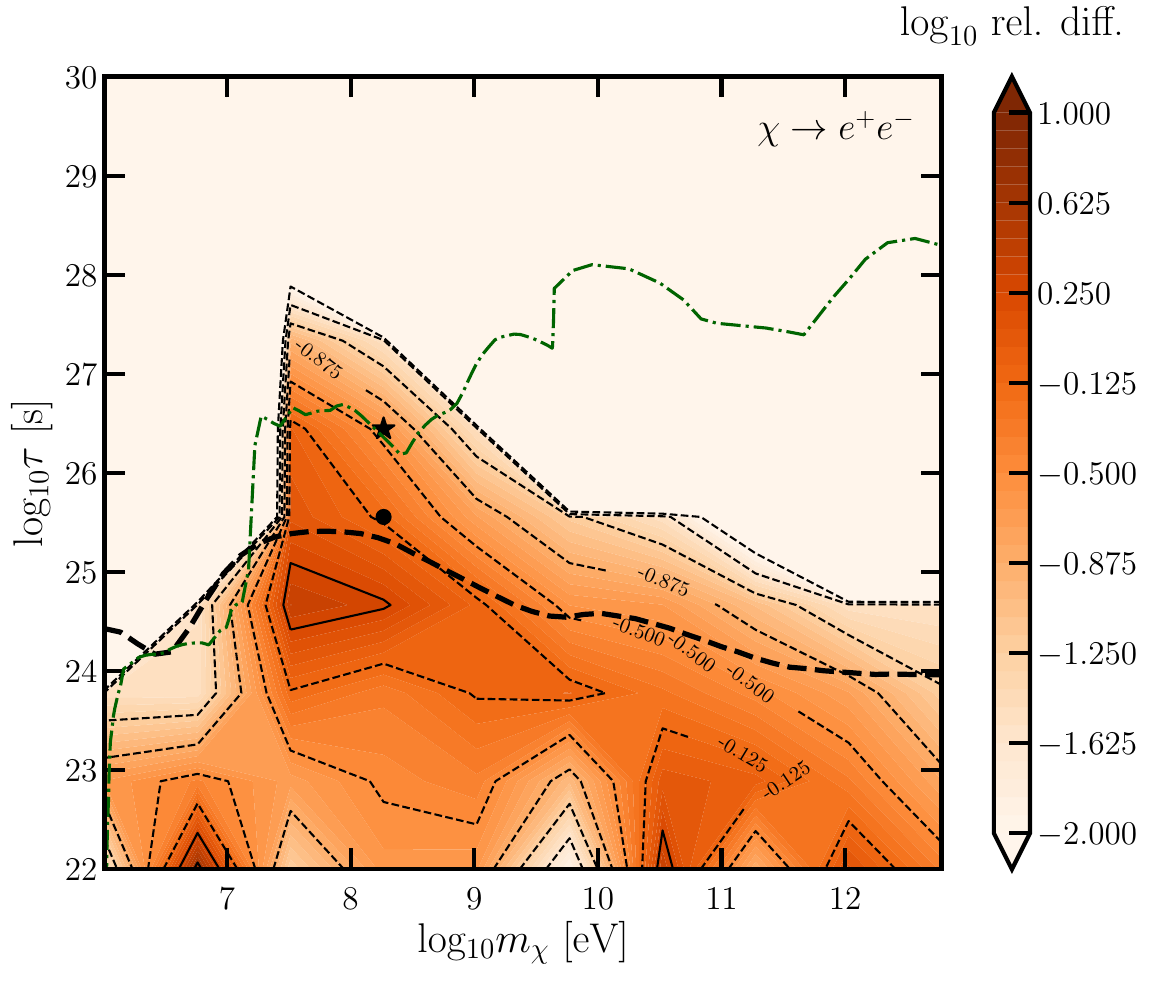} \\
    \includegraphics[scale=0.4]{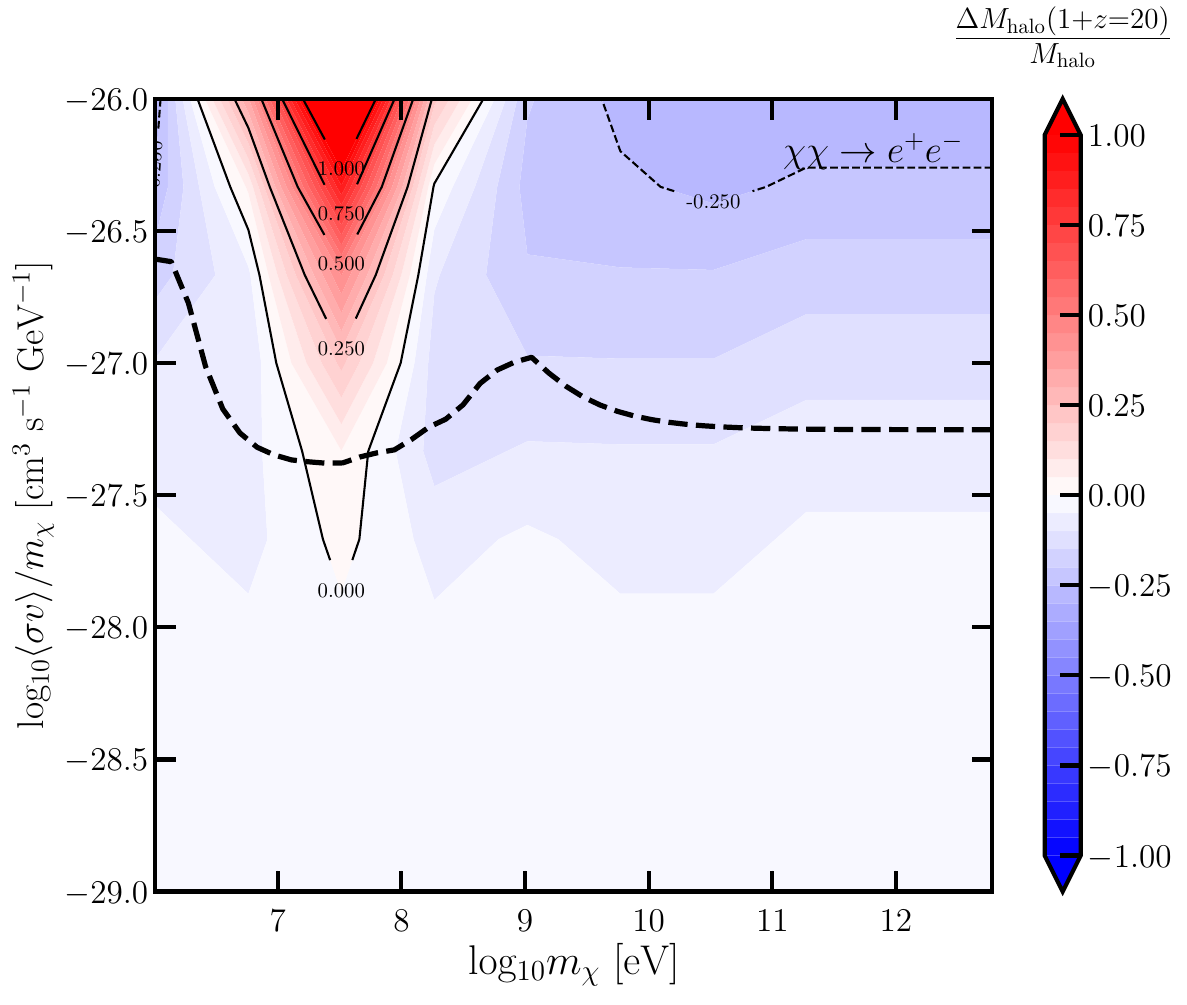}
    \includegraphics[scale=0.4]{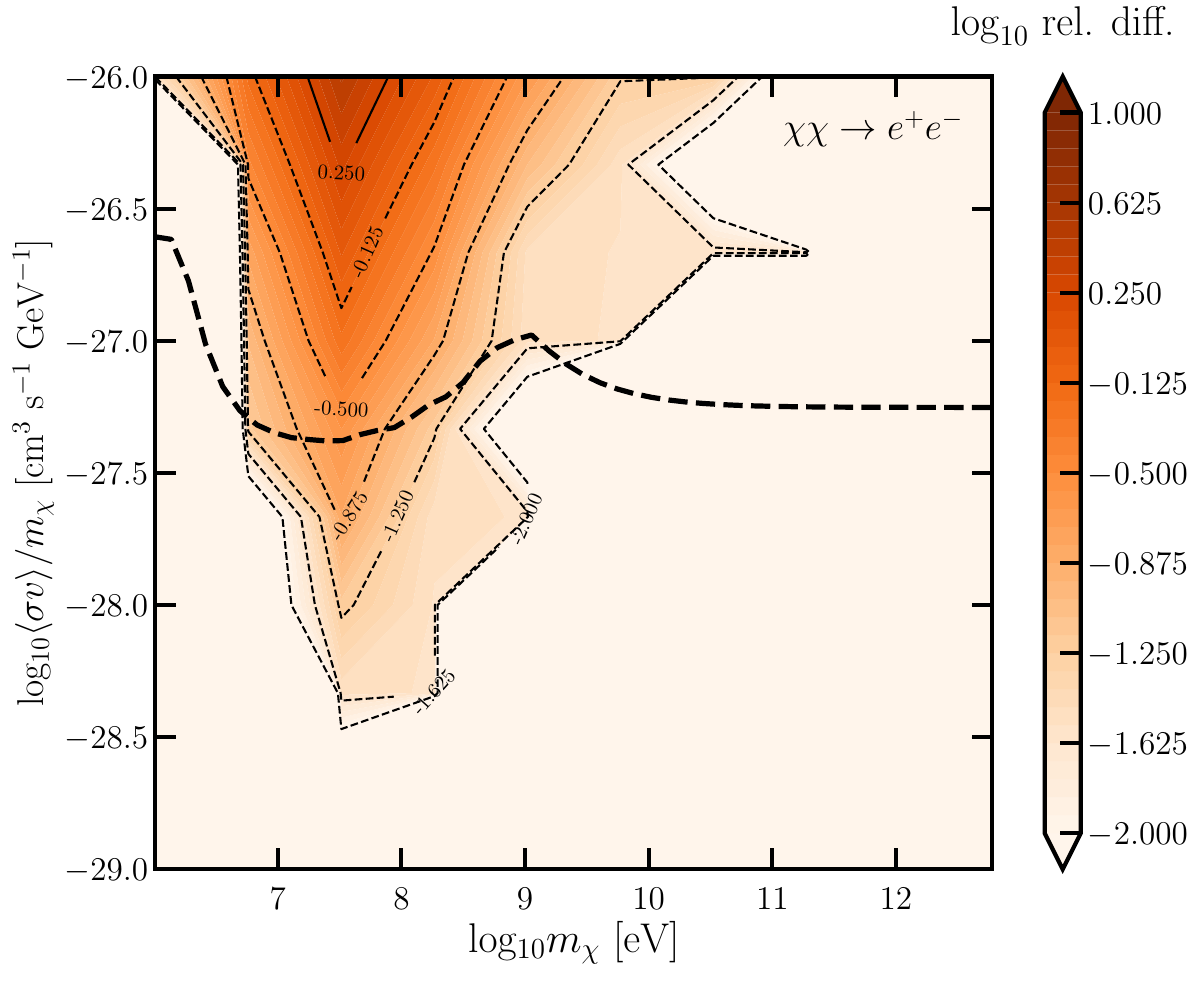}
    \caption{
    \textit{(Left)} Change to minimum halo mass $M_\mathrm{halo}$ necessary for star formation at $1+z=20$ when self-shielding is inefficient and a LW flux can have large effects.
    \textit{(Right)} $\log_{10}$ of the difference between the results for no/total self-shielding, divided by the total self-shielding results.
    In other words, the lightest contour shows where the results differ relative to each other by less than a percent.
    The top row shows the parameter space for DM decaying to $e^+ e^-$ pairs, and the bottom row is for $s$-wave annihilation to $e^+ e^-$.
    }
    \label{fig:scans_z20_LW}
\end{figure*}

In the previous subsection, we assumed that self-shielding is very efficient, such that LW photons have a negligible effect on the collapsing halo.
We now explore the opposite limit in order to bracket the effect of self-shielding.

Fig.~\ref{fig:scans_z20_LW} shows the results for decay (top row) and $s$-wave annihilation (bottom row) to $e^+ e^-$ pairs; the left panels show the same types of contours as in Fig.~\ref{fig:scans_z20} and the right shows the difference in the collapse threshold when assuming no/complete self-shielding of the halo.
See Appendix~\ref{app:LW} for discussion of other channels.
In both cases, reducing the efficacy of self-shielding raises the mass threshold for collapsing halos; the magnitude of the effect increases for larger energy injections (at even larger injections, the difference starts to decrease once the energy injection is large enough that heating matters more than the LW background).
This means that the effect on the mass threshold is increased for injection models that delay star formation, such as Model~$\bullet$, and slightly decreased for models which accelerated star formation, such as Model~$\star$.

The effect of self-shielding is most dramatic in the $s$-wave annihilation panel; whereas in the case of efficient self-shielding, all injection models shown decreased the threshold for collapse, neglecting self-shielding gives rise to a region of parameter space where it is now possible to raise the threshold.
This occurs around DM masses of tens of MeV; at this mass, the injected electrons are at the correct energy to upscatter CMB photons into the LW band through inverse Compton scattering (ICS).

Since inefficient self-shielding further slows the collapse of halos, this leads to the intriguing possibility that exotic heating and enhanced LW backgrounds may prevent gas fragmentation and subsequent star formation, but facilitate the collapse of the gas cloud directly into a black hole~\cite{Friedlander:2022ovf}.
We leave further study of the impact of DM processes on direct collapse black holes to future work.

\subsubsection{Signals in 21\,cm}
\label{sec:21cm}
\begin{figure}
    \includegraphics[scale=0.5]{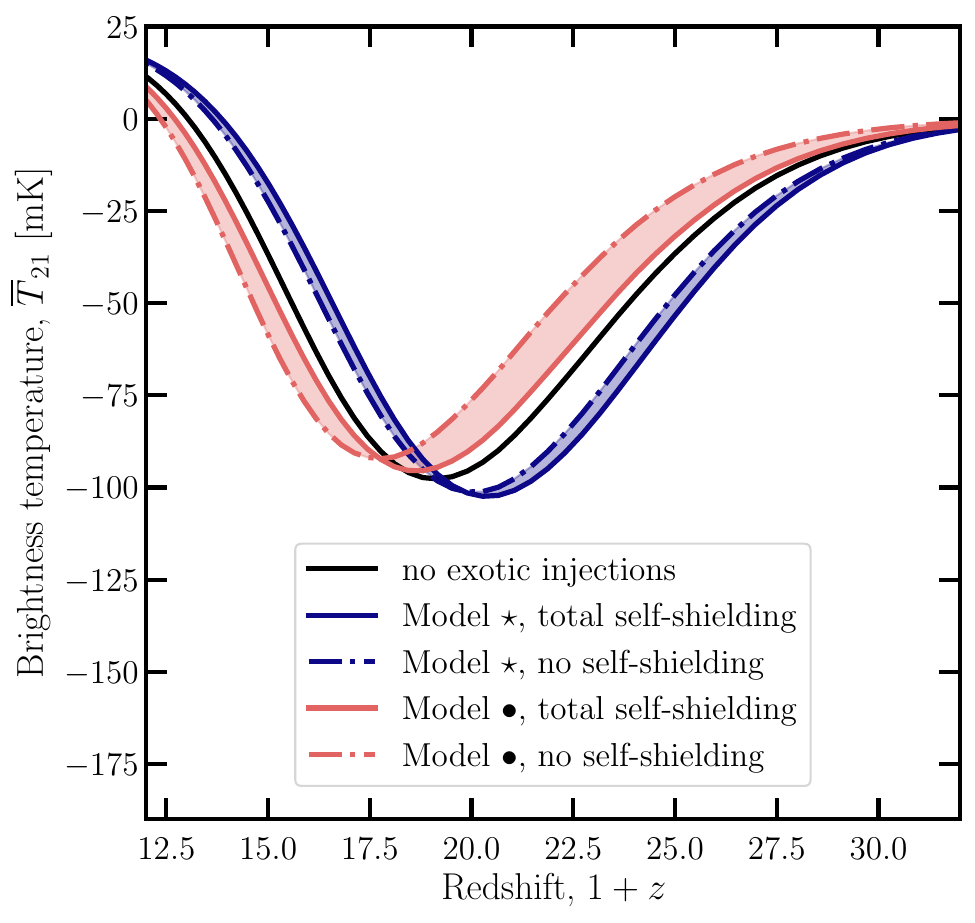}
    \includegraphics[scale=0.5]{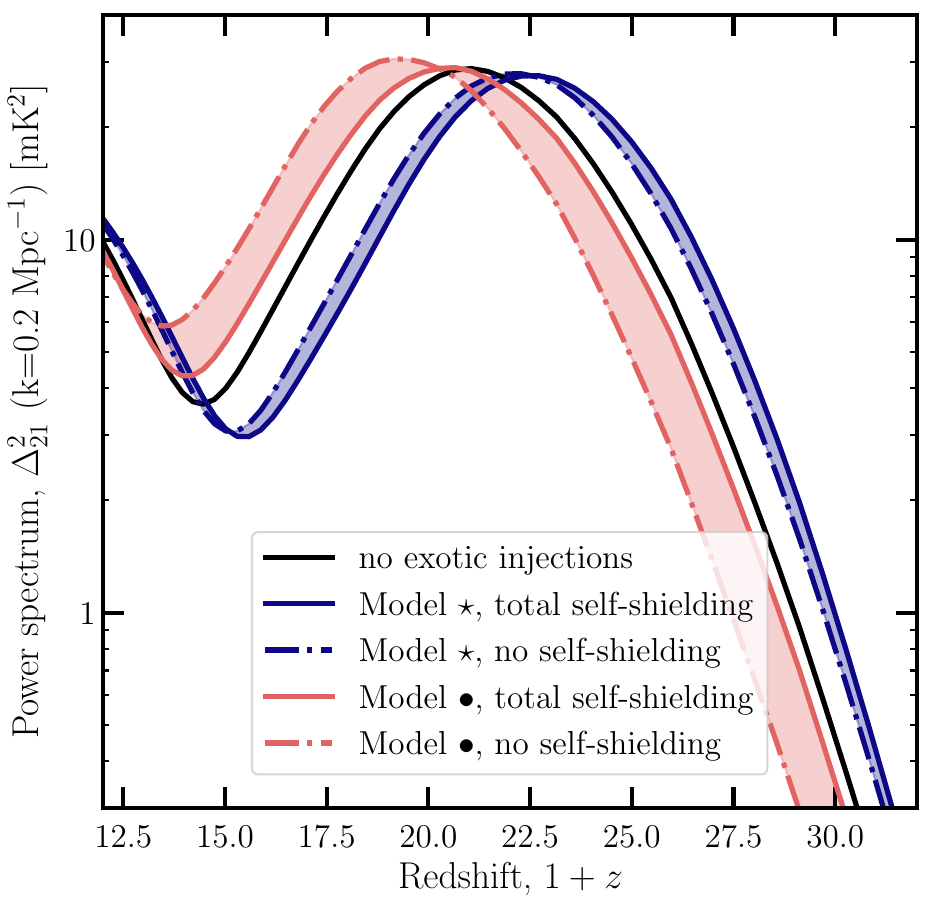}
    \caption{
    The 21\,cm global signal (top) and power spectrum at $k = 0.2$ Mpc$^{-1}$ (bottom) as a function of redshift, for our standard cosmological model and two fiducial DM models.
    The shaded contours bracket the effect of H$_2$ self-shielding.
    }
    \label{fig:21cm}
\end{figure}

One of the most promising ways to determine the timing of the first stellar formation is through the 21\,cm transition of neutral hydrogen at high redshifts (for a thorough review, see e.g.\ Ref.~\cite{Pritchard:2011xb}).
Here we briefly discuss how the DM models we have considered would affect an example 21\,cm signal.
We will focus on the timing of cosmic dawn through the enhancement or suppression of H$_2$ from exotic energy injection.
As such, we will not model other sources of feedback (e.g.\ stellar LW emission~\cite{Machacek:2000us}, DM-baryon relative velocities~\cite{Tseliakhovich:2010bj}, or their combination~\cite{Schauer:2020gvx,Kulkarni:2020ovu}, as well as heating~\cite{Valdes:2012zv,Sitwell:2013fpa,Evoli:2014pva,Liu:2018uzy}), and will focus on a few fiducial scenarios as a showcase.
Moreover, the effects of exotic energy injection may be partially degenerate with varying quantities such as the halo mass function and astrophysical parameters.
Since this is not a forecast for detectability, we defer a detailed study of the 21\,cm signal, varying astrophysical parameters and including reionization bubbles, to future work, as this will require the establishment of hydrodynamical simulations of H$_2$ formation with DM decay or annihilation.

We use the public 21\,cm code {\tt Zeus21} \githubZeus~\cite{Munoz:2023kkg}, which we modify to include stellar formation in molecular-cooling halos.
We take a toy model where we enable star formation in halos below the atomic-cooling threshold with some constant star-formation efficiency $f_*^{\rm mol}$, and we keep the baseline {\tt Zeus21} model otherwise.
That is, we take the model for the star-formation efficiency from Ref.~\cite{Munoz:2023kkg},
\begin{equation}
    f_\star \equiv \frac{\dot M_\star}{f_b\,\dot M_\mathrm{halo}},
\end{equation}
where $\dot M_*$ is the star-formation rate, $\dot M_\mathrm{halo}$ is the mass-accretion rate, and $f_b$ is the baryon fraction, which we enhance as
\begin{equation}
    \Delta f_\star(M_\mathrm{halo}) = f_*^{\rm mol} e^{-M_{\rm mol}/M_\mathrm{halo}} \, e^{-M_\mathrm{halo}/M_{\rm atom}},
\end{equation}
between the atomic- and molecular-cooling thresholds ($M_{\rm atom}$ and $M_{\rm mol}$, respectively).
The former is set by a constant virial temperature of $T_{\rm atom}=10^4$ K~\cite{Oh:2001ex}, and we obtain the latter from our results for each DM scenario (as shown in Fig.~\ref{fig:critical_collapse_Mhalo}).
Following Refs.~\cite{Qin:2021gkn,Munoz:2021psm}, we set an amplitude of star-formation efficiency of $f_*^{\rm mol}=10^{-2.5}$, in broad agreement with current constraints to reionization.
In this simple model we assume the same stellar properties (e.g.\ Lyman-$\alpha$ and X-ray photons emitted per star-forming baryon) for all galaxies, though a more realistic model would take low-mass (H$_2$-cooling) galaxies to host older (PopIII) stars, with different spectra~\cite{Bromm:2003vv}.

With these caveats in mind, we can now estimate the degree to which we expect DM decays to affect the timing of the first stellar formation.
We show the predicted 21\,cm signals from this toy model in Fig.~\ref{fig:21cm}; the top panel shows the global signal, while the bottom panel shows the amplitude of the power spectrum at $k = 0.2$ Mpc$^{-1}$, both as a function of redshift.
Since Model $\bullet$\, raises the mass threshold for stars to form, the 21\,cm signals are slightly delayed, shifting to smaller values of $z$.
Conversely, Model $\star$\, allows smaller mass halos to host stars, so cosmic star formation begins earlier and the 21\,cm signals are accelerated.
In this latter case, the signal peaks/troughs can be shifted by as much as $\Delta z \sim 2$.

The degree to which the signals are shifted also depends on the efficiency of self-shielding; we bracket this effect using shaded contours in Fig.~\ref{fig:21cm}.
The possible variation of the signal due this effect is larger for Model $\bullet$, since this model injects more energy and can therefore contribute a larger LW background.
At lower redshifts, the astrophysical contribution to the LW background will also become important; once stars begin to form, they will emit their own LW radiation, and we have not modelled this feedback here.

While a full detectability study is beyond the scope of this work, we note in passing that a fiducial signal like our case without exotic injections in Fig.~\ref{fig:21cm} is expected to be detectable by the currently operating HERA interferometer~\cite{DeBoer:2016tnn}, boasting a signal-to-noise ratio of SNR $\approx 100$~\cite{Munoz:2021psm}. 
As such, this generation of telescopes may be sensitive to the delay/acceleration of the first galaxies due to decaying DM. 

Note that we have focused on the timing of the 21\,cm signal, and not its depth. 
In a future analysis, this and other potential effects of exotic energy injection should be studied together, which will pave the way to find the full effects of DM on the cosmic dawn.

\subsection{Conclusion}
\label{sec:H2_conclusion}

We have performed an initial study of the effect of homogeneous energy deposition on early star formation.
We use \texttt{DarkHistory} to calculate how energy is deposited by decaying or annihilating DM and then track the effect of this exotic energy injection on the temperature, ionization, and H$_2$ abundance of a toy halo model.
We find that energy injection from decaying or annihilating DM can both raise and lower the critical mass/virial temperature threshold for galaxies to form, and the direction of this effect can depend on the redshift at which the halos virialize.
Hence, exotic energy injection can both accelerate and delay the onset of star formation, and this can in turn alter the timing of signals in 21\,cm cosmology.

The most interesting unconstrained parameter space comes from decaying DM, where the mass threshold for collapse can be lowered by as much as 60\%; while there are also some decaying DM models consistent with CMB anisotropy limits that can raise the mass threshold, most of these are ruled out by other indirect detection constraints.
Regarding annihilations, there is some unconstrained parameter space where $s$-wave annihilations can lower the mass threshold, although only by about 10\%; the regions where $p$-wave annihilation has any affect are ruled out by indirect detection limits.

We ignore a number of subdominant effects, including other H$_2$ formation pathways, the LW background, H$_2$ self-shielding, and baryon streaming velocities.
Moreover, although we include the boost factors for annihilation from structure formation, we neglect energy injection from within the halo itself for all channels; however, energy deposition by decays or annihilations from within a halo can also be significant, and even dominate the IGM signal in the case of annihilations.
A complete study of exotic energy injection on star formation would require modeling both contributions; this is left for future work.

Our results are obtained with a spherically collapsed model for the halo, so in order to make more precise statements, hydrodynamical simulations will be required.
However, it would be far too computationally expensive to scan over many DM models with such simulations in order to find those with the most significant effects.
In this work we have performed such a scan with a simpler semi-analytic model, and found the regions of DM parameter space that would be most interesting to simulate, including decays to photons between lifetimes of $\log_{10} (\tau / [\mathrm{s}]) = 24$ to 27, and decays to $e^+ e^-$ pairs between lifetimes of $\log_{10} (\tau / [\mathrm{s}]) = 24$ to 28, both for masses less than about 10 GeV (at larger masses, these lifetimes are excluded by indirect detection limits~\cite{Fermi-LAT:2013thd,Essig:2013goa, Massari:2015xea, Cohen:2016uyg,Laha:2020ivk,Cirelli:2020bpc,Cirelli:2023tnx,Boudaud:2016mos,Boudaud:2018oya}).
Within this range we find a very rich phenomenology, with DM both helping the formation of the first stars (by catalyzing H$_2$ formation), as well as hampering it (through heating and photodissociation).
We conclude that the cosmic-dawn era will not only teach us about the astrophysics of the first galaxies, but will also shed light onto the nature of DM.

\chapter{Production of Primordial Black Holes}
\label{sec:PBHandMFI}

Dark matter could take the form of a population of black holes that existed in the early universe well before the emission of the CMB.
Such \textit{primordial black holes} (PBHs) are relatively minimal dark matter candidates in that they do not require the existence of new degrees of freedom except in their formation mechanism, and these may be provided at high energy scales at which we already expect new physics to emerge in the early universe.

In this chapter, I will describe my work studying the formation of PBHs in well-motivated and realistic models of multifield inflation.
In Section~\ref{sec:PBHs}, I will show that the multifield model we studied is indeed capable of generating PBH dark matter.
This work is based on Ref.~\cite{Geller:2022nkr} and was performed in collaboration with Sarah Geller, Evan McDonoough, and Dave Kaiser.
Sarah Geller led this work and my primary contribution was to develop the tools which we used to analyze the multifield model and search for a suitable parameter set to create PBHs.

In Section~\ref{sec:PBH_MCMC}, I will show results from the first ever Markov Chain Monte Carlo (MCMC) study of inflationary models which can generate PBHs while remaining in compliance with observational constraints.
I will also show that future gravitational wave observatories will be able to place new constraints on such formation mechanisms.
This work is based on Ref.~\cite{Qin:2023lgo} and was done in collaboration with Sarah Geller, Shyam Balaji, Evan McDonoough, and Dave Kaiser.
I led this work and wrote the code pipeline for the MCMC, analyzed the resulting posteriors, and calculated the gravitational wave signals from these models.

\section{Primordial Black Holes from Multifield Inflation with Nonminimal Couplings}
\label{sec:PBHs}

Primordial black holes (PBHs) were first postulated more than half a century ago \cite{ZeldovichNovikov1967,Hawking1971,Carr:1974nx}, and they remain a fascinating theoretical curiosity. In recent years, many researchers have realized that PBHs provide an exciting prospect for accounting for dark matter. Rather than requiring some as-yet unknown elementary particles beyond the Standard Model, dark matter might consist of a large population of PBHs that formed very early in cosmic history. See Refs.~\cite{Carr:2020xqk,Green:2020jor,Villanueva-Domingo:2021spv} for recent reviews.

Much activity has focused on mechanisms by which PBHs could form from density perturbations that were generated during early-universe inflation. When overdensities with magnitude above some critical threshold re-enter the Hubble radius after the end of inflation, they induce gravitational collapse into black holes. Many studies have focused on specific inflationary models that can yield appropriate perturbations; PBH formation following hybrid inflation has garnered particular attention \cite{Garcia-Bellido:1996mdl,Lyth:2010zq,Bugaev:2011wy,Halpern:2014mca,Clesse:2015wea,Kawasaki:2015ppx}. Others have found clever ways to engineer
desired features of a given model so as to generate PBHs, by inserting specific features into the potential and/or non-canonical kinetic terms for the field(s) driving inflation. See, e.g., Refs.~\cite{Garcia-Bellido:2017mdw,Ezquiaga:2017fvi,Kannike:2017bxn,Germani:2017bcs,Motohashi:2017kbs,Di:2017ndc,Ballesteros:2017fsr,Pattison:2017mbe,Passaglia:2018ixg,Biagetti:2018pjj,Byrnes:2018txb,Carrilho:2019oqg,Ashoorioon:2019xqc,Aldabergenov:2020bpt,Ashoorioon:2020hln,Inomata:2021uqj,Inomata:2021tpx,Pattison:2021oen,Lin:2020goi,Palma:2020ejf,Yi:2020cut,Iacconi:2021ltm,Kallosh:2022vha,Ashoorioon:2022raz,Frolovsky:2022ewg,Aldabergenov:2022rfc}.

In this work we explore possibilities for the production of PBHs within well-motivated models of inflation that feature realistic ingredients from high-energy theory. In particular, we consider models with several interacting scalar fields, each of which includes a nonminimal coupling to the spacetime Ricci scalar. This family of models includes---but is more general than---well-known models such as Higgs inflation \cite{Bezrukov:2007ep} and $\alpha$-attractor models \cite{Kallosh:2013maa,Kallosh:2013yoa,Galante:2014ifa}. For example, the Higgs sector of the Standard Model includes four scalar degrees of freedom, all of which remain in the spectrum at high energies within renormalizable gauges \cite{Mooij:2011fi,Greenwood:2012aj}. Moreover, every candidate for Beyond Standard Model physics includes even more scalar degrees of freedom at high energies \cite{Lyth:1998xn,Mazumdar:2010sa}. Likewise, nonminimal couplings in the action of the form $\xi \phi^2 R$, where $\phi$ is a scalar field, $R$ is the spacetime Ricci scalar, and $\xi$ a dimensionless constant, are required for renormalization and, more generally, are induced by quantum corrections at one-loop order even if the couplings $\xi$ vanish at tree-level \cite{Callan:1970ze,Bunch:1980br,Bunch:1980bs,Birrell:1982ix,Odintsov:1990mt,Buchbinder:1992rb,Faraoni:2000gx,Parker:2009uva,Markkanen:2013nwa,Kaiser:2015usz}. The couplings $\xi$ generically increase with energy scale under renormalization-group flow with no UV fixed point \cite{Odintsov:1990mt,Buchbinder:1992rb}, and hence they can be large $(\vert \xi \vert \gg 1$) at the energy scales relevant for inflation. Finally, although the models we study need not make recourse to supersymmetry or supergravity, we find they can be realized in simple supergravity setups, including in models that simultaneously realize the observed cosmological constant. 

Inflationary dynamics in the family of models we consider generically yield predictions for observable quantities, such as the spectral index of primordial curvature perturbations and the ratio of power spectra for tensor and scalar perturbations, in close agreement with recent measurements \cite{Kaiser:2012ak,Kaiser:2013sna,Schutz:2013fua}. Such models also generically yield efficient post-inflation reheating, typically producing a radiation-dominated equation of state and a thermal spectrum of decay products within $N_{\rm reh} \sim {\cal O} (1)$ efolds after the end of inflation \cite{Bezrukov:2008ut,Garcia-Bellido:2008ycs,Child:2013ria,DeCross:2015uza,DeCross:2016fdz,DeCross:2016cbs,Figueroa:2016dsc,Repond:2016sol,Ema:2016dny,Sfakianakis:2018lzf,Rubio:2019ypq,Nguyen:2019kbm,vandeVis:2020qcp,Iarygina:2020dwe,Ema:2021xhq,Figueroa:2021iwm,Dux:2022kuk}. Hence such models represent an important class in which to consider PBH production.

We find that such models provide a natural framework within which PBHs could form. As in previous studies that focused on the formation of PBHs from a phase of ultra-slow-roll inflation \cite{Garcia-Bellido:2017mdw,Ezquiaga:2017fvi,Kannike:2017bxn,Germani:2017bcs,Motohashi:2017kbs,Di:2017ndc,Ballesteros:2017fsr,Pattison:2017mbe,Passaglia:2018ixg,Byrnes:2018txb,Biagetti:2018pjj,Carrilho:2019oqg,Inomata:2021tpx,Inomata:2021uqj,Pattison:2021oen}, we also find that to produce perturbation spectra relevant for realistic PBH scenarios, at least one dimensionless parameter must be highly fine-tuned. Nonetheless, we find that such models can yield accurate predictions for a significant number of observable quantities using a smaller number of relevant free parameters. In this section, we focus on the general mechanisms by which such models can produce PBHs, and defer to Section~\ref{sec:PBH_MCMC} a more thorough analysis of the full parameter space.

In Section \ref{sec:model} we introduce the family of multifield models on which we focus and identify generic features of their dynamics. Section \ref{sec:PBHform} considers the formation of PBHs after the end of inflation, including how the production of PBHs is affected by changes to various model parameters. Concluding remarks follow in Section \ref{sec:PBHdiscussion}. In Appendix \ref{appPerturbations}, we review important features of gauge-invariant perturbations in multifield models, while in Appendix \ref{appSUGRA} we demonstrate how this family of models can be realized within a supergravity framework. Appendix \ref{appTheta} includes additional details about our analytic solution for the fields' trajectory through field space during inflation.
Throughout this section we will work in terms of the reduced Planck mass, $M_{\rm pl} \equiv 1 / \sqrt{ 8 \pi G} = 2.43 \times 10^{18} \, {\rm GeV}$.

\subsection{Multifield Model and Dynamics}
\label{sec:model}

\subsubsection{Multifield Formalism}
\label{sec:multifieldformalism}

We begin with a brief review of multifield dynamics for background quantities and linearized fluctuations, following the notation of Ref.~\cite{Kaiser:2012ak}. See also Appendix \ref{appPerturbations}, Refs.~\cite{Sasaki:1995aw,Langlois:2008mn,Peterson:2010np,Gong:2011uw}, and Ref.~\cite{Gong:2016qmq} for a review of gauge-invariant perturbations in multifield models. We consider models with ${\cal N}$ scalar fields $\phi^I (x^\mu)$ with $I = 1, 2, ... , \, {\cal N}$, and work in $(3 + 1)$ spacetime dimensions. In the Jordan frame, the action may be written
\beq
\tilde{S} = \int d^4 x \sqrt{-\tilde{g}} \left[ f (\phi^I ) \tilde{R}  - \frac{1}{2} \delta_{IJ} \tilde{g}^{\mu\nu}  \partial_\mu \phi^I \partial_\nu \phi^J - \tilde{V} (\phi^I) \right] ,
\label{SJ}
\eeq
where $f (\phi^I)$ denotes the fields' nonminimal couplings and tildes indicate quantities in the Jordan frame. After performing a conformal transformation by rescaling $\tilde{g}_{\mu\nu} (x) \rightarrow g_{\mu\nu} (x) = \Omega^2 (x) \tilde{g}_{\mu\nu} (x)$ with conformal factor
\beq
\Omega^2 (x) = \frac{2}{M_{\rm pl}^2} f (\phi^I (x) ) ,
\label{Omega}
\eeq
we may write the action in the Einstein frame as \cite{Kaiser:2010ps}
\beq
S = \int d^4 x \sqrt{-g} \left[ \frac{ M_{\rm pl}^2}{2} R - \frac{1}{2} {\cal G}_{IJ} g^{\mu\nu} \partial_\mu \phi^I \partial_\nu \phi^J - V (\phi^I) \right] ,
\label{SE}
\eeq
where the potential in the Einstein frame is stretched by the conformal factor,
\beq
V (\phi^I) = \frac{ M_{\rm pl}^4}{4 f^2 (\phi^I) } \tilde{V} (\phi^I ) .
\label{VEconformal}
\eeq
The nonminimal couplings induce a curved field-space manifold in the Einstein frame with associated field-space metric
\beq
{\cal G}_{IJ} (\phi^K) = \frac{M_{\rm pl}^2}{2 f (\phi^K)} \left[ \delta_{IJ} + \frac{ 3}{f (\phi^K)} f_{, I} f_{, J} \right] ,
\label{GIJgeneral}
\eeq
where $f_{, I} \equiv \partial f / \partial \phi^I$. For ${\cal N} \geq 2$ fields with nonminimal couplings, one cannot canonically normalize all of the fields while retaining the Einstein-Hilbert form of the gravitational part of the action \cite{Kaiser:2010ps}.

We consider perturbations around a spatially flat Friedmann-Lema\^{i}tre-Robertson-Walker (FLRW) line element, as discussed further in Appendix \ref{appPerturbations}, and separate each scalar field into a spatially homogeneous vacuum expectation value and spatially varying fluctuations:
\beq
\phi^I (x^\mu) = \varphi^I (t) + \delta \phi^I (x^\mu).
\label{phivarphi}
\eeq
The equation of motion for the spatially homogeneous background fields then takes the form
\beq
{\cal D}_t \dot{\varphi}^I + 3 H \dot{\varphi}^I + {\cal G}^{IK} V_{, K} = 0 ,
\label{eomvarphi}
\eeq
where $H \equiv \dot{a} / a$ and ${\cal D}_t A^I = \dot{\varphi}^J {\cal D}_J A^I$ for any field-space vector $A^I$, and where the covariant derivative  ${\cal D}_J$ 
employs the usual Levi-Civita connection associated with the metric ${\cal G}_{IJ}$. Since we consider only linearized fluctuations in this chapter, we may set ${\cal G}_{IJ} (\phi^K) \rightarrow {\cal G}_{IJ} (\varphi^K)$, so that components of the field-space metric depend only on time. The magnitude of the background fields' velocity vector is given by
\beq
\vert \dot{\varphi}^I \vert \equiv \dot{\sigma} = \sqrt{ {\cal G}_{IJ} \, \dot{\varphi}^I \dot{\varphi}^J } ,
\label{dotsigmadef}
\eeq
in terms of which we may write the unit vector
\beq
\hat{\sigma}^I \equiv \frac{ \dot{\varphi}^I } {\dot{\sigma}} 
\label{hatsigma}
\eeq
which points along the background fields' direction of motion in field space. The quantity
\beq
\hat{s}^{IJ} \equiv {\cal G}^{IJ} - \hat{\sigma}^I \hat{\sigma}^J
\label{sIJ}
\eeq
projects onto the subspace of the field-space manifold perpendicular to the background fields' motion.

In terms of $\dot{\sigma}$, the equations of motion for background quantities may be written \cite{Kaiser:2012ak}
\beq
\begin{split}
\ddot{\sigma} &+ 3 H \dot{\sigma} + V_{, \sigma} = 0 , \\
H^2 &= \frac{1}{ 3 M_{\rm pl}^2} \left[ \frac{1}{2} \dot{\sigma}^2 + V \right] , \\
\dot{H} &= - \frac{1}{ 2 M_{\rm pl}^2} \dot{\sigma}^2 ,
\end{split}
\label{eombackground}
\eeq
where
\beq
V_{, \sigma} \equiv \hat{\sigma}^I V_{, I} .
\label{Vsigma}
\eeq
The covariant turn-rate vector is defined as \cite{Kaiser:2012ak}
\beq
\omega^I \equiv {\cal D}_t \hat{\sigma}^I = - \frac{1}{ \dot{\sigma}} V_{, K} \hat{s}^{IK} ,
\label{omegadef}
\eeq
where the last expression follows upon using Eqs.~(\ref{eomvarphi}), (\ref{sIJ}), and (\ref{eombackground}). The usual slow-roll parameter takes the form
\beq
\epsilon \equiv - \frac{ \dot{H}}{H^2} = \frac{ 1}{2 M_{\rm pl}^2} \frac{ \dot{\sigma}^2}{H^2} ,
\label{epsilon}
\eeq
where the last expression follows upon using Eq.~(\ref{eombackground}). We define the end of inflation $t_{\rm end}$ via $\epsilon (t_{\rm end}) = 1$, which corresponds to $\ddot{a} (t_{\rm end}) = 0$, the end of accelerated expansion.

In addition to $\epsilon$, we consider a second slow-roll parameter
\beq
\eta \equiv 2 \epsilon - \frac{ \dot{\epsilon}}{2 H \epsilon}.
\label{etadef}
\eeq
Using Eqs.~(\ref{eombackground}) and (\ref{epsilon}) we see that, in general,
\beq
\frac{\dot{\epsilon}}{2 H \epsilon} = \frac{\ddot{\sigma}}{H \dot{\sigma}} + \epsilon .
\label{ddotsigmaeta}
\eeq
During ordinary slow-roll evolution $\vert \ddot{\sigma} \vert \ll \vert 3 H \dot{\sigma} \vert$, and the top line of Eq.~(\ref{eombackground}) becomes $3 H \dot{\sigma} \simeq - V_{, \sigma}$. Under those conditions $\eta \sim \epsilon < 1$. However, during so-called ultra-slow-roll, the potential becomes nearly flat, $V_{, \sigma} \simeq 0$, and hence the equation of motion for the background fields becomes $\ddot{\sigma} \simeq - 3 H \dot{\sigma}$. In that case, $\epsilon$ becomes exponentially smaller than 1 and 
\beq
\eta \rightarrow 3 \quad\quad (\textrm{ultra-slow-roll}). 
\label{etaUSR}
\eeq
Eq.~(\ref{etadef}) then yields $\dot{\epsilon} + 6 H \epsilon \simeq 0$. Given $H \simeq {\rm constant}$ during ultra-slow-roll evolution (consistent with $\epsilon \ll 1$), the kinetic energy density of the background fields $\rho_{\rm kin} = \dot{\sigma}^2 / 2 = M_{\rm pl}^2 H^2 \epsilon$ rapidly redshifts as $\rho_{\rm kin} (t) \sim a^{-6} (t)$ \cite{Kinney:2005vj,Martin:2012pe,Namjoo:2012aa,Romano:2015vxz,Germani:2017bcs,Dimopoulos:2017ged,Biagetti:2018pjj,Byrnes:2018txb,Pattison:2018bct,Carrilho:2019oqg,Inomata:2021tpx,Inomata:2021uqj,Pattison:2017mbe,Pattison:2019hef,Pattison:2021oen}.

The gauge-invariant Mukhanov-Sasaki variables $Q^I$ are constructed as linear combinations of metric perturbations and the field fluctuations, as in Eq.~(\ref{QIdef}). We may project the perturbations $Q^I$ into adiabatic ($Q_\sigma$) and isocurvature ($\delta s^I$) components \cite{Gordon:2000hv,Wands:2002bn,Bassett:2005xm,Kaiser:2012ak},
\beq
Q^I = \hat{\sigma}^I Q_\sigma + \delta s^I ,
\label{QIadiso}
\eeq
where
\beq
Q_\sigma \equiv \hat{\sigma}_J Q^J , \>\> \delta s^I \equiv \hat{s}^I_{\>\> J} Q^J .
\label{Qsigmadeltasdef}
\eeq
For two-field models, as we consider below, the isocurvature perturbations are characterized by a field-space scalar $Q_s$ defined via \cite{McDonough:2020gmn}
\beq
\delta s^J = \epsilon^{IJ} \hat{\sigma}_I Q_s ,
\label{deltasQs}
\eeq
where $\epsilon^{IJ} \equiv [ {\rm det} ({\cal G}_{IJ} ) ]^{-1/2} \, \bar{\epsilon}^{IJ}$ and $\bar{\epsilon}^{IJ}$ is the usual antisymmetric Levi-Civita symbol. The equations of motion for Fourier modes of comoving $k$, $Q_\sigma (k, t)$ and $Q_s (k, t)$, are given in Eqs.~(\ref{Qsigmaeom})--(\ref{Qseom}), from which it is clear that the adiabatic and isocurvature perturbations decouple for non-turning trajectories, for which $\vert \omega^I \vert = 0$. In addition, the amplitude of isocurvature perturbations will be suppressed as $Q_s (k, t) \sim a^{-3/2} (t)$ while $\mu_s^2 / H^2 \gg 1$, where the mass of the isocurvature perturbations, $\mu_s^2$, is given in Eq.~(\ref{mus}). Hence if $\omega^2 \ll H^2$ or $\mu_s^2 / H^2 \gg 1$, or both, there will be negligible transfer of power from the isocurvature to the adiabatic modes \cite{Gordon:2000hv,Wands:2002bn,Bassett:2005xm,Kaiser:2012ak,Kaiser:2013sna,Kaiser:2015usz,Schutz:2013fua,Langlois:2008mn,Peterson:2010np,Gong:2011uw,Gong:2016qmq,McDonough:2020gmn}.

The adiabatic perturbation is proportional to the gauge-invariant curvature perturbation \cite{Kaiser:2012ak}
\beq
{\cal R} = \frac{ H}{ \dot{\sigma}} Q_\sigma = \frac{ Q_\sigma}{M_{\rm pl} \sqrt{ 2 \epsilon} } ,
\label{Rdef}
\eeq
where the last equality 
follows from Eq.~(\ref{epsilon}). To avoid confusion, we adopt the convention of 
Ref.~\cite{Iacconi:2021ltm} and denote the curvature perturbation as ${\cal R}$ and the Ricci scalar of the field-space manifold as ${\cal R}_{\rm fs}$.

We may calculate the dimensionless power spectrum for the gauge-invariant scalar curvature perturbations ${\cal R}$, defined as
\beq
{\cal P}_{\cal R} (k) \equiv \frac{ k^3}{2 \pi^2} \vert {\cal R}_k (t_{\rm end} ) \vert^2 ,
\label{PRdef}
\eeq
where $t_{\rm end}$ indicates the end of inflation. As discussed in Ref.~\cite{Geller:2022nkr}, within the family of models we are considering, the fields generically evolve within a local minimum or ``valley" of the potential in the Einstein frame, and therefore the isocurvature modes remain heavy throughout the duration of inflation, $\mu_s^2 \gg H^2$. Likewise, the covariant turn rate remains small, $\vert \omega^I \vert \ll H$. (See also Refs.~\cite{Kaiser:2012ak,Kaiser:2013sna,Schutz:2013fua,Kaiser:2015usz,DeCross:2015uza}.) Under these conditions, when the background fields $\varphi^I (t)$ undergo ordinary slow-roll evolution with $\epsilon, \vert \eta \vert \ll 1$, the power spectrum assumes the form~\cite{Gordon:2000hv,Wands:2002bn,Bassett:2005xm,Kaiser:2012ak,Geller:2022nkr}
\begin{equation}
    {\cal P}_{\cal R}^{\rm SR} (k) = \frac{ H^2 (t_k) }{8 \pi^2 M_{\rm pl}^2 \epsilon (t_k) } \left( \frac{ k}{a (t_k) H (t_k)} \right)^{3 - 2 \nu_{\rm SR}} \big[ 1 + {\cal O} (\epsilon) \big] ,
    \label{eqn:PRHepsilon}
\end{equation}
with $\nu_{\rm SR} = \frac{3}{2} + 3 \epsilon - \eta$. As discussed in Appendix \ref{app:USR}, during ordinary slow-roll the modes ${\cal R}_k (t)$ remain frozen after crossing outside the Hubble radius, $\dot{\cal R}_k \simeq 0$ for $k \ll a H$, so one may evaluate ${\cal P}_{\cal R}^{\rm SR} (k)$ for ${\cal R}_k (t_k) \simeq {\cal R}_k (t_{\rm end})$, where $t_k$ is the time when a mode of comoving wavenumber $k$ first crossed outside the Hubble radius during inflation:
\begin{align}
    k = a(t_k) H (t_k).\label{crossout}
\end{align}

Inflationary dynamics that yield a brief phase of ultra-slow-roll evolution, during which $\epsilon (t_{\rm usr}) \ll 1$, will generate a spike in the power spectrum ${\cal P}_{\cal R} (k)$ on corresponding wavenumbers $k_{\rm usr}$. Such large-amplitude perturbations, in turn, can produce PBHs upon re-entering the Hubble radius after the end of inflation \cite{Garcia-Bellido:2017mdw,Ezquiaga:2017fvi,Germani:2017bcs,Kannike:2017bxn,Motohashi:2017kbs,Di:2017ndc,Ballesteros:2017fsr,Pattison:2017mbe,Passaglia:2018ixg,Biagetti:2018pjj,Byrnes:2018txb,Carrilho:2019oqg,Figueroa:2020jkf,Inomata:2021tpx,Inomata:2021uqj,Pattison:2021oen}. The main effect from ultra-slow-roll on the amplitude of the power spectrum is captured by the usual slow-roll expression in Eq.~(\ref{eqn:PRHepsilon}), given the relationship ${\cal P}_{\cal R}^{\rm SR} (k) \propto 1 / \epsilon$. Additional growth in ${\cal P}_{\cal R} (k)$ for certain wavenumbers $k$, beyond that represented by ${\cal P}_{\cal R}^{\rm SR} (k)$, can also occur during ultra-slow-roll. As I will discuss in Section \ref{sec:data}, we have performed about 2 million simulations of the dynamics of this family of models across a broad region of parameter space. In order for this to be computationally tractable, we used the expression of Eq.~(\ref{eqn:PRHepsilon}) in our Markov Chain Monte Carlo analysis, which depends only on background-order quantities, and hence can be evaluated for a given point in parameter space very efficiently. In Appendix \ref{app:USR}, we bound the discrepancy in ${\cal P}_{\cal R} (k)$ that can arise from the ultra-slow-roll phase in this family of models within the relevant regions of parameter space. 

For discussion of possible effects from loop corrections on the power spectrum, compare Refs.~\cite{Cheng:2021lif,Kristiano_yokoyama_1,Kristiano_yokoyama_2,sayantan_1,sayantan_2,Cheng:2023ikq} with Refs.~\cite{Senatore_2010,Senatore:2012nq,Pimentel:2012tw,Senatore:2012ya,Ando:2020fjm,RiottoPBH,Firouzjahi:2023aum,Motohashi:2023syh,Firouzjahi:2023ahg,Franciolini:2023lgy,Tasinato:2023ukp}.

In order for such an ultra-slow-roll phase to produce a large spike in ${\cal P}_{\cal R} (k)$, quantum fluctuations of the fields must not whisk the system past the region of the potential in which $V_{, \sigma} \simeq 0$ too quickly, or else inflation will end before significant amplification of ${\cal P}_{\cal R} (k)$ can occur \cite{Germani:2017bcs,Motohashi:2017kbs,Di:2017ndc,Ballesteros:2017fsr,Dimopoulos:2017ged,Pattison:2017mbe,Biagetti:2018pjj,Byrnes:2018txb,Pattison:2019hef,Carrilho:2019oqg,Passaglia:2018ixg,Pattison:2021oen,Inomata:2021uqj,Inomata:2021tpx}. Backreaction from quantum fluctuations 
yields a variance of the kinetic energy density for the system \cite{Inomata:2021tpx}
\beq
\langle (\Delta K )^2 \rangle \simeq \frac{ 3 H^4}{4 \pi^2} \rho_{\rm kin} ,
\label{varianceK}
\eeq
where $\rho_{\rm kin} = \dot{\sigma}^2 / 2$ is the background fields' unperturbed kinetic energy density. Classical evolution will dominate quantum diffusion during ultra-slow-roll evolution if $\rho_{\rm kin} > \sqrt{\langle (\Delta K )^2 \rangle}$. Upon using Eq.~(\ref{epsilon}), this criterion becomes
\beq
\epsilon_{\rm usr} > \frac{3}{ 4 \pi^2} \left( \frac{ H}{ M_{\rm pl} } \right)^2 .
\label{epsUSRdominate}
\eeq
Comparing with Eq.~(\ref{PRHepsilon}), we see that Eq.~(\ref{epsUSRdominate}) is equivalent to ${\cal P}_{\cal R} (k) < 1/6$ \cite{Inomata:2021tpx}. Within the regions of parameter space that we consider in Sections \ref{sec:trajectories} and \ref{sec:PBHsUSR}, the criterion of Eq.~(\ref{epsUSRdominate}) is always satisfied, such that during ultra-slow-roll, classical evolution of the background fields continues to dominate over quantum diffusion, allowing for a robust amplification of curvature perturbations. 

In the absence of a transfer of power from isocurvature to adiabatic perturbations, predictions for observables relevant to the cosmic microwave background radiation (CMB) revert to the familiar and effectively single-field forms \cite{Kaiser:2012ak,Kaiser:2013sna}. Explicit expressions for the spectral index $n_s (k_*)$, the running of the spectral index $\alpha (k_*) \equiv (d n_s (k_*) / d {\rm ln} k)\vert_{k_*}$, and the tensor-to-scalar ratio $r (k_*)$ may be found in Eqs.~(\ref{nsdef})--(\ref{rTtoS}); here $k_* = 0.05 \, {\rm Mpc}^{-1}$ is the comoving CMB pivot scale. Likewise, inherently multifield features, such as the fraction of primordial isocurvature perturbations $\beta_{\rm iso} (k_*, t_{\rm end})$, which is defined in Eq.~(\ref{betaisodef}), and primordial non-Gaussianity $f_{\rm NL}$, defined in Eq.~(\ref{fNLSFA2}), generically remain small for multifield models in which the isocurvature modes remain heavy throughout inflation ($\mu_s^2 \gg H^2$) and the turn-rate remains negligible ($\omega^2 \ll H^2$) \cite{Kaiser:2012ak,Kaiser:2013sna,Schutz:2013fua,Kaiser:2015usz,Gordon:2000hv,Langlois:2008mn,DiMarco:2002eb,Bernardeau:2002jy,Seery:2005gb,Yokoyama:2007dw,Byrnes:2008wi,Peterson:2010mv,Chen:2010xka,Byrnes:2010em,Gong:2011cd,Elliston:2011dr,Elliston:2012ab,Seery:2012vj,Mazumdar:2012jj,Wands:2002bn,Bassett:2005xm,Peterson:2010np,Gong:2011uw,Gong:2016qmq,McDonough:2020gmn}.

\subsubsection{Supersymmetric Two-Field Models}
\label{sec:susymodels}

For the remainder of this chapter we consider supersymmetric two-field models, in which  supersymmetry is spontaneously broken. These models naturally arise in both global supersymmetry and supergravity.
Although our framework does not depend strongly on supersymmetric motivations, the supersymmetric framework provides a codex for translating a relatively large number of effective field theory parameters to a much smaller set of parameters that govern the UV completion in supergravity, which is valid at least at tree-level. The desired nonminimal couplings can then be realized in a manifestly supersymmetric manner, e.g., as in Refs.~\cite{Kallosh:2010ug,Kallosh:2013hoa}, in the superconformal approach to supergravity \cite{Freedman:2012zz}, or else generated via quantum effects once supersymmetry has been spontaneously broken. Here we provide a brief overview. Additional details may be found in Appendix \ref{appSUGRA} and Ref.~\cite{Freedman:2012zz}.

As mentioned, at the energy scales relevant for inflation, the construction yields specific arrangements among various dimensionless coupling constants, but the field operators that appear in the action include only generic dimension-4 operators that should be included in {\it any} self-consistent effective field theory for two interacting scalar fields in $(3 + 1)$ spacetime dimensions. This sort of SUSY pattern imprinted on low-energy physics has been discussed 
in the context of CMB non-Gaussianity from supersymmetric higher-spin fields \cite{Alexander:2019vtb}. 

We focus on inflation models that may be realized in the global supersymmetry limit of supergravity. The model is specified by a K\"{a}hler potential $\tilde{K}$ and superpotential $\tilde{W}$ in the Jordan frame, given by
\begin{equation}
    \tilde{K} (\Phi, \bar{\Phi} ) = - \frac{1}{2} \displaystyle \sum _{I=1} ^2 ( \Phi^I - \bar{\Phi}^{\bar{I}})^2
    \label{KahlerJordan1}
\end{equation}
and
\begin{equation}
    \tilde{W} (\Phi) =  \sqrt{2} \, \mu b_{IJ} \Phi^I \Phi^J + 2 c_{IJK} \Phi^I \Phi^J \Phi^K ,
    \label{Wtilde1}
\end{equation}
with indices $I, J, K \in \{ 1, 2 \}$. We select $\tilde{K}$ so as to provide canonical kinetic terms for the real and imaginary components of the scalar fields $\varpi^I$ associated with each chiral superfield $\Phi^I$ (as further discussed in Appendix \ref{appSUGRA}), and insert factors of $\sqrt{2}$ and $2$ in the superpotential $\tilde{W}$ to reduce clutter in the resulting equations. The coefficients $b_{IJ}$ and $c_{IJK}$ in $\tilde{W}$ are real-valued dimensionless coefficients, and repeated indices are trivially summed over. We omit possible constant and linear contributions to $\tilde{W}$, since non-renormalization of $\tilde{W}$ \cite{Grisaru:1979wc,Seiberg:1993vc} provides the freedom to do so. Expanding Eq.~(\ref{Wtilde1}), we may express $\tilde{W}$ as
\begin{equation}
    \begin{split}
    \tilde{W} &= \sqrt{2} \, b_1 \mu (\Phi_1 )^2 + \sqrt{2}  \, b_2 \mu (\Phi_2 )^2 +  2 c_1 (\Phi_1)^3  \\
   & \quad + 2 c_2 (\Phi_1)^2 \Phi_2 + 2 c_3 \Phi_1 (\Phi_2)^2 + 2 c_4 (\Phi_2 )^3 ,
    \label{Wtilde}
\end{split}
\end{equation}
where we have defined $b_1 \equiv b_{11}$, $b_2 \equiv b_{22}$, $c_1 \equiv c_{111}$, $c_2 \equiv (c_{112}+c_{121}+c_{211})$, $c_3 \equiv (c_{122}+c_{212}+c_{221})$, and $c_4 \equiv c_{222}$. We set the coupling $b_{12}$ for the quadratic cross-term $\mu \Phi_1 \Phi_2$ to zero for simplicity but without loss of generality, since this choice merely amounts to a choice of coordinates on field space.

The K\"{a}hler potential and superpotential together determine the scalar potential as
\begin{equation}
    \tilde{V} = e^{ \tilde{K}/M_{\rm pl}^2}\left( |D \tilde{W}|^2 - 3M_{\rm pl}^{-2}|\tilde{W}|^2  \right) ,
    \label{VtildeSUSY}
\end{equation}
where $D_I \equiv \partial_I + M_{\rm pl}^{-2} \tilde{K}_{, I}$ denotes a K\"{a}hler covariant derivative \cite{Freedman:2012zz}. The explicit tilde on $V$ indicates that the chiral superfields $\Phi^I$ are assumed to be nonminimally coupled to gravity, either through a manifestly supersymmetric setup or through quantum effects below the SUSY breaking scale, making the  expression for $\tilde{V}$ in Eq.~(\ref{VtildeSUSY}) the Jordan-frame potential. 

The choice of K\"{a}hler potential in Eq.~(\ref{KahlerJordan1}) guarantees that the imaginary parts of the scalar components of $\Phi^I$ are heavy during inflation, $m^2_{\psi} > H^2$, where $\Phi^I = \varpi^I + ...$ for complex scalar fields $\varpi^I$, and $\varpi^I = ( \phi^I + i \psi^I) / \sqrt{2}$, with $\phi^I$ and $\psi^I$ real-valued scalar fields. In the global supersymmetry limit ($\vert \Phi^I \vert^2 / M^2_{\rm pl} \ll 1$), the scalar potential can then be expressed as simply
\beq
\tilde{V} (\phi, \chi) \simeq \sum_I \left\vert \frac{ \partial W }{\partial \Phi^I } \right\vert^2_{\Phi^I \rightarrow \varpi^I} ,
\label{Vtildedef}
\eeq
where we label the real-valued scalar components of the chiral superfields as $\Phi_1 = \phi/\sqrt{2}$ and $\Phi_2 = \chi/\sqrt{2}$. We discuss additional details of the embedding in supergravity in Appendix \ref{appSUGRA}.

\subsubsection{The Einstein-Frame Scalar Potential}

The full form of $\tilde{V} (\phi, \chi)$ appears in Appendix \ref{appSUGRA}. For our two-field models, it is convenient to adopt polar coordinates for the field-space manifold,
\beq
\phi (t) = r (t) \, \cos \theta (t) , \>\> \chi (t) = r (t) \, \sin \theta (t) ,
\label{phichirtheta}
\eeq
with $r \geq 0$ and $0 \leq \theta \leq 2 \pi$. Then the Jordan-frame scalar potential of Eq.~(\ref{Vtildedef}) takes the form
\beq
\tilde{V} (r, \theta) = {\cal B} (\theta) \mu^2 r^2 + {\cal C} (\theta) \mu r^3 + {\cal D} (\theta) r^4
\label{Vtildertheta}
\eeq
with
\beq
\begin{split}
    {\cal B} (\theta) &\equiv 4 b_1^2 \cos^2 \theta + 4 b_2^2 \sin^2 \theta , \\
    {\cal C} (\theta) &\equiv 12 b_1 c_1 \cos^3 \theta + 4  (2b_1 + b_2) c_2 \cos^2 \theta \sin \theta \\
    &\quad + 4 (b_1 + 2 b_2) c_3 \cos \theta \sin^2 \theta + 12 b_2 c_4 \sin^3 \theta , \\
    {\cal D} (\theta) &\equiv (9 c_1^2 + c_2^2) \cos^4 \theta + 4 c_2 (3 c_1 + c_3) \cos^3 \theta \sin\theta \\
    &\quad + (4 c_2^2 + 6 c_1 c_3 + 6 c_2 c_4 + 4 c_3^2) \cos^2 \theta \sin^2\theta \\
    &\quad+ 4 c_3 (c_2 + 3 c_4) \cos \theta \sin^3 \theta + (9 c_4^2 + c_3^2) \sin^4 \theta .
\end{split}
\label{BCDdef}
\eeq
As mentioned, we consider this scalar potential in conjunction with nonminimal couplings to gravity. In a curved spacetime, scalar fields' self-interactions will generate nonminimal couplings of the form \cite{Callan:1970ze,Bunch:1980br,Bunch:1980bs,Birrell:1982ix,Odintsov:1990mt,Buchbinder:1992rb,Faraoni:2000gx,Parker:2009uva,Markkanen:2013nwa,Kaiser:2015usz}
\beq
\begin{split}
f (\phi, \chi) &= \frac{1}{2} \left[ M_{\rm pl}^2 + \xi_\phi \phi^2 + \xi_\chi \chi^2 \right] \\
&= \frac{1}{2} \left[ M_{\rm pl}^2 + r^2 \left( \xi_\phi \cos^2 \theta + \xi_\chi \sin^2 \theta \right) \right] .
\end{split}
\label{frtheta}
\eeq
Hence the action for the scalar degrees of freedom of our models takes the form of Eq.~(\ref{SJ}), with $\tilde{V} (\phi^I)$ given by Eq.~(\ref{Vtildertheta}) and $f (\phi^I)$ by Eq.~(\ref{frtheta}).

\begin{figure*}[t!]
    \centering
    \includegraphics[width=0.43\textwidth]{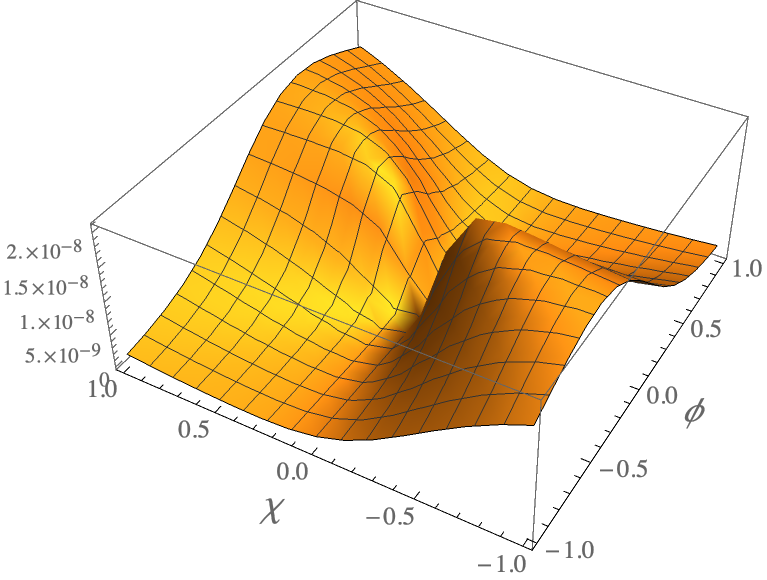} $\quad$ \includegraphics[width=0.43\textwidth]{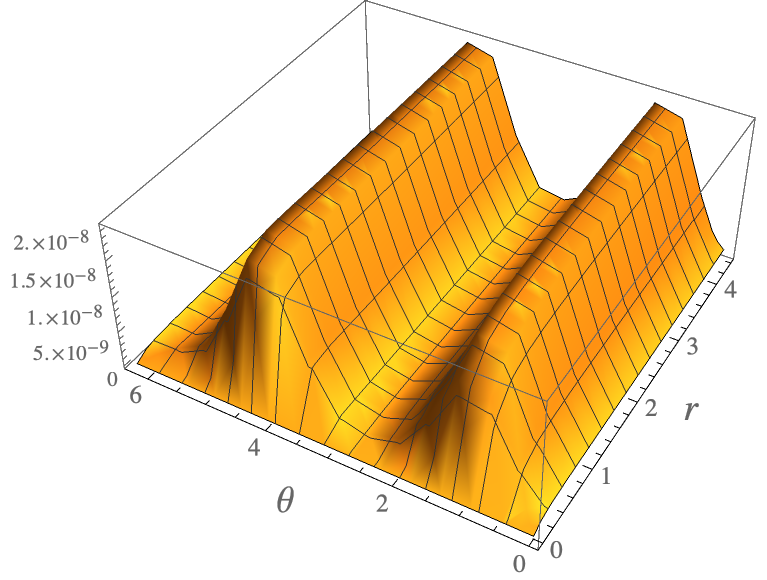}
    \caption{The scalar potential in the Einstein frame, in both $\{ \phi, \chi\}$ ({\it left}) and $\{ r, \theta \}$ ({\it right}) coordinates. Fields are shown in units of $M_{\rm pl}$. The parameters are $\mu = M_{\rm pl}$, $b_1 = b_2 = -1.8 \times 10^{-4}$, $c_1 = 2.5 \times 10^{-4}$, $c_2 = c_3 = 3.57 \times 10^{-3}$, $c_4 = 3.9 \times 10^{-3}$, and $\xi_\phi = \xi_\chi = 100$. 
    }
    \label{fig:VE}
\end{figure*}

Upon transforming to the Einstein frame, the field-space metric ${\cal G}_{IJ}$ in our $\{ r, \theta \}$ coordinates has components
\beq
\begin{split}
    {\cal G}_{rr} &= \frac{ M_{\rm pl}^2}{2f} \left[ 1 + \frac{3r^2}{f} \left( \xi_\phi \cos^2 \theta + \xi_\chi \sin^2 \theta \right)^2 \right] , \\
    {\cal G}_{r\theta} &= \frac{ M_{\rm pl}^2}{2f} \left( \frac{ 3 r^3}{f} \right) \Big[ \left( \xi_\phi \cos^2\theta + \xi_\chi \sin^2 \theta \right) \left( - \xi_\phi + \xi_\chi \right) \cos \theta \sin \theta \Big] , \\
    {\cal G}_{\theta \theta} &= \frac{ M_{\rm pl}^2}{2f} \left[ r^2 + \frac{ 3 r^4}{f} \left( - \xi_\phi + \xi_\chi \right)^2 \cos^2 \theta \sin^2 \theta \right] ,
\label{GIJcomponents}
\end{split}
\eeq
with $f (r, \theta)$ given in Eq.~(\ref{frtheta}). The potential in the Einstein frame becomes
\beq
V (r, \theta) = \frac{ M_{\rm pl}^4}{[ 2f (r, \theta) ]^2} \left[ {\cal B} (\theta) \mu^2 r^2 + {\cal C} (\theta) \mu r^3 + {\cal D} (\theta) r^4 \right] ,
\label{VErtheta}
\eeq
with the coefficients ${\cal B}, {\cal C}$, and ${\cal D}$ given in Eq.~(\ref{BCDdef}). 

The form of $V (\phi^I)$ in Eq.~(\ref{VErtheta}) has a similar structure to the single-field potential studied in Ref.~\cite{Garcia-Bellido:2017mdw}, which included both a cubic self-interaction term and the conformal factor $(M_{\rm pl}^2 + \xi \phi^2)^2$ in the denominator. The potential in Eq.~(\ref{VErtheta}) is also a natural generalization of the two-field models studied in Refs.~\cite{Kaiser:2012ak,Kaiser:2013sna,Schutz:2013fua,Kaiser:2015usz}, for which the numerator included only the term proportional to ${\cal D} (\theta)$. Much as in those multifield studies, the Einstein-frame potential of Eq.~(\ref{VErtheta}) includes local maxima and local minima (or ``ridges" and ``valleys") throughout the field space. See Fig.~\ref{fig:VE}. As we describe in Section \ref{sec:trajectories}, this structure of the potential yields strong single-field attractor behavior \cite{Kaiser:2012ak,Kaiser:2013sna,Schutz:2013fua,Kaiser:2015usz,DeCross:2015uza}: the system generically settles into a local minimum of the potential very quickly after the start of inflation and remains within that minimum for the duration of inflation.

Potentials of the form in Eq.~(\ref{VErtheta}) have very flat plateaus at large field values, of the type favored by recent measurements of CMB anisotropies \cite{Planck:2018jri}. For models in which $\xi_\phi \simeq \xi_\chi$, in the limit in which the ${\cal D} (\theta) r^4$ term dominates the numerator of $V (r, \theta)$ and $\xi_\phi r^2 \gg M_{\rm pl}^2$, the potential reduces to the simple form
\beq
V (r, \theta) \simeq \frac{ M_{\rm pl}^4 \, {\cal D} (\theta) }{\xi_\phi^2} + {\cal O} \left( \frac{ M_{\rm pl}^2}{\xi_\phi r^2} \right) .
\label{Vplateau}
\eeq
In the absence of strong turning among the background fields during inflation ($\omega^2 \ll H^2$), the upper bound on the primordial tensor-to-scalar ratio $r_{0.05} < 0.036$ at the CMB pivot scale $k_* = 0.05 \, {\rm Mpc}^{-1}$ \cite{BICEP:2021xfz} constrains $H (t_*) < 1.9 \times 10^{-5} \, M_{\rm pl}$. This constraint on $H (t_*)$ becomes more complicated for inflationary trajectories that feature strong turning before the end of inflation \cite{McDonough:2020gmn}, but is appropriate for the scenarios we consider here. Assuming that the CMB-relevant curvature perturbations crossed outside the Hubble radius while the fields were still on the large-field plateau of the potential, the constraint on $H(t_*)$ corresponds to the limit
\beq
\frac{ {\cal D} (\theta) }{\xi_\phi^2} \leq 1.1 \times 10^{-9} ,
\label{DxiCMB}
\eeq
upon relating $H$ to $V$ during slow roll. From Eq.~(\ref{BCDdef}) we see that ${\cal D} (\theta) \sim 9 c_{\rm max}^2$, where $c_{\rm max} = {\rm max} \{ c_i \}$. Hence to remain compatible with observations of the CMB, we expect the couplings to fall within a range such that
\beq
\frac{ \vert c_{\rm max} \vert}{\xi_\phi} \lesssim {\cal O} (10^{-5}) .
\label{cmax}
\eeq
As $\xi_\phi \simeq \xi_\chi$ becomes larger, the dimensionless couplings $c_i$ can likewise become larger while still remaining compatible with observations.

The Einstein-frame potential $V (r, \theta)$ of Eq.~(\ref{VErtheta}) retains the large-field plateau as in the models studied in Refs.~\cite{Kaiser:2012ak,Kaiser:2013sna,Schutz:2013fua,Kaiser:2015usz}. On the other hand, the potential of Eq.~(\ref{VErtheta}) includes modified {\it small-field} structure compared to the previous models. In particular, the coefficients ${\cal B} (\theta)$ and ${\cal C} (\theta)$ remain nonzero when at least one of the dimensionless couplings $b_i \neq 0$. These changes to the small-field structure of the potential can yield a phase of ultra-slow-roll evolution near the end of inflation, which in turn can produce PBHs.

\subsubsection{Inflationary Trajectories}
\label{sec:trajectories}

\begin{figure}[h]
    \centering
    \includegraphics[width=0.43\textwidth]{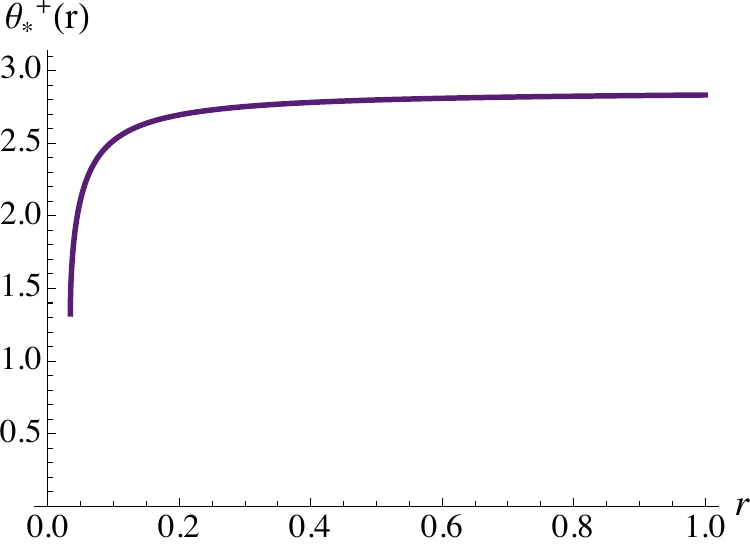}
    \caption{The angle in field space $\theta_* (r)$ along which the system evolves for the same couplings as in Fig.~\ref{fig:VE}. For this set of parameters, the local minimum of the potential lies along $\theta_*^+ (r)$, whereas $\theta_*^- (r)$ is a local maximum.}
    \label{fig:thetastar}
\end{figure}

If the dimensionless couplings that appear in Eqs.~(\ref{BCDdef})--(\ref{VErtheta}) obey additional symmetries, namely
\beq
\xi_\phi = \xi_\chi = \xi, \>\> b_1 = b_2 = b, \>\>  c_2 = c_3 ,
\label{bcxisymmetries}
\eeq
then we may find exact analytic solutions for the background fields' trajectory during inflation. In particular, if the couplings obey the relationships of Eq.~(\ref{bcxisymmetries}), then we find
\beq
V_{, \theta} (r, \theta) = \frac{ M_{\rm pl}^4 r^3}{[2 f (r) ]^2} \left[ {\cal C}' (\theta) \mu  + {\cal D}' (\theta) r \right]
\label{Vprime}
\eeq
because $f (r, \theta) \rightarrow f (r)$ and ${\cal B} (\theta) \rightarrow 4 b^2$ when $\xi_\phi = \xi_\chi$ and $b_1 = b_2 = b$. The system will evolve along a direction in field space $\theta_*$ such that $V_{, \theta} (r, \theta_*) = 0$. As shown in Appendix \ref{appTheta}, for the symmetric couplings of Eq.~(\ref{bcxisymmetries}) the extrema are given by
\beq
\theta_*^\pm (r) = {\rm arccos} (x^\pm (r))
\label{thetastar1}
\eeq
with
\beq
x^\pm (r) = \frac{ -d_1 \pm \vert d_4 \vert \sqrt{ - 1 + R^2} }{R \sqrt{ d_1^2 + d_4^2}} ,
\label{xpm}
\eeq
where
\beq
\begin{split}
    d_1 &\equiv c_1 + \frac{ c_2}{3} , \>\> d_4 \equiv c_4 + \frac{ c_2}{3} , \\
    r_{\rm imag} &\equiv \frac{ b\mu}{\sqrt{d_1^2+ d_4^2}} , \>\> R \equiv \frac{ r}{r_{\rm imag}} .
\end{split}
\label{didef}
\eeq
In the limit $b \rightarrow 0$, $x^\pm (r) \rightarrow {\rm constant}$ and hence $\dot{\theta}^\pm_* \rightarrow 0$, consistent with the non-turning attractor trajectories identified in Refs.~\cite{Kaiser:2012ak,Kaiser:2013sna,Schutz:2013fua,Kaiser:2015usz}. For $b \neq 0$, the trajectories $\theta_*^\pm (r)$ show virtually no turning until $r \ll M_{\rm pl}$, near the end of inflation. See Fig.~\ref{fig:thetastar}. The analytic solutions $\theta_*^\pm (r)$ become complex for $r < \vert r_{\rm imag} \vert$, although the fields' dynamical evolution remains smooth in the vicinity of $r \sim \vert r_{\rm imag} \vert$.

\begin{figure*}
    \centering
    \includegraphics[width=0.45\textwidth]{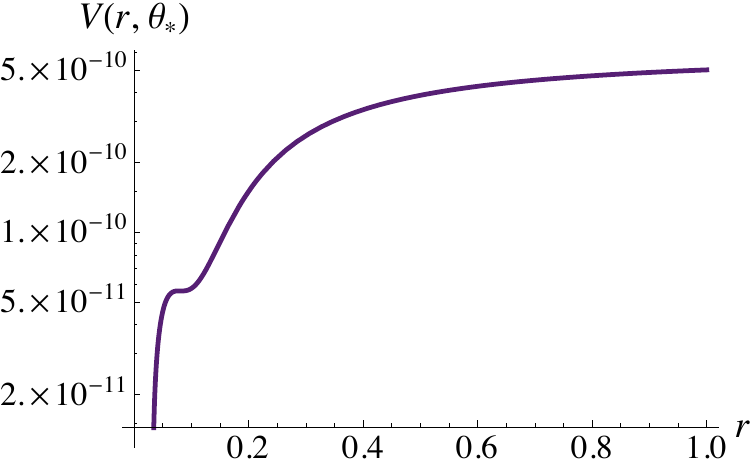} $\quad$
    \includegraphics[width=0.45\textwidth]{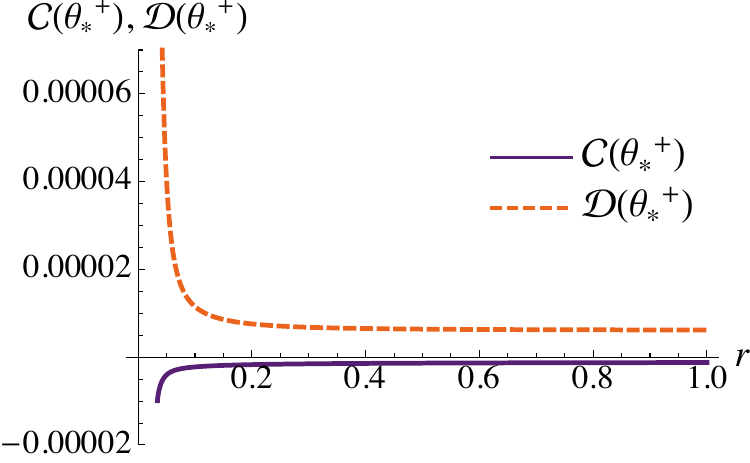}
    \caption{({\it Left}) The scalar potential in the Einstein frame $V (r, \theta_*^+)$ (in units of $M_{\rm pl}^4$) evaluated along the direction of the fields' evolution, $\theta_*^+ (r)$. ({\it Right}) The dimensionless coefficients ${\cal C} (\theta)$ (purple) and ${\cal D} (\theta)$ (orange dashed) as defined in Eq.~(\ref{BCDdef}), evaluated along the direction of the fields' evolution, $\theta_*^+ (r)$. Both plots use the same parameters as in Fig.~\ref{fig:VE}.   }
    \label{fig:VCD}
\end{figure*}

We may project the multifield potential $V (r, \theta)$ along the fields' trajectory $\theta_* (r)$, which yields $V (r, \theta_* (r))$. See Fig.~\ref{fig:VCD}. Upon including $b \neq 0$, and hence ${\cal C} \neq 0$, the potential evaluated along $\theta_* (r)$ generically develops a feature at small field values, much as in the single-field models studied in Refs.~\cite{Garcia-Bellido:2017mdw,Ezquiaga:2017fvi,Germani:2017bcs,Kannike:2017bxn}. For the example shown, the dimensionless coefficient ${\cal C} (\theta_*) < 0$ for the duration of inflation, while ${\cal B}, {\cal D} (\theta_*) > 0$ (recall that for $b_1 = b_2 = b$, ${\cal B} = 4 b^2$ is independent of $\theta$). Given the opposite signs of ${\cal C}$ and ${\cal B, D}$, the new features will emerge in $V (r, \theta_* (r))$ for field values $r$ such that $\vert {\cal C} (\theta_*)\vert \mu r \sim {\cal B} \mu^2 + {\cal D} (\theta_*) r^2$. For the parameters shown in Figs.~\ref{fig:VE}--\ref{fig:VCD}, this occurs for $r \simeq 0.1 \, \mu$.

\begin{figure*}
    \centering
    \includegraphics[width=0.48\textwidth]{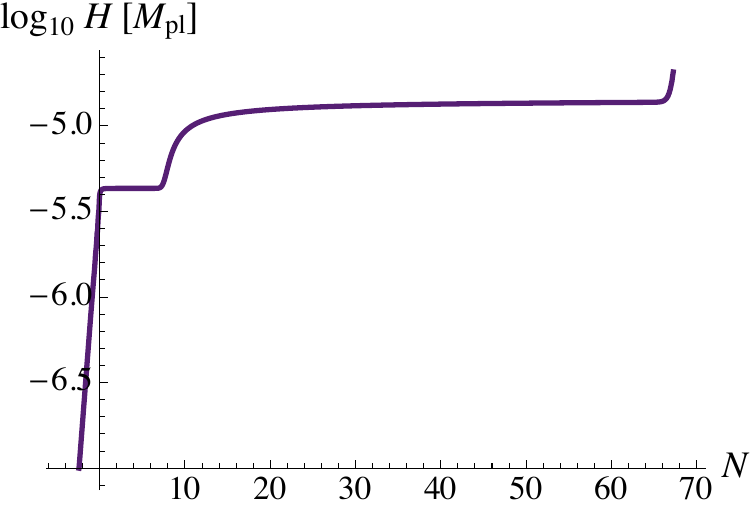} $\quad$
    \includegraphics[width=0.45\textwidth]{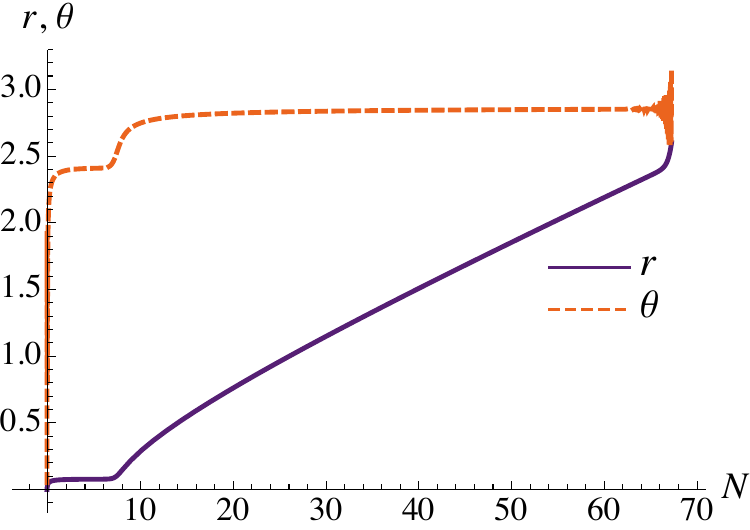}
    \caption{({\it Left}) The evolution of the Hubble parameter $H (t)$ as a function of efolds $N$ before the end of inflation ($N (t_{\rm end}) = 0$). ({\it Right}) The evolution of the fields $r (t)$ (purple, in units of $M_{\rm pl}$) and $\theta (t)$ (orange dashed) as a function of efolds $N$ before the end of inflation. Both plots use the same parameters as in Fig.~\ref{fig:VE} and initial conditions $r (t_i) = 2.6 \, M_{\rm pl}$, $\theta (t_i) = \pi - 0.02$, $\dot{r} (t_i) = - 10^{-5} \, M_{\rm pl}^2$, and $\dot{\theta} (t_i) = 4 \times 10^{-5} \, M_{\rm pl}$.  }
    \label{fig:Hfields}
\end{figure*}

\begin{figure*}
    \centering
    \includegraphics[width=0.48\textwidth]{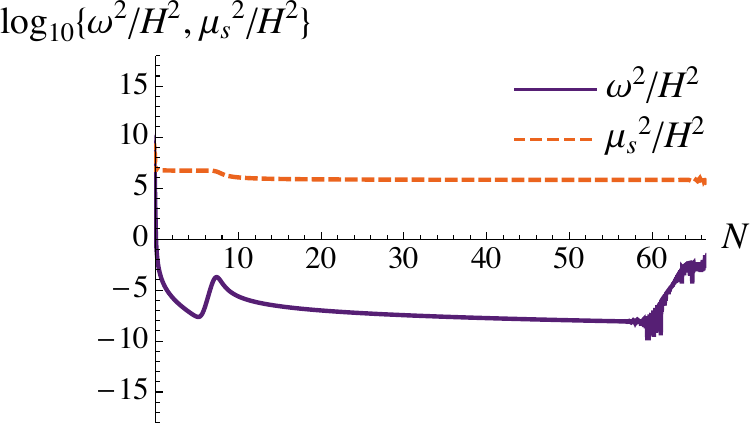} $\quad$ 
    \includegraphics[width=0.45\textwidth]{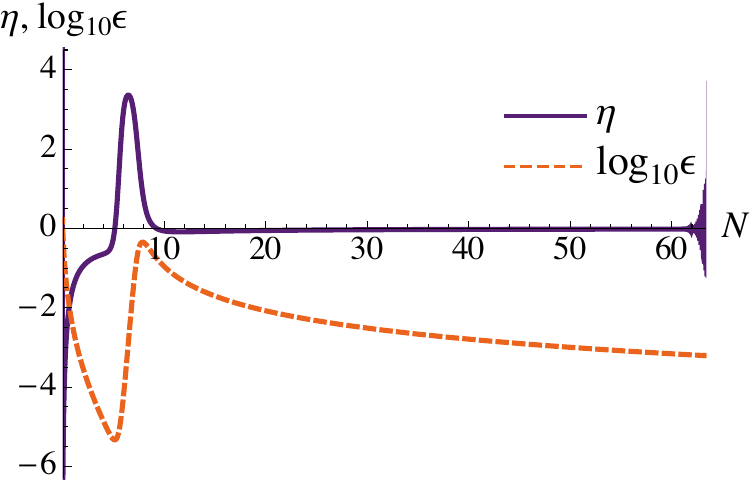}
    \caption{({\it Left}) The evolution of the covariant turn-rate $\vert \omega^I (t) \vert$ (purple) and the mass of the isocurvature modes $\mu_s (t)$ (orange dashed) as a function of efolds $N$ before the end of inflation ($N (t_{\rm end}) = 0$).  ({\it Right}) The slow-roll parameters $\eta$ (purple) and $\epsilon$ (orange dashed) as functions of efolds $N$ before the end of inflation. While the system undergoes ultra-slow-roll evolution, $\eta \rightarrow 3$ and $\epsilon \rightarrow 10^{-5}$, consistent with Eq.~(\ref{etaUSR}). Both plots use the same parameters and initial conditions as in Fig.~\ref{fig:Hfields}. 
}
    \label{fig:muomegaPR}
\end{figure*}

With fine-tuning of at least one of the couplings $\{ b_, c_i \}$, one may arrange for the small-field feature to be a quasi-inflection point, as in Refs.~\cite{Garcia-Bellido:2017mdw,Ballesteros:2017fsr,Di:2017ndc,Motohashi:2017kbs}. More generally, the projected potential will develop a local minimum along the direction $\theta_* (r)$ with a nearby local maximum, as in Ref.~\cite{Kannike:2017bxn}. When the fields encounter this small-field feature in the potential, the system enters a phase of ultra-slow-roll evolution: the fields' kinetic energy density $\rho_{\rm kin} = \dot{\sigma}^2 / 2 \rightarrow 0$ while $H \simeq {\rm constant}$, and hence $\epsilon$ falls by several orders of magnitude, given the relationship in Eq.~(\ref{epsilon}). 

We numerically solve the coupled equations of motion for the background fields $r (t), \theta (t)$ and the Hubble parameter $H (t)$ using Eqs.~(\ref{eomvarphi}) and (\ref{eombackground}). In Fig.~\ref{fig:Hfields} we plot the evolution of $H, r$ and $\theta $ for typical values of the couplings. Fig.~\ref{fig:muomegaPR} confirms that once the system settles into a local minimum of the potential in the angular direction ($V_{, \theta} (r, \theta_* (r)) = 0$), the isocurvature modes remain heavy for the duration of inflation ($\mu_s^2 \gg H^2$) and the turn-rate remains negligible ($\omega^2 \ll H^2$). When the fields encounter the small-field feature in the potential near $r \simeq 0.1 \, \mu$, the system enters a phase of ultra-slow-roll evolution, with $\eta \rightarrow 3$ and $\epsilon \rightarrow 10^{-5}$. For each of these plots, we show the evolution of the system as a function of the number of efolds $N$ before the end of inflation: $N (t) \equiv N_{\rm total} - \int_{t_i}^{t} H (t) \, dt$, where $N_{\rm total} \equiv \int_{t_i}^{t_{\rm end}} H (t) \, dt$ and $t_{\rm end}$ is determined via $\epsilon (t_{\rm end}) = 1$.

\begin{figure}[t]
    \centering
    \includegraphics[width=0.45\textwidth]{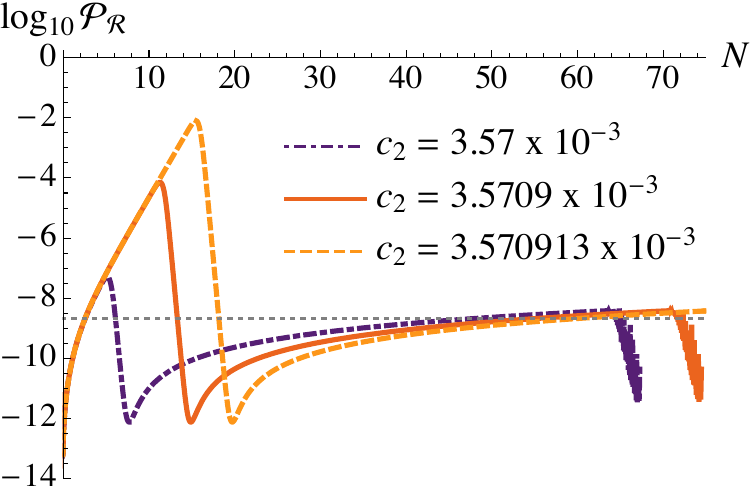}
    \caption{Fine-tuning one of the dimensionless couplings can increase the duration of the ultra-slow-roll phase. For longer periods of ultra-slow-roll, the slow-roll parameter $\epsilon$ falls to smaller values and the peak in the power spectrum ${\cal P}_{\cal R} (k)$ rises. All three curves shown here use the same parameters and initial conditions as in Figs.~\ref{fig:VE}--\ref{fig:muomegaPR}, with increasing fine-tuning of $c_2 = c_3$. The horizontal dotted line shows the COBE normalization ${\cal P}_{\cal R} (k_*) = 2.1 \times 10^{-9}$ for the CMB pivot-scale $k_* = 0.05 \, {\rm Mpc}^{-1}$.}
    \label{fig:PRtuning}
\end{figure}

Given the relationship between ${\cal P}_{\cal R} (k)$, $H$, and $\epsilon$ in Eq.~(\ref{PRHepsilon}), the power spectrum of curvature perturbations will become amplified for modes $k$ that exit the Hubble radius while the fields are in the phase of ultra-slow-roll. In general, the decrease in $\epsilon$---and hence the increase in ${\cal P}_{\cal R} (k)$---depends on the ratios of various couplings. For the parameters shown in Figs.~\ref{fig:VCD}--\ref{fig:muomegaPR}, the local maximum of the potential near $r \simeq 0.1 \, \mu$ is marginally greater than the value of the potential at the nearby local minimum, so the system spends only $\Delta N \sim 2.5$ efolds in the ultra-slow-roll phase. As shown in Fig.~\ref{fig:PRtuning}, by fine-tuning one of the dimensionless couplings, we may adjust the relative heights of the local maximum and local minimum along $\theta_* (r)$, thereby prolonging the duration over which the fields persist in the ultra-slow-roll phase and increasing the peak value of ${\cal P}_{\cal R} (k)$. Even the tallest peak of ${\cal P}_{\cal R} (k)$ shown in Fig.~\ref{fig:PRtuning} satisfies ${\cal P}_{\cal R} (k) \lesssim 10^{-2} < 1/6$, and hence the criterion of Eq.~(\ref{epsUSRdominate}) is always satisfied. In other words, even while the system undergoes ultra-slow-roll evolution, the classical evolution of the background fields dominates quantum diffusion for the  parameters considered here. 

The dynamics of the fields in the models we consider here are distinct from those recently studied in $\alpha$-attractor models \cite{Iacconi:2021ltm,Kallosh:2022vha}. In particular, we only consider positive values of the nonminimal couplings in this work, so that the conformal transformation associated with the factor $\Omega^2 (x)$ in Eq.~(\ref{Omega}) remains nonsingular. For $\xi_I > 0$, the induced field-space manifold in the Einstein frame has positive curvature, ${\cal R}_{\rm fs} > 0$, the magnitude of which falls in the limit $\xi_\phi r^2 \gg M_{\rm pl}^2$. (An explicit expression for ${\cal R}_{\rm fs}$ for these models may be found in Eq.~(115) of Ref.~\cite{Kaiser:2012ak}.) Hence curved field-space effects make fairly modest contributions to the fields' dynamics during the early stages of inflation \cite{Kaiser:2012ak,Kaiser:2013sna,Schutz:2013fua,DeCross:2015uza}. 

In $\alpha$-attractor models, on the other hand, the curvature of the field-space manifold is negative and constant, ${\cal R}_{\rm fs} = - 4 / (3 \alpha)$, with dimensionless constant $\alpha >0$. For $\alpha \sim {\cal O} (1)$, the fields' evolution will be affected by the nontrivial field-space manifold throughout the duration of inflation. Hence in $\alpha$-attractor models, the fields may ``ride the ridge," remaining on or near a local maximum of the potential for much of the duration of inflation \cite{Iacconi:2021ltm}, whereas in the family of models we consider here, the fields generically settle into a local minimum of the potential after a brief, initial transient. For the case of $\xi_I > 0$, the fields can only ``ride the ridge" of the potential for $N \gtrsim {\cal O} (1)$ efolds if the fields' initial conditions are exponentially fine-tuned \cite{Kaiser:2012ak,Schutz:2013fua,Kaiser:2013sna,DeCross:2015uza}. The fact that the fields generically settle into a local minimum of the potential in these models ensures that the isocurvature modes remain heavy throughout inflation and that the covariant turn-rate remains negligible.

\subsubsection{Scaling Relationships}
\label{sec:scaling}

As shown in Fig.~\ref{fig:PRtuning}, the evolution of perturbations is sensitive to the small-field feature in the Einstein-frame potential, which in turn depends upon ratios among the dimensionless couplings $b_i$ and $c_i$. We explore some of those relationships in this section. We first note from Eqs.~(\ref{BCDdef}) and (\ref{VErtheta}) that the mass-scale $\mu$ only appears in $V (\phi^I)$ multiplied by the $b_i$. Without loss of generality, we therefore fix $\mu = M_{\rm pl}$ and adjust the magnitude of the scalar fields' tree-level masses by changing $b_i$.

\begin{figure*}[t]
    \centering
    \includegraphics[width=0.45\textwidth]{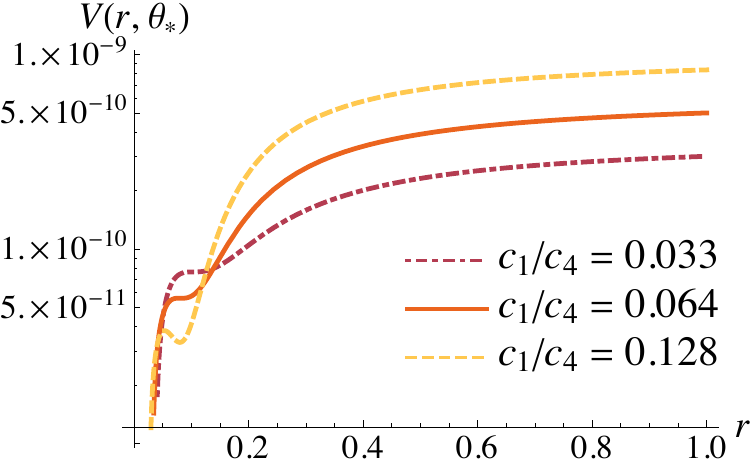} $\quad$
    \includegraphics[width=0.45\textwidth]{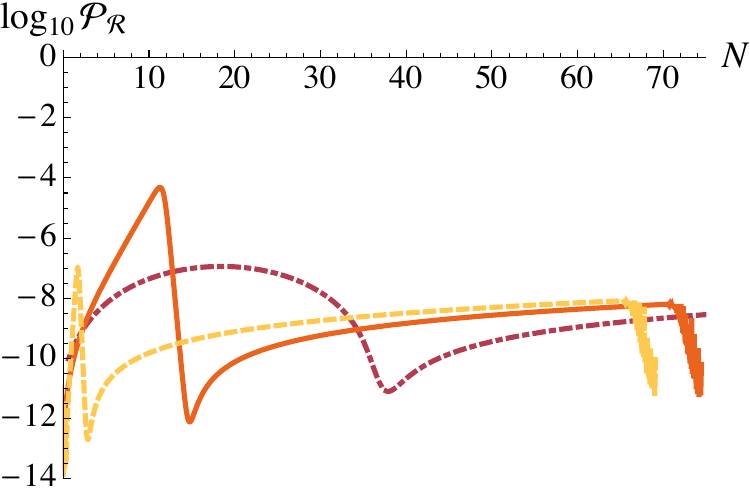}
    \caption{The potential $V (r, \theta_* (r))$ ({\it left}) and the power spectrum ${\cal P}_{\cal R} (k)$ ({\it right}) for $\mu = M_{\rm pl}$, $\xi_\phi = \xi_\chi = 100$, and $b_1 = b_2 = -1.8 \times 10^{-4}$, with varying ratio $c_1 / c_4$. In each case we keep $c_2 \sim c_4$ and fine-tune $c_2$ to a comparable degree. As the hierarchy in $V (r, \theta_*(r))$ between the large-field plateau and the small-field feature decreases, the peak in the power spectrum shifts from tall and narrow to short and wide. The curves shown here correspond to $\{ c_1, c_2, c_4 \} = \{ 1.5 \times 10^{-4}, 4.3738 \times 10^{-3}, 4.5 \times 10^{-3} \}$ (maroon dot-dashed), $\{ 2.5 \times 10^{-4}, 3.5709 \times 10^{-3}, 3.9 \times 10^{-3} \}$ (orange), and $\{ 4.1 \times 10^{-4}, 3.0879 \times 10^{-3}, 3.2 \times 10^{-3} \}$ (gold dashed).}
    \label{fig:PRcratio}
\end{figure*}

The shape of the peak in the power spectrum ${\cal P}_{\cal R}(k)$ depends on the hierarchy between the value of the potential $V (r, \theta_* (r))$ along the large-field plateau and in the vicinity of the small-field feature. This hierarchy, in turn, depends on the ratio of various coupling constants. For example, if the couplings satisfy the symmetries of Eq.~(\ref{bcxisymmetries}), we may hold $\xi$ and $b$ fixed and vary the ratio $c_1 / c_4$. If $c_1 \ll c_4$, then $V$ will develop a significant hierarchy between large and small field values, and the system will approach the small-field feature with correspondingly greater kinetic energy, much as analyzed in Ref.~\cite{Kannike:2017bxn} for similar single-field models. For $c_1 \ll c_4$, even if the value of $V$ at the local minimum is significantly lower than the value at the nearby local maximum, the system can nonetheless ``escape" to the global minimum of $V$ without lingering arbitrarily long near the small-field feature of the potential. In these scenarios, the corresponding peak in ${\cal P}_{\cal R} (k)$ is tall and narrow. In this section we 
set aside the question of whether the fields could tunnel through the local barrier more quickly than they would simply flow beyond the local maximum classically.

As the ratio $c_1 / c_4$ becomes less extreme, the small-field feature in the potential more closely resembles a quasi-inflection point, akin to those studied in Ref.~\cite{Garcia-Bellido:2017mdw}. In this case, the fields approach the small-field feature with less kinetic energy and linger longer in the ultra-slow-roll phase. The resulting feature in ${\cal P}_{\cal R} (k)$ is more rounded and wide. See Fig.~\ref{fig:PRcratio}.

\begin{figure}[h!]
    \centering
    \includegraphics[width=0.45\textwidth]{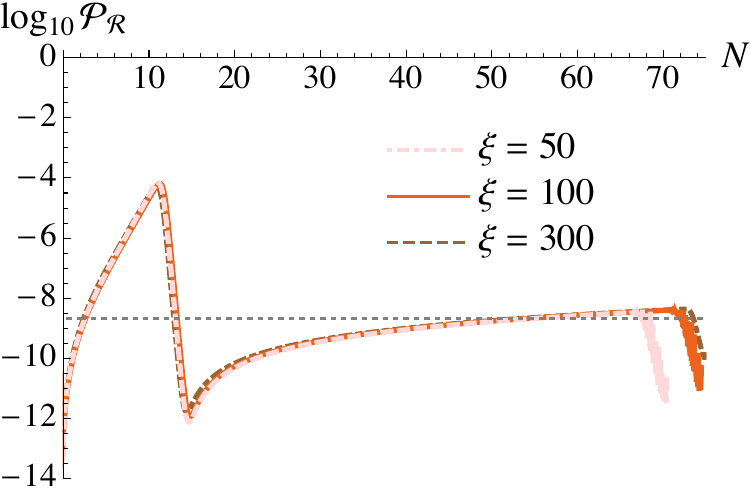}
    \caption{The power spectrum ${\cal P}_{\cal R} (k)$ for three values of the nonminimal coupling constant $\xi$, when we exploit the scaling relationships of Eq.~\eqref{eqn:scaling}. 
    For each curve we set $\mu = M_{\rm pl}$, $\hat{c}_1 = 2.5 \times 10^{-4}$, and $\hat{c}_4 = 3.9 \times 10^{-3}$. 
    For $\xi = 100$ we set $y = 1$, $\hat{b} = - 1.8 \times 10^{-4}$, and $\hat{c}_2 = 3.5709 \times 10^{-3}$.
    For the remaining curves, appropriate values of $y$ and $\hat{b}$ follow from Eq.~(\ref{eqn:scaling}); we also adjust $\hat{c}_2$ in each case by $\Delta \hat{c}_2 / \hat{c}_2 = {\cal O} (10^{-4})$, because the dynamics only become exactly invariant under the scaling of Eq.~(\ref{eqn:scaling}) in the limit $\xi \rightarrow \infty$.
    }
    \label{fig:PRscaling}
\end{figure} 


Under the symmetry of Eq.~(\ref{bcxisymmetries}), we also find that the Einstein-frame potential is invariant if we scale the parameters and fields as follows:
\beq
b = \sqrt{y} \, \hat{b}, \quad 
c_i= y\hat{c}_i, \quad
\xi = y\hat{\xi}, \quad
r = \hat{r} / \sqrt{y},
\label{eqn:scaling}
\eeq
where $y > 0$ is a real-valued constant.
As field-space scalars, the potential $V (r, \theta)$ and the angle $\theta_* (r)$ are both invariant under the rescalings of Eq.~(\ref{eqn:scaling}). The metric ${\cal G}_{IJ} (r, \theta)$, on the other hand, is a field-space tensor rather than a scalar, whose components {\it do} transform under these rescalings. The field-space curvature in these models falls as $1/\xi$~\cite{Kaiser:2012ak,McDonough:2020gmn}, and therefore the full inflationary dynamics---which depend on both $V (r, \theta)$ and ${\cal G}_{IJ} (r, \theta)$---become invariant under the rescalings of Eq.~(\ref{eqn:scaling}) in the limit $\xi \rightarrow \infty$.


The power spectrum for the set of parameters used in Fig.~\ref{fig:VE} is shown in Fig.~\ref{fig:PRscaling} as the dark orange line.
To demonstrate the scaling relation of Eq.~\eqref{eqn:scaling}, we multiply the fiducial parameter set by appropriate factors of $y$ such that we obtain the scaled parameter sets corresponding to $\xi=50$ and $300$. We further adjust $\hat{c}_2$ by $\Delta \hat{c}_2 / \hat{c}_2 \sim 1/\xi^2 \sim {\cal O} (10^{-4})$ in each case, to accommodate the non-invariance of ${\cal G}_{IJ} (r, \theta)$ under the scaling relations of Eq.~(\ref{eqn:scaling}) for finite $\xi$. The power spectra for these scaled parameters are also shown on Fig.~\ref{fig:PRscaling}; the curves are all nearly identical. Hence, given a set of parameters with some value of $\xi$, we can use the scaling relation to find a corresponding set of parameters at a different value of $\xi$ that will yield the same predictions for observables.

\subsection{PBH Formation}
\label{sec:PBHform}

PBHs can form soon after the end of inflation from large peaks in the power spectrum ${\cal P}_{\cal R} (k)$ on length-scales much shorter than those probed by the CMB. Such large perturbations cross outside the Hubble radius near the end of inflation, remain effectively frozen in amplitude while their wavelength is longer than the Hubble radius, and later re-enter the Hubble radius after the end of inflation, whereupon they can induce gravitational collapse. 

\subsubsection{Critical Collapse}
\label{sec:crit_collapse}

Upon re-entering the Hubble radius after inflation, local overdensities
\beq
\delta ({\bf x}) \equiv \frac{ \rho ({\bf x} ) - \bar{\rho} }{\bar{\rho}}
\label{deltadef}
\eeq
will induce gravitational collapse if they are of sufficient amplitude. Here $\bar{\rho} = \rho_{\rm total}$ is the energy density averaged over a Hubble volume. The collapse process is a critical phenomenon akin to other kinds of phase transitions. In particular, the masses of black holes that form at time $t_c$ follow the distribution \cite{Choptuik:1992jv,Evans:1994pj,Gundlach:2002sx,Niemeyer:1997mt,Niemeyer:1999ak,Yokoyama:1998xd,Green:1999xm,Green:2004wb,Kuhnel:2015vtw,Young:2019yug,Kehagias:2019eil,Escriva:2019phb,DeLuca:2020ioi,Musco:2020jjb,Escriva:2021aeh}
\beq
M (\delta_{\rm avg}) = {\cal K} M_H (t_c) \left( \delta_{\rm avg} - \delta_c \right)^\nu \label{Micritical}
\eeq
for overdensities $\delta_{\rm avg}$ above some threshold $\delta_c \sim {\cal O} (10^{-1})$, where $\delta_{\rm avg}$ is the spatial average of $\delta ({\bf x})$ over a region of radius $R < H^{-1}$, ${\cal K}$ is a dimensionless ${\cal O} (1)$ constant, and $\nu$ is a universal critical exponent ($\nu \simeq 0.36$ for collapse during a radiation-dominated era). The Hubble mass $M_H (t_c)$ is the mass enclosed within a Hubble sphere at time $t_c$:
\beq
\begin{split}
M_H (t_c) &\equiv \frac{ 4 \pi}{3} \rho_{\rm total} (t_c) H_c^{-3} \\
&= 4 \pi \frac{ M_{\rm pl}^2}{ H_c} ,
\end{split}
\label{MH}
\eeq
where $H_c \equiv H (t_c)$. The second line of Eq.~(\ref{MH}) follows upon using the Friedmann equation, $H^2 = \rho_{\rm total} / (3 M_{\rm pl}^2)$. Although the relationship between the threshold $\delta_c$ and the curvature perturbation ${\cal R}$ is, in general, nonlinear and depends on the spatial profile of the overdensities \cite{Young:2019yug,Kehagias:2019eil,Escriva:2019phb,DeLuca:2020ioi,Musco:2020jjb,Escriva:2021aeh}, the threshold criterion $\delta_{\rm avg} \geq \delta_c$ for the production of PBHs is typically equivalent to the threshold \cite{Young:2019yug} 
\beq
{\cal P}_{\cal R} (k_{\rm pbh}) \geq 10^{-3}, 
\label{PRthreshold}
\eeq
where ${\cal P}_{\cal R}$ is defined in Eq.~(\ref{PRdef}). The scale $k_{\rm pbh} = a (t_c) H_c$ is the comoving wavenumber of perturbations that re-enter the Hubble radius at time $t_c$ and induce collapse. 

The mass spectrum of PBHs that form via critical collapse includes a long tail for masses $M < \bar{M}$ \cite{Kuhnel:2015vtw,Young:2019yug,DeLuca:2020ioi}, though it is sharply peaked at an average value $\bar{M}$ that is remarkably close to Bernard Carr's original estimate \cite{Carr:1975qj},
\beq
\bar{M} = \gamma M_H (t_c) ,
\label{barMi}
\eeq
with dimensionless constant $\gamma \simeq 0.2$. For PBHs that form during the radiation-dominated phase, $a(t) \sim t^{1/2}$ and hence $H (t) = 1 / (2t)$, so from Eqs.~(\ref{MH}) and (\ref{barMi}) we have
\beq
\bar{M} \simeq 8.1 \times 10^{37} \, {\rm g} \left( \frac{ t_c}{1 \, {\rm s} } \right) 
\label{barMiseconds}
\eeq
upon using $\gamma = 0.2$. PBHs with average masses within the range $10^{17} \, {\rm g} \leq \bar{M} \leq 10^{22} \, {\rm g}$ could account for the entire dark-matter fraction in the observable universe today while evading various observational constraints \cite{Carr:2020xqk,Green:2020jor,Villanueva-Domingo:2021spv}; this corresponds to PBH formation times of $10^{-21} \, {\rm s} \leq t_c \leq 10^{-16} \, {\rm s}$.

We may relate the time $t_c$ to the earlier time $t_{\rm pbh}$, during inflation, when perturbations with wavenumber $k_{\rm pbh}$ first crossed outside the Hubble radius. If the first Hubble-crossing time $t_{\rm pbh}$ occurs $\Delta N$ efolds before the end of inflation, then
\beq
k_{\rm pbh} = a (t_{\rm pbh}) H (t_{\rm pbh} ) = a (t_{\rm end}) e^{- \Delta N} H (t_{\rm pbh}) ,
\label{kpbh1}
\eeq
where $t_{\rm end}$ denotes the end of inflation. As in Appendix \ref{appPerturbations}, we parameterize the post-inflation reheating phase as a brief period of matter-dominated expansion ($w_{\rm eff} \simeq 0$) which lasts $N_{\rm reh}$ efolds between the times $t_{\rm end}$ and $t_{\rm rd}$; beginning at time $t_{\rm rd}$, the universe expands with a radiation-dominated equation of state \cite{Amin:2014eta,Allahverdi:2020bys}. Then the scale factor $a (t_c)$ at the time that the perturbations of comoving wavenumber $k_{\rm pbh}$ re-enter the Hubble radius will be
\beq
a (t_c) = a (t_{\rm end}) e^{N_{\rm reh} } \left( \frac{ t_c}{t_{\rm rd } } \right)^{1/2}
\label{atc}
\eeq
and the Hubble parameter will be $H (t_c) = 1/ (2 t_c)$. Between $t_{\rm end}$ and $t_{\rm rd}$ the energy density redshits as $\rho (t_{\rm rd}) = \rho (t_{\rm end}) e^{-3 N_{\rm reh}}$, so we may write
\beq
\frac{1}{ t_{\rm rd} } = 2 H (t_{\rm end}) e^{-3 N_{\rm reh} / 2} .
\label{tRD}
\eeq
From Eqs.~(\ref{atc}) and (\ref{tRD}), we find
\beq
\begin{split}
k_{\rm pbh} &= a (t_c) H (t_c) \\
&= \frac{1}{\sqrt{2 t_c}} a (t_{\rm end}) H^{1/2} (t_{\rm end}) e^{N_{\rm reh} / 4}  .
\end{split}
\label{kpbh2}
\eeq
Equating the expressions for $k_{\rm pbh}$ in Eqs.~(\ref{kpbh1}) and (\ref{kpbh2}), we may solve for $\Delta N$:
\beq
\Delta N = \frac{1}{2} \ln \left[  \frac{ 2 H^2 (t_{\rm pbh} ) }{H (t_{\rm end} ) } e^{- N_{\rm reh} / 4}\, t_c \right] .
\label{DeltaN}
\eeq
For the parameters that we have been considering, which yield a substantial hierarchy between the values of the potential along the large-field plateau and near the small-field feature, $H (t_{\rm pbh} ) \simeq H (t_{\rm end} ) \simeq 10^{-5.4} \, M_{\rm pl}$; see the left panel of Fig.~\ref{fig:Hfields}. Previous studies of post-inflation reheating in closely related models have consistently found efficient reheating, with $N_{\rm reh} \lesssim 3$ across a wide range of parameter space \cite{DeCross:2015uza,DeCross:2016cbs,DeCross:2016fdz,Nguyen:2019kbm,vandeVis:2020qcp}; the incorporation of trilinear couplings, such as the terms proportional to the coefficient ${\cal C}$ in the effective potential of Eq.~(\ref{VErtheta}), generically increases the efficiency of reheating \cite{Bassett:1999ta,Dufaux:2006ee}. Upon taking $0 \leq N_{\rm reh} \leq 3$, we therefore find
\beq
18 \lesssim \Delta N \lesssim 25 
\label{DeltaNtarget}
\eeq
across the range of PBH formation times of interest, $10^{-21} \, {\rm s} \leq t_c \leq 10^{-16} \, {\rm s}$. 
This result is comparable to an estimate based on the discussion in Ref.~\cite{Ozsoy:2023ryl}.

\subsubsection{PBHs from Ultra-Slow-Roll Evolution in These Models}
\label{sec:PBHsUSR}

\begin{figure*}
    \centering
    \includegraphics[width=0.43\textwidth]{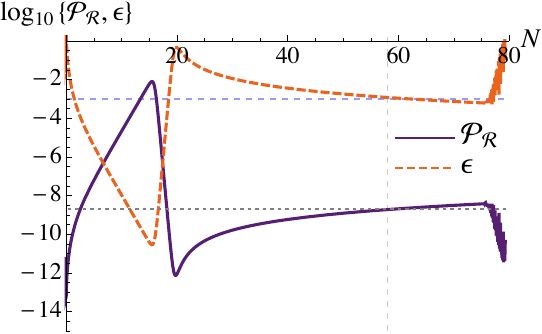} $\quad$ \includegraphics[width=0.43\textwidth]{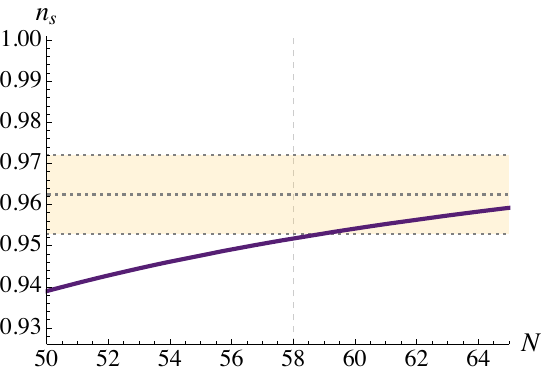} \\
    \includegraphics[width=0.43\textwidth]{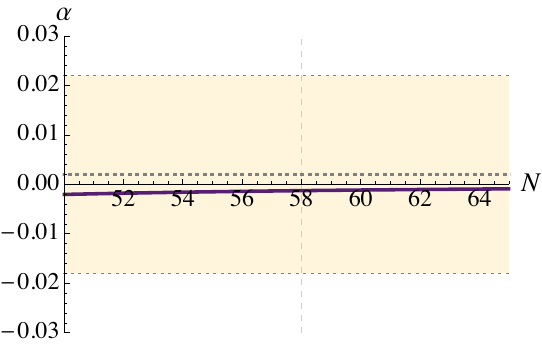} $\quad$ \includegraphics[width=0.43\textwidth]{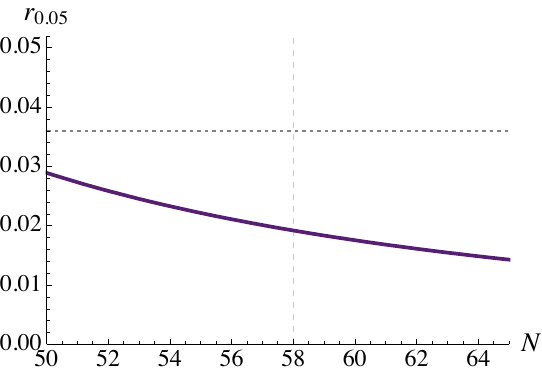}    
    \caption{Observable quantities from our two-field model with one fine-tuned parameter. For each plot, $N$ denotes the number of efolds before the end of inflation ($N (t_{\rm end}) = 0$). For the parameters chosen, the CMB pivot scale $k_* = 0.05 \, {\rm Mpc}^{-1}$ crossed outside the Hubble radius $N_* \simeq 58$ efolds before the end of inflation, and ${\cal P}_{\cal R} (k)$ first exceeded the threshold for PBH production $\Delta N = 16.3$ efolds before the end of inflation. ({\it Top left}) The power spectrum ${\cal P}_{\cal R} (k)$ (purple) and the slow-roll parameter $\epsilon$ (orange dashed). The horizontal dotted line shows the COBE normalization ${\cal P}_{\cal R} (k_*) = 2.1 \times 10^{-9}$, and the horizontal dashed blue line shows the threshold for PBH formation ${\cal P}_{\cal R} (k) = 10^{-3}$. ({\it Top right}) The spectral index $n_s (k_*)$ (purple), Planck 2018 best-fit value (dotted), and $2 \sigma$ error-bar contours \cite{Planck:2018jri}. ({\it Bottom left}). The running of the spectral index $\alpha (k_*) = (d n_s (k) / d {\rm ln} k)\vert_{k_*}$ (purple), Planck 2018 best-fit value (dotted), and $2 \sigma$ error-bar contours when the Planck analysis allows for $\alpha (k_*) \neq 0$ \cite{Planck:2018jri}. ({\it Bottom right}) The tensor-to-scalar ratio $r (k_*)$ (purple) and the 2020 Planck-BICEP/Keck upper bound (dotted) \cite{BICEP:2021xfz}. The system was evolved numerically with the same parameters and initial conditions as in Figs.~\ref{fig:VE}--\ref{fig:muomegaPR}, but with $c_2 = c_3 = 3.570913 \times 10^{-3}$ rather than $c_2 = c_3 = 3.57 \times 10^{-3}$.   }
    \label{fig:PBHrun}
\end{figure*}

As analyzed in Refs.~\cite{Byrnes:2018txb,Carrilho:2019oqg}, a rapid rise in ${\cal P}_{\cal R} (k)$ at short wavelengths $k \sim k_{\rm pbh}$, which could induce PBHs after inflation, necessarily has an impact on the long-wavelength power spectrum in the vicinity of the CMB pivot-scale $k_*$; see also Ref.~\cite{Ando:2020fjm}. Hence there is a delicate balance required to secure predictions for observables in the vicinity of the CMB pivot scale $k_*$ that remain consistent with the latest measurements \cite{Planck:2018jri,BICEP:2021xfz,Planck:2019kim} while also arranging for ${\cal P}_{\cal R} (k_{\rm pbh}) \geq 10^{-3}$. In particular, the presence of small-field features in the potential, which can yield a large peak in ${\cal P}_{\cal R} (k)$ near $k \sim k_{\rm pbh}$, tends to modestly deform the potential along the large-field plateau, relevant for ${\cal P}_{\cal R} (k_*)$. The value of the spectral index $n_s (k_*)$ is typically lower than in related models for which little or no peak appears in ${\cal P}_{\cal R} (k)$ at small scales. 

To compare with the latest observations, we must evaluate the number of efolds before the end of inflation, $N_*$, when the CMB pivot scale $k_* = 0.05 \, {\rm Mpc}^{-1}$ first crossed outside the Hubble radius. Eq.~(\ref{Nstar}) shows that $N_*$ depends weakly on the duration of reheating. Given efficient reheating in these models \cite{DeCross:2015uza,DeCross:2016cbs,DeCross:2016fdz,Nguyen:2019kbm,vandeVis:2020qcp}, we take $N_{\rm reh} \sim 0$; then Eq.~(\ref{Nstar}) yields $N_* \simeq 58$ for the parameters of interest.

The models we consider here generically induce a small but nonzero running of the spectral index, $\alpha (k_*) \equiv (d n_s (k) / d {\rm ln} k )\vert_{k_*} \sim {\cal O} (10^{-3})$. If one includes possible running $\alpha (k_*) \neq 0$ in the analysis of the latest Planck data, then the best-fit value for the spectral index is given by $n_s (k_*) = 0.9625 \pm 0.0048$, with $\alpha (k_*) = 0.002 \pm 0.010$, each at $68\%$ confidence level \cite{Planck:2018jri}. Meanwhile, the most recent combined Planck-BICEP/Keck observations constrain the tensor-to-scalar ratio at $k_* = 0.05 \, {\rm Mpc}^{-1}$ to be $r_{0.05} < 0.036$ \cite{BICEP:2021xfz}. As shown in Fig.~\ref{fig:PBHrun}, for a particular choice of parameters our two-field model yields predictions consistent with the latest observations while also producing a peak in the power spectrum that first crosses the critical threshold ${\cal P}_{\cal R} (k_{\rm pbh}) \geq 10^{-3}$ at $\Delta N = 16.3$ efolds before the end of inflation.

The timing of the peak in ${\cal P}_{\cal R} (k)$ for the set of parameters shown in Fig.~\ref{fig:PBHrun} was calculated neglecting non-Gaussian features of the probability distribution function for large-amplitude curvature perturbations, which arise from stochastic effects such as quantum diffusion and backreaction. When such effects are incorporated self-consistently, the probability distribution function typically features more power in the tails of the distribution than a simple Gaussian---meaning that large fluctuations remain rare, but much {\it less} rare than standard calculations (of the sort we incorporate here) would suggest \cite{Byrnes:2012yx,Young:2013oia,Pattison:2017mbe,Biagetti:2018pjj,Kehagias:2019eil,Ezquiaga:2019ftu,Ando:2020fjm,Tada:2021zzj,Biagetti:2021eep}. Although it remains a topic for further research, we expect that such non-Gaussian effects would likely shift $\Delta N$ by ${\cal O} (1)$ efolds, which would bring $\Delta N$ more squarely within the range of Eq.~(\ref{DeltaNtarget}) of interest for dark matter abundances.  

Even while neglecting these non-Gaussian effects, we find that the results shown in Fig.~\ref{fig:PBHrun} require a substantial fine-tuning of one of the dimensionless coupling constants: $c_2 = 3.570913 \times 10^{-3}$, rather than the more ``reasonable" value $c_2 = 3.57 \times 10^{-3}$ that was used for the plots in Figs.~\ref{fig:VE}--\ref{fig:muomegaPR}. Such substantial fine-tuning is typical among models that produce PBHs from a phase of ultra-slow-roll evolution \cite{Garcia-Bellido:2017mdw,Ezquiaga:2017fvi,Kannike:2017bxn,Germani:2017bcs,Motohashi:2017kbs,Di:2017ndc,Ballesteros:2017fsr,Pattison:2017mbe,Passaglia:2018ixg,Byrnes:2018txb,Biagetti:2018pjj,Carrilho:2019oqg,Inomata:2021tpx,Inomata:2021uqj,Pattison:2021oen}.

Although the need for fine-tuning in such models is not new, we note nevertheless that the multifield models considered here are relatively efficient. We require such models to yield accurate predictions for {\it eight} distinct quantities; our two-field model does so using {\it six} relevant free parameters. The observable quantities to match include the spatial curvature contribution to the total energy density $\Omega_K$; the spectral index $n_s (k_*)$; the running of the spectral index $\alpha (k_*)$; the tensor-to-scalar ratio $r (k_*)$; the isocurvature fraction at the end of inflation $\beta_{\rm iso} (k_*, t_{\rm end})$; the non-Gaussianity parameter $f_{\rm NL}$; the peak amplitude of the power spectrum at short scales ${\cal P}_{\cal R} (k_{\rm pbh})$; and the time $\Delta N$ when the peak in ${\cal P}_{\cal R} (k_{\rm pbh})$ first crosses the critical threshold. 

The multifield models we explore here display strong single-field attractor behavior, with negligible turning throughout the duration of inflation, $\omega^2 \ll H^2$. Such attractor behavior means that the evolution of the system---and hence predictions for observables---is sensitive to changes in {\it one} initial condition, $r (t_i)$, rather than the other $2 {\cal N} - 1$ initial conditions required in ${\cal N}$-field models. (For example, predictions for observables in the two-field case are independent of $\dot{r} (t_i)$, $\theta (t_i)$, and $\dot{\theta} (t_i)$, unless those initial conditions are exponentially fine-tuned \cite{Kaiser:2012ak,Kaiser:2013sna,Schutz:2013fua,Kaiser:2015usz,DeCross:2015uza}.) Once $r (t_i)$ is set large enough to yield sufficient inflation (with $N_{\rm total} \geq 65$ efolds), these models generically satisfy observational constraints on $\Omega_K$. Meanwhile, as emphasized above, the single-field attractor behavior generically suppresses such typical multifield phenomena as $\beta_{\rm iso} (k_*, t_{\rm end})$ and $f_{\rm NL}$, thereby easily keeping predictions consistent with observational bounds. In particular, consistent with the discussion leading to Eq.~(\ref{betaisoheavy}), we find $\beta_{\rm iso} (k_*, t_{\rm end}) < e^{-3 N_*} \sim {\cal O} (10^{-76})$ for the parameters used in Fig.~\ref{fig:PBHrun}, compared to the current Planck bound $\beta_{\rm iso} (k_*, t_{\rm end}) \leq 0.026$ \cite{Planck:2018jri}. Likewise, from the discussion leading to Eq.~(\ref{fNLSFA2}), we find $f_{\rm NL}^{\rm equil} (k_*) = - 0.019$ for the parameters used in Fig.~\ref{fig:PBHrun}, consistent with the latest measurement from Planck: $f_{\rm NL}^{\rm equil}(k_*) = -26 \pm 47$ \cite{Planck:2019kim}.

The results shown in Fig.~\ref{fig:PBHrun}, which incorporate the symmetries among coupling constants of Eq.~(\ref{bcxisymmetries}), thus reveal close agreement between predictions for $\{ \Omega_K, \beta_{\rm iso}, f_{\rm NL}, n_s (k_*), \alpha (k_*), r (k_*),$ ${\cal P}_{\cal R} (k_{\rm pbh}), \Delta N \}$ from a two-field model with six relevant free parameters: $\{ r (t_i), \xi, b, c_1, c_2, c_4 \}.$

\subsection{Discussion}
\label{sec:PBHdiscussion}

In this section we have demonstrated that inflationary models that incorporate well-motivated features from high-energy physics can produce primordial black holes (PBHs) soon after the end of inflation, of interest for present-day dark-matter abundances. In particular, we have investigated models with multiple interacting scalar fields, each with a nonminimal coupling to the spacetime Ricci curvature scalar. Our multifield models are inspired by supersymmetric constructions (with an explicit supergravity construction provided in Appendix \ref{appSUGRA}) and incorporate only generic operators in the action that would be expected in any self-consistent effective field theory treatment at high energies. 

Despite being multifield by construction, the inflationary dynamics in these models rapidly relax to effectively single-field evolution along a smooth large-field plateau in the effective potential (much as in closely related models \cite{Kaiser:2012ak,Kaiser:2013sna,Schutz:2013fua,Kaiser:2015usz}), thereby yielding predictions for primordial observables in close agreement with the latest measurements of the cosmic microwave background (CMB) radiation. Models within this family also yield efficient reheating following the end of inflation \cite{Bezrukov:2008ut,Garcia-Bellido:2008ycs,Child:2013ria,DeCross:2015uza,DeCross:2016fdz,DeCross:2016cbs,Figueroa:2016dsc,Repond:2016sol,Ema:2016dny,Sfakianakis:2018lzf,Rubio:2019ypq,Nguyen:2019kbm,vandeVis:2020qcp,Iarygina:2020dwe,Ema:2021xhq,Figueroa:2021iwm,Dux:2022kuk}. In addition, the potentials we study here include small-field features that can induce a brief phase of ultra-slow-roll evolution prior to the end of inflation, which yield sharp spikes in the power spectrum of curvature perturbations on length-scales exponentially shorter than the CMB pivot scale $k_* = 0.05 \, {\rm Mpc}^{-1}$. Upon re-entering the Hubble radius after the end of inflation, these amplified short-scale perturbations induce gravitational collapse to PBHs.

 As in previous studies of PBH formation following an ultra-slow-roll phase during inflation \cite{Garcia-Bellido:2017mdw,Ezquiaga:2017fvi,Kannike:2017bxn,Germani:2017bcs,Motohashi:2017kbs,Di:2017ndc,Ballesteros:2017fsr,Pattison:2017mbe,Passaglia:2018ixg,Byrnes:2018txb,Biagetti:2018pjj,Carrilho:2019oqg,Inomata:2021tpx,Inomata:2021uqj,Pattison:2021oen}, we find that in order to generate PBHs near the mass-range that could account for the present-day dark-matter abundance we must fine-tune one dimensionless coupling constant to several significant digits. Nonetheless, by incorporating only one fine-tuned constant, these models yield accurate predictions for eight distinct quantities---including the spectral index $n_s (k_*)$ and its running $\alpha (k_*)$, the tensor-to-scalar ratio $r (k_*)$, the isocurvature fraction $\beta_{\rm iso} (k_*, t_{\rm end})$ and primordial non-Gaussianity $f_{\rm NL}$, among others---using fewer than eight free parameters.

\section{Planck Constraints and Gravitational Wave Forecasts for Primordial Black Hole Dark Matter Seeded by Multifield Inflation
}
\label{sec:PBH_MCMC}

In Section~\ref{sec:PBHs}, we demonstrated that PBHs relevant for DM will form from a realistic class of multifield inflationary models built from well-motivated high-energy ingredients. In particular, we considered inflationary models that incorporate multiple interacting scalar fields, each with a nonminimal gravitational coupling. Such models consist of generic mass-dimension-4 operators in an effective field theory (EFT) expansion of the action at inflationary energy scales, and (as discussed in Ref.~\cite{Geller:2022nkr}) have a consistent ultraviolet completion in the context of supergravity. For various choices of model parameters, the inflationary dynamics in such models can include a phase of ultra-slow-roll evolution \cite{Garcia-Bellido:2017mdw,Ezquiaga:2017fvi,Germani:2017bcs,Kannike:2017bxn,Motohashi:2017kbs,Di:2017ndc,Ballesteros:2017fsr,Pattison:2017mbe,Passaglia:2018ixg,Biagetti:2018pjj,Byrnes:2018txb,Carrilho:2019oqg,Figueroa:2020jkf,Inomata:2021tpx,Inomata:2021uqj,Pattison:2021oen,Balaji:2022rsy,Balaji:2022zur,Kawai:2022emp,Karam:2022nym,Pi:2022ysn}, which can yield PBHs with $M_{\rm pbh}$ within the appropriate range to account for the entire DM abundance, while also matching high-precision measurements of the primordial perturbation spectrum on length-scales relevant for the cosmic microwave background radiation (CMB). In Ref.~\cite{Geller:2022nkr}, we demonstrated that eight distinct observational constraints---relating to both PBH and CMB observables---could be matched by adjusting only six free parameters in these models.

In this section, we use a Markov Chain Monte Carlo (MCMC) analysis to systematically identify the regions of parameter space for the family of models constructed in Ref.~\cite{Geller:2022nkr} that can produce relevant populations of PBHs for DM while continuing to match multiple observables related to the CMB. 
Our two-dimensional marginalized posterior distributions for pairs of parameters demonstrate that this general class of models can yield predictions for observables near the CMB pivot scale $k_* = 0.05 \, {\rm Mpc}^{-1}$ in close agreement with the latest observations \cite{Planck:2018jri,Planck:2018vyg,Planck:2019kim,BICEP:2021xfz}---including the amplitude of the primordial power spectrum $A_s (k_*)$, the spectral tilt $n_s (k_*)$, and the ratio of power in tensor to scalar modes $r (k_*)$---while also producing PBHs with masses $M_{\rm pbh}$ within the range for which they could account for the entire DM abundance. Such regions of parameter space also yield predictions for related observables that are easily compatible with the latest observational bounds, such as the running of the spectral index $\alpha_s (k_*)$, the primordial isocurvature fraction $\beta_{\rm iso} (k_*)$, and local primordial non-Gaussianity $f_{\rm NL}$. 

In Ref.~\cite{Geller:2022nkr}, much as in previous studies of PBH formation from ultra-slow-roll evolution \cite{Garcia-Bellido:2017mdw,Ezquiaga:2017fvi,Germani:2017bcs,Kannike:2017bxn,Motohashi:2017kbs,Di:2017ndc,Ballesteros:2017fsr,Pattison:2017mbe,Passaglia:2018ixg,Biagetti:2018pjj,Byrnes:2018txb,Carrilho:2019oqg,Figueroa:2020jkf,Inomata:2021tpx,Inomata:2021uqj,Pattison:2021oen}, we found that one model parameter needed to be highly fine-tuned in order to match predictions for all eight relevant observables. As we will discuss here, with the aid of the MCMC we identify a degeneracy direction in parameter space such that shifts up to $\sim 10\%$ of any particular model parameter can be compensated by comparable shifts among the other parameters while preserving a close fit with observations. The overall tuning of each parameter required to match all the observables of interest, as measured by the posterior distributions for various {\it ratios} of parameters, is at the percent level, driven largely by the sub-percent-level accuracy to which the spectral index $n_s$ has been measured.

In addition to studying predictions from these models for CMB observables and PBH formation, we also analyze predictions for the amplification of primordial gravitational waves (GWs). GWs provide a tantalizing means to test the physics of the early universe. (For recent reviews, see Refs.~\cite{Guzzetti:2016mkm,Caprini:2018mtu,Domenech:2021ztg}.) In the context of PBH formation, the amplified spectrum of scalar curvature perturbations---necessary to induce gravitational collapse into PBHs---will source tensor modes beyond linear order in perturbation theory. Therefore PBH formation should be accompanied by a contribution to a stochastic GW background (SGWB) with a particular spectral shape \cite{Saito:2008jc,Saito:2009jt,Assadullahi:2009jc,Bugaev:2009zh,Bugaev:2010bb,Inomata:2018epa}. 

Whereas the primordial GW spectrum is tightly constrained on scales near the CMB pivot scale $k_*$, it is mostly unconstrained on the much shorter length-scales relevant for PBH formation. We calculate the expected contribution to the SGWB from PBH formation in our models, and find that the signal overlaps significantly with the (projected) integrated sensitivity curves for several next-generation detectors, including Advanced LIGO-Virgo (LIGO A+) \cite{KAGRA:2021kbb}, LISA \cite{Barausse:2020rsu}, the Einstein Telescope (ET) \cite{Maggiore:2019uih}, Cosmic Explorer (CE)~\cite{Reitze:2019iox}, and DECIGO \cite{Yuan:2021qgz,Kawamura:2020pcg}. These results suggest the exciting possibility that the production of DM in the form of PBHs from multifield models could soon be testable. 

This section is organized as follows. 
We will use the same multifield model and formalism described in Section~\ref{sec:model}.
In \S~\ref{sec:observables}, we discuss relevant physical observables predicted by the model. 
In \S~\ref{sec:constraints}, we discuss how we constrain the allowed model parameter space.
In \S~\ref{sec:results} we present our results and finally we conclude with further discussion in \S~\ref{sec:conclusion}. We discuss the effects of a phase of ultra-slow-roll evolution on the power spectrum of scalar curvature perturbations in Appendix \ref{app:USR}, and present additional details on the calculation of the induced GW spectrum in Appendix \ref{app:tensor_spectrum}.

\subsection{Observables}
\label{sec:observables}

Upon imposing the symmetries between couplings given by Eq.~(\ref{bcxisymmetries}), the two-field models under consideration are specified by five free dimensionless parameters, $\{ \xi, b, c_1, c_2, c_4 \}$, and the fields' initial value $r (t_0)$. Our aim is to determine how generically such models will satisfy CMB constraints, produce PBHs that could account for the DM abundance, and produce detectable GW signals.

We do so by determining the regions of parameter space that yield predictions that are consistent with both the empirical constraints and meet the criteria for producing PBHs. The latter---which yield a population of PBHs within the mass range of interest---are more restrictive, since by slightly relaxing these PBH constraints, the model generally remains in compliance with CMB observational constraints. In this section, we identify specific observables of interest and consider how model predictions for these observables vary with parameters.

\subsubsection{Cosmic Microwave Background }
\label{CMBsubsection}

The dimensionless power spectrum for the gauge-invariant curvature perturbations, ${\cal P}_{\cal R} (k)$, defined in Eq.~(\ref{PRdef}), is central to the consideration of CMB constraints. In particular, predictions from this model must be consistent with the latest high-precision measurements of several quantities related to ${\cal P}_{\cal R} (k)$ in the vicinity of the CMB pivot scale $k_* = 0.05 \, {\rm Mpc}^{-1}$ \cite{Planck:2018jri,Planck:2018vyg,Planck:2019kim,BICEP:2021xfz}, including the amplitude (or COBE normalization)
\beq
A_s \equiv {\cal P}_{\cal R} (k_*)  
\label{AsPlanck}
\eeq
and the spectral index
\begin{align}
    n_s (k_*) &\equiv 1 + \left( \frac{d\, {\rm ln} {\cal P}_{\cal R} (k) }{ d \,{\rm ln} k} \right) \Big\vert_{k_*} \n 
    &\simeq 1 - 6 \epsilon (t_*) + 2 \eta (t_*) ,
    \label{nsdef}
    \end{align}
where the second line holds to first order in slow-roll parameters, $\epsilon$ and $\eta$ are defined in Eqns.~\eqref{epsilon} and \eqref{etadef}, and $t_*$ is the time when $k_*$ crosses outside of the Hubble radius. We also consider the running of the spectral index,
\begin{equation}
\alpha (k_*) \equiv \left( \frac{ d n_s (k) }{ d \, {\rm ln} k} \right) \Big\vert_{k_*} \simeq \left( \frac{ \dot{n}_s (k)}{H} \right) \Big \vert_{k_*}.
\label{alphadef}
\end{equation}
Observables related to the CMB may be calculated using the expression in Eq.~(\ref{PRHepsilon}) in our model, across all the regions of parameter space under study here.
    
As noted in the previous section, within these models the fields evolve along single-field attractors during inflation, with exponentially suppressed turning within field space. In the limit $\vert \omega^I \vert \ll H$, the tensor-to-scalar ratio for our multifield models reverts to its usual single-field form \cite{Bassett:2005xm,Gong:2016qmq,Kaiser:2012ak,Geller:2022nkr}
\begin{equation}
    r (k_*) = 16 \epsilon (t_*) .
    \label{rTtoS}
\end{equation}
Given that the isocurvature perturbations remain heavy and the turn rate remains suppressed in these models, we also find that typical multifield features, such as primordial isocurvature perturbations $\beta_{\rm iso} (k_*)$ and primordial non-Gaussianity (parameterized by various dimensionless coefficients $f_{\rm NL}$, corresponding to different shape functions for the bispectrum) remain exponentially suppressed \cite{Kaiser:2013sna,Kaiser:2015usz,Geller:2022nkr}, easily consistent with the latest observations \cite{Planck:2018jri,Planck:2019kim}.

We compare predictions from our model with the {\it Planck} 2018 results (when the spectral index is allowed to run with wavenumber)~\cite{Planck:2018jri} and the {\it Planck}-BICEP/Keck 2021 constraint on the tensor-to-scalar ratio \cite{BICEP:2021xfz}:
\beq
\begin{split}
    \ln (10^{10} A_s ) &= 3.044 \pm 0.014 , \\
    n_s (k_*) &= 0.9625 \pm 0.0048 , \\
    \alpha_s (k_*) &= 0.002 \pm 0.010 , \\
    r (k_*) &< 0.036 ,
    \label{eqn:PlanckCMBconstraints}
\end{split}
\eeq
where the reported error bars correspond to $68\%$ confidence intervals.

\subsubsection{Primordial Black Holes }
\label{sec:PBHconstraints}

The primordial power spectrum ${\cal P}_{\cal R} (k)$ must exceed some threshold on appropriate scales $k_{\rm pbh}$ in order for the curvature perturbations to seed primordial overdensities that will ultimately undergo gravitational collapse when these perturbations re-enter the Hubble radius after the end of inflation. For the models under consideration, this threshold is achieved for modes with $k=k_{\text{pbh}}$ that cross outside the Hubble radius during the transient period of ultra-slow roll.
Typical estimates suggest a threshold for PBH formation of ${\cal P}_{\cal R} (k_{\rm pbh}) \geq 10^{-3}$, about six orders of magnitude greater than the amplitude around the CMB pivot scale $k_*$, ${\cal P}_{\cal R}  (k_*) = A_s = 2.1 \times 10^{-9}$ \cite{Kuhnel:2015vtw,Young:2019yug,Kehagias:2019eil,Escriva:2019phb,DeLuca:2020ioi,Musco:2020jjb,Escriva:2021aeh}. 

Various effects beyond linear order in perturbation theory, such as stochastic dynamics and quantum diffusion, typically yield a non-Gaussian probability distribution function for curvature perturbations of various amplitudes, increasing the likelihood of large-amplitude perturbations compared to the Gaussian approximation. Such effects, in turn, can reduce the required threshold on ${\cal P}_{\cal R} (k_{\rm pbh})$ by one to two orders of magnitude \cite{Ezquiaga:2019ftu,Figueroa:2020jkf,Tada:2021zzj,Biagetti:2021eep,Animali:2022otk,Gow:2022jfb,Ferrante:2022mui}. Nevertheless, in this work we use the threshold ${\cal P}_{\cal R} (k_{\rm pbh}) \geq 10^{-3}$; this is conservative in that relaxing this threshold would only lead to a larger region of parameter space that would be consistent with observations.
For computational tractability, we impose this conservative threshold via ${\cal P}_{\cal R}^{\rm SR} (k_{\rm pbh}) \geq 10^{-3}$, where the slow-roll expression for the power spectrum ${\cal P}_{\cal R}^{\rm SR} (k)$ is given in Eq.~\eqref{eqn:PRHepsilon}, given that ${\cal P}_{\cal R} (k) \geq {\cal P}_{\cal R}^{\rm SR} (k)$ when ultra-slow-roll effects are taken into account, as discussed further in Appendix \ref{app:USR}.

In addition to the peak height of the power spectrum, the PBHs that form after inflation are also sensitive to the time, during inflation, when the large-amplitude perturbations were first amplified and crossed outside the Hubble radius. We denote this time as 
\begin{align}
\Delta N \equiv N_{\rm pbh} - N_{\rm end},
\end{align}
where $\Delta N$ is the number of $e$-folds before the end of inflation when $k_{\rm pbh}$ crossed outside the Hubble radius.

In Section~\ref{sec:crit_collapse}, we determined that the correct range of $\Delta N$ to produce asteroid mass primordial black holes is $18 \lesssim \Delta N \lesssim 25$.
As discussed in Appendix~\ref{app:USR}, corrections to the power spectrum under the slow-roll approximation, ${\cal P}_{\cal R}^{\rm SR} (k)$, indicate that the full power spectrum peaks $N_{\rm usr}$ efolds earlier than ${\cal P}_{\cal R}^{\rm SR} (k)$.
In the family of models that we consider here, we typically find $N_{\rm usr} \simeq 3.5$.
Given this typical value of $N_{\rm usr}$, and also taking into account the residual uncertainty in the duration of reheating, we set our lower-bound for $\Delta N = 14$, ensuring that the region of our resulting parameter space that passes the threshold will produce PBHs large enough to avoid evaporation bounds.


\subsubsection{Parameter Dependence \& Degeneracies}
\label{sec:degen1}
%
\begin{figure*}[t!]
    \includegraphics[width=0.49\textwidth]{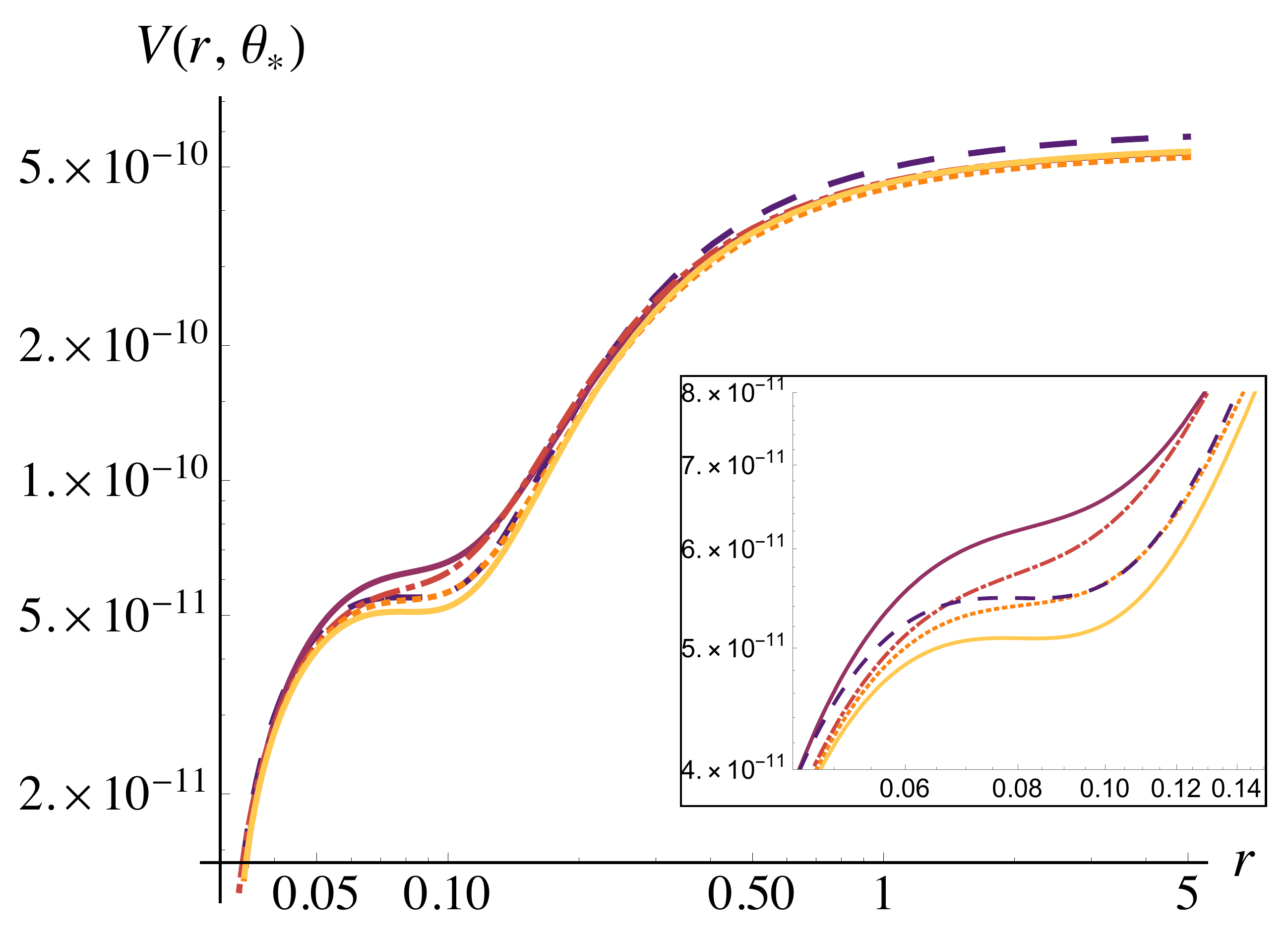}\quad
    \includegraphics[width=0.48\textwidth]{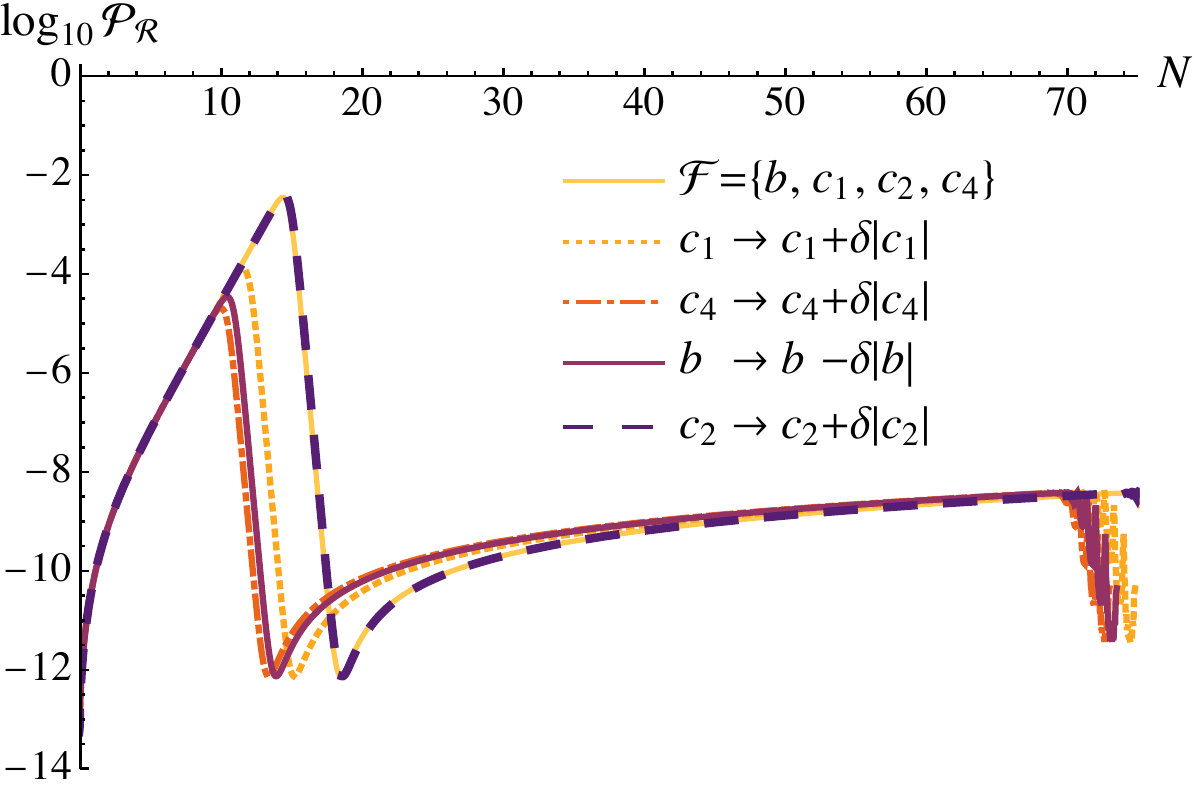}
    \caption{
    The potential (\emph{left panel}) and power spectrum (\emph{right panel}) are plotted for parameter set $\mathcal{F}=\{b, c_1, c_2, c_4\}$ in yellow. We then vary one parameter at a time, cumulatively, to obtain each of the other curves: first we increase $c_1$ (orange dotted curve), then increase $c_4$  (red dot-dashed curve), then decrease $b$ (magenta curve), and finally increase $c_2$ to obtain a degenerate parameter set (dashed purple curve), $\mathcal{F}'=\{b-\delta|b|, c_1+\delta|c_1|, c_2+\delta|c_2|, c_4+\delta|c_4|\}$. 
    The power spectrum is much more sensitive to the step size along the degeneracy direction than the potential; hence, the parameter variations in the left panel are of order $\delta\simeq 10^{-2}$, whereas in the right panel they are of order $\delta\simeq 10^{-6}$.
    }
    \label{fig:V_PR_degen}
\end{figure*}

Once the nonminimal coupling $\xi$ is fixed, the parameter space is described by the four remaining free parameters, $b, c_1, c_2, c_4$. Varying these parameters changes the behavior of the potential and thus also changes predictions for $n_s$, $\mathcal{P}_{\mathcal{R}}$, and other observables in characteristic ways.

Measurements of CMB observables are sensitive to physics at the pivot scale, thus they will be affected by changes to the potential at \textit{large} field values, that is, around $r(t_*)$. Meanwhile, the PBH constraints are largely sensitive to changes in the potential at \textit{small} field values, around $r(t_{\text{pbh}})$, corresponding to the period of ultra-slow roll during which the modes with $k\sim k_{\text{pbh}}$ first exit the Hubble radius. The tension is thus between tuning the small-field features to get a sufficiently large spike in $\mathcal{P}_{\mathcal{R}}$ to seed PBH formation without compromising the large-field dynamics. 
(See also Refs.~\cite{Geller:2022nkr,Byrnes:2018txb,Carrilho:2019oqg,Ando:2020fjm}.)

In general, the longer the period of ultra-slow roll, that is, the larger the relative depth of the local minimum at small field values as compared to the local maximum and the large-field plateau, the larger the spike in the primordial power spectrum $\mathcal{P}_{\mathcal{R}}$ will be. There is a limit to this trend, however. If the relative depth of the local minimum is too large, the characteristic time for the fields to quantum tunnel out of the local minimum becomes comparable to the time for classical transit through the ultra-slow-roll region. In this instance, we cannot ignore the effects of quantum diffusion on the dynamics, hence, there is a range of parameters for which the fields undergo ultra-slow roll evolution long enough for PBHs to form post-inflation, but not so long that quantum effects become dominant. 
We can determine whether a potential will have a small-field feature that falls within this range by considering the magnitude of kinetic energy for the fields as they enter the ultra-slow-roll region. If the kinetic energy is too high, the fields will roll past the region of the potential for which $V_{,\sigma}\simeq 0$ too quickly for ultra-slow roll to yield a sufficient spike in $\mathcal{P}_{\mathcal{R}}$, whereas if the kinetic energy is too low, the fields will become classically ``stuck'' in the local minimum. 
This is discussed in more detail in our previous work \cite{Geller:2022nkr} (see the discussion around Eqn.~\eqref{varianceK}). 

Varying each of the parameters individually affects both the shape of the minimum/maximum feature at small field values as well as the slope of the potential between the large-field plateau and the local minimum, which will change the kinetic energy of the fields as they approach the region of the potential for which $V_{,\sigma}\simeq 0$, as follows: 

\begin{itemize}
    \item Increasing $|b|$ while keeping $b<0$ will both increase the relative depth of the local minimum while \textit{decreasing} the slope of the potential as the fields approach the region with $V_{,\sigma}\simeq 0$. The overall effect is to \emph{decrease} the kinetic energy of the fields as they enter the ultra-slow roll regime.
    
    \item Increasing $c_2=|c_2|$ also increases the relative depth of the local minimum, but \emph{increases} the slope of the potential as the fields approach $V_{,\sigma}\simeq 0$. The latter effect dominates, so the net result is to \textit{increase} the kinetic energy of the fields. 
    
    \item Increasing both $c_1=|c_1|$ and $c_4=|c_4|$ will \textit{increase} the kinetic energy of the fields as they approach the ultra-slow roll region. 
\end{itemize}

Thus the effect of parameter variations is such that increasing the magnitude of $b$ has the opposite effect to increasing the magnitudes of the $c_i$. We find that the interplay of parameters is such that a certain sequence of small parameter variations will lead to a degenerate potential and power spectrum. An example of this is shown in Fig.~\ref{fig:V_PR_degen}. We begin at the fiducial parameter set $\vec{\mathcal{F}}$ and take a small step $\delta$ in the degeneracy direction given by the unit vector $\hat{n}$ to a degenerate parameter set $\vec{\mathcal{F}}^{'}=\vec{\mathcal{F}}+\hat{n}\delta$, where $\hat{n}\delta=(-|b|,|c_1|,| c_2|, |c_4|)\delta$. The power spectrum is much more sensitive to the step size along the degeneracy direction than is the potential, so the right panel, for ${\cal P}_{\cal R} (k)$, uses $\delta\simeq 10^{-6}$, whereas the left panel, for $V (r, \theta_*)$, uses a step size of $\delta\simeq 10^{-2}$ so that the effects of varying parameters can be more readily seen.

\subsubsection{Gravitational Waves}
\label{sec:GWs}

In this work we also consider a complementary observable in inflationary PBH models, namely GWs sourced by the amplified curvature perturbation.

At second order in cosmological perturbation theory, scalar modes can source a SGWB. The induced GWs are usually described by the energy density $\rho_{\rm GW}$ per logarithmic frequency interval, normalized by the critical density (see, e.g., Ref.~\cite{Domenech:2021ztg})
\begin{equation}
    \Omega_{\textrm{GW,0}} h^2 = \frac{h^2}{3M_{\rm Pl}^2 H_0^2} \frac{d\rho_{\rm GW}}{d \textrm{ln} k} .
\end{equation}
Assuming that the GWs were induced by modes that crossed back inside the Hubble radius at a temperature $T_c$ during the radiation-dominated epoch, the fractional energy density today can be written as
\begin{equation}
    \Omega_{\textrm{GW,0}} h^2 = \Omega_{r0}h^2 \left( \frac{g_* (T_c)}{g_{*,0}} \right) \left( \frac{g_{*s} (T_c)}{g_{*s,0}} \right)^{-4/3} \Omega_{\textrm{GW,c}} , 
\end{equation}
where $H_0$ is the present value of the Hubble constant, $h = H_0 / (100 \,\mathrm{km}\, \mathrm{s}^{-1}\, \mathrm{Mpc}^{-1})$, and $\Omega_{\textrm{GW,c}}$ is the GW spectral density at the time the waves were induced. The quantities
$g_* (T)$ and $g_{*s} (T)$ are the effective number of degrees of freedom for the radiation energy density and entropy; today, their values are equal to $g_{*,0} = 3.36$ and $g_{*,0} = 3.91$.
Using $\Omega_{r0}h^2 = 4.18 \times 10^{-5}$ \cite{Planck:2018vyg} and $g_* (T_c) \approx g_{*s} (T_c) \approx 106.75$, this becomes 
\begin{align}
\label{eq:dimensionlessGWspectraldensity}
    \Omega_{\textrm{GW,0}} h^2 \approx 1.62 \times 10^{-5} \, \Omega_{\textrm{GW,c}}.
\end{align}

The dimensionless spectral density when the modes are contained within the Hubble radius during the radiation-dominated epoch is given by
\begin{align}
    \label{eq:GWspectrum}
    \Omega_{\textrm{GW,c}}(k,\tau)&=\frac{1}{24}\left(\frac{k}{aH}\right)^2 \overline{P_h(k,\tau)},
\end{align}
where the conformal time is defined as $\tau = (aH)^{-1}$ at horizon reentry in the radiation-dominated era, and the two respective polarization modes of GWs have been summed over. $P_h$ is the power spectrum of the induced tensor-mode perturbation sourced by linear scalar-mode perturbations at second order given by Eq.~ \eqref{def_P_h}, which can be solved via the Green's function method \cite{Ananda:2006af,Baumann:2007zm} as
\begin{align}\label{sol_h_Green}
    h_\lambda(\vec{k},\tau) = 4 \int^{\tau} d\tau_1 G_{\vec{k}}(\tau;\tau_1) \frac{a(\tau_1)}{a(\tau)} S_\lambda (\vec{k},\tau_1),
\end{align}
where $\lambda = +, \times$ are the two polarizations, $G_{\vec{k}}(\tau;\tau_1) =\frac{1}{k}\sin(k(\tau -\tau_1))$ is the Green's function in radiation domination, and $S_\lambda$ is the source term; detailed information is provided in Appendix~\ref{app:tensor_spectrum}.
The overline in Eq.~\eqref{eq:GWspectrum} denotes an average over a few wavelengths for time oscillations led by the Green's function.
The GW spectrum induced by curvature perturbations is given by~\cite{Kohri:2018awv,Domenech:2021ztg}
\begin{equation}
    P_h (\tau, k) = 2 \int_0^\infty \,dt \int_{-1}^1 \,ds \left[ \frac{t (2+t) (s^2-1)}{(1-s+t) (1+s+t)} \right]^2 I^2 (v, u, x) \mathcal{P_R}(kv) \mathcal{P_R}(ku),
    \label{eqn:Ph}
\end{equation}
where $u = (t+s+1)/2$, $v = (t-s+1)/2$, $x = k\eta$, and the appropriate kernel $I (v, u, x)$ is given in Appendix~\ref{app:tensor_spectrum}.

\subsection{Planck Constraints on PBH-seeding Multifield Inflation}
\label{sec:constraints}

\subsubsection{Data and Likelihood}
\label{sec:data}

To constrain the model presented above, we use data from the {\it Planck} 2018 CMB temperature and polarization anisotropies and lensing spectra, and enforce a minimal requirement that the model can produce PBHs that could comprise all DM.
We incorporate the CMB data using Gaussian priors corresponding to the {\it Planck} 2018 constraints on the $\Lambda$CDM cosmological model. Specifically, we use measurements of the spectral index $n_s (k_*)$, the amplitude $\ln\left[10^{10} A_s(k_*)\right]$, and the running of the spectral index  $\alpha (k_*)$, corresponding to the marginalized parameter constraints in the context of the $\Lambda$CDM model. We also enforce a one-sided Gaussian constraint on the tensor-to-scalar ratio $r (k_*)$ corresponding to the combined {\it Planck}-BICEP/Keck observations. The best fit values for these quantities and their error bars are given in Eq.~\eqref{eqn:PlanckCMBconstraints}.

In order to translate these constraints to an inflation model, one must assume a reheating scenario; the CMB measurements are reported at the pivot scale $k_*$, and calculating the time during inflation when this mode crossed the Hubble radius requires knowing the time spent during the reheating phase~\cite{Dodelson:2003vq,Liddle:2003as,Amin:2014eta,Martin:2021frd}.
If we assume that reheating is efficient and lasts for much less than one efold, then the CMB pivot scale for this model with typical parameters corresponds to $N_* \simeq 58$~\cite{Geller:2022nkr}; however, the longer the reheating phase lasts, the smaller $N_*$ is. 
In inflationary models similar to the one that we consider, post-inflation reheating is typically efficient and lasts for $N_\text{reh} \sim \mathcal{O} (1)$ $e$-folds \cite{Bezrukov:2008ut,Garcia-Bellido:2008ycs,Child:2013ria,DeCross:2015uza,DeCross:2016fdz,DeCross:2016cbs,Figueroa:2016dsc,Repond:2016sol,Ema:2016dny,Sfakianakis:2018lzf,Nguyen:2019kbm,vandeVis:2020qcp,Iarygina:2020dwe,Ema:2021xhq,Figueroa:2021iwm,Dux:2022kuk}.

In our analysis, we allow $N_*$ to take on values within the range typically considered \cite{Planck:2018jri}, $N_* = 55 \pm 5$, and fix $N_*$ to optimize our reheat history. In other words, we choose $N_*$ to be the value between 50 and 60 such that the CMB observables at that scale most closely match the measurements listed in Eq.~(\ref{eqn:PlanckCMBconstraints}). An alternative approach would be to marginalize over the possible reheat histories; however, since this would make our MCMC computationally expensive, we choose to fix $N_*$ using this simpler procedure, wherein $N_*$ is treated as a derived parameter parameterizing the optimal reheating scenario. We leave a dedicated study of reheating dependence, e.g., in analogy to Ref.~\cite{PhysRevD.93.103532}, to future work.

To summarize, we take the model likelihood to be the following:
\begin{enumerate}
    \item A Gaussian over the {\it Planck} and BICEP/Keck observables, $\{ \ln(10^{10} A_s), n_s (k_*), \alpha (k_*), r(k_*) \}$, corresponding to the {\it Planck} constraints on each of these quantities.
    
    \item A uniform likelihood for the peak of the power spectrum in the restricted range ${\cal P}_{\cal R} (k_{\rm pbh}) \geq 10^{-3}$, namely the threshold to form PBHs, and zero likelihood for the peak falling below this.
    
    \item A uniform likelihood for the {\it position} of the peak of the power spectrum, in the restricted range $14 \leq \Delta N \leq 25$, corresponding to the mass window where PBHs can comprise an $\mathcal{O}(1)$ fraction of DM, and zero likelihood outside of this range.
\end{enumerate}

Next, in order to determine the observability of the induced GWs from curvature perturbations, we compare our predicted signals to the sensitivity curves from LISA~\cite{Schmitz:2020syl}, LIGO A+~\cite{Cahillane:2022pqm}, the ET~\cite{Maggiore:2019uih}, CE~\cite{Reitze:2019iox}, and DECIGO~\cite{Yagi:2011wg,Kawamura:2020pcg}.
\begin{figure}
	\centering
	\includegraphics[scale=0.5]{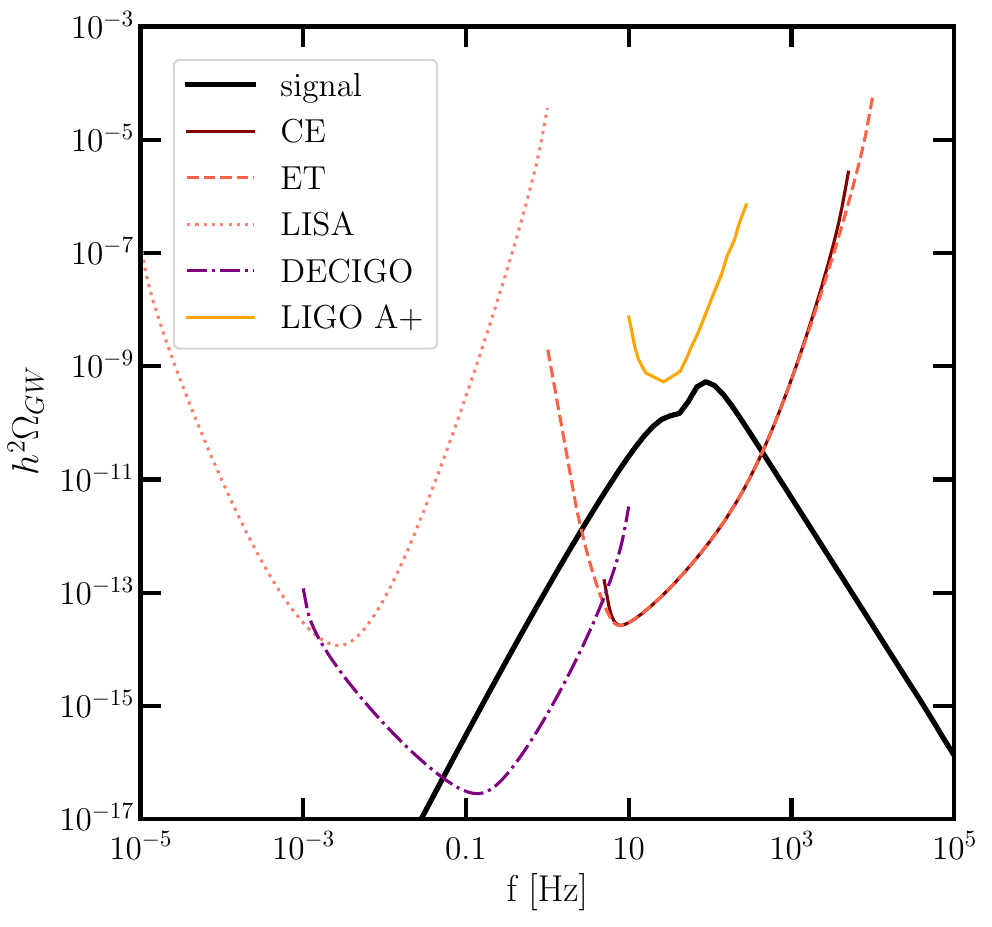}
	\caption{
	The spectral density of gravitational waves from our model with parameters $\xi=100$, $b=-1.8 \times 10^{-4}$, $c_1=2.5 \times 10^{-4}$, $c_2=3.570913 \times 10^{-3}$, and $c_4=3.9 \times 10^{-3}$.
	For comparison, we also plot the power-law integrated sensitivity curves for LIGO A+~\cite{Cahillane:2022pqm}, LISA~\cite{Schmitz:2020syl}, ET~\cite{Maggiore:2019uih}, CE~\cite{Reitze:2019iox}, and DECIGO~\cite{Yagi:2011wg,Kawamura:2020pcg}.}
	\label{fig:SGWB}
\end{figure}
In Fig.~\ref{fig:SGWB}, we show an example of a GW signal from our model with a particular set of parameters against these sensitivity curves.\footnote{Note that for plotting experimental sensitivities, we use power-law integrated sensitivity curves~\cite{Thrane:2013oya}. 
The power-law integrated curves will overlap with a SGWB when the signal-to-noise ratio (SNR) is greater than 1. Hence they are a better tool for visualizing the observability of a signal than the typically reported noise spectra.}
Given the total observation time $t_\text{obs}$ and noise spectrum of an experiment $\Omega_\text{noise} (f)$, we can calculate the signal-to-noise ratio (SNR) of a background of GWs with power spectral density $\Omega_\text{GW,0} (f)$:
\begin{equation}
\label{eq:SNR}
    \rho = \sqrt{2 t_\text{obs} \int_{f_\text{min}}^{f_\text{max}} df \, \left(\frac{\Omega_\text{GW,0} (f)}{\Omega_\text{noise} (f)}\right)^2 }
\end{equation}
The SNR for this signal is maximal for CE at $\rho=1811$ and also quite large for ET at $\rho=667$, whereas $\rho\simeq0$ in LIGO A+ and LISA, and finally in DECIGO the SNR is sizeable at $\rho=243$. 

The frequency limits for integration in the above expression, denoted $[f_\textrm{min},f_\textrm{max}]$, define the bandwidth of the detector. 
Eq.~(\ref{eq:SNR}) therefore represents the total broadband SNR, integrated over both time and frequency. 
It can be computed as the expected SNR of a filtered cross-correlation. 
Here, in general one assumes that the SGWB is sufficiently described by a power-law of the from $\Omega_\textrm{GW}$ = $\Omega_\beta (f /f_\textrm{ref})^\beta$ where $\beta$ is the spectral index and $f_\textrm{ref}$ is the reference frequency, which we set for example to $100$ Hz for ground-based detectors over the sensitivity region of interest. We set the observation times to the duration of data taking by the experiment. We can then use Eq.~\eqref{eq:SNR} to compute the value of GW amplitude required to reach a target SNR. In order to determine the detectability of the SGWB signal, we consider the spectrum from a particular inflationary model to be observable if it gives an SNR of $\rho \geq 1$.

\subsubsection{Results}
\label{sec:results}

We perform an MCMC analysis \cite{2013PASP..125..306F} of our multifield inflation model fit to cosmological data as described in Sec.~\ref{sec:data}. The posterior sampling is performed using an ensemble sampler \cite{2010CAMCS...5...65G} implemented in the Python package \texttt{emcee} \cite{2013PASP..125..306F}, with 200 walkers. We use the Python package \texttt{Corner} \cite{corner} for plotting results.

Given the scaling relationship in Eq.~\eqref{eqn:scaling}, we choose to fix $\xi =100$ and allow the remaining parameters $b$, $c_1$, $c_2$, and $c_4$ to vary.\footnote{Whereas the dynamics of these models become independent of $\xi$ in the limit $\xi \rightarrow \infty$, we expect that data would not be able to constrain $\xi$ due to the scaling relations of Eq.~(\ref{eqn:scaling}), and the relative constraints on the other parameters would be comparable for any fixed value of $\xi$. Hence we choose to fix $\xi$ to 100.}
We take broad uniform priors on the model parameters given by $b=[-10^{-3},-10^{-4}]$, $c_1 = [10^{-4},10^{-3}]$, $c_2 = [10^{-3},10^{-2}]$, $c_4 = [10^{-3}, 10^{-2}]$.

We assess convergence of our MCMC chains by a combination of the autocorrelation time~\cite{2010CAMCS...5...65G,2013PASP..125..306F} and stability of marginalized parameter constraints.
The \texttt{emcee} documentation recommends running an analysis for 50 autocorrelation times to ensure convergence; however, this would be prohibitively computationally expensive for our case. On the other hand, as also noted in Ref.~\cite{2013PASP..125..306F}, an accurate approximation to marginalized parameter constraints can be realized with significantly fewer samples.  
In total, we include approximately 1,300,000 samples for the final analysis, corresponding to an estimated 11 autocorrelation times. We find that as we vary the number of samples included in the analysis by 10\%, the marginalized parameter constraints (central value and error bars) vary at the sub-percent level.

\begin{table}[htb!]
	\centering
    Constraints from requiring PBH DM and satisfying \emph{Planck} 2018 data \\
    \renewcommand{\arraystretch}{1.5}
    \begin{tabular}{|l|c|}
    \hline\hline
    Parameter & Constraint \\ 
    \hline \hline
    
    $\mathbf{b}$ & $-1.87\, (-1.73)_{-0.11}^{+0.09} \times 10^{-4}$  \\
    $\mathbf{c_1}$ & $2.61 \,(2.34) _{-0.17}^{+0.24} \times 10^{-4}$  \\
    $\mathbf{c_2}$ & $3.69\, (3.42) _{-0.16}^{+0.22} \times 10^{-3}$ \\
    $\mathbf{c_4}$ & $4.03\, (3.75) _{-0.17}^{+0.24} \times 10^{-3}$ \\

    \hline
    
    $n_s (k_*)$ & $0.952 \, (0.956) _{-0.003}^{+0.002}$   \\
    $\ln (10^{10} A_s)$ & $3.049 \, (3.048) _{-0.001}^{+0.001}$  \\
     $N_*$ & $58.8\, (60.0) _{-2.2}^{+1.2}$  \\
     $\alpha (k_*)$ & $-0.0012\, (-0.0010) _{-0.0002}^{+0.0001}$  \\
     $r (k_*)$ & $0.019 \, (0.016) _{-0.001}^{+0.002}$ \\

    \hline

    $b/c_2$ & $-5.04 (-5.05) _{-0.05}^{+0.03} \times 10^{-2}$  \\
    $c_1/c_2$ & $7.07 (6.84) _{-0.26}^{+0.32} \times 10^{-2}$  \\
    $c_4/c_2$ & $1.091 (1.096) _{-0.008}^{+0.009}$  \\

    \hline
  \end{tabular} 
  \caption{
  The mean (best-fit) $\pm1\sigma$ constraints on parameters and derived observables of our multifield inflation model. Sampled parameters are in boldface.
  To generate these constraints, we fix $\xi=100$, require that the peak of the curvature power spectrum satisfies the requirements to produce PBH DM, and also include \emph{Planck} 2018 measurements on $A_s$, $n_s$, $\alpha$ and $r$.
  The last three rows of the table show the constraints on ratios of model parameters, which are subject to fewer degeneracies.
  }
  \label{tab:best_fit_params}
\end{table}

\begin{figure*}
	\centering
    \includegraphics[width=0.8\textwidth]{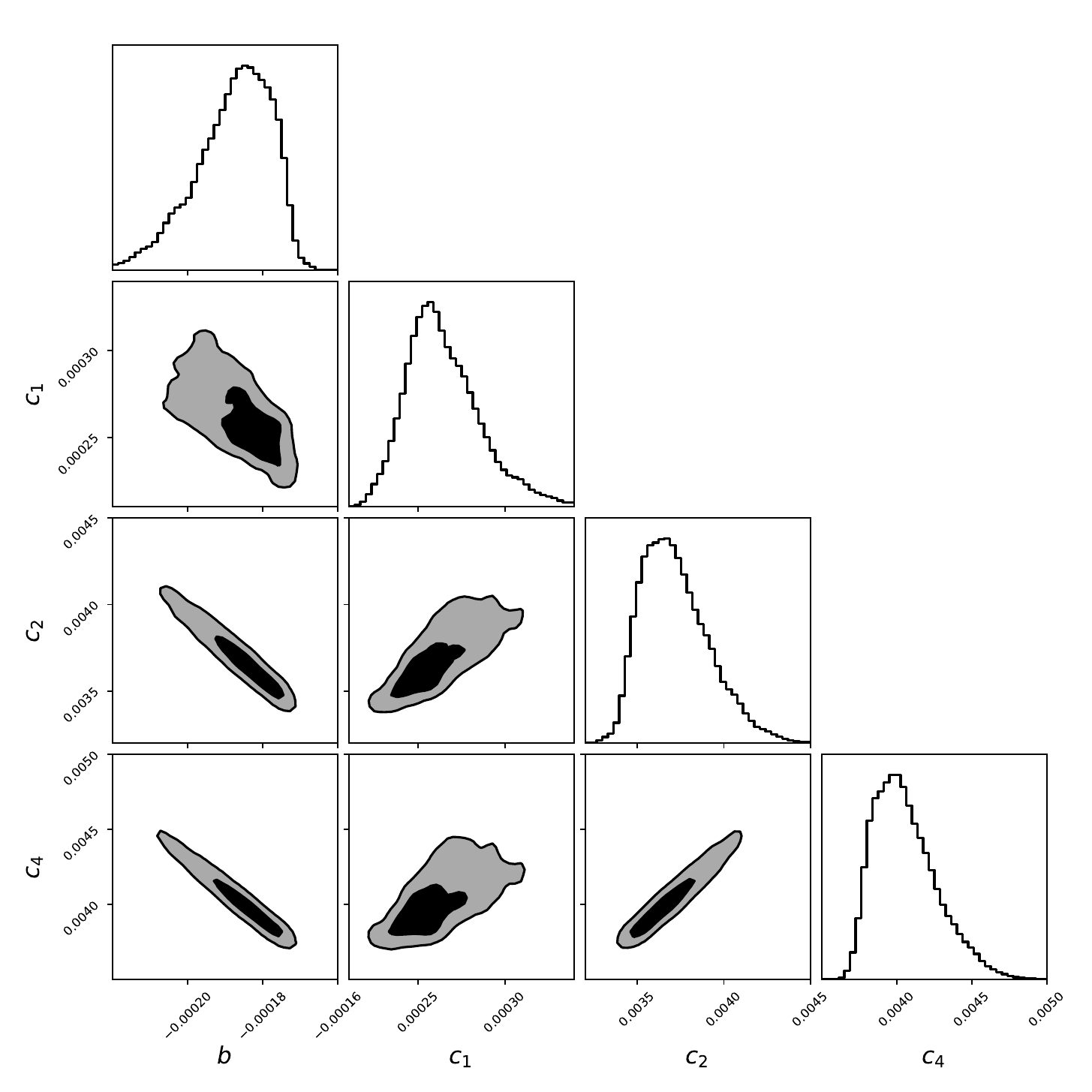}
    \\
    \includegraphics[width=0.6\textwidth]{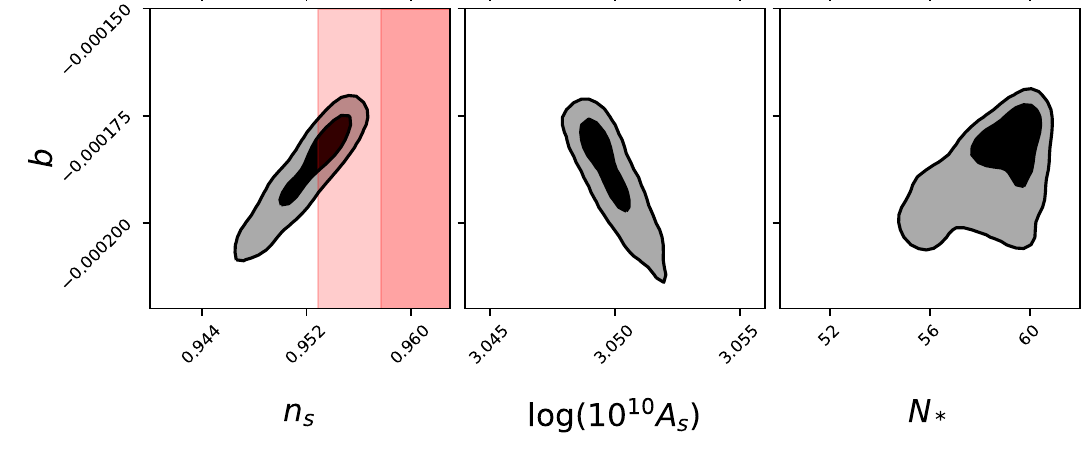}
    \caption{
    Posterior distributions in the fit of the two-field inflation model given by Eqs.~\eqref{Vtildertheta} and \eqref{bcxisymmetries} to CMB data from {\it Planck} 2018, with a prior that the model produce PBHs which could comprise all of DM.
    The last row also shows the posteriors between $b$ and derived observables, which are optimized over possible reheating histories.
    The black and grey filled contours correspond to the 68\% and 95\% deviations from the distribution means. 
    The red shaded regions show the 68\% and 95\% CL for $n_s (k_*)$ from {\it Planck} 2018.
    }
    \label{fig:MCMC_corner}
\end{figure*}
\begin{figure*}
	\centering
    \includegraphics[scale=0.5]{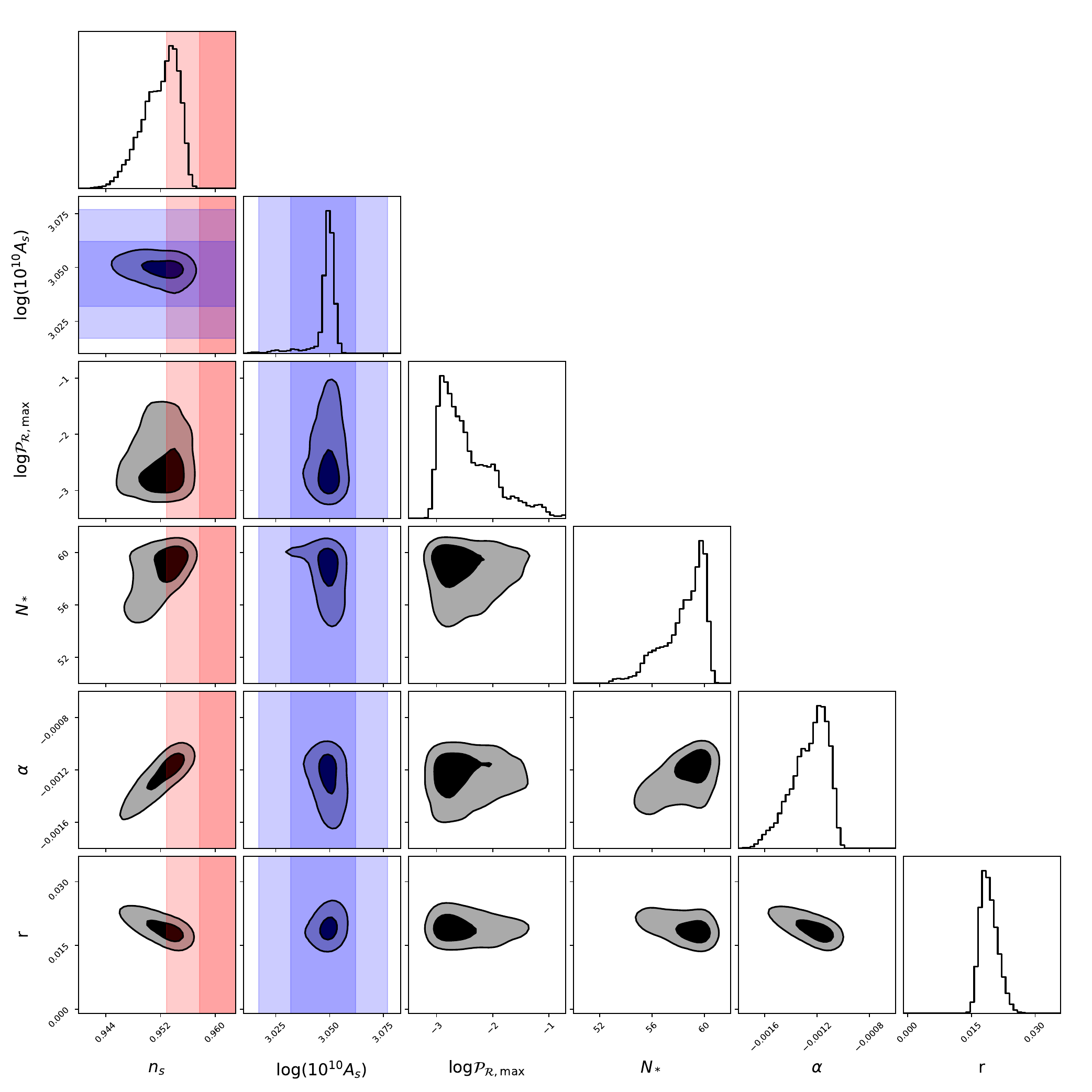}
    \caption{
    Posterior distributions on $n_s (k_*)$, $A_s (k_*)$, the peak of $\mathcal{P}_\mathcal{R}$, $N_*$, $\alpha (k_*)$, and $r(k_*)$; all of these quantities are optimized over possible reheating histories.
    The red shaded regions show the 68\% and 95\% CL for $n_s (k_*)$ from {\it Planck} 2018, and the blue shaded regions show the 68\% and 95\% CL for $A_s (k_*)$ from {\it Planck} 2018.
    There is a correlation between $n_s (k_*)$ and $N_*$. 
    At the high-end cutoff of $N_*=60$, the contours show that higher values of $n_s (k_*)$ are preferred; as $N_*$ decreases, $n_s (k_*)$ decreases as well, eventually falling well outside of the {\it Planck} 2018 constraints.
    Whereas there is also a clear correlation between $n_s (k_*)$ and $A_s (k_*)$, the variations in $A_s (k_*)$ are much less than the {\it Planck} 2018 1$\sigma$ errors, hence $A_s (k_*)$ is not a key constraint.
    }
    \label{fig:obs_corner}
\end{figure*}

The main results of this analysis are shown in Table~\ref{tab:best_fit_params} and Figs.~\ref{fig:MCMC_corner} and \ref{fig:obs_corner}.
Table~\ref{tab:best_fit_params} shows the marginalized posterior means, best-fit values, and corresponding error bars on model parameters, as well as the constraints on their ratios. The maximum likelihood model has $b=-1.73 \times 10^{-4}$, $c_1 = 2.34 \times 10^{-4}$, $c_2 = 3.42 \times 10^{-3}$, and $c_4 = 3.75 \times 10^{-3}$, and yields predictions for CMB observables $n_s (k_*) = 0.9560$, $\alpha (k_*)=-0.001$, $r (k_*)=0.016$, and $\log(10^{10} A_s)=3.048$, in excellent compliance with {\it Planck} constraints. 
This demonstrates that there is a region of parameter space in our model that is both compatible with {\it Planck} constraints and can produce PBH DM.

The posterior distributions on the model parameters are shown in Fig.~\ref{fig:MCMC_corner}, and posterior distributions for derived (cosmological) parameters are shown in Fig.~\ref{fig:obs_corner}. 
Derived parameters are analyzed in post-processing of the MCMC chains, for a subset of $\approx 4 \times 10^4$ samples. Consistent with \texttt{emcee} documentation \cite{2013PASP..125..306F}, we find the marginalized constraints on the model parameters $\{ b,c_1,c_2,c_4\}$ from this subset of steps are near-identical to those from the full MCMC chains, thus validating our use of a subset of points for constraints on derived parameters, such as $n_s$ and $A_s$.

From Fig.~\ref{fig:MCMC_corner}, we can clearly see the degeneracies discussed in \S~\ref{sec:degen1}.
As expected, $b$ is anticorrelated with the other $c_i$'s, while all the $c_i$ parameters are positively correlated with one another.
Moreover, at the larger end of the posterior distribution for $b$, we see a sharp cutoff, whereas toward smaller values there is a more gradual tail.
Due to the anticorrelation, this behavior is reversed for the $c_i$; the posteriors exhibit cutoffs at small values and a tail at larger values.

We can understand this behavior if we look at the marginalized posteriors for $b$ and the cosmological observables, a subset  of which is shown in the last row of Fig.~\ref{fig:MCMC_corner}.
The parameter $b$ shows a clear positive correlation with $n_s (k_*)$; as we move along the contour to more negative values for $b$, $n_s (k_*)$ decreases past the $2\sigma$ {\it Planck} 2018 error bars.
Hence, towards smaller values of $b$ (larger values of $c_i$), the posteriors show a tail corresponding to the Gaussian prior on the value of $n_s (k_*)$.
There is also a correlation between $b$ and $N_*$: larger values of $b$ produce models that prefer a larger $N_*$. Hence at a large enough value for $b$, we are eventually constrained by the requirement that $N_*$ be less than 60.
This explains the sharp cut-offs observed in the posteriors.

We reiterate that the distributions of the $n_s$, $A_s$, $N_*$, and the other observables shown in Figs.~\ref{fig:MCMC_corner} and \ref{fig:obs_corner} are the {\it optimal} values for a given set of model parameters. 
That is, for a given set of model parameters, we consider the optimal reheat history, rather than marginalizing over possible reheat histories. 
From Fig.~\ref{fig:obs_corner}, one may appreciate that in comparison with the {\it Planck} measurement of $A_s$, the (optimal) $A_s$ is essentially fixed in our model. Underlying this, however, is a delicate $A_s$-$n_s$ compensation in the choice of $N_*$, which in turn generates a spread of $N_*$ values (see, e.g., the $n_s-N_*$ posterior), in contrast to what one might expect if $n_s$ were solely driving the constraints on the models (which would lead to $N_*=60$). 
The optimization of reheat history is discussed further in Sec.~\ref{sec:deg}. 

In addition, although the resulting PBH masses tend to populate the lower end of the allowed DM window, $ 10^{-16}\lesssim M_{\textrm{PBH}}/M_\odot \lesssim 10^{-11}$, we find regions of parameter space that yield PBHs within that mass range. The tendency to produce lower mass black holes is driven by the {\it Planck} 2018 data; compliance with measurements at the pivot scale drives the parameters towards models with $\Delta N \lesssim 14$, which is consistent with the results of Ref.~\cite{Geller:2022nkr}. (See also Refs.~\cite{Byrnes:2018txb,Carrilho:2019oqg,Ando:2020fjm}.)
However, we emphasize that the estimation of the required $\Delta N$ range to obtain black holes in this window neglected the effects of non-Gaussianity in large-amplitude curvature perturbations, which would enhance power in large fluctuations and hence yield a higher probability of producing larger black holes~\cite{Byrnes:2012yx,Young:2013oia,Pattison:2017mbe,Biagetti:2018pjj,Kehagias:2019eil,Ezquiaga:2019ftu,Ando:2020fjm,Tada:2021zzj,Biagetti:2021eep,Ferrante:2022mui}.
We leave incorporating these effects to future studies.

Finally, while the model parameters are constrained at the $\sim 10 \%$ level, this is deceptive because of the strong degeneracies in parameter space.
If we instead examine the ratios of the parameters, as shown in the bottom rows of Table~\ref{tab:best_fit_params}, we see that these are constrained at the percent level. Hence, the parameters must be tuned to less than a percent in order for the model to produce an appropriate population of PBHs while also remaining in compliance with existing measurements.

\subsubsection{Degeneracy directions and compensations}
\label{sec:deg}

As mentioned in Sec.~\ref{sec:degen1}, we identify degeneracy directions in parameter space: directions along which parameter variations will leave the potential and power spectrum, and thus predictions for CMB observables, unchanged. In this section we study this quantitatively. 

\begin{figure}[h]   
    \centering
    \includegraphics[width=.5\textwidth]{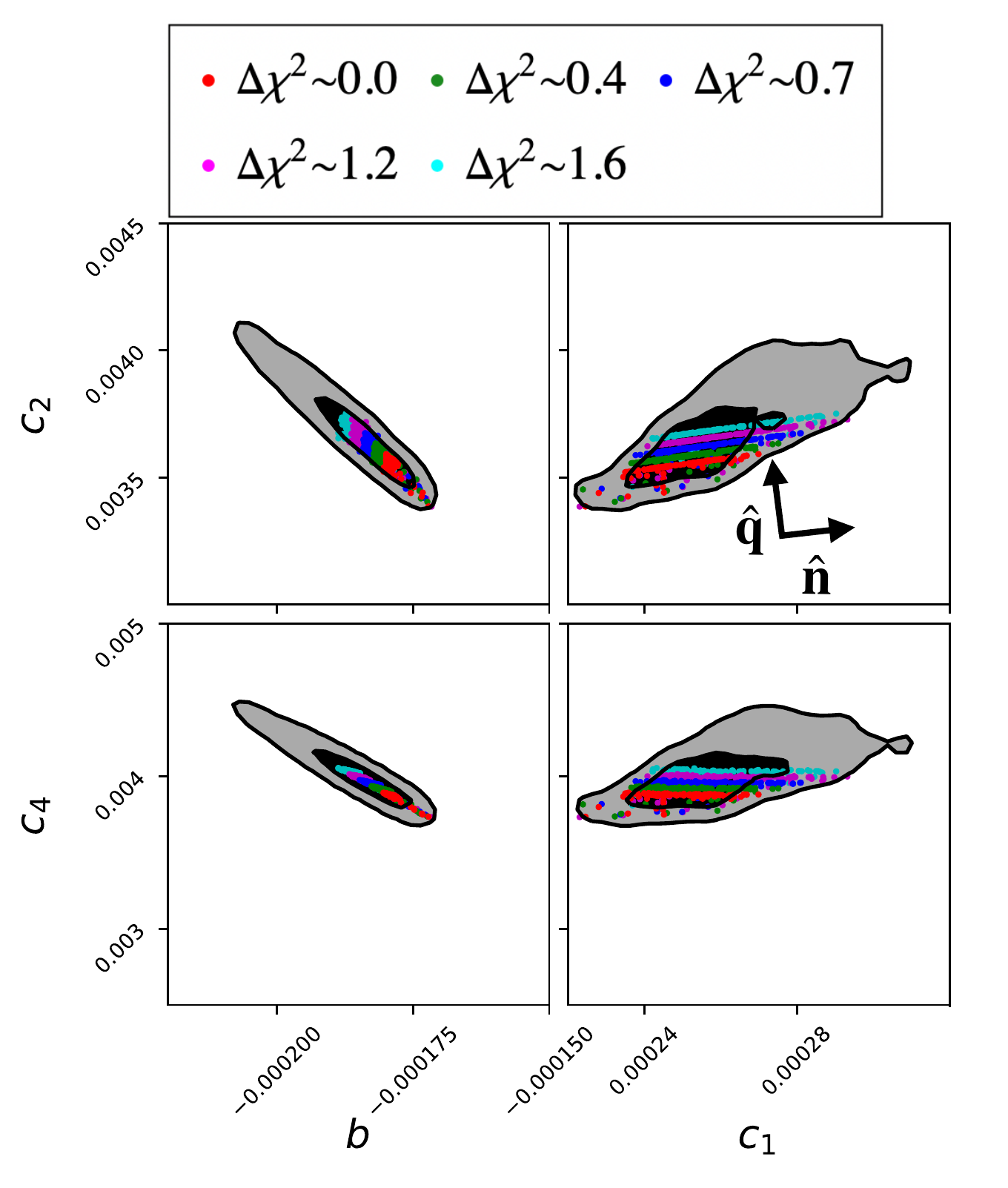}
    \caption{Four of the panels from the corner plot in Fig.~\ref{fig:MCMC_corner}, overlaid with a scatter plot of five sets of parameter sets. Points within a single color region yield nearly degenerate overall fits to the data, with differences of their total $\chi^2$ within $0.01$.
    The fiducial parameter set ${\cal F}$, for which the potential and power spectrum are plotted in Fig.~\ref{fig:V_PR_degen}, is one point within the red set of points. One example of a degeneracy direction is indicated in the top right panel by the unit vector $\hat{n}$, with orthogonal direction indicated by $\hat{q}$. Moving along $\hat{n}$ means moving along one color (with fixed $\chi^2$), whereas moving along $\hat{q}$ means moving from one color to another (changing $\chi^2$). The legend shows the approximate difference in total $\chi^2$ between each (degenerate) set of points relative to the fiducial parameter set ${\cal F}$. 
    } 
    \label{fig:bands}
\end{figure}

We consider two parameter sets to be degenerate if the difference of their total $\chi^2$ values is less than $0.01$. We define five super-sets of degenerate points (parameter sets) within the parameter space, as illustrated in the scatter plots in Fig.~\ref{fig:bands} superposed upon four of the two-dimensional posteriors, in which each super-set is assigned a color: red, green, blue, magenta, and cyan. Parameter sets within a given color yield nearly identical values of $\chi^2$; parameter sets belonging to different colors yield different values of $\chi^2$. The fiducial parameter set ${\cal F}$, used in the plots in Fig.~\ref{fig:V_PR_degen}, belongs to the set of red points in Fig.~\ref{fig:bands}.

\begin{figure}[h]   
    \centering
    \includegraphics[width=.48\textwidth]{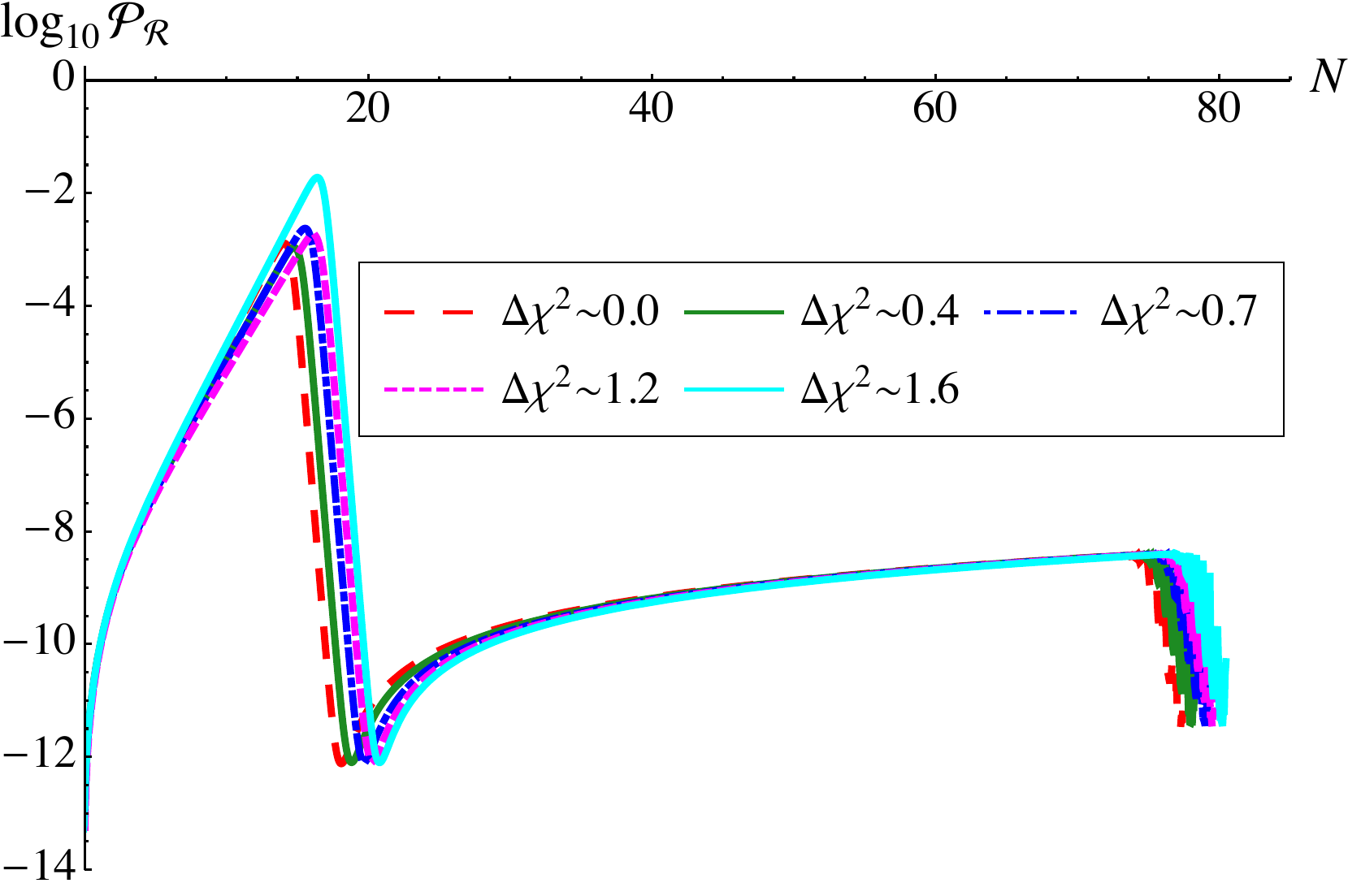}
    \caption{The power spectra for representative points in parameter space drawn from each of the five super-sets shown as distinct color bands in Fig.~\ref{fig:bands}. As in Fig.~\ref{fig:bands},
    $\Delta\chi^2$ for each parameter set is calculated relative to the $\chi^2$ of the fiducial parameter set $\mathcal{F}$ (red).}
     \label{fig:pr_rainbow}
\end{figure}

By computing the total $\Delta \chi^2$ relative to the red points, 
we can locally identify two directions within the parameter space. Along the degenerate direction $\hat{n}$, $\Delta\chi^2$ remains effectively constant (to within $0.01$). Along the orthogonal direction $\hat{q}$, $\Delta\chi^2$ changes appreciably, as shown in the legend of Fig.~\ref{fig:bands}. The power spectra for five choices of parameters are plotted in Fig.~\ref{fig:pr_rainbow}, where each curve represents a single point from the super-set of corresponding color shown in Fig.~\ref{fig:bands}. 

\begin{figure}[h]
	\centering
    \includegraphics[width=.48\textwidth]{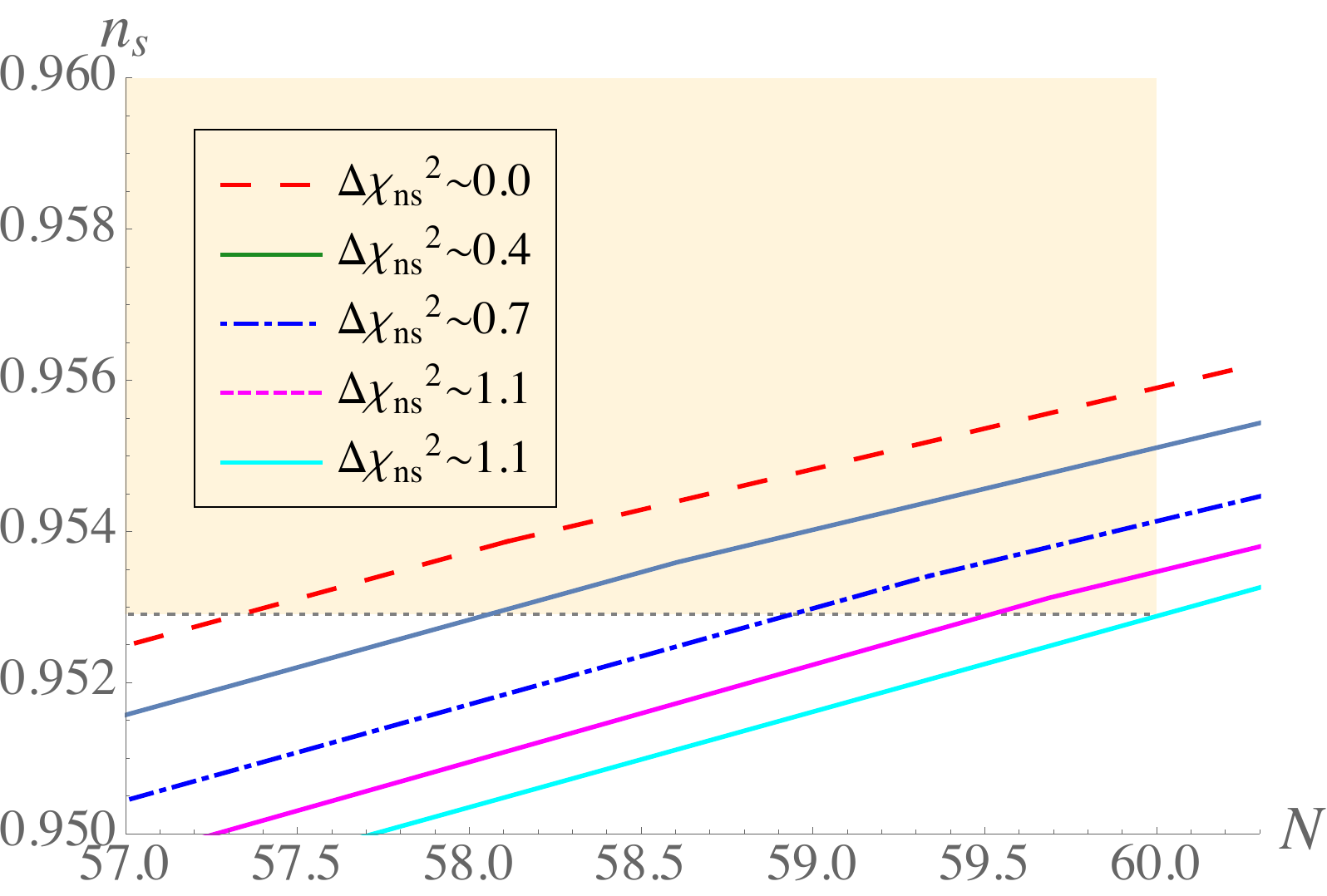}
    \caption{The spectral index $n_s$ for the same five sets of parameters used in Fig.~\ref{fig:pr_rainbow}, each drawn from the corresponding color bands in Fig.~\ref{fig:bands}.
    The legend shows differences in $\chi^2_{\rm ns}$ for each parameter set relative to the fiducial parameter set ${\cal F}$ (red). Moving along the direction $\hat{q}$ orthogonal to the degeneracy direction corresponds to moving from red to green, blue, magenta, and cyan, with increasing $\Delta \chi^2_{\rm ns}$.
    } 
    \label{fig:ns_bands}
\end{figure}

Our analysis allows $N_{*}$ to vary within a range of $(50,60)$ $e$-folds, and for a given parameter set, the ultimate fit to data is performed for the value of $N_{*}$ which minimizes $\chi^2$; we call this the \textit{optimal} $N_*$ value. From the $N_{*}-n_s$ posterior in Fig.~\ref{fig:obs_corner}, we see that a larger value of $n_s$, approaching the central {\it Planck} value, favors a smaller range of values for $N_{*}$, whereas a smaller value of $n_s$, moving away from the central {\it Planck} value, allows for a larger range of $N_{*}$. Whereas we might expect the opposite behavior due to the scaling of the running $\alpha$ with the size of $n_s$, this behavior is in fact explained by looking at the trends in $\Delta\chi^2_{n_s}$ and $\Delta\chi^2_{A_s}$ as we move from the fiducial (red) set along the orthogonal direction $\hat{q}$. Here we use the notation $\chi^2_{y_i}$ to mean the normalized $\chi^2$ given by the square of the difference between the {\it Planck} and model value divided by the $\sigma_{y_i}^2$ for observable $y_i$. 

The value of $\Delta\chi^2_{n_s}$, where the difference is calculated relative to the fiducial parameter set $\mathcal{F}$, changes as: $\Delta\chi^2_{n_s}=0$ (red), $0.4$ (green), $0.7$ (blue), $1.1$ (magenta), and $1.1$ (cyan). These values almost exactly track the overall differences $\Delta\chi^2$, for all but the cyan parameter set, showing that the change in $\Delta\chi^2$ is driven primarily by the fit to $n_s$ data. However, the optimal choice of $N_{*}$ is also somewhat driven by $A_s$. At larger values of $n_s$, i.e. closer to the \textit{Planck} central value, the compensatory behavior of $A_s$ pushes it further from its \textit{Planck} value, which results in an optimal choice of $N_*$ slightly below the value of 60 $e$-folds that would minimize $\chi^2_{n_s}$ alone. This compensation between $n_s$ and $A_s$ results in a wider spread of optimal $N_*$ values for larger $n_s$ (closer to the \textit{Planck} value) and a narrower spread for smaller $n_s$ (further from the \textit{Planck} value). At smaller $n_s$, $A_s$ is closer to the \textit{Planck} value, and $\Delta\chi^2_{A_s}$ remains small for a wider range of $N_*$ values. 

This interplay between values of $A_s$ and $n_s$ and the optimal value of $N_*$ can be seen by comparing the cyan and magenta curves in Fig.~\ref{fig:ns_bands}, which were chosen to have the same value of $n_s$ and thus equivalent $\Delta\chi^2_{n_s}$, but which have different values of $A_s$: $\Delta\chi^2_{A_s}\sim 0.4$ for cyan and $\Delta\chi^2_{A_s}\sim 0.008$ for magenta. As a result, the favored value $N_*$ is lower for the magenta parameter set (59.5) than for the cyan parameter set (60.0), indicating that the length of time for which the field experiences ultra slow-roll is longer for the cyan parameter set by about $0.5$ $e$-folds. This is consistent with the fact that the cyan parameter set shown in Figs.~\ref{fig:pr_rainbow}-\ref{fig:ns_bands} has a value of the coupling $b$ with larger magnitude $|b|$ than does the magenta parameter set. As we saw in Sec.~\ref{sec:degen1}, increasing the magnitude of $|b|$ increases the depth of the local minimum in the small-field feature of the potential, and thus lengthens the duration of ultra-slow roll. This results in a modest but noticeable increase in the 
height of the peak in $\mathcal{P}_{\mathcal{R}}$ for cyan relative to magenta, as can be seen in Fig.~\ref{fig:pr_rainbow}. The interplay between $A_s, n_s$ and the optimal choice of $N_*$ thus connects the small-field physics to the CMB-scale physics.

\subsection{Gravitational Wave Forecasts}

%
\begin{figure*}
	\centering
    \includegraphics[scale=0.54]{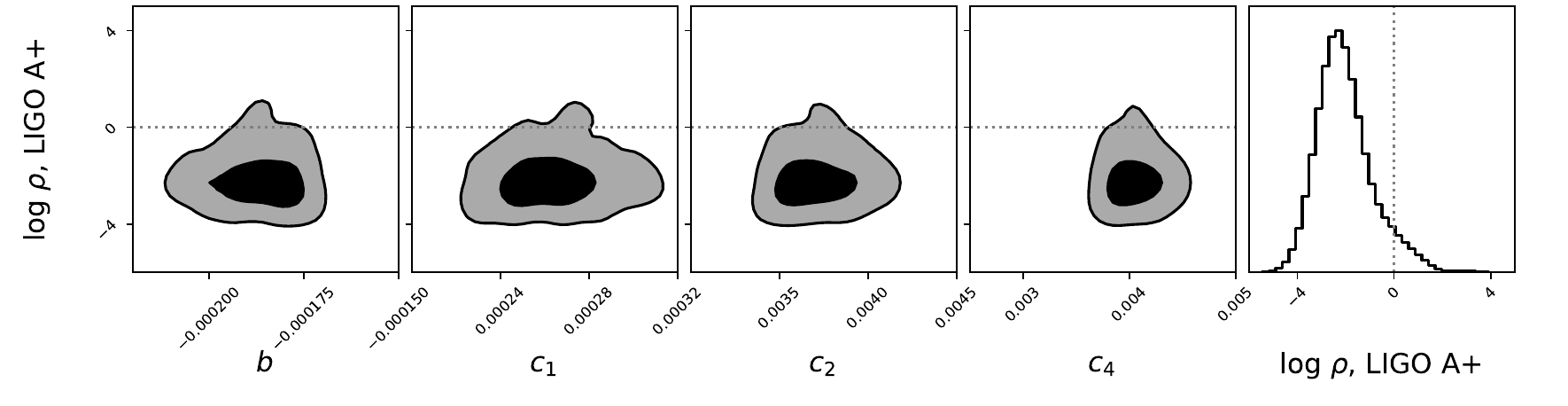}
    \includegraphics[scale=0.54]{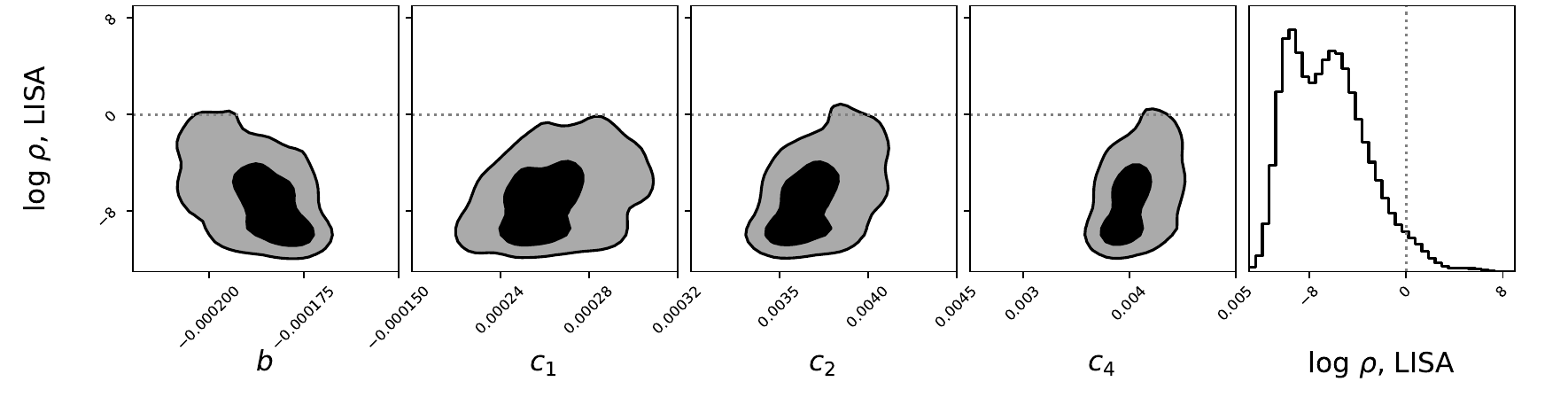}    \includegraphics[scale=0.54]{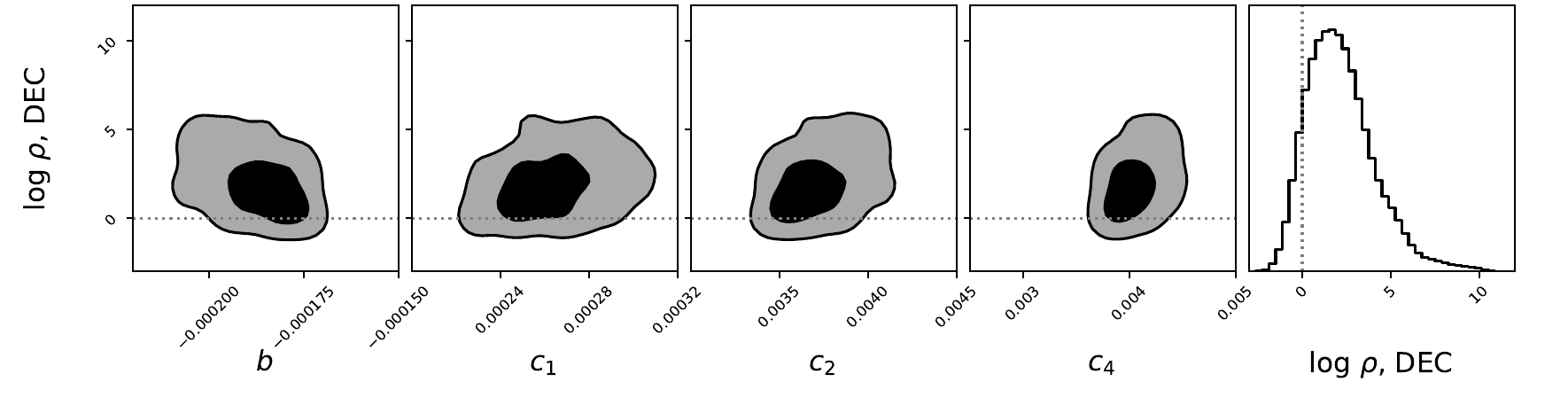}
    \includegraphics[scale=0.54]{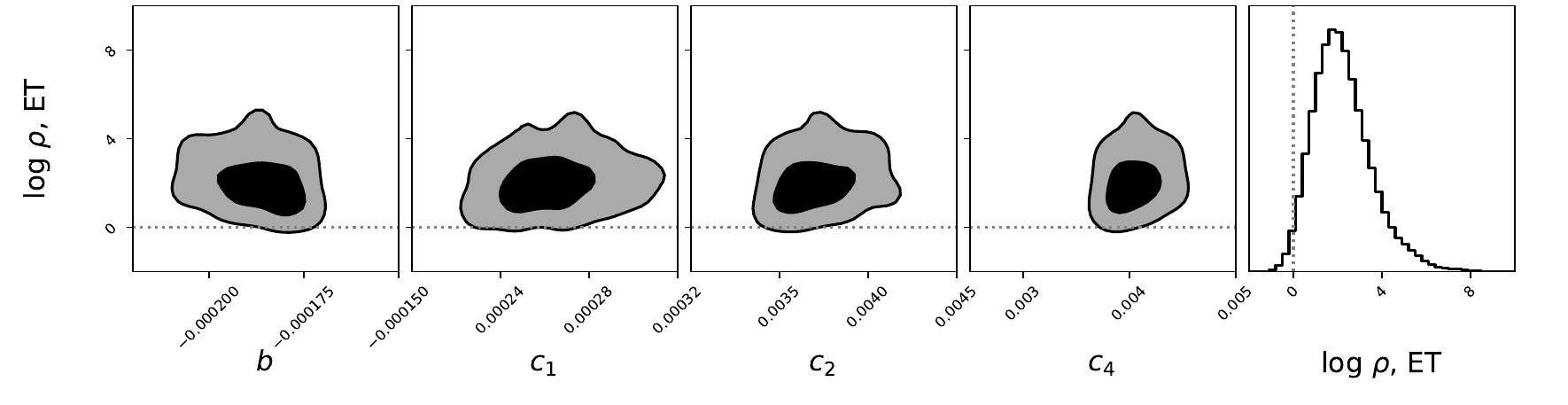}
    \includegraphics[scale=0.54]{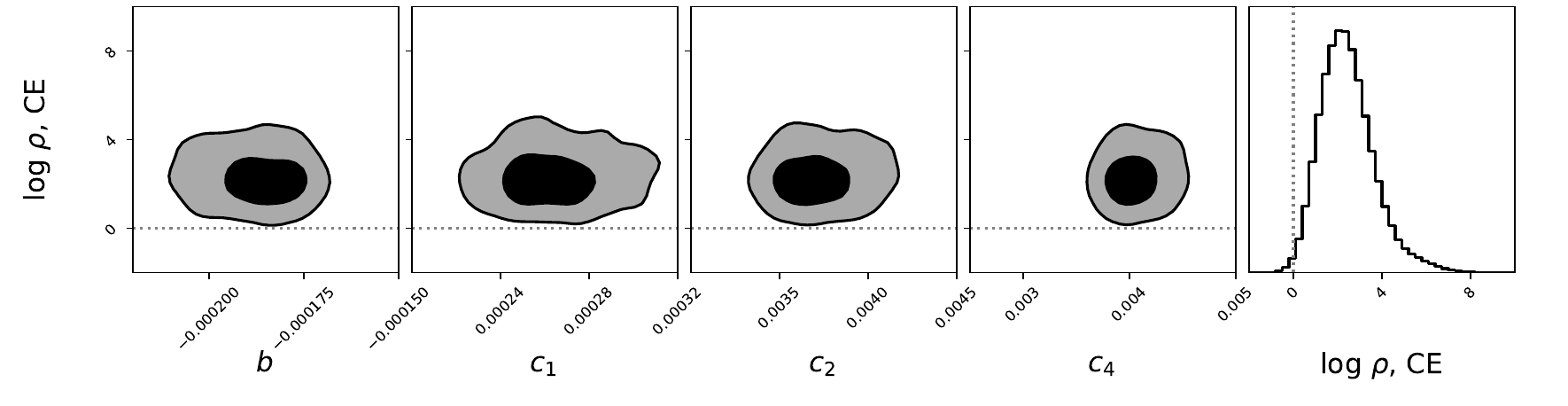}
    \caption{
    \footnotesize
    Signal-to-noise ratio ($\rho$) of gravitational waves from this multifield inflation model in LIGO A+, LISA, DECIGO, ET, and CE, shown from top to bottom respectively. Here, we use a subset of 13400 of the data points sampled by the MCMC.
    For all experiments, we find models in some parameter region with $\rho > 1$. 
    The sensitivity to these induced gravitational waves is greatest in ET and CE.
    }
    \label{fig:SNR_all}
\end{figure*}

In the previous sections, we have discussed the 68\% and 95\% CL parameter regions for our multifield inflation model in terms of the parameters $b, c_1, c_2, c_4$ and $\xi$, to obtain PBHs as all DM. 
This culminated in the marginalized constraints shown in \S~\ref{sec:results}. 
Since inflationary models with large amplifications of scalar perturbations can also produce secondary tensor perturbations, we compute the GW spectra that result from these tensor perturbations in the present universe, using the allowed parameter space in \S~\ref{sec:results}. 
We use the formulation described in \ref{sec:GWs} to compute the dimensionless present-day spectral density of GW modes contained within the Hubble radius, $\Omega_{\textrm{GW},0}h^2$, from Eq.~\eqref{eq:dimensionlessGWspectraldensity} and the corresponding SNR, $\rho$, as per Eq.~\eqref{eq:SNR} for various experiments. 
We describe in further detail the experiments and their sensitivities below.

At high frequencies, CE and ET are sensitive to the range $10$-$10^3$ Hz. 
At slightly lower frequencies, between e.g. $10^{-5}$-$10$ Hz, we also expect LISA and DECIGO to have some sensitivity to the models we consider. 
It should be noted that at lower frequencies, data from the International Pulsar Timing Array (IPTA), which comes from a combination of three pulsar timing array experiments, can search for GWs in the frequency band $10^{-9}$-$10^{-7}$ Hz. 
However, the multifield inflation models which produce PBHs that can comprise all of DM do not produce strong signals in this frequency range, hence we do not consider IPTA.

In Fig.~\ref{fig:SNR_all}, we show the SNR for the GWs from the sampled models in ET, CE, DECIGO, LISA, and LIGO A+. 
The dashed lines demarcate the threshold for observability where $\rho \geq 1$.  
We also list the mean $\rho$ for each experiment, along with 68\% and 95\% CL regions, in Table~\ref{tab:SNR}. 
ET, CE, and DECIGO are all very sensitive to the induced GWs from the sampled parameter region in Fig.~\ref{fig:MCMC_corner}, with most points yielding SNRs that are much larger than unity. 
For all three experiments, the centers of the posteriors lie at around $\rho\simeq 10^2$.
However, LISA and LIGO A+ have much less sensitivity for the favoured parameter region of Fig.~\ref{fig:MCMC_corner}. 
In the case of LISA, we see that only a very small part of the 95\% confidence region lies above $\rho=1$ and for LIGO A+, we see that there is only a slightly larger, albeit small region lying above the threshold.

The peak sensitivity for each experiment can be understood in terms of the comoving wavenumbers that exit the Hubble radius during inflation, as well as the GW frequencies. 
For ET and CE, the peak sensitivity is around $k_{\rm p}/k_{\rm eq}\simeq 10^{18}$, where $k_p$ is the position of the hypothetical peak of the primordial power spectrum and $k_{\rm eq}=0.01\,\textrm{Mpc}^{-1}$ is the comoving wavenumber that corresponds matter-radiation equality~\cite{Planck:2018vyg}.
For DECIGO the peak lies at around $k_{\rm p}/k_{\rm eq}\simeq 10^{16}$ and for LISA it is smaller still, at around $k_{\rm p}/k_{\rm eq}\simeq 10^{14}$. 
In our case, since producing PBHs as DM is an important physical requirement, we demand $k_{\rm p}=k_{\rm pbh}$, which we recall from \S~\ref{sec:PBHconstraints}. 
For the resulting marginalized posterior distribution favoured for this model, we see that the  $10^{16}\lesssim k_{\rm p}/k_{\rm eq}\lesssim 10^{18}$ is preferred, culminating in GWs favoured with frequencies lying in the mHz to kHz range; this is optimally situated for observation by ET, CE, and DECIGO. It should be noted that a very small region of the GW spectrum predicted by the posterior distribution is in the LIGO band. It has been shown that Gaussian density perturbations producing secondary GWs in the LIGO band produce PBHs that would have evaporated \cite{Kapadia:2020pnr}. However, since there is an underestimation of $\Delta N$ as described at the end of \S~\ref{sec:PBHconstraints} due to simplified assumptions, we confirm that the PBH mass would surpass the evaporation bound when propagating these effects. We intend to perform a detailed study of the PBH mass spectrum of which we leave to future work.

\begin{table}[htb!]
	\centering
    \renewcommand{\arraystretch}{1.5}
    \begin{tabular}{|l|c|c|c|}
    \hline\hline 
    Experiment & $\log_{10} \, \rho$ & $68\%$ CL & $95\%$ CL \\
    \hline \hline
    LIGO A+ & $-2.25$ & ${}^{+1.18}_{-0.74}$ & ${}_{-1.34}^{+3.02}$ \\

    LISA & $-6.52$ & ${}_{-3.33}^{+3.28}$ & ${}_{-3.47}^{+6.99} $\\

    ET & 2.04 & ${}_{-0.95}^{+1.35}$ & ${}_{-1.58}^{+3.23}$ \\

    DECIGO & 1.91 & ${}_{-1.49}^{+1.91} $&$ {}_{-2.22}^{+4.49}$ \\

    CE & 2.32 & ${}_{-0.99}^{+1.17}$ & ${}_{-1.55}^{+3.15}$ \\

    \hline
  \end{tabular} 
  \caption{
    The signal-to-noise ratio ($\rho$) for relevant experiments with upper and lower $68\%$ and $95\%$CL bounds. 
    The central values and bounds are shown in log scale.
    }
  \label{tab:SNR}
\end{table}

In summary, we find tantalizing GW phenomenology predicted for the posterior distribution of model parameters in our multifield inflation model. 
The correspondence between PBHs as DM and primordial GWs remains a compelling prospect for future ground- and space-based experiments.

\subsection{Conclusions}
\label{sec:conclusion}

We have performed an MCMC analysis of a simple yet generic multifield inflation model characterized by two fields coupled to each other and nonminimally coupled to gravity. This model was fit to {\it Planck} 2018 data, parametrized by measurements of the spectral parameters $A_s$, $n_s$, $\alpha$, and $r$, and with a prior that the primordial power spectrum should lead to PBH production in the ultralight asteroid-mass range, where constraints still allow for PBHs to account for all of DM. We find a nontrivial region of parameter space in our model that is both compatible with {\it Planck} data and can produce PBH DM. The constraints on allowable regions of parameter space are driven in particular by $n_s$ and $N_*$.

There are a number of reasons why we choose to focus on this multifield model.
The Standard Model includes multiple scalar fields, and extensions to the Standard Model typically feature many more. Therefore, embeddings of inflationary dynamics within realistic models of high-energy physics are likely to involve multiple interacting scalar fields. Given that several types of
single-field inflation models have successfully yielded predictions for PBHs as DM while remaining in compliance with CMB data~\cite{Carr:2020gox,Carr:2020xqk,Green:2020jor,Escriva:2022duf,Escriva:2022duf,Ozsoy:2023ryl}, it seems natural to study whether multifield models can accomplish the same, while also reducing the necessary amount of fine-tuning to produce black holes. The family of models we consider is a natural generalization of multifield models that have been studied extensively in the literature~\cite{Kaiser:2012ak,Kaiser:2013sna,Schutz:2013fua,Kaiser:2015usz}, and is closely related to well-known examples such as Higgs inflation \cite{Bezrukov:2007ep,Greenwood:2012aj,Kawai:2014gqa,Kawai:2015ryj} and $\alpha$-attractors \cite{Kallosh:2013hoa,Kallosh:2013maa,Galante:2014ifa}. 

The results of our MCMC show that there is a robust region of parameter space for which this family of models can produce PBHs in the appropriate mass range to comprise all of DM, while also remaining in compliance with {\it Planck} data.
In particular, the posteriors on all parameters show a Gaussian-like tail at one end that is controlled by the measurement of $n_s (k_*)$, and a sharp cutoff at the other end from our requirement that $N_*$ remain in the range $55 \pm 5$.
Due to our procedure for optimizing over possible reheat scenarios, we find that the optimal $N_*$ typically fixes $A_s$ to the central value of the {\it Planck} measurement.

Through this analysis, we found that whereas the parameters of the model are constrained at around a 10\% level, there is a degeneracy direction in the parameter space that leads to fine-tuned ratios at the percent level. 
It is possible that the actual required level of fine-tuning is  
greater than this, given that studies of single-field inflation models have found that model parameters typically need to be fine-tuned to as much as one part in $10^7$ in order to give rise to enhancements to the power spectrum that are sufficiently large to produce PBH DM~\cite{Garcia-Bellido:2017mdw,Ezquiaga:2017fvi,Kannike:2017bxn,Germani:2017bcs,Motohashi:2017kbs,Di:2017ndc,Ballesteros:2017fsr,Pattison:2017mbe,Passaglia:2018ixg,Byrnes:2018txb,Biagetti:2018pjj,Carrilho:2019oqg,Inomata:2021tpx,Inomata:2021uqj,Pattison:2021oen}. 
On the other hand, our results suggest that our multifield model may require less fine-tuning than some well-studied single-field models.

Upon fixing $\xi$, predictions from our multifield model depend on only four free parameters. Moreover, we can shift any one of these parameters at the 10\% level, and the constraints on the parameter ratios require the remaining three to be tuned at the 1\% level, yielding a total degree of fine-tuning of approximately $10^{-7}$. However, since the most constraining quantity is $n_s$, which has error bars at the 1\% level, the relative amount of fine-tuning needed to produce PBHs is $10^{-5}$. While there exist more rigorous measures of fine-tuning~\cite{Athron:2007ry,Fowlie:2014xha}, we leave such a quantitative analysis to future work.

Furthermore, the allowed parameter region in this model produces observable GW signals in frequency ranges that future experiments such as LIGO A+ and Virgo, ET, CE, DECIGO and LISA are projected to be sensitive to. 
The observational prospects for DECIGO, ET, and CE are particularly good for this model; the latter two experiments have a central region of the MCMC posterior distribution with signal-to-noise ratio of $\rho > 100$. 
This result suggests that this inflation model is a viable and well-motivated candidate to explain both the observed DM and to generate observationally relevant primordial GWs.
 
Given these results, there are a number of interesting directions for future work. First, since we find that the spectral index $n_s$ is especially constraining for these models, it may be instructive to look at forecasts for constraints by CMB-S4~\cite{Abazajian:2019eic,CMB-S4:2016ple}. Moreover, improving measurements on the running of the spectral index $\alpha (k_*)$ could also play an important role in helping to distinguish among such models. In addition, predictions for the tensor-to-scalar ratio, $r (k_*) \simeq 0.016$, are approximately a factor of two below current observational bounds \cite{BICEP:2021xfz}, whereas similar models (with multiple interacting scalar fields and nonminimal couplings, but with no small-field features that could yield PBHs) tend to predict considerably smaller values, $r (k_*) \simeq 0.004$ \cite{Kaiser:2013sna}. Thus we expect measurements of $r(k_*) $ to play a key role in future tests of this model. Along these same lines, it would be interesting to consider multifield models that combine the suppression of $r(k_*)$ found in Ref.~\cite{McDonough:2020gmn} with the enhancement on small scales studied here. We leave this for future work.

Second, in this work, we examined the SGWB resulting from inflation models that can produce PBHs as DM. It would also be interesting to perform a dedicated MCMC analysis to find the regions of parameter space for such models that produce detectable GWs, regardless of the implications for DM. 

Third, given the discussion above of the amount of fine-tuning in our model, another direction one could pursue would be a Bayesian comparison of our model versus similar single-field models, or to further consider more rigorous quantitative measures of fine-tuning in these classes of models. (See, e.g., Ref.~\cite{Cole:2023wyx}.)

Finally, a more systematic analysis of uncertainties associated with the post-inflation reheating phase \cite{Martin:2021frd} would be valuable for considering how the domain of viable parameter space for the multifield models studied here compares with those of other types of inflationary models. It is likely that if we were to sample $N_*$ within the range $55 \pm 5$ and then marginalize over it, that would increase the error bars in the estimates of our parameter constraints, which in turn would reduce the implied degree of fine-tuning for such models. Such questions remain a topic for further research.

\chapter{New Analytic Tools for 21\,cm Cosmology}
\label{sec:EFTof21cm}

21\,cosmology is an exciting new probe of the early universe that will allow us to see the cosmos at redshifts around the epoch of reionization and cosmic dawn, which have not yet been directly observed.
Radio interferometers such as HERA~\cite{DeBoer:2016tnn,HERA:2021bsv} are currently searching for fluctuations in the 21\,cm line and are likely to deliver the first 21\,cm power spectrum measurement within the decade.
The theory of the 21\,cm signal has traditionally been driven by computationally expensive methods such as hydrodynamic or seminumerical simulations.
This is mainly because reionization is a relatively rapid and patchy process; hence this regime was considered to be too nonlinear to study with analytic methods.
However, at redshifts and length scales that will be observable by HERA and future radio telescopes, the signal is in fact only mildly nonlinear and perturbative enough to be treated with e.g. effective field theory.

In this chapter, I will discuss my work to establish an analytic description of the 21\,cm signal from the early universe using techniques from effective field theory.
We include realistic observational effects such as redshift space distortions and also validate our description against \texttt{THESAN}, a suite of state-of-the-art radiation-magneto-hydrodynamic simulations of the epoch of reionization.
This chapter is based on Ref.~\cite{Qin:2022xho} and was performed in collaboration with Katelin Schutz, Aaron Smith, Enrico Garaldi, Rahul Kannan, Tracy Slatyer, and Mark Vogelsberger. 
I led this work and performed the majority of the calculations, including renormalizing the contributions from redshift space distortions and analyzing the output of the simulations.

\section{An Effective Bias Expansion for 21 cm Cosmology in Redshift Space}

The 21\,cm transition of neutral hydrogen provides a promising avenue for mapping out large scale structure (LSS) and testing cosmological theories at redshifts where there are few or no other detectable luminous tracers of the underlying matter field. 
Most empirical cosmological information either comes from measurements of the cosmic microwave background (CMB), which was emitted around the time of recombination $z \approx 1100$, or from surveys of tracers like galaxies at lower redshifts. 
To better understand how structure in our universe evolved at intermediate redshifts, we need observations of the diffuse neutral hydrogen gas from immediately after recombination through to the Epoch of Reionization (EoR).

Several experiments are already actively attempting to map the cosmological 21\,cm signal from the EoR, both at the level of the global signal (monopole) using experiments like EDGES~\cite{Monsalve:2016xbk}, LEDA~\cite{2018MNRAS.478.4193P}, PRI$^Z$M~\cite{2019JAI.....850004P}, and SARAS~\cite{Singh:2017syr}, as well as the fluctuations in the 21\,cm signal using interferometric experiements like PAPER~\cite{2010AJ....139.1468P}, the MWA ~\cite{2013PASA...30....7T,2013PASA...30...31B}, LOFAR~\cite{2013A&A...556A...2V}, HERA~\cite{DeBoer:2016tnn}, and the upcoming Square Kilometre Array (SKA)~\cite{Weltman:2018zrl}; there are also a number of post-reionization intensity mapping efforts such as CHIME~\cite{2014SPIE.9145E..22B}, HIRAX~\cite{Newburgh:2016mwi}, and CHORD~\cite{2019clrp.2020...28V}.
There has already been a tentative detection of the global signal from cosmic dawn in the form of a deep absorption trough at $z\sim 17$ made by the EDGES collaboration~\cite{Bowman:2018yin}, although this interpretation is in strong tension with observations from SARAS~3~\cite{Singh:2022}.
Further study of this feature is a major goal of 21\,cm experiments moving forward, while at the same time there is a push towards measuring the 21\,cm power spectrum from the EoR and eventually performing full tomographic mapping.

One complication of measuring the 21\,cm power spectrum is redshift space distortions (RSDs), which are contributions to the observed redshift that arise due to the peculiar velocities of neutral hydrogen rather than Hubble expansion. 
In other words, using the redshift of an observed line emission to infer a distance without accounting for line-of-sight peculiar velocities will yield the wrong distance. 
One can only directly measure distances in this illusory ``redshift space'' since there is no other independent way of inferring the peculiar velocity of the gas; 
thus, we are faced with the problem of extracting information about real space cosmology from redshift space observables. 
This is particularly relevant for interferometric measurements of the 21\,cm EoR signal, since substantial foregrounds that contaminate the signal lie in a ``wedge'' in $k_\parallel$ vs. $k_\perp$, where $k_\parallel$ and $k_\perp$ denote the line-of-sight and transverse components of Fourier modes, respectively. 
Modes with even moderate projections onto the $k_\perp$ direction will be within the foreground wedge whereas modes with larger projections onto the line-of-sight direction will be less contaminated by foregrounds~\cite{2010ApJ...724..526D,2012ApJ...745..176V,2012ApJ...752..137M,2012ApJ...756..165P,Trott:2012md,2013ApJ...768L..36P,2013ApJ...770..156H,Thyagarajan:2013eka,Liu:2014bba,Liu:2014yxa,2015ApJ...804...14T,2015ApJ...807L..28T,2016ApJ...833..242L,2016MNRAS.458.2928C,2018MNRAS.476.3051A}. 
In addition to evading foregrounds, from an instrumental perspective the high-$k$ modes that are nearly parallel to the line of sight are more readily observable due to the ease of attaining high spectral resolution as opposed to angular resolution. 
Experiments like HERA~\cite{DeBoer:2016tnn} therefore predominantly observe modes that are nearly parallel to the line of sight, and these are precisely the modes that will be most affected by RSDs~\cite{2016MNRAS.456...66J}.
For useful reviews on 21\,cm foreground mitigation, see Refs.~\cite{2019arXiv190912369C} and \cite{2020PASP..132f2001L}.

In this chapter, we parametrize the effects of RSDs on the 21\,cm field using techniques inspired by effective field theory (EFT)~\cite{Baumann:2010tm,Carrasco:2012cv}. 
In recent years, EFT techniques have become a powerful tool for studying large scale structure~\cite{Carrasco:2013sva,Carrasco:2013mua,Carroll:2013oxa,Senatore:2014via,Baldauf:2015zga,Foreman:2015lca,Baldauf:2015aha,Cataneo:2016suz,Lewandowski:2017kes,Konstandin:2019bay,DAmico:2019fhj,Ivanov:2019pdj}. 
As structure formation progresses, nonlinear effects at a given scale become increasingly important; in other words, while density perturbations in the recombination epoch can be accurately described purely by linear theory, perturbations in the EoR cannot. With EFT techniques, one can systematically treat mildly nonlinear effects to increasingly high accuracy, up to some cutoff scale where structure formation becomes fully nonlinear. 
More specifically, we use EFT-inspired methods to treat the feedback of small-scale nonlinear effects on the larger scales of interest; this procedure is analogous to renormalization~\cite{Assassi:2014fva}.

\begin{figure*}
	\centering
    \begin{tabular}{cc}
    \includegraphics[width=0.35\textwidth]{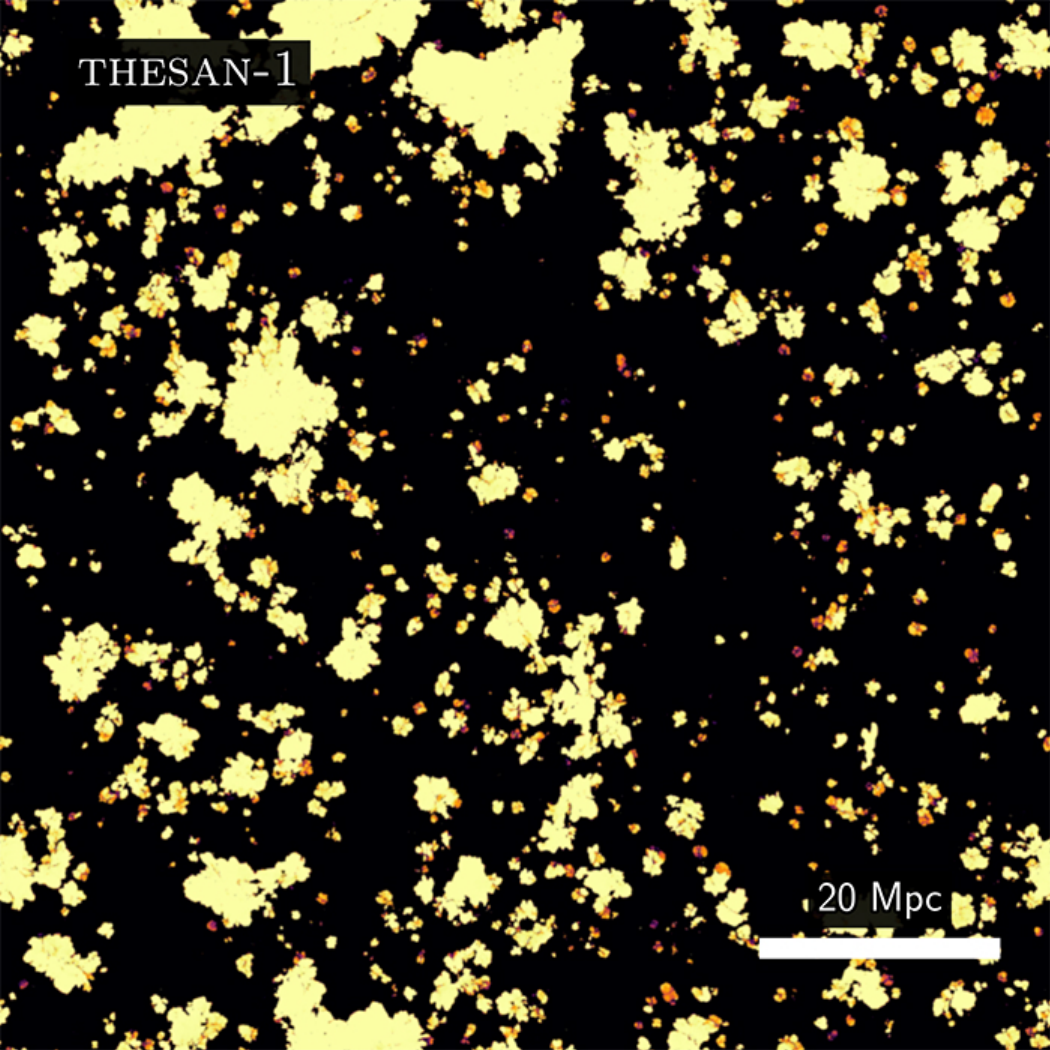} & \quad \includegraphics[width=0.35\textwidth]{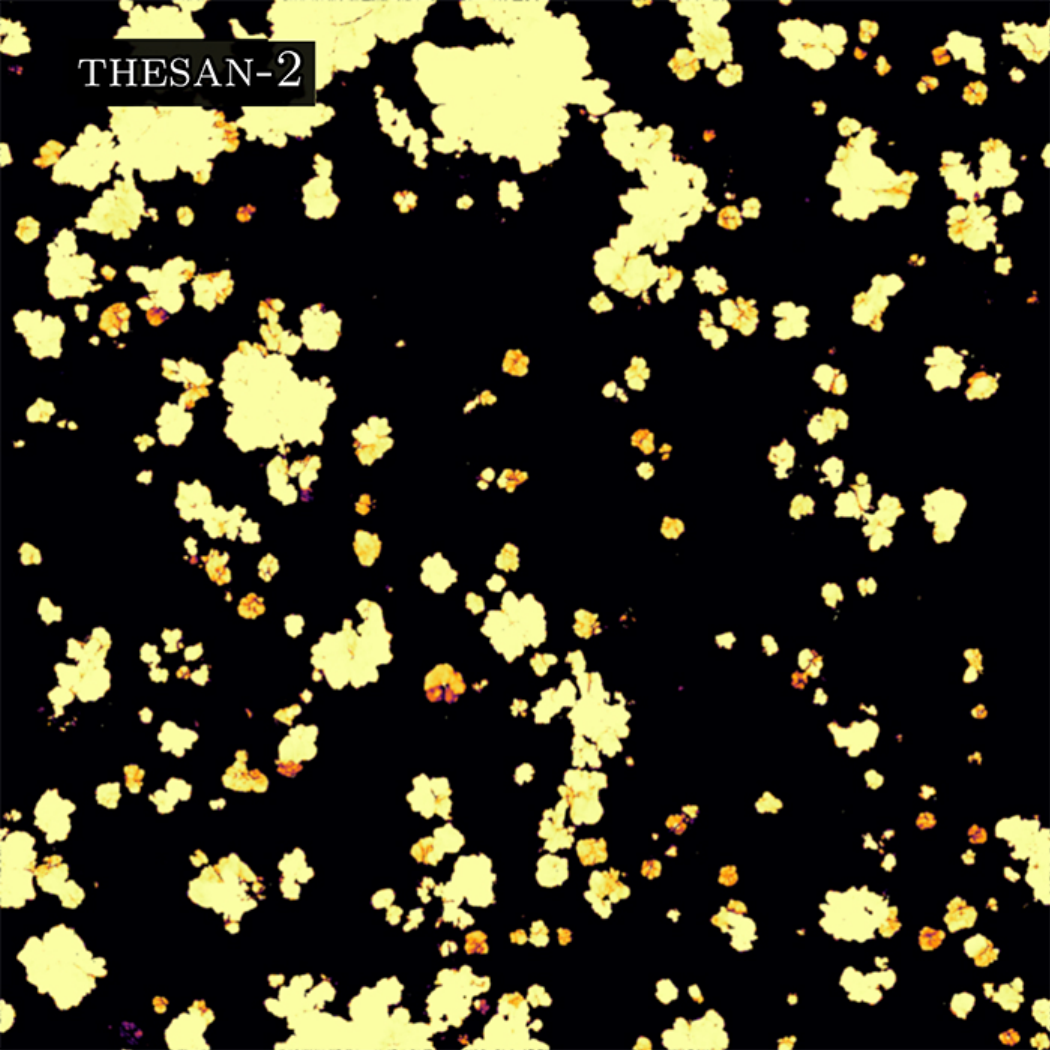} \\ \\
    \includegraphics[width=0.35\textwidth]{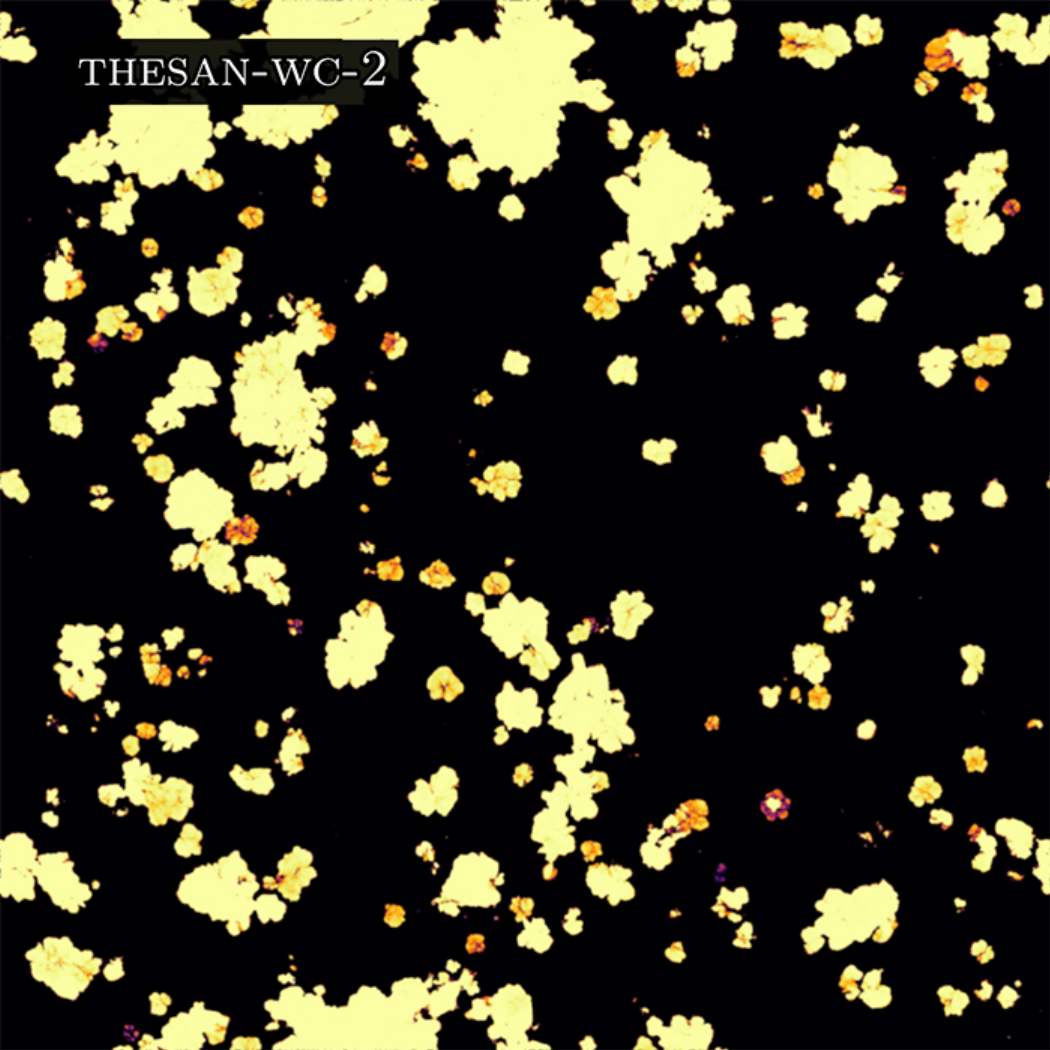} & \quad \includegraphics[width=0.35\textwidth]{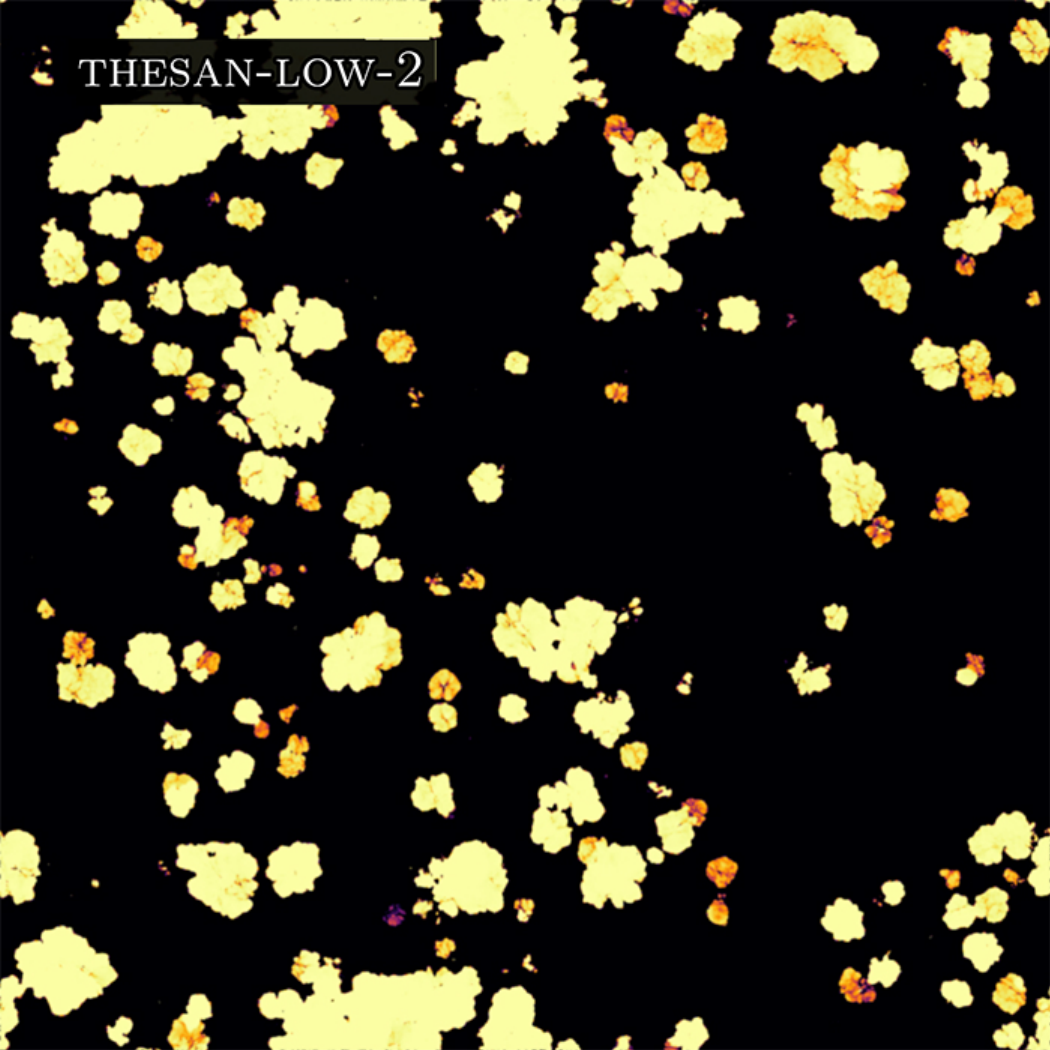} \\ \\
    \includegraphics[width=0.35\textwidth]{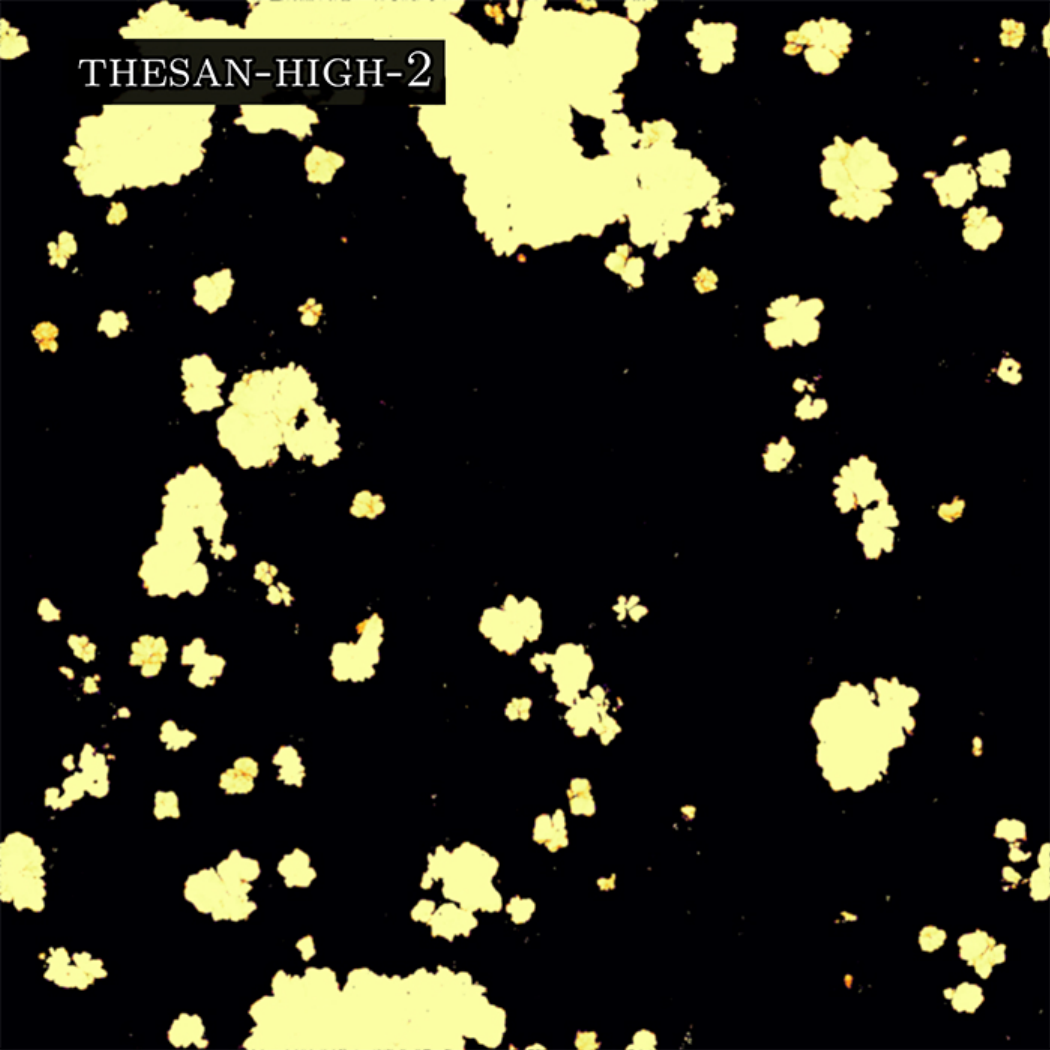} & \quad \includegraphics[width=0.35\textwidth]{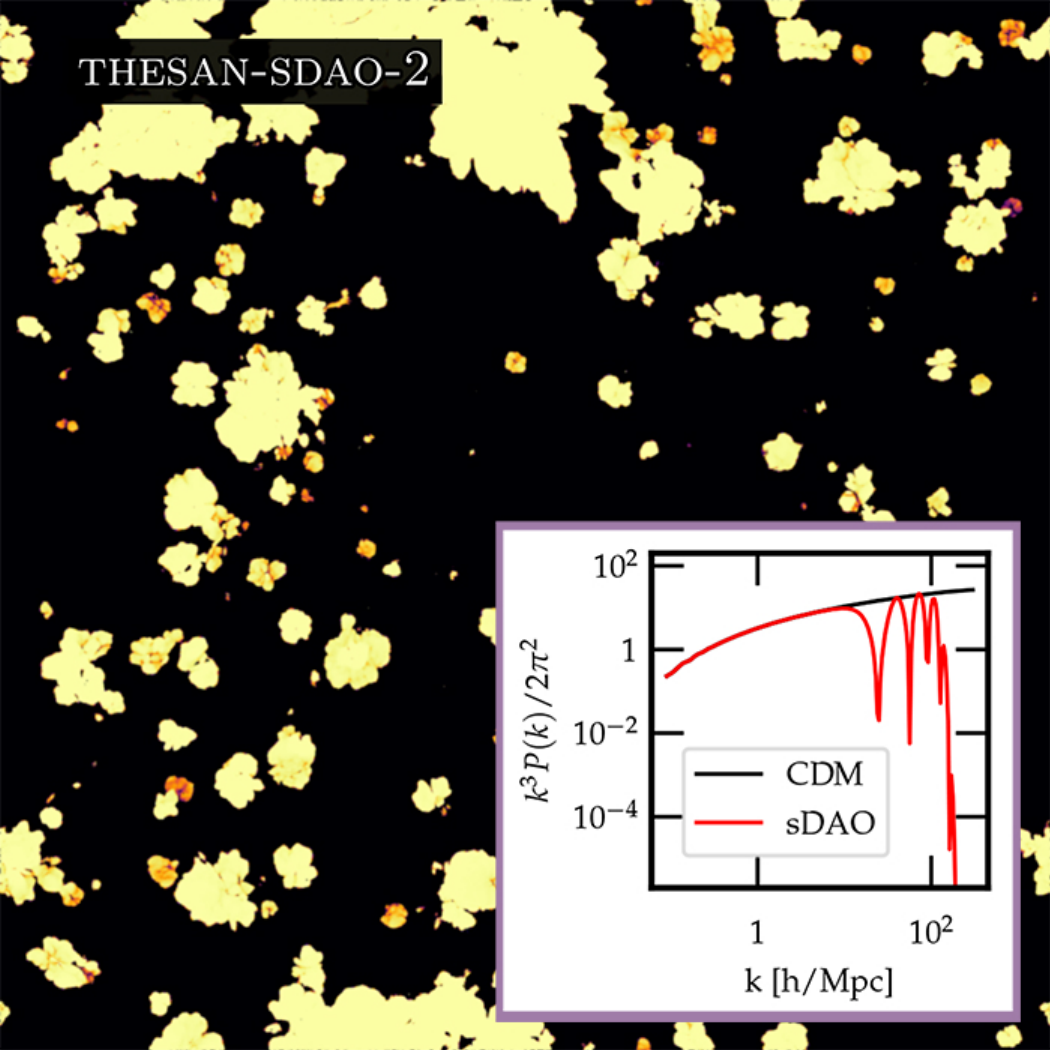}
    \end{tabular}
    \caption{
    \footnotesize
    Maps of the ionized bubble distribution in the different \thesan simulations, at $x_\mathrm{HI} \sim 0.7$.
    The bubbles are projected from a thick `slice' of the simulation spanning 8\% of the box volume.
    \thesan-1 is the highest resolution simulation in the \thesan suite.
    \thesan-2 is a medium resolution simulation that is otherwise the same as \thesan-1; \thesan-\textsc{wc}-2 has a slightly higher escape fraction to compensate for the lower star formation in \thesan-2 compared to \thesan-1.
    \thesan-\textsc{low}-2 is the same as \thesan-2 except that only halos below $10^{10}\,\Msun$ contribute to reionization, whereas in \thesan-\textsc{high}-2, only halos above $10^{10}\,\Msun$ contribute.
    \thesan-\textsc{sdao}-2 is the same as \thesan-2 but uses a non-standard dark matter model that effectively cuts off the linear matter power spectrum at small scales.
    The inset plot shows the linear matter power spectrum for cold dark matter and the dark acoustic oscillation model.
    As expected, \thesan-\textsc{high}-2 exhibits the largest ionized bubbles.
    Small ionized bubbles are also less abundant in the \thesan-\textsc{high}-2 and \thesan-\textsc{sdao}-2 simulations, compared to the others.}
    \label{fig:slices}
\end{figure*}

The application of EFT-inspired techniques to the EoR 21\,cm intensity field has only been studied relatively recently, as it was previously thought that the 21\,cm signal was nonperturbative in the wavenumber range probed by telescopes due to the presence of large ionized structures. 
To date, most of the theoretical analysis of the 21\,cm signal has been driven by computationally expensive radiative transfer simulations~\cite{gnedin1997reionization,ciardi2003simulating,McQuinn:2006et, iliev2006simulating,trac2007radiative,gnedin2014cosmic,pawlik2017aurora}, or by semi-analytic models such as 21\textsc{cmfast}~\cite{2011MNRAS.411..955M,2020JOSS....5.2582M} that can survey a wide range of theories of reionization with $\mathcal{O}(10\%)$ level agreement with simulation~\cite{zahn2007simulations,zahn2011comparison,majumdar2014use} (although the level of agreement depends sensitively on the ability of simulations to resolve self-shielded Lyman limit systems~\cite{kaurov2016cosmic}).
There have also been studies using phenomenological models that model the distribution of bubble sizes~\cite{2018ApJ...860...55R,2019ApJ...876...56R,Mirocha:2022}, parameterized models tuned to radiative transfer simulations~\cite{2013ApJ...776...81B}, models that match to a given mass-weighted ionization fraction~\cite{2022ApJ...927..186T}, and hybrid numerical methods that simulate the distribution of the first stars~\cite{2012Natur.487...70V,2014MNRAS.437L..36F}.

However, perturbative methods have gradually been developed with increasing success to study the process of reionization.
Linear perturbation theory can qualitatively reproduce many of the features of the EoR~\cite{2007MNRAS.375..324Z,2015PhRvD..91h3015M}, and theories including quadratic bias can match semi-analytic models to the level of tens of percent on very large scales~\cite{2019MNRAS.487.3050H}.
Ref.~\cite{McQuinn:2018zwa} pioneered the use of an effective bias expansion together with large-scale reionization simulations to show that the signal is in fact only mildly nonlinear on observable scales and that the field can be described accurately in real space with a small number of free parameters. 

In this chapter, we extend the perturbative description of the 21\,cm signal to include RSDs, the effects of which have previously been encapsulated in EFT treatments of the density field and halos~\cite{Senatore:2014vja,Lewandowski:2015ziq,Perko:2016puo}. 
In particular, we find that the RSDs give rise to terms that are not multiplied by any bias coefficients; therefore, the contribution of these terms to observable quantities is fixed and does not add any degrees of freedom when fitting to measurements or simulations.
Previous works have studied these terms using linear theory, and found that these terms can enhance the power spectrum up to a factor of $\sim 2$~\cite{Barkana:2004zy} and that the size of the resulting anisotropies varies with redshift~\cite{2015PhRvL.114j1303F}.
Moreover, these contributions contain information about the underlying cosmological density field that is free from astrophysical influence. We test the validity of our theoretical approach using \thesan, a suite of state-of-the-art radiation hydrodynamic simulations~\cite{thesan1, thesan2, thesan3, thesan4, thesan5, thesan6}.
Figure \ref{fig:slices} shows example slices from the simulations included in the suite.

All distances in this chapter are reported in terms of comoving units.
Readers mainly interested in the results of fitting the effective field theory expansion to simulations may choose to skip Section~\ref{sec:review}, which reviews perturbative methods for studying cosmological density fields, including standard perturbation theory (SPT), EFT, and the renormalization of local composite operators that appear in bias expansions.
In Section~\ref{sec:21cmrad}, we apply these perturbative techniques to the 21\,cm brightness temperature in redshift space. In Section~\ref{sec:thesan}, we introduce the \thesan simulations and describe our method for fitting the coefficients in the theory expansion to the simulations, which is done at the level of the cosmological \textit{fields}, instead of directly to the power spectrum to mitigate the possibility of overfitting. 
Section~\ref{sec:discussion} elaborates on the physical interpretation of the bias parameters and compares the fit parameters for simulations run with different physics, including a simulation with interacting dark matter that exhibits strong dark acoustic oscillations (sDAOs). 
Finally, we summarize our findings and outline some future directions in Section~\ref{sec:EFT_conclusion}.

\subsection{Review of Cosmological perturbation theories}
\label{sec:review}

In this section, we review results from SPT and the EFT of LSS, as well as effective renormalization in bias expansions. 
We begin by deriving the equations of motion for density and velocity perturbations and showing the perturbative solutions in SPT. 
We then review some results from EFT and introduce a diagrammatic language to help organize calculations involving higher-order terms and composite operators. 
See also Ref.~\cite{Desjacques:2016bnm} for a comprehensive review of cosmological perturbation theory, particularly in the context of galaxy bias.

\subsubsection{Standard Perturbation Theory}
\label{sec:spt}

Given a phase space distribution of collisionless particles $f (\tau, \boldsymbol{x}, \boldsymbol{p})$, the Boltzmann equation in an expanding universe is~\cite{Dodelson:2003ft}
\begin{equation}
    \frac{\mathrm{d}f}{\mathrm{d}t} = \frac{1}{a} \frac{\partial f}{\partial \tau} + \frac{\boldsymbol{p}}{a^2 m} \cdot \frac{\partial f}{\partial \boldsymbol{x}} - m \frac{\partial f}{\partial \boldsymbol{p}} \cdot \frac{\partial \phi}{\partial \boldsymbol{x}} = 0 .
\end{equation}
Here, $t$ denotes cosmic time, and is related to conformal time $\tau$ via $dt = a d\tau$, where $a$ is the scale factor, 
$\boldsymbol{x}$ and $\boldsymbol{p}$ are the comoving positions and momenta, and $\phi$ is the gravitational potential.
In this and following equations, we use boldface type to represent spatial 3-vectors; however, we occasionally use Einstein index notation to avoid ambiguities.

The first three moments of $f (\tau, \boldsymbol{x}, \boldsymbol{p})$ correspond to the comoving mass density, momentum density, and velocity dispersion:
\begin{align}
    \rho (\tau, \boldsymbol{x}) &\equiv m \int \dbar^3 p \, f (\tau, \boldsymbol{x}, \boldsymbol{p}) , \\
    \boldsymbol{\pi} (\tau, \boldsymbol{x}) &\equiv \int \dbar^3 p \, f (\tau, \boldsymbol{x}, \boldsymbol{p}) \, \boldsymbol{p} , \\
    \sigma^{ij} (\tau, \boldsymbol{x}) &\equiv \frac{1}{m^2} \int \dbar^3 p \, f (\tau, \boldsymbol{x}, \boldsymbol{p}) \, p^i p^j - \frac{\pi^i \pi^j}{m\rho} .
\end{align}
In these definitions, we denote $\dbar^3 p = d^3 p / (2\pi)^3$.
The first two moments of the Boltzmann equation correspond to the continuity equation
\begin{equation}
    0 = \partial_\tau \rho + \frac{1}{a} \boldsymbol{\nabla} \cdot \boldsymbol{\pi}
    \label{eqn:continuity}
\end{equation}
and the Euler equation
\begin{equation}
    0 = \partial_\tau \pi_i + \frac{1}{a} \partial_j \left( \frac{\pi_i \pi_j}{\rho} \right) + a \rho \nabla_i \phi .
    \label{eqn:euler}
\end{equation}
In the present formulation, the fluid equations contain no terms corresponding to shear forces, viscosity, or heat conduction; our collisionless particles therefore constitute a perfect fluid.

We solve these fluid equations perturbatively, defining $\delta = \rho/\bar{\rho} - 1$, where $\bar{\rho}$ is the mean density, and noting that the momentum density can be rewritten in terms of the physical peculiar velocity $\boldsymbol{v}$ as $\boldsymbol{\pi} = \rho a \boldsymbol{v}$; this is not the velocity of individual particles, but the bulk velocity of the field, i.e. averaged velocity of the particles in a region. 
Since the mean velocity of a homogeneous universe vanishes, $\boldsymbol{v}$ is perturbatively small. 
In terms of $\delta$ and $\boldsymbol{v}$, the continuity and Euler equations become
\begin{align}
    0 &= \partial_\tau \delta + \boldsymbol{\nabla} \cdot [ (1+\delta) \boldsymbol{v}] \\
    0 &= \partial_\tau \boldsymbol{v} + \mathcal{H} \boldsymbol{v} + (\boldsymbol{v} \cdot \boldsymbol{\nabla}) \boldsymbol{v} + \boldsymbol{\nabla} \phi
\end{align}
Here, $\mathcal{H} = \partial_\tau a / a$ is the conformal Hubble parameter.
We also include the Poisson equation in comoving coordinates as an equation of motion; since we are dealing with scales much smaller than the Hubble length, gravity can be treated as Newtonian:
\begin{equation}
    \partial^2 \phi = \frac{3}{2} \mathcal{H}^2 \Omega_m \delta.
\end{equation}
Above, $\Omega_m$ is the mass density in units of the critical density.
We hereafter set $\Omega_m=1$ since reionization occurs deep in the matter-dominated era.
The velocity can be further decomposed in terms of its divergence, $\theta = \boldsymbol{\nabla} \cdot \boldsymbol{v}$, and curl or vorticity, $\boldsymbol{\omega} = \boldsymbol{\nabla} \times \boldsymbol{v}$.
However, at leading order in SPT, any initial vorticity decays linearly with the expansion of the Universe; therefore, we neglect the contribution to the velocity field coming from $\boldsymbol{\omega}$.
\footnote{At higher order in SPT, there can be growing vorticity modes; however, the sources always contain powers of the vorticity at linear order, and are therefore still suppressed relative to the growing modes of $\delta$ and $\theta$.
Vorticity can also matter in EFT at third order, because the stress tensor and heat conduction terms source a non-decaying contribution~\cite{Bertolini:2016bmt}.}
With this velocity decomposition, in Fourier space the continuity and Euler equations are
\begin{align}
	\partial_\tau \delta_{\boldsymbol{k}} + \theta_{\boldsymbol{k}} &= - \int \dbar^3 q \, \frac{\boldsymbol{q} \cdot \boldsymbol{k}}{\boldsymbol{q}^2} \, \theta_{\boldsymbol{q}} \delta_{(\boldsymbol{k}-\boldsymbol{q})} , \label{eqn:S_a} \\
	\partial_\tau \theta_{\boldsymbol{k}} + \mathcal{H} \theta_{\boldsymbol{k}} + \frac{3}{2} \mathcal{H}^2 \delta_{\boldsymbol{k}} &= - \int \dbar^3 q \, \left( \frac{k^2 [\boldsymbol{q} \cdot (\boldsymbol{k} - \boldsymbol{q})]}{2 q^2 (\boldsymbol{k} - \boldsymbol{q})^2} \right) \theta_{\boldsymbol{q}} \theta_{(\boldsymbol{k}-\boldsymbol{q})} . \label{eqn:S_b}
\end{align}
Above, we've used bold subscripts to denote Fourier transformed quantities, e.g. $\delta_{\boldsymbol{k}} = \int d^3 x \, \delta (\boldsymbol{x}) e^{-i \boldsymbol{k} \cdot \boldsymbol{x}}$.
It will also be useful to convert the time derivatives into derivatives with respect to scale factor using $\partial_\tau = \mathcal{H} a \partial_a$ and the fact that $\mathcal{H} \propto 1 / \sqrt{a}$ during matter domination.

The right hand sides of Eqns. \eqref{eqn:S_a} and \eqref{eqn:S_b} mix the $\delta$ and $\theta$ modes.
To solve these coupled fluid equations, one can adopt the perturbative ansatz
\begin{gather}
    \delta_{\boldsymbol{k}} (\tau) = \sum_{n=1}^\infty D^n (\tau) \delta^{(n)}_{\boldsymbol{k}} , \\
    \theta_{\boldsymbol{k}} (\tau) = - \mathcal{H}(\tau) f(\tau) \sum_{n=1}^\infty D^n (\tau) \theta^{(n)}_{\boldsymbol{k}}
\end{gather}
where $\delta^{(n)}$ and $\theta^{(n)}$ are $\mathcal{O}(\delta^{(1)})^n$ and where $D(\tau)$ and $f(\tau) = d \ln D(\tau) / \mathcal{H} d\tau $ are the linear and logarithmic growth functions. Since reionization occurs deep in the matter-dominated epoch, we set $D(\tau) = a (\tau)$ and $f(\tau) = 1$ in this work; to include the effect of a dark energy component, one can substitute the appropriate growth factors~\cite{Bernardeau:2001qr}. Given the form of the ansatz, the solution to Eqns. \eqref{eqn:S_a} and \eqref{eqn:S_b} can be expressed as
\begin{align}
	\delta^{(n)}_{\boldsymbol{k}} &= \int \dbar^3 q_1 \dots \int \dbar^3 q_n \, (2\pi)^3 \delta^D \left( \boldsymbol{k} - \sum_{i=1}^n \boldsymbol{q}_i \right) F_n (\boldsymbol{q}_1, \dots, \boldsymbol{q}_n) \delta^{(1)}_{\boldsymbol{q}_1} \dots \delta^{(1)}_{\boldsymbol{q}_n} , \label{eqn:dn} \\
	\theta^{(n)}_{\boldsymbol{k}} &= \int \dbar^3 q_1 \dots \int \dbar^3 q_n \, (2\pi)^3 \delta^D \left( \boldsymbol{k} - \sum_{i=1}^n \boldsymbol{q}_i \right) G_n (\boldsymbol{q}_1, \dots, \boldsymbol{q}_n) \delta^{(1)}_{\boldsymbol{q}_1} \dots \delta^{(n)}_{\boldsymbol{q}_n}  \label{eqn:tn}
\end{align}
where the mode coupling kernels $F_n$ and $G_n$ have well-known recursion relations~\cite{Goroff:1986ep,Jain:1993jh,Bernardeau:2001qr}.
The first few kernels, symmetrized over permutations of the momenta, are
\begin{gather}
    F_1 = G_1 = 1 , \\
    F_2(\boldsymbol{q}_1, \boldsymbol{q}_2) = \frac{5}{7} + \frac{2}{7} \frac{(\boldsymbol{q}_1 \cdot \boldsymbol{q}_2)^2}{\boldsymbol{q}_1^2 \boldsymbol{q}_2^2} + \frac{\boldsymbol{q}_1 \cdot \boldsymbol{q}_2}{2} \left( \frac{1}{\boldsymbol{q}_1^2} + \frac{1}{\boldsymbol{q}_2^2} \right) , \\
    G_2(\boldsymbol{\boldsymbol{q}}_1, \boldsymbol{\boldsymbol{q}}_2) = \frac{3}{7} + \frac{4}{7} \frac{(\boldsymbol{q}_1 \cdot \boldsymbol{q}_2)^2}{\boldsymbol{q}_1^2 \boldsymbol{q}_2^2} + \frac{\boldsymbol{q}_1 \cdot \boldsymbol{q}_2}{2} \left( \frac{1}{\boldsymbol{q}_1^2} + \frac{1}{\boldsymbol{q}_2^2} \right) .
\end{gather}
One can calculate correlation functions of these fields using a diagrammatic representation.
The diagram rules are:
\begin{enumerate}
    \item Each $\delta^{(n)}_{\boldsymbol{k}}$ and $\theta^{(n)}_{\boldsymbol{k}}$ corresponds to a vertex with one external leg of wavenumber $\boldsymbol{k}$ and $n$ internal legs representing the factors of $\delta^{(1)}_{\boldsymbol{q}_i}$.
    The vertex couples the $n$ modes of the internal legs, and therefore corresponds to $F_n (\boldsymbol{q}_1, \dots, \boldsymbol{q}_n)$ or $G_n (\boldsymbol{q}_1, \dots, \boldsymbol{q}_n)$ depending on which field is involved. In analogy to conservation of momentum, wavenumber is conserved so each vertex also carries a factor of $(2\pi)^3 \delta^D \left( \boldsymbol{k} - \sum_{i=1}^n \boldsymbol{q}_i \right)$.
    We used filled dots to represent the density field and open dots to represent the velocity field.
    \begin{equation}
        \delta^{(n)}_{\boldsymbol{k}} \quad \rightarrow \quad
    	\begin{tikzpicture}[baseline=(current bounding box.center)]
    	\begin{feynman}
        	\vertex (i);
        	\vertex [right=1.5cm of i, scale=1.5, dot] (j) {};
        	
        	\vertex [above right=2cm of j] (a) {};
        	\vertex [below=0.5cm of a] (b) {};
        	\vertex [below right=2cm of j] (c) {};
        	
        	\vertex [below=1cm of b, scale=0.3, dot] (l) {};
        	\vertex [below=0.2cm of l, scale=0.3, dot] (m) {};
        	\vertex [below=0.2cm of m, scale=0.3, dot] (n) {};
        	
        	\diagram*{
        		(i) -- [edge label=\(\boldsymbol{k}\)] (j) ,
        		(j) -- [scalar, edge label=\(\boldsymbol{q_1}\)] {(a)},
        		(j) -- [scalar] {(b)},
        		(j) -- [scalar, edge label'=\(\boldsymbol{q_n}\)] {(c)}
        	};
    	\end{feynman}
    	\end{tikzpicture}
    	\quad = \quad (2\pi)^3 \delta^D \left( \boldsymbol{k} - \sum_{i=1}^n \boldsymbol{q}_i \right) F_n (\boldsymbol{q}_1, \dots, \boldsymbol{q}_n)
    \end{equation}
    \begin{equation}
        \theta^{(n)}_{\boldsymbol{k}} \quad \rightarrow \quad
    	\begin{tikzpicture}[baseline=(current bounding box.center)]
    	\begin{feynman}
        	\vertex (i);
        	\vertex [right=1.5cm of i, scale=1.5, empty dot] (j) {};
        	
        	\vertex [above right=2cm of j] (a) {};
        	\vertex [below=0.5cm of a] (b) {};
        	\vertex [below right=2cm of j] (c) {};
        	
        	\vertex [below=1cm of b, scale=0.3, dot] (l) {};
        	\vertex [below=0.2cm of l, scale=0.3, dot] (m) {};
        	\vertex [below=0.2cm of m, scale=0.3, dot] (n) {};
        	
        	\diagram*{
        		(i) -- [edge label=\(\boldsymbol{k}\)] (j) ,
        		(j) -- [scalar, edge label=\(\boldsymbol{q_1}\)] {(a)},
        		(j) -- [scalar] {(b)},
        		(j) -- [scalar, edge label'=\(\boldsymbol{q_n}\)] {(c)}
        	};
    	\end{feynman}
    	\end{tikzpicture}
    	\quad = \quad (2\pi)^3 \delta^D \left( \boldsymbol{k} - \sum_{i=1}^n \boldsymbol{q}_i \right) G_n (\boldsymbol{q}_1, \dots, \boldsymbol{q}_n)
    \end{equation}
    \item To compute a correlation function, draw all connected diagrams that can be made by contracting the internal $\delta^{(1)}$ legs. 
    Since we are using symmetrized $F_n$ and $G_n$ kernels, permuting the $\delta^{(1)}$ legs on each $\delta^{(n)}$ vertex will give rise to a symmetry factor of $n!$, and loops introduce additional combinatoric factors to prevent double-counting.
    \item For each internal leg carrying wavenumber $\boldsymbol{p}$, write down a factor of $P_L (\boldsymbol{p})$, the linear matter power spectrum. This is analogous to the propagator of the linear, ``free'' density fields of the internal legs, since the power spectrum is related to the two-point correlation function.
    \begin{equation}
        \begin{tikzpicture}
    	\begin{feynman}
    	\vertex [dot, scale=1.5] (i) {};
    	\vertex [right=2cm of i, empty dot, scale=1.5] (j) {};
    	\diagram*{
    		(i) --[scalar, momentum=\(\boldsymbol{p}\)] (j)
    	};
    	\end{feynman}
    	\end{tikzpicture}
    	\quad = \quad P_L (\boldsymbol{p})
    \end{equation}
    The vertices on the ends of the propagator can correspond to both $\delta$, both $\theta$, or one of each; the factor of $P_L (\boldsymbol{p})$ associated with the propagator is the same in any case, since $\delta^{(1)}$ and $\theta^{(1)}$ are spatially the same up to time-dependent factors.
    \item Integrate over the wavenumber $\boldsymbol{q}$ of each loop with $\int \dbar^3 q$.
\end{enumerate}

\subsubsection{Effective Field Theory}
\label{sec:eft}

For small-wavelength modes, perturbations will have collapsed enough to have become nonlinear and be outside of the regime of validity of the present perturbative theoretical treatment. 
These non-linearities can also affect large scales, since modes of different scales are coupled by the vertex kernels of Eqns.~\eqref{eqn:dn} and \eqref{eqn:tn} and since integrals over loops formally run over all wavenumbers.
This motivates introducing a smoothed version of the fields, where we convolve the densities or velocities with a windowing function $W_\Lambda$ of characteristic length scale $1/\Lambda$. 
We can apply this smoothing to the equations of motion e.g. Eqns. \eqref{eqn:continuity} and \eqref{eqn:euler}; however, smoothed composite operators can not be straightfowardly expressed as a product of smoothed fields, e.g. $(\delta v)_\text{smooth} \neq \delta_\text{smooth} \,v_\text{smooth}$. 
In order to express the equations of motion in terms of smoothed fields, one can express the smoothed composite operators as a product of smoothed fields after introducing additional correction terms~\cite{Baumann:2010tm,Carrasco:2012cv,Pajer:2013jj,Mercolli:2013bsa,Abolhasani:2015mra}. 
These terms are unknown \emph{a priori} but can be constructed from the bottom-up from all terms consistent with the symmetries (e.g. Galilean invariance). 
These terms take the form of an effective stress tensor for the long-wavelength fluid and the sensitivity to unknown behaviour of small-scale modes is parameterized as an effective speed of sound, viscosity, shear, etc. 
In other words, the smoothed density field is not a perfect fluid because of the feedback from small-scale modes.

The new terms in the effective stress tensor can be constructed order by order, by expanding the stress tensor in terms of convective time derivatives (co-moving with fluid elements) of local operators~\cite{Bertolini:2016bmt}.
As a result of this change to the equations of motion, we must modify the perturbative ansatz to include additional counterterms, which we denote by $\tilde{\delta}^{(n)}$ and $\tilde{\theta}^{(n)}$,
\begin{gather}
    \delta_{\boldsymbol{k}} (\tau) = \sum_{n=1}^\infty \left(  a(\tau)^n \delta^{(n)}_{\boldsymbol{k}} + \epsilon a(\tau)^{n+2} \tilde{\delta}^{(n)}_{\boldsymbol{k}} \right) \label{eqn:EFT_delta} \\
    \theta_{\boldsymbol{k}} (\tau) = - \mathcal{H}(\tau)  \sum_{n=1}^\infty \left( a(\tau)^n \theta^{(n)}_{\boldsymbol{k}} + \epsilon a(\tau)^{n+2} \tilde{\theta}^{(n)}_{\boldsymbol{k}} \right). \label{eqn:EFT_theta}
\end{gather}
Here, $\epsilon$ is a parameter that allows us to keep track of the EFT power counting. The counterterms come with an additional factor of $a (\tau)^2$ compared to the SPT terms because the EFT terms must have the same time dependence as loop contributions from SPT in order to correct them. 
The EFT kernels $\tilde{F}_n$ and $\tilde{G}_n$ are analogously defined relative to the SPT kernels as
\begin{align}
	\tilde{\delta}^{(n)}_{\boldsymbol{k}} &= \int \dbar^3 q_1 \dots \int \dbar^3 q_n \, (2\pi)^3 \delta^D \left( \boldsymbol{k} - \sum_{i=1}^n q_i \right) \tilde{F}_n (\boldsymbol{q}_1, \dots, \boldsymbol{q}_n) \delta^{(1)}_{\boldsymbol{q}_1} \dots \delta^{(1)}_{\boldsymbol{q}_n} , \\
	\tilde{\theta}^{(n)}_{\boldsymbol{k}} &= \int \dbar^3 q_1 \dots \int \dbar^3 q_n \, (2\pi)^3 \delta^D \left( \boldsymbol{k} - \sum_{i=1}^n q_i \right) \tilde{G}_n (\boldsymbol{q}_1, \dots, \boldsymbol{q}_n) \delta^{(1)}_{\boldsymbol{q}_1} \dots \delta^{(1)}_{\boldsymbol{q}_n} .
\end{align}
The forms of the EFT kernels, $\tilde{F}_n$ and $\tilde{G}_n$, are derived in Ref.~\cite{Bertolini:2016bmt} and listed up to $n=3$. 
Correlation functions can then be computed and are robust to the effects of non-linearities affecting the results at the level of the fluid equations; i.e. the large perturbative scales are less affected by the uncertainties of small scale physics.

\subsubsection{Renormalized bias}
\label{sec:renorm}

Much of the formalism for cosmological effective field theories has been developed in the context of large scale structure and the matter density field.
However, EFT techniques can also be extended to study biased tracers of the matter field, such as galaxies and halos~\cite{2009JCAP...08..020M,Assassi:2014fva,Senatore:2014eva,Angulo:2015eqa,Fujita:2016dne,Perko:2016puo,Nadler:2017qto,Donath:2020abv}. 
Cosmological 21\,cm radiation is also a biased tracer of the underlying matter field on large scales.
When expressing the 21\,cm intensity field as a local bias expansion in terms of $\delta$, one needs to include all operators that respect homogeneity and isotropy; specifically, the theoretical 21\,cm field is built up only from operators that obey these symmetries, and should generically include contributions from all such operators,
\begin{equation}
    (\delta_{21})_{\boldsymbol{k}} = b_1 \delta_{\boldsymbol{k}} - b_{\nabla^2} k^2 \delta_{\boldsymbol{k}} + b_2 \left(\delta^2\right)_{\boldsymbol{k}} + b_{\mathcal{G}2} (\mathcal{G}_2)_{\boldsymbol{k}} + \cdots
    \label{eqn:bias}
\end{equation}
Note that the momentum subscript denotes a Fourier transformation over the entire operator, e.g. $\left(\delta^2\right)_{\boldsymbol{k}} = \int \dbar^3 k \, \delta^2 (\boldsymbol{x}) e^{-i \boldsymbol{k} \cdot \boldsymbol{x}} \neq \left(\delta_{\boldsymbol{k}}\right)^2$.
In the above equation, $\mathcal{G}_2$ is the second Galileon or tidal operator, defined in configuration space as
\begin{equation}
    \mathcal{G}_2 = (\nabla_i \nabla_j \phi) (\nabla^i \nabla^j \phi) - (\nabla^2 \phi)^2 .
\end{equation}
The bias coefficients $b$ for the various operators in Eq.~\eqref{eqn:bias} are not known \emph{a priori} and must be determined from real or simulated data; in fact, the field on the left-hand side of Eqn.~\eqref{eqn:bias} can be replaced with any biased tracer of the underlying matter field, e.g. halos or galaxies, and the inferred coefficients will differ depending on the physics of the particular tracer in question.

From Eqn.~\eqref{eqn:bias}, we see that composite operators such as $\delta^2 (\boldsymbol{x})$ appear in the configuration space picture. 
Diagramatically, we represent this composite operator in Fourier space with the vertex
\begin{equation}
	\begin{tikzpicture}[baseline=(current bounding box.center)]
	\begin{feynman}
	\vertex (i);
	\vertex [right=1cm of i, small, blob] (a) {};
	\vertex [above right=1cm of a, dot] (b) {};
	\vertex [below right=1cm of a, dot] (c) {};
	\diagram*{
		(i) -- (a) ,
		(a) -- {(b),(c)}
	};
	\end{feynman}
	\end{tikzpicture}
\end{equation}
where the ``blob'' indicates a convolution in Fourier space and where the two legs ending in solid dots represent $\delta^{(n)}$ component fields entering the convolution.
Because these operators are local in configuration space, in Fourier space the convolution includes all wavenumbers.
Therefore, these composite operators contain contributions from small-scale modes that are non-linear, in analogy to the previous subsection. These non-linear contributions to the bias expansion are not removed by the EFT formalism described above, because the counterterms in Eqns. \eqref{eqn:EFT_delta} and \eqref{eqn:EFT_theta} only correct the non-linearities that affect the equations of motion for matter.
We follow the renormalization procedure in Ref.~\cite{Assassi:2014fva} to remove the small-scale or UV-dependence of composite operators order by order. 

To renormalize an operator $f$, we take correlation functions of $f$ with factors of the linear density field $\delta^{(1)}$ and add counterterms that cancel UV-sensitive loop contributions to these correlation functions in the zero-mode limit; this leaves only the tree-level (or zero-loop) contribution.
In other words, our renormalization condition is
\begin{align}
	\braket{[f_{\boldsymbol{k}}] \delta^{(1)}_{\boldsymbol{q}_1} \cdots \delta^{(1)}_{\boldsymbol{q}_n}} &= \braket{f_{\boldsymbol{k}} \delta^{(1)}_{\boldsymbol{q}_1} \cdots \delta^{(1)}_{\boldsymbol{q}_n}}_\mathrm{tree} \, \mathrm{for} \, \boldsymbol{q}_i = 0 , \, \forall \, i.
	\label{eqn:renorm_condition}
\end{align}
where the square brackets denote the renormalized operator $[f] = f + \sum_\mathcal{O} Z^f_\mathcal{O} \mathcal{O}$ such that the sum over all counterterm operators $\mathcal{O}$,  $\sum_\mathcal{O} Z^f_\mathcal{O} \mathcal{O}$ cancel the loop contributions. 
We evaluate the renormalization conditions at zero wavenumber, since this is the limit where the theory is most perturbative.
In evaluating loops, we only include diagrams where the loops connect multiple component fields of the convolution vertex; such diagrams are sometimes called ``one particle irreducible" or 1PI due to their similarity with such diagrams from quantum field theory~\cite{Assassi:2014fva,Abolhasani:2015mra}, but we stress that these definitions are not exactly the same.
The diagrams we include capture the additional mixing between small and large scale modes in the convolution that we are concerned with, as opposed to the mixing that arises from the equations of motion. We now make these definitions more explicit:
\begin{itemize}
    \item 1PI diagrams are diagrams that cannot be separated into two valid, disconnected diagrams by cutting a single internal line.
    The following graph is an example of a fully 1PI diagram.
    \begin{equation}
    \begin{tikzpicture}[baseline=(current bounding box.center)]
	\begin{feynman}
	\vertex (i);
	\vertex [right=1cm of i, small, blob] (a) {};
	\vertex [above right=1cm of a, dot] (b1) {};
	\vertex [right=1cm of a, dot] (b2) {};
	\vertex [below right=1cm of a, dot] (b3) {};
	\vertex [right=1cm of b1, dot] (c1) {};
	\vertex [right=1cm of c1] (d1);
	\vertex [right=1cm of b3, dot] (c2) {};
	\vertex [right=1cm of c2] (d2);
	\diagram*{
		(i) -- (a) ,
		(a) -- {(b1),(b2),(b3)},
		(b1) -- [scalar] (c1),
		(b1) -- [scalar] (b2),
		(c1) -- (d1),
		(b3) -- [scalar] (c2),
		(b3) -- [scalar] (b2),
		(c2) -- (d2),
	};
	\end{feynman}
	\end{tikzpicture}
	\nonumber
	\end{equation}
    Since the momenta in the external legs coming out of the composite operator are related via loops, this diagram involves mode mixing, so we include it in the renormalization procedure.
    Note that this example is a two-loop diagram, so we do not include this for calculating the one-loop power spectrum.
    
    \item We also include diagrams of the following type.
    \begin{equation}
    \begin{tikzpicture}[baseline=(current bounding box.center)]
	\begin{feynman}
	\vertex (i);
	\vertex [right=1cm of i, small, blob] (a) {};
	\vertex [above right=1cm of a, dot] (b1) {};
	\vertex [right=1cm of a, dot] (b2) {};
	\vertex [below right=1cm of a, dot] (b3) {};
	\vertex [right=1cm of b1, dot] (c1) {};
	\vertex [right=1cm of c1] (d1);
	\vertex [right=1cm of b3, dot] (c2) {};
	\vertex [right=1cm of c2] (d2);
	\diagram*{
		(i) -- (a) ,
		(a) -- {(b1),(b2),(b3)},
		(b1) -- [scalar] (c1),
		(b1) -- [scalar] (b2),
		(c1) -- (d1),
		(b3) -- [scalar] (c2),
		(c2) -- (d2),
	};
	\end{feynman}
	\end{tikzpicture}
	\nonumber
	\end{equation}
	This is not 1PI in the conventional sense, since it can be separated into two valid diagrams by cutting the bottom leg coming out of the convolution vertex.
	However, we still include it for the purpose of renormalization, since there is a loop that relates the momenta of the other two legs.
	Such diagrams have been termed ``partially 1PI" in the literature~\cite{Assassi:2014fva}.
    
    \item Below is an example of a diagram that is neither fully 1PI nor partially 1PI.
    \begin{equation}
    \begin{tikzpicture}[baseline=(current bounding box.center)]
	\begin{feynman}
	\vertex (i);
	\vertex [right=1cm of i, small, blob] (a) {};
	\vertex [above right=1cm of a, dot] (b1) {};
	\vertex [right=1cm of a, dot] (b2) {};
	\vertex [below right=1cm of a, dot] (b3) {};
	\vertex [right=1cm of b1, dot] (c1) {};
	\vertex [right=1cm of c1] (d1);
	\vertex [right=1cm of b2, dot] (c2) {};
	\vertex [right=1cm of c2] (d2);
	\vertex [right=1cm of b3, dot] (c3) {};
	\vertex [right=1cm of c3] (d3);
	\diagram*{
		(i) -- (a) ,
		(a) -- {(b1),(b2),(b3)},
		(b1) -- [scalar] (c1),
		(c1) -- (d1),
		(b2) -- [scalar] (c2),
		(c2) -- (d2),
		(b3) -- [scalar] (c3),
		(c3) -- (d3),
	};
	\draw[dashed] (b1) arc [start angle=270, end angle=-180, radius=0.3cm];
	\end{feynman}
	\end{tikzpicture}
	\nonumber
	\end{equation}
	We do not include this for bias renormalization, since the momenta running through the external legs of the composite operator vertex do not mix.
\end{itemize}

As an example, we show the calculation of the first few of counterterms for $\delta^2$. 
The first counterterm cancels UV sensitivity from the expectation value of $\delta^2$,
\begin{align}
    \braket{\left(\delta^2\right)_{\boldsymbol{k}}} \quad &= \quad
	\begin{tikzpicture}[baseline=(current bounding box.center)]
	\begin{feynman}
	\vertex (i);
	\vertex [right=1cm of i, small, blob] (a) {};
	\vertex [above right=1cm of a, dot] (b) {};
	\vertex [below right=1cm of a, dot] (c) {};
	\diagram*{
		(i) -- [momentum=\(\boldsymbol{k}\)] (a) ,
		(a) -- {(b),(c)},
		(b) -- [scalar, half left, momentum=\(\boldsymbol{p}\)] (c)
	};
	\end{feynman}
	\end{tikzpicture}
	\n
	&= \quad
	\int_0^\Lambda \frac{\mathrm{d}p}{2\pi^2} p^2 P_L (\boldsymbol{p})
	\equiv \sigma^2 (\Lambda).
\end{align}
Thus the lowest order counterterm is $-\sigma^2 (\Lambda)$.
By subtracting tadpole diagrams involving operators that contribute to Eqn.~\eqref{eqn:bias}, we ensure that the expectation value of the biased tracer vanishes at the one-loop level, $\langle \delta_{21} \rangle = 0$. The next counterterm cancels the UV sensitivity of
\begin{align}
    \braket{\left(\delta^2\right)_{\boldsymbol{k}} \delta^{(1)}_{\boldsymbol{q}}} \quad &= \quad 2 \times
	\begin{tikzpicture}[baseline=(current bounding box.center)]
	\begin{feynman}
	\vertex (i);
	\vertex [right=1cm of i, small, blob] (a) {};
	\vertex [above right=1cm of a, dot] (b) {};
	\vertex [below right=1cm of a, dot] (c) {};
	\vertex [right=1cm of b, dot] (d) {};
	\vertex [right=1cm of d] (f);
	\diagram*{
		(i) -- (a) ,
		(a) -- {(b),(c)},
		(b) -- [scalar] (c),
		(b) -- [scalar] (d),
		(d) -- (f),
	};
	\end{feynman}
	\end{tikzpicture}
	\n
	&= \quad
	P_L (\boldsymbol{k}) \int \dbar^3 p \, F_2(\boldsymbol{k}, \boldsymbol{p}) P_L (\boldsymbol{p}) \n
	&= \frac{68}{21} \sigma^2 (\Lambda) P_L (\boldsymbol{k}) .
\end{align}
Since the corresponding counterterm must be proportional to $P_L (\boldsymbol{k})$ when correlated with a factor of $\delta^{(1)}$, the counterterm must be proportional to the operator $\delta$.
Thus, the first few terms of the renormalized $\left(\delta^2\right)_{\boldsymbol{k}}$ are
\begin{equation}
    [\delta^2] = \delta^2 - \sigma^2 (\Lambda) - \frac{68}{21} \sigma^2 (\Lambda) \delta .
\end{equation}
Note that this equation is written in configuration space.
In order to include corrections from higher-order counterterms, we would continue to calculate higher point correlation functions. In this way, we renormalize all the composite operators that appear in our bias expansion. 

The Galilean operator, $\mathcal{G}_2$, is not renormalized at leading order in derivatives~\cite{Assassi:2014fva}.
Even if we had decided to calculate these higher derivative counterterms, the limit $q_i \rightarrow 0$ in the renormalization conditions ensures that these counterterms vanish anyway; thus, within this framework, $\mathcal{G}_2$ is not renormalized.

\subsection{The 21 cm radiation field in redshift space}
\label{sec:21cmrad}

The 21\,cm differential brightness temperature is a biased tracer of the underlying matter density and can be written as~\cite{Zaroubi_notes,Furlanetto:2019jso}
\begin{align}
	\delta T_b \approx& \,28 (1+\delta) x_\mathrm{HI} \left( 1 - \frac{T_\mathrm{CMB} (\nu)}{T_\mathrm{spin}} \right) \left(\frac{\Omega_b h^2}{0.0223}\right) \n
	&\times \sqrt{\left(\frac{1+z}{10}\right) \left(\frac{0.24}{\Omega_m}\right)} \left(\frac{H(z)/(1+z)}{\mathrm{d}v_\parallel / \mathrm{d}r_\parallel}\right) \,\mathrm{mK} .
	\label{eqn:dTb}
\end{align}
In this expression, $x_\mathrm{HI}$ is the fraction of hydrogen that is neutral, $T_\mathrm{CMB} (\nu)$ is the CMB brightness temperature, $\Omega_b$ is the baryon density in units of the critical density, $h$ is the Hubble constant in units of $100 \,\mathrm{km}\, \mathrm{s}^{-1} \mathrm{Mpc}^{-1}$, $\Omega_m$ is the mass density in units of the critical density, $H(z)$ is the Hubble expansion at $z$, and $dv_\parallel / dr_\parallel$ is the gradient of the proper velocity along the line of sight. 
The spin temperature $T_\mathrm{spin}$ is defined in terms of the ratio of the occupancy of the spin-1 and spin-0 ground states of hydrogen.
\begin{equation}
    \frac{n_1}{n_0} = 3 \exp(-T_*/T_{\mathrm{spin}}), \label{eqn:spintemp}
\end{equation}
Here, $T_* = 0.0681$ K is the temperature corresponding to the 21\,cm wavelength.
The spin temperature varies throughout space and even throughout individual clumps of neutral hydrogen.
Since we are studying redshifts well into the EoR, we assume $T_\mathrm{spin} \gg T_\mathrm{CMB}$, so the factor of $\left( 1 - \frac{T_\mathrm{CMB} (\nu)}{T_\mathrm{spin}} \right)$ in Eqn. \eqref{eqn:dTb} becomes saturated and the effect of spatial fluctuations in the spin temperature is negligible.
Henceforth, we neglect spin temperature fluctuations. 
This is a common simplification; however there are also a number of studies that do not assume $T_\mathrm{spin} \gg T_\mathrm{CMB}$ ~\cite{2017MNRAS.464.1365M,2018MNRAS.478.5591M}.
In particular, for the higher redshifts where we expect the 21\,cm radiation field to be more perturbative and the EFT framework to in principal be a better descriptor of the signal, the assumption of spin temperature saturation may hinder the accuracy of this method.
Extending our formalism to higher redshifts relevant for cosmic dawn will require that spin temperature fluctuations be taken into account, and will be the subject of future work.

To be explicit, we define $\delta_{21} = (\delta T_b - \overline{\delta T_b}) / \overline{\delta T_b}$ to be the \textit{fluctuations} in the brightness temperature, and not $\delta T_b$ itself.
Then, by comparing Eqns. \eqref{eqn:bias} and \eqref{eqn:dTb}, we see that measuring the bias coefficients gives us information about the distribution and ionization of the intervening hydrogen, as well as cosmological parameters.

\subsubsection{From real space to redshift space}
\label{sec:RSDs}

Observations of the 21\,cm radiation field are complicated by the fact that neutral hydrogen has a peculiar velocity which give rise to RSDs. In other words, the measured redshift of the 21\,cm line cannot be attributed purely to the expansion of the Universe.
The distances mapped out from the redshift and ignoring the peculiar velocity form a distorted ``redshift space'' and the coordinates $\boldsymbol{x}_r$ in this space are related to real space coordinates $\boldsymbol{x}$ by
\begin{equation}
	\boldsymbol{x}_r = \boldsymbol{x} + \frac{\boldsymbol{\hat{n}} \cdot \boldsymbol{v}_\mathrm{pec}}{\mathcal{H}} \boldsymbol{\hat{n}} .
\end{equation}
Here, $\boldsymbol{\hat{n}}$ is the line-of-sight direction and $\boldsymbol{v}_\mathrm{pec}$ is the peculiar velocity at the location indicated by the real space coordinate.

The effect of RSDs has been accounted for in effective field theory descriptions of LSS and biased tracers of LSS~\cite{Matsubara:2007wj,Senatore:2014vja,Lewandowski:2015ziq,Perko:2016puo,2018JCAP...12..035D,Donath:2020abv,2020PhRvD.102b3515G,2022MNRAS.513..117L}. RSDs have also been treated perturbatively for the 21\,cm signal~\cite{2012MNRAS.422..926M}, but without including a fully systematic treatment of small-scale nonlinearities as described in the previous section.
To derive the effect of RSDs, we can use the above relationship to transform between the real and redshift space density contrast. 
If the density in real space is $\rho$ and in redshift space is $\rho_r$, then conservation of mass implies $\rho_r (\boldsymbol{x}_r) d^3 x_r = \rho (\boldsymbol{x}) d^3 x$. 
Then, expanding in terms of the density constrast $\delta = \rho/\bar{\rho} - 1$, we find
\begin{equation}
	\delta_r (\boldsymbol{x}_r) = (1 + \delta (\boldsymbol{x})) \left| \frac{\partial \boldsymbol{x}_r}{\partial \boldsymbol{x}} \right|^{-1} -1 .
\end{equation}
Fourier transforming this relation yields
\begin{equation}
    (\delta_r)_{\boldsymbol{k}} = \delta_{\boldsymbol{k}} + \int d^3 x \, e^{-i\boldsymbol{k} \cdot \boldsymbol{x}} \left( \exp\left[ -i \frac{k_\parallel v_\parallel}{\mathcal{H}} \right] - 1 \right) (1 + \delta (\boldsymbol{x})) .
\end{equation}
Here, we have defined $v_\parallel \equiv \boldsymbol{\hat{n}} \cdot \boldsymbol{v}_\mathrm{pec}$ and $k_\parallel \equiv \boldsymbol{\hat{n}} \cdot \boldsymbol{k}$. 

The quantity $k_\parallel v_\parallel$ can be thought of as the rate at which modes of length scale $1/k$ are changing along the line of sight, due to peculiar velocities. 
For the modes of interest, this rate is quite small compared to the expansion rate of the universe because the peculiar velocities are very nonrelativistic, hence $k_\parallel v_\parallel / \mathcal{H}$ is a small quantity. Taylor expanding in this parameter then gives another series expansion,
\begin{align}
	(\delta_r)_{\boldsymbol{k}} =& \delta_{\boldsymbol{k}} 
	-i \frac{k_\parallel}{\mathcal{H}} (v_{\parallel})_{\boldsymbol{k}} 
	-i \frac{k_\parallel}{\mathcal{H}} \left({\delta v_{\parallel}}\right)_{\boldsymbol{k}}
	- \frac{1}{2} \left(\frac{k_\parallel}{\mathcal{H}} \right)^2 \left({v_\parallel^2}\right)_{\boldsymbol{k}} \n
	&- \frac{1}{2} \left(\frac{k_\parallel}{\mathcal{H}} \right)^2 \left({\delta v_\parallel^2}\right)_{\boldsymbol{k}}
	+ \frac{i}{6} \left(\frac{k_\parallel}{\mathcal{H}} \right)^3 \left({v_\parallel^3}\right)_{\boldsymbol{k}}
	+ \cdots
	\label{eqn:RSD_expansion}
\end{align}
We see that we now have new operators that include factors of the velocity field which will also need to be renormalized following the prescription of Sec.~\ref{sec:renorm}.

Note that the $v_\parallel$ that appears in Eqn.~\eqref{eqn:RSD_expansion} is a projection of the \textit{baryon} velocity, while the velocity that appears throughout Section~\ref{sec:review} is the \textit{matter} velocity.
Relative velocities between baryons and dark matter can affect the formation and distribution of the first bound objects, leaving imprints on the matter power spectrum and galaxy bispectrum~\cite{Tseliakhovich:2010bj,2011JCAP...07..018Y}, as well as affecting the Lyman-$\alpha$ forest~\cite{2020PhRvD.102b3515G}, reionization, and the 21\,cm signal~\cite{Munoz:2019fkt,Munoz:2019rhi,Cain:2020npm,Park:2020ydt,2022MNRAS.513..117L}.
However, during the EoR, when the first collapsed objects have already formed, the effect of these relative velocities is negligible~\cite{Stacy:2010gg}.
In the simulations we use, the difference between the two velocities is less than 2\% in the vast majority of the volume and this approximation is also justified and used in other studies~\cite{2012MNRAS.422..926M}. 
Hence, for our purposes, we take the velocity of the neutral hydrogen and matter to be the same and leave the inclusion of a relative velocity term for future study.

\subsubsection{The effective 21 cm field}
\label{sec:field}

The steps to building up our effective field theory are:
\begin{enumerate}
    \item Use standard perturbation theory to treat the evolution of the matter density field, $\delta$.
    See Section~\ref{sec:spt} for a review of SPT.
    \item Include a bias expansion to write the 21\,cm field $\delta_{21}$ in terms of $\delta$.
    See also Ref.~\cite{Desjacques:2016bnm} for a comprehensive review of cosmological perturbation theory for biased tracers.
    \item Include an RSD expansion in $k_\parallel v_\parallel / \mathcal{H}$, in order to write the redshift space field $\delta_{21,r}$ in terms of the real space field $\delta_{21}$.
    \item Smooth over non-perturbative modes in the field using some wavenumber $\Lambda$ and renormalize the composite operators that appear.
    See Section~\ref{sec:renorm} for more details.
\end{enumerate}
Putting together the bias and RSD expansions, we obtain
\begin{align}
    (\delta_{21,r})_{\boldsymbol{k}} = &b_{1} \delta_{\boldsymbol{k}} - b_{\nabla^2} k^2 \delta_{\boldsymbol{k}} + b_{2} {\left(\delta^2\right)_{\boldsymbol{k}}} + b_{G2} (\mathcal{G}_2)_{\boldsymbol{k}} \n
    &-i \frac{k_\parallel}{\mathcal{H}} \left[ (v_\parallel)_{\boldsymbol{k}} + b_1 ({\delta \,v_\parallel})_{\boldsymbol{k}} - b_{\nabla^2} k^2 ({\delta\, v_\parallel})_{\boldsymbol{k}} \right] - \frac{1}{2} \left(\frac{k_\parallel}{\mathcal{H}} \right)^2 ({v_\parallel^2})_{\boldsymbol{k}} + \cdots
    \label{eqn:d21_bare}
\end{align}
There are terms in Eqn.~\eqref{eqn:d21_bare} that are not multiplied by any bias coefficients; thus, when fitting the theory to data or simulations, the size of these terms cannot be adjusted. 
We have checked that in the linear limit, the bias-independent term in Eqn. \eqref{eqn:d21_bare} resulting from RSDs matches the result used in Ref~\cite{Barkana:2004zy}.
RSDs therefore give rise to a bias-independent contribution to the power spectrum, which can enhance the 21\,cm power spectrum relative to the matter spectrum by a factor of up to $\sim 2$~\cite{Barkana:2004zy}.
The measurability of these contributions depends on redshift~\cite{2015PhRvL.114j1303F} and the angular dependence of these terms can also be used to distinguish the contributions to the 21\,cm power spectrum due to density fluctuations from ionization fluctuations~\cite{2012MNRAS.422..926M}.

We can now apply the procedure outlined in section \ref{sec:renorm} to renormalize the operators appearing in Eqn.~\eqref{eqn:d21_bare}.
As discussed above, $\mathcal{G}_2$ is not renormalized at leading order in derivatives~\cite{Assassi:2014fva}, and furthermore $\left(\delta\, v_\parallel\right)_{\boldsymbol{k}}$ receives no extra counterterms because the momentum $\pi_i \propto (1 + \delta) v_i$ is automatically renormalized through the continuity equation~\cite{Senatore:2014vja,Perko:2016puo}.
\footnotemark[2]
Then for the remaining operators, we find
\begin{align}
	\left[\delta^2\right] &= \delta^2 - \sigma^2 (\Lambda) \left( 1 + \frac{68}{21} \delta + \frac{8126}{2205} \delta^2 + \frac{254}{2205} \mathcal{G}_2 \right) + \cdots \n
	\left[ v_\parallel^2 \right] &= v_\parallel^2 - \mathcal{H}^2 \varsigma^2 (\Lambda) \left[ \frac{1}{3} + \frac{2}{105} \left( 24 + 23 \frac{k_{\parallel}^2}{k^2} \right) \delta + v^{(2)}_{ct} \right] + \cdots
\end{align}
where $\varsigma^2 (\Lambda) = \int_0^\Lambda \frac{\mathrm{d}p}{2\pi^2} P_L (\boldsymbol{p})$ and $v^{(2)}_{ct}$ is given in Fourier space by 
\begin{align}
    \left( v^{(2)}_{ct} \right)_{\boldsymbol{k}} &= \frac{\delta^D (\boldsymbol{q}_1 + \boldsymbol{q}_2 - \boldsymbol{k})}{10290} \left[ 996 + 2041 \left( \frac{q_{1,\parallel}^2}{q_1^2} + \frac{q_{2,\parallel}^2}{q_2^2} \right) - 2142 \frac{q_{1,\parallel} q_{2,\parallel}}{q_1 q_2} \right. \n
    &\left.\qquad + \frac{\boldsymbol{q}_{1} \cdot \boldsymbol{q}_{2}}{q_1 q_2} \left( 1071 \frac{q_{1,\parallel}^2}{q_1^2} + 1071 \frac{q_{2,\parallel}^2}{q_2^2} - 948 \frac{\boldsymbol{q}_{1} \cdot \boldsymbol{q}_{2}}{q_1 q_2} + 2844 \frac{q_{1,\parallel} q_{2,\parallel}}{q_1 q_2} \right) \right] \delta_{\boldsymbol{q_1}} \delta_{\boldsymbol{q_2}} .
\end{align}
More details about the derivation of the counterterms for $v_\parallel^2$ can be found in Appendix~\ref{sec:v2ct}.
Re-expressing Eqn.~\eqref{eqn:d21_bare} in terms of these renormalized operators, we obtain
\begin{align}
    (\delta_{21,r})_{\boldsymbol{k}} = &b_{1}^{(R)} \delta_{\boldsymbol{k}} - b_{\nabla^2} k^2 \delta_{\boldsymbol{k}} + b_{2}^{(R)} \left[ \delta^2 \right]_{\boldsymbol{k}} + b_{G2}^{(R)} (\mathcal{G}_2)_{\boldsymbol{k}} \n
    &-i \frac{k_\parallel}{\mathcal{H}} \left[ (v_\parallel)_{\boldsymbol{k}} + b_1 (\delta v_\parallel)_{\boldsymbol{k}} - b_{\nabla^2} k^2 (\delta v_\parallel)_{\boldsymbol{k}} \right] - \frac{1}{2} \left(\frac{k_\parallel}{\mathcal{H}} \right)^2 \left[ v_\parallel^2 \right]_{\boldsymbol{k}} + \cdots
    \label{eqn:d21_renorm}
\end{align}
where the renormalized bias coefficients are given by
\begin{align}
    b_1^{(R)} &= b_{1} + \sigma^2 (\Lambda) \left( \frac{34}{21} b_2 \right) - \frac{2}{420} \left( 24 + 23 \frac{k_{\parallel}^2}{k^2} \right) k_\parallel^2 \varsigma^2 (\Lambda) \n
    b_2^{(R)} &= b_{2} + \frac{8126}{2205} \sigma^2(\Lambda) b_2 - \frac{1}{2} k_\parallel^2 \varsigma^2 (\Lambda) v^{(2)}_{ct} \n
    b_{G2}^{(R)} &= b_{G2} + \frac{254}{2205} \sigma^2 (\Lambda) b_2 .
\end{align}
In these expansions, we have only gone to second order in fields.
This is because renormalized operators that start at third order in $\delta^{(1)}$ do not contribute to the one-loop power spectrum.
For example, consider the bare operator $\delta^3$.
This operator's only contribution to the one-loop power spectrum is through the correlation function with the linear density field $\braket{(\delta)^3 \delta^{(1)}}$, which begins at one-loop order and has no tree-level component. To build the renormalized operator $[\delta^3]$, the renormalization condition in Eqn. \eqref{eqn:renorm_condition} requires that the correlation function between $[\delta^3]$ and factors of the linear density field equal the tree-level contribution, which is zero.
Hence, these third order operators have no contribution at one-loop order. This is in contrast to the bare operator $\delta^2$, which contributes to the one-loop power spectrum through $\braket{(\delta)^2 \delta^{(2)}}$; the renormalization procedure does not null out this contribution unlike for the case of $\delta^3$.
\footnote{To be more explicit, the momentum can be decomposed into gradient and curl components as $\pi^i = a \bar{\rho} \left( \frac{\partial^i}{\partial^2} \pi_s + \epsilon^{ijk} \frac{\partial_j}{\partial^2} \pi_{v,k} \right)$. 
The scalar potential $\pi_s$ is related to the density by the continuity equation, $\pi_s = - \dot{\delta}$, and so receives the same counterterms as $\delta$. 
The vector potential $\pi_v$ does not need to be renormalized since it first appears at third order, which is all we need for the one-loop power spectrum, and receives no counterterms at this order. 
Thus, $\pi_i$ requires no additional counterterms. Since $\pi_i \propto (1 + \delta) v_i$, and $v_i$ is already renormalized through Eqn.\eqref{eqn:EFT_theta}, then $\delta v_i$ also has no additional counterterms.}

Finally, we use the SPT ansatzes for $\delta$ and $\theta$ to write the field in terms of the linear density perturbations $\delta^{(1)}$.
We substitute
\begin{align}
	\delta &= \delta^{(1)} + \delta^{(2)} + \delta^{(3)}, \\
	\delta^2 &= (\delta^{(1)})^2 + 2 \delta^{(1)} \delta^{(2)} , \\
	\mathcal{G}_2 &= \left( \frac{\nabla_i \nabla_j}{\nabla^2} \delta^{(1)} \right)^2 + 2 \left( \frac{\nabla_i \nabla_j}{\nabla^2} \delta^{(1)} \right) \left( \frac{\nabla^i \nabla^j}{\nabla^2} \delta^{(2)} \right) \n
	&\qquad - (\delta^{(1)})^2 - 2 \delta^{(1)} \delta^{(2)} , \\
	v_\parallel &= - \mathcal{H} \frac{\nabla_\parallel}{\nabla^2} (\theta^{(1)} + \theta^{(2)} + \theta^{(3)}), \\
	\delta v_\parallel &= - \mathcal{H} \left( \delta^{(1)} \frac{\nabla_\parallel}{\nabla^2} \theta^{(1)} + \delta^{(1)} \frac{\nabla_\parallel}{\nabla^2} \theta^{(2)} + \delta^{(2)} \frac{\nabla_\parallel}{\nabla^2} \theta^{(1)} \right) , \\
	v_\parallel^2 &= \mathcal{H}^2 \left( \frac{\nabla_\parallel}{\nabla^2} \theta^{(1)} \frac{\nabla_\parallel}{\nabla^2} \theta^{(1)} + 2 \frac{\nabla_\parallel}{\nabla^2} \theta^{(1)} \frac{\nabla_\parallel}{\nabla^2} \theta^{(2)} \right) .
\end{align}
As we explain in the next section, we also use the second-order approximation $\delta^2 = (\delta^{(1)})^2$, as we find the $\delta^{(1)} \delta^{(2)}$ term is very noisy and affects the quality of the fit to simulations.

\subsection{Fitting to the \thesan simulations}
\label{sec:thesan}

To validate our EFT calculation of the 21\,cm power spectrum, we fit Eqn.~\eqref{eqn:d21_renorm} to simulations, emphasizing that this procedure is performed at the field level.
For this study, we use the recently developed \thesan simulations~\cite{thesan1,thesan2,thesan3}.

\thesan is a new suite of radiation-magneto-hydrodynamical simulations designed to simultaneously capture the complex physics of cosmic reionization and high-redshift galaxy formation. 
These simulations combine a large comoving box size of $95.5$\,Mpc, high resolution (sufficient to model the formation of atomic cooling halos, the smallest structures significantly contributing to the reionization process), a wide range of self-consistent realistic prescriptions for high-redshift physics (built on top of the successful IllustrisTNG galaxy formation model, described in Refs.~\cite{Weinberger+2017,Pillepich+2018}), and the approach to initial conditions production described in Ref.~\cite{Angulo&Pontzen2016}, which significantly reduces the effect of sample variance, increasing the statistical fidelity of the simulations.
As a reminder, all distances and wavenumbers are reported in comoving units.

The simulations are performed using the code \mbox{\textsc{arepo-rt}} \cite{Arepo, ArepoRT}. 
Radiation-magneto-hydrodynamics equations are solved on a mesh, built from a set of mesh-generating points that approximately follow the gas flow as their Voronoi tessellation. 
This approach ensures a natural increase of resolution in the high-density regions, where it is needed. Gravity is instead computed using an hybrid Tree-PM approach, where long-range forces are computed using a particle mesh algorithm and short-range ones are calculated using a hierarchical oct-tree \cite{Barnes&Hut86}.
The photon production rate for star is computed using the BPASS \cite{BPASS2017, BPASS2018} library. 

Among other observables, the \thesan simulations have been shown to reproduce realistic realizations of the reionization history of the Universe, IGM temperature evolution, optical depth to the CMB, $z \geq 6$ UV luminosity function \cite{thesan1}, photo-ionization rate, mean free path of ionising photons, IGM opacity and temperature-density relation \cite{thesan2}.

The different simulations that make up the \thesan suite are described in Ref.~\cite{thesan1} and shown in Figure \ref{fig:slices}.
Here, we briefly summarize the properties of the simulations we include in this study.
\begin{itemize}
    \item \thesan-1: The highest resolution simulation with $2100^3$ dark matter particles of mass $3.12 \times 10^6\,\Msun$ and $2100^3$ gas particles of mass $5.82 \times 10^5\,\Msun$.
    
    \item \thesan-2: A medium resolution simulation with $1050^3$ dark matter particles of mass $2.49 \times 10^7\,\Msun$ and $1050^3$ gas particles of mass $4.66 \times 10^6\,\Msun$.
    This simulation is the same as \thesan-1, but the spatial resolution has been lowered by a factor of 2 (i.e. the particles in the simulations have been coarse-grained to be more massive by a factor of 8).
    
    \item \thesan-\textsc{wc}-2: Same as \thesan-2, but the birth cloud escape fraction is slightly higher to compensate for lower star formation in the medium resolution runs. 
    The total integrated number of photons emitted in \thesan-1 and \thesan-\textsc{wc}-2 are the same.
    
    \item \thesan-\textsc{high}-2: Same as \thesan-2, but using a halo-mass-dependent escape fraction, with only halos \textit{above} $10^{10}\,\Msun$ contributing to reionization.
    
    \item \thesan-\textsc{low}-2: Same as \thesan-2, but using a halo-mass-dependent escape fraction, with only halos \textit{below} $10^{10}\,\Msun$ contributing to reionization.
    
    \item \thesan-\textsc{sdao}-2: Same as \thesan-2, but using a non-standard dark matter model that includes couplings to relativistic particles. 
    The effect of these new interactions is to give rise to sDAOs and cut off the linear matter power spectrum at small scales~\cite{Cyr-Racine:2013fsa}.
    This difference is quantified through a transfer function, which is defined as the square root of the ratio between the DAO matter power spectrum and the standard cold dark matter power spectrum~\cite{Bohr:2020yoe}.
\end{itemize}

For this study, we use each simulation on a $128^3$ grid.
Higher resolutions are not necessary since our methods are only relevant on the largest scales.
To simulate the redshift space distortions, we create a mock observer and adjust the particle data according to their peculiar velocities.
Due to the $128^3$ render grid that we use, we can only resolve RSDs corresponding to peculiar velocities greater than $\Delta v = H(z) \Delta r = 39 \,\mbox{km/s} \times \sqrt{\frac{1+z}{7}} \frac{128}{N_\mathrm{pix}}$, where we have used Hubble's law, $\Delta r$ is the smallest change in distance we can resolve on the grid, and $N_\mathrm{pix}$ is the number of pixels along one dimension of the grid.

\subsubsection{Fitting the bias parameters}
\label{sec:fitting}

\begin{figure}
	\centering
	\includegraphics{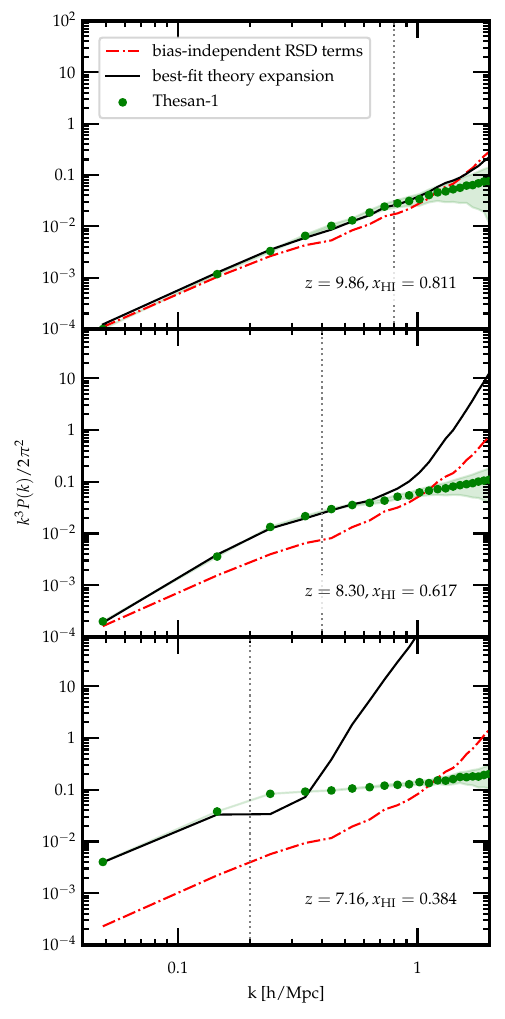}
	\caption{21\,cm power spectra in redshift space at various redshifts and values for the neutral hydrogen fraction.
	Green dots indicate the binned power spectrum from the \thesan-1 simulation; the shaded regions indicate the shot noise error.
	The black dashed line shows the best fit from the effective field theory. 
	The fits are performed at the level of the fields, rather than at the power spectrum level.
	The red dash-dotted line shows the relative contribution of the terms with no bias coefficients. 
	The vertical gray dotted line is at $k_\mathrm{NL}$, which is the maximum wavenumber that we fit up to.
	For earlier redshifts, our theory expansion remains a good fit to \thesan at slightly larger wavenumbers than $k_\mathrm{NL}$, indicating that the 21\,cm intensity field is still perturbative at these times and our theory still has predictive power for smaller scales than the ones that we fit.
	}
	\label{fig:power_spectra}
\end{figure}

To test our perturbative expansion, we use the same method as in Ref.~\cite{McQuinn:2018zwa} and fit to simulations at the level of the fields, instead of the power spectrum itself.
Since the power spectrum has a broad shape and our bias expansion has many parameters, fitting directly at the level of the power spectrum could be subject to overfitting. 
Instead, we fit the \textit{fields} at every mode with wavenumber less than $k_\mathrm{NL}$, where for practical purposes we define $k_\mathrm{NL}$ to be the scale at which the simulated 21~cm field smoothed over $k_\mathrm{NL}$ has a maximum value of $|\delta_\mathrm{sim}| = 0.8$ in redshift space. 
In other words, $k_\mathrm{NL}$ is analogous to the smoothing scale $\Lambda$ we introduced in Section~\ref{sec:eft}.
In principle, one should choose the $\Lambda$ to be much less than $k_\mathrm{NL}$; however, the number of modes available to fit drastically decreases as we lower the wavenumber cutoff, from 6043 modes at $\Lambda = 1.1$ h/Mpc to 7 modes at $\Lambda = 0.1$ h/Mpc.
Hence, we find that using a smaller $\Lambda$ only increases the fit error, without substantially changing the best fit parameters, so we choose to use $\Lambda = k_\mathrm{NL}$.
In addition, one could choose a different criterion with which to define the nonlinear wavenumber, but we find that varying this threshold between 0.6 and 1.0 also does not substantially change the range of wavenumbers that we fit.

In fitting at the field level, we take the error power spectrum,
\begin{equation}
    P_\mathrm{err} (\boldsymbol{k}) = \frac{|(\delta_\mathrm{sim})_{\boldsymbol{k}} - (\delta_\mathrm{EFT})_{\boldsymbol{k}}|^2}{V} ,
\end{equation}
where $V$ is the simulation volume, and minimize the value of $P_\mathrm{err} (\boldsymbol{k})$ over all modes up to $k_\mathrm{NL}$.
We emphasize that this quantity is \textit{not} the same as the error on the power spectrum; it is the power spectrum of errors at the field level.
Thus, our cost function is
\begin{equation}
    \mathcal{A} = \sum_{\boldsymbol{k}} w_{\boldsymbol{k}} P_\mathrm{err} (\boldsymbol{k}),
\end{equation}
where the sum is over every \textit{mode} and not just every wavenumber value, $w_{\boldsymbol{k}}$ is the weight that we assign to each mode and quantifies how we smooth the fields, $\delta_\mathrm{sim}$ is the simulation field we want to fit to, and $\delta_\mathrm{EFT}$ is the theory expansion.
In this study, $\delta_\mathrm{sim}$ describes perturbations to the redshifted 21\,cm brightness temperature, neglecting fluctuations in spin temperature since we are assuming the $T_\mathrm{spin} \gg T_\mathrm{CMB}$ limit, and $\delta_\mathrm{EFT}$ is given by Eqn.~\eqref{eqn:d21_renorm}. 
The error power spectrum is also sometimes referred to as ``stochasticity'' and is commonly used to quantify the error in estimators of fields~\cite{Seljak:2004ni,Bonoli:2008rb,2010PhRvD..82d3515H,2011MNRAS.412..995C,Baldauf:2013hka,Modi:2016dah}.
If we were able to perfectly construct the 21\,cm field using our model, we would expect $P_\mathrm{err} (\boldsymbol{k}) = 0$.
However, there is an irreducible shot noise contribution to the error due to the discreteness of the simulation particles, which is given by~\cite{2018MNRAS.475..676S}
\begin{equation}
    P_\mathrm{shot} = \frac{V}{N_\mathrm{eff}},
    \label{eqn:Pshot}
\end{equation}
where $N_\mathrm{eff}$ is the effective number of neutral hydrogen tracers, given by
\begin{equation}
    N_\mathrm{eff} = \frac{M^2}{\braket{m^2}}.
    \label{eqn:Neff}
\end{equation}
In this expression, $M = \sum_i m_i$ is the total mass of the tracers and $\braket{m^2} = \left( \sum_i m_i^2 \right) / N$ their mean squared mass, with $N$ being the total number of tracers.

Since we are only fitting modes with wavenumber less than $k_\mathrm{NL}$,
\begin{equation}
    w_{\boldsymbol{k}} = \begin{cases} 1, k < k_\mathrm{NL}, \\ 0, k > k_\mathrm{NL}. \end{cases}
\end{equation}
This choice of weights corresponds to performing least-squares regression. Instead of implementing a sharp cutoff such as this, we could choose to fit the simulation using a smoother filter, such as a Gaussian that down-weights the relative importance of modes closer to the nonlinear scale in determining the fit parameters. 
However, we find that the best-fit parameters are robust to the choice of filter, and $\mathcal{A}$ is smaller for the sharp cutoff filter compared to the Gaussian filter by about 25-30\%.

To calculate the operators appearing in Eqn.~\eqref{eqn:d21_renorm}, we take $\delta^{(1)}$ to be the initial conditions of the simulation, which are seeded at $1+z = 50$ when the perturbations on the scales of interest should still be in the linear regime.
We calculate $\delta^{(2)}$ and $\delta^{(3)}$ using equivalent methods from Lagrangian perturbation theory, since the Lagrangian theory displacements are easier to compute~\cite{Scoccimarro:1997gr,Baldauf:2015zga}.
The velocity factors can then be calculated using
\begin{align}
    \theta^{(1)} &= \delta^{(1)} , \nonumber \\
    \theta^{(2)} &= \delta^{(2)} + \frac{2}{7} \mathcal{G}_2^{(2)} , \nonumber \\
    \theta^{(3)} &= \delta^{(3)} + \frac{2}{9} \left[\mathcal{G}_{2,v} + \frac{1}{7} \nabla^2 \left( \nabla_i \phi \frac{\nabla^i}{\nabla^2} \mathcal{G}_2 \right) \right]^{(3)} .
\end{align}
Above, we denote $\mathcal{G}_{2,v} = \nabla_i \nabla_j \phi \nabla^i \nabla^j \phi_v - \nabla^2 \phi \nabla^2 \phi_v$, where $\phi_v = \theta / \nabla^2$ is the velocity potential.
The superscript 3 at the end of the brackets indicates that we are only keeping terms up to third order in $\delta^{(1)}$.
See Appendix~\ref{sec:theta} for a derivation of these relationships.
Finally, we find that using the second-order approximation for the term $\delta^2 = (\delta^{(1)})^2$ leads to a better fit at most redshifts, see Appendix~\ref{sec:d2_or_d3} for details.
Hereafter, we only show fits using the second-order approximation for $\delta^2$.

Figure~\ref{fig:power_spectra} shows the resulting power spectrum that comes from fitting Eqn.~\eqref{eqn:d21_renorm} to the \thesan-1 simulation \emph{fields} at redshifts of $z =$ 9.86, 8.30, and 7.16.
This corresponds to neutral hydrogen fractions of $x_\mathrm{HI} =$ 0.811, 0.617, and 0.384, respectively.
The red dash-dotted lines show the relative contribution of the bias-independent terms that arise from the RSD expansion, namely $-i \frac{k_\parallel}{\mathcal{H}} (v_\parallel)_{\boldsymbol{k}} - \frac{1}{2} \left(\frac{k_\parallel}{\mathcal{H}} \right)^2 ({v_\parallel^2})_{\boldsymbol{k}}$.
In the first panel, which is the earliest redshift at $z=9.86$, we see that although we only fit up to a maximum wavenumber of $k_\mathrm{NL} = 0.8$ h/Mpc, the series expansion power spectrum smoothly diverges from that of the simulation at wavenumbers above $k_\mathrm{NL}$.
We find similar behavior at $z=8.30$, where we instead fit up to $k_\mathrm{NL} = 0.4$ h/Mpc.
In this case, the theory continues to fit the simulation power spectrum up to a wavenumber of about 0.7 h/Mpc.
This is encouraging, as it indicates that our effective theory has some predictive power past the wavenumbers that we fit. 
By the time the simulation reaches $z = 7.16$, reionization has nearly concluded and the the neutral fraction is much smaller compared to the other redshifts we show.
As a result, $\delta_{21,r}$ is becoming nonperturbative even on the largest scales, since the ionized bubbles have grown quite large as well. 
This degrades the quality of the fit in the last panel relative to the previous redshifts.
For this redshift, we use $k_\mathrm{NL} = 0.2$ h/Mpc and the series expansion is not at all predictive above $k_\mathrm{NL}$.

\begin{figure}
	\centering
	\includegraphics{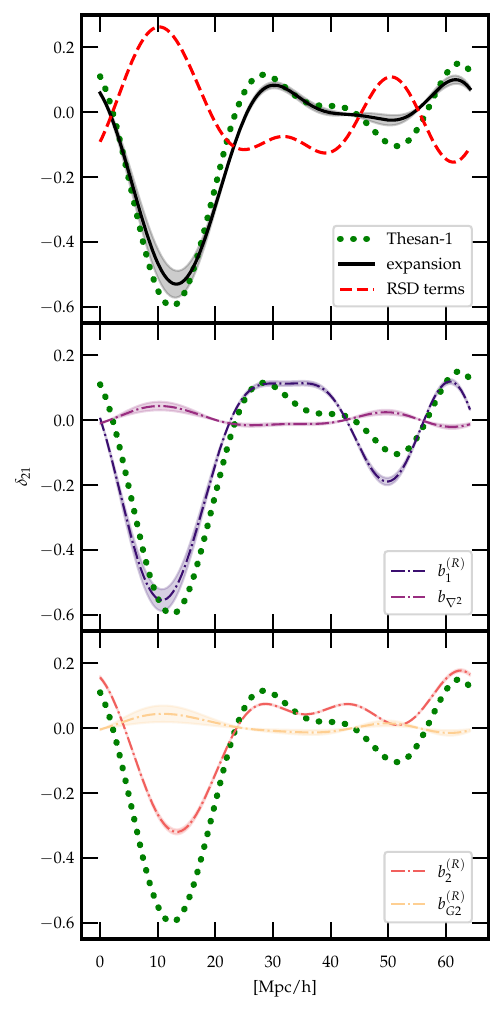}
	\caption{The redshift space 21\,cm differential brightness temperature along a line through the simulation volume at $z = 8.30$, $x_\mathrm{HI} = 0.617$, smoothed over $k_\mathrm{NL} = 0.4$ h/Mpc. 
	The green dots show the signal from the \thesan-1 simulation, the thick black line in the first panel is the best fit theory expansion. 
	For comparison, we also show the contributions of each of the bias parameters to the black line in the other panels, as well as the bias independent contribution (thick red dashed).
	The filled contours show the 68\% confidence intervals on the fitted coefficients.
	}
	\label{fig:1d-slice}
\end{figure}

To see how well the bias expansion fits at the field level, we show in Figure \ref{fig:1d-slice} the fluctuations in the redshift space 21\,cm differential brightness temperature along a line through the simulation volume at $z = 8.30$ and $x_\mathrm{HI} = 0.617$ (in Appendix~\ref{sec:fit_grid}, we show fluctuations in the differential brightness temperature along several lines through the volume).
The fields are smoothed over $k_\mathrm{NL} = 0.4$ h/Mpc.
The green dots show the signal from \thesan-1 and the solid black line in the top panel is the signal from the best fit theory expansion.
We also show the contributions of each term to the best fit theory expansion in the other panels.
Along this particular line, we see that the shapes that dominate the fit are the terms multiplying the $b_1^{(R)}$ coefficient.
Moreover, some of curves show a degree of degeneracy with each other.
Past studies have dealt with such degeneracies using a Gram-Schmidt process to orthogonalize the shapes~\cite{Schmittfull:2018yuk}; we leave an exploration of such a procedure on the 21\,cm field in redshift space to future work. 

To quantify the level of agreement between the simulation and best-fit 21\,cm fields in configuration space, we take the root mean square of fluctuations in the brightness temperature, as well as their difference.
We find the simulation box has a root mean square fluctuation of $0.134$, the best-fit theory field is $0.133$, and their root mean square difference is $0.050$. 
Thus, the disagreement at the \textit{field level} is about $\sim 30\%$. 
We note that the differences appear greatest at the field's extrema.
In the EFT of LSS, comparisons are not typically done at the level of fields in configuration space, but at the level of the power spectrum.
We emphasize that the level of agreement between the power spectra is still percent-level; however, if we were to use our effective field theory description to ``paint on'' the 21\,cm field over a linear initial density field, the root-mean-square difference with the simulated 21\,cm field would likely be degraded at the level of $\sim \mathcal{O}(10\%)$.

\begin{figure}
	\centering
	\includegraphics{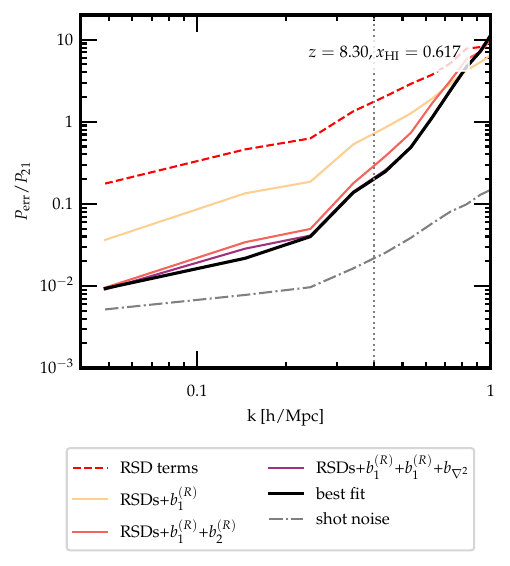}
	\caption{The error power spectrum for the best fit theory field to the \thesan-1 simulation at $z = 8.30$ and $x_\mathrm{HI} = 0.617$.
	Starting with the contribution of the bias-independent RSD terms, the error spectrum is reduced as we add terms one by one, starting with the most dominant terms.
	Shown in the grey dash-dotted line is the contribution from shot noise.
	}
	\label{fig:error}
\end{figure}

We can also assess goodness of fit by looking at the error power spectrum with the best fit coefficients.
Figure~\ref{fig:error} shows the error power spectrum at $z = 8.30$ for the best fit to \thesan-1.
The red dashed line shows the contribution of the bias-independent RSD terms; as the most dominant terms in the theory are added one by one, the error power spectrum is reduced.
For comparison, we also show the curve for shot noise (grey dash-dotted), which is calculated using Eqns.~\eqref{eqn:Pshot} and \eqref{eqn:Neff}.

Our method for fitting the bias parameters takes up to a few minutes to run on a standard laptop and can be further optimized for speed.
This highlights the promise of effective field theory methods for studying 21\,cm cosmology, as this runtime is a mere drop in the bucket compared to the e.g. 30 million CPU hours required to generate the \thesan simulation suite~\cite{thesan_web}.

\subsection{Discussion}
\label{sec:discussion}

\begin{figure}
	\centering
	\includegraphics{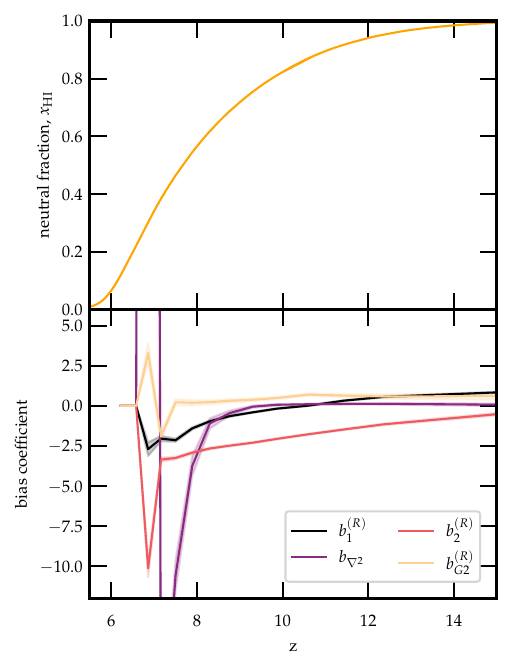}
	\caption{Evolution of the neutral hydrogen fraction and bias coefficients for \thesan-1.
	The top panel shows the reionization history of \thesan-1.
	The bottom panel shows how the bias coefficients change between $z \in [6,15]$.
	The filled contours show the 68\% confidence intervals on the fitted coefficients.
	The evolution of the coefficient is quite smooth down to redshift $z \sim 7.5$, but evolves rapidly thereafter, signalling the breakdown of the perturbation theory.}
	\label{fig:evo}
\end{figure}

Figure~\ref{fig:evo} shows the evolution of the neutral hydrogen fraction and best-fit bias coefficients for \thesan-1.
While the evolution of the bias coefficients is smooth prior to $z \sim 7.5$, at later times the evolution is very rapid, signalling that our perturbative treatment is breaking down as the universe becomes very ionized.

The bias parameters have natural physical interpretations:
\begin{itemize}
    \item $b_1^{(R)}$ is the linear bias and measures how well the 21\,cm field traces the underlying linear matter density.
    
    \item $b_{\nabla^2}$ is related to the effective size of the ionization bubbles $R_\mathrm{eff}$, as argued in Ref.~\cite{McQuinn:2018zwa},
    $$b_{\nabla^2} = \frac{1}{3} b_1^{(R)} R_\mathrm{eff}^2.$$
    As we would expect, this quantity is small at the beginning of reionization, but grows larger with time.
    Once this quantity becomes very large, we expect the ionization field to be quite nonperturbative, hence our formalism will no longer apply.
    
    \item $b_2^{(R)}$ is the quadratic bias, and therefore related to nonlinearities in the 21\,cm field. 
    
    \item $b_{G2}^{(R)}$ is the coefficient for the tidal field, which captures the effects of local anisotropies in the matter field. 
    The length scales of these anisotropies are small compared to the ionization bubbles, hence on the large scales we consider, this term should be subdominant. 
    In addition, the galaxies that source ionizing radiation are highly biased tracers, hence the quadratic bias terms will dominate over the tidal term~\cite{McQuinn:2018zwa}. 
\end{itemize}

\begin{figure*}
	\includegraphics[scale=0.83]{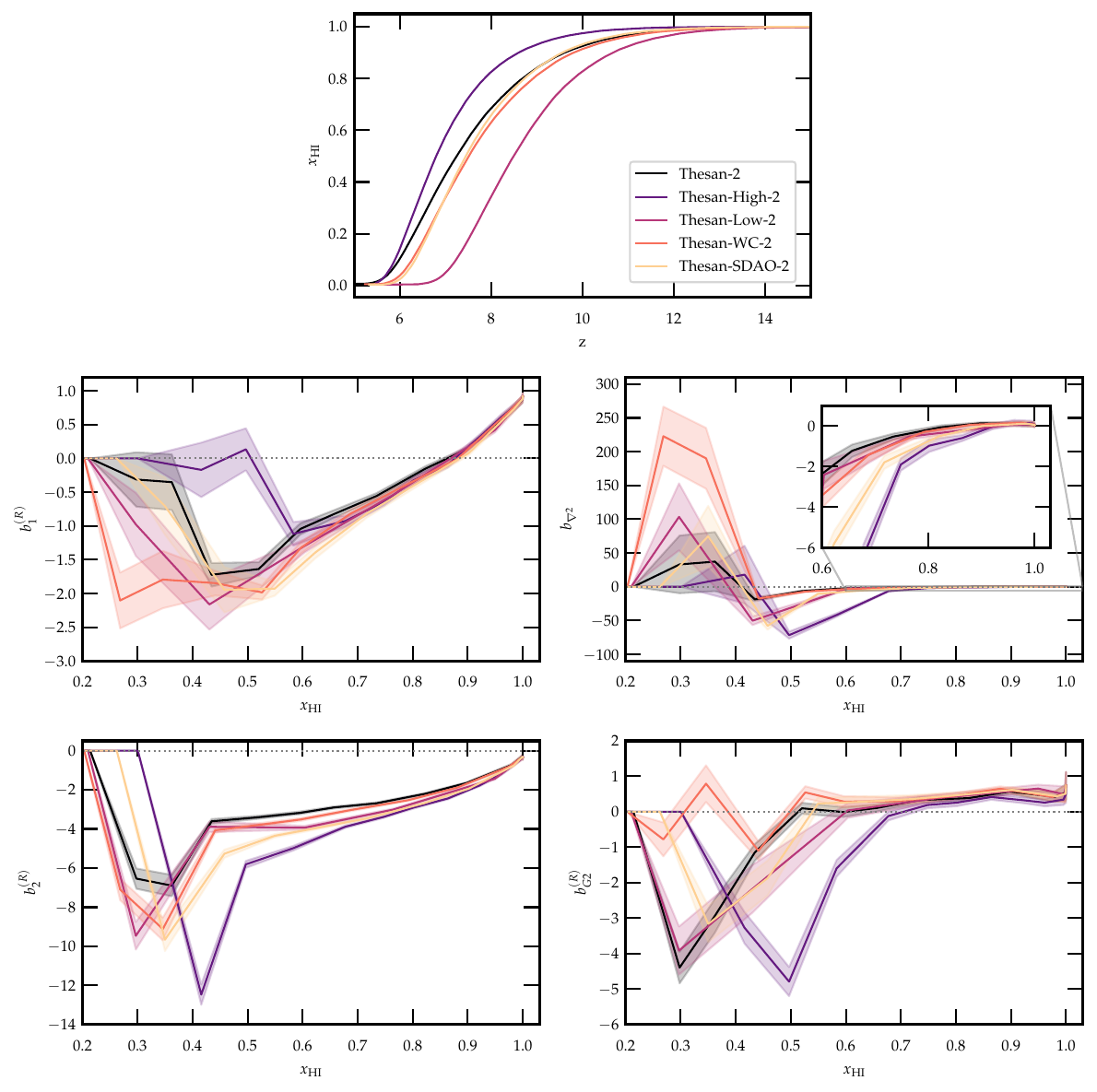}
	\caption{Evolution of the neutral hydrogen fraction and bias coefficients for the various \thesan-2 simulations.
	The filled contours show the 68\% confidence intervals on the fitted coefficients.
	For these fits, we continue to use the approximation $\delta^2 = (\delta^{(1)})^2$, which yields a slightly better fit for many redshifts compared to the third-order expression.
	The top panel shows the neutral fraction as a function of redshift for the different simulations---since the ionization histories vary widely, it is more appropriate to compare simulations at similar values of $x_\mathrm{HI}$, instead of redshift.
	The remaining panels show the best-fit bias coefficients as a function of $x_\mathrm{HI}$.
	At early times, i.e. high neutral fractions, the parameters evolve relatively smoothly; the curves begin to diverge at different values of the neutral fraction, indicating when the perturbative expansion breaks down for each simulation.}
	\label{fig:compare_bias}
\end{figure*}

In order to verify these interpretations of the bias parameters, we can look at how the fitted parameters change between simulations run with different physical models.
Figure~\ref{fig:compare_bias} compares the ionization histories and bias parameters for the five different versions of \thesan-2.
Since the ionization histories can vary widely depending on the physics, fair comparisons of the simulations should be done at similar values of the neutral hydrogen fraction, instead of similar redshifts.
Hence, the remaining panels of Figure~\ref{fig:compare_bias} show the evolution of the bias parameters as a function of $x_\mathrm{HI}$.
At early times and high neutral fractions, the evolution of the bias parameters is relatively smooth.
As reionization progresses and $x_\mathrm{HI}$ decreases, the curves begin to diverge from each other and evolve more rapidly, indicating when the perturbative expansion breaks down for each of the simulations.
At late times, the fitted values of these coefficients should not be trusted; however, it is not surprising that they tend towards zero at the very end of reionization, since the 21\,cm signal should vanish as the neutral fraction goes to zero.

For all of the simulations, $b_{\nabla^2}$ is initially small, then blows up after some critical value for the neutral fraction.
Note that since $b_1^{(R)}$ is negative at the redshifts where significant bubble growth occurs, we expect $b_{\nabla^2}$ to also be negative according to its relation with $R_\mathrm{eff}$.
$b_{\nabla^2}$ diverges earliest for \thesan-\textsc{high}-2 and \thesan-\textsc{sdao}-2, as we would expect since these simulation source the largest bubbles; the coefficients exceed $b_{\nabla^2} < -5$ before reionization has even reached the halfway point.
In contrast, the other three simulations evolve quite similarly at high neutral fractions, with $b_{\nabla^2} > -4$ up to $x_\mathrm{HI} = 0.6$. In addition, the coefficient $b_2^{(R)}$ starts near zero at the beginning of reionization and grows in magnitude with time.
This indicates the growing bias of the signal over time; since $b_2^{(R)}$ is consistently most negative for \thesan-\textsc{high}-2 and thesan-\textsc{sdao}-2 at early times, the 21\,cm signal is more highly biased in these models.
Again, this is not surprising, since reionization is driven by the largest halos in these simulations, and such halos form in the largest overdensities. Finally, $b_{G2}^{(R)}$ is relatively small at early times for all the simulations, with values ranging between about 0 and 0.5 for $x_\mathrm{HI} > 0.7$.
This is in line with our argument that contributions to the 21\,cm power spectrum from anisotropies should be small.

\begin{figure}
	\centering
	\includegraphics{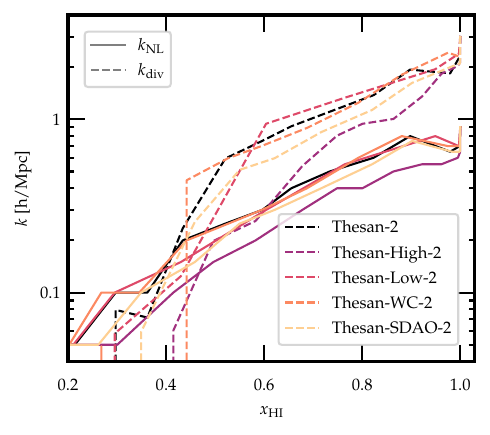}
	\caption{$k_\mathrm{NL}$ (solid lines) and $k_\mathrm{div}$ (dashed lines) as a function of redshift for the various \thesan-2 simulations.
	Regions where $k_\mathrm{div} > k_\mathrm{NL}$ indicate that our theory has predictive power past the wavenumbers that we fit, demonstrating that perturbative methods are valid at high enough values for the neutral fraction, i.e. early enough in the process of reionization.}
	\label{fig:compare_ks}
\end{figure}

To estimate the validity of our EFT methods beyond the range of modes we fit to, we define $k_\mathrm{div}$ as the wavenumber at which the power spectrum of the perturbative expansion with the best-fit bias parameters diverges from the simulation power spectrum by a factor of two.
Figure~\ref{fig:compare_ks} shows $k_\mathrm{NL}$ (dashed lines) and $k_\mathrm{div}$ (solid lines) as a function of $x_\mathrm{HI}$.
At early times or high enough neutral fractions, $k_\mathrm{div} > k_\mathrm{NL}$, which indicates the theory has predictive power even at scales smaller than those that we fit to.
This indicates that our EFT method is valid earlier on in reionization.
Notice that $k_\mathrm{div}$ falls below $k_\mathrm{NL}$ at the highest value of $x_\mathrm{HI}$ for \thesan-\textsc{high}-2 compared to the other simulations, again demonstrating that this simulation field is the least perturbative, due to the large size of the ionized bubbles.

\subsubsection{Observational limits}
\label{sec:experiment}

\begin{figure}
	\centering
	\includegraphics{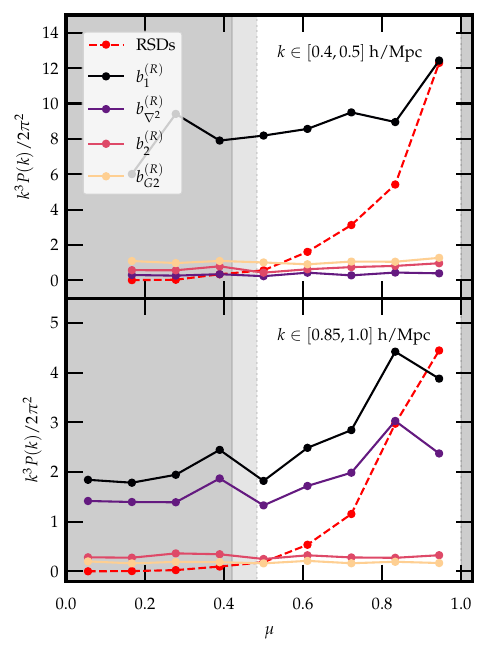}\vspace{-0.4cm}
	\caption{Power in the different operators appearing in the bias expansion, binned over the angle from the line of sight $\mu = \cos\theta$.
	For the top panel, we only average over modes with wavenumbers $k \in [0.4,0.5]$ h/Mpc.
	For the lower panel, we only average over modes with wavenumbers $k \in [0.85,1.0]$ h/Mpc.
	The windows between the light gray (dark gray) regions show the range of $\mu$ that HERA could probe for Band 1 (Band 2), if the foreground wedge could be mitigated.
	We see that all the shapes are relatively flat, except for the term multiplying $b_{\nabla^2}$ on certain scales and the terms appearing due to RSDs.
	The top panel is missing the lowest bin because the simulation did not have modes at those angles for the given range of $k$, due to the lower resolution of the $128^3$ grid used for our analysis.}
	\label{fig:mu_shapes}
\end{figure}

Experiments such as HERA will soon have measurements of the 21\,cm brightness temperature field.
Already, HERA has set upper limits on the 21\,cm power spectra in the spectral windows spanning 117.1--132.6\,MHz (Band 1) and 150.3--167.8\,MHz (Band 2)~\cite{HERA:2021bsv}.
The central frequencies of these bands are 124.8 and 159.0\,MHz, corresponding to redshifts of 10.4 and 7.9, respectively.
HERA and other instruments primarily observe modes along the line of sight; in other words, if we define $\mu = \cos\theta$, where $\theta$ is the angle from the line of sight, then typically instruments probing the 21\,cm power spectrum are most sensitive to larger values of $\mu$. 
Furthermore, due to the chromatic response of interferometers, foregrounds in 21\,cm intensity mapping will leak into the so-called ``foreground wedge"~\cite{2010ApJ...724..526D,2012ApJ...745..176V,2012ApJ...752..137M,Trott:2012md,Thyagarajan:2013eka,Liu:2014bba,Liu:2014yxa}; cuts to avoid foregrounds thus reduce the observable range of $\mu$ even further~\cite{DeBoer:2016tnn}.

If the shapes of the terms that appear in the bias expansion power spectrum look similar across the range of angles that HERA can probe, then varying the bias coefficients of these terms will have the same effect, and a measurement of the power spectrum will not allow us to distinguish the different operators.
To investigate which of the terms in our bias expansion are effectively degenerate, we plot the power in the terms multiplying each bias coefficient in Figure~\ref{fig:mu_shapes} as a function of $\mu$, averaging over $k \in [0.4,0.5]$ h/Mpc and $k \in [0.85,1.0]$ h/Mpc.
The windows between the light gray (dark gray) regions show the range of $\mu$ that HERA could probe in Band 1 (Band 2), if one could mitigate foregrounds within the aforementioned wedge.
From these figures, we see that the power in most of the terms is relatively flat as a function of $\mu$. 
The exceptions are the terms multiplying $b_1$ and $b_{\nabla^2}$ on small scales and the bias-independent terms arising as a result of RSDs.
Thus, without signal contamination from the wedge, HERA could distinguish a few types of shapes in the power spectrum: the relatively flat contribution coming from terms multiplying $b_2^{(R)}$ and $b_{G2}^{(R)}$; the contribution from $b_1^{(R)}$ and $b_{\nabla^2}$; and the contribution from RSDs, which have the strongest scaling with $\mu$ in the relevant range of angles.
In reality, the data cuts that HERA uses to avoid foregrounds limits the observable range to $\mu \gtrsim 0.98$, i.e. nearly directly along the line of sight.
This foreground window may differ between experiments, e.g. LOFAR can probe $\mu \gtrsim 0.97$~\cite{Mertens:2020llj}, and SKA predicts an observable window of $\mu \gtrsim 0.67$~\cite{Wolz:2015sqa}; however, it is clear that the capacity for experiments to distinguish between different shapes in the bias expansions would be greatly improved if one could recover information inside the foreground wedge.

Using the upper limits set by HERA at 95\% confidence level, it is also possible to constrain the possible range of bias parameters. 
A simple method to estimate the constraints would be to set all parameters to zero except the one of interest; we then vary the parameter until we find the values where the power spectrum of the bias expansion lies just under the HERA power spectrum upper limits.
While this is neither the most conservative nor accurate method for constraining the bias parameters, we expect the true values of the bias coefficients to have absolute values much smaller than these limits anyways, since the current upper limits on the power spectrum as measured by HERA are largely set by instrumental systematics and thermal noise~\cite{HERA:2021bsv}.
We find that the estimated HERA constraints on the bias parameters are about one to two orders of magnitude larger than the values fit from simulations. As 21~cm experiments continue to peel away instrumental systematics and take more data, these constraints will shrink and give a more meaningful estimate for the bias parameters.

\subsection{Conclusions}
\label{sec:EFT_conclusion}

In this study, we have incorporated renormalized bias and redshift space distortions into an EFT-inspired description for the 21\,cm brightness temperature.
Using the \thesan simulations, we have shown that these perturbative techniques are valid for describing the behavior of large scales ($k \lesssim 0.8$ h/Mpc) early in reionization ($x_\mathrm{HI} \gtrsim 0.4$). In particular, we can achieve percent-level agreement at the level of the power spectrum and $\mathcal{O}(10\%)$ level agreement at the field level on large scales. We have given physical interpretations for the bias parameters and used simulations run with different physics to test these interpretations.
Since the simulations have very different ionization histories, we have compared them at the same values for ionization and found that the \thesan-\textsc{high}-2 simulation is perturbative for a smaller range of $x_\mathrm{HI}$ due to the larger sizes of the ionization bubbles, while the behavior of \thesan-\textsc{sdao}-2 lies between \thesan-\textsc{high}-2 and the other simulations.

Finally, we have drawn connections between our work and interferometry experiments by showing which shapes in the power spectrum HERA will be able to distinguish in the range of angles that they are sensitive to. We have also estimated how the HERA upper limits on the 21\,cm power spectrum constrain our parameters.

There are many directions that can be taken to further develop this perturbative treatment of the 21\,cm intensity field.
For example, we found some of the terms in the theory are partially degenerate in describing the power spectrum. To reduce such degeneracies and better understand the true degrees of freedom involved, one could apply a method to orthogonalize the shapes appearing in the effective field theory, as was done in Ref.~\cite{Schmittfull:2018yuk}.
In addition, we ignored spin temperature fluctuations in this work; however, this will be an important effect to include if we want to extend this description to describe redshifts where $T_\mathrm{spin} < T_\mathrm{CMB}$.
Such improvements will be critical for using these methods to extract astrophysical and cosmological information from the 21\,cm power spectrum, as we gain important new insights into the EoR in the next few years.

\chapter{Conclusions}
\label{sec:thesis-conclusion}

The work I have done over the course of pursuing my degree represents a few small steps to push forward the frontiers of dark matter phenomenology and cosmology.
In this thesis, I discussed my research studying multiple cosmological probes of exotic energy injection, exploring a multifield model of inflation that successfully produces PBH dark matter, and developing an analytic description of 21\,cm signals that will be imminently measured by radio interferometers.

For each chapter, there are a number of interesting future directions to pursue.
Recent combined observations by the James Webb Space Telescope and the Chandra X-ray Observatory have identified~\cite{Bogdan:2023ilu,Natarajan:2023rxq} a high-redshift quasar that appears to host a surprisingly large black hole with a mass of $\sim 4 \times 10^7 M_\odot$.
One hypothesis for how such a black hole could form is directly from the collapse of a gas cloud that is cooling too slowly to fragment into stars.
Hence, sources of exotic heat such as decaying or annihilating dark matter could not only modify early star formation, but facilitate the formation of these direct collapse black holes.
Another hypothesis is that the observed quasar could have started as a heavy PBH seed with mass of about $10^4 M_\odot$.
Such an object would have to accrete three more orders of magnitude of mass, but this is still a more realistic accretion scenario compared to the required accretion when starting with a Pop III stellar remnant, and therefore is in some ways a more plausible explanation of high-redshift overmassive black holes.

Regarding the multifield model described in Chapter~\ref{sec:PBHandMFI}, 
the NANOGrav 15-year dataset shows strong evidence for a gravitational wave background~\cite{NANOGrav:2023gor,NANOGrav:2023hvm}.
Such a signal could be generated by models like the one studied in this thesis.
In addition, fields that would normally be considered spectators during inflation (e.g. that do not contribute to the evolution of background dynamics) could have a significant impact on the USR phase that produces PBHs, due to the necessary fine-tuning to achieve a sufficiently large enhancement of the power spectrum.
Studying how the addition of a spectator impacts this particular model would be a first step towards showing the effect of spectators generally on such PBH production mechanisms.

Finally, the EFT of 21\,cm signals is still in the very early stages of its development.
There are a number of ways that this formalism could be extended, such as incorporating spin temperature fluctuations or orthogonalizing the operators that appear in the theory to reduce the large degeneracies between terms.
I am also looking into applying this method to mode reconstruction.
Since large scale modes can induce correlations among small-scale modes, it may be possible to reconstruct modes within the 21\,cm foreground wedge using the modes that are directly observed~\cite{Darwish:2020prn}.

Many of the key developments of modern physical cosmology were established in the past few decades.
With the ongoing transition into the age of precision cosmology and the plethora of upcoming datasets, the next few decades also hold great promise to illuminate the remaining mysteries of our cosmos, such as the unknown nature of dark matter.
I hope the ideas presented in this thesis will prove to be useful steps towards advancing our collective understanding of the early universe, and I aim to continue exploring the fields of dark matter and cosmology and help uncover the answers to these long-standing questions.
\appendix
\chapter{Supplementary material for Section~\ref{sec:Lya}}
\label{app:Lya_supp}

\section{Terms in the Evolution Equations}
\label{app:rates}

In this appendix we provide explicit expressions for the terms appearing in Eq.~\eqref{eq:ionization_diff_eq} and~\eqref{eq:temp_diff_eq} and explicitly write down the helium ionization evolution equations.  Starting with the non-DM temperature sources,
\begin{alignat}{1}
\dot{T}_\text{adia} &= -2 H T_\text{m} \,,\nonumber \\ 
\dot{T}_\text{C} &= -\Gamma_C (T_\text{CMB} - T_\text{m}) \, ,
\end{alignat}
where $H$ is the Hubble parameter, $T_\text{CMB}$ is the temperature of the CMB, and $\Gamma_C$ is the Compton cooling rate 
\begin{align}
\Gamma_C = \frac{x_\text{e}}{1 +\chi + x_\text{e}} \frac{8 \sigma_T a_r T_\text{CMB}^4}{3 m_\text{e}} \,.
\end{align}
Here, $\sigma_T$ is the Thomson cross section, $a_r$ is the radiation constant, and $m_\text{e}$ is the electron mass.
The DM temperature source is given by
\begin{alignat}{1}
\dot{T}_\text{DM} &= \frac{2 f_\text{heat}(z, \mathbf{x})}{3(1 + \chi + x_\text{e}) n_\text{H}} \left(\frac{dE}{dV\, dt}\right)^\text{inj}\,
\end{alignat}
where $f_\text{heat}(z, \mathbf{x})$ is the deposition efficiency fraction into heating of the IGM as a function of redshift $z$ and a vector, ${\bf x}$, storing the ionization levels of HI and HeII, which is computed by \texttt{DarkHistory}.  
$\left(\frac{dE}{dV\, dt}\right)^\text{inj}$ is the total amount of energy injected per volume per time through DM decays or annihilations.
Finally, $\dot{T}_\text{atom}$ is given by the sum of the recombination, collisional ionization, collisional excitation, and bremsstrahlung cooling rate fitting functions given in Appendix B4 of Ref.~\cite{Bolton:2006pc}. 
In Fig.~\ref{fig:rates}, we plot these rates for a model of DM decaying to photons with a lifetime of $\SI{2e22}{\s}$ and $m_\chi = \SI{800}{\mega\eV}$.
We set $x_\text{e}^\text{Pl}$ to Planck's latest FlexKnot ionization history and use the `conservative' treatment for the photoheating term. 
Fig.~\ref{fig:rates} demonstrates that in a hot and reionized universe, cooling processes that were once negligible become important and possibly dominant.

\begin{figure}[h!]
	\centering
	\includegraphics[scale=0.48]{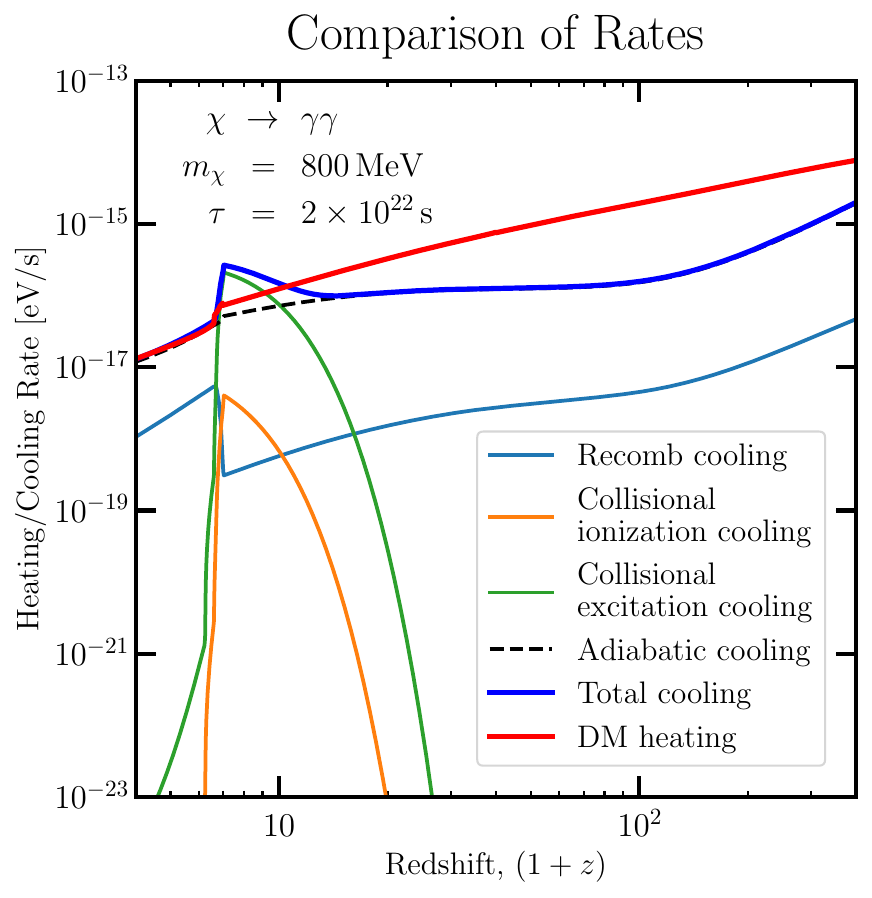}
	\caption{The absolute value of the atomic cooling rates included in $\dot{T}_\text{atom}$, the adiabatic cooling rate, and the DM heating rate. We assume a model of DM decaying to photons with a lifetime of $\SI{2e22}{\s}$ and $m_\chi = \SI{800}{\MeV}$.  The blue line corresponds to the sum of all cooling rates while the red corresponds to the DM heating rate, the only source of heating in the `conservative' treatment.
	}
	\label{fig:rates}
\end{figure}

Moving on to the ionization equations, we write down the helium version of Eq.~\eqref{eq:ionization_diff_eq},
\begin{alignat}{1}
\dot{x}_\text{HeII} & = \dot{x}_\text{HeII}^\text{atom} + \dot{x}_\text{HeII}^\text{DM} + \dot{x}_\text{HeII}^\star \, , \nonumber \\
x_\text{HeIII} & = 0 \,,
\label{eq:He_ionization_diff_eq}
\end{alignat}
where $x_\text{HeII} \equiv n_\text{HeII}/n_\text{H}$ is the density of singly-ionized helium atoms in the IGM normalized to the density of hydrogen atoms, and $x_\text{HeIII}$ is defined similarly. As explained above, the second of these two equations reflects the fact that there are negligibly few fully ionized helium atoms in the IGM over the redshifts under consideration in our analysis. Therefore we only need to keep track of the relative levels of HeI and HeII using the first equation.
Similarly to the $\dot{x}_\text{HII}^\star$ term, we have engineered the astrophysical reionization source term to turn off for $z > z^\star$ and produce a helium ionization curve that is equal to $\frac{\chi}{1+\chi} x_\text{e}^\text{Pl}(z)$ for $z < z^\star$. In other words,
\begin{alignat}{1}
\begin{dcases}
\dot{x}_\text{HeII} = \dot{x}_\text{HeII}^\text{atom} + \dot{x}_\text{HeII}^\text{DM} \,, & z > z^\star \,, \\
x_\text{HeII} = \frac{\chi}{1+\chi} x_\text{e}^\text{Pl}(z) \,, & z < z^\star \,.
\end{dcases}
\label{eq:He_photoionization_rate}
\end{alignat}
Notice that we do not need to know the explicit form of $\dot{x}^\star_\text{HeII}$
in contrast to $\dot{x}^\star_\text{HII}$, which we need to compute to evaluate $\dot{T}^\star$ in Eq.~\eqref{eq:modeled_term}.
Due to this simplified treatment, $x_\text{HeII}$ can be discontinuous at $z^\star$; we have tested alternative prescriptions and found negligible effects on our constraints.

The atomic sources contain a contribution from photoionization and a contribution from recombination.
For $z>z^\star$,
we assume a case-B scenario~\cite{Seager:1999bc, Seager:1999km, Wong:2007ym},
\begin{alignat}{3}
&\dot{x}_\text{HII}^\text{atom} &=& \; 4 \,  \mathcal{C}_\text{H} \, \left[ (1 - x_\text{HII}) \, \beta^B_\text{H} e^{-E_\text{H}/T_\text{CMB}}
-n_\text{H} \, x_\text{e} \, x_\text{HII} \, \alpha^B_\text{H} \right] \nonumber \\
&\dot{x}_\text{HeII}^\text{atom} &=& \; 4 \,  \sum_s \mathcal{C}_{\text{HeII}, s} \, \Big[ g_s (\chi - x_\text{HeI}) \, \beta^B_{\text{HeI}, s} e^{-E_{\text{HeI}, s}/T_\text{CMB}} - \, n_\text{H} \, x_\text{e} \, x_\text{HeII} \, \alpha^B_{\text{HeI}, s} \Big] \, ,
\end{alignat}
where $E_i$, $\beta^B_i$, $\alpha^B_i$, and $\mathcal{C}_i$ are, respectively,  the binding energy, case-B photoionization coefficient (including the gaussian fudge factor used in RECFAST v1.5.2~\cite{Seager:1999bc, Seager:1999km}), case-B recombination coefficient, and Peebles $\mathcal{C}_i$ factor for species $i \in \{ \text{H}; \text{HeI, singlet}; \text{HeI, triplet} \}$ \cite{Peebles:1968ja}.
Notice, there is a sum over both spin states of the two electrons in the excited HeI atom. 
For the spin singlet, $g_1 = 1$ and $E_{\text{HeI}, 1} = \SI{20.616}{\eV}$, while for the spin triplet state $g_3 = 3$ and $E_{\text{HeI}, 3} = \SI{19.820}{\eV}$.

When $z < z^\star$, we assume a case-A scenario, which is applicable during reionization~\cite{Bolton:2006pc}:
\begin{alignat}{1}
\dot{x}_\text{HII}^\text{atom} = & \; n_\text{H}  \, (1-x_\text{HII}) \, x_\text{e}\, \Gamma_\text{eHI} - n_\text{H} \, x_\text{e} \, x_\text{HII} \, \alpha^A_\text{HII} \, .
\label{eqn:caseA}
\end{alignat}
The collisional ionization rate, $\Gamma_\text{eHI}$, and case-A recombination coefficient, $\alpha^A_\text{HII}$, can be found in Ref.~\cite{Bolton:2006pc}. Notice that the case-A photoionization term from CMB photons is not included because it is exponentially suppressed at these low redshifts, and that the photoionization term from astrophysical reionization sources is already accounted for in $\dot{x}^\star_\text{HII}$.
Additionally, we do not need the analogous HeII version of Eq.~\eqref{eqn:caseA} since at these redshifts we have assumed $x_\text{HeII} = \chi \, x_\text{HII}$.

\begin{figure*}[t!]
	\begin{tabular}{c}
		\includegraphics[width=0.95\textwidth]{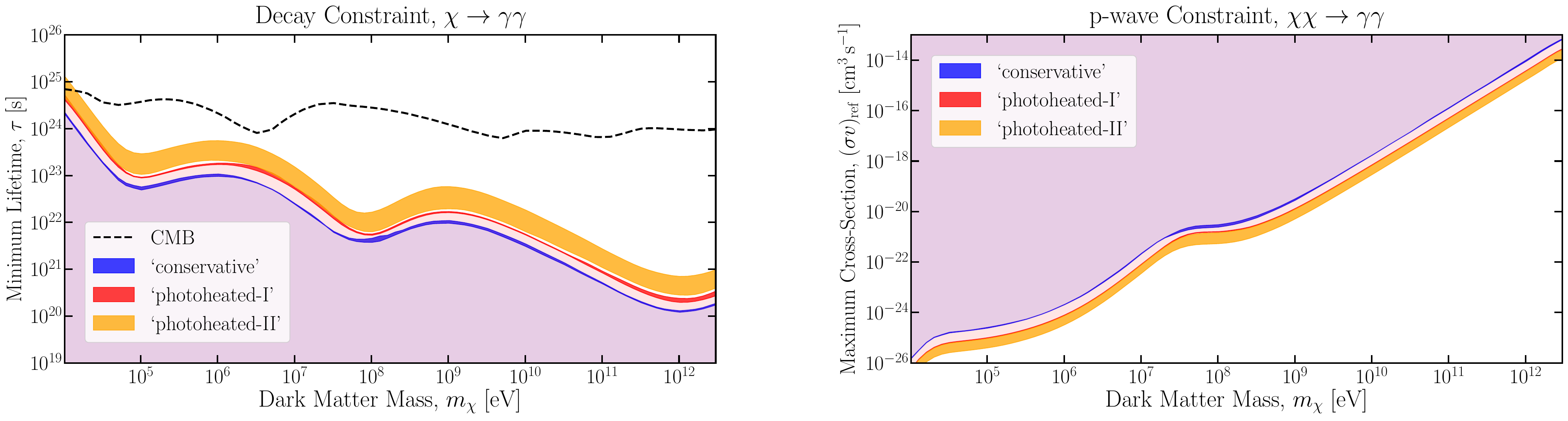}
	\end{tabular}
	\caption{Constraints for decay (left) or $p$-wave annihilation (right) to $\gamma\gamma$ pairs with $v_\text{ref} = \SI{100}{\km \, \s^{-1}}$.  We show our constraints using the `conservative' (blue band), `photoheated-I' (red band), and `photoheated-II' (orange band) treatments. We also include the CMB constraint for decay~\cite{Slatyer:2016qyl} (dashed-black). Telescope constraints~\cite{Ackermann:2012qk,Boddy:2015efa,Archambault:2017wyh,Abdallah:2018qtu,Acciari:2018sjn} are many orders of magnitude stronger than ours, and are not shown for clarity.
	}
	\label{fig:constraints_gg}
\end{figure*}

The DM ionization source terms are given by
\begin{alignat}{1}
\dot{x}_\text{HII}^\text{DM}   &= \left[\frac{f_\text{H ion}(z, \mathbf{x})}{E_\text{H} n_\text{H}} + \frac{(1 - \mathcal{C_\text{H}}) f_\text{exc} (z, \mathbf{x})}{0.75 E_\text{H} n_\text{H}}\right] \left(\frac{dE}{dV \, dt}\right)^\text{inj} \,, \nonumber \\
\dot{x}_\text{HeII}^\text{DM} &= \frac{f_\text{He ion}(z, \mathbf{x})}{E_\text{HeI} n_\text{He}} \left(\frac{dE}{dV \, dt}\right)^\text{inj} \,, \nonumber \\ 
\dot{x}_\text{HeIII}^\text{DM} & = 0 \,,
\label{eqn:TLA_DM_sources}
\end{alignat}
where $f_\text{H ion}(z, \mathbf{x})$, $f_\text{He ion}(z, \mathbf{x})$, $f_\text{exc}(z, \mathbf{x})$ are the deposition efficiency fractions into hydrogen ionization, single neutral helium ionization, and hydrogen excitation calculated by \texttt{DarkHistory}. 

\section{Other Final States}
\label{app:phot_constr}
In this appendix we provide constraints for DM decay and $p$-wave annihilation to $\gamma\gamma$, $\mu^+\mu^-$, $\pi^+\pi^-$, and $\pi^0\pi^0$.

\subsection{Photons}
Fig.~\ref{fig:constraints_gg} shows decay and annihilation constraints for $\gamma\gamma$ final states using the `conservative' (blue), `photoheated-I' (red), or `photoheated-II' treatments (orange).
As in the main text, the $p$-wave annihilation cross-section is defined by $\sigma v = (\sigma v)_\text{ref} \times (v/v_\text{ref})^2$ with $v_\text{ref} = \SI{100}{\kilo\meter\per\second}$ and we use the NFW boost factor for $p$-wave annihilation calculated in Ref.~\cite{Liu:2016cnk}, which accounts for enhanced annihilation due to increased DM density and dispersion velocity in halos. 
Just as in the main text, the darkly shaded blue, red, and orange bands show the variation of our constraints as we vary $x_\text{e}^\text{Pl}$ in Eq.~\eqref{eq:photoionization_rate} over the 95\% confidence region of Planck's FlexKnot and Tanh late-time ionization curves.
As before, the `conservative' and `photoheated-I' bands are narrow, demonstrating an insensitivity to the precise form of the reionization curve, while the `photoheated-II' curve is broader for the reasons discussed in the main text.

The photon final state constraints are less competitive with existing constraints than are the $e^+e^-$ constraints. 
For example, CMB constraints~\cite{Slatyer:2016qyl} are stronger for all masses in the decay channel. 
Additionally, telescope constraints (see e.g. Refs.~\cite{Ackermann:2012qk,Boddy:2015efa,Archambault:2017wyh,Abdallah:2018qtu,Acciari:2018sjn}) 
are many orders of magnitude stronger than ours since telescopes can search directly for the produced photons, in contrast to our temperature constraints that indirectly look for the effects that these photons have on the IGM.

Our $\gamma\gamma$ constraints are weaker than our $e^+e^-$ constraints because the photoionization probability is small (equivalently, the path length is long) for the redshifts and photon energies of interest. In contrast, electrons can efficiently heat the gas either through direct Coulomb interactions (for non-relativistic and mildly relativistic electrons) or through inverse Compton scattering that produces efficiently-ionizing photons (for higher-energy electrons).

\subsection{Muons and Pions}
While we could also consider any other Standard Model particle final state, 
our results from the previous section and the main text indicate that
our constraints are most competitive at masses below $\SI{10}{GeV}$. 
Therefore, we consider 
some of the most important final states that are available to sub-GeV DM besides those already considered: muons, charged pions, and neutral pions. 
To compute the final spectra of $e^+e^-$ and $\gamma$ produced by the decay of pions or muons, we use the PPPC4DMID for DM masses above $\SI{10}{\GeV}$.

For DM masses below $\SI{10}{\GeV}$, we follow the method described in Ref.~\cite{Cirelli:2020bpc}. We start with the spectrum of electrons in the muon rest frame, which is given by
\begin{equation}
\frac{d N^{\mu \rightarrow e\nu\bar{\nu}}_e}{d E_e} = \frac{4 \sqrt{\xi^2 - 4 \varrho^2}}{m_\mu} [ \xi (3 - 2 \xi) + \varrho^2 (3 \xi - 4)]
\end{equation}
between energies of $m_e < E_e < (m_\mu^2 + m_e^2) / (2 m_\mu)$ and is otherwise zero.
In this equation, $\xi = 2 E_e/m_\mu$ and $\varrho = m_e/m_\mu$.
For a particle $A$ with mass $m_A$ decaying with some spectrum $dN/dE'$ in its rest frame, the spectrum $dN/dE$ in an arbitrary frame where $A$ has energy $E_A$ is given by
\begin{equation}
\frac{dN}{dE} = \frac{1}{2 \beta \gamma} \int^{E'_\mathrm{max}}_{E'_\mathrm{min}} \frac{dE'}{p'} \frac{dN}{dE'} ,
\end{equation}
where $\gamma = E_A/m_A$ is the Lorentz factor, $\beta = \sqrt{1 - \gamma^{-2}}$, and $E'_{\mathrm{max}/\mathrm{min}} = \gamma (E \pm \beta p)$.
In the case of decay to muons, we can use this equation to boost from the muon frame to the dark matter frame, where the muon has energy $m_\chi$ for annihilations or $m_\chi/2$ for decays.
For decay to pions, we first boost to the pion rest frame where the muon has energy $(m_\pi^2 + m_\mu^2) / (2 m_\pi)$, and then the dark matter frame where the pion similarly has energy $m_\chi$ for annihilations or $m_\chi/2$ for decays.

\begin{figure*}[t!]
	\begin{tabular}{c}
		\includegraphics[width=0.95\textwidth]{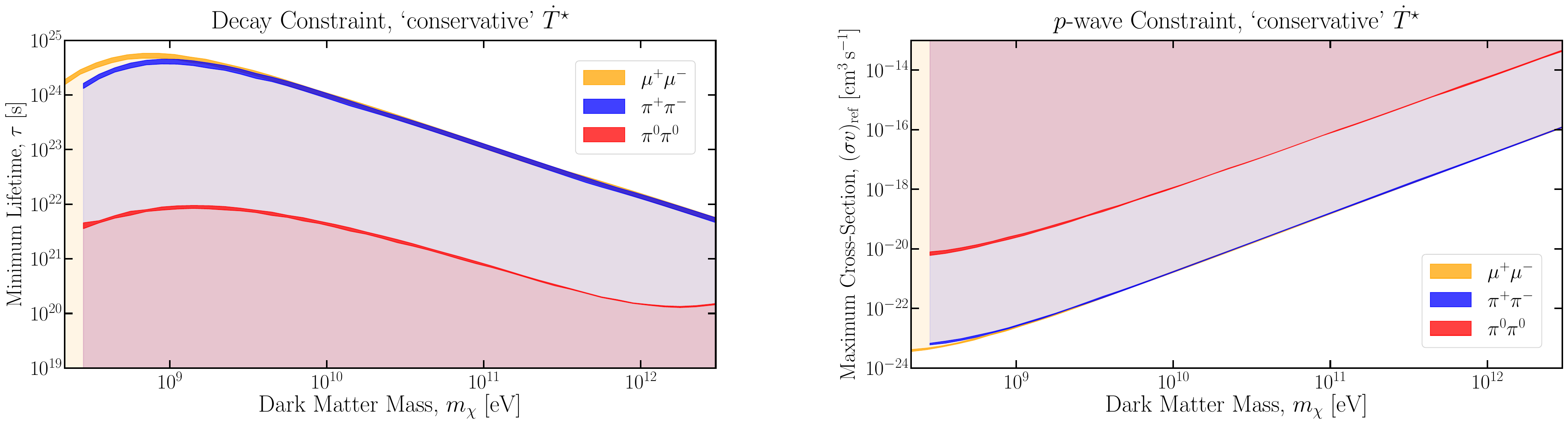}
	\end{tabular}
	\caption{Constraints for decay (left) or $p$-wave annihilation (right) to $\mu^+\mu^-$ (yellow), $\pi^+\pi^-$ (blue), and $\pi^0\pi^0$ (red) pairs with $v_\text{ref} = \SI{100}{\km \, \s^{-1}}$.  
		We show our constraints only using the `conservative' treatments. 
	}
	\label{fig:constraints_mu_pi}
\end{figure*}

We plot our constraints in Fig.~\ref{fig:constraints_mu_pi}. 
$\mu^+\mu^-$ and $\pi^+\pi^-$ ultimately decay to $e^+e^-$ and neutrinos, meaning that these constraints are comparable to
the $e^+e^-$ constraints, though somewhat weaker because the produced electrons share at most an $\mathcal{O}(1)$ 
fraction of the total DM injected energy with the other neutrinos. 
$\pi^0\pi^0$ decays almost exclusively to 4$\gamma$, so the photons carry half the energy as compared to photons that result from $\chi \rightarrow \gamma\gamma$ decays. Thus, the pion constraints look exactly like the $\gamma\gamma$ constraints shifted by a factor of 2 to the left.


%
%


\section{Cross Checks}
\label{app:xchecks}
Here, we provide cross checks to validate the assumptions we made in our analysis. 
First, we will validate maintaining $x_\text{HeIII} = 0$ after H and HeI reionization despite DM injecting HeII ionizing photons. 
Second, we will check that our $p$-wave constraints are insensitive to the uncertainty in the halo boost factor coming from the halo profile. Finally, we will validate our use of ionization histories that feature significant ionization levels prior to reionization, by checking that they do not violate constraints on the total $z < 50$ optical depth.

\subsection{Treatment of HeIII}
\label{sec:HeIII}
In calculating the constraints shown in Fig.~\ref{fig:constraints}, we assume that there is no ionization of HeII to HeIII---i.e. $x_\text{HeIII}=0$ --- consistent with the assumptions that went into the making of \texttt{DarkHistory}'s transfer functions. 
We still account for energy deposition through ionization of HeII by
allowing photons with energies $E_\gamma > \SI{54.4}{\eV}$ to be absorbed by HeII atoms, thus producing electrons of energy $E_\gamma - \SI{54.4}{\eV}$ that thermalize with the IGM.
This is not entirely self-consistent because these photoionization events would gradually increase the fraction of HeII atoms as they convert into HeIII atoms.  
Having fewer HeII atoms could then affect our constraints by decreasing the heating deposition fraction, since fewer photoionized electrons could be produced and thermalize with the IGM.

We test our sensitivity of our constraints to this approximation by adding a new $\dot{x}_\text{HeIII}$ source term and accounting for recombination photons once HI/HeI reionization is complete.
We restrict this correction to after HI/HeI reionization
because it is expected to make the biggest difference in the heating rate then, since HeII atoms are the only possible source of photoionized electrons at this point, and because the temperature data we use are primarily in this redshift range.

To apply our correction,
we first modify Eq.~\eqref{eq:He_ionization_diff_eq} to track the fully ionized helium fraction,
\begin{alignat}{1}
\dot{x}_\text{HeIII} = & \frac{f_\text{He ion}(z, \mathbf{x})}{4 E_\text{H} n_\text{He}} \left(\frac{dE}{dV \, dt}\right)^\text{inj} + n_\text{H}  \, (\chi-x_\text{HeIII}) \, x_\text{e}\, \Gamma_\text{eHeII} - n_\text{H} \, x_\text{e} \, x_\text{HeIII} \, \alpha^A_\text{HeIII} \, ,
\end{alignat}
where the deposition fraction $f_\text{He ion}(z, \mathbf{x})$ computed by \dhis accounts for the total energy deposited into HeII ionization
and $\Gamma_\text{eHeII}$ is the collisional ionization rate of HeII~\cite{Bolton:2006pc}. 
We then compute the fraction of HeIII atoms that recombines within a timestep of the code, $\Delta t$,
\begin{alignat}{1}
f^\text{HeIII}_\text{recomb} = 1 - e^{-\alpha^{\text{A}}_{\text{HeIII}} x_\text{HeIII} n_e \Delta t} \, .
\end{alignat}
We convert this fraction to the number of $\SI{54.4}{\eV}$ photons per baryon emitted by HeIII atoms in this time step
\begin{alignat}{1}
N^\text{HeIII}_\text{recomb} = f^\text{HeIII}_\text{recomb} n_\text{HeIII}/n_\text{B} \, ,
\end{alignat}
then add these photons to \texttt{DarkHistory}'s low energy photon spectrum within that time step.

\begin{figure}[h]
	\centering
	\includegraphics[scale=0.38]{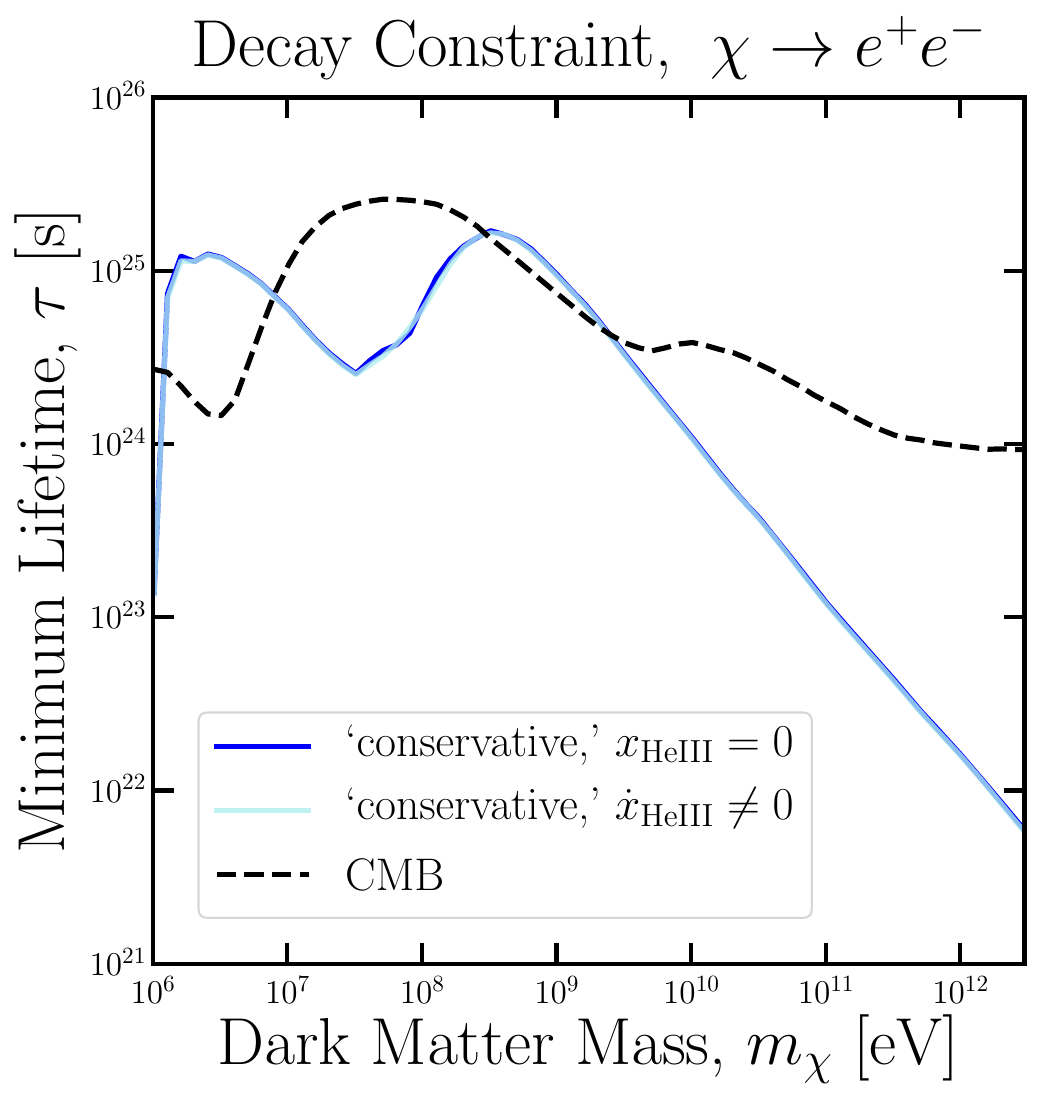}
	\caption{Comparison of `conservative' constraints for decay to electrons, including and not including the effects of HeII ionization after HI/HeI reionization. Both constraints were generated assuming Planck's earliest Tanh reionization history.
	}
	\label{fig:HeIII}
\end{figure}

Fig.~\ref{fig:HeIII} shows a comparison of constraints for dark matter decaying to electrons, where the two curves either allow for a non-zero HeIII fraction (light blue) or do not (blue). 
The difference in constraints is always less than 1\%, and so is not an important source of error in our analysis.

\subsection{Boost factor for $p$-wave annihilation}
\label{sec:pwave_boost}
The boost factor due to enhanced density and velocity dispersion in halos depends on the halo profile chosen. However, in Ref.~\cite{Liu:2016cnk}, the boost factor was found to be highly robust to this choice, since the main contribution to the boost factor comes from the largest halos, which are fully resolved in $N$-body simulations.
We find that the difference in our constraints made by using the Einasto $p$-wave boost factor rather than the NFW $p$-wave boost factor from Ref.~\cite{Liu:2016cnk} is negligibly small, resulting in a modification of no more than 0.5\% to our constraints.
Notice that the two boost factors only vary over the halo mass function and halo profile, and do not include uncertainties due to mergers, asphericity, etc.  

\subsection{Optical depth}
\label{sec:opdepth}

In this appendix, we discuss the relation between temperature and ionization constraints, focusing in particular on the complementarity of these constraints. One might worry that scenarios excluded by excess heating of the IGM are strictly a subset of those excluded by the ionization history.
In some cases, the DM contribution to the optical depth $\tau$ before reionization, combined with one of the Planck reionization models, can exceed the Planck limit on $\tau$. DM energy injection starts to increase the ionization fraction and temperature immediately after recombination, and so our computed ionization histories will always be in excess of Planck's reionization curves at early enough redshifts.

To some extent, these worries have already been addressed by the fact that the temperature constraints can sometimes be stronger than the CMB power spectrum constraints for DM decays as derived in Refs.~\cite{Slatyer:2016qyl, Poulin:2016anj}, which account for the effect of excess ionization on the full multipole structure of the CMB power spectrum. For simplicity, however, we would like to compare the IGM temperature constraints derived in the main body with limits on the ionization history coming simply from the Planck upper limit on $\tau$. 

Given an ionization history $x_\text{e}(z)$, the optical depth is
\begin{equation}
\tau = n_{\text{H,0}}  \sigma_T \int_0^{z_\mathrm{max}} dz \, x_\text{e}(z) \frac{(1+z)^2}{H(z)},
\label{eqn:opt_dpth}
\end{equation}
where $\sigma_T$ is the Thomson cross-section and $z_\mathrm{max}$ is set to 50, as is done in Ref.~\cite{Planck2018}. The 68\% upper bound on the optical depth from Planck assuming a tanh function reionization history is $\tau = 0.0549$~\cite{Planck2018}. To derive a constraint, 
we compute an ionization history in the presence of DM energy injection 
and exclude it if the history's optical depth is greater than $0.0549$.

Clearly, these optical depth constraints will be highly sensitive to the reionization curve we choose. 
For example, if we were to use the earliest Tanh reionization curve that already saturates the optical depth bound we would rule out all DM models since they all increase $\tau$.
On the other hand, we saw that our temperature constraints were very weakly dependent on the choice of reionization curve.
For a fair comparison, we choose a reionization history with the smallest optical depth.
While we could choose the latest Tanh reionization curve, we instead follow the instantaneous reionization method described in Ref.~\cite{Liu:2016cnk} so that we can compare to older optical depth constraints.
We will assume an instantaneous HI/HeI reionization at $z=6$, then an instantaneous HeII reionization at $z=3$, but no other sources of reionization other than DM for $z>6$.
The optical depth contributed by the range $0 < z < 6$ is $0.384$,\footnote{This is nearly equivalent to using the earliest Tanh reionization history, which has an optical depth contribution of $0.383$ to the same redshift range.}
meaning that DM models that contribute more than $\delta \tau = 0.0165$ to the optical depth within the range $6 < z < 50$ will be ruled out.

\begin{figure}[h]
	\centering
	\includegraphics[scale=0.48]{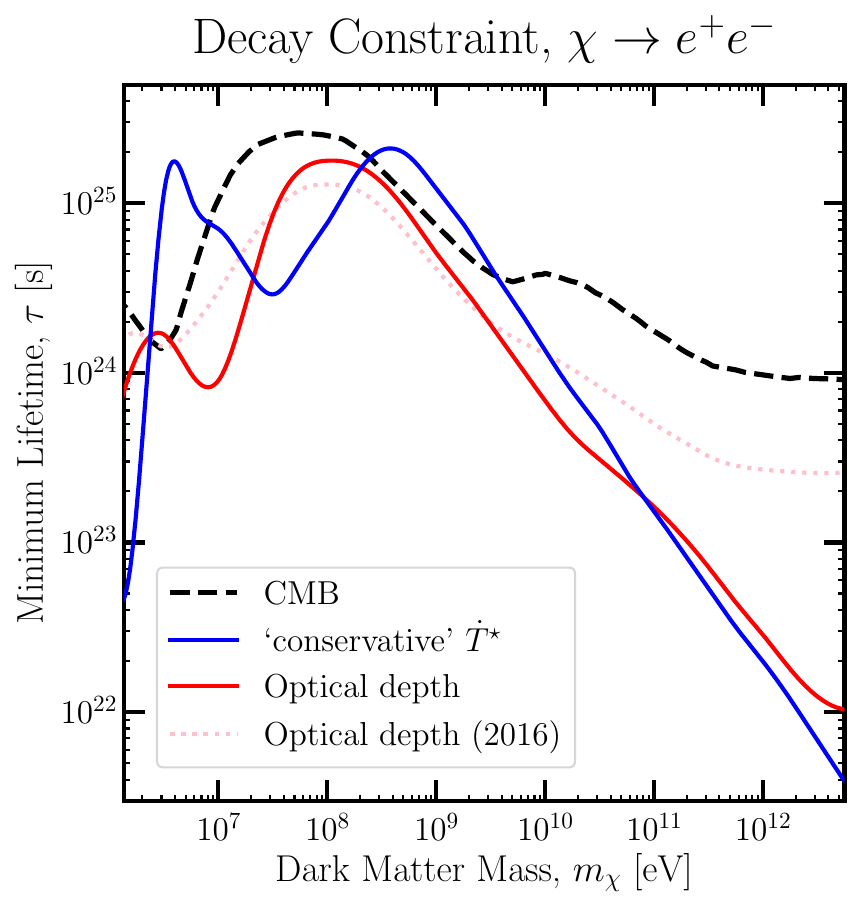}
	\caption{Constraints obtained from the IGM temperature (red) and optical depth (blue, solid), as well as previous bounds derived in Ref.~\cite{Liu:2016cnk} from the optical depth (pink, dashed). The black line shows the constraints derived using a principal component analysis of CMB data~\cite{Slatyer:2016qyl}. 
	}
	\label{fig:opdepth}
\end{figure}

Fig.~\ref{fig:opdepth} shows a comparison of the optical depth constraint in blue to the IGM temperature constraint in red for a model of DM decay to $e^+e^-$, as well as to a previous constraint made with Planck intermediate results ~\cite{Liu:2016cnk}, which measured $\tau = 0.058 \pm 0.012$ and is represented by the dashed curve~\cite{Adam:2016hgk}. We see that across most of the mass range, the two methods of constraining dark matter parameters are comparable, but there is a large range of DM masses over which the temperature constraints do better than the optical depth limits. To summarize, since the IGM temperature constraints are insensitive to the exact ionization history during reionization, they probe a different aspect of energy injection from DM that is distinct from ionization-based constraints like optical depth and the CMB power spectrum. Finally, we show for reference an older optical depth constraint across all masses~\cite{Liu:2016cnk}, which calculated $\delta \tau$ by integrating over the excess ionization fraction over the standard three-level atom result up to recombination, following the method in Ref.~\cite{Cirelli:2009bb}. 

Fig.~\ref{fig:opdepth} shows a comparison of the optical depth constraint in blue to the IGM temperature constraint in red for a model of DM decay to $e^+e^-$. 
For reference, we show the CMB power spectrum constraint in dashed-black~\cite{Slatyer:2016qyl}, 
and also show an older optical depth constraint made with Planck intermediate results~\cite{Adam:2016hgk} in dotted-pink~\cite{Liu:2016cnk}. 
This latter constraint calculated $\tau = 0.058 \pm 0.012$ by following the method in Ref.~\cite{Cirelli:2009bb} and integrating over the excess ionization fraction over the standard three-level atom result up to recombination. 
We see that across most of the mass range, the two methods of constraining dark matter lifetimes are comparable, but there is a large range of DM masses over which the temperature constraints do better than the optical depth limits. 
Had we considered a model of p-wave annihilation to $e^+ e^-$ instead, we would find that the heating constraints are generically stronger than those from the optical depth since, as compared to the decay case, the power from p-wave annihilation is enhanced at low redshifts where temperature limits are most sensitive.
To summarize, since the IGM temperature constraints are insensitive to the exact ionization history during reionization, they probe a different aspect of energy injection from DM that is distinct from ionization-based constraints like optical depth and the CMB power spectrum. 

\section{Test Statistics}
\label{app:stats}

In this section, we derive the distribution of the modified $\chi^2$-like test statistic (TS) that we use in conjunction with the `conservative' treatment of the  $\dot{T}^\star$ photoheating term (i.e. $\dot{T}^\star=0$). We are working in a frequentist framework, so we wish to evaluate the probability distribution for the TS defined in Eq.~\eqref{eq:one_sided_TS}, when assuming a certain pattern of heating due to DM energy injection. We can then say that this scenario is excluded if the TS observed in the real data is sufficiently unlikely. We make the assumption that the data points in different redshift bins are independent and Gaussian distributed.

Suppose that there are $N$ redshift bins, and in the $i$th bin the temperature value $T_{i,\text{data}}$ is drawn from a Gaussian distribution with mean $T_{i,\text{pred}}$ and standard deviation $\sigma_{i,\text{data}}$. There is then a $50\%$ chance that $T_{i,\text{data}} > T_{i,\text{pred}}$, so the probability distribution for $\text{TS}_i$ as defined in Eq.~\eqref{eq:one_sided_TS} is:
\begin{align} 
f(\text{TS}_i | T_{i,\text{pred}})& = \frac12 \delta(\text{TS}_i) + P(T_{i,\text{data}}) \frac{d(T_{i,\text{data}})}{d(\text{TS}_i)}  \nonumber \\
& = \frac12 \delta(\text{TS}_i) + \frac{1}{2 \sqrt{2 \pi}} \text{TS}_i^{-1/2} \exp(-\text{TS}_i/2) \,,
\end{align}
where $\delta$ is the Dirac delta function. Let the $\chi^2$ probability distribution function with $j$ degrees of freedom be denoted by $f_{\chi^2}(\text{TS}; j)$. Then one can rewrite this distribution in terms of the $\chi^2$ distribution with one degree of freedom:
\begin{align} 
f(\text{TS}_i|T_{i,\text{pred}}) & = \frac12 \delta(\text{TS}_i) + \frac12 f_{\chi^2}(\text{TS}_i; 1) \, .
\end{align}

Now we want to know the distribution for the total TS value from combining the bins (assuming uncorrelated data), $\text{TS} \equiv \sum_i \text{TS}_i$. We can write:
\begin{alignat}{2}
f(\text{TS}|\{T_{i,\text{pred}} \}) &=&& \left[ \prod_{i=1}^N \int_0^\infty d\text{TS}_i \,f(\text{TS}_i | T_{i,\text{pred}}) \right] \delta (\text{TS} - \sum_{j=1}^N \text{TS}_j )  \nonumber \\
&=&& \left[ \prod_{i=1}^N \int_0^\infty d\text{TS}_i \, \left[ \delta(\text{TS}_i) + f_{\chi^2} (\text{TS}_i;1) \right] \right]  \frac{1}{2^N}  \delta (\text{TS} - \sum_{j=1}^N \text{TS}_j ) \,.
\end{alignat}

Expanding the product inside the integrals gives a sum of terms, which each consist of a product of delta-functions and $f_{\chi^2}$ functions. For a term with $n$ delta-functions, the delta-functions can be used to do $n$ of the integrals, resulting in a term of the form:
\begin{equation} 
\left[ \prod_{j=n+1}^N \int_0^\infty d \text{TS}_{i_j}  f_{\chi^2}(\text{TS}_{i_{j}}; 1)\right] \delta(\text{TS} - \sum_{k=n+1}^N \text{TS}_{i_k})\,,
\end{equation}
where $n$ can take on values from $0$ to $N$, and $i_{n+1}, i_{n+2}, \cdots, i_N$ are a collection of indices between 1 and $N$ for this particular term. However, this is exactly the standard probability distribution function for the sum of the $\chi^2$ test statistic over $N-n$ bins, so we can write it as $f_{\chi^2}(\text{TS}; N-n)$.

The coefficient of each such term will be the number of ways of choosing which $n$ indices correspond to $\delta$-function terms as opposed to the $N-n$ indices labeling $f_{\chi^2}(\text{TS}_{i}; 1)$ contributions -- which is the binomial coefficient $\binom{N}{n}$. Since $\binom{N}{n}$ = $\binom{N}{N-n}$, we can write:
\begin{equation} 
f(\text{TS}|\{T_{i,\text{pred}}\})  =  \frac1{2^N} \sum_{n=0}^N \binom{N}{n} f_{\chi^2}(\text{TS}; n) \, .
\end{equation}
This completes the proof of Eq.~\eqref{eq:TS_pdf}.

Note that this expression integrates correctly to 1, as $\int d \text{TS} f_{\chi^2}(\text{TS}; n)= 1$ and $\sum_{n=0}^N \binom{N}{n} = 2^N$.
The largest binomial coefficients $\binom{N}{n}$ will occur for $n\approx N/2$, and so we may approximate the distribution as a $\chi^2$ distribution with $N/2$ degrees of freedom. However, for the actual constraints in the main text we use the full distribution, rather than this approximation.

We can also understand this distribution by thinking of $\text{TS}$ as the standard $\chi^2$ test statistic, in the presence of a model for the data where each redshift bin contains an irreducible (dark matter) contribution plus a non-negative but otherwise arbitrary increase to the temperature from photoheating. If we profile over the nuisance parameters describing the unknown astrophysics, we see that the minimum $\chi^2$ will be attained when:
\begin{itemize}
	\item in bins where the irreducible contribution from dark matter already exceeds the measured temperature, extra contributions from photoheating are set to zero; the contribution to the TS is the usual $\chi^2$ computed using the irreducible model and the data,
	\item in bins where the irreducible contribution from dark matter does not exceed the measured temperature, the additional photoheating contribution is chosen to precisely match the data, and consequently the contribution to the TS is zero.
\end{itemize}
This is exactly the prescription for our modified TS, Eq.~\eqref{eq:one_sided_TS}. 

Because this is a standard $\chi^2$ test, just with a flexible background model, the probability distribution for the TS should follow that of a $\chi^2$ distribution with $N-m$ degrees of freedom, where $m$ is the number of floated parameters in the fit. The number of floated parameters for this signal model is the number of bins where the data is greater than the irreducible model, which can vary from 0 to $N$; thus the full probability distribution is obtained as a linear combination of $\chi^2$ distributions with degrees of freedom varying from $0$ to $N$.


\chapter{Supplemental Material for Sections~\ref{sec:DHv2_tech}-\ref{sec:DHv2_apps} }

\section{Collisional Ionization and Excitation Rates}
\label{app:collisional_rates}

For collisional ionization, we adopt the results of Ref.~\cite{Kim_Rudd}, using the binary-encounter-Bethe model for HI shown in Eq.~(57)---which shows excellent agreement with experimental results between \SI{13.6}{\eV} and \SI{3}{\kilo\eV}---and the binary-encounter-dipole model for HeI and HeII, shown in Eq.~(55). All quantities required for the cross section are tabulated in Table~I of the same paper. Note that these cross-section fits are not expected to hold when the incoming electron is relativistic; however, this does not affect our results significantly, since relativistic electrons in the early universe lose their energy predominantly through ICS, with ionization being a small contribution to the total energy loss. 

For collisional excitation rates, we rely on the tabulated cross sections in Ref.~\cite{Stone_Kim_Desclaux} for hydrogen $np$ states and HeI excitation (we only track excitation up to the $2p$ state) in the energy range of \SI{10}{\eV}--\SI{3}{\kilo\eV}. For all other hydrogen states, we use the data provided by the CCC database~\cite{CCC}, which gives the cross sections of excitations from the ground state of hydrogen up to and including the $4f$ state, between \SI{14}{\eV} and \SI{999}{\eV}. 

For energies higher than those that are tabulated above, we use the Bethe approximation~\cite{RevModPhys.43.297}, which expands the excitation cross section as a function of $\mathcal{R}/E' \ll 1$ for incoming electron energies above \SI{1}{\kilo\eV}. Ref.~\cite{Stone_Kim_Desclaux} provides a nonrelativistic Bethe approximations---suitable for $E' < \SI{10}{\kilo\eV}$---for the excitation cross sections of hydrogen $np$ states and HeI excitation of the form 
\begin{alignat}{1}
\sigma_{np} (E') = \frac{4 \pi a_0^2 \mathcal{R}}{T + B + E_{1s \to np}} \frac{f_\text{accu}}{f_\text{sc}} \left[a_{np} \log \left(\frac{E'}{\mathcal{R}}\right) + b_{np} + c_{np} \frac{\mathcal{R}}{E'}\right] \,,
\end{alignat}
where $a_0$ is the Bohr radius, $B$ is the binding energy of the ground state electron, $E_{1s \to np}$ is the excitation energy of the $np$ state, and $a_{np}$, $b_{np}$ and $c_{np}$ are fit coefficients. $f_\text{accu} / f_\text{sc}$ is 1 for hydrogen, and is a correction factor applied to atoms with more than one electron in the ground state. All unknown values in this expression are tabulated in Ref.~\cite{Stone_Kim_Desclaux}. For all other hydrogen states, we perform a fit to the three data points with the highest energies in the CCC database with the following functional form, which is appropriate for transitions which are optically forbidden~\cite{RevModPhys.43.297}: 
\begin{alignat}{1}
\sigma_{nl}(E') = \frac{4 \pi a_0^2}{E' / \mathcal{R}} \left(\beta_{nl} + \frac{\gamma_{nl}}{E' / \mathcal{R}}\right) \,,
\end{alignat}
where $\beta_{nl}$ and $\gamma_{nl}$ are fit coefficients for each $nl$ state, and use this asymptotic form between \SI{1}{\kilo\eV} and \SI{10}{\kilo\eV}. 

Above \SI{10}{\kilo\eV}, relativistic corrections start to become important. In this regime, we switch to the relativistic version of the Bethe approximation. For optically allowed transitions, this is of the form~\cite{RevModPhys.43.297}
\begin{alignat}{1}
\sigma_{np} = \frac{8 \pi a_0^2}{m_e \beta^2 / \mathcal{R}} \left( M_{np}^2 \left[\log \left(\frac{\beta^2}{1 - \beta^2}\right) - \beta^2\right] + C_{np} \right) \,,
\end{alignat}
where $\beta$ is the electron velocity, $M_{np}^2 = a_{np}$, $C_{np} = a_{np} \log (2 m_e \zeta_{np} / \mathcal{R})$, and $\zeta_{np} = (\mathcal{R} / 4) \exp (b_{np} / a_{np})$; $a_{np}$ and $b_{np}$ are the same coefficients used in the nonrelativistic Bethe approximation. Similarly, for optically forbidden transitions, we have
\begin{alignat}{1}
\sigma_{nl}(E') = \frac{8 \pi a_0^2}{\mathcal{R} / (m_e \beta^2)} \beta_{nl} \,,
\end{alignat}
where once again $\beta_{nl}$ is the same coefficient as above. 

For HeII excitation, we continue using the cross section provided in Ref.~\cite{PhysRevA.55.329}.

\section{Derivation of $y$-parameter from heating}
\label{app:y_deriv}

As mentioned in Section~\ref{sec:y-type}, at early enough times, one can write the $y$-parameter as an integral over the heating rate.
Here we present the full details of the derivation.

First, one can rewrite Eq.~\eqref{eq:general_Tm_evolution} in terms of $T_m / T_\text{CMB}$ as
\begin{equation}
\frac{d}{dt} \left( \frac{T_m}{T_\text{CMB}} \right) + H (1+J) \frac{T_m}{T_\text{CMB}} = JH + \frac{\dot{T}_m^\text{inj}}{T_\text{CMB}} .
\end{equation}
We can now write $T_m = T_m^{(0)} + \Delta T$, and note that $T_m^{(0)}$ solves the differential equation above without $\dot{T}_m^\text{inj}$, i.e. 
\begin{alignat}{1}
\frac{d}{dt} \left(\frac{T_m^{(0)}}{T_\text{CMB}}\right) = JH \left(1 - \frac{T_m^{(0)}}{T_\text{CMB}}\right) - H \frac{T_m^{(0)}}{T_\text{CMB}} \,.
\end{alignat}
Since $J \gg 1$ prior to $1+z \approx 500$, the first term on the right-hand side drives $T_m^{(0)} \to T_\text{CMB}$ until the two terms are roughly equal; in other words, any large deviations of $T_m^{(0)}$ from $T_\text{CMB}$ are erased on a timescale much faster than the Hubble timescale, leaving the right-hand side at a value close to zero. This means that for $J \gg 1$, 
\begin{alignat}{1}
T_m^{(0)} \approx \left(1 - \frac{1 }{J}\right) T_\text{CMB} \,.
\label{eqn:analytic_Tm}
\end{alignat}
The temperature evolution equation also gives
\begin{alignat}{1}
\frac{d}{dt} \left(\frac{\Delta T}{T_\text{CMB}}\right) = - H (1 + J) \left(\frac{\Delta T}{T_\text{CMB}} \right) + \frac{\dot{T}_m^\text{inj}}{T_\text{CMB}} \,.
\end{alignat}
Applying the same argument as before, we can see that $\Delta T \to 0$ until the two terms on the right-hand side are roughly equal, with any large deviations in $\Delta T$ from zero being erased well before a Hubble time. This then gives 
\begin{alignat}{1}
\Delta T \approx \frac{\dot{T}_m^\text{inj}}{H J} = \frac{3 (1 + \chi + x_e) m_e \dot{T}_m^\text{inj}}{8 \sigma_T u_\text{CMB} x_e} \,.
\end{alignat}
for $J \gg 1$. 

With this expression, it is easy to see from Eq.~\eqref{eqn:y_DM} that the $y$-parameter due to energy injection is
\begin{align}
y_\text{inj} &\approx \int_0^t dt\, \frac{3 n_\text{H} (1 + \chi + x_e) \dot{T}_m^\text{inj}}{8 u_\text{CMB}} = \frac{1}{4} \int_0^t dt\, \frac{\dot{Q}}{\rho_\text{CMB}} \,.
\end{align}
%

\section{Atomic Transition Rates}
\label{app:atomic_rates}

To calculate the bound-free and bound-bound transition rates, we follow the method outlined in Ref.~\cite{AliHaimoud:2010dx}.
Starting with the bound-free rates, we calculate the recombination rate to state $nl$ and the photoionization rate from state $nl$ using~\cite{Burgess1965}
\begin{alignat}{1}
\alpha_{nl} & = \left( \frac{2\pi}{\mu_e T_m} \right)^{3/2}
\int_0^{\infty} e^{-\mathcal{R} \kappa^2/T_m} \gamma_{nl} [1+f^\text{CMB}+\Delta f] d(\kappa^2) , \nonumber \\
\beta_{nl} &= \int_{\omega_{nl}}^{\mathcal{R}} d\omega [f^\text{CMB}+\Delta f] a_{nl}(k^2) \, .
\label{eqn:bound_free_rates}
\end{alignat}
Above, $\mu_e$ is the reduced mass of the electron and proton, $\kappa = p_e a_0$ is the momentum of the unbound electron in units of the Bohr radius, $a_0$, and $\omega_{nl}$ is the energy required to photoionize a hydrogen atom in the $nl$ state.
In the expression for $\beta_{nl}$, the integral is cut off at $\mathcal{R}$ because photons with energy above this are assumed to ionize the 1s state. 
$\gamma_{nl}$ and $a_{nl}(k^2)$ are defined by
\begin{alignat}{1}
\gamma_{nl} &= \frac{2}{3n^2} \frac{\mathcal{R}}{2\pi} (1+n^2 \kappa^2)^3 \sum_{l' = l\pm1}\text{max}(l,l') g(n,l,\kappa,l')^2 \nonumber \\
a_{nl}(k^2) &= \left( \frac{4\pi \alpha a_0^2}{3}\right) n^2 (1+n^2 \kappa^2) \sum_{l' = l\pm1}\frac{\text{max}(l,l')}{2l+1} g(n,l,\kappa,l')^2
\label{eqn:gamma_and_a}
\end{alignat}
$g(n,l,\kappa,l')$ is proportional to a matrix element of the dipole transition operator, 
and we calculate it using an iterative procedure described in Ref.~\cite{Burgess1965}.

Turning to the bound-bound rates, we calculate dipole up-transitions and down-transitions using
\begin{alignat}{1}
R_{nl\to n'l'} &= A_{nl\to n'l'} [1+f^\text{CMB}+\Delta f] \;\;\;\; E_n > E_{n'} , \\
R_{nl\to n'l'} &= \frac{g_{l'}}{g_l} A_{n'l'\to nl} [f^\text{CMB}+\Delta f] \;\;\;\; E_n < E_{n'}  \, .
\label{eqn:bound_bound_rates}
\end{alignat}
Here, $A_{nl\to n'l'}$ is the Einstein A-coefficient, which we calculate using an iterative procedure described in Ref.~\cite{Hey_2006} (See their Eqs.~(52)-(53)), and $g_l$ or $g_i$ is the degeneracy of the corresponding energy level. 

When added to the MLA equations, Eq.~\eqref{eq:general_xi_evolution}, the rates $R_{1s\to np}$ and $R_{np\to 1s}$ need to be treated with more care. 
While it is technically true that one could solve the MLA using the $R_{1s\to np}$ rates as defined above, 
one would have to use a step size smaller than the fastest timescale, $R_{1s\to 2p}^{-1}$, 
to be able to resolve the frequent emission and absorption of Lyman-series photons. 
Instead, the standard method is to use a much larger stepsize and replace $R_{1s\to np}$ and $R_{np\to 1s}$ by smaller effective rates. 
These effective rates only keep track of the transitions that produce or absorb a Lyman-series photon that is not instantly absorbed, 
and is able to redshift out of the resonant energy line~\cite{Seager:1999km}.
To calculate these modified rates we first calculate the Sobolev optical depth and then the probability that a photon will redshift out of the resonant energy line,
\begin{alignat}{1}
\tau_{ij} &= \frac{A_{ji} \lambda_{ij}^3 [n_i (g_j/g_i) - n_j]}{8 \pi H(z)} \\
p_{ij} &= \frac{1-\exp (-\tau_{ij})}{\tau_{ij}}  \, .
\label{eqn:Sobolev}
\end{alignat}
where $\lambda_{ij}$ is the line photon's wavelength.

In addition to the dipole transition rates, we include the most important quadrupole transition, $1s \leftrightarrow 2s$.
The transition rate from the $2s$ to $1s$ state in the presence of a background radiation field with occupation number $f(\omega)$ is given by~\cite{Chluba:2005uz}
\begin{equation}
A_{2s1s} = \frac{A_0}{2} \int_0^1 \phi(y) \left[ 1 + f^\gamma (\omega) \right] \left[ 1 + f^\gamma (E_\alpha - \omega) \right] \,dy ,
\label{eqn:A2s1s}
\end{equation}
where $A_0 =  \SI{4.3663}{\s^{-1}}$, $y = \omega / E_\alpha$, 
and $\phi (y)$ is proportional to the probability density for emitting two photons at frequencies $\omega$ and $E_\alpha - \omega$.
The factor of $1/2$ is required since by integrating $y$ from 0 to 1, we count each photon twice.
We use the analytic fit for $\phi(y)$ given in Refs.~\cite{1984A&A...138..495N,Chluba:2005uz}.
\begin{equation}
\phi(y) = C [w(1-4^{c_3} w^{c_3}) + c_1 w^{{c_2} + {c_3}} 4^{c_3}] ,
\end{equation}
where $w = y(1-y)$, $C = 46.26$, $c_1 = 0.88$, ${c_2} = 1.53$, and ${c_3} = 0.8$.
One can check that in the absence of a radiation field, i.e. $f^\gamma (\omega) = 0$, Eq.~\eqref{eqn:A2s1s} yields the transition rate in vacuum, $A_{2s1s} = \SI{8.22}{\s^{-1}}$.
The reverse rate is
\begin{equation}
A_{1s2s} = \frac{A_0}{2} \int_0^1 \phi(y) f^\gamma (\omega) f^\gamma (E_\alpha - \omega) \,dy .
\end{equation}

We would like to obtain the spectrum of photons resulting from this transition.
If $n_{2s}$ and $n_{1s}$ are the number of atoms in the $2s$ and $1s$ states, respectively, 
then the net change in number of photons per unit time and volume in the energy bin containing $y_i$ with width $dy_i$ is given by
\begin{align}
n_B \frac{d N_\omega}{dt} &= 2 \,dy_i \left( n_{2s} \frac{d A_{2s1s}}{dy} - n_{1s} \frac{d A_{1s2s}}{dy} \right) \n
&= A_0 \phi(y_i) \,dy_i \times \left\{ n_{2s} \left[ 1 + f^\gamma (\omega_i) \right] \left[ 1 + f^\gamma (E_\alpha - \omega_i) \right] - n_{1s} f^\gamma (\omega_i) f^\gamma (E_\alpha - \omega_i) \right\} .
\end{align}
The factor of 2 in the first line accounts for the fact that there will be a contribution from transitions corresponding to photons with energy $\omega_i$, as well as $E_\alpha - \omega_i$.

\section{Solving for excited states}
\label{app:matrix_method}

In Section~\ref{sec:steady_state}, we described our simplified evolution equations under the steady state approximation and calculated the rates $\tilde{\alpha}_\text{B}$
, $\tilde{\beta}_\text{B}$, and $\dot{x}_\text{inj}$ in terms of the quantity $Q_k$.
The derivation outlined there required only one matrix inversion; however, since $Q_k$ is nearly equal to the identity, calculating $1-Q_k$ to adequate precision is very slow.
Our code is based upon the following procedure, which is numerically faster.

As noted in Section~\ref{sec:steady_state}, under the steady-state approximation, determining the hydrogen level populations amounts to solving the matrix equation
\begin{alignat}{1}
x_k = \sum_{l > 1s} M^{-1}_{kl} (b_l^{1s} + b_l^\text{rec} + b_l^\text{inj}) .
\end{alignat}
In the code, we use a slightly different normalization for $M_{kl}$ and the $b_l^i$ terms such that
\begin{gather}
M_{kl} = \delta_{kl} - \frac{\tilde{R}_{l \rightarrow k}}{\tilde{R}_k^\text{out}} , \\
b_l^{1s} = x_{1s} \frac{\tilde{R}_{1s \rightarrow l}}{\tilde{R}_l^\text{out}} , \quad
b_l^\text{rec} = x_e^2 n_\text{H} \frac{\alpha_l}{\tilde{R}_l^\text{out}}, \quad
b_l^\text{inj} = \frac{\dot{x}^\text{exc}_{\text{inj}, l}}{\tilde{R}_l^\text{out}} .
\label{eq:M_and_b_code}
\end{gather}

Using these expressions, we can again simplify Eq.~\eqref{eq:simplified_intermediate_xe} to obtain 
\begin{alignat}{1}
\dot{x}_e = - x_e^2 n_\text{H} \tilde{\alpha}_\text{B} + x_{1s} \tilde{\beta}_\text{B} + \dot{x}_\text{inj} \,,
\end{alignat}
where we now have
\begin{align}
\tilde{\alpha}_\text{B} &= \sum_{k > 1s} \alpha_k - \beta_k M_{kl}^{-1} b_l^\text{rec} , \\
\tilde{\beta}_\text{B} &= \beta_k M^{-1}_{kl} b^{1s}_l , \\
\dot{x}_\text{inj} &= \beta_k M^{-1}_{kl} b^\text{inj}_l + \dot{x}_\text{inj}^\text{ion} \, .
\end{align}

\section{Derivation of the Three-Level Atom Model}
\label{app:derivation_TLA}

We begin with the multi-level atom model discussed in Eq.~\eqref{eq:final_xe}, given by 
\begin{alignat}{1}
\dot{x}_e = - x_e^2 n_\text{H} \tilde{\alpha}_\text{B} + x_{1s} \tilde{\beta}_\text{B} + \dot{x}_\text{inj} \,,
\label{eq:xe_evolution_appendix}
\end{alignat}
where we have defined the following objects:
\begin{alignat}{1}
\tilde{\alpha}_\text{B} &= \sum_{k > 1s}  Q_k \alpha_k \,, \nonumber \\
\tilde{\beta}_\text{B} &= \sum_{k > 1s} (1 - Q_k) \tilde{R}_{1s \to k} \,, \nonumber \\
\dot{x}_\text{inj} &= \sum_{k > 1s} (1 - Q_k) \dot{x}_{\text{inj}, k}^\text{exc} + \dot{x}_\text{inj}^\text{ion} \,, \nonumber \\
Q_k &= \sum_{l > 1s} M_{lk}^{-1} \tilde{R}_{l \to 1s} \,, \nonumber \\
M_{kl} &= \delta_{kl} \tilde{R}_k^\text{out} - \tilde{R}_{l \to k} \,. 
\end{alignat}
It is also useful to recall the relation 
\begin{alignat}{1}
\beta_k = \sum_{l > 1s} M_{lk} - \tilde{R}_{k \to 1s}
\end{alignat}

The first assumption to derive the TLA is that all $n \geq 2$ states are in Boltzmann equilibrium with each other. Under this assumption, there are no net bound-bound transitions. Furthermore, since $g_i \exp(-\omega_i / T) \tilde{R}_{i \to j} = g_j \exp(-\omega_j / T) \tilde{R}_{j \to i}$ for any two states $i$ and $j$ by detailed balance, we note that
\begin{alignat}{1}
g_l e^{-\omega_l / T} M_{kl} = g_k e^{- \omega_k / T} M_{lk} \,, \qquad g_l e^{- \omega_l / T} M_{kl}^{-1} = g_k e^{- \omega_k / T} M_{lk}^{-1} \,.
\label{eq:symmetry_of_M}
\end{alignat}
We also have the following detailed balance relation between photoionization and recombination coefficients, obtained using the Saha relation:
\begin{alignat}{1}
\alpha_k = \frac{\lambda_\text{th}^3}{2} g_k e^{- \omega_k / T} \beta_k \,,
\label{eq:thermal_eq_alpha_beta_relation}
\end{alignat}
where $\lambda_\text{th} \equiv \sqrt{2 \pi / (m_e T_\text{CMB})}$ is the thermal de Broglie wavelength of an electron. Now, revisiting the expression for $x_k$, under the assumption of Boltzmann equilibrium for all excited states, we can write with the help of Eqs.~\eqref{eq:symmetry_of_M} and~\eqref{eq:thermal_eq_alpha_beta_relation} 
{\small
\begin{alignat}{1}
x_k &= \sum_{l > 1s} M_{kl}^{-1} \left( x_{1s} \tilde{R}_{1s \to l} + x_e^2 n_\text{H} \alpha_l + \dot{x}_{\text{inj}, l}^\text{exc} \right) \nonumber \\
&= \sum_{l > 1s} M_{kl}^{-1} \left( x_{1s} \tilde{R}_{1s \to l} + x_e^2 n_\text{H} \frac{\lambda_\text{th}^3}{2} g_l e^{-\omega_l / T} \beta_l + \dot{x}_{\text{inj},l}^\text{exc} \right) \nonumber \\
&= \sum_{l > 1s} M_{kl}^{-1} \left[ \frac{x_{1s}}{2} e^{\omega_1/T} g_l e^{-\omega_l / T} \tilde{R}_{l \to 1s} + x_e^2 n_\text{H} \frac{\lambda_\text{th}^3}{2} g_l e^{- \omega_l / T} \left( \sum_{p > 1s} M_{pl} - \tilde{R}_{l \to 1s} \right) + \dot{x}_{\text{inj},i}^\text{exc} \right] \nonumber \\
&= \frac{1}{2} \left(x_{1s} e^{\omega_1 / T} - x_e^2 n_\text{H} \lambda_\text{th}^3 \right) \sum_{l > 1s}  g_l e^{- \omega_l / T} M_{kl}^{-1} \tilde{R}_{l \to 1s} + \frac{1}{2} x_e^2 n_\text{H} \lambda_\text{th}^3 \sum_{l > 1s} \sum_{q > 1s} g_l e^{- \omega_l / T} M_{kl}^{-1} M_{ql} + \sum_{l > 1s} M_{kl}^{-1} \dot{x}_{\text{inj}, l}^\text{exc} \nonumber \\
&\approx \frac{1}{2} \left(x_{1s} e^{\omega_1 / T} - x_e^2 n_\text{H} \lambda_\text{th}^3 \right) g_k e^{- \omega_k / T} \sum_{l > 1s}   M_{lk}^{-1} \tilde{R}_{l \to 1s} + \frac{1}{2} x_e^2 n_\text{H} \lambda_\text{th}^3 g_k e^{- \omega_k / T} \sum_{l > 1s} \sum_{q > 1s} M_{lk}^{-1} M_{ql} + \sum_{l > 1s} M_{kl}^{-1} \dot{x}_{\text{inj}, l}^\text{exc} \nonumber \\
&= \frac{1}{2} g_k e^{- \omega_k / T} \left[ \left(x_{1s} e^{\omega_1/T} - x_e^2 n_\text{H} \lambda_\text{th}^3 \right) Q_k + x_e^2 n_\text{H} \lambda_\text{th}^3 \right] + \sum_{l > 1s} M_{kl}^{-1} \dot{x}_{\text{inj},l}^\text{exc} \,. 
\end{alignat}
}
At this point, in order to derive expressions previously used in the literature, we must make the further approximation that $\dot{x}_{\text{inj},l}^\text{exc}$ is small, and can be dropped; we will return to this point later in the section. Doing so, and applying the assumption about a Boltzmann distribution of states, we find
\begin{alignat}{1}
\frac{g_k e^{-\omega_k / T}}{2 e^{- \omega_2 / T}} x_{2s} \approx \frac{1}{2} g_k e^{- \omega_k / T} \left[ \left(x_{1s} e^{\omega_1/T} - x_e^2 n_\text{H} \lambda_\text{th}^3 \right) Q_k + x_e^2 n_\text{H} \lambda_\text{th}^3 \right] \,,
\end{alignat}
or
\begin{alignat}{1}
Q_k \approx \frac{x_{1s} e^{\omega_1/T} - x_e^2 n_\text{H} \lambda_\text{th}^3}{x_{2s} e^{\omega_2 / T} - x_e^2 n_\text{H} \lambda_\text{th}^3} \equiv Q \,.
\end{alignat}
Note that the right-hand side of this expression is independent of $k$, i.e.\ $Q_k$ takes the same approximate value $Q$ for all states $k$. 

In fact, $Q$ can be written purely in terms of atomic transition rates, independent of the population of the various hydrogen states and the free electron fraction. Beginning from the definition of $Q_k$, we can write the sum
\begin{alignat}{1}
\sum_{l > 1s} \sum_{k > 1s}  g_k e^{-\omega_k / T}  M_{lk} Q_k &= \sum_{l > 1s} \sum_{k > 1s}  g_k e^{-\omega_k / T} M_{lk} \sum_{p > 1s} M_{pk}^{-1} \tilde{R}_{p \to 1s} \nonumber \\
&\approx \sum_{l > 1s} \sum_{k > 1s} g_l e^{- \omega_l / T} M_{kl} \sum_{p > 1s} M_{pk}^{-1} \tilde{R}_{p \to 1s} \nonumber \\
&= \sum_{l > 1s} g_l e^{-\omega_l / T} \tilde{R}_{l \to 1s} 
\end{alignat}
On the other hand, 
\begin{alignat}{1}
\sum_{l > 1s} \sum_{k > 1s} g_k e^{-\omega_k / T} M_{lk} Q_k &= \sum_{k > 1s} g_k e^{-\omega_k / T} Q_k \sum_{l > 1s} M_{lk} \nonumber \\
&= \sum_{k > 1s} g_k e^{-\omega_k / T} Q_k (\beta_k + \tilde{R}_{k \to 1s})
\end{alignat}
Given the fact that $Q_k$ is approximately constant, we can put these two expression together to find
\begin{alignat}{1}
Q \approx \frac{\sum_{k > 1s} g_l e^{-\omega_k / T} \tilde{R}_{k \to 1s}}{\sum_{k > 1s} g_k e^{-\omega_k / T} \beta_k + \sum_{k > 1s} g_k e^{-\omega_k / T} \tilde{R}_{k \to 1s}} \,.
\end{alignat}
The final approximation that we make is that recombination to the ground state from states $n > 2$ is negligible compared to the $n = 2$ states, which is an excellent approximation assuming the excited states are Boltzmann distributed~\cite{AliHaimoud:2010dx}. Defining the case-B photoionization rate $\beta_\text{B} = (1/8)\exp(\omega_2 / T) \sum_{k > 1s} g_k \exp(-\omega_k / T) \beta_k$, this gives us our final expression for $Q$, 
\begin{alignat}{1}
Q &\approx \frac{2 e^{-\omega_2 / T} (\tilde{R}_{2s \to 1s} + 3 \tilde{R}_{2p \to 1s}) }{8 e^{-\omega_2 / T} \beta_\text{B} +  2 e^{-\omega_2 / T} (\tilde{R}_{2s \to 1s} + 3 \tilde{R}_{2p \to 1s})} = \frac{\tilde{R}_{2s \to 1s} / 4 + 3 \tilde{R}_{2p \to 1s} / 4}{\beta_\text{B} + \tilde{R}_{2s \to 1s} / 4 + 3 \tilde{R}_{2p \to 1s} / 4} = \mathcal{C} \,,
\end{alignat}
where in the last step we note that the expression is exactly to the Peebles-$\mathcal{C}$ factor. With this approximation, we find
\begin{alignat}{1}
\tilde{\alpha}_\text{B} &\approx Q \sum_{k > 1s} \alpha_k = \mathcal{C} \alpha_\text{B} \,, \nonumber \\
\tilde{\beta}_\text{B} &\approx (1 - \mathcal{C}) \sum_{k > 1s} \frac{g_k e^{-\omega_k / T}}{2 e^{-\omega_1 / T}} \tilde{R}_{k \to 1s} \approx (1 - \mathcal{C}) e^{-(\omega_2 - \omega_1)/T} (\tilde{R}_{2s \to 1s} + 3 \tilde{R}_{2p \to 1s}) = 4 \mathcal{C} e^{-(\omega_2 - \omega_1) / T} \beta_\text{B} \,, \nonumber \\
\dot{x}_\text{inj} &\approx (1 - \mathcal{C}) \sum_{k > 1s}\dot{x}_{\text{inj}, k}^\text{exc} + \dot{x}_\text{inj}^\text{ion} \,,
\end{alignat}
where $\alpha_\text{B}$ is the case-B recombination coefficient. Upon substitution into Eq.~\eqref{eq:xe_evolution_appendix}, this finally leads to
\begin{alignat}{1}
\dot{x}_e = \mathcal{C} \left(- x_e ^2 n_\text{H} \alpha_\text{B} + 4 \beta_\text{B} e^{-(\omega_2 - \omega_1) / T} x_{1s} \right) + (1 - \mathcal{C}) \sum_{k > 1s} \dot{x}_{\text{inj}, k}^\text{exc} + \dot{x}_\text{inj}^\text{ion} \,,
\end{alignat}
the evolution equation for the three-level atom. In summary, the assumptions made are that \textit{1)} all excited states follow a Boltzmann distribution with respect to each other, \textit{2)} injected photons play a small role in setting the occupation number of the excited states, \textit{3)} recombination proceeds primarily through the $n = 2$ states, which is a consequence of the Boltzmann suppression of the population in higher level states made in the first assumption. Of these assumptions, the second assumption is exactly true for standard recombination, but may be false in the presence of exotic energy injection---in fact, injected low-energy photons that excite hydrogen atoms can easily dominate the spectrum, since the CMB blackbody distribution is exponentially suppressed for such energies. With the proper inclusion of injected photons, $Q_k$ becomes $k$-dependent, and the effective recombination rate after recombining to state $k$ is no longer independent of $k$, and cannot be simply expressed as a single factor $\mathcal{C}$; the usual way of including excitations through a term proportional to $(1 - \mathcal{C})$~\cite{Chen:2003gz,DarkHistory} is therefore not correct. Of course, the presence of nonthermal photons already breaks the first assumption to begin with, necessitating the full multi-level treatment described in Section~\ref{sec:evolution}.

\section{Additional Cross Checks}
\label{app:cross_checks}

In this appendix, we validate our improvements to the \dhis code by comparing against results in the literature.

\subsection{Comparison to \dhis \texttt{v1.0}}

%
\begin{figure*}
	\includegraphics[width=\textwidth]{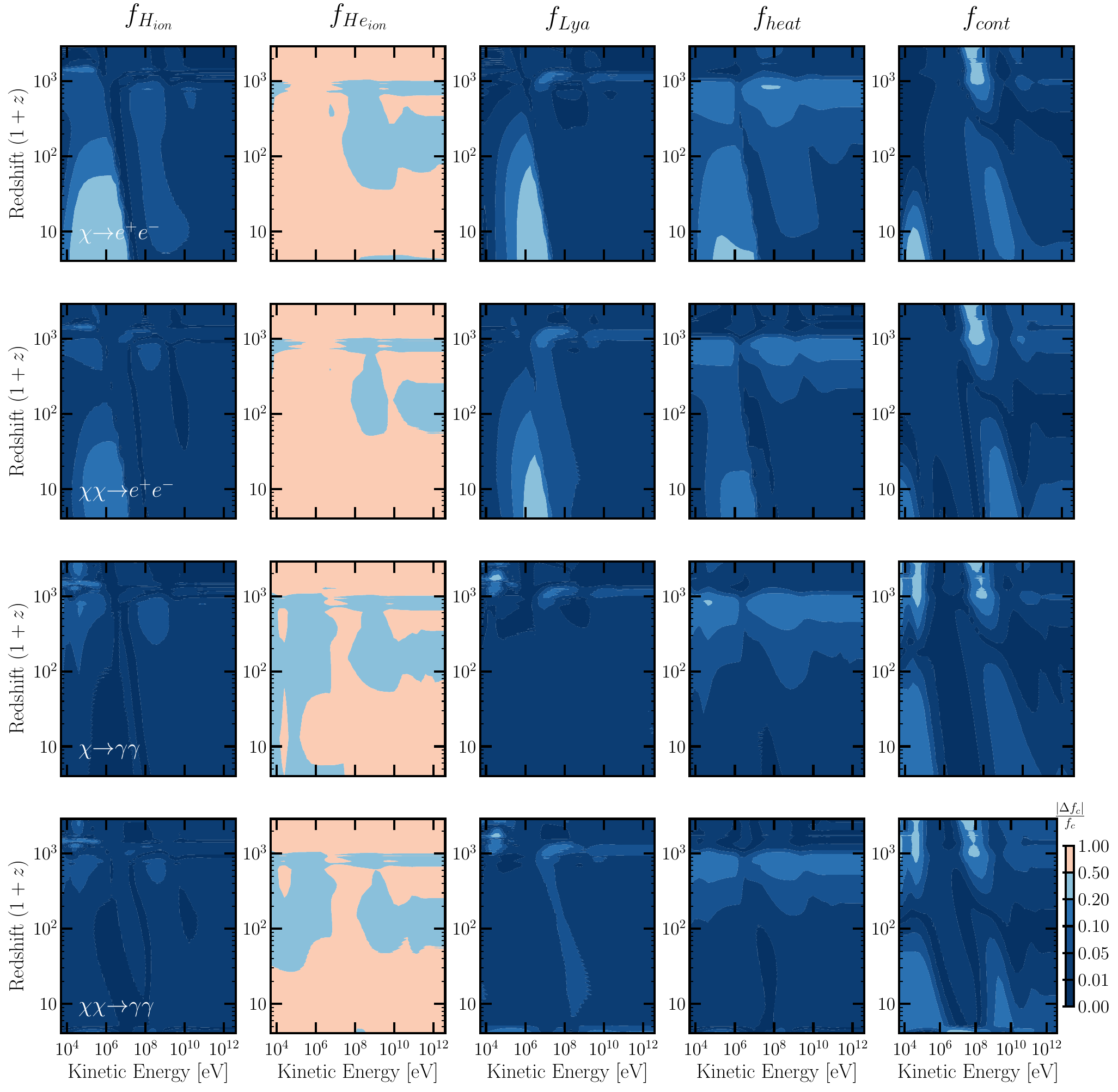}
	\caption{Comparison against the $f_c$ values calculated from \dhis \texttt{v1.0}.
		Each row shows either decay or annihilation to $e^+ e^-$ or photon pairs, and each column shows a different energy deposition channel.}
	\label{fig:f_heatmap}
\end{figure*}
In Fig.~\ref{fig:f_heatmap}, we show the difference between the $f_c$'s calculated using the updated treatment of low-energy electrons (using $n_\text{max}=10$ and only one iteration) and \dhis \texttt{v1.0}, as a function of redshift and the kinetic energy of the injected electron.
The channels are defined as follows.
\begin{itemize}
	\item H ion: energy deposited by photoionization and collisional ionization of hydrogen.
	
	\item He ion: energy deposited by photoionization and collisional ionization of helium.
	
	\item Ly-$\alpha$: energy deposited by photons and electrons into $1s \rightarrow 2p$ excitations. This was previously labeled as the `exc' channel, since this was the only excitation that we tracked.
	In the new method, since we can track an arbitrary number of excited states, then the number of Lyman-$\alpha$ photons emitted depends on the probability that an excited states cascades to the ground state by first deexciting to $2p$; as mentioned in Section~\ref{sec:atom-cross-checks}, this can be calculating using the $R_{i \to j}$ transition rates.
	
	\item heat: energy deposited by injected electrons into internal energy of the IGM.
	
	\item cont: energy deposited into photons with energy less than $E_\alpha$.
	Again, since the new method can track excited states other than $2p$, the continuum channel includes contributions from deexcitations of excited states to $2p$, as well as deexcitations to $2s$ and the two photon transition from $2s$ to $1s$.
\end{itemize}
With the exception of helium ionization and certain regions of continuum deposition, the difference in $f_c$ for all the channels is under 10\% for most redshifts and energies.
As discussed in Section~\ref{sec:lowengelec}, the helium ionization channel in MEDEA is rather noisy due to the Monte Carlo procedure they employ; this explains the large relative difference in $f_{\text{He ion}}$.

We also expect some differences in $f_\text{cont}$ since we are including a new contribution: the spectrum of photons upscattered by ICS off low-energy electrons and photoionized electrons. 
This new component is important at high redshifts, which is also where we find the largest discrepancies.
There are also a number of larger discrepancies at low redshifts in channels which primarily produce $e^+ e^-$ pairs; these are due to the fact that we are including new excitation states and are using different cross-sections from before.
We find that if we set these contributions to zero, agreement with \dhis \texttt{v1.0} is restored at the level of about 10\%.

In addition, we include an option to calculate an `effective' $f_\text{exc}$ such that if one uses the \dhis \texttt{v1.0} TLA evolution equation,
\begin{gather}
\dot{x}_e = - \mathcal{C} \left[ n_\text{H} x_e x_\text{HII} \alpha_\text{B} - 4 (1 - x_\text{HII}) \beta_\text{B} e^{-E_{21} / T_\text{CMB}} \right] + \left[ \frac{f_\text{ion}}{\mathcal{R} n_\text{H}} + \frac{f_\text{exc}}{E_\alpha n_\text{H}} \right] \left( \frac{dE}{dV \, dt} \right)^\text{inj} + \dot{x}^\text{re} ,
\label{eq:TLA_xe}
\end{gather}
together with the newly defined $f_\text{exc}$, then one obtains the same histories as that calculated using the MLA.
Note that compared to the Eq. (5) in Ref.~\cite{DarkHistory}, we have absorbed a factor of $(1-\mathcal{C})$ into $f_\text{exc}$ for numerical stability.

To summarize, with this version of \texttt{DarkHistory}, we include the option \texttt{elec\_method} in \texttt{main.evolve()}, which allows one to calculate $f_c$ by one of three methods depending on if the option is set to \texttt{`old'}, \texttt{`new'}, or \texttt{`eff'}.
\begin{itemize}
	\item \texttt{`old'}: calculate the $f_c$'s as in \dhis \texttt{v1.0}, using MEDEA for the energy deposition of electrons with energy $<3$ keV.
	
	\item \texttt{`new'}: calculate the $f_c$'s without separating low-energy electrons from high-energy electrons.
	
	\item \texttt{`eff'}: calculate the $f_c$'s without separating low-energy electrons from high-energy electrons, and also output an effective $f_\text{exc}$ that can be plugged into the TLA equations.
\end{itemize}

\subsection{Spectral distortions from ICS and heating}

As mentioned in Section~\ref{sec:y-validation}, a key difference between our work and Ref~\cite{Acharya:2018iwh} is that they treat the IGM as completely ionized.
At the early redshifts they are interested in, this is a good assumption.
However, between the redshift of $1+z=3000$ (the default initial redshift for \dhis) and recombination, there is a small but non-negligible amount of neutral hydrogen.
Secondary electrons resulting from ionization of this hydrogen can contribute to heating and therefore contribute to a $y$-type distortion.
\begin{figure*}
	\includegraphics[width=\textwidth]{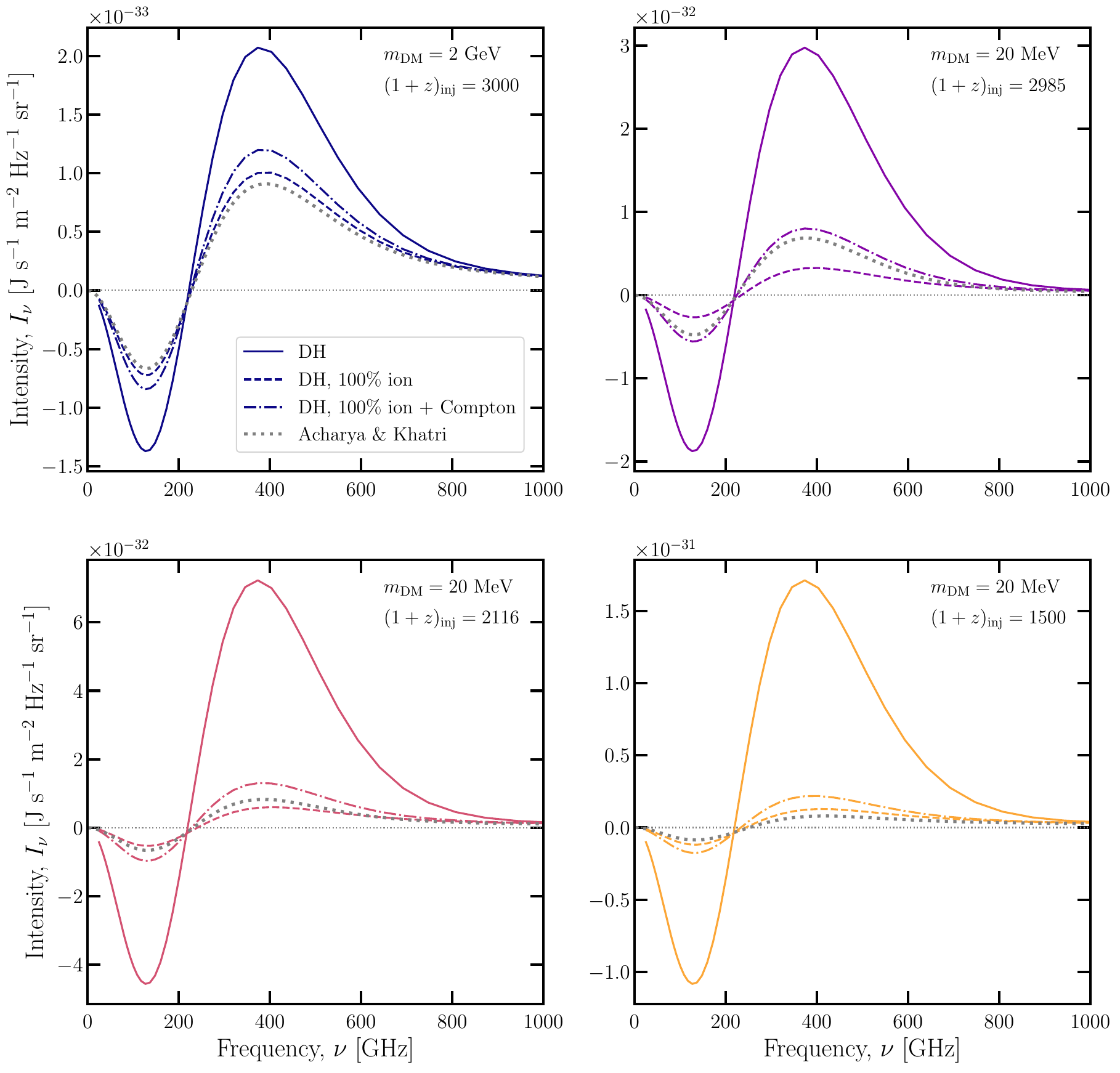}
	\caption{Spectral distortions from dark matter decaying to $e^+ e^-$ pairs, at various dark matter masses and injection redshifts. The solid lines shows the distortions contributed by low-energy photons in \texttt{DarkHistory}; the dashed lines neglect energy deposition by secondary low-energy electrons, which may result from photoionizations; the dot-dashed curves make the same assumptions as the dashed, but furthermore include the effects of heating from Compton scattering (see text for details). For comparison, the Green's functions from Ref.~\cite{Acharya:2018iwh} are shown in the gray dashed line.
		The dashed and dot-dashed curves bracket, or nearly bracket the Green's functions; this is as expected, since Ref.~\cite{Acharya:2018iwh} assumes 100\% ionization and accounts for the heating component.}
	\label{fig:greens_fncs}
\end{figure*}

Fig.~\ref{fig:greens_fncs} shows the Green's functions from Ref~\cite{Acharya:2018iwh} for dark matter of different masses decaying to $e^+ e^-$ pairs at redshift $1+z_\text{inj}$, compared to the same distortions generated by \dhis using a few different methods.
The solid line shows the component from only summing over the low-energy photon spectra; we see that as a consequence of tracking the ionization level consistently in \texttt{DarkHistory}, the solid line is always larger in amplitude than the Green's functions of Ref.~\cite{Acharya:2018iwh}-- that is to say, assuming full ionization may underestimate spectral distortions from the epoch prior to recombination.

To check consistency with Ref.~\cite{Acharya:2018iwh} under matched assumptions, we can try to turn off the contribution to heating from secondary electrons produced by ionization.
The dashed line shows the predictions of \dhis without including the heating from ionized secondary electrons in the module for low energy deposition.
While modifying the low energy deposition accounts for most of the energy going into heat, this does not fully bracket the contribution from photoionization.
Our high energy deposition transfer functions do not extend to a value of $x_e = 1$, since at the highest redshifts we consider, there is still a small but non-negligible amount of neutral hydrogen.
Hence, for this cross-check, we cannot truly set $x_e = 1$, so some ionizations are necessarily included and the secondary electrons from this channel can propagate into the low-energy electrons and contribute to heating.
Hence, while the dashed curves are lower than the Green's function from Ref~\cite{Acharya:2018iwh} for $m_\text{DM} = 20$ MeV and injection redshifts 2985 and 2116, they are slightly above for the other panels.

Finally, this comparison also omits a contribution that {\it is} included in Ref.~\cite{Acharya:2018iwh}, where photons heat the free electrons through Compton scattering (since they are not absorbed via photoionization).
Thus, the dot-dashed curves include this heating from Compton scattering, calculated using Equation (B.2) in Ref.~\cite{Acharya:2018iwh}.
These and the dashed curves fully bracket or nearly bracket the Green's functions from Ref~\cite{Acharya:2018iwh} in Fig.~\ref{fig:greens_fncs}.

Thus, we achieve reasonable agreement with Ref.~\cite{Acharya:2018iwh} when we do not include photoionizations from low-energy photons, but do include Compton scattering from these photons that would realistically photoionize at these redshifts.

\subsection{MLA treatment validation}
\label{sec:MLA_vs_TLA}

\begin{figure}[t]
	\centering
	\includegraphics[width=0.5\textwidth]{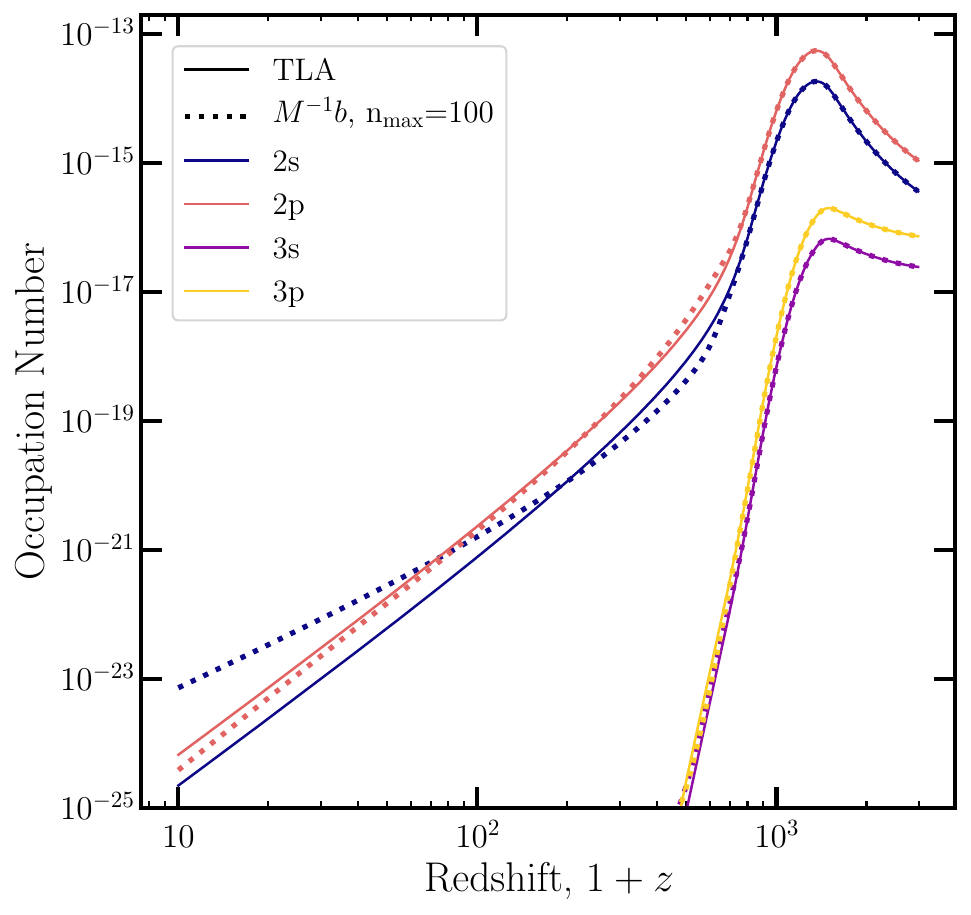}
	\caption{Occupation number of hydrogen levels under both the MLA and TLA treatments.}
	\label{fig:MLA_vs_TLA}
\end{figure}

At high enough redshifts, when the density of hydrogen is large enough that the TLA assumptions hold, the TLA and MLA treatments should yield the same results.
Fig.~\ref{fig:MLA_vs_TLA} shows the evolution of the occupation levels for the lowest hydrogen levels calculated using Eq.~\eqref{eq:matrix_MLA} and with the TLA method; both are calculated without including any sources of exotic energy injection.
The two methods agree well at redshifts above a few hundred.
%

\section{Spectral distortions from other redshifts}
\label{app:other_rs}

In Section~\ref{sec:dist_results}, we showed the spectral distortions resulting from energy injection between redshifts of $3000 > 1+z > 4$.
By default, \dhis only treats this redshift range, since $1+z = 3000$ is the highest redshift for which we have photon cooling transfer functions, and at lower redshifts one has to treat helium reionization.
However, energy injection from other redshift ranges can have contributions to the distortion that are comparable in amplitude.

\begin{figure*}
	\includegraphics[width=\textwidth]{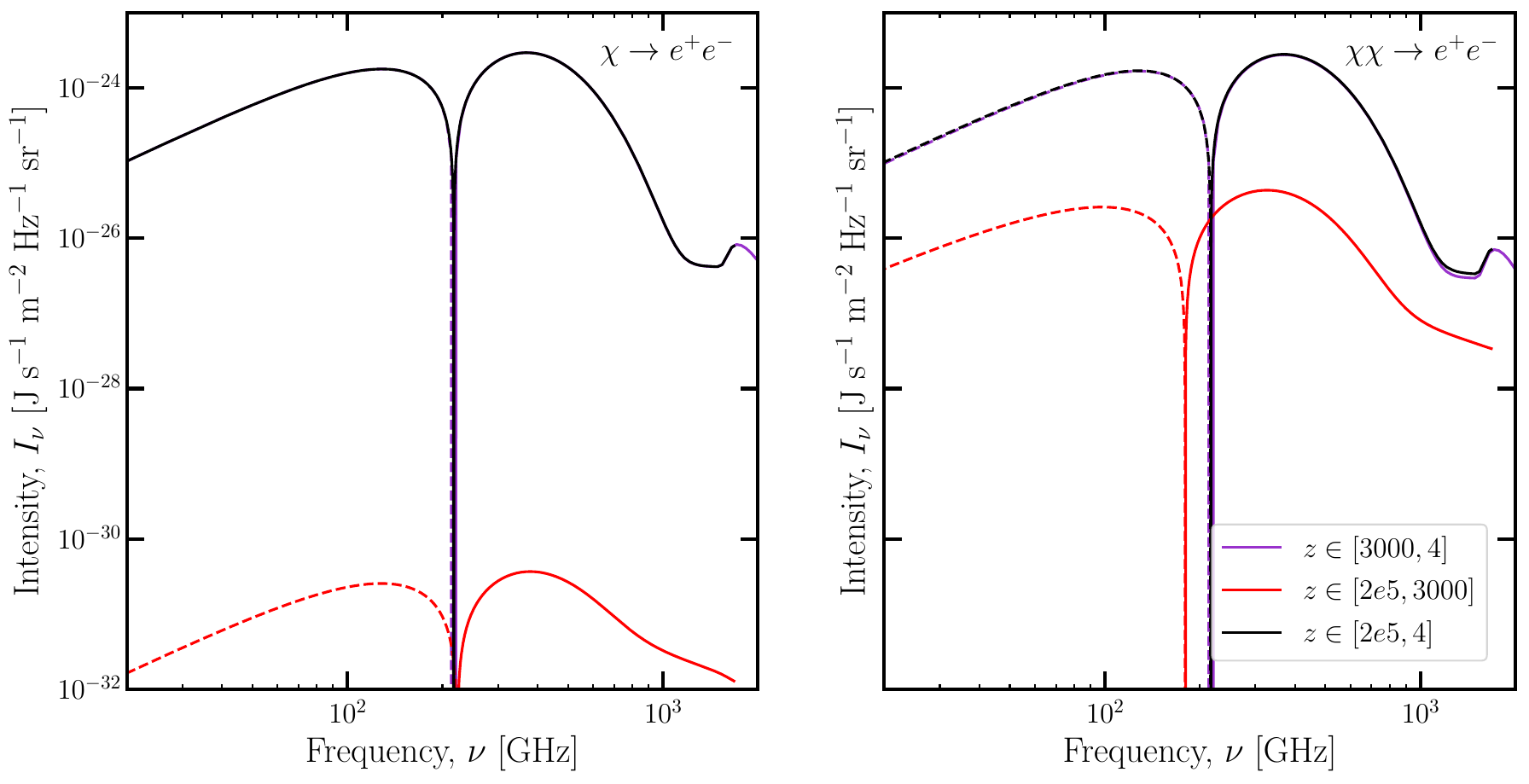}
	\caption{
		Contribution to the spectral distortion from high redshifts. 
		The DM masses are chosen such that the injected electrons and positrons have kinetic energy of 1 GeV; for decaying DM, the lifetime is $10^{25}$ s and for annihilation, the cross-section is $10^{-26}$ cm$^3$ s$^{-1}$.
		Red shows the contribution from redshifts $2 \times 10^5 > 1+z > 3000$ calculated using the Green's functions described in Ref.~\cite{Acharya:2018iwh}.
		Purple is the contribution from \texttt{DarkHistory}, while black shows the sum of the two contributions.
		Note that the $x$-axis range here is much narrower than for other figures.
	}
	\label{fig:high_rs}
\end{figure*}

For redshifts between $2 \times 10^5 > 1+z > 3000$, one can determine the spectral distortion contribution using the Green's functions calculated in Ref.~\cite{Acharya:2018iwh}.
We find that the early redshift contribution is subdominant for the masses we have tested, as shown in Fig.~\ref{fig:high_rs}; the left panel shows decaying DM and the right shows annihilating DM.
The mass of the DM particle is chosen such that the injected electrons and positrons have a kinetic energy of 1 GeV, so that we can use the Green's functions presented in Ref.~\cite{Acharya:2018iwh}.
The red line shows the early contribution and the purple line shows the contribution calculated by \texttt{DarkHistory}; black shows the sum of the two contributions, i.e. the total spectral distortion contributed by energy injection between $2 \times 10^5 > 1+z > 4$.

The relative size of the contributions can be roughly understood by considering the redshift dependence of the energy injection rate.
For decays, the energy injected per log redshift interval as a fraction of the CMB energy density is proportional to $\rho_\chi / H / u_\text{CMB} (T)$, where $H$ is the Hubble parameter and $u_\text{CMB}(T)$ is the blackbody energy density with temperature $T$.
Hence, this quantity depends on redshift as $(1+z)^{-2.5}$ during matter domination and $(1+z)^{-3}$ during radiation domination.
Then integrating this over the appropriate log redshift intervals, we estimate that the contribution from $3000 > 1+z >4$ should be larger than that of $2 \times 10^5 > 1+z > 3000$ by about seven orders of magnitude, which is roughly consistent with Fig.~\ref{fig:high_rs}.
Similarly for annihilation, the energy injected per Hubble time as a fraction of the CMB energy density is $\rho_\chi^2 / H / u_\text{CMB} (T)$, which goes as $(1+z)^{0.5}$ during matter domination and is constant during radiation domination.
Integrating this over the two redshift ranges, we find that two contributions are nearly equal, with the later contribution being smaller by a factor of only about 2.
However, as we found in Section~\ref{sec:DHv2_tech}, when the universe is fully ionized, i.e. prior to $z \sim 1100$, the conversion of energy injection to spectral distortions is less efficient; the reason is there are no photoionizations, and hence no secondary electrons, which can efficiently contribute to heating.
Hence, this suppresses the distortion from earlier redshifts by a couple orders of magnitude.
Note that these estimates are independent of dark matter mass, so we expect it to be generally true that the contribution from $3000 > 1+z > 4$ dominates over the contribution from higher redshifts.

\begin{figure*}
	\includegraphics[width=\textwidth]{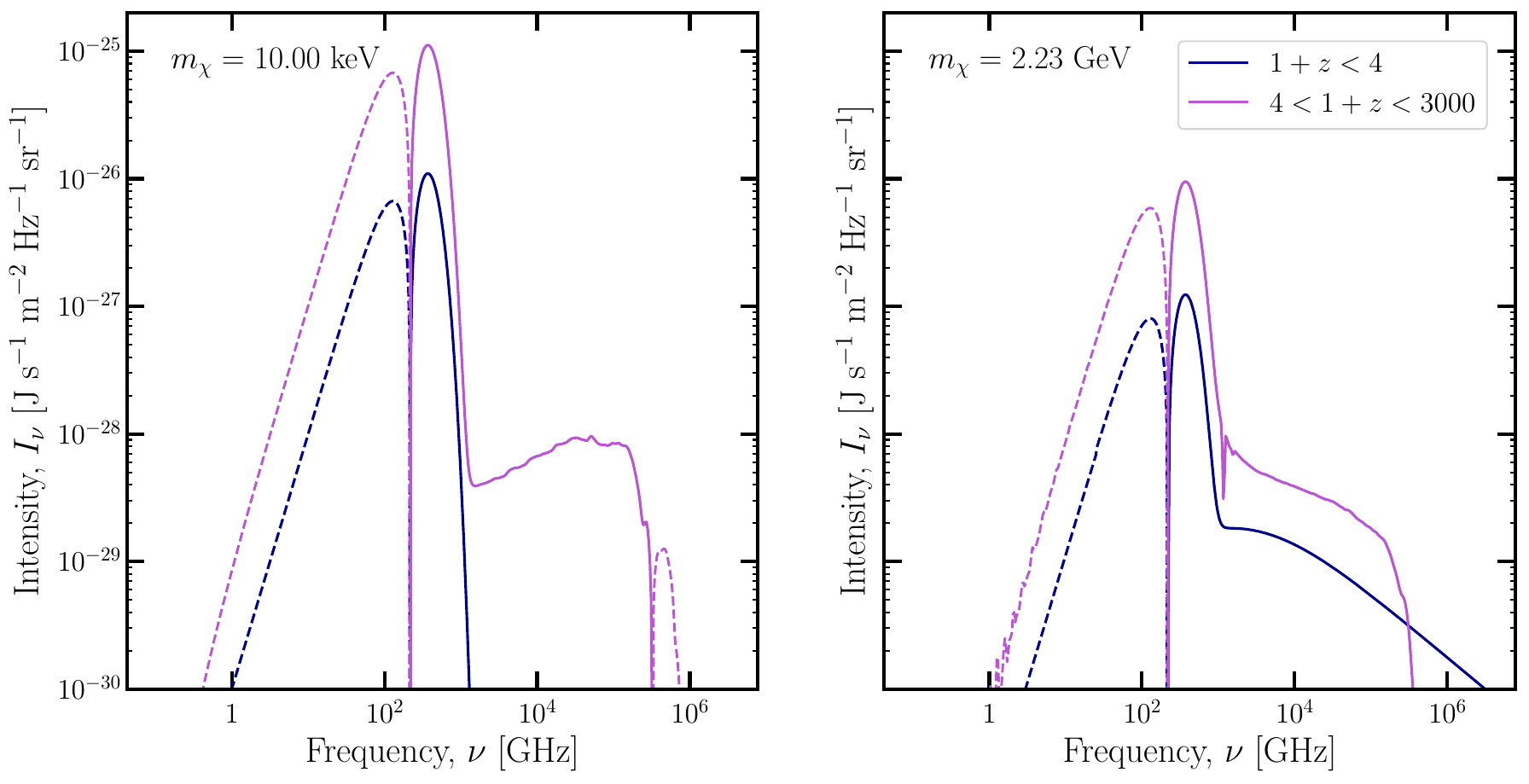}
	\caption{
		Estimate of the contribution to the spectral distortions from low redshifts.
		Here, we assume DM decays to photons with the lifetime taken to be at CMB constraints~\cite{Slatyer:2016qyl}.
		Purple shows the early contribution from \texttt{DarkHistory}, while blue shows the late time contribution.
	}
	\label{fig:low_rs}
\end{figure*}
We can also estimate the contribution to the distortion during the latest redshifts, $1+z < 4$.
At these redshifts, we can approximate the universe as completely ionized, hence the only cooling mechanisms available to photons are Compton scattering, pair production, and redshifting; the particle cascades resulting from these processes will eventually contribute to a spectral distortion through heating or ICS.
For photon energies much smaller than the electron mass $m_e$, the photons lose energy to electrons at the rate~\cite{1990ApJ...349..415S,Acharya:2018iwh}.
\begin{equation}
\frac{d \omega}{dt} = n_e \sigma_T \frac{\omega^2}{m_e} ,
\end{equation}
where $n_e$ is the number density of electrons and $\sigma_T$ is the Thomson cross-section.
Then, based on the timescales for electron cooling processes given in Ref~\cite{2010MNRAS.404.1869F}, the dominant cooling process for electrons at energies less than a few MeV is heating, as opposed to ICS.
At energies much larger than this, we can no longer use the previous energy loss formula and we must also take ICS into account.
Hence, we can estimate the low-redshift spectral distortion in this regime using three steps:
\begin{enumerate}
	\item We calculate the rate at which electrons are upscattered by Compton scattering.
	
	\item We assume the electrons gain most of the injected photon's energy; this can be seen by averaging the Compton scattering formula
	\begin{equation}
	\omega' = \frac{\omega}{1 + \frac{\omega}{m_e} (1-\cos\theta)}
	\end{equation}
	over the angle $\theta$, where $\omega'$ is the energy of the photon after scattering.
	We find that
	\begin{equation}
	\frac{\langle \omega' \rangle}{\omega} = \frac{m_e}{\omega} \log \left( 1 + \frac{\omega}{m_e} \right).
	\end{equation}
	In the limit that $\omega \gg m_e$, this asymptotes to zero, hence the photon loses most of its energy to the electron upon scattering.
	
	\item We process the electrons through the electron cooling module of \dhis to determine the spectrum of secondary photons from ICS, as well as how much energy was deposited into heating.
\end{enumerate}

Fig.~\ref{fig:low_rs} compares the spectral distortion calculated by \texttt{DarkHistory}, shown in purple, and the missing contribution from late times, shown in blue.
We assume DM decays to photons with a mass of 10 keV or 2.23 GeV and lifetime taken to be at the CMB bounds~\cite{Slatyer:2016qyl}; these two models are representative of results for energy deposition by low and high energy particles.
The amplitude of the late-time distortion is smaller than the contribution from $3000 > 1+z > 4$ by about an order of magnitude in the models we consider. 
Again, we can estimate the relative size of these contributions using the argument that we outlined above for high redshifts.
Using the redshift dependence for energy injection by decaying dark matter, one might guess that the late time contributions from $1+z < 4$ should be larger in amplitude than that of $3000 > 1+z > 4$ by a factor of about 10. However, since the universe is also fully ionized at these late times, the amplitude of the resulting spectral distortion is suppressed and turns out to be smaller than the distortion from $3000 > 1+z > 4$ by an order of magnitude for the two masses that we have tested.

Moreover, if the late time spectral distortion does not have significant ICS or atomic line contributions, then it is degenerate with the distortions from reionization and structure formation, since all of these are $y$-type distortions sourced by heating.
If ICS is significant, then the distortion will have a high energy tail that can be distinguished from other sources of spectral distortions, as shown in the right panel of Fig.~\ref{fig:low_rs}.

\begin{figure*}
	\includegraphics[width=\textwidth]{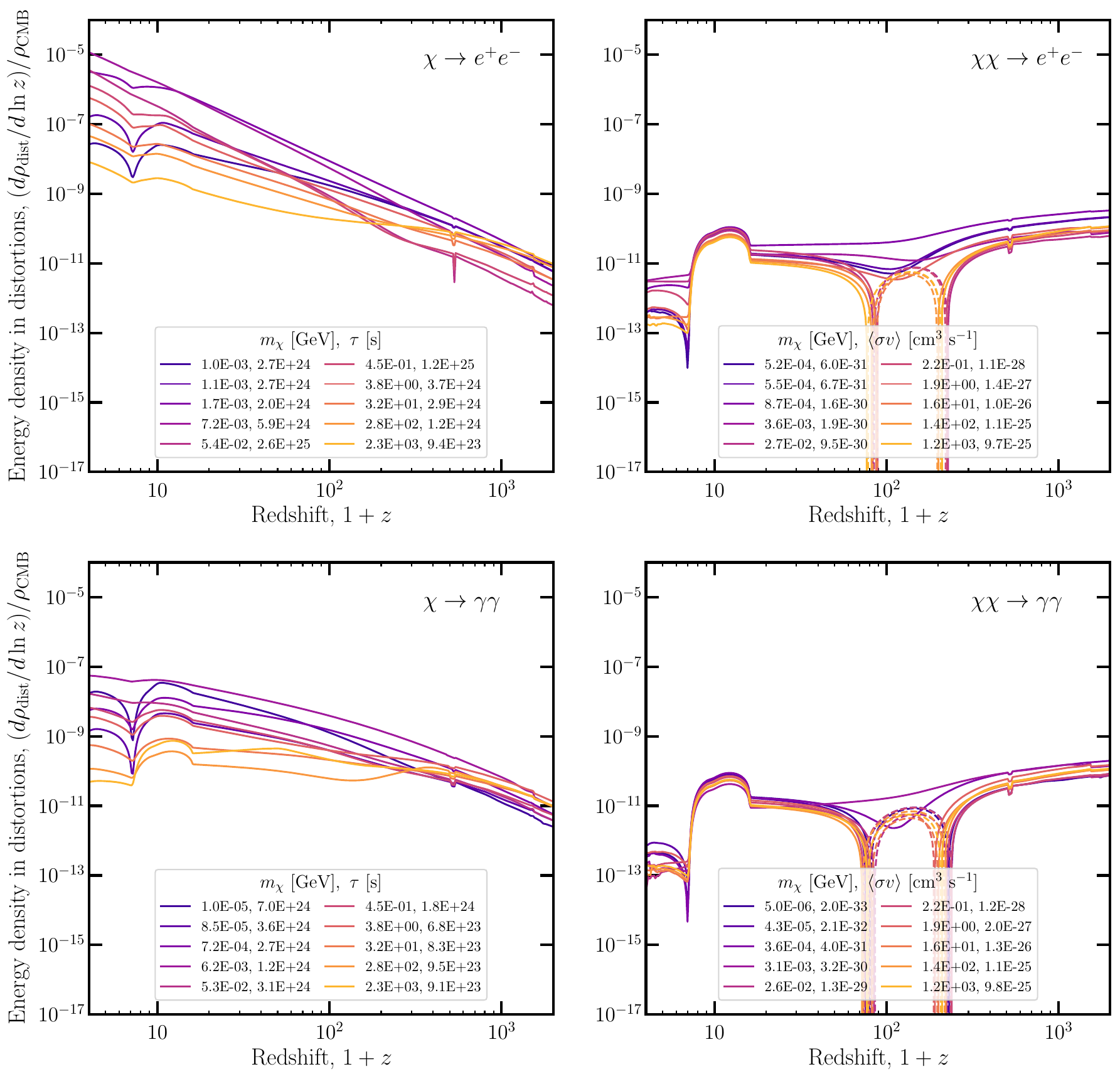}
	\caption{
		Rate of change in the energy of the spectral distortion relative to $\Lambda$CDM for the same models as in  Fig.~\ref{fig:dist_grid}.
		In other words, this figure shows how the energy in the spectra shown in Fig.~\ref{fig:dist_grid_noLCDM} evolves with time.
	}
	\label{fig:eng_rate_grid_noLCDM}
\end{figure*}
\begin{figure}
	\centering
	\includegraphics[width=0.5\columnwidth]{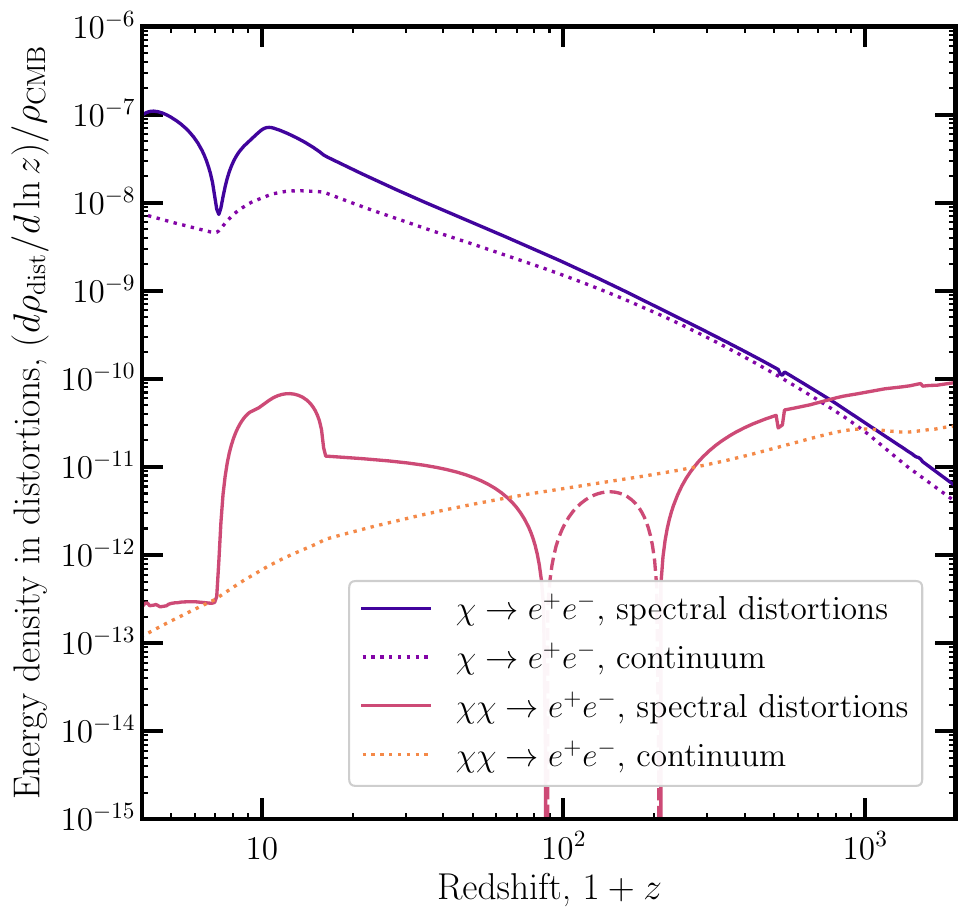}
	\caption{
		Comparison of the rate of change in the energy of the spectral distortion to the rate of energy deposited in the continuum channel for two of the models shown in Fig.~\ref{fig:eng_rate_grid_noLCDM}.
	}
	\label{fig:dist_vs_cont}
\end{figure}
%

\section{Energy density in spectral distortions}
\label{app:eng_rate}
%
Here, we validate our results by examining how much energy is deposited into the spectral distortions as a function of redshift.
Fig.~\ref{fig:eng_rate_grid_noLCDM} shows the rate of change in the energy of the spectral distortion, not including the contributions to the distortion from $\Lambda$CDM processes.
In other words, this figure shows how the energy in the spectra shown in Fig.~\ref{fig:dist_grid_noLCDM} evolves with time.

For comparison, in Fig.~\ref{fig:dist_vs_cont}, we show two of the models in Fig.~\ref{fig:dist_grid_noLCDM}, one for decay to $e^+ e^-$ and one for annihilation to $e^+ e^-$, and also show an estimate for the rate at which energy is deposited into the continuum for those same models, computed as discussed in Section~\ref{sec:DHv2_tech}. 
This calculation largely tracks the $f_\text{cont}$ calculation in the original version of \texttt{DarkHistory}, albeit with improvements to the treatment of low-energy electrons; it was intended as a calorimetric estimate of the total power into spectral distortions. 
The full spectral distortion calculation includes effects that the continuum calculation does not---in particular, $y$-type distortions from heating are not included in the contributions to the continuum channel---and provides the spectrum itself rather than simply an integrated quantity. 
This comparison can be viewed as a cross-check on our previous estimate of the integrated energy, and we do find that the shape and normalization of the curves is generally similar; however, detailed sensitivity estimates should use the spectral distortion results.

We see that the general shape of all the curves is consistent with the scaling for the energy injection rate discussed in App.~\ref{app:other_rs}: for decays, this rate scales as $(1+z)^{-2.5}$ during matter domination and $(1+z)^{-3}$ during radiation domination and for annihilations, the rate goes as $(1+z)^{0.5}$ during matter domination and is constant during radiation domination.
There are obvious exceptions to this trend.
For example, when reionization begins, there is an increase to the rate of energy deposited in spectral distortions due to photoheating and emission from excited hydrogen atoms.
There is also a small artifact around $1+z \sim 500$ due to a change in the way we calculate the $y$ parameter after this redshift, see Section~\ref{sec:DHv2_tech} for details.

In addition, for $s$-wave annihilation, there is a negative feature around $1+z \sim 100$.
The main contributions to spectral distortions at this time are atomic line emission and a negative amplitude $y$-type distortion. 
This $y$-distortion is present because the matter is always slightly colder than the CMB; the matter temperature would cool faster than the CMB if their temperatures were not coupled at early times.
The atomic lines are a positive contribution to the energy density in spectral distortions, whereas the $y$-distortion is a negative contribution.
Both contributions are enhanced in the presence of exotic energy injection due to the larger residual ionization at these times; hence the contribution to the energy density caused by exotic energy injection can be negative when the enhancement of the $y$-distortion dominates.

\section{Comparison to \texttt{Recfast}}
\label{app:recfast_xcheck}

%
\begin{figure*}
	\includegraphics[width=\textwidth]{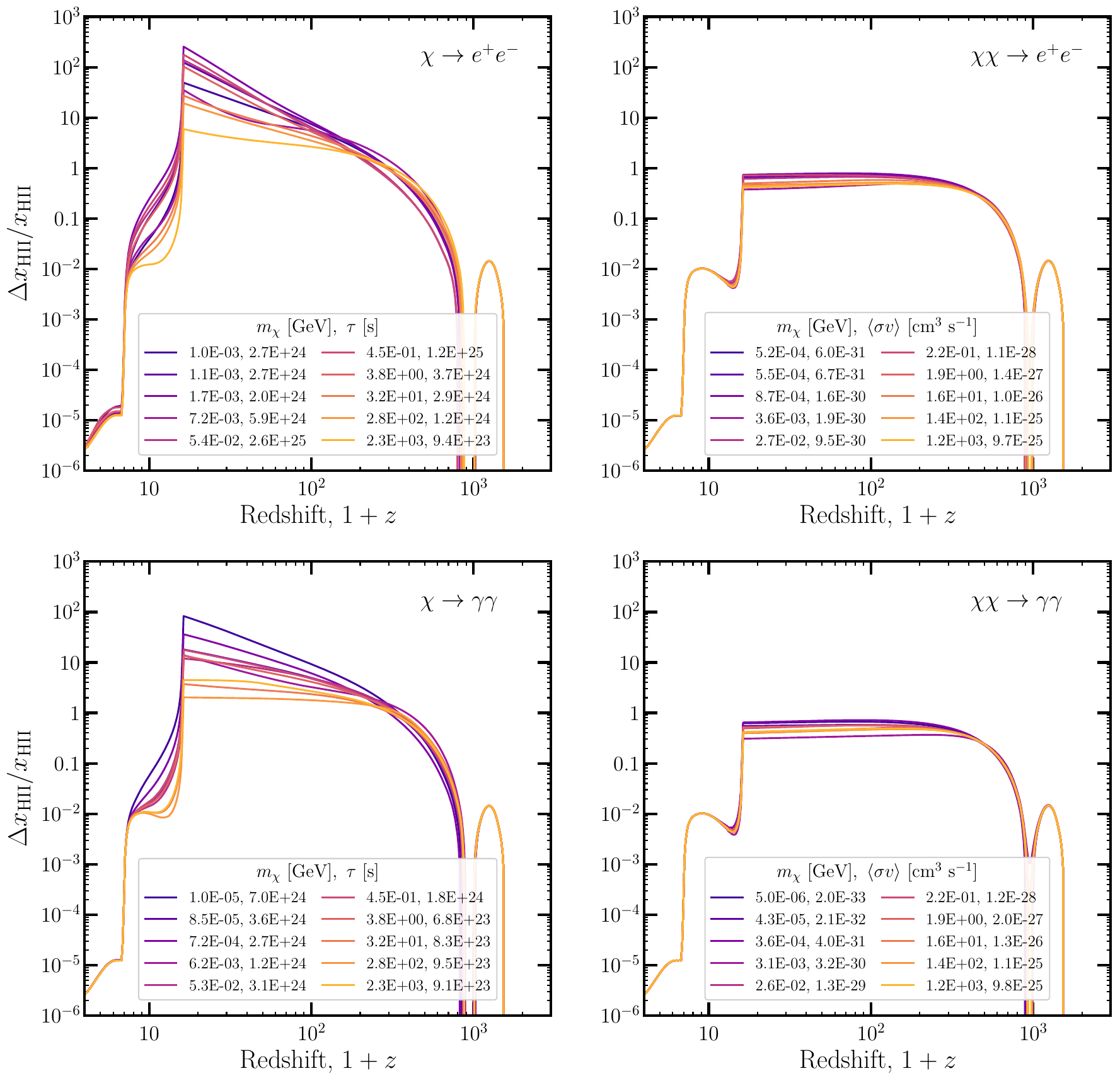}
	\caption{Difference in the ionization history relative to that calculated using \texttt{Recfast}.
		In other words, this is the difference in ionization between the two methods, divided by the history calculated with  \texttt{Recfast}.
		While the differences in ionization between histories with/without exotic energy injection is $\mathcal{O}(1)$ for annihilating DM, the relative change can be as large as a factor of a few hundred for decaying models.}
	\label{fig:delta_xe_grid_recfast}
\end{figure*}

In Section~\ref{sec:old_ion}, we showed that the difference in the ionization histories calculated using the method described in Section~\ref{sec:DHv2_tech} and \dhis \texttt{v1.0} were not significant.
Hence, a comparison of the ionization histories calculated with the new method against \texttt{Recfast} is not technically new, but we show the results here for completion.

Fig.\ref{fig:delta_xe_grid_recfast} shows the relative difference in $x_\text{HII} (z)$ calculated using the upgrades from Section~\ref{sec:DHv2_tech} and \texttt{Recfast}; in other words, Fig.\ref{fig:delta_xe_grid_recfast} shows the change in the ionization when we include the effect of exotic energy injection.
The signal is largest for decaying dark matter models, where the ionization history can be enhanced by as much as $\mathcal{O} (100)$ around $1+z \sim 20$.

\clearpage
\newpage

\chapter{Supplementary material for Section~\ref{sec:first_stars}}

\section{Comparison of IGM and halo contributions}
\label{app:IGM_vs_halo}

In this work, we have assumed that energy deposited per particle in the halo can be approximated by the energy deposited per particle from decays and annihilations in the IGM.
A natural question to ask is how the homogeneous contribution (i.e. from the IGM) compares to 
to the contributions of decays and annihilations within the halo itself.
For example, for Milky Way-like halos, the high density within the halo means that the annihilations within the halo typically contribute much more to indirect-detection signals than the IGM, whereas for decays the contributions are comparable~\cite{Ibarra:2009nw,Essig:2013goa}.
However, for earlier halos which are much smaller than the Milky Way halo, it is not as clear which contribution will dominate.

In the case where particles propagate with long path lengths, one can estimate the intensity of particles sourced by exotic energy injections within a system by calculating the $J$-factor for annihilations or the $D$-factor for decays.
For a spherically-symmetric system, these are given by
\begin{gather}
    J = \int d\ell \, \rho^2 (r (\ell, \theta, \psi)), \\
    D = \int d\ell \, \rho (r (\ell, \theta, \psi)) ,
\end{gather}
where $r$ denotes the distance from the center of the density profile, $\ell$ is the line-of-sight distance, and $\theta$ and $\psi$ specify the angle of the source relative to the observer.
These expressions can be derived by calculating the flux from annihilating or decaying DM and isolating the factors that depend on astrophysics~\cite{Lisanti:2016jxe,Slatyer:2021qgc}.

The effect of annihilations within early halos has been studied in e.g.\ Refs.~\cite{Schon:2014xoa,Schon:2017bvu}, hence we will compare to typical halos used in their work.
Consider a halo at redshift $1+z \sim 30$ with a mass of about $10^6 M_\odot$.
We will assume it has an Einasto density profile,
\begin{equation}
    \rho_\mathrm{Ein} = \rho_0 \exp \left\{ -\frac{2}{\alpha} \left[ \left( \frac{r}{r_0} \right)^\alpha - 1 \right] \right\}
\end{equation}
with $\alpha = 0.17$ and concentration parameter of approximately $c \sim 7$.
Given $M_\mathrm{halo}$, one can infer the virial radius $r_\mathrm{vir}$, which is related to the Einasto scale radius by $r_\mathrm{vir} = c r_0$; by calculating the mass within $r_\mathrm{vir}$, one can also determine the correct normalization $\rho_0$ for the density profile.
With these parameters, we find that at the center of the halo, $J \sim 10^{30}$ GeV$^2$ cm$^{-5}$ and $D \sim 10^{24}$ GeV cm$^{-2}$.

Turning to the IGM contribution, the meaning of the $D$ and $J$-factors becomes somewhat ambiguous, since when integrating out to cosmological distances, the expression for flux cannot be so easily factored into ``particle physics" and ``astrophysics" contributions due to redshifting of the emitted spectrum.
However, for the sake of an estimate, we can follow the discussion in Section 3.3 of Ref.~\cite{Slatyer:2021qgc} to effectively factor out the model-dependent terms, modifying this derivation appropriately for decays.
Then the $D$-factor from a homogeneous universe is
\begin{equation}
    D_\mathrm{IGM} = (1+z_\mathrm{obs})^3 \frac{\rho_\mathrm{DM,0}}{H_0} \int_{z_\mathrm{obs}}^\infty \frac{1}{\sqrt{\Omega_m (1+z)^3 + \Omega_\Lambda}} \, dz ,
\end{equation}
where $\rho_\mathrm{DM,0}$ is the average energy density of DM today and $\Omega_\Lambda$ is the density parameter for dark energy.
At $1+z_\mathrm{obs} \sim 30$, we find $D \sim 10^{26}$ GeV cm$^{-2}$; hence for early halos at this redshift, the IGM contribution can dominate by a factor up to a few orders of magnitude.

For annihilations, there is an extra factor of $\overline{\rho} (1+z)^3$; however, this integral will not converge if we integrate out to $1+z \rightarrow \infty$.
One could introduce an effective cutoff that accounts for the redshifting of the spectrum, as well as the potential absorption of emitted particles.
\begin{equation}
    J_\mathrm{IGM} = (1+z_\mathrm{obs})^6 \frac{\rho_\mathrm{DM,0}^2}{H_0} \int_{z_\mathrm{obs}}^{z_\mathrm{cut}} \frac{(1+z)^3}{\sqrt{\Omega_m (1+z)^3 + \Omega_\Lambda}} \, dz .
\end{equation}
However, this integral is likely an overestimate since the biggest contributions to $J_\mathrm{IGM}$ come from high redshifts, when densities are higher and secondary particles are more likely to be absorbed, or the emitted photons get redshifted out of observable wavelengths.
Given these uncertainties, it is less clear for the case of annihilations whether or not the contributions of the IGM and halo are comparable.

\section{Impact of the halo on dark matter energy deposition}
\label{app:fs_halo}

In reality, not all particles have long path lengths relative to the halo, and we also need to account for the enhanced gas density of the halo to understand how particles deposit their energy within it. Eqns.~\eqref{eqn:T_inj} and~\eqref{eqn:x_inj} describe the effect from DM energy deposition on the ionization and temperature within the halo. In this work, we have made the simplifying assumption that the per-baryon effect of DM is identical to that expected assuming homogeneity, allowing us to make use of results computed using \texttt{DarkHistory}~\cite{DarkHistory}. 
In this appendix, we will argue that this assumption can be justified throughout most of our parameter space of interest, including for the two fiducial models examined in detail in the main body. 
Even though the enhanced density of the halo leads to an increase in the rate of dark matter processes in the halo, and also provides additional targets for cascading particles to scatter off and cool, in most cases, we will show that the intensity of energy-depositing particles seen by targets inside the halo is essentially identical to the intensity of particles present in the homogeneous IGM. 
This means that the energy deposited \emph{per particle} is similar in both the halo and in the homogeneous IGM, justifying the use of the expressions shown in Eqns.~\eqref{eqn:T_inj} and~\eqref{eqn:x_inj}. 

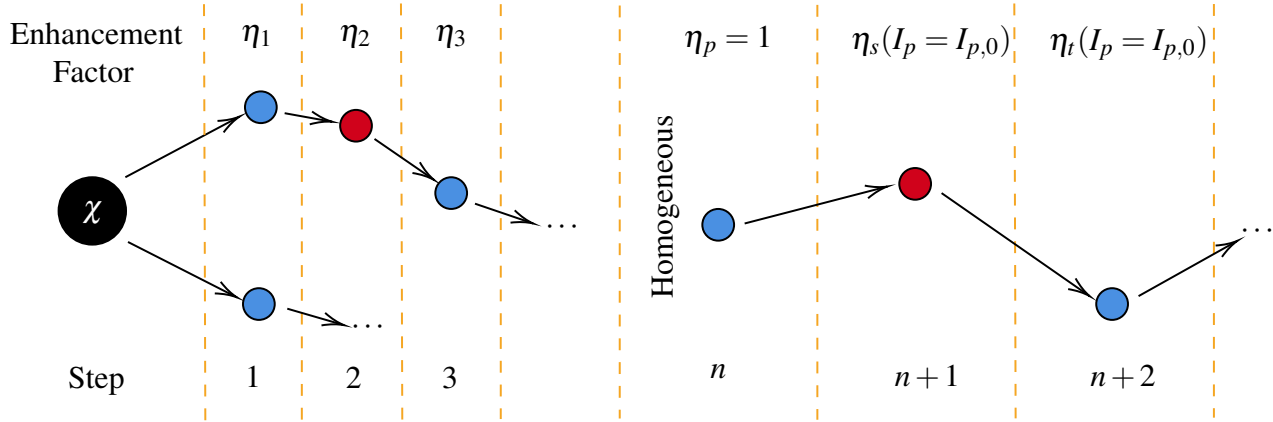
\begin{figure}[t]

    \tikzset{every picture/.style={line width=0.75pt}} 
    
    \begin{tikzpicture}[x=0.75pt,y=0.75pt,yscale=-1,xscale=1]
    
    \draw  [fill={rgb, 255:red, 0; green, 0; blue, 0 }  ,fill opacity=1 ] (36,155.33) .. controls (36,145.94) and (43.61,138.33) .. (53,138.33) .. controls (62.39,138.33) and (70,145.94) .. (70,155.33) .. controls (70,164.72) and (62.39,172.33) .. (53,172.33) .. controls (43.61,172.33) and (36,164.72) .. (36,155.33) -- cycle ;
    \draw    (69.67,138.33) -- (125.57,108.61) ;
    \draw [shift={(127.33,107.67)}, rotate = 152] [color={rgb, 255:red, 0; green, 0; blue, 0 }  ][line width=0.75]    (10.93,-3.29) .. controls (6.95,-1.4) and (3.31,-0.3) .. (0,0) .. controls (3.31,0.3) and (6.95,1.4) .. (10.93,3.29)   ;
    \draw    (71,171) -- (124.22,198.09) ;
    \draw [shift={(126,199)}, rotate = 206.98] [color={rgb, 255:red, 0; green, 0; blue, 0 }  ][line width=0.75]    (10.93,-3.29) .. controls (6.95,-1.4) and (3.31,-0.3) .. (0,0) .. controls (3.31,0.3) and (6.95,1.4) .. (10.93,3.29)   ;
    \draw [color={rgb, 255:red, 245; green, 166; blue, 35 }  ,draw opacity=1 ] [dash pattern={on 4.5pt off 4.5pt}]  (109.33,50.98) -- (110,261.33) ;
    \draw  [fill={rgb, 255:red, 74; green, 144; blue, 226 }  ,fill opacity=1 ] (130,103) .. controls (130,98.58) and (133.58,95) .. (138,95) .. controls (142.42,95) and (146,98.58) .. (146,103) .. controls (146,107.42) and (142.42,111) .. (138,111) .. controls (133.58,111) and (130,107.42) .. (130,103) -- cycle ;
    \draw  [fill={rgb, 255:red, 74; green, 144; blue, 226 }  ,fill opacity=1 ] (129,202.33) .. controls (129,197.92) and (132.58,194.33) .. (137,194.33) .. controls (141.42,194.33) and (145,197.92) .. (145,202.33) .. controls (145,206.75) and (141.42,210.33) .. (137,210.33) .. controls (132.58,210.33) and (129,206.75) .. (129,202.33) -- cycle ;
    \draw [color={rgb, 255:red, 245; green, 166; blue, 35 }  ,draw opacity=1 ] [dash pattern={on 4.5pt off 4.5pt}]  (158.67,50.67) -- (158.67,260.7) ;
    \draw    (151.33,204.67) -- (178.08,212.44) ;
    \draw [shift={(180,213)}, rotate = 196.21] [color={rgb, 255:red, 0; green, 0; blue, 0 }  ][line width=0.75]    (10.93,-3.29) .. controls (6.95,-1.4) and (3.31,-0.3) .. (0,0) .. controls (3.31,0.3) and (6.95,1.4) .. (10.93,3.29)   ;
    \draw [color={rgb, 255:red, 245; green, 166; blue, 35 }  ,draw opacity=1 ] [dash pattern={on 4.5pt off 4.5pt}]  (208.67,50.63) -- (209.33,260.67) ;
    \draw    (150,106) -- (172.03,109.98) ;
    \draw [shift={(174,110.33)}, rotate = 190.23] [color={rgb, 255:red, 0; green, 0; blue, 0 }  ][line width=0.75]    (10.93,-3.29) .. controls (6.95,-1.4) and (3.31,-0.3) .. (0,0) .. controls (3.31,0.3) and (6.95,1.4) .. (10.93,3.29)   ;
    \draw  [fill={rgb, 255:red, 208; green, 2; blue, 27 }  ,fill opacity=1 ] (178,112.33) .. controls (178,107.92) and (181.58,104.33) .. (186,104.33) .. controls (190.42,104.33) and (194,107.92) .. (194,112.33) .. controls (194,116.75) and (190.42,120.33) .. (186,120.33) .. controls (181.58,120.33) and (178,116.75) .. (178,112.33) -- cycle ;
    \draw [color={rgb, 255:red, 245; green, 166; blue, 35 }  ,draw opacity=1 ] [dash pattern={on 4.5pt off 4.5pt}]  (259,50.97) -- (259,260.37) ;
    \draw    (195.33,119) -- (223.69,138.53) ;
    \draw [shift={(225.33,139.67)}, rotate = 214.56] [color={rgb, 255:red, 0; green, 0; blue, 0 }  ][line width=0.75]    (10.93,-3.29) .. controls (6.95,-1.4) and (3.31,-0.3) .. (0,0) .. controls (3.31,0.3) and (6.95,1.4) .. (10.93,3.29)   ;
    \draw  [fill={rgb, 255:red, 74; green, 144; blue, 226 }  ,fill opacity=1 ] (226,146.33) .. controls (226,141.92) and (229.58,138.33) .. (234,138.33) .. controls (238.42,138.33) and (242,141.92) .. (242,146.33) .. controls (242,150.75) and (238.42,154.33) .. (234,154.33) .. controls (229.58,154.33) and (226,150.75) .. (226,146.33) -- cycle ;
    \draw    (246,151.67) -- (273.44,161.02) ;
    \draw [shift={(275.33,161.67)}, rotate = 198.82] [color={rgb, 255:red, 0; green, 0; blue, 0 }  ][line width=0.75]    (10.93,-3.29) .. controls (6.95,-1.4) and (3.31,-0.3) .. (0,0) .. controls (3.31,0.3) and (6.95,1.4) .. (10.93,3.29)   ;
    \draw [color={rgb, 255:red, 245; green, 166; blue, 35 }  ,draw opacity=1 ] [dash pattern={on 4.5pt off 4.5pt}]  (319,51.65) -- (319,260.1) ;
    \draw  [fill={rgb, 255:red, 74; green, 144; blue, 226 }  ,fill opacity=1 ] (360.67,162.33) .. controls (360.67,157.92) and (364.25,154.33) .. (368.67,154.33) .. controls (373.08,154.33) and (376.67,157.92) .. (376.67,162.33) .. controls (376.67,166.75) and (373.08,170.33) .. (368.67,170.33) .. controls (364.25,170.33) and (360.67,166.75) .. (360.67,162.33) -- cycle ;
    \draw [color={rgb, 255:red, 245; green, 166; blue, 35 }  ,draw opacity=1 ] [dash pattern={on 4.5pt off 4.5pt}]  (418.67,51.33) -- (418.67,260.73) ;
    \draw [color={rgb, 255:red, 245; green, 166; blue, 35 }  ,draw opacity=1 ] [dash pattern={on 4.5pt off 4.5pt}]  (518,50.63) -- (518.67,260.03) ;
    \draw    (382,161.67) -- (452.06,143.5) ;
    \draw [shift={(454,143)}, rotate = 165.47] [color={rgb, 255:red, 0; green, 0; blue, 0 }  ][line width=0.75]    (10.93,-3.29) .. controls (6.95,-1.4) and (3.31,-0.3) .. (0,0) .. controls (3.31,0.3) and (6.95,1.4) .. (10.93,3.29)   ;
    \draw  [fill={rgb, 255:red, 208; green, 2; blue, 27 }  ,fill opacity=1 ] (460,141.67) .. controls (460,137.25) and (463.58,133.67) .. (468,133.67) .. controls (472.42,133.67) and (476,137.25) .. (476,141.67) .. controls (476,146.08) and (472.42,149.67) .. (468,149.67) .. controls (463.58,149.67) and (460,146.08) .. (460,141.67) -- cycle ;
    \draw [color={rgb, 255:red, 245; green, 166; blue, 35 }  ,draw opacity=1 ] [dash pattern={on 4.5pt off 4.5pt}]  (618.67,50.9) -- (618.67,259.67) ;
    \draw    (481.33,146.33) -- (554.36,197.19) ;
    \draw [shift={(556,198.33)}, rotate = 214.85] [color={rgb, 255:red, 0; green, 0; blue, 0 }  ][line width=0.75]    (10.93,-3.29) .. controls (6.95,-1.4) and (3.31,-0.3) .. (0,0) .. controls (3.31,0.3) and (6.95,1.4) .. (10.93,3.29)   ;
    \draw  [fill={rgb, 255:red, 74; green, 144; blue, 226 }  ,fill opacity=1 ] (559.33,202.33) .. controls (559.33,197.92) and (562.92,194.33) .. (567.33,194.33) .. controls (571.75,194.33) and (575.33,197.92) .. (575.33,202.33) .. controls (575.33,206.75) and (571.75,210.33) .. (567.33,210.33) .. controls (562.92,210.33) and (559.33,206.75) .. (559.33,202.33) -- cycle ;
    \draw    (580.67,198.33) -- (628.92,171.31) ;
    \draw [shift={(630.67,170.33)}, rotate = 150.75] [color={rgb, 255:red, 0; green, 0; blue, 0 }  ][line width=0.75]    (10.93,-3.29) .. controls (6.95,-1.4) and (3.31,-0.3) .. (0,0) .. controls (3.31,0.3) and (6.95,1.4) .. (10.93,3.29)   ;

    \draw (46.67,149.07) node [anchor=north west][inner sep=0.75pt]  [color={rgb, 255:red, 255; green, 255; blue, 255 }  ,opacity=1 ]  {$\chi $};
    \draw (128.67,232.07) node [anchor=north west][inner sep=0.75pt]    {$1$};
    \draw (39.33,232.33) node [anchor=north west][inner sep=0.75pt]   [align=left] {{Step}};
    \draw (182,208.73) node [anchor=north west][inner sep=0.75pt]    {$\cdots $};
    \draw (179.33,232.4) node [anchor=north west][inner sep=0.75pt]    {$2$};
    \draw (228,232.4) node [anchor=north west][inner sep=0.75pt]    {$3$};
    \draw (280,158.73) node [anchor=north west][inner sep=0.75pt]    {$\cdots $};
    \draw (10,50.33) node [anchor=north west][inner sep=0.75pt]   [align=left] {\begin{minipage}[lt]{63.16pt}\setlength\topsep{0pt}
    \begin{center}
    {Enhancement Factor}
    \end{center}
    
    \end{minipage}};
    \draw (126.33,59.73) node [anchor=north west][inner sep=0.75pt]    {$\eta _{1}$};
    \draw (349.67,60.4) node [anchor=north west][inner sep=0.75pt]    {$\eta _{p} =1$};
    \draw (176,59.73) node [anchor=north west][inner sep=0.75pt]    {$\eta _{2}$};
    \draw (224.33,59.73) node [anchor=north west][inner sep=0.75pt]    {$\eta _{3}$};
    \draw (630.67,160.68) node [anchor=north west][inner sep=0.75pt]    {$\cdots $};
    \draw (363.33,231.4) node [anchor=north west][inner sep=0.75pt]    {$n$};
    \draw (456.33,231.4) node [anchor=north west][inner sep=0.75pt]    {$n+1$};
    \draw (555,231.73) node [anchor=north west][inner sep=0.75pt]    {$n+2$};
    \draw (324.95,200.86) node [anchor=north west][inner sep=0.75pt]  [rotate=-269.85] [align=left] {\begin{minipage}[lt]{63.67pt}\setlength\topsep{0pt}
    \begin{center}
    {Homogeneous}
    \end{center}
    
    \end{minipage}};
    \draw (434,60.07) node [anchor=north west][inner sep=0.75pt]    {$\eta _{s}( I_{p} =I_{p,0})$};
    \draw (534,60.73) node [anchor=north west][inner sep=0.75pt]    {$\eta _{t}( I_{p} =I_{p,0})$};

    \end{tikzpicture}
    \caption{Cartoon illustrating a particle cascade and the two scenarios that we consider in order to calculate enhancement factors $\eta$: \emph{(Left)} $\eta_1$ for particles directly emitted by DM, followed by the secondary $\eta_2$ and tertiary $\eta_3$, as well as \emph{(Right)} uniform intensity of primaries at step $n$ ($I_p = I_{p,0}$ and $\eta_p = 1$), from which we can calculate the enhancement factor of step $n+1$ secondaries $\eta_s(I_p = I_{p,0})$ and step $n+2$ tertiaries $\eta_t(I_p = I_{p,0})$.}
    \label{fig:cascade_cartoon}
    
\end{figure}

A full and precise treatment of this problem would be challenging, requiring tracking the 4D evolution of the secondary particle cascade that leads to energy deposition.  In this appendix, we instead perform a simplified analysis using the following assumptions: 
\begin{enumerate}
    \item We model the halo as a simple tophat of mass $M_\mathrm{halo}$, and radius $r_\mathrm{vir}$, with a density $\Delta$ times larger than the mean matter energy density. 
    \item The intensity of particles seen by targets in the halo is approximated by the intensity of particles seen at the center of this halo, which is spherically symmetric about this point. 
    \item All spectra produced are treated as monochromatic, with particles entirely scattering into the peak of their scattered spectra; for example, every electron with Lorentz factor $\gamma$ that inverse Compton scatters against the CMB is assumed to produce photons with energy given by the mean photon energy $2 \pi^4 \gamma^2 T_\mathrm{CMB} / [45 \zeta(3)]$ only.
    \item All scattering processes occur only in the forward direction. 
\end{enumerate}
In general, the particle cascade can undergo many steps before producing particles that are rapidly absorbed as ionization and heating, or which free-stream (in the case of sub-\SI{13.6}{\eV} photons).
Our goal is to determine the intensity of particles in the last step of the cascade, which either deposit their energy directly into ionization or heating of the gas. 
We will do this by calculating the intensity of appropriately chosen intermediate steps.
Consider a particle at step $n$ of the particle cascade that we will call the `primary' particle, cascading into a `secondary' particle at step $n+1$ and a `tertiary' particle at step $n+2$.
Under the assumptions listed above, we will find it useful to obtain the steady-state intensity of secondary and tertiary particles under the following conditions, both illustrated in Fig.~\ref{fig:cascade_cartoon}:  
\begin{enumerate}
    \item If $n = 1$, the primary particles are particles emitted directly by the DM process, and we can compute the intensity of the $n = 2$ secondary and $n = 3$ tertiary particles; 
    \item For any $n$, if the intensity of the primary particle $I_p$ is equal to its intensity expected in the homogeneous limit $I_{p,0}$,  we can obtain an expression for the intensities of $n+1$ secondary particles, denoted $I_s(I_p = I_{p,0})$, and $n+2$ tertiary particles $I_t(I_p = I_{p,0})$.
\end{enumerate}
To characterize these intensities, we define the quantity $\eta_n \equiv I_n / I_{n,0}$, i.e.\ the ratio of the intensity at step $n$ in the halo to the homogeneous, steady-state intensity that exists deep in the homogeneous IGM.
$\eta_n = 1$ means that the intensity of particles in step $n$ within the halo is given by the homogeneous, steady-state intensity, while $\eta_n \neq 1$ represents either an enhancement or a suppression inside the halo.

With a combination of the two scenarios summarized in Fig.~\ref{fig:cascade_cartoon}, we are able to determine the intensity of particles in the final step for DM decay into $e^+e^-$ and photon pairs for $m_\chi < \SI{10}{\giga\eV}$. For example, starting from particles emitted directly from the DM, we can first find that $\eta_3 \approx 1$, and then use $\eta_s(I_p = I_{p,0})$ to calculate $\eta_4(I_3 = I_{3,0})$, finding that $\eta_4 \approx 1$, allowing us to proceed down the cascade. 
$\eta_3$ and $\eta_t(I_p = I_{p,0})$ are computed so that we can skip intermediate steps that have short path length, which may result in $\eta \neq 1$. This will ultimately enable us to estimate the impact of the halo on energy deposition. 
In general, cascades will pass through a series of steps with long path lengths that have intensities that are close to the homogeneous limit; the impact of the halo is strong only when the last or last few steps have short path lengths. 
These results can easily be extended to annihilations as well.
We leave a more thorough understanding of how the intensity of particles from DM processes in a halo differs from the homogeneous intensity to future work. 

In this appendix, we will mainly focus on ionization and heating, and only note here that LW photons can also be affected by the presence of the halo. 
To model the intensity of LW photons within the halo, we would have to go significantly beyond the tophat model discussed here, since it depends sensitively on the H$_2$ abundance within the halo. 
Even within $\Lambda$CDM, the amount of self-shielding or equivalently the path length of LW photons is still somewhat unclear. 
If self-shielding of LW photons is not significant, then we expect LW photons to have path lengths a factor of a few times smaller than the Hubble radius, since the H$_2$ fraction in the IGM is negligible, and the path length is by definition much longer than the halo itself. 
We would therefore expect the LW intensity to be similar to the homogeneous, steady-state intensity in the IGM, loosely justifying our use of the LW intensity in the IGM in the limit of no self-shielding. 
For now, we set aside the question of the effect of the halo on LW photons originating from DM, leaving it to simulations to address in detail.

\subsection{Particles Directly Emitted from the Dark Matter Process}
\label{app:primaries}
We first begin by understanding the intensity of $n=1$ particles emitted directly from the DM process, and received at the center of the halo. 
We start from the radiative transport equation applied to lines pointing radially outward from the center of the halo, 
\begin{alignat*}{1}
    \frac{dI_1}{ds} = j_1(s) - \alpha_1(s) I_1(s)  \,,
\end{alignat*}
which relates the intensity of particles emitted at a distance $s$ from the halo center to the emission coefficient $j$ and the extinction coefficient $\alpha$.
We will neglect redshifting throughout this appendix for simplicity.
The emission coefficient is the energy emitted per volume, time, frequency and solid angle by the medium, which for DM processes is 
\begin{alignat*}{1}
    j_1 \equiv  \frac{dE_\omega}{dV \, dt \, d\omega \, d\Omega} = \frac{1}{4\pi} \left( \frac{dE}{dV \, dt} \right)^\mathrm{inj} \frac{d \overline{N}_1}{d \omega} \,,
\end{alignat*}
where $d \overline{N}_1 / d \omega$ is the spectrum of particles directly emitted per DM process. 
We have also made use of the isotropy of the DM process in writing down this expression. 
Outside of the halo, we denote the energy injection rate to be $(dE / dV \, dt)^\mathrm{inj}_0$, the usual injection rate under the homogeneous assumption. 
Inside the halo, however, we have
\begin{alignat*}{1}
    j_1(s < r_\mathrm{vir}) = \frac{\Delta^\beta}{4 \pi} \left( \frac{dE}{dV \, dt} \right)^\mathrm{inj}_0 \frac{d \overline{N}_1}{d\omega} \,,
\end{alignat*}
where $\beta = 1$ for decay and $\beta = 2$ for annihilation. 
The extinction coefficient is related to the optical depth of the directly emitted particle via the usual relation
\begin{alignat*}{1}
    \tau_1(s, s') = \int_s^{s'} dx \, \alpha_1(x) \,,
\end{alignat*}
where $\tau_1(s, s')$ is the optical depth between points $s$ and $s'$, with $s < s'$. 
$\alpha_1^{-1}(s)$ is the local mean free path of the directly emitted particle at $s$. 
Under the homogeneous assumption, the extinction coefficient for the particle takes on some constant value $\alpha_{1,0}$; inside the halo, however, the extinction coefficient is enhanced by a factor $\Delta_1$, the enhancement in density of the targets of the primary in the halo. 
In some cases, e.g.\ ICS of electrons off the CMB, the density of targets presented to the particle is not enhanced by the halo. 
For simplicity, we only consider the dominant process responsible for scattering of the directly emitted particle. 
The optical depth between the origin and some point $s$ away from the origin can therefore be written as
\begin{alignat}{1}
    \tau_1(0, s) = \begin{cases}
        \Delta_1 \alpha_{1,0} s \,, & s < r_\mathrm{vir} \,, \\
        \Delta_1 \alpha_{p,0} r_\mathrm{vir} + \alpha_1(s - r_\mathrm{vir})  \,, & s \geq r_\mathrm{vir}
    \end{cases}
\end{alignat}
under our simple tophat approximation. 

The radiative transfer equation can be integrated radially inward to obtain
\begin{alignat}{1}
    I_1(s) = \int_s^\infty ds'\, j_p(s') e^{-\tau_p(s, s')}  \,.
    \label{eqn:gen_pri_intensity_expr}
\end{alignat}
Under the homogeneous assumption, the intensity of directly emitted particles at all points in space is
\begin{alignat}{1}
    I_{1,0} \equiv \frac{1}{4\pi \alpha_0} \left( \frac{dE}{dV \, dt} \right)^\mathrm{inj}_0 \frac{d \overline{N}}{d\omega}  \,.
\end{alignat}

We now define the quantity $\eta_1 \equiv I_1(s = 0)/I_{1,0}$ as a measure of the impact of the halo on the intensity of directly emitted particles received inside the halo. 
If $\eta_1 \sim 1$, then the intensity of these particles is essentially equal to the intensity expected under the assumption of homogeneity. 
Otherwise, the halo plays a significant role in determining the intensity received. 
Under our tophat assumption, we can perform the integral Eqn.~\eqref{eqn:gen_pri_intensity_expr} to obtain 
\begin{alignat}{1}
    \eta_1 = \frac{\Delta^\beta}{\Delta_1} \left(1 - e^{-\Delta_1 \alpha_{1,0} r_\mathrm{vir}} \right) + e^{-\Delta_1 \alpha_{1,0}  r_\mathrm{vir}} \,.
    \label{eqn:eta_1}
\end{alignat}

Let us consider the following limits: 
\begin{itemize}
    \item $\Delta_1 \alpha_{1,0} r_\mathrm{vir} \ll 1$: since $\Delta_1 \alpha_{1,0}$ is the inverse of the local mean free path of the directly emitted particle in the halo, this corresponds to the limit where the local mean free path in the halo is much longer than the halo itself, i.e.\ the halo is optically thin. 
    We find $\eta(\Delta_1 \alpha_{1,0} r_\mathrm{vir} \ll 1) \approx 1 + \Delta^\beta \alpha_{1,0} r_\mathrm{vir} \approx 1$ for $n = 1$. 
    This demonstrates one of the key takeaways of this appendix: \emph{the center of the halo is illuminated with the same intensity of particles as in the homogeneous IGM, as long as the mean free path of the particles involved is sufficiently long.}
    
    \item $\Delta_1 \alpha_{1,0} r_\mathrm{vir} \gg 1$: the halo is optically thick to directly emitted particles, and we therefore find $\eta(\Delta_1 \alpha_{1,0} r_\mathrm{vir} \gg 1) \approx \Delta^\beta / \Delta_1$. 
    The intensity is enhanced by the $\Delta^\beta$ enhancement in the DM process rate within the halo, but is shielded by the enhanced density of targets surrounding the center of the halo. In particular, \emph{if primaries sourced by DM decays deposit their energy promptly into ionization and heating via any process with a rate that scales as $\Delta$, e.g.\ atomic processes, the intensity within the halo remains at the homogeneous value.} An alternative way to see this result is to consider the total power injected in the halo, and divide by the number of gas particles in the halo, under the assumption that all the injected power is promptly deposited. If the both the injected power and the gas particle density are enhanced by the same factor, the effect cancels out in the ratio.
\end{itemize}
We observe that at least for the case of DM decay, these nominally opposite limits actually lead to the same behavior; this is a first hint that this behavior ($\eta\approx 1$) will be common.

\subsection{Secondaries}
\label{app:secondaries}

In most cases, particles from DM processes cool by scattering into other particles, which can further undergo subsequent interactions. 

We will first consider how to determine the intensity of secondary particles (at step $n+1$) in the limit where \emph{1)} the primary (at step $n = 1$) producing this particle is sourced directly by the DM process, with $j_1(s \leq r_\mathrm{vir}) = \Delta^\beta j_{1,0}$ and $j_1(s > r_\mathrm{vir}) = j_{1,0}$, and \emph{2)} the primary (at step $n$ for any $n$) has constant intensity $I_p = I_{p,0}$. These are two limits that we will frequently encounter in a DM process particle cascade. 
Throughout this section, we use subscript `$p$' to denote both $n = 1$ and more general primaries, and `1' only when discussing $n = 1$ directly emitted primaries.
Our goal is to compute the intensity of secondary particles at step $n+1$.

Under the simplifying assumption that all particles scatter only in the forward direction, the intensity of the primaries $I_p$ acts as a source of emission for the secondaries, i.e.\ along radial paths pointing out from the center of the halo,
\begin{alignat}{1}
    j_s(s) = \int d \omega_p \, \alpha_p(s) \frac{I_p(s)}{\omega_p} \cdot \omega_s \frac{d \overline{N}_s}{d \omega_s}(\omega_p) \,,
    \label{eqn:sec_emission_coeff}
\end{alignat}
where $\alpha_p$ is the extinction coefficient of the primaries, and $d \overline{N}_s / d \omega_s$ is the spectrum of secondaries produced per primary scattering event. Note that $\alpha_p$, $I_p$ and $d \overline{N}_s / d \omega_s$ depend on $\omega_p$, while $j_s$ depends also on $\omega_s$.
\footnote{An additional integral over $\Omega_p$ would appear without the assumption of forward scattering.}
Integrating the radiation transfer equation for secondaries gives
\begin{alignat}{1}
    I_s(s) = \int d \omega_p \, \frac{\omega_s}{\omega_p} \frac{d \overline{N}_s}{d \omega_s}(\omega_p) \int_s^\infty ds' \, \alpha_p(s') I_p(s') e^{-\tau_s(s, s')}
    \label{eqn:gen_sec_intensity_expr}
\end{alignat}

For a homogeneous medium, we have
\begin{alignat*}{1}
    I_{s,0} = \int d\omega_p \, \frac{\omega_s}{\omega_p} \frac{d \overline{N}_s}{d\omega_s} (\omega_p) I_{p,0} \frac{\alpha_{p,0}}{\alpha_{s,0}} \,,
\end{alignat*}
where $I_{p,0}$ is the homogeneous primary intensity.

First, let us consider the case where the primary intensity is given by its homogeneous value, i.e.\ $I_p = I_{p,0}$. From Eqn.~\eqref{eqn:gen_sec_intensity_expr}, the intensity of the secondaries is given by
\begin{alignat*}{2}
    I_s(I_p = I_{p,0}) &=&& \int d \omega_p \, \frac{\omega_s}{\omega_p} \frac{d \overline{N}_s}{d\omega_s} (\omega_p) I_{p,0} \int_s^\infty ds' \, \alpha_{p}(s') e^{-\tau_s(s,s')} \\
    &=&& \int d \omega_p \, \frac{\omega_s}{\omega_p}  \frac{d \overline{N}_s}{d\omega_s} (\omega_p) I_{p,0} \frac{\alpha_{p,0}}{\alpha_{s,0}} \left[ e^{-\Delta_s \alpha_{s,0} r_\mathrm{vir}} + \frac{\Delta_p}{\Delta_s} \left(1 - e^{-\Delta_s \alpha_{s,0} r_\mathrm{vir}} \right) \right] \,,
\end{alignat*}
Yet again, under the assumption of a monochromatic primary spectrum, we obtain
\begin{alignat}{1}
    \eta_s(I_p = I_{p,0}) = e^{-\Delta_s \alpha_{s,0} r_\mathrm{vir}} + \frac{\Delta_p}{\Delta_s} \left(1 - e^{-\Delta_s \alpha_{s,0} r_\mathrm{vir}} \right) \,.
\end{alignat}
Note that in general $\Delta_p \neq \Delta_s$; for example, primary electrons that undergo ICS into photoionizing photons have $\Delta_p = 1$, since ICS occurs off CMB photons which are not enhanced within a halo, but $\Delta_s = \Delta$, since photoionization occurs off neutral atoms which are enhanced in a halo. 
The limits of interest are: 
\begin{itemize}
    \item $\Delta_s \alpha_{s,0} r_\mathrm{vir} \ll 1$: this corresponds to the limit where  the halo is optically thin to secondaries. 
    We find $\eta_s(I_p = I_{p,0}) \approx 1 + \Delta_p \alpha_{s,0} r_\mathrm{vir}$, i.e. we get a potential enhancement if the primary path length is sufficiently short, due to the fact that we become dominated by primaries scattering inside the halo (and not in the homogeneous IGM), which comes with a $\Delta_p$ enhancement; 
    \item $\Delta_s \alpha_{s,0} r_\mathrm{vir} \gg 1$: the halo is optically thick to secondaries, leading to $\eta_s(I_p = I_{p,0}) \approx \Delta_p / \Delta_s$. 
    In this limit, the dominant contribution to the intensity comes from primaries scattering close to the center of the halo. 
    The scattering rate is therefore enhanced by $\Delta_p$, but suppressed by $\Delta_s$ due to the shielding that the halo overdensity provides to the secondaries. 
\end{itemize}
Another way to think about the $\Delta_p$ halo enhancement is to consider primaries with short path length. 
In this case, the assumption of a homogeneous intensity for the primaries implies that the primaries are more efficiently {\it injected} in regions of high $\Delta_p$ (where they are also more efficiently depleted). 
Since the primaries convert promptly to secondaries, the production of secondaries is similarly enhanced.

Next, we examine the case where the primaries are sourced by DM processes, i.e.\ $n = 1$ and $j_1(s \leq r_\mathrm{vir}) = \Delta^\beta j_{1,0}$, and $j_1(s > r_\mathrm{vir}) = j_{1,0}$, where $j_{1,0}$ is some constant emission coefficient. 
Substituting Eqn.~\eqref{eqn:gen_pri_intensity_expr} into Eqn.~\eqref{eqn:gen_sec_intensity_expr}, we find
\begin{alignat*}{1}
    I_2(s) = \int d \omega_1 \, \frac{\omega_2}{\omega_1} \frac{d \overline{N}_2}{d \omega_2} (\omega_1) \int_s^\infty ds' \alpha_1(s') e^{-\tau_2(s, s')} \int_{s'}^\infty ds'' j_1(s'') e^{-\tau_1(s', s'')} \,.
\end{alignat*}
The structure of this result can be understood as follows: the intensity of secondaries at $s$ is given by the sum intensity of primaries in shells of width $ds'$, multiplied by $\alpha_1$ to obtain the intensity into secondaries, and finally multiplied by the survival probability of secondaries traveling from $s'$ to $s$. 
Once again, for a monochromatic primary spectrum, we can define $\eta_2 \equiv I_2(s = 0) / I_{2,0}$, and using the fact that $I_{1,0} = j_{1,0} / \alpha_{1,0}$, we obtain
\begin{multline}
    \eta_2 = \alpha_{2,0} \left[ \int_0^{r_\mathrm{vir}} ds\, \alpha_1(s) e^{-\tau_2(0,s)} \left( \Delta^\beta \int_s^{r_\mathrm{vir}} ds' e^{-\tau_1(s, s')} + \int_{r_\mathrm{vir}}^\infty ds' \, e^{-\tau_1(s, s')} \right) \right. \\
    \left. + \int_{r_\mathrm{vir}}^\infty ds \, \alpha_1(s) e^{-\tau_2(0,s)} \int_s^\infty ds' \, e^{-\tau_1(s, s')} \right] \,.
\end{multline}
This can be evaluated with our tophat model, giving 
\begin{alignat*}{1}
    \eta_2 = e^{- \Delta_2 \alpha_{2,0} r_\mathrm{vir}} + \frac{\Delta^\beta}{\Delta_2} \left(1 - e^{-\alpha_{2,0} \Delta_2 r_\mathrm{vir}} \right) - \alpha_{2,0} (\Delta^\beta - \Delta_1) \frac{e^{- \Delta_2 \alpha_{2,0} r_\mathrm{vir}} - e^{-\Delta_1 \alpha_{1,0} r_\mathrm{vir}}}{\Delta_1 \alpha_{1,0} - \Delta_2 \alpha_{2,0}}  \,.
\end{alignat*}
Let us consider the following limits: 
\begin{itemize}
    \item $\alpha_{2,0} \to \infty$: this corresponds to a secondary with extremely short path length. 
    In this limit, the center of the halo only receives secondaries that are produced near the center.
    We find $\eta_2 = (\Delta_1 / \Delta_2) \eta_1$, i.e.\ we get an enhancement from the primary intensity itself being larger, and the fact that there can be more targets for primaries to scatter off in the halo; on the other hand, we receive a suppression due to screening of the secondaries by the dense halo; 
    \item $\alpha_{2,0} \to 0$: in this limit, secondaries have a very long path length, and the halo is optically thin. 
    One finds that also assuming $\alpha_{1,0} \to 0$, i.e.\ the halo is optically thin also to primaries, we obtain $\eta_2 = 1 + \Delta_1 \alpha_{2,0} r_\mathrm{vir}$, which is the same result as $I_s(I_p = I_{p,0})$, since $\eta_p \to 1$ as $\alpha_{p,0} \to 0$. 
    On the other hand, for $\alpha_{1,0} \to \infty$, both primaries and secondaries have a short path length. 
    We find $\eta_2 \approx 1 + \Delta^\beta \alpha_{2,0} r_\mathrm{vir}$, which is similar to the result for $\eta_1$---if the directly emitted particles have a sufficiently short path length, then $n = 2$ particles can be treated as the directly emitted particles instead. 
\end{itemize}

\subsection{Tertiaries}
\label{app:tertiaries}

Intensities of tertiaries and subsequent particles can be calculated iteratively, with increasingly more complicated integrals to perform. For our purposes, as with secondary particles, we only need to determine the intensity of tertiaries under the two assumptions of \emph{1)} the primary producing the tertiary is sourced directly by the DM process, with $j_1(s \leq r_\mathrm{vir}) = \Delta^\beta j_{1,0}$ and $j_1(s > r_\mathrm{vir}) = j_{1,0}$, and \emph{2)} the primary has constant steady-state intensity $I_p = I_{p,0}$. 

Following the same procedure as before, we can recursively obtain the intensity of tertiaries as
\begin{alignat*}{1}
    I_t(s) = \int d \omega_s \, \frac{\omega_t}{\omega_s} \frac{d \overline{N}_t}{d \omega_t} (\omega_s) \int d \omega_p \, \frac{\omega_s}{\omega_p} \frac{d \overline{N}_s}{d \omega_s} (\omega_p) \int_s^\infty ds' \, \alpha_s (s') e^{-\tau_t(s, s')}  \int_{s'}^\infty ds'' \, \alpha_p (s'') e^{-\tau_s(s', s'')} I_p(s'') \,.
\end{alignat*}
Once again, we evaluate the expected intensity in the homogeneous limit, which is
\begin{alignat*}{1}
    I_{t,0} = \int d \omega_s \, \frac{\omega_t}{\omega_s} \frac{d \overline{N}_t}{d \omega_t} (\omega_s) \int d \omega_p \, \frac{\omega_s}{\omega_p} \frac{d \overline{N}_s}{d \omega_s} (\omega_p) \frac{\alpha_{p,0}}{\alpha_{t,0}} I_{p,0} \,.
\end{alignat*}
For $n = 1$ and $j_1(s \leq r_\mathrm{vir}) = \Delta^\beta j_{1,0}$ and $j_1(s > r_\mathrm{vir}) = j_{1,0}$, assuming all cascades are monochromatic, we can again define $\eta_3 = I_3(s = 0) / I_{3,0}$ and perform the integrals over the various domains. This ultimately gives
\begin{multline}
    \eta_3 = e^{- \Delta_3 \alpha_{3,0} r_\mathrm{vir}} + \frac{\Delta^\beta}{\Delta_3} \left(1 - e^{-\Delta_3 \alpha_{3,0} r_\mathrm{vir}} \right) - \frac{\Delta_2(\Delta^\beta - \Delta_1) \alpha_{2,0} \alpha_{3,0}}{\Delta_1 \alpha_{1,0} - \Delta_1 \alpha_{1,0}} \frac{e^{-\Delta_3 \alpha_{3,0} r_\mathrm{vir}} - e^{- \Delta_1 \alpha_{1,0} r_\mathrm{vir}}}{\Delta_3 \alpha_{3,0} - \Delta_1 \alpha_{1,0}} \\ + \frac{\Delta_2 ( \Delta_1 - \Delta_2) \alpha_{2,0} \alpha_{3,0} - \Delta_1 ( \Delta^\beta - \Delta_2) \alpha_{1,0} \alpha_{3,0}}{\Delta_1 \alpha_{1,0} - \Delta_2 \alpha_{2,0}} \frac{e^{-\Delta_3 \alpha_{3,0} r_\mathrm{vir}} - e^{- \Delta_2 \alpha_{2,0} r_\mathrm{vir}}}{\Delta_2 \alpha_{2,0} - \Delta_3 \alpha_{3,0}} \,.
\end{multline}

For any $n$, in the limit where $I_p = I_{p,0}$, we can likewise define $\eta_t(I_p = I_{p,0}) \equiv I_t(s = 0) / I_{t,0}$, which is given by
\begin{multline}
    \eta_t(I_p = I_{p,0}) = e^{- \Delta_t \alpha_{t,0} r_\mathrm{vir}} + \frac{\alpha_{s,0} \Delta_p \Delta_s}{\Delta_t} \frac{1 - e^{- \Delta_t \alpha_{t,0} r_\mathrm{vir}}}{\Delta_s \alpha_{s,0} - \Delta_t \alpha_{t,0}} \\
    - \alpha_{t,0} \Delta_p \frac{1 - e^{-\Delta_s \alpha_{s,0} r_\mathrm{vir}}}{\Delta_s \alpha_{s,0} - \Delta_t \alpha_{t,0}} - \alpha_{t,0} \Delta_s \frac{e^{- \Delta_s \alpha_{s,0} r_\mathrm{vir}} - e^{- \Delta_t \alpha_{t,0} r_\mathrm{vir}}}{\Delta_s \alpha_{s,0} - \Delta_t \alpha_{t,0}} \,.
\end{multline}
This result is finite as $\Delta_s \alpha_{s,0} \to \Delta_t \alpha_{t,0}$. 
In the limit where $\alpha_{s,0} \to \infty$, i.e.\ the secondaries have an extremely short path length, we find $\eta_t \approx e^{-\Delta_t \alpha_{t,0} r_\mathrm{vir}} + (\Delta_p / \Delta_t) (1 - e^{- \Delta_t \alpha_{t,0} r_\mathrm{vir}})$. 
Comparing this with the result for the intensity of a daughter particle originating from a mother particle with homogeneous intensity, this result shows that we can simply skip the secondaries step in the cascade if the path length is sufficiently short, which matches our intuitive expectations.

\subsection{Effect of the Halo}
\label{app:effect_of_the_halo}

%
\begin{figure}
    \centering
    \includegraphics[scale=0.5]{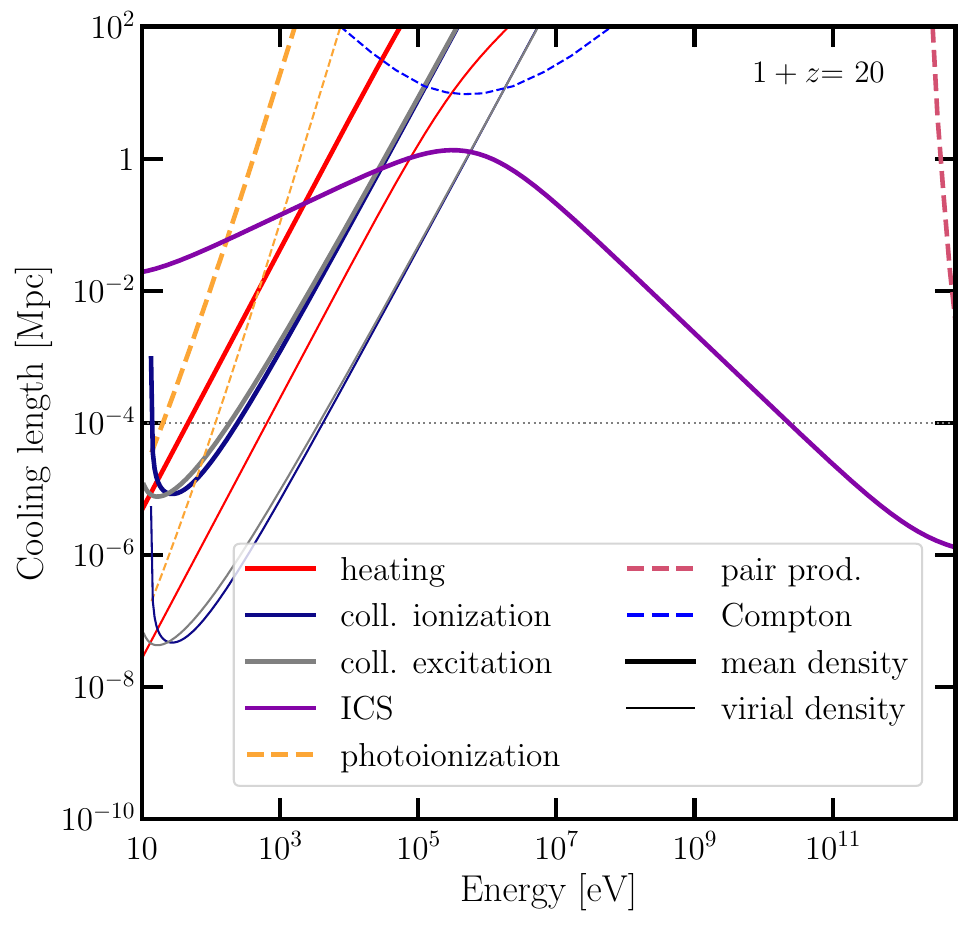}
    \caption{
    Path lengths for particles to lose a significant amount of energy by various processes.
    For cooling lengths that depend on the gas density, we show results for the mean density in solid lines and virialized halo density in dashed lines.
    The horizontal dotted line marks 0.1 kpc, which is the approximate size of $r_\mathrm{vir}$ for halos of mass $10^6 M_\odot$ virializing at $1+z=20$.
    }
    \label{fig:cooling_length}
\end{figure}
\begin{figure}
    \centering
    \includegraphics[scale=0.48]{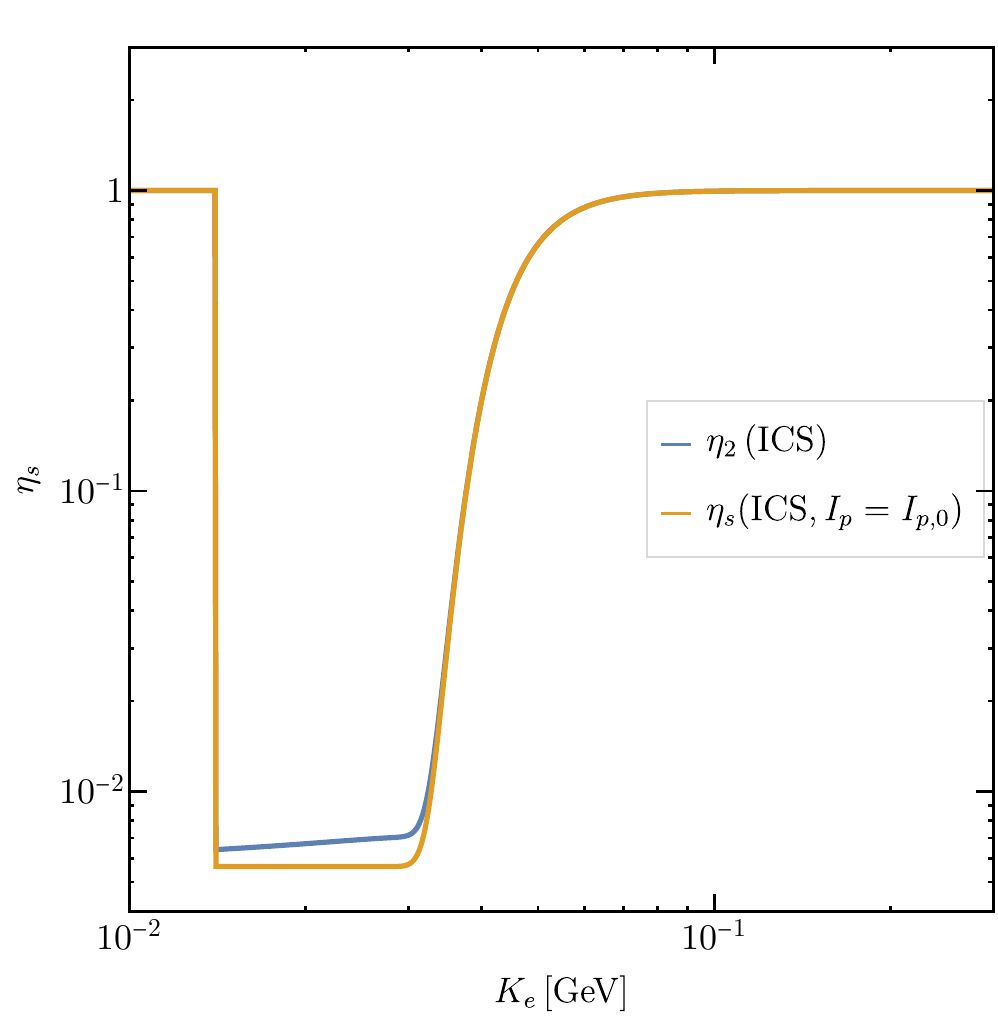}
    \includegraphics[scale=0.48]{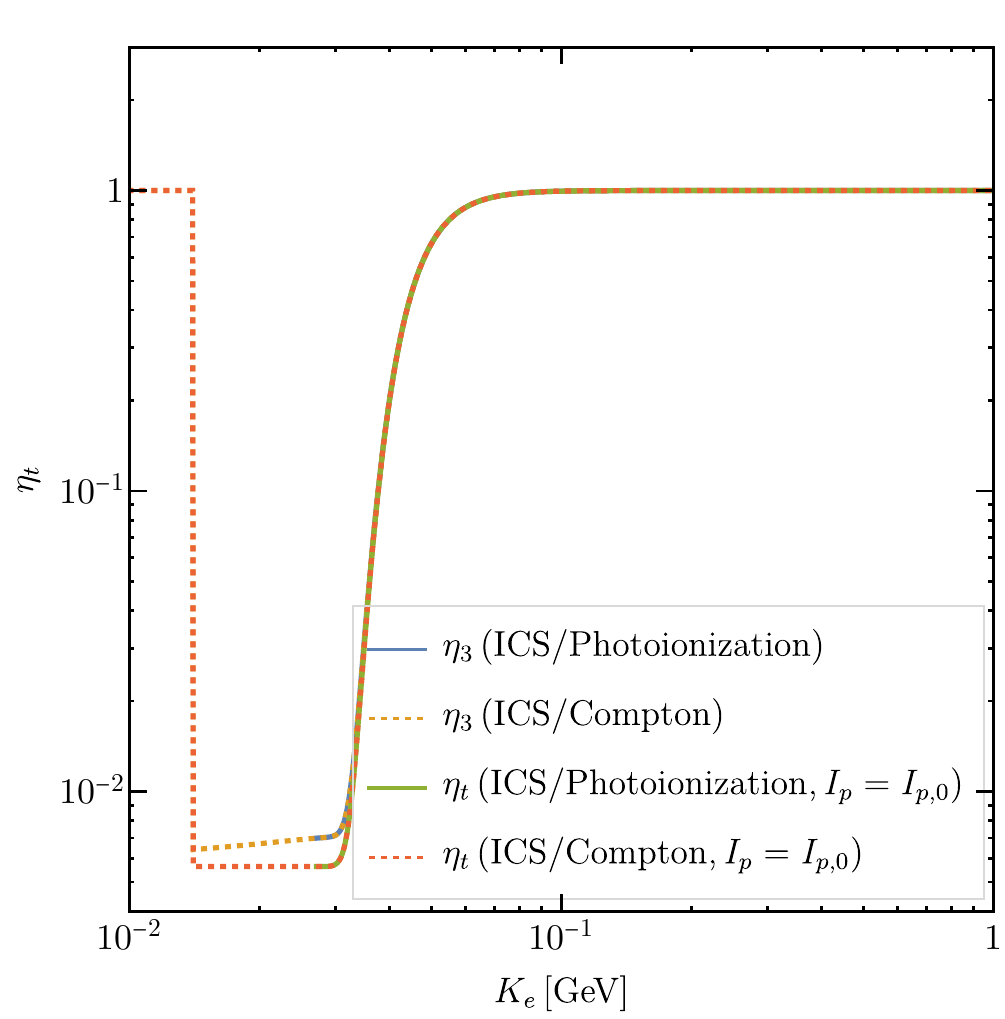} \\
    \includegraphics[scale=0.48]{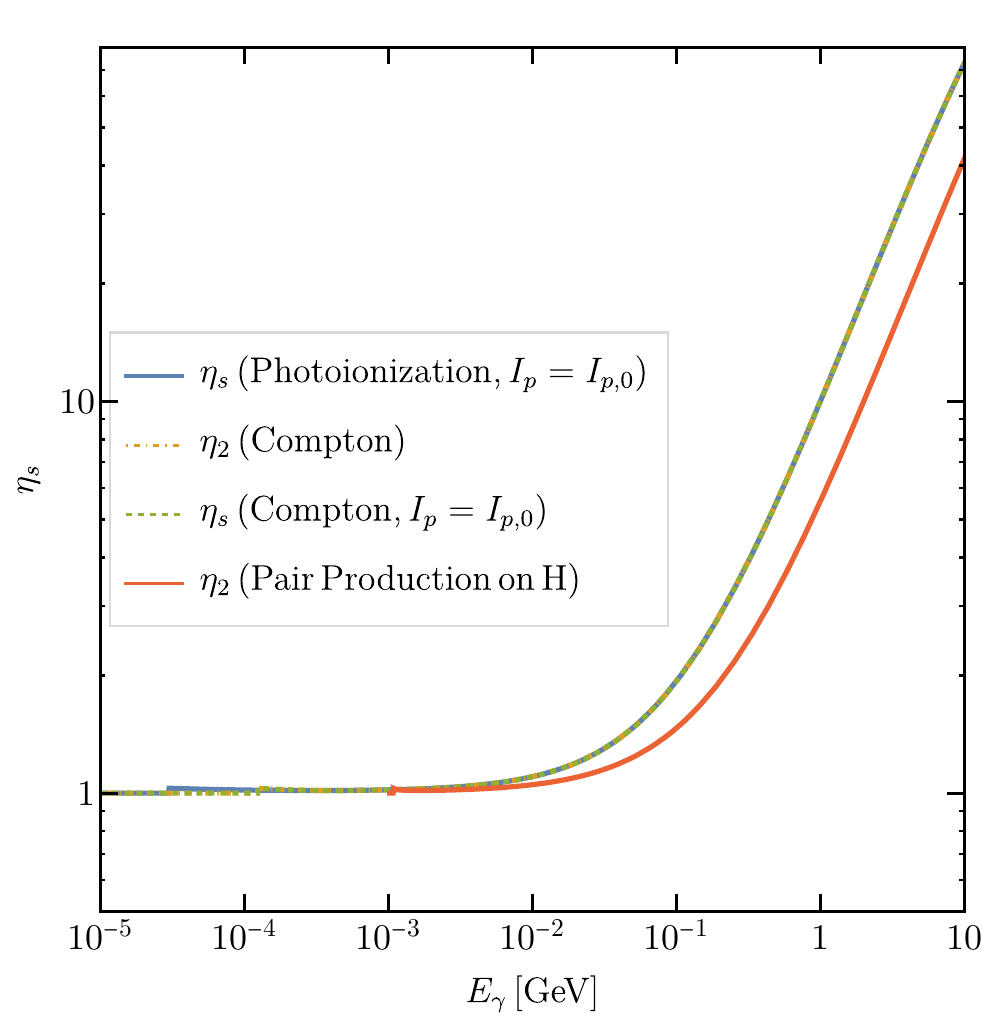}
    \includegraphics[scale=0.48]{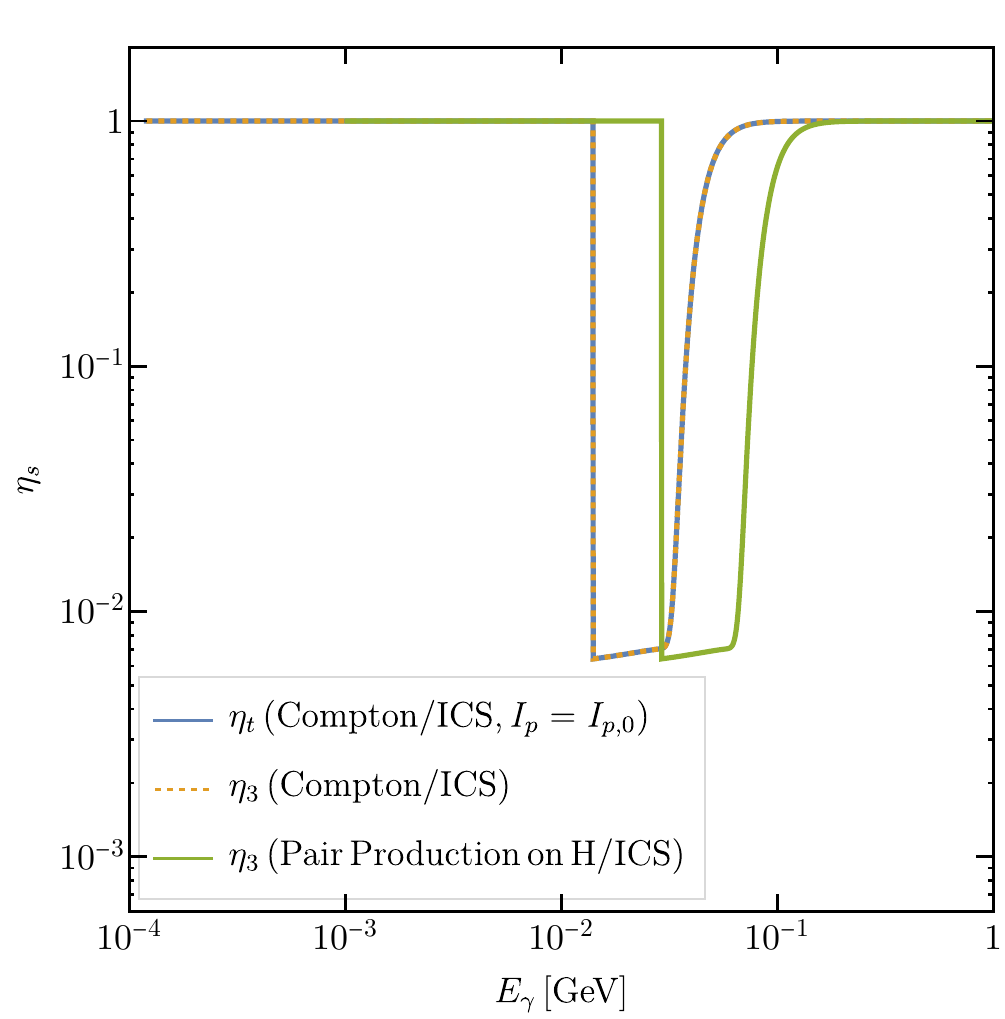} 
    \caption{
    \footnotesize
    The halo enhancement factor calculated for DM decay, under different assumptions as a function of primary particle energy, including \emph{(top left)} $\eta_2$ for secondaries produced by the ICS of electrons sourced by DM decay (blue), or $\eta_s(I_p = I_{p,0})$ for secondaries by primaries with the homogeneous, steady-state intensity (orange); \emph{(top right)} $\eta_3$ for tertiaries produced by ICS of electrons sourced by DM decay followed by photoionization of the secondary photons (blue), or Compton scattering (dashed orange), as well as $\eta_t(I_p = I_{p,0})$ for tertiaries produced by ICS of electrons with the homogeneous, steady-state intensity followed by photoionization (green), or Compton scattering (dashed red); \emph{(bottom left)} $\eta_s(I_p = I_{p,0})$ for secondaries produced by photoionization with the homogeneous, steady-state primaries (blue), $\eta_2$ for Compton scattering of photons sourced by DM decay (orange, dot-dashed), $\eta_s(I_p = I_{p,0})$ for Compton scattering of primary photons with the homogeneous, steady-state intensity (green, dashed), and $\eta_2$ for pair production of photons sourced by DM decay (red), and \emph{(bottom right)} $\eta_t(I_p = I_{p,0})$ for primary photons with the homogeneous, steady-state intensity undergoing Compton scattering, producing electrons that undergo ICS (blue), and $\eta_3$ for photons from DM decays undergoing either Compton scattering (orange, dashed) or pair production on neutral hydrogen (green), followed by ICS of the secondary electrons.
    }
    \label{fig:eta}
\end{figure}

We are now ready to examine the effect of the enhanced density in the halo on energy deposition with in the halo. 
We focus only on decaying DM in this section, since this is the process that has largest impact on star formation given existing experimental constraints.
First, we note that the energy deposited per volume per time $(dE / dV \, dt)_\mathrm{dep} \propto \int d \omega_f \alpha_f I_f$, where $f$ labels the last step in the cascade in the halo. 
Therefore, if the halo receives the same intensity as in the homogeneous limit, the energy deposited per volume per time increases by a factor of $\Delta_f$ relative to the homogeneous limit, where $\Delta_f$ is the overdensity of targets for the last step of the cascade, since $\alpha_f = \Delta_f \alpha_{f,0}$, where $\alpha_{f,0}$ is the extinction coefficient of the last step in the cascade in the homogeneous limit. 
Equivalently, the energy deposited \emph{per baryon} per time is enhanced by $\Delta_f / \Delta$, which is 1 for the final step in all cascades, since ionization and heating occurs through scattering with atoms or free electrons. In other words, receiving a homogeneous intensity at the center of the halo typically implies the same energy deposited per baryon per time as in the homogeneous limit. 

To determine what effect the halo has on the intensity of the particles in the final step, we estimate $\eta_f = I_f(s = 0) / I_{f,0}$ by obtaining $\eta$ for some of the intermediate steps, using either our ability to calculate $\eta_2$ for secondaries and $\eta_3$ for tertiaries from particles emitted directly from the DM process, or by making use of $\eta_n \approx 1$ for some intermediate step in the cascade, allowing us to calculate $\eta_{n+1}$ or $\eta_{n+2}$.
We present results for $1+z = 20$, which is the more experimentally accessible redshift, and for $M_\text{halo} = 10^6 M_\odot$, which is close to the critical halo mass at that redshift including DM effects (see Fig.~\ref{fig:critical_collapse_Mhalo}). 
Our results are relatively insensitive to halo masses within an order of magnitude of $M_\text{halo} = 10^6 M_\odot$, since the relevant parameter is $r_\text{vir} \propto M_\text{halo}^{1/3}$.

The extinction coefficients for the cooling processes in the cascade of $e^+e^-$ pairs and photons are given by the inverse of the energy loss path length, i.e. $\alpha = (1/v) (d \log E/ dt)$. 
The energy loss path lengths $\alpha^{-1}$ for relevant processes are shown in Fig.~\ref{fig:cooling_length} for $1+z = 20$ and $1+z = 100$, and are discussed and derived in detail in Refs.~\cite{Slatyer:2009yq,Liu:2023fgu}.

Tables~\ref{tab:cascade_elec} and~\ref{tab:cascade_phot} shows the approximate monochromatic cascade produced by an electron or positron, and a photon respectively, as a function of initial energy. 
Each row corresponds to a range of energies over which the cascade goes through the same processes and intermediate states, differing only in energy of the intermediate states. 
The dominant process taking particles from step $n$ to step $n+1$ in the cascade are shown under $n \to n+1$. 
The final step by which energy is deposited directly into ionization, heating or low-energy photons is shown in bold.

In order to obtain $\eta_f$ for all relevant final states, we need the following results, summarized in Fig.~\ref{fig:eta}: 
\begin{itemize}
    \item $\eta_2$ and $\eta_s(I_p = I_{p,0})$ for secondary photons produced by primary electrons undergoing ICS, assumed to either be sourced by DM decay for $\eta_2$, or to have the homogeneous, steady-state intensity, $I_p = I_{p,0}$ for $\eta_s(I_p = I_{p,0})$ (Fig.~\ref{fig:eta} top left); 
    \item $\eta_3$ and $\eta_t(I_p = I_{p,0})$ for tertiary electrons of primary electrons undergoing ICS, producing secondary photons, which subsequently produce tertiary electrons through either photoionization or Compton scattering (Fig.~\ref{fig:eta} top right); 
    \item $\eta_s(I_p = I_{p,0})$ for secondary electrons produced by primary photons undergoing photoionization or Compton scattering, and $\eta_2$ for secondary electrons produced by Compton scattering or pair production on neutral hydrogen of primary photons sourced by DM decay (Fig.~\ref{fig:eta} bottom left), and
    \item $\eta_t(I_p = I_{p,0})$ for tertiary photons produced by the ICS of secondary electrons, which is in turn produced by  primary photons undergoing Compton scattering, as well as $\eta_3$ for tertiary photons produced by the ICS of secondary electrons, coming from primary photons sourced by DM decay undergoing pair production (Fig.~\ref{fig:eta} bottom right). 
\end{itemize}

\renewcommand{\arraystretch}{1.8}

\begin{table*}[!t]
\footnotesize
\begin{center}
\begin{tabular}{C{2cm}|C{1cm}C{2cm}C{2.2cm}C{2cm}C{1.5cm}C{1.5cm}C{1.6cm}}
\textbf{Electron Kinetic Energy} & $1 \to 2$  & $2$ & $2 \to 3$ & $3$ & $3 \to 4$ & $4$ & $4 \to 5$ \\
\Xhline{1\arrayrulewidth}
\SIrange{1}{14}{\mega\eV} & ICS & $< \SI{13.6}{\eV}$ $\gamma$ \tiny{\textcolor{red}{(ICS,2)}} & \bf{No Ionization /Heating} & -- & --  &-- &--\\
\SIrange{14}{60}{\mega\eV} & ICS & \SIrange{13.6}{230}{\eV} $\gamma$ & Photoionization & \SIrange{0}{215}{\eV} $e^-$ \tiny{\textcolor{red}{(ICS,3)}} & \textcolor{blue}{$e^-$ \bf{Atomic}} &-- &--\\
\SIrange{60}{350}{\mega\eV} & ICS & \SIrange{0.23}{8}{\kilo\eV} $\gamma$ & Photoionization & \SIrange{0.215}{8}{\kilo\eV} $e^-$ \tiny{\textcolor{red}{(ICS,3)}} & $e^-$ \bf{Atomic} &-- &-- \\
\SIrange{0.35}{1.37}{\giga\eV} & ICS & \SIrange{8}{120}{\kilo\eV} $\gamma$ & Compton & \SIrange{0.125}{30}{\kilo\eV} $e^-$ \tiny{\textcolor{red}{(ICS,3)}} & $e^-$ \bf{Atomic} &-- &--\\
\SIrange{1.37}{10}{\giga\eV} & ICS & \SIrange{0.12}{6.4}{\mega\eV} $\gamma$ \tiny{\textcolor{red}{(ICS,2)}} & Compton & \SIrange{0.03}{1.8}{\mega\eV} $e^-$ & ICS & $<$ \SI{13.6}{\eV} $\gamma$ \tiny{\textcolor{red}{($\gamma$,t)}}& \bf{No Ionization /Heating} \\
\end{tabular}
\end{center}
\caption{Dominant cascade for primary electrons/positrons with energies between \SI{1}{\mega\eV} and \SI{10}{\giga\eV}. The steps in the cascade and the dominant process between steps are shown in each column. The channel through which energy is deposited in the last step of the cascade is written in bold. The relevant method to calculate $\eta$ and determine the intensity at each step in the cascade is shown in red, with `ICS' and `$\gamma$' referring to the electron ICS and photon $\eta$ results, `2' and `3' indicating the use of $\eta_2$ and $\eta_3$, while `$s$' and `$t$' indicating the use of $\eta_s(I_p = I_{p,0})$ and $\eta_t(I_p = I_{p,0})$ respectively. Final intensities which are suppressed relative to the homogeneous intensity are shown in blue.}
\label{tab:cascade_elec}
\end{table*}    

With this information, we can explain the estimate for $\eta_f$ for each type of cascade as a function of the primary particle energy.

\vspace{2mm}

\underline{$\chi \to e^+e^-$}:

\begin{itemize}
    \item \SIrange{1}{14}{\mega\eV}: $e^+e^-$ pairs mainly undergo ICS into sub-\SI{13.6}{\eV} photons with long path lengths,\footnote{\SIrange{10.2}{13.6}{\eV} photons scatter rapidly, but elastically, which we consider as having a long path length.} giving $\eta_f \approx 1$ (see Fig.~\ref{fig:eta} top left). 
    \item \SIrange{14}{60}{\mega\eV}: The primary particles undergo ICS into photons just above the ionization threshold of hydrogen, and have very short path lengths. These photons then produce low-energy electrons, which again undergo collisional ionization with short path lengths. By considering $\eta_3$ for ICS (Fig.~\ref{fig:eta} top right), we find that $\eta_f \approx 1/\Delta$;
    \item \SIrange{60}{350}{\mega\eV}: This energy range results in the same cascade as the previous range, but the photoionizing photons are of sufficiently high energy that their path lengths are much longer than the virial radius of the halo. We find $\eta_3 \approx 1$ for ICS into photons, which then photoionize neutral atoms to produce the final, tertiary final low-energy electrons (Fig.~\ref{fig:eta} top right);
    \item \SIrange{0.35}{1.37}{\giga\eV}: The photons from ICS now cool mainly by Compton scattering, and not photoionization, but $\eta_f \approx 1$ remains true in this regime, again by considering $\eta_3$ of for primary electrons undergoing ICS, producing photons that Compton cool (Fig.~\ref{fig:eta} top right);
    \item \SIrange{1.37}{10}{\giga\eV}: For this energy range, the cascade becomes longer, with the secondary photons Compton scattering into electrons that predominantly undergo ICS instead of atomic processes. We use the fact that $\eta_2 \approx 1$ for ICS into photons to show that these secondary photons have the homogeneous, steady-state intensity (Fig.~\ref{fig:eta} top left), and then use $\eta_t(I_p = I_{p,0})$ for photons that undergo Compton cooling to electrons that cool mainly via ICS to show that ultimately, $\eta_f \approx 1$ (Fig.~\ref{fig:eta} bottom right). 
\end{itemize}

\begin{table*}[!t]
\scriptsize
\begin{center}
\begin{tabular}{C{1.4cm}|C{1.3cm}C{1.8cm}C{.7cm}C{1.5cm}C{1.2cm}C{1.7cm}C{.8cm}C{1.2cm}C{1.4cm}}
\textbf{Photon Energy} & $1 \to 2$  & $2$ & $2 \to 3$ & $3$ & $3 \to 4$ & $4$ & $4 \to 5$ & $5$ & $5 \to 6$ \\
\Xhline{1\arrayrulewidth}
\SIrange{10}{120}{\kilo\eV} & Compton & \SIrange{0.125}{30}{\kilo\eV} $e^-$ \tiny{\textcolor{red}{($\gamma$,2)}} & $e^-$ \bf{Atomic} & -- & --  &-- &--\\
\SIrange{0.12}{14}{\mega\eV} & Compton & \SIrange{0.03}{14}{\mega\eV} $e^-$ & ICS & $<$\SI{13.6}{\eV} $\gamma$ \tiny{\textcolor{red}{($\gamma$,3)}} & \bf{No Ionization /Heating} &-- &-- \\
\SIrange{14}{60}{\mega\eV} & Compton & \SIrange{14}{60}{\mega\eV} $e^-$ \tiny{\textcolor{red}{($\gamma$,2)}} & ICS & \SIrange{13.6}{230}{\eV} $\gamma$ & Photo- ionization & \SIrange{0}{215}{\eV} $e^-$ \tiny{\textcolor{red}{(ICS,t)}} & \textcolor{blue}{$e^-$ \bf{Atomic}} \\
\SIrange{60}{120}{\mega\eV} & H Pair Production & \SIrange{30}{60}{\mega\eV} $e^-$ \tiny{\textcolor{red}{($\gamma$,2)}} & ICS & \SIrange{58}{230}{\eV} $\gamma$ & Photo- ionization & \SIrange{43}{215}{\eV} $e^-$ \tiny{\textcolor{red}{(ICS,t)}} & \textcolor{blue}{$e^-$ \bf{Atomic}} \\
\SIrange{120}{700}{\mega\eV} & H Pair Production & \SIrange{60}{350}{\mega\eV} $e^-$ & ICS & \SIrange{0.145}{8}{\kilo\eV} $\gamma$ \tiny{\textcolor{red}{($\gamma$,3)}} & Photo- ionization & \SIrange{0.13}{8}{\kilo\eV} $e^-$ \tiny{\textcolor{red}{($\gamma$,s)}} & $e^-$ \bf{Atomic} \\
\SIrange{0.7}{2.8}{\giga\eV} & H Pair Production & \SIrange{0.35}{1.4}{\giga\eV} $e^-$ & ICS & \SIrange{8}{120}{\kilo\eV} $\gamma$ \tiny{\textcolor{red}{($\gamma$,3)}} & Compton & \SIrange{0.125}{30}{\kilo\eV} $e^-$ \tiny{\textcolor{red}{($\gamma$,s)}} & $e^-$ \bf{Atomic} \\
\SIrange{2.8}{10}{\giga\eV} & H Pair Production & \SIrange{1.4}{5}{\giga\eV} $e^-$ & ICS & \SIrange{120}{450}{\kilo\eV} $\gamma$ \tiny{\textcolor{red}{($\gamma$,3)}} & Compton & \SIrange{30}{400}{\kilo\eV} $e^-$ & ICS & $<$ \SI{10.2}{\eV} $\gamma$ \tiny{\textcolor{red}{($\gamma$,t)}} & \bf{No Ionization /Heating}\\
\end{tabular}
\end{center}
\caption{Dominant cascade for primary photons with energies between \SI{1}{\mega\eV} and \SI{10}{\giga\eV}. The steps in the cascade and the dominant process between steps are shown in each column. The channel through which energy is deposited in the last step of the cascade is written in bold. The relevant method to calculate $\eta$ and determine the intensity at each step in the cascade is shown in red, with `ICS' and `$\gamma$' referring to electron ICS and photon $\eta$ results, `2' and `3' indicating the use of $\eta_2$ and $\eta_3$, and `$s$' and `$t$' indicating the use of $\eta_s(I_p = I_{p,0})$ and $\eta_t(I_p = I_{p,0})$ respectively. Final intensities which are suppressed relative to the homogeneous intensity are shown in blue.}
\label{tab:cascade_phot}
\end{table*}    

Note that our benchmark Models $\bullet$\, and $\star$\, directly emit \SI{92}{\mega\eV} electrons and positrons, and therefore have $\eta_f \approx 1$.

\vspace{2mm}

\underline{$\chi \to \gamma \gamma$}:

\begin{itemize}
    \item \SIrange{10}{120}{\kilo\eV}: The primary photons Compton scatter to produce low-energy electrons, which also rapidly lose all their energy through atomic processes. We find $\eta_f \approx 1$ by calculating $\eta_2$ of primary photons which Compton scatter (Fig.~\ref{fig:eta} bottom left), finding that the enhancement of $\Delta$ due to enhanced production of primaries in the halo is exactly canceled by the shielding of secondaries due to the same enhancement in density; 
    \item \SIrange{0.12}{14}{\mega\eV}: Primary photons Compton scatter once again, but produce electrons that predominantly cool via ICS into sub-\SI{13.6}{\eV} photons. $\eta_f \approx 1$ by considering $\eta_t$ for photons that Compton scatter (Fig.~\ref{fig:eta} bottom right); 
    \item \SIrange{14}{60}{\mega\eV}: Once again, we have Compton scattering into electrons ($\eta_2 \approx 1$ from Fig.~\ref{fig:eta} bottom left, so these electrons have homogeneous intensity) that ICS into photons, that are now just above the hydrogen ionization threshold; they photoionize hydrogen, producing low-energy electrons with a short path length. We find $\eta_f \approx 1 / \Delta$ by looking at $\eta_t(I_p = I_{p,0})$ for electrons undergoing ICS and subsequent photoionization (Fig.~\ref{fig:eta} bottom right); 
    \item \SIrange{60}{120}{\mega\eV}: Similar to the above, except the first step is pair production on neutral hydrogen of the primary photons. $\eta_f \approx 1 / \Delta$; 
    \item \SIrange{120}{700}{\mega\eV}: Similar to the above, except that the photoionizing photons in the 3rd step of the cascade have a long path length. Because of this, $\eta_f \approx 1$. We can deduce this by starting with $\eta_3$ for photons to show that the intensity of tertiary photons is the homogeneous intensity (Fig.~\ref{fig:eta} bottom right), and then looking at $\eta_s(I_p = I_{p,0})$ for photoionizing photons (Fig.~\ref{fig:eta} bottom left); 
    \item \SIrange{0.7}{2.8}{\giga\eV}: Similar to the above, except that the tertiary photons cool by Compton scattering, which also has a long path length, leading to $\eta_f \approx 1$ as before; 
    \item \SIrange{2.8}{10}{\giga\eV}: Finally, in this energy range, photons undergo pair production on neutral hydrogen, and follows the same cascade as \SIrange{1.4}{5}{\giga\eV} electrons do as described above. We find that $\eta_f \approx 1$ by using $\eta_3 \approx 1$ of Compton scattering photons going into electrons that ICS, and then applying $\eta_t(I_p = I_{p,0})$ on these tertiary photons (which produces electrons that again ICS into $<\SI{13.6}{\eV}$ photons) (see Fig.~\ref{fig:eta} bottom right). 
\end{itemize}

In summary, we find that $\eta_f \approx 1$ across most of the parameter space of interest for DM decays into $e^+e^-$ and $\gamma \gamma$, except for $\SI{28}{\mega\eV} \lesssim m_\chi \lesssim \SI{120}{\mega\eV}$ for $\chi \to e^+e^-$, and $\SI{28}{\mega\eV} \lesssim m_\chi \lesssim \SI{240}{\mega\eV}$ for $\chi \to \gamma \gamma$, where $\eta_f \approx 1 / \Delta$. 
This implies a reduction of $1/\Delta$ with respect to the homogeneous $f_c(z)$ calculated in \texttt{DarkHistory} for these narrow ranges of parameter space. 
One way to summarize the physical origin of this reduction is that in all these cases, near the end of the cascade, homogenized electrons with a relatively long path length undergo ICS to produce photons that promptly deposit their energy. 
The production of these photons is not enhanced by the presence of the halo, and their energy must be divided between the larger density of halo particles, leading to the observed suppression in deposited power per particle. 
\emph{In all other regimes, we have argued that $\eta_f \approx 1$ implies $f_c / n_\mathrm{H}$ in both the IGM and the halo are approximately equal, justifying our approach in Section~\ref{sec:first_stars}.}

Although the same formalism applies to annihilation, we do not perform the same in-depth analysis, since the potential effect of DM annihilation on the formation of the first stars appears to be strongly constrained by existing experimental probes such as the CMB power spectrum. 
We note briefly that in the case of annihilation, we can get both enhancement and suppression of the intensity of particles at the final step, likely violating the assumption that the same energy is deposited per baryon per time in the halo as in the homogeneous IGM over a broader swathe of parameter space. 

\begin{figure}
    \centering
    \includegraphics[scale=0.5]{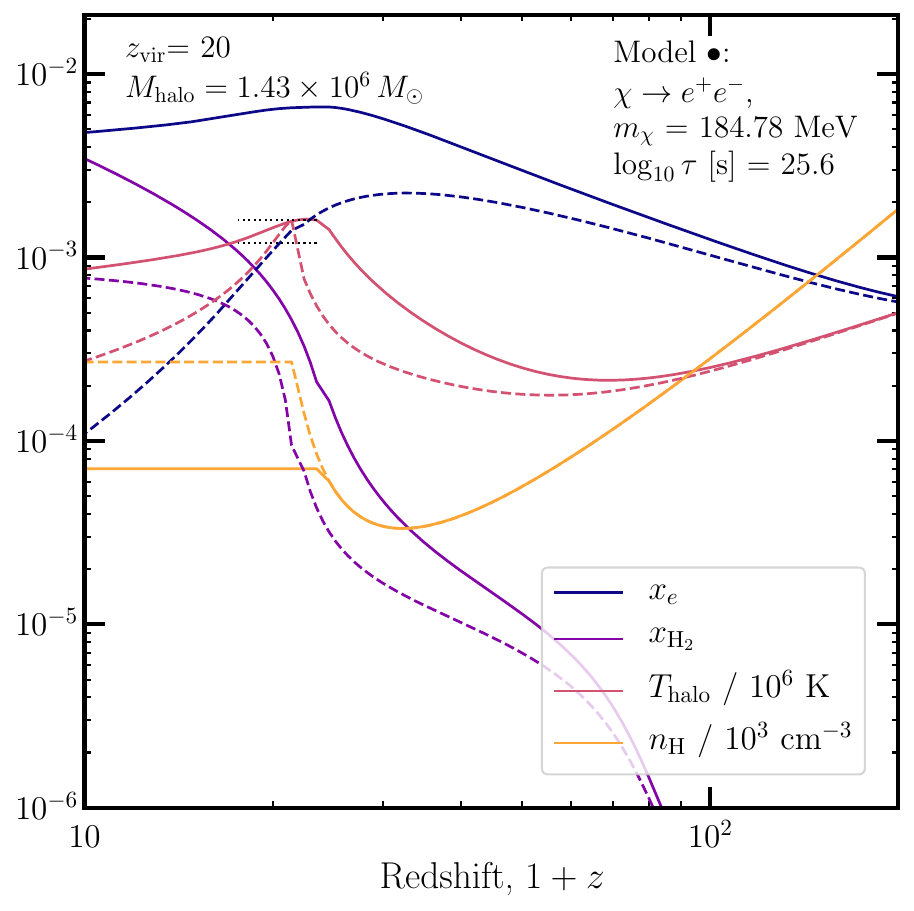}
    \hspace{5mm}
    \includegraphics[scale=0.5]{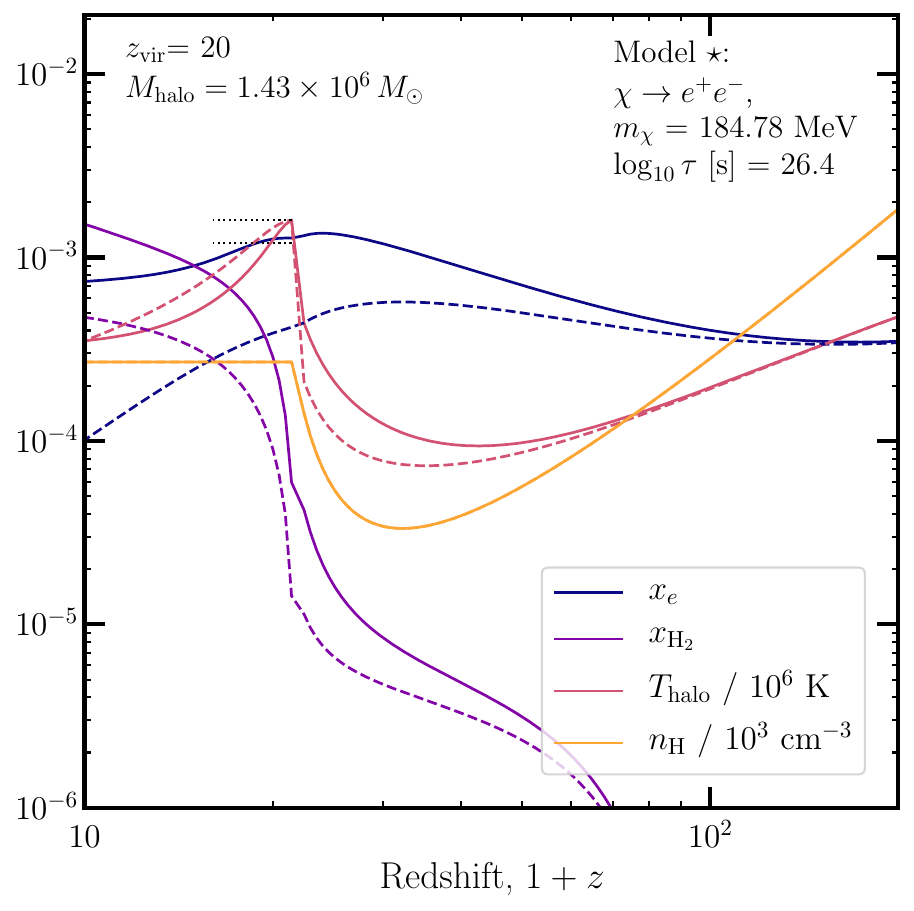}
    \caption{
    Halo evolution for Model $\bullet$\, (left) and Model $\star$\, (right), with the same halo as in Fig.~\ref{fig:halo_examples}.
    Solid lines indicate results assuming the same $f_c$'s as calculated in the IGM; dashed lines show results where the $f_c$'s are suppressed by a factor of the halo number density, i.e. the worst case scenario for energy deposition.
    The horizontal line segments on each panel indicate $T_\mathrm{vir}$ and $0.75 T_\mathrm{vir}$, and span the redshift range $(z_\mathrm{vir}, 0.75 z_\mathrm{vir})$; if the temperature curve crosses the lower line segment after virialization, the halo passes the criterion for collapsing and forming stars.
    }
    \label{fig:halo_f_suppressed}
\end{figure}

Fig.~\ref{fig:halo_f_suppressed} shows the difference in the halo evolution when we assume that the $f_c$'s are the same as in the IGM, or when we assume the $f_c$'s is suppressed by an additional factor of the overdensity $\Delta$.
For Model $\bullet$, this mild suppression of energy deposition reduces the effect of heating, such that the halo is no longer pressure supported when it virializes and is able to reach the virial density.
After virialization, the reduced heating rate means the halo cools much faster and is now able to pass the star-forming condition.
For Model $\star$, where the cooling rate of the halo is enhanced compared to the case with no exotic injection, reducing energy deposition causes the halo to cool more slowly.

From examining these two models, we can infer the ways in which Figs.~\ref{fig:scans_z20} and \ref{fig:scans_z100} would change if the $f_c$'s were suppressed for certain dark matter masses.
Overall, the contours would shift towards smaller lifetimes/larger cross-sections, since in order to have the same effect on a halo, one would have to dial up the rate of energy injection to counteract the reduced deposition rate.

\section{Bracketing Lyman-Werner effects for other channels}
\label{app:LW}

%
\begin{figure*}
    \includegraphics[scale=0.4]{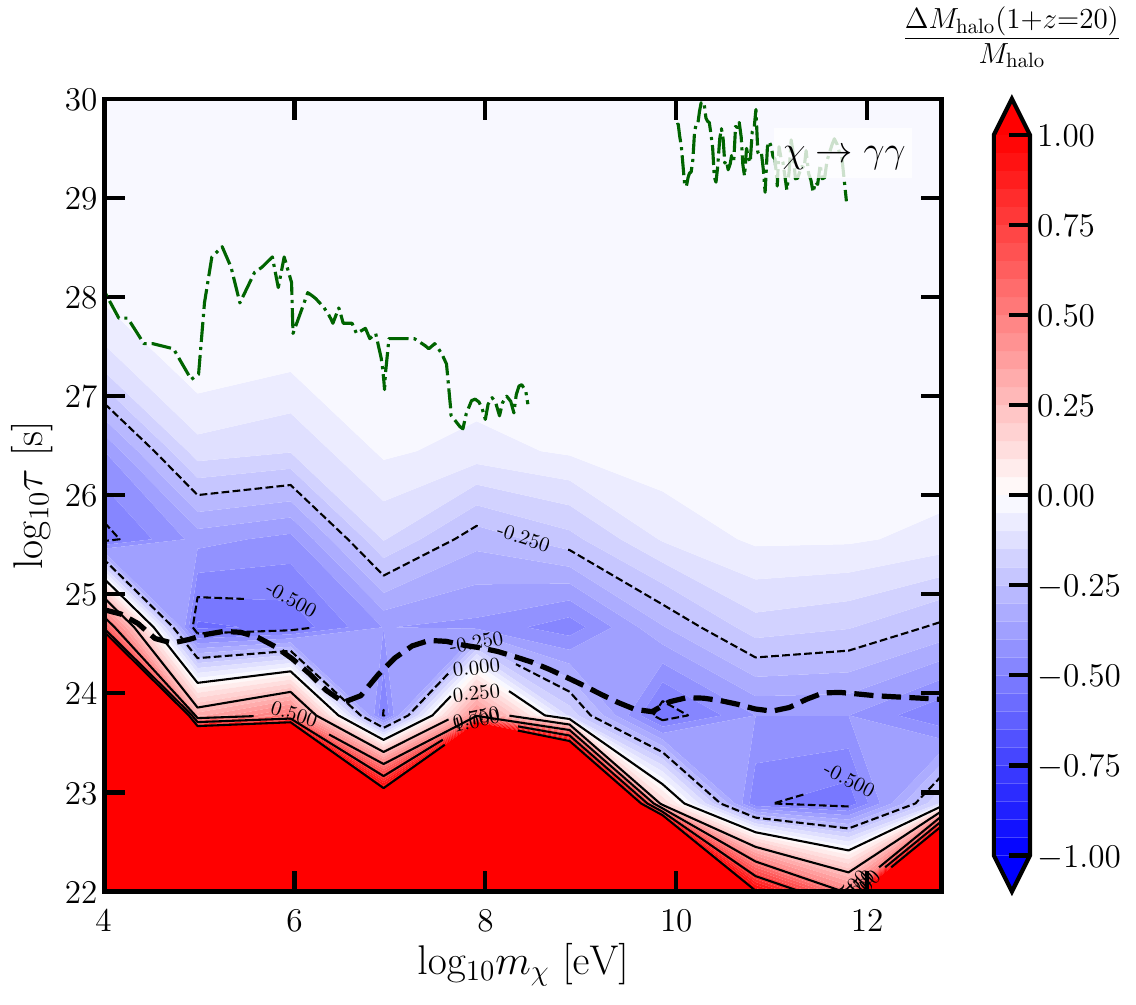}
    \includegraphics[scale=0.4]{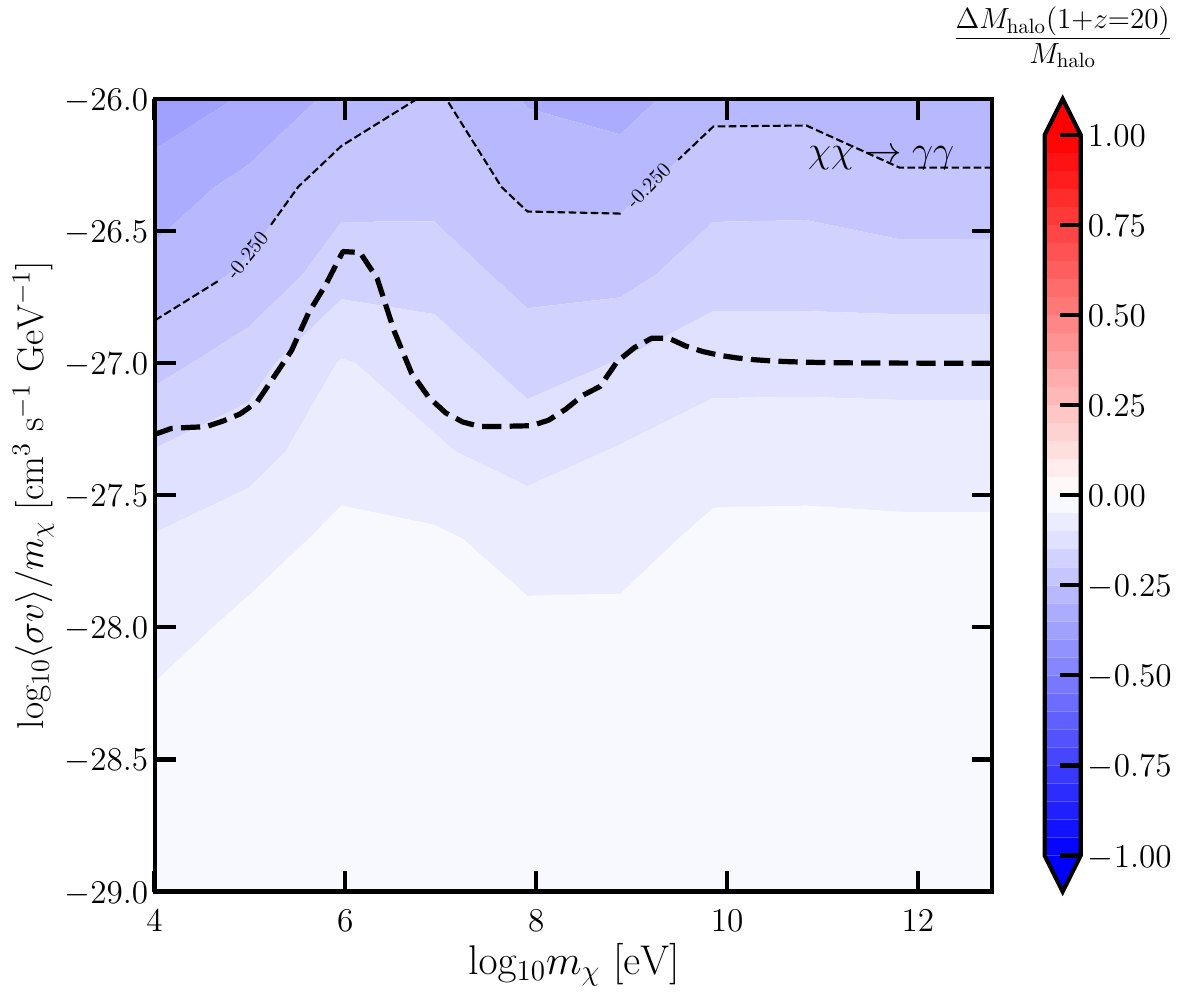}
    \includegraphics[scale=0.4]{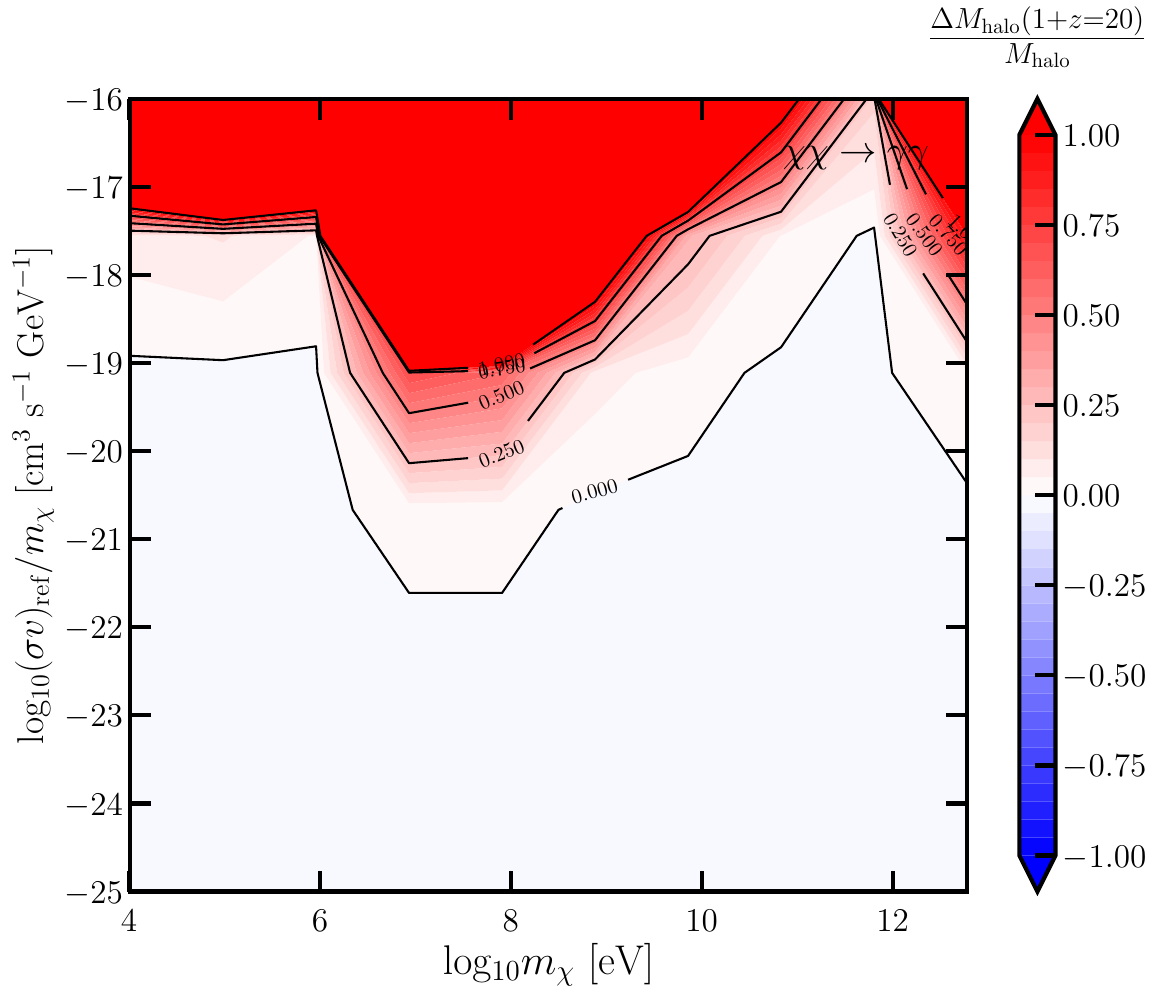}
    \includegraphics[scale=0.4]{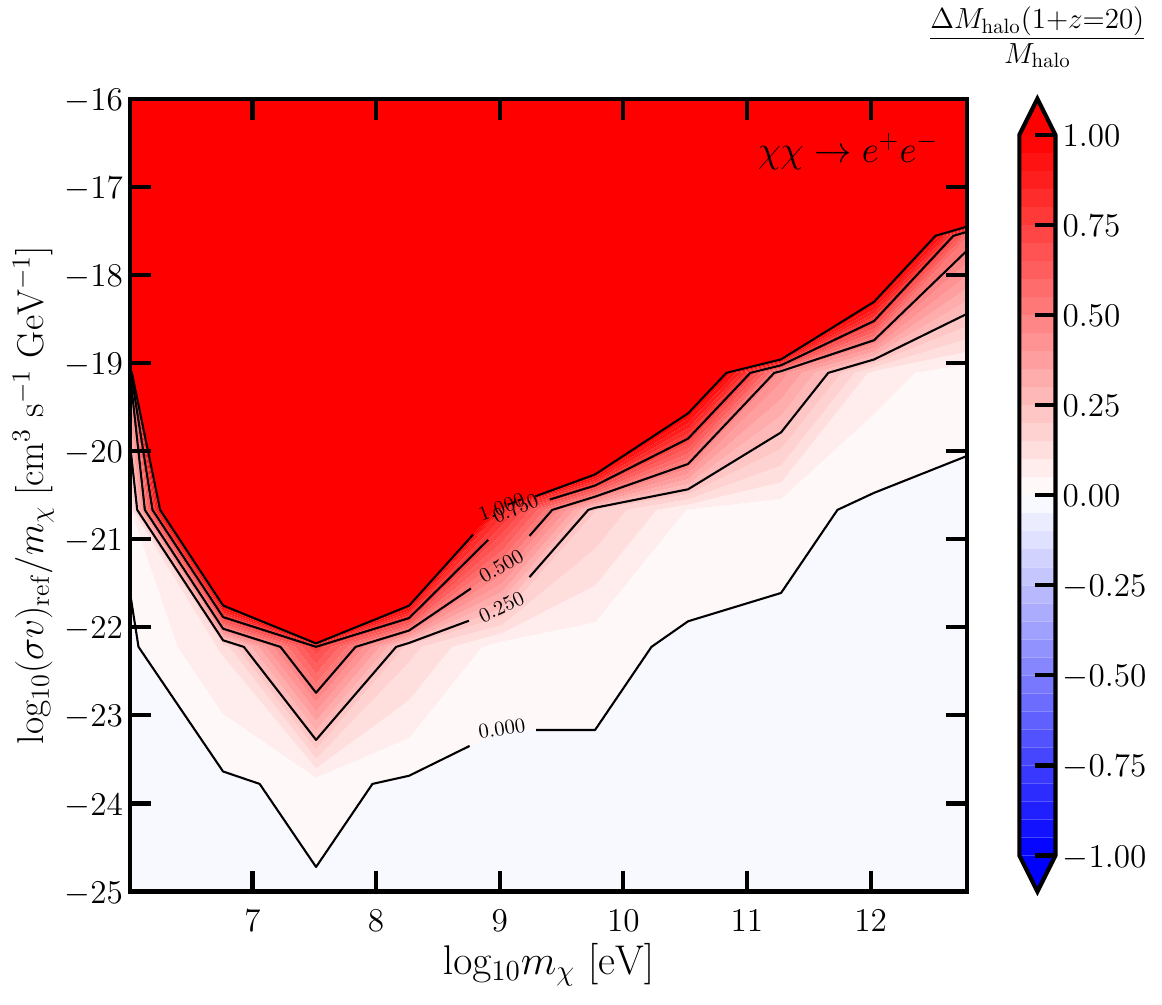}
    \caption{
    Same as Fig.~\ref{fig:scans_z20}, but assuming inefficient H$_2$ self-shielding.
    Clockwise from the top left, each panel shows the parameter space for decay to photons, $s$-wave annihilation to photons, $p$-wave annihilation to $e^+ e^-$ pairs, and $p$-wave annihilation to photons.
    }
    \label{fig:other_scans_z20_LW}
\end{figure*}

In Sec.~\ref{sec:shielding}, we showed the changes to the critical mass threshold in the parameter space for decay and $s$-wave annihilation to $e^+ e^-$ pairs, when self-shielding of the halo is inefficient.
Here, we discuss the other channels.
Fig.~\ref{fig:other_scans_z20_LW} shows the same results as in Fig.~\ref{fig:scans_z20_LW}, but for decay to photons, $s$-wave annihilation to photons, and $p$-wave annihilation to both $e^+ e^-$ pairs and photons.

For decay to photons, the depth of the blue contours is slightly reduced relative to the results assuming strong self-shielding, with the largest differences reaching to about 20\%.
The $s$-wave annihilation to photon results are only marginally impacted by self-shielding assumptions and differ by less than a percent in most of the parameter space shown.

For both $p$-wave annihilation channels, the depth of the red contours is significantly increased such that there is much more parameter space where we would likely see a significant delay to star formation.
We see that that $p$-wave to $e^+ e^-$ results are especially enhanced at masses of tens of MeV.
This is for the same reason as discussed in Sec.~\ref{sec:shielding} for the $s$-wave results; the primary electrons are injected at the right energy to upscatter CMB photons through ICS into the LW band.
However, as mentioned in Sec.~\ref{sec:results}, our assumption that energy deposition is very similar between the IGM and the halo is most likely to break for $p$-wave annihilation, where the energy injection is dominated by the largest halos with the highest velocity dispersions---we leave a more accurate calculation of the $p$-wave results to future study.

\chapter{Supplementary Material for Section~\ref{sec:PBHs}}

\section{Perturbations in Multifield Models}
\label{appPerturbations}

We consider scalar perturbations around a spatially flat Friedmann-Lema\^{i}tre-Robertson-Walker (FLRW) line element,
\beq
ds^2 = - (1 + 2A) dt^2 + 2a (t) (\partial_i B) dt dx^i + a^2 (t) \left[ (1 - 2 \psi) \delta_{ij} + 2 \partial_i \partial_j E \right] dx^i dx^j .
\label{ds}
\eeq
Gauge freedom means that only two of the four metric functions $A$, $B$, $\psi$, and $E$ in Eq.~(\ref{ds}) are independent. The field fluctuations $\delta \phi^I$ introduced in Eq.~(\ref{phivarphi}) are also gauge-dependent. We construct the gauge-invariant Mukhanov-Sasaki variables as linear combinations of field fluctuations and metric perturbations \cite{Kaiser:2012ak,Bassett:2005xm,Gong:2016qmq},
\beq
Q^I \equiv \delta \phi^I + \frac{ \dot{\varphi}^I }{ H} \psi ,
\label{QIdef}
\eeq
and project the perturbations $Q^I$ into adiabatic ($Q_\sigma$) and isocurvature ($Q_s$) components as in Eqs.~(\ref{QIadiso})--(\ref{deltasQs}). The equations of motion for modes $Q_\sigma (k, t)$ and $Q_s (k, t)$ then take the form \cite{Kaiser:2012ak}
\beq
\begin{split}
    \ddot{Q}_\sigma + 3 H \dot{Q}_\sigma + &\left[ \frac{ k^2}{a^2} + {\cal M}_{\sigma\sigma} - \omega^2 - \frac{1}{ M_{\rm pl}^2 a^3} \frac{d}{dt} \left( \frac{ a^3 \dot{\sigma}^2}{H} \right) \right] Q_\sigma \\
    &= 2 \frac{d}{dt} \left( \omega Q_s \right) - 2 \left( \frac{ V_{, \sigma}}{\dot{\sigma}} + \frac{ \dot{H}}{H} \right) \left( \omega Q_s \right) 
    \label{Qsigmaeom}
\end{split}
\eeq
and
\beq
\ddot{Q}_s + 3 H \dot{Q}_s + \left[ \frac{k^2}{a^2} + \mu_s^2 \right] Q_s 
= 4 M_{\rm pl}^2 \frac{ \omega}{\dot{\sigma}} \frac{ k^2}{a^2} \Psi ,
\label{Qseom}
\eeq
where $\omega \equiv \epsilon^{IJ} \hat{\sigma}_I \omega_J = \pm \vert \omega^I \vert$ is the scalar turn rate \cite{Achucarro:2016fby,McDonough:2020gmn}. The gauge-invariant Bardeen potential $\Psi \equiv \psi + a^2 H (\dot{E} - B  a^{-1})$ may be related to $Q_\sigma$ and $Q_s$ via the $00$ and $0i$ components of the Einstein field equations \cite{Kaiser:2012ak}; the form of Eq.~(\ref{Qseom}) is particularly convenient for understanding the behavior of the isocurvature modes $Q_s (k, t)$ in the long-wavelength limit, $k \ll a H$. The mass matrix for the perturbations is given by
\beq
{\cal M}^I_{\>\> J} \equiv {\cal G}^{IK} \left( {\cal D}_J {\cal D}_K V \right) - {\cal R}^I_{\>\> LMJ} \dot{\varphi}^L \dot{\varphi}^M 
\label{MIJdef}
\eeq
with the projections
\beq
{\cal M}_{\sigma \sigma} \equiv \hat{\sigma}_I \hat{\sigma}^J {\cal M}^I_{\>\> J}, \>\> {\cal M}_{ss} \equiv \hat{s}_I^{\>\> J} {\cal M}^I_{\>\> J}
\label{Mprojections}
\eeq
and the mass of the isocurvature perturbations is
\beq
\mu_s^2 \equiv {\cal M}_{ss} + 3 \omega^2 .
\label{mus}
\eeq
In Eq.~(\ref{MIJdef}), ${\cal R}^I_{\>\> LMJ}$ is the Riemann tensor for the field-space manifold. 

When the isocurvature modes remain heavy ($\mu_s^2 \gg H^2$) and/or the turn-rate remains negligible ($\omega^2 \ll H^2$), the predictions for CMB observables revert to covariant versions of the familiar single-field forms \cite{Kaiser:2012ak,Kaiser:2013sna}. In particular, if the adiabatic perturbations remain light during inflation and we initialize the gauge-invariant perturbations in the usual Bunch-Davies vacuum state, then at Hubble crossing, solutions of Eq.~(\ref{Qsigmaeom}) will have amplitude \cite{Gordon:2000hv,Wands:2002bn,Bassett:2005xm}
\beq
\vert Q_\sigma (k_*, t_*) \vert = \frac{ H (t_*) }{\sqrt{ 2 k_*^3}} 
\label{Qsigmastar}
\eeq
up to an irrelevant phase, where $t_*$ is the time when $k_* = a (t_*) H (t_*)$ during inflation. Then Eqs.~(\ref{Rdef}) and (\ref{PRdef}) yield
\beq
{\cal P}_{\cal R} (k_*) = \frac{ H^2 (t_*) }{8 \pi^2 M_{\rm pl}^2 \epsilon (t_*) } .
\label{PRHepsilon}
\eeq
The spectral index $n_s (k_*)$ at some pivot scale $k_*$ is given by \cite{Kaiser:2012ak}
\beq
n_s (k_*) \equiv 1 + \left( \frac{d\, {\rm ln} {\cal P}_{\cal R} (k) }{ d \,{\rm ln} k} \right) \Big\vert_{k_*} \simeq 1 - 6 \epsilon (t_*) + 2 \eta (t_*) 
\label{nsdef}
\eeq
to first order in slow-roll parameters, where $\epsilon (t)$ and $\eta (t)$ are defined in Eqs.~(\ref{epsilon}) and (\ref{etadef}). 
The expression for $n_s (k_*)$ in Eq.~(\ref{nsdef}) is easiest to derive by using the usual slow-roll relation $(dx / d {\rm ln} k )\vert_{k_*} \simeq \dot{x} / H (t_*)$ at Hubble crossing \cite{Bassett:2005xm}. Likewise, the running of the spectral index is given by
\beq
\alpha (k_*) \equiv \left( \frac{ d n_s (k) }{ d \, {\rm ln} k} \right) \Big\vert_{k_*} \simeq \left( \frac{ \dot{n}_s (k)}{H} \right) \Big \vert_{k_*}.
\label{alphadef}
\eeq
The tensor-to-scalar ratio is given by \cite{Kaiser:2013sna,Bassett:2005xm,Gong:2016qmq}
\beq
r (k_*) = 16 \epsilon (t_*) .
\label{rTtoS}
\eeq

For multifield models, we may compare the power spectra of curvature and isocurvature perturbations. If we adopt the conventional normalization \cite{Kaiser:2012ak,Gordon:2000hv,Bassett:2005xm,Gong:2016qmq}
\beq
{\cal S} \equiv \frac{ Q_s }{M_{\rm pl} \sqrt{2 \epsilon}},
\label{calSdef}
\eeq
then the dimensionless isocurvature power spectrum may be written
\beq
{\cal P}_{\cal S} (k) \equiv \frac{ k^3}{2 \pi^2} \vert {\cal S}_k \vert^2 .
\label{PSdef}
\eeq
The isocurvature fraction $\beta_{\rm iso} (k_*, t)$ is defined as
\beq
\beta_{\rm iso} (k_*, t) \equiv \frac{ {\cal P}_{\cal S} (k_*, t) }{ \left[ {\cal P}_{\cal R} (k_*, t) + {\cal P}_{\cal S} (k_*, t) \right]} .
\label{betaisodef}
\eeq
For inflationary trajectories along which the isocurvature modes remain heavy, $\mu_s^2 \gg H^2$ (as in Fig.~\ref{fig:muomegaPR}), the amplitude of isocurvature perturbations falls as $Q_s (k_*, t) \simeq Q_s (k_* , t_*) [ a (t_*) / a (t)]^{3/2}$ for times $t > t_*$. If $\vert Q_s (k_*, t_*) \vert = H (t_*) / \sqrt{ 2 k_*^3}$, akin to Eq.~(\ref{Qsigmastar}), then the amplitude of the mode ${\cal S} (k_*, t)$ will evolve for times $t > t_*$ as 
\beq
\vert {\cal S} (k_*, t) \vert \simeq \frac{ H (t_*) e^{- 3 (N_* - N(t))/2 }}{ 2 M_{\rm pl} \sqrt{ k_*^3 \, \epsilon (t) }} ,
\label{Skstar}
\eeq
where $N(t) \leq N_*$ is the number of efolds before the end of inflation. Then 
\beq
{\cal P}_{\cal S} (k_*, t) \simeq \frac{ H^2 (t_*)}{8 \pi^2 M_{\rm pl}^2 \epsilon (t) } e^{-3 (N_* - N (t))} .
\label{PSheavy}
\eeq
Meanwhile, for $\omega^2 \ll H^2$, the amplitude of the mode ${\cal R} (k_*, t)$ remains frozen for $t > t_*$, so ${\cal P}_{\cal R} (k_*, t) = {\cal P}_{\cal R} (k_*, t_*)$, with magnitude given in Eq.~(\ref{PRHepsilon}). In that case, ${\cal P}_{\cal S} (k_*, t) \ll {\cal P}_{\cal R} (k_*, t)$ for $t > t_*$, and we find
\beq
\beta_{\rm iso} (k_*, t) \simeq \frac{  \epsilon (t_*) }{ \epsilon (t) } e^{-3 (N_* - N(t))} .
\label{betaisoheavy}
\eeq
For $\mu_s^2 \gg H^2$ and $\omega^2 \ll H^2$, the isocurvature fraction is therefore exponentially suppressed by the end of inflation, $\beta_{\rm iso} (k_*, t_{\rm end}) \simeq \epsilon (t_*) e^{-3 N_*} \ll 1$  \cite{Schutz:2013fua,Gordon:2000hv,Wands:2002bn,Bassett:2005xm,DiMarco:2002eb,Peterson:2010np}. 

Similarly, for heavy isocurvature modes ($\mu_s^2 \gg H^2$) and weak turning ($\omega^2 \ll H^2$), the non-Gaussianity also behaves much as in single-field models. In particular, for multifield models with curved field-space manifolds, the dimensionless coefficient $f_{\rm NL}$ may be written \cite{Seery:2005gb,Langlois:2008mn,Gong:2011cd,Elliston:2012ab,Kaiser:2012ak}
\beq
f_{\rm NL} = - \frac{5}{6} \frac{ N^{, A} N^{, B} {\cal D}_A {\cal D}_B N}{ (N_{, I} N^{, I} )^2} - \frac{5}{6} \frac{ N_{, A} N_{, B} N_{, C} \, {\cal A}^{ABC} (k_1, k_2, k_3)}{(N_{, I} N^{, I} )^2 \sum k_i^2 },
\label{fNLdef}
\eeq
where $N = {\rm ln} \, a (t_{\rm end}) - {\rm ln} \, a (t_*)$ is the number of efolds before the end of inflation when the mode with comoving wavenumber $k_*$ first crossed outside the Hubble radius. The term ${\cal A}^{ABC} (k_i)$ vanishes for flat field-space manifolds, ${\cal G}_{IJ} = \delta_{IJ}$; for the curved field-space manifold we consider here, most contributions to ${\cal A}^{ABC}$ vanish identically for equilateral configurations ($k_1 = k_2 = k_3 = k_*$), and (for arbitrary shape functions) the terms proportional to ${\cal A}^{ABC}$ remain subdominant to the contributions arising from the first term in Eq.~(\ref{fNLdef}) \cite{Kaiser:2012ak}. In addition, if the isocurvature modes remain heavy during inflation, then the dominant contribution to the bispectrum arises from variations of $N$ due to fluctuations along the fields' direction of motion. In that case, Eq.~(\ref{fNLdef}) reduces to
\beq
f_{\rm NL} \simeq - \frac{5}{6} \frac{ \hat{\sigma}^A \hat{\sigma}^B {\cal D}_A {\cal D}_B N }{ (\hat{\sigma}^I {\cal D}_I N )^2 } .
\label{fNLSFA1}
\eeq
Recall that $\dot{\varphi}^I {\cal D}_I A^J = {\cal D}_t A^J$ is the covariant directional derivative of vector $A^J$ in the field space. Hence for the term in the denominator of Eq.~(\ref{fNLSFA1}), we may write
\beq
\hat{\sigma}^I {\cal D}_I N = \frac{1}{ \dot{\sigma} } {\cal D}_t N = - \frac{ H}{\dot{\sigma} } .
\label{fNLdenom}
\eeq
For the numerator of Eq.~(\ref{fNLSFA1}), we may write
\beq
\hat{\sigma}^A \hat{\sigma}^B {\cal D}_A {\cal D}_B N = \hat{\sigma}^A {\cal D}_A \hat{\sigma}^B {\cal D}_B N - \frac{1}{\dot{\sigma}}\omega^B {\cal D}_B N ,
\label{fNLnum1}
\eeq
upon using the definition of the turn-rate vector $\omega^I$ in Eq.~(\ref{omegadef}). We note that 
\beq
{\cal D}_B N = -\frac{ H}{\dot{\varphi}^B} = -\frac{ H \hat{\sigma}_B}{\dot{\sigma}} ,
\eeq
and hence the term proportional to $\omega^B$ in Eq.~(\ref{fNLnum1}) vanishes, given the orthogonality of $\omega^B$ and $\hat{\sigma}^B$. Again using $\dot{\varphi}^I {\cal D}_I A^J = {\cal D}_t A^J$, we then have
\beq
\begin{split}
\hat{\sigma}^A {\cal D}_A \hat{\sigma}^B {\cal D}_B N &= \left( \frac{ H^2}{ \dot{\sigma}^2} \right) \left[ - \frac{  \dot{H}}{H^2} + \frac{ \ddot{\sigma}}{H \dot{\sigma}} \right] \\
&=  \left( \frac{ H^2}{ \dot{\sigma}^2} \right) \left( 2 \epsilon - \eta  \right) ,
\end{split}
\label{fNLnum2}
\eeq
upon using the definitions of $\epsilon$ in Eq.~(\ref{epsilon}), $\eta$ in Eq.~(\ref{etadef}), and the relationship in Eq.~(\ref{ddotsigmaeta}). Combining Eqs.~(\ref{fNLdenom})--(\ref{fNLnum2}), we then find for Eq.~(\ref{fNLSFA1})
\beq
f_{\rm NL} \simeq \frac{5}{6} \left( \eta - 2 \epsilon \right) + {\cal O} \left(\frac{ \omega^2 }{ H^2} \right) + {\cal O} \left( \frac{ H^2}{\mu_s^2} \right).
\label{fNLSFA2}
\eeq
For ordinary slow-roll evolution within a single-field attractor, we therefore find that the coefficients for equilateral, orthogonal, and local configurations of the bispectrum will each generically remain small, $\vert f_{\rm NL} \vert \lesssim {\cal O} (10^{-2})$. During ultra-slow-roll, when $\eta \rightarrow 3$, the non-Gaussianity will rise to be ${\cal O} (1)$ \cite{Kaiser:2012ak,Kaiser:2013sna,Kaiser:2015usz,Langlois:2008mn,Bernardeau:2002jy,Seery:2005gb,Yokoyama:2007dw,Byrnes:2008wi,Peterson:2010mv,Chen:2010xka,Byrnes:2010em,Gong:2011cd,Elliston:2011dr,Elliston:2012ab,Seery:2012vj,Mazumdar:2012jj,Peterson:2010np,Gong:2011uw,Gong:2016qmq}.

The comoving CMB pivot scale $k_* = 0.05 \, {\rm Mpc}^{-1}$ first crossed outside the Hubble radius $N_* \equiv N (k_*)$ efolds before the end of inflation \cite{Dodelson:2003vq,Liddle:2003as}
\beq
\begin{split}
N_* &= 67 - {\rm ln} \left( \frac{ k_*}{a_0 H_0} \right) + \frac{1}{4} {\rm ln} \left( \frac{ V^2 (t_*)}{ M_{\rm pl}^4 \rho (t_{\rm end}) } \right) + \frac{1 - 3 w_{\rm eff} }{12 (1 + w_{\rm eff} ) } {\rm ln} \left( \frac{ \rho (t_{\rm rd}) }{\rho (t_{\rm end} ) } \right) \\
&\simeq 62 + \frac{1}{4} {\rm ln} \left( \frac{ V^2 (t_*)}{3 M_{\rm pl}^6 H^2 (t_{\rm end}) } \right) - \frac{ N_{\rm reh}}{4} ,
\end{split}
\label{Nstar}
\eeq
where the subscript $0$ denotes present-day values, $t_*$ is the time when $k_* = a (t_*) H (t_*)$ during inflation, $t_{\rm end}$ is the time at which inflation ends, and $t_{\rm rd}$ is the time when the universe first attains a radiation-dominated equation of state after the end of inflation. In the second line, we assume that the reheating epoch persists for $N_{\rm reh}$ efolds after the end of inflation, during which the universe expands with a matter-dominated equation of state $w_{\rm eff} \simeq 0$ \cite{Amin:2014eta,Allahverdi:2020bys}.

\section{Realization in Supergravity}
\label{appSUGRA}

For a textbook review of supergravity, we refer the reader to Ref.~\cite{Freedman:2012zz}. For a concise review, we refer the reader to the appendices of Ref.~\cite{Kolb:2021xfn}.

The potential in Eq.~\eqref{Vtildertheta} is realized within the framework of $\mathcal{N}=1$ supergravity in $d=4$ dimensions. We take two chiral superfields $\Phi^{I}$, with $I=\{1,2\}$, with field content 
\begin{align}
    \Phi^{I}(x,\theta)=\varpi^{I}+\sqrt{2}\theta\eta^{I}+\theta\theta F^{I} ,
\end{align}
where each $\varpi^I$ (for $I \in \{ 1, 2 \}$) is a complex scalar field, each $\eta^{I}$ is a two-component Weyl spinor, $\theta$ is the fermionic coordinate on superspace, and $F^{I}$ are non-dynamical auxiliary fields; $\bar{\Phi}^{\bar{I}}$ denotes the corresponding anti-chiral superfields. Each complex scalar field $\varpi^I$ can be written in terms of its real and imaginary parts as 
\begin{align}
    \varpi^{I}=\frac{1}{\sqrt{2}}(\phi^{I}+i\psi^{I}) .
\end{align}
Our model is specified in the Jordan frame by a superpotential  $\tilde{W}(\Phi^I)$ and  K\"{a}hler potential $\tilde{K}(\Phi^I,\bar{\Phi}^{\bar{I}})$. The kinetic terms of the scalar components are given by
\begin{equation}
    \mathcal{L}_{\rm kinetic} = - \tilde{\mathcal{G}}_{I \bar{J}} \tilde{g}^{\mu \nu } \partial_{\mu} \varpi^{I} \partial_{\nu} \bar{\varpi}^{\bar{J}} ,
\end{equation}  
with field-space metric 
\begin{align}
    \tilde{\mathcal{G}}_{I\bar{J}}=\frac{\partial}{\partial\Phi^{I}}\frac{\partial}{\partial\bar{\Phi}^{\bar{J}}}\tilde{K}(\Phi^J,\bar{\Phi}^{\bar{J}})_{\Phi^I\rightarrow \varpi^I, \bar{\Phi}^{\bar{I}} \rightarrow \bar{\varpi}^{\bar{I}} }.
    \label{GtildeIJ}
\end{align}
The scalar potential in the Jordan frame is given by
\begin{equation}
     \tilde{V} = \bigg\{ e^{\tilde{K}/M_{\rm pl}^2}\left[ |D\tilde{W} |^2 - 3 M_{\rm pl}^{-2}|\tilde{W} |^2\right]\bigg\}_{\Phi^I\rightarrow \varpi^I, \bar{\Phi}^{\bar{I}} \rightarrow \bar{\varpi}^{\bar{I}} } ,
\end{equation}
where $D_I \equiv \partial_I + M_{\rm pl}^{-2} \tilde{K}_{, I}$.

We select the K\"{a}hler potential to be
\begin{equation}
    \tilde{K} = -\frac{1}{2} \displaystyle \sum _{I=1} ^2 ( \Phi^I - \bar{\Phi}^{\bar{I} })^2
    \label{Kahler1}
\end{equation}
and work with the generic superpotential
\begin{equation}
    \tilde{W}=  \sqrt{2} \mu b_{IJ} \Phi^I \Phi^J + 2 c_{IJK} \Phi^I \Phi^J \Phi^K ,
\end{equation}
where $\mu$ is a mass-scale. Given Eqs.~(\ref{GtildeIJ}) and (\ref{Kahler1}), the field-space metric in the Jordan frame is flat,
\beq
\tilde{\cal G}_{I \bar{J}} = \delta_{I \bar{J}} .
\label{Gtildeflat}
\eeq

For $\tilde{K}$ given in Eq.~(\ref{Kahler1}), we find $\tilde{K}\rightarrow  \sum_I (\psi^I)^2$ upon projecting $\{ \Phi^I, \bar{\Phi}^{\bar{I}} \} \rightarrow \{ \varpi^I, \bar{\varpi}^{\bar{I}} \}$; hence the imaginary components $\psi^I$ of each scalar field $\varpi^I$ become heavy, due to the exponential dependence of $\tilde{V}$ on the K\"{a}hler potential. In particular, it is straightforward to show that $m_\psi^2 \simeq 10 H^2$ (in the Einstein frame), which allows us to integrate out the imaginary components $\psi^I$ during inflation. The resulting scalar potential for the real components $\varpi^1 \equiv \phi / \sqrt{2}$ and $\varpi^2 \equiv \chi / \sqrt{2}$ is given by
\begin{equation}
    \begin{split}
    \tilde{V}(\phi,\chi) = & \> 4 b_1^2 \mu ^2 \phi ^2-\frac{3 b_1^2 \mu ^2 \phi ^4}{2 M_{\rm pl}^2} -\frac{3 b_1 b_2 \mu ^2 \chi ^2 \phi ^2}{M_{\rm pl}^2}+12 b_1 c_1 \mu  \phi ^3-\frac{3 b_1 c_1 \mu  \phi ^5}{M_{\rm pl}^2}  \\
    &+8 b_1 c_2 \mu  \chi  \phi ^2-\frac{3 b_1 c_2 \mu  \chi  \phi ^4}{M_{\rm pl}^2}+4 b_1 c_3 \mu  \chi ^2 \phi -\frac{3 b_1 c_3 \mu  \chi ^2 \phi ^3}{M_{\rm pl}^2}-\frac{3 b_1 c_4 \mu  \chi ^3 \phi ^2}{M_{\rm pl}^2}+4 b_2^2 \mu ^2 \chi ^2  \\
    & -\frac{3 b_2^2 \mu ^2 \chi ^4}{2 M_{\rm pl}^2}-\frac{3 b_2 c_1 \mu  \chi ^2 \phi ^3}{M_{\rm pl}^2}+4 b_2 c_2 \mu  \chi  \phi ^2 -\frac{3 b_2 c_2 \mu  \chi ^3 \phi ^2}{M_{\rm pl}^2}+8 b_2 c_3 \mu  \chi ^2 \phi -\frac{3 b_2 c_3 \mu  \chi ^4 \phi }{M_{\rm pl}^2} \\
    & +12 b_2 c_4 \mu  \chi ^3-\frac{3 b_2 c_4 \mu  \chi ^5}{M_{\rm pl}^2}-\frac{3 c_1^2 \phi ^6}{2 M_{\rm pl}^2}+9 c_1^2 \phi ^4 -\frac{3 c_1 c_2 \chi  \phi ^5}{M_{\rm pl}^2}+12 c_1 c_2 \chi  \phi ^3-\frac{3 c_1 c_3 \chi ^2 \phi ^4}{M_{\rm pl}^2}  \\
    & +6 c_1 c_3 \chi ^2 \phi ^2 -\frac{3 c_1 c_4 \chi ^3 \phi ^3}{M_{\rm pl}^2}-\frac{3 c_2^2 \chi ^2 \phi ^4}{2 M_{\rm pl}^2} +4 c_2^2 \chi ^2 \phi ^2+c_2^2 \phi ^4-\frac{3 c_2 c_3 \chi ^3 \phi ^3}{M_{\rm pl}^2}  \\
    & +4 c_2 c_3 \chi  \phi ^3+4 c_2 c_3 \chi ^3 \phi -\frac{3 c_2 c_4 \chi ^4 \phi ^2}{M_{\rm pl}^2} +6 c_2 c_4 \chi ^2 \phi ^2-\frac{3 c_3^2 \chi ^4 \phi ^2}{2 M_{\rm pl}^2}+c_3^2 \chi ^4+4 c_3^2 \chi ^2 \phi ^2 \\
    & -\frac{3 c_3 c_4 \chi ^5 \phi }{M_{\rm pl}^2}+12 c_3 c_4 \chi ^3 \phi -\frac{3 c_4^2 \chi ^6}{2 M_{\rm pl}^2}+9 c_4^2 \chi ^4  ,
\end{split}
\label{Vtildefull}
\end{equation}
where, as noted below Eq.~(\ref{Wtilde}), we define $b_1 \equiv b_{11}$, $b_2 \equiv b_{22}$, $c_1 \equiv c_{111}$, $c_2 \equiv (c_{112}+c_{121}+c_{211})$, $c_3 \equiv (c_{122}+c_{212}+c_{221})$, and $c_4 \equiv c_{222}$. If one considers inflationary models with $\xi \gg 1$, the perturbation modes accessible to observation correspond to the those that exited the Hubble radius when $\phi,\chi \ll M_{\rm pl}$. Taking the $\phi,\chi \ll M_{\rm pl}$ limit, Eq.~(\ref{Vtildefull}) simplifies to
\begin{eqnarray}
    \tilde{V}(\phi,\chi) =&& 4 b_1^2 \mu ^2 \phi ^2+12 b_1 c_1 \mu  \phi ^3+8 b_1 c_2 \mu  \chi  \phi ^2 \\
    && +4 b_1 c_3 \mu  \chi ^2 \phi +4 b_2^2 \mu ^2 \chi ^2+4 b_2 c_2 \mu  \chi  \phi ^2 \nonumber \\
    && +8 b_2 c_3 \mu  \chi ^2 \phi +12 b_2 c_4 \mu  \chi ^3+9 c_1^2 \phi ^4+12 c_1 c_2 \chi  \phi ^3 \nonumber \\
    && +6 c_1 c_3 \chi ^2 \phi ^2+4 c_2^2 \chi ^2 \phi ^2+c_2^2 \phi ^4+4 c_2 c_3 \chi  \phi ^3 \nonumber \\
    && +4 c_2 c_3 \chi ^3 \phi +6 c_2 c_4 \chi ^2 \phi ^2+c_3^2 \chi ^4+4 c_3^2 \chi ^2 \phi ^2 \nonumber \\
    && +12 c_3 c_4 \chi ^3 \phi +9 c_4^2 \chi ^4 + \mathcal{O}\left(\phi^5/M_{\rm pl}, \chi^5/M_{\rm pl} \right) \nonumber .
\end{eqnarray}
We note that the benchmark value of $\xi$ in Higgs inflation is ${\cal O}(10^4)$ \cite{Bezrukov:2007ep}, and further note that our model can accommodate $\xi$ over many orders of magnitude, via the rescaling of Eq.~\eqref{eqn:scaling}. Finally, translating to polar coordinates, we arrive at Eq.~\eqref{Vtildertheta}.

These models can easily be unified with the current epoch of cosmic acceleration and the observed cosmological constant. This is done by introducing an additional superfield $S$ which satisfies a nilpotency constraint,
\begin{equation}
    S(x,\theta)^2  =0 .
\end{equation}
This condition projects out the scalar component of $S$ from the bosonic sector of the theory. The cosmological applications of the nilpotent superfields were developed in, e.g., Refs.~\cite{Ferrara:2014kva,McDonough:2016der,Kallosh:2017wnt}. The simplest model is given by, 
\begin{equation}
    W = M S   \,\, , \,\, K = S \bar{S},
\end{equation}
leading to a scalar potential which is simply a cosmological constant
\begin{equation}
    V  = M^2 .
\end{equation}
Inflation and dark energy can be realized in this context either by promoting $M$ to a function of fields, or else through field-dependent corrections to the K\"{a}hler potential such as \cite{McDonough:2016der},
\begin{equation}
  \delta K =  f(\Phi,\bar{\Phi}) S \bar{S}  .
\end{equation}
In both cases the scalar potential is simply,
\begin{equation}
    V =  G^{S \bar{S}} \partial_{S}W \partial_{\bar{S}}\bar{W}  .
\end{equation}

We may easily combine the nilpotent superfield models with the inflation models proposed in this study. For example, we may consider,
\begin{equation}
\begin{split}
     \tilde{W} &=  M S + \tilde{W}_{\rm infl}(\Phi^I) , \\ 
     \tilde{K} &= S \bar{S} + \tilde{K}_{\rm infl}(\Phi^I,\bar{\Phi}^{\bar{I}}) ,
\end{split}
\end{equation}
where $\tilde{W}_{\rm infl}$ and $\tilde{K}_{\rm infl}$ refer to the Jordan-frame $W$ and $K$ of our multifield inflation model. The resulting (Jordan-frame) scalar potential is given by,
\begin{equation}
    \tilde{V} = M^2 + \tilde{V}_{\rm infl}(\phi,\chi) ,
\end{equation}
where $\tilde{V}_{\rm infl}$ is the Jordan frame inflationary potential of our two-field model. This approach allows for additional spectator fields during inflation, simply by promoting $M$ to a function of fields, or by corrections to $\tilde{K}$ \cite{McDonough:2016der}.

Finally, nonminimal couplings of the superfields $\Phi^I$ to gravity, in a manifestly supersymmetric form, can be accomplished following the procedure of Ref.~\cite{Kallosh_2013}, slightly generalized from one inflaton to two.

\section{Analytic Solution for the Background Fields' Trajectory}
\label{appTheta}

As noted in Section \ref{sec:trajectories}, if the dimensionless couplings obey the symmetries of Eq.~(\ref{bcxisymmetries}), then we may solve analytically for the background fields' trajectory during inflation. We identify local minima of the potential in the angular direction by calculating
\beq
\begin{split}
    V_{, \theta} (r, \theta) &= \frac{ M_{\rm pl}^4}{[2 f (r) ]^2} \left[ {\cal C}' (\theta) \mu r^3 + {\cal D}' (\theta) r^4 \right] \\
    &= F( r) \, G (r, \theta) ,
\end{split}
\label{Vcommatheta1}
\eeq
where $F (r)$ is some function independent of $\theta$, and
\beq
G (r, \theta) \equiv {\cal C}' (\theta) \mu + {\cal D}' (\theta) r .
\label{Grthetadef}
\eeq
The system will evolve along local minima $\theta_*$ such that $V_{, \theta} (r, \theta_*) = 0$, which corresponds to $G(r, \theta_*)$ = 0. Given the definitions of ${\cal C} (\theta)$ and ${\cal D} (\theta)$ in Eq.~(\ref{BCDdef}), the terms that appear in $G (r, \theta)$ may be written
\beq
\begin{split}
    {\cal C}' (\theta) &= - 18 b c_1 \sin (2 \theta) \left[ \cos \theta - \left( \frac{ c_4 }{c_1} \right) \sin \theta \right] + 12 b c_2 g_1 (\theta) , \\
    {\cal D}' (\theta) &= - 18 c_1^2 \sin (2\theta) \left[ \cos^2 \theta - \left( \frac{c_4 }{c_1} \right)^2 \sin^2 \theta \right] + 4 c_2 g_2 (\theta)
\end{split}
\label{CprimeDprime}
\eeq
with
\beq
\begin{split}
    g_1 (\theta) &\equiv \cos^3 \theta + \sin (2\theta) \left( \cos \theta - \sin \theta \right) - \sin^3 \theta , \\
    g_2 (\theta) &\equiv (3 c_1 + c_2 ) \cos^4 \theta \\
    &\quad + \frac{3}{2} (c_1 + c_2 + c_4) \sin (2 \theta) \left( \cos^2 \theta - \sin^2 \theta \right) \\
    &\quad - 9 (c_1 - c_4) \cos^2 \theta \sin^2 \theta - (3 c_4 + c_2 ) \sin^4 \theta.
\end{split}
\label{g1g2}
\eeq
Closed-form solutions to the equation $G (r, \theta_*) = 0$ may then be found by using the substitution $\theta_* (r) = {\rm arccos} (x (r))$, resulting in the expression for $x^\pm (r)$ given in Eq.~(\ref{xpm}). 
\chapter{Supplemental Material for Section~\ref{sec:PBH_MCMC}}              

\section{The power spectrum of scalar curvature perturbations during ultra-slow-roll} 
\label{app:USR}

We may expand the spacetime line element to linear order in scalar metric perturbations around a spatially flat FLRW metric as in Ref.~\cite{Bassett:2005xm}
\beq
ds^2 = - (1 + 2 A) dt^2 + 2 a \left( \partial_i B \right) dt \, dx^i + a^2 \left[ \left( 1 - 2 \psi \right) \delta_{ij} + 2 \partial_i \partial_j E \right] dx^i dx^j .
\label{dsFLRWscalar}
\eeq
The metric functions $A, B, \psi$, and $E$ are each gauge dependent, as are the field fluctuations $\delta \phi^I$ identified in Eq.~(\ref{phivarphi}). We may construct the gauge-invariant curvature perturbation ${\cal R}$ as the linear combination \cite{Bassett:2005xm}
\beq
{\cal R} \equiv \psi - \frac{ H}{\rho + p} \delta q ,
\label{Rdefpsi}
\eeq
where $\rho = \frac{1}{2} \dot{\sigma}^2 + V$, $p = \frac{1}{2} \dot{\sigma}^2 - V$, and
\beq
\delta q = - {\cal G}_{IJ} \dot{\varphi}^I \delta \phi^J = - \dot{\sigma} \hat{\sigma}_J \delta \phi^J ,
\label{deltaq}
\eeq
with $\dot{\sigma}$ and $\hat{\sigma}^I$ defined in Eqs.~(\ref{dotsigmadef})--(\ref{hatsigma}). In a multifield model we may project onto the hypersurface of the field-space manifold that is orthogonal to the direction of the background fields' motion via \cite{Kaiser:2012ak}
\beq
\hat{s}^{IJ} \equiv {\cal G}^{IJ} - \hat{\sigma}^I \hat{\sigma}^J ,
\label{hatsIJ}
\eeq
in terms of which we may define the ${\cal N} - 1$ remaining gauge-invariant scalar perturbations
\beq
\delta s^I \equiv \hat{s}^I_{\>\> J} \delta \phi^J .
\label{sIdef}
\eeq
Although the field-space vector $\delta s^I$ includes ${\cal N}$ components, only ${\cal N} - 1$ are linearly independent. Moreover, where the field fluctuations $\delta \phi^I$ are gauge dependent, the perturbations $\delta s^I$ are gauge independent, up to linear order in fluctuations \cite{Kaiser:2012ak}. 

In terms of these quantities, the derivative with respect to cosmic time $t$ of a mode ${\cal R}_k (t)$ is given by \cite{Kaiser:2012ak}
\beq
\dot{\cal R}_k = \frac{ 2 H}{\dot{\sigma}} \left( \omega_J \delta s_k^J \right) + \frac{ H}{\dot{H}} \frac{ k^2}{a^2} \Psi_k .
\label{dotR}
\eeq
Here $\omega^I$ is the covariant turn-rate vector defined in Eq.~(\ref{omegadef}) and $\Psi$ is the gauge-invariant Bardeen potential \cite{Bassett:2005xm}
\beq
\Psi \equiv \psi + a^2 H \left( \dot{E} - \frac{B}{a} \right) .
\label{PsidefBardeen}
\eeq
Upon using Eq.~(\ref{eombackground}) for $\dot{H}$ and Eq.~(\ref{epsilon}) for $\epsilon$, Eq.~(\ref{dotR}) is equivalent to
\beq
\frac{ \dot{\cal R}_k}{H} = \frac{1}{ M_{\rm pl} H} \sqrt{ \frac{2}{\epsilon}} \left( \omega_J \delta s_k^J \right) - \frac{1}{ \epsilon} \left(\frac{ k}{aH} \right)^2 \Psi_k .
\label{dotR2}
\eeq
During ordinary slow-roll evolution of the background fields $\varphi^I (t)$, with $\epsilon \sim 10^{-3} - 10^{-2}$, we find the usual result that modes ${\cal R}_k (t)$ will remain conserved outside the Hubble radius, $\dot{\cal R}_k \simeq 0$, in effectively single-field models, for which $\vert \omega_J \delta s_k^J \vert / (M_{\rm pl} H) \sim 0$. Yet even in such (effectively) single-field scenarios, modes ${\cal R}_k (t)$ may undergo significant growth outside the Hubble radius during a phase of ultra-slow-roll evolution, as $\epsilon$ falls exponentially, $\epsilon (t_{\rm usr}) \ll 1$. (See, e.g., Refs.~\cite{Kinney:2005vj,Martin:2012pe}.)

In the limit $\vert \omega_J \delta s^J \vert / (M_{\rm pl} H) \sim 0$, the equation of motion for modes ${\cal R}_k$ assumes the simple form \cite{Bassett:2005xm}
\beq
\frac{1}{ a^3 \epsilon} \frac{d}{dt} \left( a^3 \epsilon \dot{\cal R}_k \right) + \frac{ k^2}{a^2} {\cal R}_k = 0 .
\label{eomR}
\eeq
Using the definitions of the slow-roll parameters $\epsilon$ and $\eta$ in Eqs.~(\ref{epsilon}) and \eqref{etadef}, this is equivalent to
\beq
\ddot{\cal R}_k + 3 H \left( 1 + \frac{4}{3} \epsilon - \frac{2}{3} \eta \right) \dot{\cal R}_k + \frac{ k^2}{a^2} {\cal R}_k = 0 .
\label{eomR2}
\eeq
During ultra-slow-roll, the slow-roll parameter $\epsilon$ falls rapidly, $\epsilon \rightarrow 0^+$. In that limit, the amplitude of modes ${\cal R}_k$ that have crossed outside the Hubble radius (with $k \ll aH$) will grow whenever $\eta > 3/2$ \cite{Kinney:2005vj,Martin:2012pe,Byrnes:2018txb,Liu:2020oqe,Cole:2022xqc}. Near an inflection point of the potential, for which $V_{, \sigma} \simeq 0$, the background equations of motion yield $\epsilon \sim a^{-6}$ and hence $\eta \rightarrow  3$. For small-field features in the potential of the sort we analyze here, however, $V_{, \sigma}$ need not vanish identically during the entire duration of ultra-slow-roll, and $\eta$ can exceed 3. More generally, $\eta (t)$ need not remain constant during the ultra-slow-roll phase.

Given that the amplitude of modes can grow outside the Hubble radius while $\eta > 3/2$, departures from the typical slow-roll evolution are characterized by the quantity
\beq
{\cal U} \equiv \int_{t_s}^{t_e} dt \left( \eta(t) - \frac{3}{2} \right) H (t) = \left( \bar{\eta} - \frac{3}{2} \right) N_{\rm usr} ,
\label{Udef}
\eeq
where $t_s$ and $t_e$ indicate the start and end of the ultra-slow-roll phase, each determined by the times when $\eta$ crosses $3/2$, and $\bar{\eta}$ denotes the average of $\eta (t)$ between $t_s$ and $t_e$. Given that $\epsilon \rightarrow 0^+$ during ultra-slow-roll, $\dot{H} \sim 0$ and the duration of ultra-slow-roll (in $e$-folds) may be approximated as $N_{\rm usr} \simeq H (t_e - t_s)$. 

The rapid fall of $\epsilon$ after the start of ultra-slow-roll means that some modes with wavenumber $k < k_s$ will become amplified after crossing outside the Hubble radius, beginning with wavenumber $k_{\rm min}$ given by \cite{Liu:2020oqe}
\beq
k_{\rm min} = k_s \exp \left[ - {\cal U} \right] \simeq k_s e^{- (\bar{\eta} - 3/2 ) N_{\rm usr} } ,
\label{kmin}
\eeq
where $k_s = a (t_s) H (t_s)$ is the wavenumber of the mode that first exits the Hubble radius at the start of ultra-slow-roll. The value of $k_{\rm min}$ comes from balancing the fall in amplitude of the ``decaying" contribution to ${\cal R}_k$ between the time that mode crosses outside the Hubble radius ($t_k$) and the onset of ultra-slow-roll ($t_s$), with the growth of that same term during the ultra-slow-roll phase ($t_s \leq t \leq t_e$) \cite{Liu:2020oqe}.

For modes with $k_{\rm min} \leq k \leq k_s$, which cross outside the Hubble radius prior to the onset of ultra-slow-roll, the power spectrum will be modified compared to the slow-roll expression ${\cal P}_{\cal R}^{\rm SR} (k)$ of Eq.~(\ref{PRHepsilon}) as \cite{Liu:2020oqe} (see also Refs.~\cite{Martin:2012pe,Byrnes:2018txb,Cole:2022xqc})
\beq
{\cal P}_{\cal R} (k) \simeq \left( \frac{ k}{k_s} \right)^4 \exp\left[ 4 \, {\cal U} \right] \, {\cal P}_{\cal R}^{\rm SR} (k ) 
\simeq \left( \frac{ k}{k_s} \right)^4 {\cal P}_{\cal R}^{\rm SR} (k)\, e^{( 4 \bar{\eta} - 6 ) N_{\rm usr}} \quad {\rm for} \> \, k_{\rm min} \leq k \leq k_s .
\label{PRusrkmin}
\eeq
The steep growth $\sim k^4$ begins at $k_{\rm min}$ and reaches a peak at $k_s$. For modes that exit the Hubble radius during ultra-slow-roll, with $k_e \leq k \leq k_s$, the power spectrum falls from its peak at ${\cal P}_{\cal R} (k_s)$, since those modes merely oscillate prior to exiting the Hubble radius and then experience a shorter duration ($N < N_{\rm usr}$) of growth outside the Hubble radius during the remainder of the ultra-slow-roll phase. For modes that cross outside the Hubble radius later than $t_e$, the background system again undergoes ordinary slow-roll evolution, so the standard expression for ${\cal P}_{\cal R}^{\rm SR} (k)$ of Eq.~(\ref{PRHepsilon}) applies for $k > k_e$ \cite{Martin:2012pe,Liu:2020oqe,Byrnes:2018txb,Cole:2022xqc}.

Incorporating the added growth of various modes during ultra-slow-roll, the dimensionless power spectrum reaches its maximum value at $k_s$ \cite{Martin:2012pe,Liu:2020oqe,Byrnes:2018txb,Cole:2022xqc}:
\beq
{\cal P}_{\cal R, \rm max} (k)  = {\cal P}_{\cal R} (k_s) \simeq \exp \left[ 4 \, {\cal U} \right] {\cal P}_{\cal R}^{\rm SR} (k_s)  .
\label{PRmaxUSR}
\eeq
The usual slow-roll expression for ${\cal P}_{\cal R}^{\rm SR} (k)$ in Eq.~(\ref{PRHepsilon}) already incorporates parametric growth comparable to the term $\exp [ 4 \, {\cal U} ]$, given that ${\cal P}_{\cal R}^{\rm SR} (k) \propto 1 / \epsilon$ and $\epsilon \sim a^{-6} \sim \exp [6 N_{\rm usr}]$ during ultra-slow-roll. This amplification matches the ultra-slow-roll amplification factor $\exp [ 4 \, {\cal U} ]$ if $\bar{\eta} = 3$ during ultra-slow-roll. The main impacts of the ultra-slow-roll phase, compared to the usual slow-roll expression, are then twofold: the peak value ${\cal P}_{\cal R, \rm max}$ can exceed the value represented by ${\cal P}_{\cal R, \rm max}^{\rm SR}$ if $4 \, {\cal U} > 6 N_{\rm usr}$; and the wavenumber corresponding to the peak in ${\cal P}_{\cal R} (k)$ shifts to shorter values compared to ${\cal P}_{\cal R}^{\rm SR} (k)$, due to the added growth of some modes after they cross outside the Hubble radius. In particular, whereas ${\cal P}_{\cal R}^{\rm SR} (k)$ typically reaches its {\it minimum} value at $k_s$ and its maximum value at $k_e$, the full expression for ${\cal P}_{\cal R}$ reaches its {\it maximum} value at $k_s \simeq k_e \exp[ - N_{\rm usr}]$ \cite{Liu:2020oqe}.

We may consider the impact of the ultra-slow-roll phase on ${\cal P}_{\cal R} (k)$ for our family of models in typical regions of parameter space. We select a point among the red regions of Fig.~\ref{fig:bands}, near our fiducial parameter set. For this set of parameters, the system enters ultra-slow-roll evolution (with $\eta > 3/2$) at $N_s = 18.12$ 
$e$-folds before the end of inflation and exits ultra-slow-roll at $N_e = 14.59$, 
for a total duration $N_{\rm usr} = 3.53$ $e$-folds of ultra-slow-roll. 

If we neglect the effects of ultra-slow-roll on the power spectrum, then the slow-roll approximation to the power spectrum reaches a peak value ${\cal P}_{\cal R, \rm max}^{\rm SR} = {\cal P}_{\cal R}^{\rm SR} (k_e) = 1.18 \times 10^{-3}$ at $N_e$, consistent with the behavior of most regions of parameter space under study here (see the posterior distribution of ${\cal P}_{\cal R, \rm max}$ in Fig.~\ref{fig:obs_corner}). Given ${\cal P}_{\cal R}^{\rm SR} (k_s) = 7.14 \times 10^{-13}$, this means that the slow-roll approximation to the power spectrum grows by a factor $1.66 \times 10^9$ during the ultra-slow-roll phase, consistent with the parametric growth noted above: ${\cal P}_{\cal R}^{\rm SR} (k) \propto 1 / \epsilon \sim \exp [ 6 N_{\rm usr} ]$. For these parameters, meanwhile, the ultra-slow-roll amplification factor ${\cal U}$ defined in Eq.~(\ref{Udef}) is ${\cal U} = 5.70$. We thus find $4 \, {\cal U} = 22.80 > 6 N_{\rm usr} = 21.18$, which indicates that ${\cal P}_{\cal R, \rm max} > {\cal P}_{\cal R, \rm max}^{\rm SR}$. In particular, the additional growth during the ultra-slow-roll phase yields ${\cal P}_{\cal R, \rm max} = \exp \left[ 4 \, {\cal U} \right] {\cal P}_{\cal R}^{\rm SR} (k_s) = 5.70 \times 10^{-3}$, a factor $\sim 5$ greater than the peak predicted by the slow-roll approximate form ${\cal P}_{\cal R}^{\rm SR} (k)$. 

Although details of the impact of ultra-slow-roll on ${\cal P}_{\cal R} (k)$ depend on the model and parameter set under consideration, the growth of ${\cal P}_{\cal R, \rm max}$ compared to ${\cal P}_{\cal R, \rm max}^{\rm SR}$ that we find here is consistent with previous studies, which typically find ${\cal P}_{\cal R, \rm max} \sim {\cal O} (10) \times {\cal P}_{\cal R, \rm max}^{\rm SR}$ \cite{Kinney:2005vj,Martin:2012pe,Byrnes:2018txb,Liu:2020oqe,Cole:2022xqc}. Given that the majority of points in parameter space that we sampled in our MCMC analysis yield ${\cal P}_{\cal R, \rm max}^{\rm SR} \sim 10^{-3}$, we therefore conclude that additional growth from the ultra-slow-roll phase in this family of models is consistent with ${\cal P}_{\cal R, \rm max} \lesssim 10^{-2}$ across most regions of parameter space, and hence should evade constraints (not incorporated here) based on overproducing PBHs, producing excessive spectral $\mu$-distortions, and similar small-scale effects. (Evaluating such constraints typically requires moving beyond the approximation of a Gaussian probability distribution function for the scalar perturbations, and hence remains beyond the scope of our present analysis. See, e.g., Refs.~\cite{Gow:2020bzo,Gow:2022jfb}.)

\section{The power spectrum of induced tensor perturbations}
\label{app:tensor_spectrum}

We provide the complete equations for the computation of the tensor power spectrum $P_h$ used in Eq.~\eqref{eq:GWspectrum}. 
Let us begin with the definition of linear perturbations in the conformal Newtonian gauge for the metric of the form
\begin{align}
ds^2 = -a^2(\tau)(1+2\Phi) d\tau^2 +  a^2(\tau) \left[(1-2\Psi)\delta_{ij} + \frac{1}{2} h_{ij} \right] dx^i dx^j, 
\end{align}
where $\Phi = A - \partial_t [ a^2 (\dot{E} - B / a) ]$ is the Newtonian potential and $\Psi$ is the Bardeen curvature potential defined in Eq.~(\ref{PsidefBardeen}), while $A, B$, and $E$ are defined in Eq.~(\ref{dsFLRWscalar}). We define
\begin{align}
h_{ij}(\vec{x},\tau) = \int \frac{d^3k}{(2\pi)^{3/2} } [e_{ij}^+(\vec{k})h_+(\vec{k},\tau) + e_{ij}^\times (\vec{k})h_\times(\vec{k},\tau)] e^{i\vec{k}\cdot\vec{x}},
\end{align}
which is the linear tensor perturbation including the two polarization modes. The transverse-traceless polarization tensors are
\begin{align}
e_{ij}^+(\vec{k}) &= \frac{[e_i^1(\vec{k})e_j^1(\vec{k}) - e_i^2(\vec{k})e_j^2(\vec{k})]}{\sqrt{2}},& 
e_{ij}^\times(\vec{k}) = \frac{[e^1_i(\vec{k}) e^2_j(\vec{k}) + e^2_i(\vec{k}) e^1_j(\vec{k})]}{\sqrt{2}},
\end{align}
which are expressed in terms of orthonormal basis vectors $\textbf{e}^1$ and $\textbf{e}^2$ orthogonal to $\vec{k}$.

Keeping the tensor perturbation at linear order and the linear scalar perturbations up to second order, one can obtain the equation of motion for each polarization $h_\lambda$ from the Einstein equation as
\begin{align}\label{eom_h}
h_\lambda^{\prime\prime}(\vec{k},\tau) + 2\mathcal{H} h_\lambda^\prime (\vec{k},\tau)+ k^2 h_\lambda(\vec{k},\tau) = 4 S_\lambda (\vec{k},\tau),
\end{align} 
where second-order perturbations are projected away in the transverse-traceless decomposition \cite{Baumann:2007zm} and we have neglected the anisotropic stress in the energy momentum tensor, so that $\Psi = \Phi$, and thus
\begin{align}
S_\lambda(\vec{k},\tau) =& \int \frac{d^3q}{(2\pi)^{3/2}} e_{ij}^\lambda(\vec{k}) q^iq^j \psi_{\vec{p}}\psi_{\vec{p}} f(p,q,\tau), \\
f(p,q,\tau) =& 2T(p\tau)T(q\tau) + \frac{4}{3(1+w)} 
\left[\frac{T^\prime(p\tau)}{\mathcal{H}} + T(p\tau)\right] \left[\frac{T^\prime(q\tau)}{\mathcal{H}} + T(q\tau)\right],
\end{align}
where $\mathcal{H} = aH = 2/ [(1+3w)\tau]$ and $w$ determines the equation of state of the fluid that fills the universe, $P=w\rho$. The two internal momenta are given by $p\equiv \vert\vec{p}\vert$ and $q\equiv \vert\vec{q}\vert$, and $\vec{k} = \vec{p} +\vec{q}$. 
The time evolution of the scalar potential is described by $\Psi_{\vec{k}}(\tau) = T(k\tau)\psi_{\vec{k}}$ with respect to the primordial value $\psi_{\vec{k}}$, where the transfer function $T(k\tau)$ in the radiation-dominated universe is given in Ref.~\cite{Domenech:2021ztg}.

The primordial Newtonian potential $\psi_{\vec{k}}$ well outside the Hubble radius is related to the (gauge-invariant) curvature perturbation 
${\cal R}$ as $\psi_{\vec{k}} = [(3 + 3 w) / (5 + 3 w) ] {\cal R}$, which yields
\begin{align}
\left\langle \psi_{\vec{k}}\psi_{\vec{K}}\right\rangle = \delta^{(3)}\left(\vec{k} +\vec{K}\right) \frac{2\pi^2}{k^3} \left(\frac{3+3w}{5+3w}\right)^2 \mathcal{P}_{\cal R}(k).
\end{align}
This is where parameters of the inflationary scenario given in \S~\ref{sec:model} enter the power spectrum of the tensor perturbation. 

Solving the equation of motion of Eq.~\eqref{eom_h} by virtue of the Green's function method of Eq.~\eqref{sol_h_Green}, we can compute the total power spectrum of the tensor perturbation as
\begin{align}\label{def_P_h}
\delta^{(3)}(\vec{k}+\vec{K}) P_h(k, \tau) 
=& \frac{k^3}{2\pi^2} \sum_\lambda^{+,\times} \left\langle h_\lambda(\vec{k},\tau)h_\lambda(\vec{K},\tau)\right\rangle, \nonumber\\
=& \frac{k^3}{2\pi^2} \int^\tau d\tau_1 G_{\vec{k}}(\tau ;\tau_1) \frac{a(\tau_1)}{a(\tau)}
\int^\tau d\tau_2 G_{\vec{K}}(\tau;\tau_2)   \frac{a(\tau_2)}{a(\tau)} \nonumber\\
&\qquad\times \sum_\lambda^{+,\times} \left\langle S_\lambda(\vec{k},\tau_1)S_\lambda(\vec{K},\tau_2)\right\rangle.
\end{align}
It is convenient to use the dimensionless variables $u \equiv p/k$, $v\equiv q/k$ and $z \equiv k\tau$ to rewrite the tensor spectrum as
\begin{align}\label{eq:P_h_uv_variable}
P_h(k, z) &= 4\int_{0}^{\infty} dv\int_{\vert 1-v \vert}^{1+v} du 
\left[\frac{v}{u} - \frac{(1-u^2+v^2)}{4uv}\right]^2 I^2(u,v,z) \mathcal{P}_{\cal R}(ku) \mathcal{P}_{\cal R}(kv), \\
I(u,v,z) &= \frac{9(1+w)^2}{(5+3w)^2}\int_{0}^z dz_1 \frac{a(z_1)}{a(z)} kG_{\vec{k}} (z,z_1) f(u,v,z), 
\end{align}
where our definition of $I(u,v,z)$ coincides with that defined in Ref.~\cite{Kohri:2018awv}. Note that the projection of momentum under polarization tensors can be found in the Appendix B of Ref.~\cite{Atal:2021jyo}, where 
\begin{align}
(e_{ij}^+ q_iq_j)^2 + (e_{ij}^\times q_iq_j)^2 = k^4v^4\left[1-\frac{(1-u^2+v^2)^2}{(2v)^2}\right]^2.   
\end{align}
For numerical evaluation, we adopt new variables $t = u+v-1$, $s =u-v$ introduced in Ref.~\cite{Kohri:2018awv}, where $u=(t+s+1)/2$, $v=(t-s+1)/2$ and the tensor spectrum now reads
\begin{align}
\label{eq:powerspectrum}
P_h(k, z) =& 2\int_{0}^{\infty} dt \int_{-1}^{1} ds \left[ \frac{t (2 + t) (s^2 - 1)}{(1 - s + t) (1 + s + t)}\right]^2 \\\nonumber
&\quad \times {\cal P}_{\cal R}\left(\frac{k (t+s+1)}{2}\right) {\cal P}_{\cal R}\left(\frac{k (t-s+1)}{2}\right) I_{\rm RD}^2(s,t, z).
\end{align}
In the late-time limit of the radiation-dominated universe, that is, for $\tau\rightarrow\infty$ and $z \gg 1$, we have the oscillation-averaged result from Ref.~\cite{Kohri:2018awv}:
\begin{align}
\label{eq:GWtf}
&\overline{I^2_{\textrm{RD}}(s,t, k\tau \to \infty)} =  \frac{288 (-5 + s^2 + t (2 + t))^2}{z^2(1 - s + t)^6 (1 + s + t)^6} \times
\\
&\qquad\Bigg\{ \frac{\pi^2}{4} \left(-5 + s^2 + t (2 + t)\right)^2 \Theta\left(t-(\sqrt{3}-1)\right) +
\nonumber \\\nonumber
&\qquad \left[-(t - s + 1) (t + s + 1) + 
\frac{1}{2} (-5 + s^2 + t (2 + t)) \ln\left|\frac{(-2 + t (2 + t))}{3 - s^2}\right|\right]^2\Bigg\},
\end{align}
where $\Theta$ is the usual Heaviside theta function. Hence the averaged analytical transfer function during radiation domination in Eq.~\eqref{eq:GWtf} can be substituted into Eq.~\eqref{eq:powerspectrum}. The resulting integral will yield the oscillation-averaged power spectrum $\overline{P_h(k,\eta)}$ which can be substituted into Eq.~\eqref{eq:GWspectrum}, and the dimensionless GW background to be compared with experimental limits or signals can be determined. We have the calculation by direct comparison with the scale-invariant power spectrum normalized to unity ($A_{\cal R} = 1$), which yields a dimensionless gravitational wave spectrum of $\Omega_{\textrm{GW}}/A_{\cal R}^2 = 0.822$ as expected in Ref.~\cite{Kohri:2018awv}.
\chapter{Supplemental material for Chapter~\ref{sec:EFTof21cm}}
\label{app:EFT_supp}

\section{Counterterms for the $v_\parallel^2$ operator}
\label{sec:v2ct}

The renormalization of $v_\parallel^2$ is very similar to the renormalization of $\delta^2$, as described in Section~\ref{sec:renorm}.
However, the vertices corresponding to the component $\theta^{(n)}_{\boldsymbol{p}}$ fields will come with $G_n$ kernels instead of $F_n$, as well as a factor of $p_\parallel / p = \cos\theta$ for projecting the velocity onto the line of sight.

The zeroth order counterterm is given by
\begin{align}
    \braket{(v_\parallel^2)_{\boldsymbol{k}}} \quad &= \quad
	\begin{tikzpicture}[baseline=(current bounding box.center)]
	\begin{feynman}
	\vertex (i);
	\vertex [right=1cm of i, small, blob] (a) {};
	\vertex [above right=1cm of a, empty dot] (b) {};
	\vertex [below right=1cm of a, empty dot] (c) {};
	\diagram*{
		(i) -- [momentum=\(\boldsymbol{k}\)] (a) ,
		(a) -- {(b),(c)},
		(b) -- [scalar, half left, momentum=\(\boldsymbol{p}\)] (c)
	};
	\end{feynman}
	\end{tikzpicture}
	\n
	&= \quad
	\mathcal{H}^2 \int \dbar^3 p \, P(\boldsymbol{p}) \left( \frac{p_\parallel}{p^2} \right)^2 = \frac{\mathcal{H}^2}{(2\pi)^3} \int d p \, P(\boldsymbol{p}) \times 2\pi \int d \cos\theta \, \cos^2 \theta = \frac{\mathcal{H}^2}{6\pi^2} \int d p \, P(\boldsymbol{p}) .
\end{align}

At $n=1$, there are two identical diagrams that contribute, one each from $\delta^{(1)}$ contracting with one of the component $\theta^{(n)}$ fields.
Each of these diagrams also comes with a symmetry factor of 2, from permuting the two linear legs emerging from $\theta^{(2)}$.
\begin{align}
    \braket{(v_\parallel^2)_{\boldsymbol{q}} \delta^{(1)}_{\boldsymbol{q}}} 
    \quad &= \quad 2 \times
	\begin{tikzpicture}[baseline=(i.base)]
	\begin{feynman}
	\vertex (i);
	\vertex [right=1cm of i, small, blob] (a) {};
	\vertex [above right=1cm of a, empty dot] (b) {};
	\vertex [below right=1cm of a, empty dot] (c) {};
	\vertex [right=1cm of b, dot] (d) {};
	\vertex [right=1cm of d] (f);
	\diagram*{
		(i) -- (a) ,
		(a) -- {(b),(c)},
		(b) -- [scalar, reversed momentum=\(\boldsymbol{p}\)] (c),
		(b) -- [scalar, momentum=\(\boldsymbol{q}\)] (d),
		(d) -- (f),
	};
	\end{feynman}
	\end{tikzpicture}
	\n
    &= - 4 \mathcal{H}^2 P(\boldsymbol{q}) \int \dbar^3 p \, P(\boldsymbol{p}) \frac{p_\parallel}{p^2} \frac{(\boldsymbol{q} - \boldsymbol{p})_\parallel}{(\boldsymbol{q} - \boldsymbol{p})^2} G_2(-\boldsymbol{p}, \boldsymbol{q}) \n
	&= \frac{\mathcal{H}^2}{105} \left( 71 + 23 \frac{q_{\parallel}^2 - q_{\perp}^2}{q^2} \right) P(\boldsymbol{q}) \varsigma^2 (\Lambda)
\end{align}

At $n=2$, there are three diagrams to consider.
\begin{equation}
    \braket{(v_\parallel^2)_{\boldsymbol{q}_1 + \boldsymbol{q}_2} \delta^{(1)}_{\boldsymbol{q}_1}  \delta^{(1)}_{\boldsymbol{q}_2}}
    =
	\begin{tikzpicture}[baseline=(current bounding box.center)]
	\begin{feynman}
	\vertex (i);
	\vertex [right=1cm of i, small, blob] (a) {};
	\vertex [above right=1cm of a, empty dot] (b) {};
	\vertex [below right=1cm of a, empty dot] (c) {};
	\vertex [right=1cm of b, dot] (d) {};
	\vertex [right=1cm of c, dot] (e) {};
	\vertex [right=1cm of d] (f);
	\vertex [right=1cm of e] (g);
	\diagram*{
		(i) -- (a) ,
		(a) -- {(b),(c)},
		(b) -- [scalar] (c),
		(b) -- [scalar] (d),
		(c) -- [scalar] (e),
		(d) -- (f),
		(e) -- (g)
	};
	\end{feynman}
	\end{tikzpicture}
	+ 2 \times
	\begin{tikzpicture}[baseline=(current bounding box.center)]
	\begin{feynman}
	\vertex (i);
	\vertex [right=1cm of i, small, blob] (a) {};
	\vertex [above right=1cm of a, empty dot] (b) {};
	\vertex [below right=1cm of a, empty dot] (c) {};
	\vertex [right=1cm of b, dot] (d) {};
	\vertex [right=1cm of c, dot] (e) {};
	\vertex [right=1cm of d] (f);
	\vertex [right=1cm of e] (g);
	\diagram*{
		(i) -- (a) ,
		(a) -- {(b),(c)},
		(b) -- [scalar] (c),
		(b) -- [scalar] (d),
		(b) -- [scalar] (e),
		(d) -- (f),
		(e) -- (g)
	};
	\end{feynman}
	\end{tikzpicture}
	+
	\begin{tikzpicture}[baseline=(current bounding box.center)]
	\begin{feynman}
	\vertex (i);
	\vertex [right=1cm of i, crossed dot] (a) {};
	\vertex [right=1cm of a, empty dot] (b) {};
	\vertex [above right=1cm of b, dot] (c) {};
	\vertex [below right=1cm of b, dot] (d) {};
	\vertex [right=1cm of c] (f);
	\vertex [right=1cm of d] (g);
	\diagram*{
		(i) -- (a) ,
		(a) -- (b),
		(b) -- [scalar] {(c),(d)},
		(c) -- (f),
		(d) -- (g)
	};
	\end{feynman}
	\end{tikzpicture}
\end{equation}
The crossed dot on the third diagram represents the $n=1$ counterterm.
We label these terms $\mathcal{M}_1$, $\mathcal{M}_2$, and $\mathcal{M}_\mathrm{c.t.}$, respectively.
The first diagram should have a symmetry factor of 8; $2$ from each of the $G_2$ vertices, and another factor of 2 from choosing which of the linear legs contracts with which vertex. 
There will be two diagrams of the second type with identical contributions; each will have a symmetry factor of $3!$ on the $G_3$ kernel. 
The third diagram will again have a symmetry factor of 2 from permuting the external legs or the $G_2$ vertex.
\begin{align}
	\mathcal{M}_1 &= - 2^3 \,\mathcal{H}^2 P(\boldsymbol{q}_1) P(\boldsymbol{q}_2) \int \dbar^3 p \, P(\boldsymbol{p}) G_2 (\boldsymbol{q}_1, -\boldsymbol{p}) G_2 (\boldsymbol{q}_2, \boldsymbol{p}) \frac{(\boldsymbol{q}_1 - \boldsymbol{p})_\parallel}{(\boldsymbol{q}_1 - \boldsymbol{p})^2} \frac{(\boldsymbol{q}_2 + \boldsymbol{p})_\parallel}{(\boldsymbol{q}_2 + \boldsymbol{p})^2} \n
	\mathcal{M}_2 &= 2\times 3!\,  \mathcal{H}^2 P(\boldsymbol{q}_1) P(\boldsymbol{q}_2) \int \dbar^3 p \, P(\boldsymbol{p}) G_3 (\boldsymbol{q}_1, \boldsymbol{q}_2, \boldsymbol{p}) \frac{p_\parallel}{p^2} \frac{(\boldsymbol{q}_1 + \boldsymbol{q}_2 + \boldsymbol{p})_\parallel}{(\boldsymbol{q}_1 + \boldsymbol{q}_2 + \boldsymbol{p})^2} \n
	\mathcal{M}_\mathrm{c.t.} &= - \frac{\mathcal{H}^2}{105} \left( 71 + 23 \frac{(\boldsymbol{q}_1 + \boldsymbol{q}_2)_{\parallel}^2 - (\boldsymbol{q}_1 + \boldsymbol{q}_2)_{\perp}^2}{(\boldsymbol{q}_1 + \boldsymbol{q}_2)^2} \right) \varsigma^2 (\Lambda) \times 2 P(\boldsymbol{q}_1) P(\boldsymbol{q}_2) F_2 (\boldsymbol{q}_1, \boldsymbol{q}_2)
\end{align}
Summing these together, integrating, and keeping only the contribution that is nonzero when $\{q_1, q_2\} \rightarrow 0$, we find
\begin{align}
	\braket{(v_\parallel^2)_{\boldsymbol{q}_1 + \boldsymbol{q}_2} \delta^{(1)}_{\boldsymbol{q}_1}  \delta^{(1)}_{\boldsymbol{q}_2}} 
	&= \mathcal{M}_1 + \mathcal{M}_2 + \mathcal{M}_3 \n
	&= \frac{\mathcal{H}^2}{5145} P(\boldsymbol{q}_1) P(\boldsymbol{q}_2) \varsigma^2 (\Lambda) \left[ 996 + 2041 \left( \frac{q_{1,\parallel}^2}{q_1^2} + \frac{q_{2,\parallel}^2}{q_2^2} \right) - 2142 \frac{q_{1,\parallel} q_{2,\parallel}}{q_1 q_2} \right. \n
	&\qquad\left. + \frac{\boldsymbol{q}_{1} \cdot \boldsymbol{q}_{2}}{q_1 q_2} \left( 1071 \frac{q_{1,\parallel}^2}{q_1^2} + 1071 \frac{q_{2,\parallel}^2}{q_2^2} - 948 \frac{\boldsymbol{q}_{1} \cdot \boldsymbol{q}_{2}}{q_1 q_2} + 2844 \frac{q_{1,\parallel} q_{2,\parallel}}{q_1 q_2} \right) \right] .
\end{align}

\section{Perturbative velocity divergences from perturbative densities}
\label{sec:theta}

Here, we derive some relationships between the lowest order $\theta^{(n)}$'s and the operators $\delta^{(1)}$, $\delta^{(2)}$, $\delta^{(3)}$, and $\mathcal{G}_2$.
At first order, we see from the form of the perturbative ansatzes that $\theta^{(1)} = \delta^{(1)}$. 
Deriving the second order relationship is also fairly straightforward.
For compactness, we denote integrals in momentum or wavenumber space by $\int_q = \int \dbar^3 q$.
\begin{align}
\theta^{(2)}_{\boldsymbol{q}} &= \int_{q_1} \int_{q_2} (2\pi)^3 \delta^D (\boldsymbol{q}_1 + \boldsymbol{q}_2 - \boldsymbol{q}) G_2 (\boldsymbol{q}_1, \boldsymbol{q}_2) \delta^{(1)}_{\boldsymbol{q}_1} \delta^{(1)}_{\boldsymbol{q}_2} \n
&= \int_{q_1} \int_{q_2} (2\pi)^3 \delta^D (\boldsymbol{q}_1 + \boldsymbol{q}_2 - \boldsymbol{q}) F_2 (\boldsymbol{q}_1, \boldsymbol{q}_2) \delta^{(1)}_{\boldsymbol{q}_1} \delta^{(1)}_{\boldsymbol{q}_2} \n 
&\qquad\qquad + \frac{2}{7} \int_{q_1} \int_{q_2} (2\pi)^3 \delta^D (\boldsymbol{q}_1 + \boldsymbol{q}_2 - \boldsymbol{q}) \left[ \frac{(\boldsymbol{q}_1 \cdot \boldsymbol{q}_2)^2}{q_1^2 q_2^2} - 1 \right] \delta^{(1)}_{\boldsymbol{q}_1} \delta^{(1)}_{\boldsymbol{q}_2} \n
&= \delta^{(2)}_{\boldsymbol{q}} + \frac{2}{7} \int_{q_1} \int_{q_2} (2\pi)^3 \delta^D (\boldsymbol{q}_1 + \boldsymbol{q}_2 - \boldsymbol{q}) \left[ \frac{(\boldsymbol{q}_1 \cdot \boldsymbol{q}_2)^2}{q_1^2 q_2^2} - 1 \right] \delta^{(1)}_{\boldsymbol{q}_1} \delta^{(1)}_{\boldsymbol{q}_2}
\end{align}
In configuration space, the above expression becomes
$$ \theta^{(2)} = \delta^{(2)} + \frac{2}{7} \mathcal{G}_2 ^{(2)}. $$
Since the perturbative ansatz includes a factor of $- \mathcal{H}$, the velocity is given by $v_i = \frac{\partial_i}{\partial^2} \theta = - \mathcal{H} \frac{\partial_i}{\partial^2} (\theta_1 + \theta_2 + \cdots)$.

The process is analogous for the third order term.
Using the recursive relations given in Refs.~\cite{Goroff:1986ep,Jain:1993jh,Bernardeau:2001qr}, we obtain the following expressions for $\theta^{(3)}$:
\begin{align}
\theta^{(3)}_{\boldsymbol{q}} &= \int_{q_1} \int_{q_2} \int_{q_3} (2\pi)^3 \delta^D (\boldsymbol{q}_1 + \boldsymbol{q}_2 + \boldsymbol{q}_3 - \boldsymbol{q}) G_3 (\boldsymbol{q}_1, \boldsymbol{q}_2, \boldsymbol{q}_3) \delta^{(1)}_{\boldsymbol{q}_1} \delta^{(1)}_{\boldsymbol{q}_2} \delta^{(1)}_{\boldsymbol{q}_3} \\
&= \delta^{(3)}_{\boldsymbol{q}} - \frac{2}{9} \int_{q_1} \int_{q_2} \int_{q_3} (2\pi)^3 \delta^D (\boldsymbol{q}_1 + \boldsymbol{q}_2 + \boldsymbol{q}_3 - \boldsymbol{q}) \n 
&\qquad\qquad \left[ \alpha(\boldsymbol{q}_1, \boldsymbol{q}_2 + \boldsymbol{q}_3) F_2(\boldsymbol{q}_2, \boldsymbol{q}_3) - \beta(\boldsymbol{q}_1, \boldsymbol{q}_2 + \boldsymbol{q}_3) G_2(\boldsymbol{q}_2, \boldsymbol{q}_3) \right. \nonumber \\
&\qquad\qquad
\left. + \alpha(\boldsymbol{q}_1 + \boldsymbol{q}_2, \boldsymbol{q}_3) G_2(\boldsymbol{q}_1, \boldsymbol{q}_2) - \beta(\boldsymbol{q}_1 + \boldsymbol{q}_2, \boldsymbol{q}_3) G_2(\boldsymbol{q}_1, \boldsymbol{q}_2) \right] \delta^{(1)}_{\boldsymbol{q}_1} \delta^{(1)}_{\boldsymbol{q}_2} \delta^{(1)}_{\boldsymbol{q}_3}.
\end{align}
The $\alpha(\boldsymbol{q}_1, \boldsymbol{q}_2)$ and $\beta(\boldsymbol{q}_1, \boldsymbol{q}_2)$ kernels are given by
\begin{equation}
    \alpha(\boldsymbol{k}_1, \boldsymbol{k}_2) = \frac{\boldsymbol{k}_1 \cdot (\boldsymbol{k}_1 + \boldsymbol{k}_2)}{k_1^2} , \qquad \beta(\boldsymbol{k}_1, \boldsymbol{k}_2) = \frac{(\boldsymbol{k}_1 + \boldsymbol{k}_2)^2 \boldsymbol{k}_1 \cdot \boldsymbol{k}_2}{2 k_1^2 k_2^2} .
\end{equation}
Introducing the combinations $\boldsymbol{m} = \boldsymbol{q}_2 + \boldsymbol{q}_3$ and $\boldsymbol{n} = \boldsymbol{q}_1 + \boldsymbol{q}_2$, we find
{\small
\begin{align}
\theta^{(3)}_{\boldsymbol{q}} &= \delta^{(3)}_{\boldsymbol{q}} - \frac{2}{9} \int_{m} \int_{q_1} (2\pi)^3 \delta^D (\boldsymbol{q}_1 + \boldsymbol{m} - \boldsymbol{q}) \alpha(\boldsymbol{q}_1, \boldsymbol{m}) \int_{q_2} \int_{q_3} (2\pi)^3 \delta^D (\boldsymbol{q}_2 + \boldsymbol{q}_3 - \boldsymbol{m}) F_2(\boldsymbol{q}_2, \boldsymbol{q}_3) \delta^{(1)}_{\boldsymbol{q}_1} \delta^{(1)}_{\boldsymbol{q}_2} \delta^{(1)}_{\boldsymbol{q}_3} \nonumber \\
&\qquad\quad
+ \frac{2}{9} \int_{m} \int_{q_1} (2\pi)^3 \delta^D (\boldsymbol{q}_1 + \boldsymbol{m} - \boldsymbol{q}) \beta(\boldsymbol{q}_1, \boldsymbol{m}) \int_{q_2} \int_{q_3} (2\pi)^3 \delta^D (\boldsymbol{q}_2 + \boldsymbol{q}_3 - \boldsymbol{m}) G_2(\boldsymbol{q}_2, \boldsymbol{q}_3) \delta^{(1)}_{\boldsymbol{q}_1} \delta^{(1)}_{\boldsymbol{q}_2} \delta^{(1)}_{\boldsymbol{q}_3} \nonumber \\
&\qquad\quad
- \frac{2}{9} \int_{n} \int_{q_3} (2\pi)^3 \delta^D (\boldsymbol{n} + \boldsymbol{q}_3 - \boldsymbol{q}) \alpha(\boldsymbol{n}, \boldsymbol{q}_3) \int_{q_1} \int_{q_2} (2\pi)^3 \delta^D (\boldsymbol{q}_1 + \boldsymbol{q}_2 - \boldsymbol{n}) G_2(\boldsymbol{q}_1, \boldsymbol{q}_2) \delta^{(1)}_{\boldsymbol{q}_1} \delta^{(1)}_{\boldsymbol{q}_2} \delta^{(1)}_{\boldsymbol{q}_3} \nonumber \\
&\qquad\quad
+ \frac{2}{9} \int_{n} \int_{q_3} (2\pi)^3 \delta^D (\boldsymbol{n} + \boldsymbol{q}_3 - \boldsymbol{q}) \beta(\boldsymbol{n}, \boldsymbol{q}_3) \int_{q_1} \int_{q_2} (2\pi)^3 \delta^D (\boldsymbol{q}_1 + \boldsymbol{q}_2 - \boldsymbol{n}) G_2(\boldsymbol{q}_1, \boldsymbol{q}_2) \delta^{(1)}_{\boldsymbol{q}_1} \delta^{(1)}_{\boldsymbol{q}_2} \delta^{(1)}_{\boldsymbol{q}_3} \nonumber \\
&= \delta^{(3)}_{\boldsymbol{q}} - \frac{2}{9} \int_{m} \int_{q_1} (2\pi)^3 \delta^D (\boldsymbol{q}_1 + \boldsymbol{m} - \boldsymbol{q}) \alpha(\boldsymbol{q}_1, \boldsymbol{m})  \delta^{(1)}_{\boldsymbol{q}_1} \delta^{(2)}_{\boldsymbol{m}} 
+ \frac{2}{9} \int_{m} \int_{q_1} (2\pi)^3 \delta^D (\boldsymbol{q}_1 + \boldsymbol{m} - \boldsymbol{q}) \beta(\boldsymbol{q}_1, \boldsymbol{m}) \delta^{(1)}_{\boldsymbol{q}_1} \theta^{(2)}_{\boldsymbol{m}} \nonumber \\
&\qquad\quad
- \frac{2}{9} \int_{n} \int_{q_3} (2\pi)^3 \delta^D (\boldsymbol{n} + \boldsymbol{q}_3 - \boldsymbol{q}) \alpha(\boldsymbol{n}, \boldsymbol{q}_3) \theta^{(2)}_{\boldsymbol{n}} \delta^{(1)}_{\boldsymbol{q}_3} 
+ \frac{2}{9} \int_{n} \int_{q_3} (2\pi)^3 \delta^D (\boldsymbol{n} + \boldsymbol{q}_3 - \boldsymbol{q}) \beta(\boldsymbol{n}, \boldsymbol{q}_3) \theta^{(2)}_{\boldsymbol{n}} \delta^{(1)}_{\boldsymbol{q}_3}.
\end{align}
}
Upon relabeling the wavenumber that is being integrated over, this expression becomes:
{\small
\begin{align}
\theta^{(3)}_{\boldsymbol{q}} &= \delta^{(3)}_{\boldsymbol{q}} - \frac{2}{9} \int_{q_1} \int_{q_2} (2\pi)^3 \delta^D (\boldsymbol{q}_1 + \boldsymbol{q}_2 - \boldsymbol{q}) \left\{ \alpha(\boldsymbol{q}_1, \boldsymbol{q}_2) \delta^{(1)}_{\boldsymbol{q}_1} \delta^{(2)}_{\boldsymbol{q}_2} + \left[ \alpha(\boldsymbol{q}_2, \boldsymbol{q}_1) - \beta(\boldsymbol{q}_1, \boldsymbol{q}_2) - \beta(\boldsymbol{q}_2, \boldsymbol{q}_1) \right] \delta^{(1)}_{\boldsymbol{q}_1} \theta^{(2)}_{\boldsymbol{q}_2} \right\} \nonumber \\
&= \delta^{(3)}_{\boldsymbol{q}} - \frac{2}{9} \int_{q_1} \int_{q_2} (2\pi)^3 \delta^D (\boldsymbol{q}_1 + \boldsymbol{q}_2 - \boldsymbol{q}) \left[ \alpha(\boldsymbol{q}_1, \boldsymbol{q}_2) - \alpha(\boldsymbol{q}_2, \boldsymbol{q}_1) + \beta(\boldsymbol{q}_1, \boldsymbol{q}_2) + \beta(\boldsymbol{q}_2, \boldsymbol{q}_1) \right] \delta^{(1)}_{\boldsymbol{q}_1} \delta^{(2)}_{\boldsymbol{q}_2} \nonumber \\
&\qquad\quad - \frac{2}{9} \int_{q_1} \int_{q_2} (2\pi)^3 \delta^D (\boldsymbol{q}_1 + \boldsymbol{q}_2 - \boldsymbol{q}) \left[ \alpha(\boldsymbol{q}_2, \boldsymbol{q}_1) - \beta(\boldsymbol{q}_1, \boldsymbol{q}_2) - \beta(\boldsymbol{q}_2, \boldsymbol{q}_1) \right] \left( \theta^{(1)}_{\boldsymbol{q}_1} \delta^{(2)}_{\boldsymbol{q}_2} + \delta^{(1)}_{\boldsymbol{q}_1} \theta^{(2)}_{\boldsymbol{q}_2} \right). \nonumber \\
\end{align}
}
Since the choice of momentum labeling was arbitrary, the expressions should be symmetrized over $q_1$ and $q_2$.
\begin{align}
\theta^{(3)}_{\boldsymbol{q}} &= \delta^{(3)}_{\boldsymbol{q}} - \frac{2}{9} \int_{q_1} \int_{q_2} (2\pi)^3 \delta^D (\boldsymbol{q}_1 + \boldsymbol{q}_2 - \boldsymbol{q}) \beta(\boldsymbol{q}_1, \boldsymbol{q}_2) \times 2 \delta^{(1)}_{\boldsymbol{q}_1} \delta^{(2)}_{\boldsymbol{q}_2} \nonumber \\
&\qquad\quad - \frac{2}{9} \int_{q_1} \int_{q_2} (2\pi)^3 \delta^D (\boldsymbol{q}_1 + \boldsymbol{q}_2 - \boldsymbol{q}) \left[ \frac{\alpha(\boldsymbol{q}_1, \boldsymbol{q}_2) + \alpha(\boldsymbol{q}_2, \boldsymbol{q}_1)}{2} - 2 \beta(\boldsymbol{q}_1, \boldsymbol{q}_2) \right] \left( \theta^{(1)}_{\boldsymbol{q}_1} \delta^{(2)}_{\boldsymbol{q}_2} + \delta^{(1)}_{\boldsymbol{q}_1} \theta^{(2)}_{\boldsymbol{q}_2} \right) \nonumber \\
&= \delta^{(3)}_{\boldsymbol{q}} - \frac{2}{9} \int_{q_1} \int_{q_2} (2\pi)^3 \delta^D (\boldsymbol{q}_1 + \boldsymbol{q}_2 - \boldsymbol{q}) \beta(\boldsymbol{q}_1, \boldsymbol{q}_2) \times 2 \delta^{(1)}_{\boldsymbol{q}_1} \delta^{(2)}_{\boldsymbol{q}_2} \nonumber \\
&\qquad\quad + \frac{2}{9} \int_{q_1} \int_{q_2} (2\pi)^3 \delta^D (\boldsymbol{q}_1 + \boldsymbol{q}_2 - \boldsymbol{q}) \left[ \frac{(\boldsymbol{q}_1 \cdot \boldsymbol{q}_2)^2}{q_1^2 q_2^2} - 1 + \beta(\boldsymbol{q}_1, \boldsymbol{q}_2) \right] \left( \theta^{(1)}_{\boldsymbol{q}_1} \delta^{(2)}_{\boldsymbol{q}_2} + \delta^{(1)}_{\boldsymbol{q}_1} \theta^{(2)}_{\boldsymbol{q}_2} \right) \nonumber \\
&= \delta^{(3)}_{\boldsymbol{q}} + \frac{2}{9} \left(\mathcal{G}_{2,v}^{(3)}\right)_{\boldsymbol{q}} - \frac{2}{9} \int_{q_1} \int_{q_2} (2\pi)^3 \delta^D (\boldsymbol{q}_1 + \boldsymbol{q}_2 - \boldsymbol{q}) \beta(\boldsymbol{q}_1, \boldsymbol{q}_2) \times 2 \delta^{(1)}_{\boldsymbol{q}_1} \delta^{(2)}_{\boldsymbol{q}_2} \nonumber \\
&\qquad\qquad\qquad + \frac{2}{9} \int_{q_1} \int_{q_2} (2\pi)^3 \delta^D (\boldsymbol{q}_1 + \boldsymbol{q}_2 - \boldsymbol{q}) \beta(\boldsymbol{q}_1, \boldsymbol{q}_2) \left( \theta^{(1)}_{\boldsymbol{q}_1} \delta^{(2)}_{\boldsymbol{q}_2} + \delta^{(1)}_{\boldsymbol{q}_1} \theta^{(2)}_{\boldsymbol{q}_2} \right) \nonumber \\
\end{align}
As a reminder, we denote $\mathcal{G}_{2,v} = \nabla_i \nabla_j \phi \nabla^i \nabla^j \phi_v - \nabla^2 \phi \nabla^2 \phi_v$, where $\phi_v = \theta / \nabla^2$ is the velocity potential.
Then in configuration space, we can write this as
\begin{align}
    \theta_3 &= \delta_3 + \frac{2}{9} \left[\mathcal{G}_{2,v} + \frac{\nabla^2}{2} \left( \nabla_i \phi \nabla^i \phi_v - \nabla_i \phi \nabla^i \phi \right) \right]_3 \nonumber \\
    &= \delta_3 + \frac{2}{9} \left[\mathcal{G}_{2,v} + \frac{1}{7} \nabla^2 \left( \nabla_i \phi \frac{\nabla^i}{\nabla^2} \mathcal{G}_2 \right) \right]_3
\end{align}
The subscript 3 at the end of the brackets indicates that we are only keeping terms up to third order in $\delta_1$.

\section{Noisiness of the $\delta^2$ operator}
\label{sec:d2_or_d3}

\begin{figure}
	\centering
	\includegraphics{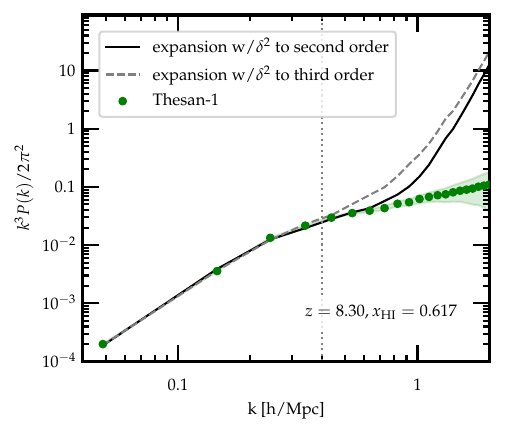}
	\caption{The 21\,cm power spectrum at $z=8.30$ ($x_\mathrm{HI} = 0.617$).
	Green dots indicate the binned power spectrum from the \thesan-1 simulation; the shaded regions indicate the shot noise error.
	The grey dashed line is the theory expansion fit to the simulations using the third-order expression for $\delta^2$, while the black dashed line shows the best fit from the effective field theory, using the approximation $\delta^2 = (\delta^{(1)})^2$.
	The vertical dotted line shows $k_\mathrm{NL}$, the maximum wavenumber that we fit up to.
	Due to the noisiness of the $\delta^2$ operator, we find that the lower-order approximation provides a slightly better fit to the simulation than the full expression, and has greater predictive power for small, mildly nonlinear scales.
	}
	\label{fig:power_spectrum_d2}
\end{figure}

When fitting our theoretical expansion for the 21\,cm field to simulations, we find that using the second-order approximation for the term $\delta^2 = (\delta^{(1)})^2$ leads to a better fit at most redshifts compared to using the full third-order expression.
An example of this is shown in Figure~\ref{fig:power_spectrum_d2}.
The grey dashed line shows the best fit using the third-order expression; the black line uses the second-order expression; the vertical dotted line shows $k_\mathrm{NL}$, the maximum wavenumber that we fit up to.
Not only is the line with the second-order approximation a better fit, it has better predictive power at wavenumbers above those that we fit to.

The degradation of the fit as we go to higher order can be attributed to the fact that $\delta^2$ captures the effects of nonlinear bias and therefore has a large shot noise contribution; including higher order terms then makes this expansion a noisier template when fitting to the simulation field~\cite{McQuinn:2018zwa}.
Hence, the main results of this study use the second-order approximation.

\section{Best fit at the field level}
\label{sec:fit_grid}

\begin{figure*}
	\includegraphics{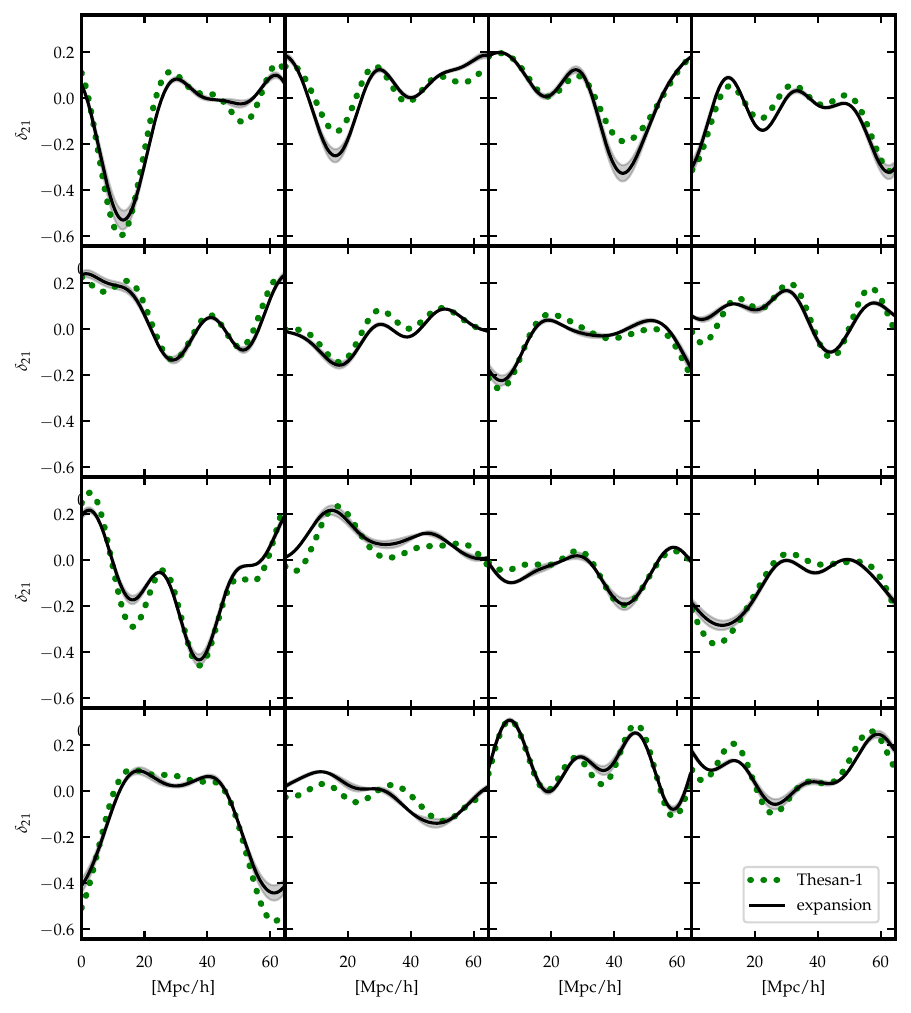}
	\caption{Examples of fluctuations in the redshift space 21\,cm differential brightness temperature along evenly spaced lines through the simulation volume at $z = 8.30$, $x_\mathrm{HI} = 0.617$, smoothed over $k_\mathrm{NL} = 0.4$ h/Mpc. 
	The green dots show the signal from the \thesan-1 simulation and the thick black line in the first panel is the best fit theory expansion.
	The filled contours show the 68\% confidence intervals on the best fit curves.
	}
	\label{fig:1d-slice-grid}
\end{figure*}

Figure~\ref{fig:1d-slice-grid} shows examples of fluctuations in the redshift space 21\,cm differential brightness temperature along several different lines through the simulation volume at $z = 8.30$, $x_\mathrm{HI} = 0.617$, smoothed over $k_\mathrm{NL} = 0.4$ h/Mpc. 
The lines are chosen to be evenly spaced along the $x$ and $y$ coordinates of the simulation volume.

\chapter{Useful Estimates}

In this appendix, I will sketch out some estimates and back-of-the-envelope calculations that have been useful throughout my Ph.D for studying exotic energy injection.

\section{Shape of cosmological constraints on energy injection}
\label{app:chi-ee-shape}

Cosmological constraints on dark matter energy injection as a function of the dark matter mass typically show a few different broad features.
Their shapes can be understood by knowing the dominant cooling processes in each mass range.

As a concrete example, I will look at CMB anisotropy constraints on the decay channel $\chi \rightarrow e^+ e^-$ (see e.g. Figs.~\ref{fig:constraints} and \ref{fig:scans_z20}).
In order for dark matter to be able to decay to these pairs, $m_\chi > 2 m_e$, hence the threshold is about 1 MeV.
Right at this threshold, all of the dark matter mass-energy is converted into electron and positron mass-energy, and since the $e^+ e^-$ pair have no kinetic energy to deposit, they quickly annihilate into photons.
Hence, lifetime constraint at $m_\chi = 2 m_e$ should be equal to the lifetime constraint for $\chi \rightarrow \gamma \gamma$ at the same mass.

Just above this threshold, the constraint will rise slightly as there is now some kinetic energy available to deposit into heating.
However, inverse Compton scattering quickly becomes the dominant cooling process for the electrons and positrons, hence they transfer their energy to photons.
An estimate for the amount of energy transferred to the photon is given by $\epsilon' = \gamma_e^2 \epsilon$, where $\epsilon$ is the initial photon energy, $\gamma_e$ is the electron Lorentz boost factor, and $\epsilon'$ is the photon energy after scattering.

For small masses, the energy of the photons will be too low for other cooling processes to be significant, so energy will be carried away in the form of low-energy photons, which does not show up in the CMB anisotropies, hence the constraint will weaken somewhat.
As we increase the dark matter mass, these photons become energetic enough that they are able to efficiently ionize hydrogen and thus deposit their energy, so the constraint will again become stronger.
We expect this turnover to occur when $\epsilon' = 13.6$ eV.
If the initial photon is a CMB photon at redshift $1+z=1100$, then $\epsilon \approx 0.22$ eV, which means that $\gamma_e \approx 8$ and $m_\chi = 2 \gamma_e m_e = 8$ MeV, which is not too far off.
For the Lyman-$\alpha$ forest constraints described in Sec.~\ref{sec:Lya}, the measurements of the IGM temperature come from $1+z \approx 6$.
In this case, the CMB photons have an energy of about $\epsilon \approx 0.0012$ eV, so the dark matter mass at which the constraints begin to strengthen again is about $m_\chi = 100$ MeV.

At even higher masses, the photons are too energetic for photoionization to be efficient and the constraints weaken from there.
If an observational probe is sensitive to later redshifts, when the universe is less dense and more transparent, then the rate of energy deposition falls off more quickly (see Fig.~17 in Ref.~\cite{DarkHistory}). 
Therefore, the constraints also weaken more quickly for late-redshift probes such as Lyman-$\alpha$ and early star formation, compared to constraints set using the CMB.

\section{CMB constraints on the lifetime of dark matter}
\label{app:lifetime_est}

Now that we know how the shape of these `calorimetric' constraints arises, we would like to estimate their normalization.
We start by noting that there is about fives times as much energy in density in dark matter as there is in baryons, $\rho_\chi = 5 \rho_b$.
If we also make the simplifying approximation that all baryons are in the form of hydrogen (which is good up to some factor of order unity), that means that there is 5 GeV of energy in dark matter per baryon in our universe.
\begin{equation}
	\frac{\rho_\chi}{n_b} = \frac{5 \rho_b}{\rho_b / m_\mathrm{H}} = 5 m_\mathrm{H} \approx 5 \,\mathrm{GeV} 
\end{equation}

The energy required to ionize a hydrogen atom is merely 13.6 eV.
That means if dark matter transfers just $\frac{13.6}{5 \times 10^9} \approx 3 \times 10^{-9}$ of its energy to baryons, this would be enough to \textit{ionize the entire universe}.
That is an effect that we would definitely be able to observe.

Hence, if dark matter can only release this much energy over the age of the universe, then for decaying dark matter, this means its lifetime is constrained to be greater than
\begin{align}
	\tau &\gtrsim 3 \times 10^{8} \times \mbox{lifetime of the universe} \n
	&\approx 3 \times 10^{8} \times (13.8 \times 10^9 \,\mathrm{yrs}) \times (\pi \times 10^7 \,\mathrm{s} / \mathrm{yr}) \n
	&\approx 10^{26} \,\mathrm{s} .
\end{align}

There can be some additional orders of magnitude to consider here based on how much of the injected energy ends up in ionization and how precisely an experiment like \textit{Planck} can measure the ionization history, but these factors mostly cancel out so I do not consider them.
One can also modify this estimate by assuming only a fraction of the dark matter can decay.

\section{Model building long-lived dark matter}
\label{app:model-building}

The estimate above gives a constraint on the lifetime for decaying dark matter.
One might then ask what kinds of models can accommodate such an incredibly long lifetime.

If the dark matter decay process is sourced by a dimension-$N$ operator, then the interaction strength goes as $\Lambda^{4-N}$, where $\Lambda$ is the scale at which new physics appears. 
For example, the grand unification energy is predicted to be at $\Lambda_\mathrm{GUT} \sim 10^{16}$ GeV and the Planck scale is approximately $\Lambda_\mathrm{Pl} \sim 10^{19}$ GeV.
Assuming that the dark matter mass is only other relevant energy scale, then the decay lifetime goes as
\begin{equation}
	\tau \sim
	\left|
	\begin{tikzpicture}[baseline=(current bounding box.center)]
	\begin{feynman}
	\vertex (a) {\(\chi\)};
	\vertex [right=1.4cm of a] (b);
	\vertex [above right=1cm and 1cm of b] (f1) {\(\psi_1\)};
	\vertex [above right=.3cm and 1cm of b] (f2) {\(\psi_2\)};
	\vertex [below right=1cm and 1cm of b] (f3) {\(\psi_j\)};
	\diagram* {
		(a) -- (b),
		(b) -- (f1),
		(b) -- (f2),
		(b) -- (f3),
	};
	\node at ($(f2)!0.5!(f3)$) {\(\vdots\)};
	\end{feynman}
	\end{tikzpicture}
	\right|^{-2}
	\sim \frac{1}{m_\chi} \left( \frac{\Lambda}{m_\chi}  \right)^{2 (N-4)} .
\end{equation}
The field content of $\psi_1$, $\psi_2$, ..., $\psi_j$ is arbitrary so long as the operator has mass dimension of $N$.

Two more useful numbers to keep in mind are $\hslash = 6.58 \times 10^{-16} \,\mbox{eV s} \approx 10^{-15} \,\mbox{eV s}$ and the lower bound on the dark matter mass from requiring its de Broglie wavelength to be smaller than a dwarf galaxy, $m_\chi \geq 10^{-22}$ eV~\cite{Hui:2021tkt}.

The lowest order operators that give rise to decay are dim-3 operators (for example, a 3-scalar interaction).
In this case, $\tau \sim m_\chi / \Lambda^2$, or
\begin{equation}
	m_\chi \sim \tau \Lambda^2 \sim 10^{91} \,\mathrm{eV} \left( \frac{\tau}{10^{26} \,\mathrm{s}} \right) \left( \frac{\Lambda}{10^{16} \,\mathrm{eV}} \right)^2 .
\end{equation}
$10^{91}$ eV is equivalent to $10^{55}$ kg or $10^{25} \,M_\odot$, which is more than the \textit{mass of the observable universe}.
Hence, from a dim-3 operator, there is no way to have the dark matter be a fundamental particle and be long-lived.

For dim-4 operators, $\tau \sim m_\chi^{-1}$.
This means that
\begin{equation}
	\tau < \frac{\hslash}{10^{-22} \,\mathrm{eV}} \sim 10^7 \,\mathrm{s} \sim 1 \,\mathrm{year}.
\end{equation}
This is considerably shorter than the age of the universe, so the decaying particle would not be cosmologically stable and is therefore unsuitable as a dark matter candidate.

For dim-5 operators, $\tau \sim \Lambda^2 / m_\chi^3$.
Rewriting this as an expression for the dark matter mass, we find
\begin{equation}
	m_\chi \sim \left( \frac{\Lambda^2}{\tau} \right)^{1/3} \sim 10^5 \,\mathrm{eV} \, \left( \frac{10^{26} \mathrm{s}}{\tau} \right)^{1/3} \left( \frac{\Lambda}{10^{19} \,\mathrm{GeV}} \right)^{2/3} .
\end{equation}
Hence, dim-5 operators are a viable decay channel if the dark matter has a mass of less than 100 keV, which is lighter than classical WIMP dark matter but otherwise allowed.

The last example we will consider is dim-6 and is also discussed in e.g. Ref.~\cite{Arvanitaki:2009yb}.
In this case,
\begin{equation}
	\tau \sim \frac{\Lambda^4}{m_\chi^5} \sim 10^{25} \,\mathrm{s} \left( \frac{1 \,\mathrm{TeV}}{m_\chi} \right)^5 \left( \frac{\Lambda}{10^{16} \,\mathrm{GeV}} \right)^4 .
\end{equation}
Hence, at this order, WIMP dark matter at the GUT scale gives a lifetime near existing constraints.

Examples of dark matter models with decay channels from dim-5 and dim-6 operators are given in Refs.~\cite{Cirelli:2005uq} and \cite{Arvanitaki:2009yb}.

\clearpage
\newpage

\begin{singlespace}
{\small \bibliography{main}}
\bibliographystyle{apsrev4-1+title}
\end{singlespace}

\end{document}